\documentclass[11pt,twoside,openany]{book}
\usepackage[a4paper,width=150mm,top=25mm,bottom=25mm,bindingoffset=5mm]{geometry}
\usepackage[utf8]{inputenc}
\usepackage[english]{babel}
\usepackage{lipsum}
\usepackage{amsmath, amsfonts, amssymb, amsthm}
\usepackage{mathrsfs}
\usepackage{tensor}
\usepackage{graphicx}
\graphicspath{{Figures/}}
\usepackage[labelsep=period,font={small}]{caption}
\usepackage{subfigure}
\usepackage{datetime}
\usepackage{hyperref}
\usepackage{enumerate}
\usepackage{fancyhdr}
\usepackage[page,toc]{appendix}
\usepackage{cite}
\usepackage{xcolor}

\hypersetup{colorlinks, linkcolor=blue, citecolor=blue, urlcolor=blue, linktoc=page}

\newdateformat{monthyeardate}{\monthname[\THEMONTH] \THEYEAR}

\let\origdoublepage\cleardoublepage
\newcommand{\clearemptydoublepage}{\clearpage{\pagestyle{empty}\origdoublepage}}

\newcommand{\ed}{\mathrm{d}}
\newcommand{\eD}{\mathrm{D}}
\newcommand{\bnab}{\boldsymbol{\nabla}}
\newcommand{\on}[1]{\operatorname{#1}}
\newcommand{\ind}{\indices}
\newcommand{\ol}{\overline}
\newcommand{\wt}{\widetilde}
\newcommand{\wh}{\widehat}
\newcommand{\vp}{\vphantom}
\newcommand{\mqolegup}[2]{\widetilde{\lambda}_{#1}\tensor{\vphantom{\lambda}}{^{#2}}}
\newcommand{\mqolegdn}[2]{\widetilde{\lambda}_{#1}\tensor{\vphantom{\lambda}}{_{#2}}}
\newcommand{\mlegup}[2]{\lambda_{#1}\tensor{\vphantom{\lambda}}{^{#2}}}

\newcommand{\nps}[3]{{#1}\ind{^{a}} {#2}\ind{_{a ; b}} {#3}\ind{^{b}}}

\DeclareMathAlphabet{\mathsfit}{T1}{\sfdefault}{\mddefault}{\sldefault}

\usepackage{array}
\newcolumntype{C}[1]{>{\centering\arraybackslash$}p{#1}<{$}}

\newcommand{\bmyi}{\begin{itemize} \color{red} \sf \item[] My to do list:}
\newcommand{\emyi}{\end{itemize}}

\begin{document}


\pagestyle{empty}
\pagenumbering{roman}

\begin{titlepage}

\topskip0pt
\vspace*{\fill}

\begin{center}
\Huge\textbf{Strong-field gravitational lensing by black holes}\par

\vspace{1cm}

{\LARGE Jake O.~Shipley}

\vspace{1cm}

\begin{figure}[h]
\centering
\includegraphics[height=4cm]{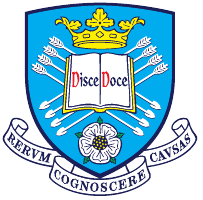}
\end{figure}

\vspace{1cm}

\Large{
A thesis submitted for the degree of

\emph{Doctor of Philosophy}}\par

\vspace{1cm}

\Large{
School of Mathematics and Statistics

University of Sheffield}\par

\vspace{1cm}

\Large{\monthyeardate\today}

\vspace*{\fill}

\end{center}

\end{titlepage}

\clearemptydoublepage
\pagestyle{plain}



\chapter*{Summary} \label{chap:summary}

In this thesis we study aspects of strong-field gravitational lensing by black holes in general relativity, with a particular focus on the role of integrability and chaos in geodesic motion.

We begin with a review of theoretical aspects of Einstein's theory of general relativity, as well as important concepts and techniques from the fields of non-linear dynamics and chaos theory. Next, we review the topic of gravitational lensing. The history of gravitational lensing and the perturbative lensing formalism are discussed. We then present an overview of gravitational lensing from a spacetime perspective, including the leading-order geometric optics approximation for high-frequency electromagnetic waves on curved spacetime; the role of (unstable) photon orbits; and black hole shadows.

We investigate binary black hole shadows using the Majumdar--Papapetrou static binary black hole (or di-hole) solution. We demonstrate that the propagation of null geodesics on this spacetime background is a natural example of chaotic scattering. The role of unstable photon orbits is discussed. We develop a symbolic dynamics to describe null geodesics and to understand the structure of one-dimensional binary black hole shadows. We demonstrate that, in situations where chaotic scattering is permitted, the shadows exhibit a self-similar fractal structure akin to the Cantor set. A gallery of two-dimensional binary black hole shadows, realised using backwards ray-tracing, is presented and analysed in detail.

Next, we use techniques from the field of non-linear dynamics to quantify fractal structures in the shadows of binary black holes, using the static Majumdar--Papapetrou di-hole as a toy model. Our perspective is that binary black hole shadows can be viewed as the exit basins of an open Hamiltonian system with three escapes. We compute the uncertainty exponent -- a quantity related to the fractal dimension -- for one-dimensional binary black hole shadows. Using a recently developed numerical algorithm, called the merging method, we demonstrate that parts of the Majumdar--Papapetrou di-hole shadow may possess the Wada property: any point on the boundary of one basin is on the boundary of at least two additional basins.

We study the existence, stability and phenomenology of circular photon orbits in stationary axisymmetric four-dimensional spacetimes in general relativity. We use a Hamiltonian formalism to describe null geodesics of the Weyl--Lewis--Papapetrou geometry. Using the Einstein--Maxwell equations, we demonstrate that generic stable photon orbits are forbidden in pure vacuum, but may arise in electrovacuum. As a case study, we consider stable photon orbits around the Reissner--Nordstr\"{o}m family of static di-holes. We examine the onset of chaos in the motion of bounded null geodesics using Poincar\'{e} sections.

In the final chapter, we apply a higher-order geometric optics formalism to describe the propagation of electromagnetic waves on Kerr spacetime. Our principal motivation is to calculate the sub-dominant correction to the electromagnetic stress--energy tensor. We use a complex self-dual bivector, built from the closed conformal Killing--Yano tensor and its Hodge dual, to construct a complex null tetrad which is parallel-propagated along null geodesics. We introduce a system of transport equations to calculate certain Newman--Penrose and higher-order geometric optics quantities. We derive generalised power series solutions to these transport equations through sub-leading order in the neighbourhood of caustic points. Finally, we introduce a practical method which may be used to evolve transport equations for divergent quantities through caustic points. 
\clearemptydoublepage

\chapter*{Acknowledgements} \label{chap:acknowledgements}

I would like express my sincere gratitude to Dr.~Sam Dolan for his support, guidance and encouragement throughout the last four years. I could not have wished for a more enthusiastic, knowledgeable and patient supervisor; it has been a pleasure and a privilege to learn from him and work with him. I would especially like to thank Sam for going above and beyond the role of supervisor. No matter how busy he is, his door is always open; he is a constant source of advice and inspiration; and he is always the first to buy a round at the pub! I am extremely fortunate to be able to call Sam my friend, as well as my supervisor.

Chapter \ref{chap:fractal_structures} of this thesis is the result of a thoroughly enjoyable collaboration with Professor Miguel Sanju\'{a}n and Dr.~\'{A}lvar Daza. I thank them both for their warm welcome and generous hospitality during my visit to Madrid in December 2016, and for teaching me with such enthusiasm about non-linear dynamics and chaos theory.

I acknowledge financial support from the University of Sheffield's Harry Worthington Scholarship. This funding has allowed me to pursue my research interests for four years, and it has provided me with the opportunity to visit a host of amazing places -- including Paris, Aveiro, Dublin, Madrid and Rome -- all in the name of research.

It has been a pleasure to be part of the Cosmology, Relativity and Gravitation research group for the last four years. I would like to acknowledge the kindness and friendship of all members past and present. Special thanks go to Carl Kent, Jack Morrice, David Dempsey, Tom Morley, Tom Stratton, Jake Percival, Vis Balakumar, Elisa Maggio, Mohamed Ould El Hadj, Carsten van de Bruck and Elizabeth Winstanley.
%

Lastly, I would like to thank my parents, sisters and Hannah for their encouragement, support and unconditional love throughout the last four years and beyond. It would take an acknowledgements section the size of this thesis to even begin to thank you all properly. 
\clearemptydoublepage


\setcounter{tocdepth}{1}
\tableofcontents

\clearemptydoublepage
\pagenumbering{arabic}
\pagestyle{fancy}



\chapter{Introduction}
\label{chap:introduction}

On 10 April 2019, the Event Horizon Telescope (EHT) collaboration reported the first image of a black hole \cite{EHTC2019a, EHTC2019b, EHTC2019c, EHTC2019d, EHTC2019e, EHTC2019f}. Using very-long-baseline interferometry (VLBI), the EHT -- a global network of telescopes observing at millimetre wavelengths -- captured exquisite-resolution images of radio emission from the supermassive black hole candidate M87$^{\ast}$, which is believed to lie at the heart of the giant elliptical galaxy Messier 87 (M87). A key feature of the EHT's images is a bright asymmetric ring which surrounds a dark central region -- the \emph{black hole shadow}. (See Figure 3 of \cite{EHTC2019a}, for example.) Overall, the observed images are consistent with expectations for a rotating Kerr black hole, as predicted by Einstein's general theory of relativity. This ground-breaking observation confirms the existence of black holes -- a key prediction of general relativity -- and provides a new way to test Einstein's theory in its most extreme limit.

The observation of M87$^{\ast}$ is in fact a detection of a \emph{gravitational lensing} effect, i.e., the deflection of light by gravity \cite{Perlick2004}. According to general relativity, black holes (and all other massive bodies) generate spacetime curvature, which leads to the deviation in the paths of photons as they trace out null geodesics on curved spacetime. The outline of the black hole shadow observed by the EHT is associated with the black hole's unstable \emph{light-ring}, where spacetime curvature is so strong that light is able to orbit the black hole. Beyond the light-ring, radially infalling photons are doomed to plunge into the black hole, crossing its \emph{event horizon} -- a one-way causal boundary in spacetime beyond which nothing can escape to infinity.

The shadow observed by the Event Horizon Telescope encodes important information about the black hole and the spacetime geometry close to the event horizon. Comparing the observed shadow of M87$^{\ast}$ with a library of ray-traced general-relativistic magnetohydrodynamic simulations, the EHT collaboration have been able to infer a mass of $M = (6.5 \pm 0.7) \times 10^{9} M_{\odot}$ \cite{EHTC2019a, EHTC2019f}. Moreover, the observations are consistent with the black hole's spin axis being oriented at $17$ degrees from the line of sight, with the black hole rotating in the clockwise direction (i.e., the spin axis points away from us) \cite{EHTC2019a}. It is hoped that future high-resolution VLBI imaging of M87$^{\ast}$ and other supermassive black hole candidates will allow scientists to test the spin and inclination of the black holes \cite{PsaltisNarayanFishEtAl2014}, and to continue to probe general relativity in the strong-field regime \cite{JohannsenPsaltis2010, BroderickJohannsenLoebEtAl2014, Johannsen2016, RicarteDexter2014, JohannsenBroderickPlewaEtAl2016}.

In 2015, the Laser Interferometer Gravitational-Wave Observatory (LIGO) collaboration detected a gravitational-wave signal emitted by a pair of merging stellar-mass black holes \cite{Abbottothers2016a}. The characteristic ``chirp'' profile of the gravitational-wave signal, dubbed GW150914, is consistent with the inspiral, merger and ringdown phases of waveform templates generated by numerical relativity simulations. As well as providing compelling evidence for the existence of both gravitational waves and binary black holes in nature, this detection signalled the birth of gravitational-wave astronomy as an observational science.

Two years later, in 2017, a gravitational-wave signal from a pair of coalescing neutron stars (GW170817) was observed by the LIGO--Virgo collaboration \cite{Abbottothers2017d}. This signal was accompanied, approximately $1.7$ seconds later, by a gamma-ray burst; and its optical transient was detected via a host of observations across the electromagnetic spectrum \cite{Abbottothers2017e}. This was evidence of another important prediction of Einstein's theory of general relativity: gravitational waves propagate outwards from their source at the speed of light. Moreover, GW170817 was the first gravitational-wave signal which was observed alongside an electromagnetic counterpart, marking an exciting breakthrough in the field of multi-messenger astronomy.

In total, LIGO--Virgo observed ten gravitational-wave signals during its first two observing runs, of which nine are consistent with the gravitational-wave signal generated by merging black holes, and the other with a binary neutron star merger \cite{AbbottAbbottAbbottEtAl2016, Abbottothers2018}. The LIGO--Virgo collaboration began their third observing run on 1 April 2019, and a host of candidate gravitational-wave signals have since been detected from compact binary coalescences. With the improvements made to the LIGO--Virgo detectors for the third observing run, there is an exciting possibility that astronomers may detect several binary neutron star mergers, in addition to one or more neutron star--black hole mergers \cite{LSCVDAC2018}.

The ground-breaking discoveries of gravitational waves and the observation of M87$^{\ast}$ have reignited interest in Einstein's theory of relativity and gravitational physics. With further ground- and space-based gravitational-wave missions (e.g.~KAGRA \cite{AsoMichimuraSomiyaEtAl2013}, LISA \cite{Amaro-SeoaneAudleyBabakEtAl2017} and others \cite{Hough2011}) on the horizon, and images of Sagittarius A$^{\ast}$ -- the supermassive black hole at the centre of the Milky Way -- expected soon \cite{Psaltis2018}, we are on the cusp of an exciting new era of gravitational astronomy and high-precision experimental tests of extreme gravity.

Whilst astrophysical black holes have only recently been observed directly (in both electromagnetic- and gravitational-wave channels), the theoretical properties of black holes have been studied for decades.

In 1916, just one year after the publication of Einstein's general theory of relativity, Schwarzschild \cite{Schwarzschild1916} found an exact solution to the vacuum field equations which describes the spacetime geometry of a static spherically symmetric gravitational source. In fact, Schwarzschild's solution describes a black hole, although it took a range of detailed analyses performed over a number of years to build up a more complete understanding of the solution's properties. A generalisation of the Schwarzschild solution, in which the gravitational source is endowed with an electric charge, was revealed independently by Reissner \cite{Reissner1916} and Nordstr\"{o}m \cite{Nordstroem1918} between 1916 and 1918. The Reissner--Nordstr\"{o}m solution is an exact solution to the Einstein--Maxwell field equations of general relativity and electromagnetism which exhibits the same spacetime symmetries as the uncharged Schwarzschild solution. In 1923, Birkhoff \cite{BirkhoffLanger1923} proved that the Schwarzschild spacetime is the unique spherically symmetric solution of the Einstein field equations in pure vacuum, a result now known as \emph{Birkhoff's theorem}. (In fact, Birkhoff's theorem was first published in 1921 by Jebsen \cite{Jebsen1921}.) High levels of symmetry also play an important role in geodesic motion on Schwarzschild spacetime: in spherical coordinates, which are well-adapted to the spacetime symmetries, geodesic motion is separable, and therefore integrable (in the sense of Liouville) \cite{MisnerThorneWheeler1973}.

Interest in the properties of very compact gravitational sources continued throughout the 1930s. Key contributions included Chandrasekhar's work on white dwarf stars, and the work of Landau \cite{Landau1932} and others \cite{BaadeZwicky1934, Zwicky1938, OppenheimerVolkoff1939} on neutron stars. At the end of the decade, Oppenheimer and Snyder \cite{OppenheimerSnyder1939} described the gravitational collapse of massive stars, and the subsequent formation of black holes.

Over the next few decades, the field of black hole physics went through a period of relative dormancy, until Kerr \cite{Kerr1963} discovered an exact solution to Einstein's field equations describing a rotating black hole in vacuum in 1963. This marked the beginning of the so-called golden age of black hole physics, and a range of influential perspectives on the theoretical properties of black holes followed. These included the generalisation of Kerr's solution (to include electric charge) by Newman and collaborators \cite{NewmanCouchChinnaparedEtAl1965, NewmanJanis1965}; the formulation and proof of uniqueness and ``no-hair'' theorems for Kerr black holes \cite{Israel1967, Israel1968, Robinson1975}; the proposal of the Penrose process, a mechanism which allows for energy extraction from rotating black holes \cite{Penrose2002}; an improved understanding of singularities in general relativity, and the proposal of the cosmic censorship conjecture \cite{Penrose2002, HawkingPenrose1970}; the formulation of the laws of black hole mechanics \cite{BardeenCarterHawking1973} and a development of the concept of black hole entropy \cite{Bekenstein1973}; and the prediction of Hawking radiation and black hole evaporation \cite{Hawking1974}.

%
In Boyer--Lindquist coordinates \cite{BoyerLindquist1967}, the stationarity and axial symmetry of the spacetime are manifest: the corresponding isometries are encoded in the existence of a pair of commuting Killing vectors. However, it is not immediately obvious that any further symmetries exist. In 1968, Carter \cite{Carter1968b} demonstrated that there exists a ``hidden'' symmetry of Kerr spacetime -- a fourth integral of motion which permits the separability of the geodesic equations on the spacetime background. This conserved quantity is known as the \emph{Carter constant}. Later, Floyd \cite{Floyd1973} and Penrose \cite{Penrose1973} demonstrated that the Kerr spacetime admits a Killing--Yano tensor, whose ``square'' is the Killing tensor. Subsequent investigations demonstrated that there is a deep geometrical reason behind the existence of the Carter constant: the Kerr spacetime admits a rank-two closed conformal Killing--Yano tensor (or \emph{principal tensor}), which gives rise to a family of Killing tensors. These Killing tensors may then be used to generate the full set of explicit and hidden symmetries on Kerr spacetime. For a comprehensive overview of this topic, see the review by Frolov \emph{et al.} \cite{FrolovKrtousKubiznak2017} and references therein.
%

The existence of the principal tensor and its associated ``hidden'' symmetry is implicated in a number of key results. For example, parallel-transport along geodesics is straightforward \cite{Marck1983a, Marck1983b}; gravitational Faraday rotation of the polarisation plane of light is trivial \cite{Penrose1973}; the Hamilton--Jacobi, Schr\"{o}dinger \cite{Carter1968b}, Klein--Gordon \cite{BrillChrzanowskiPereiraEtAl1972}, Dirac \cite{Unruh1973, Chandrasekhar1976}, Rarita--Schwinger \cite{Gueven1980} and Proca \cite{FrolovKrtousKubiznakEtAl2018} equations are all separable in Boyer--Lindquist coordinates; and the Maxwell equations \cite{Teukolsky1972} and those governing gravitational perturbations \cite{Teukolsky1973} are separable for the Maxwell and Weyl scalars of extreme spin weight in a certain complex null tetrad.

On Kerr spacetime, the high levels of symmetry ensure the integrability of geodesic motion, which means that trajectories are highly ``ordered'' in phase space. Moreover, the existence of the principal tensor underpins a range of important results relating to particle motion and wave propagation on Kerr spacetime (as described above). In general stationary axisymmetric spacetimes, however, such results do not hold. On one hand, this renders the study of particle motion and wave propagation a more technically demanding task; on the other, the lack of symmetry gives rise to the possibility of rich chaotic motion, which is typically associated with a range of distinct phenomena, particularly in the field of strong gravitational lensing by black holes and other (ultra-)compact objects \cite{CunhaHerdeiro2018}.

As well as marking the announcement of the first direct detection of a black hole using electromagnetic telescopes by the EHT \cite{EHTC2019a}, the year 2019 was the centenary of the \emph{first} detection of a gravitational lensing event. On 29 May 1919, two expeditions -- one to Pr\'{i}ncipe, an island of the west coast of Africa, and the other to Sobral, a city in north-east Brazil -- carried out experiments during a total solar eclipse to measure the deflection of starlight by our own Sun \cite{Eddington1924}. (For a recent review, see the article by Crispino and Kennefick \cite{CrispinoKennefick2019}.) The detection of gravitational light deflection provided experimental verification of Einstein's then novel theory of gravitation and was influential evidence for its superiority over the Newtonian theory \cite{Will1988}. Since then, the field of gravitational lensing has undergone a series of interesting developments -- both theoretically and observationally.

In the two decades that followed the 1919 detection, various gravitational lensing phenomena -- such as multiple images and Einstein rings -- were proposed by a range of authors \cite{Chwolson1924, Einstein1936}. In the 1960s, interest in the field was rekindled after the development of the quasi-Newtonian (or \emph{perturbative}) lensing formalism by Refsdal \cite{Refsdal1964a} and others \cite{Klimov1963, Liebes1964}. This was followed, in 1978, by the discovery of the multiply imaged quasar Q0957+561 \cite{WalshCarswellWeymann1979} -- the first experimental detection of a gravitational lensing effect since the observation of light deflection by the Sun almost sixty years earlier. To date, a host of gravitational lensing phenomena have been observed, including multiply imaged sources, Einstein rings, giant luminous arcs, image distortion, and galactic microlensing; for a review, see \cite{Wambsganss1998}.

In many cases of interest, gravitational lensing effects are well-described by the perturbative lensing formalism, which is based on a first-order post-Newtonian approximation to general relativity \cite{Dodelson2017}. For example, even in the formation of spectacular lensing phenomena, such as giant luminous arcs, photons are deflected by no more than a few arcseconds \cite{Wambsganss1998}. However, in many other situations (e.g.~in the strong-field regions around black holes or other compact objects), the quasi-Newtonian formalism breaks down, and a more careful treatment is required. In order to adequately describe the strong-field lensing effects associated with extreme compact objects, one must employ \emph{non-perturbative lensing} (also known as \emph{lensing from a spacetime perspective}), using the full theory of general relativity \cite{Perlick2004}. In this approach, photons propagate along the null geodesics of a (four-dimensional) Lorentzian spacetime, which is typically assumed to be a solution to Einstein's field equations of general relativity (or some alternative theory of gravity).

The non-perturbative lensing formalism is particularly well-equipped to deal with the study of strong-field gravitational lensing by black holes. Interest in the theoretical aspects of gravitational lensing by black holes dates back to the 1960s. In a seminal paper, Synge \cite{Synge1966} calculated the angular radius of the shadow of a Schwarzschild black hole, as seen by a distant observer. Then, in 1979, Luminet \cite{Luminet1979} determined the apparent optical image of a Schwarzschild black hole illuminated by a distant light source, as well as the astrophysically interesting case of the image of a Schwarzschild black hole surrounded by an emitting accretion disk. The simulated images obtained by Luminet bear a striking resemblance to the (real) black hole shadow images captured by the EHT \cite{EHTC2019a}.

Since the early work of Synge and Luminet, there has been much interest in the analysis of the strong-field gravitational lensing effects of black holes (and other compact objects). For example, theorists have built up an understanding of the geodesic dynamics on black hole spacetimes \cite{Bardeen1973, Sharp1979, Teo2003}; explored the existence, stability and phenomenology of light-rings \cite{Liang1974, BalekBicakStuchlik1989, CunhaHerdeiroRadu2017}; and investigated the structure of black hole shadows \cite{FalckeMeliaAgol1999, JohannsenPsaltis2010, CunhaHerdeiro2018}. Taking into account the continuing efforts to deepen our understanding of the theoretical aspects of strong-field gravitational lensing, and an ever-increasing ability to test the theoretical predictions of general relativity at exquisite levels of precision, it is clear that gravitational lensing will either play a key role confirming Einstein's theory as a fundamental law of nature, or perhaps open the door to exciting new physics.

\section*{Outline}

In this thesis, we analyse theoretical aspects of gravitational lensing by black holes in general relativity, with a particular focus on the role of (non-)integrability, order and chaos. Chapters \ref{chap:general_relativity} and \ref{chap:gravitational_lensing} are devoted to a review of key concepts and techniques from the fields of general relativity, dynamical systems, chaos theory and gravitational lensing. The remaining chapters present new work carried out by the author and collaborators on the broad theme of strong-field gravitational lensing by black holes. In Chapters \ref{chap:binary_black_hole_shadows} and \ref{chap:fractal_structures}, we investigate chaotic scattering and fractal structures in binary black hole shadows. Chapter \ref{chap:stable_photon_orbits} studies the existence and phenomenology of stable photon orbits in stationary axisymmetric spacetimes. Finally, in Chapter \ref{chap:geometric_optics_kerr}, we study the propagation of electromagnetic radiation on Kerr spacetime by applying an extended geometric optics formalism. We conclude the present chapter with a more detailed chapter-by-chapter account of the work presented in this thesis.

\subsection*{Chapter \ref{chap:general_relativity}. Dynamics in general relativity}

In Chapter \ref{chap:general_relativity}, we review a range of important mathematical tools required for the study of general relativity. In particular, we introduce some aspects of the theory of differential geometry, including differentiable manifolds; vectors, one-forms and tensors; the metric tensor; covariant differentiation and parallel transport; Lie differentiation; Killing vectors; stationarity and staticity; and curvature. We then proceed to look at geodesics, presenting the Lagrangian and Hamiltonian formalisms for geodesic motion on curved spacetime, and deriving the geodesic deviation equation. Next, we present Einstein's field equations of general relativity, and the Einstein--Maxwell equations of gravity and electromagnetism. Key solutions are reviewed. We then discuss black holes and introduce the Schwarzschild, Kerr and Kerr--Newman geometries. We give a brief overview of the tetrad formalism, before looking at a special case -- the Newman--Penrose formalism -- in more detail. We conclude the chapter with a discussion of themes which are central to this thesis: integrability and chaos in dynamical systems, with an emphasis on geodesic motion in general relativity.

\subsection*{Chapter \ref{chap:gravitational_lensing}. Gravitational lensing}

Chapter \ref{chap:gravitational_lensing} is devoted to a review of gravitational lensing, which accounts for all effects of a gravitational field on the propagation of electromagnetic radiation. We divide the field into two subfields: perturbative lensing, which uses quasi-Newtonian approximations to describe weak gravitational fields; and non-perturbative lensing (or gravitational lensing from a spacetime perspective), in which light propagates along null geodesics in a four-dimensional Lorentzian spacetime which is a solution to the field equations of general relativity. We begin with a discussion of perturbative lensing. This includes a review of the history of the field as an observational science, beginning with the detection of gravitational light deflection by the Sun in 1919; a discussion of gravitational lensing phenomena which have been observed to date; and a brief account of the mathematical formalism of perturbative lensing, including an illustration of how this can be used to calculate the deflection angle of light due to a static quasi-Newtonian gravitational field. We then proceed to a discussion of gravitational lensing from a spacetime perspective. The theory of electromagnetism in curved spacetime is introduced in a fully covariant manner. We then review the leading-order geometric optics approximation for electromagnetism, which relies of the fundamental assumption that the wavelength (and inverse frequency) is significantly shorter than all other characteristic length (and time) scales, such as the scale set by the spacetime curvature. This scheme reduces the problem of solving wave equations (i.e., Maxwell's equations) on curved spacetime to one of solving transport equations along the rays (i.e., null geodesics) of the geometry. The application of geometric optics to the study of gravitational lensing phenomena is discussed. We conclude the chapter with an introduction to strong-field gravitational lensing effects associated with black holes, including (unstable) photon orbits and black hole shadows.

\subsection*{Chapter \ref{chap:binary_black_hole_shadows}. Binary black hole shadows and chaotic scattering}

In Chapter \ref{chap:binary_black_hole_shadows}, we investigate the qualitative features of binary black hole shadows using the Majumdar--Papapetrou binary black hole (or \emph{di-hole}) solution to the Einstein--Maxwell equations, which describes a pair of extremally charged black holes in static equilibrium. We advance the view that the propagation of null geodesics on a binary black hole spacetime is a natural example of chaotic scattering. We find that the existence of two or more dynamically connected fundamental photon orbits gives rise to an uncountable infinity of non-escaping null orbits (comprising the countable set of periodic orbits and the uncountable set of aperiodic orbits), which generate scattering singularities in the initial data. Using the Gaspard--Rice three-disc scatterer as a guide, we develop an appropriate symbolic dynamics to describe null geodesics. We compare and contrast our approach -- referred to here as \emph{decision dynamics} -- with an existing symbolic dynamics for the Majumdar--Papapetrou di-hole, which we refer to as \emph{collision dynamics}. We then demonstrate that our symbolic dynamics may be used to construct a one-dimensional binary black hole shadow on initial data, using an iterative procedure akin to the construction of the Cantor set; this argument demonstrates that the one-dimensional binary black hole shadow is self-similar. We then proceed by analysing non-planar null geodesics, aiming to quantify the effect of varying the (conserved) azimuthal angular momentum on the existence and properties of the fundamental null orbits. Using the Hamiltonian formalism for null geodesics, we introduce an effective potential which is independent of the photon's orbital parameters (i.e., energy and angular momentum); we use this effective potential to understand and classify the null geodesics of the Majumdar--Papapetrou di-hole geometry. In our analysis, we uncover an unexpected feature: the existence of stable bounded photon orbits around Majumdar--Papapetrou di-holes separated by dimensionless coordinate distance $\frac{d}{M} \in \left( \sqrt{\frac{16}{27}}, \sqrt{\frac{32}{27}} \right)$. (This is explored in greater detail in Chapter \ref{chap:stable_photon_orbits}.) We then use ray-tracing to generate two-dimensional binary black hole shadow images for the equal-mass Majumdar--Papapetrou di-hole. We find that a two-dimensional binary black hole shadow image can be decomposed into one-dimensional shadow images which can exhibit regular or Cantor-like structure. We propose a method for tracking null rays through the event horizons of the two black holes in the maximally extended Majumdar--Papapetrou di-hole spacetime. We conclude this chapter with a discussion of related work on black hole shadows and chaotic gravitational lensing signatures. An algorithm for passing between the symbolic codes discussed in this chapter is given in Appendix \ref{chap:appendix_c}. Supplementary calculations on the existence of circular photon orbits around the Majumdar--Papapetrou di-hole can be found in Appendix \ref{chap:appendix_a}.

\subsection*{Chapter \ref{chap:fractal_structures}. Fractal structures in binary black hole shadows}

Chapter \ref{chap:fractal_structures} extends on the work presented in Chapter \ref{chap:binary_black_hole_shadows}. In particular, we employ techniques from the field of non-linear dynamics to characterise the fractal structures which arise in the shadows of the Majumdar--Papapetrou binary black hole system. We review the scattering of null geodesics in the Majumdar--Papapetrou di-hole spacetime from the perspective of Hamiltonian dynamics. We construct the exit basins in phase space, and highlight qualitative similarities between the Majumdar--Papapetrou system and the H\'{e}non--Heiles Hamiltonian system, a paradigmatic model of two-dimensional time-independent chaotic scattering in Hamiltonian mechanics. (A pedagogical review of the H\'{e}non--Heiles system is given in Appendix \ref{chap:appendix_b}.) We also discuss the structure of black hole shadows, which are viewed as exit basin diagrams on the image plane of a distant observer. We review the uncertainty exponent and present a numerical method to calculate this quantity. We test and calibrate our method by applying it to a simple model -- the Cantor basins -- for which exact results are known. We then calculate the uncertainty exponent numerically for Majumdar--Papapetrou di-hole shadows. This successfully differentiates between fractal and regular (i.e., non-fractal) regions of the black hole shadow, and agrees with the theoretical predictions of Chapter \ref{chap:binary_black_hole_shadows}. Next, we apply a recently developed algorithm -- the merging method -- to demonstrate that parts of the Majumdar--Papapetrou di-hole shadow possess the Wada property: any point on the boundary of one basin is on the boundary of at least two additional basins. The algorithm is able to successfully distinguish between the Wada and non-Wada parts of the binary black hole shadow.

\subsection*{Chapter \ref{chap:stable_photon_orbits}. Stable photon orbits in stationary axisymmetric spacetimes}

In Chapter \ref{chap:stable_photon_orbits}, we explore the existence and phenomenology of \emph{stable} photon orbits in four-dimensional stationary axisymmetric electrovacuum solutions to the Einstein--Maxwell equations. We review the Kerr--Newman solution, and give an overview of the classification of its equatorial circular photon orbits in the charge--spin parameter space. Using a Hamiltonian formalism for rays, we demonstrate that the null geodesics of the stationary axisymmetric Weyl--Lewis--Papapetrou spacetime in four dimensions may be understood by introducing a pair of two-dimensional effective potentials. The fixed points of these potentials correspond to the circular null orbits of the geometry. Restricting attention to the electrovacuum case, we employ a subset of the Einstein--Maxwell equations to classify the fixed points of the effective potentials. We arrive at the following key result for four-dimensional stationary axisymmetric spacetimes: generic stable photon orbits are not permitted in pure vacuum, but may arise in electrovacuum. We investigate the existence and stability of photon orbits around Reissner--Nordstr\"{o}m static di-holes, a two-parameter subfamily of the general Bret\'{o}n--Manko--Aguilar di-hole (reviewed in Appendix \ref{chap:appendix_e}). The Reissner--Nordstr\"{o}m static di-hole family includes the uncharged Weyl--Bach di-hole and the extremal Majumdar--Papapetrou di-hole as special cases. In scenarios with high levels of symmetry (e.g.~the equal-mass Majumdar--Papapetrou di-hole), we derive closed-form expressions for stable photon orbit existence regions in parameter space. In cases with less symmetry, we employ a numerical method to search for stable photon orbits in parameter space. Finally, using Poincar\'{e} sections, we explore the transition from order to chaos for null rays which are bounded in a toroidal region around the black holes in the Majumdar--Papapetrou di-hole geometry. Intriguingly, we find that the Poincar\'{e} sections and bounded trajectories of the equal-mass Majumdar--Papapetrou di-hole with dimensionless coordinate separation parameter $\frac{d}{M} = 1$ share many qualitative features with those of the H\'{e}non--Heiles Hamiltonian system (cf.~Chapter \ref{chap:fractal_structures} and Appendix \ref{chap:appendix_b} for the case of \emph{unbounded} trajectories).

\subsection*{Chapter \ref{chap:geometric_optics_kerr}. Higher-order geometric optics on Kerr spacetime}

In Chapter \ref{chap:geometric_optics_kerr}, we apply an extended geometric optics formalism to understand aspects of strong-field gravitational lensing on Kerr spacetime, with the principal aim of computing the sub-dominant correction to the electromagnetic stress--energy tensor. We begin with a review of leading-order geometric optics for the electromagnetic field on an arbitrary curved spacetime, before reviewing the higher-order extension to the geometric optics formalism, originally presented by Dolan \cite{Dolan2018}. We discuss the geometry of Kerr spacetime, with an emphasis on the closed conformal Killing--Yano tensor, related Killing objects, explicit and hidden symmetries on spacetime and in phase space, and the implications for the separability and complete integrability of null geodesic motion. We construct a complex null tetrad using a self-dual bivector built from the closed conformal Killing--Yano tensor and its Hodge dual. Performing a Lorentz transformation, we are able to transform this tetrad to a new one which is parallel-transported along null geodesics. We discuss the Newman--Penrose formalism from the perspective of (higher-order) geometric optics on a Ricci-flat spacetime. In particular, we present a closed system of transport equations for the complex Newman--Penrose scalars, and a system of transport equations for certain directional derivatives of Newman--Penrose scalars. We calculate the complex Weyl curvature scalars and their directional derivatives in the parallel-transported complex null tetrad. We determine the far-field behaviour of the Weyl scalars, Newman--Penrose scalars and higher-order geometric optics quantities as generalised power series in $r$, the radial Boyer--Lindquist coordinate. We discuss wavefronts in geometric optics, and describe caustics, where neighbouring rays cross. The transport equations for Newman--Penrose quantities break down at caustic points. We derive near-caustic solutions to these transport equations as generalised power series in the affine parameter through sub-leading order. Finally, we present a practical method which may be employed to evolve transport equations for divergent quantities through caustic points. We comment on this method and its numerical implementation. Some explicit calculations for this chapter are contained in Appendix \ref{chap:appendix_d}.

\section*{Notation and conventions}

In this thesis, we adopt the sign conventions of Misner, Thorne and Wheeler \cite{MisnerThorneWheeler1973}. The four-dimensional spacetime metric has Lorentzian signature $(-, +, +, +)$. The Riemann tensor and Ricci tensor are defined in Chapter \ref{chap:general_relativity}. The Einstein summation convention for repeated indices is assumed throughout. Spacetime indices are denoted by Latin letters from the beginning of the alphabet (e.g.~$a, b, \ldots$); spatial indices are denoted by Latin letters from the middle of the alphabet (e.g.~$i, j, \ldots$). We employ geometrised units in which the speed of light $c$ and the gravitational constant $G$ are set to unity. Occasionally, we reinsert dimensional constants to aid physical interpretation. All other notation and conventions will be introduced as required.
%

\chapter{Dynamics in general relativity}
\label{chap:general_relativity}

\section{Geometry of spacetime}

General relativity describes space and time as a four-dimensional continuum known as \emph{spacetime}. The mathematical machinery required to describe spacetime is the differential geometry of pseudo-Riemannian manifolds. This section is devoted to a review of elements of this theory which are important for the study of general relativity. A more exhaustive treatment of topics covered in this section can be found in \cite{MisnerThorneWheeler1973, Wald1984, StephaniKramerMacCallumEtAl2003}, for example.

\subsection{Differentiable manifolds}

Intuitively, an $n$-dimensional manifold $\mathcal{M}$ is a space which looks locally like Euclidean space $\mathbb{R}^{n}$, but which may have different global properties. Before formally defining a manifold, it will be necessary to introduce some preliminary definitions and terminology.

The \emph{open ball} in $\mathbb{R}^{n}$ of radius $r$ centred on the point $p$, denoted $b_{r}(p)$, is the collection of points $x \in \mathbb{R}^{n}$ such that $\left| x - p \right| < r$, where $\left| \cdot \right|$ denotes the Euclidean norm on $\mathbb{R}^{n}$.

A subset $U \subset \mathbb{R}^{n}$ is called \emph{open} if it can be expressed as a union of open balls. Equivalently, the set $A$ is open if, for every point $p \in U$, there exists some real number $\varepsilon > 0$ such that $b_{\varepsilon}(p) \subset U$.

A \emph{manifold} is a set $\mathcal{M}$ such that any point $p \in \mathcal{M}$ has a neighbourhood $\mathcal{U} \subset \mathcal{M}$ which is homeomorphic to the interior of the $n$-dimensional unit ball. In order to give a more precise mathematical definition of a manifold, we first require some additional terminology.

On a manifold $\mathcal{M}$, a \emph{chart} $(\mathcal{U}, \psi)$ consists of a subset $\mathcal{U} \subset \mathcal{M}$ (called a \emph{chart neighbourhood}) and a bijection $\psi \colon \mathcal{U} \to U \subseteq \mathbb{R}^{n}$ (called a \emph{chart map}). The chart map $\psi$ assigns to each point $p \in \mathcal{M}$ an $n$-tuple of real variables $\left( x\ind{^{1}}, \ldots, x\ind{^{n}} \right)$, which are called \emph{local coordinates}.

Suppose that $(\mathcal{U}, \psi)$ and $(\mathcal{U}^{\prime}, \psi^{\prime})$ are two charts for $\mathcal{M}$ such that $\mathcal{U} \cap \mathcal{U}^{\prime} \neq \emptyset$. The \emph{transition map} from $\psi \left( \mathcal{U} \cap \mathcal{U}^{\prime} \right) \subset U$ to $\psi^{\prime} \left( \mathcal{U} \cap \mathcal{U}^{\prime} \right) \subset U^{\prime}$ is the map defined by $\psi^{\prime} \circ \psi^{-1}$. The charts $(\mathcal{U}, \psi)$ and $(\mathcal{U}^{\prime}, \psi^{\prime})$ are said to be \emph{compatible} if the transition map $\psi^{\prime} \circ \psi^{-1}$ is a homeomorphism (i.e., a continuous bijection with a continuous inverse).

An \emph{atlas} is a collection of compatible charts $\left\{ (\mathcal{U}_{\alpha}, \psi_{\alpha}) \right\}$ which cover the manifold $\mathcal{M}$. That is, each point $p \in \mathcal{M}$ is in at least one of the chart neighbourhoods $\mathcal{U}_{\alpha}$. (Here, $\alpha$ takes values from some indexing set.)

An $n$-dimensional \emph{topological manifold} consists of a space $\mathcal{M}$ with an atlas $\left\{ (\mathcal{U}_{\alpha}, \psi_{\alpha}) \right\}$ on $\mathcal{M}$. The manifold is said to be a \emph{$C^{k}$-differentiable manifold} if the transition maps $\psi_{i} \circ \psi_{j}^{-1}$ are not only continuous but $C^{k}$-differentiable. If the transition maps are infinitely differentiable, the manifold is said to be $C^{\infty}$ (or \emph{smooth}). For differentiable manifolds, the coordinates are related by $n$ differentiable functions with non-vanishing Jacobian at each point of the overlap, i.e., $x\ind{^{i^{\prime}}} = x\ind{^{i^{\prime}}}(x\ind{^{j}})$ with $\det{\left( \frac{ \partial x\ind{^{i^{\prime}}} }{\partial x\ind{^{j}}} \right)} \neq 0$ at all points in $\mathcal{U}_{i} \cap \mathcal{U}_{j}$.

In addition to the properties listed above, definitions of manifolds often feature additional topological restrictions, such as \emph{Hausdorffness} (i.e., any two distinct points in $\mathcal{M}$ have disjoint neighbourhoods) and \emph{paracompactness} (i.e.,  every open cover of $\mathcal{M}$ has a locally finite open refinement). A detailed consideration of these properties is not necessary here.

Given two manifolds $\mathcal{M}$ and $\mathcal{M}^{\prime}$ of dimension $n$ and $n^{\prime}$, respectively, the $(n + n^{\prime})$-dimensional product space $\mathcal{M} \times \mathcal{M}^{\prime}$ consisting of all pairs $(p, p^{\prime})$ with $p \in \mathcal{M}$ and $p^{\prime} \in \mathcal{M}^{\prime}$ can be made into a manifold in a natural way. If $\psi_{\alpha} \colon \mathcal{U}_{\alpha} \to U_{\alpha}$ and ${\psi^{\prime}}_{\beta} \colon \mathcal{U}^{\prime}_{\beta} \to U^{\prime}_{\beta}$ are charts on $\mathcal{M}$ and $\mathcal{M}^{\prime}$, respectively, then one can define a chart on the product $\mathcal{M} \times \mathcal{M}^{\prime}$ as $\psi_{\alpha \beta} \colon \mathcal{U}_{\alpha} \times \mathcal{U}^{\prime}_{\beta} \to U_{\alpha} \times U^{\prime}_{\beta} \subseteq \mathbb{R}^{n + n^{\prime}}$, by taking $\psi_{\alpha \beta}(p, p^{\prime}) = \left( \psi_{\alpha}(p), {\psi^{\prime}}_{\beta}(p^{\prime}) \right)$. The family of charts $\left\{ \psi_{\alpha \beta} \right\}$ satisfies the properties required to define a manifold structure on the product $\mathcal{M} \times \mathcal{M}^{\prime}$.

Having defined a manifold, whose structure is given by charts $\left\{ \psi_{\alpha} \right\}$, we may now define the notions of smoothness and differentiability for maps between manifolds. Let $\mathcal{M}$ and $\mathcal{M}^{\prime}$ be manifolds, and let $\{ \psi_{\alpha} \}$ and $\{ {\psi^{\prime}}_{\beta} \}$ denote their respective chart maps. A \emph{map} $f \colon \mathcal{M} \to \mathcal{M}^{\prime}$ is said to be $C^{\infty}$ (or \emph{smooth}) if, for each pair $(\alpha, \beta)$, the map ${\psi^{\prime}}_{\beta} \circ f \circ \psi_{\alpha}^{-1}$ from $\mathbb{R}^{n}$ to $\mathbb{R}^{n^{\prime}}$ is $C^{\infty}$.

If a map $f \colon \mathcal{M} \to \mathcal{M}^{\prime}$ is a smooth bijection and has a smooth inverse, then $f$ is called a \emph{diffeomorphism}, and the manifolds $\mathcal{M}$ and $\mathcal{M}^{\prime}$ are said to be \emph{diffeomorphic}.

A \emph{smooth curve} $\gamma(\lambda)$ in $\mathcal{M}$ is defined to be a smooth map from an interval of $\mathbb{R}$ to $\mathcal{M}$, where $\lambda \in \mathbb{R}$ is a parameter along the curve.

\subsection{Vectors, one-forms and tensors}
\label{sec:vectors_one_forms_tensors}

\subsubsection{Vectors}

Consider an $n$-dimensional manifold $\mathcal{M}$. A \emph{tangent vector} $v$ at a point $p \in \mathcal{M}$ is a linear functional $v \colon C^{\infty}(\mathcal{M}, \mathbb{R}) \to \mathbb{R}$, where $C^{\infty}(\mathcal{M}, \mathbb{R})$ denotes the set of $C^{\infty}$ functions from $\mathcal{M}$ to $\mathbb{R}$. The tangent vector is linear and satisfies the Leibniz property:
\begin{enumerate}[(i)]
\item $v ( c_{1} f_{1} + c_{2} f_{2} ) = c_{1} v( f_{1} ) + c_{2} v( f_{2} )$, for all $f_{1}, f_{2} \in C^{\infty}(\mathcal{M}, \mathbb{R})$, and $c_{1}, c_{2} \in \mathbb{R}$;
\item $v( f_{1} f_{2} ) = v( f_{1} ) f_{2} + f_{1} v( f_{2} )$.
\end{enumerate}
It follows from axioms (i) and (ii) that if $f \in C^{\infty}(\mathcal{M}, \mathbb{R})$ is a constant function, i.e., $f(p) = \textrm{constant}$ for all $p \in \mathbb{R}$, then $v(f) = 0$.

A tangent vector is a \emph{directional derivative} along a smooth curve $\gamma(\lambda)$ which passes through $p$. One may demonstrate by performing a Taylor series expansion of a function $f$ at $p$ and using the axioms (i) and (ii) above that a tangent vector at $p$ may be expressed in the form
\begin{equation}
v = v\ind{^{a}} \frac{\partial}{\partial x\ind{^{a}}}.
\end{equation}
The real coefficients $v\ind{^{a}}$ are called the \emph{components} of $v$ at the point $p$, with respect to the local coordinate system $\{ x\ind{^{a}} \}$ in a neighbourhood of $p$. The directional derivatives $\frac{\partial}{\partial x\ind{^{a}}}$ along coordinate curves at $p$ form a basis of an $n$-dimensional vector space whose elements are the tangent vectors at $p$. This vector space is called the \emph{tangent space} (at the point $p$), and is denoted $T_{p} \mathcal{M}$. The basis $\left\{ \frac{\partial}{\partial x\ind{^{a}}} \right\}$ is called a \emph{coordinate basis}; we often write its elements as $\frac{\partial}{\partial x\ind{^{a}}} = \partial\ind{_{a}}$. In the coordinate basis, the action of a basis vector on a function is written in the form $f\ind{_{, a}} = \partial\ind{_{a}} f = \frac{\partial f}{\partial x\ind{^{a}}}$. One may express $v$ in terms of some other basis $\left\{ \frac{\partial}{\partial {x^{\prime}}\ind{^{a}}} \right\}$. The old basis can be expressed in terms of the new basis as
\begin{equation}
\frac{\partial}{\partial x\ind{^{a}}} = \frac{\partial {x^{\prime}}\ind{^{b}}}{\partial x\ind{^{a}}} \frac{\partial}{\partial {x^{\prime}}\ind{^{b}}}.
\end{equation}
Moreover, the components ${v^{\prime}}\ind{^{a}}$ of $v$ in terms of the new basis are related to those of the old basis by
\begin{equation}
\label{eqn:vector_tranformation_law}
{v^{\prime}}\ind{^{a}} = \frac{\partial {x^{\prime}}\ind{^{a}}}{\partial x\ind{^{b}}} v\ind{^{b}}.
\end{equation}
This is known as the vector transformation law.

One may express a tangent vector in terms of a \emph{general basis} $\{ e\ind{_{a}} \}$, which is a collection of $n$ linearly independent vectors $e\ind{_{a}}$ at $p$. Any vector $v \in T_{p} \mathcal{M}$ can then be written in the form
\begin{equation}
v = v\ind{^{a}} e\ind{_{a}}.
\end{equation}
A coordinate basis is simply a special choice of general basis. Frequently, we will perform calculations in a coordinate basis; however, there will be a number of occasions when it is preferable to use a general basis.

The disjoint union of all tangent spaces $T_{p} \mathcal{M}$ at points $p \in \mathcal{M}$ forms the \emph{tangent bundle} $T \mathcal{M}$. In local coordinates, the elements of $T \mathcal{M}$ are the $2n$-tuples $(x\ind{^{a}}, v\ind{^{a}})$. The tangent bundle is then a $2n$-dimensional manifold. Moreover, if $\mathcal{M}$ is $C^{k}$, then $T \mathcal{M}$ is $C^{k - 1}$.

One may now construct a \emph{vector field} $v(p)$ on $\mathcal{M}$ by assigning to each $p \in \mathcal{M}$ a vector $v \in T_{p} \mathcal{M}$ such that the components $v\ind{^{a}}$ are differentiable functions of the local coordinates $x\ind{^{a}}$. A vector field $v(p)$ may then be regarded as a smooth map from $\mathcal{M}$ to $T \mathcal{M}$.

Let $X \in T \mathcal{M}$ be a vector field on $\mathcal{M}$, and consider a point $p \in \mathcal{M}$. Let $J \subseteq \mathbb{R}$ be an open interval which contains the point $0$. A smooth curve $\gamma \colon J \to \mathcal{M}$ is called an \emph{integral curve} of $X$ passing through $p$ if it satisfies the initial value problem
\begin{equation}
\label{eqn:integral_curve_ivp}
\dot{\gamma}(\lambda) = \left. X \right|_{\gamma}, \qquad \gamma(0) = p,
\end{equation}
where an overdot denotes differentiation with respect to the parameter $\lambda$. In local coordinates $\{ x\ind{^{a}} \}$, the problem of finding such curves reduces to solving the system
\begin{equation}
\frac{\ed}{\ed \lambda} x\ind{^{a}} (\lambda) = X\ind{^{a}} \left( x\ind{^{1}}(\lambda), \ldots, x\ind{^{n}}(\lambda) \right) ,
\end{equation}
where $X\ind{^{a}}$ denotes the $a$th component of the vector field $X$ in the coordinate basis $\left\{ \partial\ind{_{a}} \right\}$. Given a starting point $p$ at $\lambda = 0$, such a system of ordinary differential equations has a unique solution, so every smooth vector field $X$ has a unique family of integral curves \cite{Wald1984}.

Let $\mathcal{D} = \left\{ (\lambda, p) \, | \, \lambda \in J, \, p \in \mathcal{M} \right\}$. The \emph{flow of $X$} is the smooth map $\varphi_{X} \colon \mathcal{D} \to \mathcal{M}$ defined by $\varphi_{X} (\lambda, p) = \gamma(\lambda)$, where $\gamma \colon J \to \mathcal{M}$ is the unique solution to \eqref{eqn:integral_curve_ivp}.

\subsubsection{One-forms}

A \emph{one-form} (or \emph{dual vector}) $\sigma$ maps a vector $v \in T_{p} \mathcal{M}$ into a real number, which is the \emph{contraction} of $\sigma$ and $v$, denoted $\langle \sigma , v \rangle$. This mapping is linear, i.e., $\langle \sigma , c_{1} v_{1} + c_{2} v_{2} \rangle = c_{1} \langle \sigma , v_{1} \rangle + c_{2} \langle \sigma , v_{2} \rangle$, for all $v_{1}, v_{2} \in T_{p} \mathcal{M}$, and $c_{1}, c_{2} \in \mathbb{R}$. Linear combinations of one-forms are then defined by the linearity property $\langle c_{1} \sigma_{1} + c_{2} \sigma_{2} , v \rangle = c_{1} \langle \sigma_{1} , v \rangle + c_{2} \langle \sigma_{2} , v \rangle$, for all $c_{1}, c_{2} \in \mathbb{R}$.

The set of $n$ linearly independent one-forms $\{ \omega\ind{^{a}} \}$, which are determined by $\langle \omega\ind{^{a}} , e\ind{_{b}} \rangle = \delta\ind{^{a}_{b}}$, form a basis of the \emph{dual space} $T_{p}^{\ast} \mathcal{M}$ to the tangent space $T_{p} \mathcal{M}$. This is a vector space known as the \emph{cotangent space}. The basis $\{ \omega\ind{^{a}} \}$ is said to be dual to the basis $\{ e\ind{_{a}} \}$. A one-form $\sigma \in T_{p}^{\ast} \mathcal{M}$ can be expressed in terms of the basis $\{ \omega\ind{^{a}} \}$ as
\begin{equation}
\sigma = \sigma\ind{_{a}} \omega\ind{^{a}}.
\end{equation}
The contraction of any one-form $\sigma \in T_{p}^{\ast} \mathcal{M}$ and any vector $v \in T_{p} \mathcal{M}$ can be expressed as
\begin{equation}
\langle \sigma , v \rangle = \sigma\ind{_{a}} v\ind{^{a}},
\end{equation}
with respect to the bases $\{ \omega\ind{^{a}} \}$ and $\{ e\ind{_{a}} \}$.

The \emph{differential} $\ed f$ of a function $f$ is a  one-form, with the defining property $\langle \ed f , v \rangle = v(f) = v\ind{^{a}} e\ind{_{a}}(f)$. Taking $f$ to be the local coordinate functions $\{ x\ind{^{a}} \}$ and $v$ to be the coordinate basis, the previous definition gives $\langle \ed x\ind{^{a}} , \partial\ind{_{b}} \rangle = \frac{\partial x\ind{^{a}}}{\partial x\ind{^{b}}} = \delta\ind{^{a}_{b}}$, which implies that the basis $\{ \ed x\ind{^{a}} \}$ of the cotangent space is dual to the coordinate basis $\{ \partial\ind{_{a}} \}$ of the tangent space. Of course, any one-form $\sigma \in T_{p}^{\ast} \mathcal{M}$ can then be written in terms of the basis $\{ \ed x\ind{^{a}} \}$ as
\begin{equation}
\sigma = \sigma\ind{_{a}} \ed x\ind{^{a}}.
\end{equation}
In local coordinates, the differential $\ed f$ is simply $\ed f = f\ind{_{, a}} \ed x\ind{^{a}}$.

If $\sigma\ind{_{a}}$ denote the components of a one-form with respect to the dual basis $\{ \ed x\ind{^{a}} \}$, then it follows from $\langle \ed x\ind{^{a}} , \partial\ind{_{b}} \rangle = \delta\ind{^{a}_{b}}$ and the vector transformation law \eqref{eqn:vector_tranformation_law} that
\begin{equation}
{\omega^{\prime}}\ind{_{a}} = \frac{\partial x\ind{^{b}}}{\partial {x^{\prime}}\ind{^{a}}} \omega\ind{_{b}}.
\end{equation}
This is the transformation law for one-form components.

In a fashion analogous to the construction of the tangent bundle and vector fields, one may construct the \emph{cotangent bundle} $T^{\ast} \mathcal{M}$, which is the disjoint union of cotangent spaces $T_{p}^{\ast} \mathcal{M}$. A \emph{one-form field} $\sigma(p)$ is then constructed by assigning to each $p \in \mathcal{M}$ a one-form $\sigma \in T_{p}^{\ast} \mathcal{M}$, such that the components $\sigma\ind{_{a}}$ are differentiable functions of the local coordinates. The components $\sigma\ind{_{a}}$ are often referred to as the covariant components of a vector.

\subsubsection{Tensors}

If $V$ and $W$ are vector spaces, then the \emph{tensor product} of $V$ and $W$, denoted $V \otimes W$, is also a vector space. There is a standard bilinear map $V \times W \to V \otimes W$, denoted by $(v , w) \mapsto v \otimes w$, which satisfies the following axioms:
\begin{enumerate}[(i)]
\item $\left( c_{1} v_{1} + c_{2} v_{2} \right) \otimes w = c_{1} v_{1} \otimes w + c_{2} v_{2} \otimes w$;
\item $v \otimes \left( c_{1} w_{1} + c_{2} w_{2} \right) = c_{1} v \otimes w_{1} + c_{2} v \otimes w_{2}$;
\end{enumerate}
for all $v_{1}, v_{2} \in V$, $w_{1}, w_{2} \in W$ and $c_{1}, c_{2} \in \mathbb{R}$. The vector space $V \otimes W$ is then the space of all finite linear combinations of formal symbols of the form $v \otimes w$ for $v \in V$ and $w \in W$.

A \emph{tensor} $T$ of type $(r ,s)$ at $p$ is a multilinear map from the tensor product of $r$ copies of the tangent space at $p$ with $s$ copies of the cotangent space at $p$ to the real numbers:
\begin{equation}
T \colon T_{p} \mathcal{M} \otimes \ldots \otimes T_{p} \mathcal{M} \otimes T_{p}^{\ast} \mathcal{M} \otimes \ldots \times T_{p}^{\ast} \mathcal{M} \to \mathbb{R}.
\end{equation}
The tensor $T$ maps any ordered set of $r$ one-forms and $s$ vectors into a real number.

An arbitrary type-$(r, s)$ tensor $T$ can be expressed in terms of the bases $\{ e\ind{_{a}} \}$ and $\{ \omega\ind{^{b}} \}$ as
\begin{equation}
T = T\ind{^{a_{1} \ldots a_{r}}_{b_{1} \ldots b_{s}}} e\ind{_{a_{1}}} \otimes \ldots \otimes e\ind{_{a_{r}}} \otimes \omega\ind{^{b_{1}}} \otimes \ldots \otimes \omega\ind{^{b_{s}}}.
\end{equation}
The \emph{components} of the tensor $T$ with respect to $\{ e\ind{_{a}} \}$ and $\{ \omega\ind{^{b}} \}$ are the real coefficients $T\ind{^{a_{1} \ldots a_{r}}_{b_{1} \ldots b_{s}}}$. In general, the components of a tensor $T$ of type $(r ,s)$ transform as
\begin{equation}
\label{eqn:tensor_transformation_law}
{T^{\prime}} \ind{^{a_{1} \ldots a_{r}}_{b_{1} \ldots b_{s}}} = \frac{\partial {x^{\prime}}\ind{^{a_{1}}}}{\partial x\ind{^{c_{1}}}} \ldots \frac{\partial {x^{\prime}}\ind{^{a_{r}}}}{\partial x\ind{^{c_{r}}}} \frac{\partial x\ind{^{d_{1}}}}{\partial {x^{\prime}}\ind{^{b_{1}}}} \ldots \frac{\partial x\ind{^{d_{s}}}}{\partial {x^{\prime}}\ind{^{b_{s}}}} T\ind{^{c_{1} \ldots c_{r}}_{d_{1} \ldots d_{s}}}.
\end{equation}
The generalisation of a tensor to a \emph{tensor field} is straightforward: a tensor field is a choice of a tensor at each point $p \in \mathcal{M}$ that varies smoothly with any coordinates (i.e., the components of the tensor are smooth functions of the local coordinates). In particular, type-$(1, 0)$ tensors are vector fields, type-$(0, 1)$ tensors are one-form (or covector) fields, and type-$(0, 0)$ tensors are defined to be functions. For simplicity, we hereafter refer to tensor fields as tensors.


\subsubsection{Exterior calculus}

At this stage we introduce some notation for the totally symmetric and totally antisymmetric parts of tensors. For a tensor $T$ of type $(0, p)$ with components $T\ind{_{a_{1} \ldots a_{p}}}$, we define its totally symmetric and totally antisymmetric parts, respectively, as
\begin{align}
T\ind{_{( a_{1} \ldots a_{p} )}} &= \frac{1}{p!} \sum_{\pi} T\ind{_{ a_{\pi(1)} \ldots a_{\pi(p)} }}, \\
T\ind{_{[ a_{1} \ldots a_{p} ]}} &= \frac{1}{p!} \sum_{\pi} \on{sgn}(\pi) T\ind{_{ a_{\pi(1)} \ldots a_{\pi(p)} }},
\end{align}
where the summation is taken over all permutations $\pi$ of the ordered set $\{ 1, \ldots, p \}$, and $\on{sgn(\pi)}$ is the sign of the permutation, which takes the value $+1$ ($-1$) for even (odd) permutations of $\{ 1, \ldots, p \}$. (Equivalent definitions apply for the symmetrisation and anti-symmetrisation of contravariant indices.)

A \emph{$p$-form} $\alpha$ is a totally antisymmetric tensor of type $(0, p)$. The set of all $p$-forms on a manifold $\mathcal{M}$ is a vector space, denoted $\Omega^{p}(\mathcal{M})$. A smooth function $f$ on $\mathcal{M}$ is a $0$-form, $f \in \Omega^{0}(\mathcal{M})$.

The \emph{exterior product} (or \emph{wedge product}) of a $p$-form $\alpha$ with a $q$-form $\beta$ is denoted by $\alpha \wedge \beta$. It is a totally antisymmetric tensor of type $(0, p + q)$, i.e., a $(p + q)$-form, whose components are given (up to normalisation) by the anti-symmetrisation of the tensor product of $\alpha$ and $\beta$:
\begin{equation}
(\alpha \wedge \beta)\ind{_{a_{1} \ldots a_{p} b_{1} \ldots b_{q}}} = \frac{(p + q)!}{p! q!} \alpha\ind{_{ [ a_{1} \ldots a_{p}}} \beta\ind{_{ b_{1} \ldots b_{q} ]}}.
\end{equation}
The exterior product obeys the property $\alpha \wedge \beta = (-1)^{p q} \beta \wedge \alpha$.

For one-forms $\alpha$ and $\beta$, their \emph{symmetric product} is defined in terms of the tensor product as
\begin{equation}
\label{eqn:symmetric_product_one_forms}
\alpha \beta = \frac{1}{2} \left( \alpha \otimes \beta + \beta \otimes \alpha \right).
\end{equation}

The \emph{exterior derivative} $\ed$ is a map $\ed \colon \Omega^{p}(\mathcal{M}) \to \Omega^{p + 1}(\mathcal{M})$, which is completely determined by the axioms
\begin{enumerate}[(i)]
\item $\ed (\alpha + \beta) = \ed \alpha + \ed \beta$;
\item $\ed (\alpha \wedge \beta) = \ed \alpha \wedge \beta + (-1)^{p} \alpha \wedge \ed \beta$;
\item $\ed^{2} f = \ed (\ed f) = 0$;
\item $\ed f = f\ind{_{, a}} \ed x\ind{^{a}}$.
\end{enumerate}
Property (iv) says that the exterior derivative maps a function $f$ (i.e., a $0$-form) to its differential.

A differential $p$-form $\alpha$ is \emph{closed} if its exterior derivative vanishes $(\ed \alpha = 0)$. A differential $p$-form $\alpha$ is \emph{exact} if it can be expressed as the exterior derivative of another differential $(p-1)$-form $\beta$ ($\alpha = \ed \beta$). The form $\beta$ is called a \emph{potential form} for $\alpha$. The potential form $\beta$ is non-unique: $\beta^{\prime} = \beta + \ed \gamma$, where $\gamma$ is any $(p - 2)$-form, is also a potential form for $\alpha$, since $\ed^{2} \gamma = 0$.

By property (iii) from the above list of properties for the exterior derivative operator, any exact form is necessarily closed. The question of whether the converse of this statement is true depends on the topology of the domain of interest. On a contractible domain, every closed form is exact by the Poincar\'{e} lemma.


\subsection{Metric tensor}
\label{sec:metric_tensor}

A \emph{pseudo-Riemannian manifold} $(\mathcal{M}, g)$ is an $n$-dimensional smooth manifold $\mathcal{M}$ endowed with a tensor $g$ of type $(0, 2)$, called the \emph{metric tensor}, such that, at each point $p \in \mathcal{M}$, $g$ is a symmetric non-degenerate bilinear quadratic form. In a coordinate basis, one can write the metric as
\begin{equation}
g = g\ind{_{a b}} \ed x\ind{^{a}} \otimes \ed x\ind{^{b}},
\end{equation}
Often, we use the notation $\ed s^{2}$ to denote the \emph{line element}, writing
\begin{equation}
\label{eqn:line_element}
\ed s^{2} = g\ind{_{a b}} \ed x\ind{^{a}} \ed x\ind{^{b}},
\end{equation}
where $\ed x\ind{^{a}} \ed x\ind{^{b}}$ denotes the symmetric product of the basis one-forms $\ed x\ind{^{a}}$ and $\ed x\ind{^{b}}$; see \eqref{eqn:symmetric_product_one_forms}. The notation \eqref{eqn:line_element} is consistent with the intuitive notion that the metric represents an infinitesimal squared distance on the manifold $\mathcal{M}$. A \emph{Riemannian manifold} is equipped with a metric of signature $(+, +, \ldots, +)$; whereas a \emph{Lorentzian manifold} is endowed with a metric of signature $(-, +, \ldots, +)$.

Einstein's theory of general relativity is based on the concept of \emph{spacetime}, a four-dimensional continuum which unifies the three spatial dimensions and one temporal dimension. In this work, a \emph{spacetime} is a four-dimensional Lorentzian manifold $\mathcal{M}$, equipped with a metric $g$ of signature $(-, +, +, +)$.

The scalar product of two vectors $v$ and $u$ is given by the contraction $v \cdot u = g\ind{_{a b}} v\ind{^{a}} u\ind{^{b}}$. Two vectors are said to be \emph{orthogonal} if their scalar product vanishes. A vector $v$ is said to be \emph{timelike}, \emph{null} or \emph{spacelike} if $g\ind{_{a b}} v\ind{^{a}} v\ind{^{b}}$ is negative, zero or positive, respectively.

The contravariant components $g\ind{^{a b}}$ form the matrix inverse of the metric tensor $g\ind{_{a b}}$; the ``inverse metric'' $g\ind{^{a b}}$ is a type-$(2, 0)$ tensor. One may use the metric and its inverse to raise and lower indices in the standard fashion; for example, $v\ind{_{a}} = g\ind{_{a b}} v\ind{^{b}}$ and $v\ind{^{a}} = g\ind{^{a b}} v\ind{_{b}}$. This means that the vector field $v\ind{^{a}} e\ind{_{a}}$ and the one-form field $v\ind{_{a}} \omega\ind{^{a}}$ represent the same geometrical object. These will be used interchangeably, and we will often denote a vector (one-form) field by its components, i.e., $v\ind{^{a}}$ ($v\ind{_{a}}$).

Let $\Omega > 0$ be a smooth function. A \emph{conformal transformation} of the metric is a mapping of the form $g\ind{_{a b}} \mapsto \tilde{g}\ind{_{a b}} = \Omega^{2} g\ind{_{a b}}$. The inverse metrics are related by $\tilde{g}\ind{^{a b}} = \frac{1}{\Omega^{2}} g\ind{^{a b }}$. Conformal transformations arise in a range of contexts in general relativity; we will see that they are particularly important in the treatment of geodesics (Section \ref{sec:geodesics}).

On an $n$-dimensional pseudo-Riemannian manifold $\mathcal{M}$, the \emph{Hodge dual} of a $p$-form $\alpha$ is a $(n - p)$-form $\vp{\alpha}^{\star}\alpha$, whose components are defined by
\begin{equation}
(\vp{\alpha}^{\star}\alpha)\ind{_{a_{p + 1} \ldots a_{n}}} = \alpha\ind{^{a_{1} \ldots a_{p}}} \epsilon\ind{_{ a_{1} \ldots a_{p} a_{p + 1} \ldots a_{n}}},
\end{equation}
where $\epsilon$ is the Levi-Civita tensor (or totally anti-symmetric tensor). On four-dimensional spacetime, the Hodge dual of a $p$-form is a $(4 - p)$-form, and the Levi-Civita tensor is $\epsilon\ind{_{a b c d}} = \sqrt{- \det{g}} \left[ a b c d \right]$, where $\det{g}$ denotes the determinant of the metric tensor, and $\left[ a b c d \right]$ is the fully anti-symmetric Levi-Civita symbol with $\left[ 0 1 2 3 \right] = 1$.

\subsection{Covariant differentiation and parallel transport}
\label{sec:covariant_differentiation_parallel_transport}

A \emph{covariant derivative} operator $\nabla$ is a map which takes a smooth type-$(r, s)$ tensor field to a smooth type-$(r, s + 1)$ tensor field. For a tensor $T$ with components $T\ind{^{a_{1} \ldots a_{r}}_{b_{1} \ldots b_{s}}}$, the action of the covariant derivative on $T$ is denoted $\nabla\ind{_{c}} T\ind{^{a_{1} \ldots a_{r}}_{b_{1} \ldots b_{s}}}$ in index notation; we employ the standard notation $\nabla\ind{_{a}}$, attaching an index to the covariant derivative operator.

On an $n$-dimensional smooth pseudo-Riemannian manifold $(\mathcal{M}, g)$, the covariant derivative satisfies the following conditions.
\begin{enumerate}[(i)]
\item Linearity: $\nabla\ind{_{c}} \left(\alpha X\ind{^{a_{1} \ldots a_{r}}_{b_{1} \ldots b_{s}}} + \beta Y\ind{^{a_{1} \ldots a_{r}}_{b_{1} \ldots b_{s}}} \right) = \alpha \nabla\ind{_{c}} X\ind{^{a_{1} \ldots a_{r}}_{b_{1} \ldots b_{s}}} + \beta \nabla\ind{_{c}} Y\ind{^{a_{1} \ldots a_{r}}_{b_{1} \ldots b_{s}}}$, for all tensors $X$, $Y$ of type $(r, s)$, and all $\alpha, \beta \in \mathbb{R}$.
\item Leibniz rule for the tensor product of two tensors: $\nabla\ind{_{e}} \left( X\ind{^{a_{1} \ldots a_{r}}_{b_{1} \ldots b_{s}}} Y\ind{^{c_{1} \ldots c_{r^{\prime}}}_{d_{1} \ldots d_{s^{\prime}}}} \right) = \left( \nabla\ind{_{e}} X\ind{^{a_{1} \ldots a_{r}}_{b_{1} \ldots b_{s}}} \right) Y\ind{^{c_{1} \ldots c_{r^{\prime}}}_{d_{1} \ldots d_{s^{\prime}}}} + X\ind{^{a_{1} \ldots a_{r}}_{b_{1} \ldots b_{s}}} \left( \nabla\ind{_{e}} Y\ind{^{c_{1} \ldots c_{r^{\prime}}}_{d_{1} \ldots d_{s^{\prime}}}} \right)$, for all tensors $X$ of type $(r, s)$ and tensors $Y$ of type $(r^{\prime}, s^{\prime})$.
\item Commutativity with contraction: $\nabla\ind{_{c}} \left(T\ind{^{a_{1} \ldots d \ldots a_{r}}_{b_{1} \ldots d \ldots b_{s}}} \right) = \nabla\ind{_{c}} T\ind{^{a_{1} \ldots d \ldots a_{r}}_{b_{1} \ldots d \ldots b_{s}}}$, for all tensors $T$ of type $(r, s)$ (so that the parentheses are not necessary).
\item Consistency with tangent vectors as directional derivatives of scalar functions: $v(f) = v\ind{^{a}} \nabla\ind{_{a}} f$, for all smooth functions $f \in C^{\infty}(\mathcal{M}, \mathbb{R})$ and all tangent vectors $v \in T \mathcal {M}$.
\item Torsion-free: $\nabla\ind{_{a}} \nabla\ind{_{b}} f = \nabla\ind{_{b}} \nabla\ind{_{a}} f$, for all smooth functions $f \in C^{\infty}(\mathcal{M}, \mathbb{R})$.
\end{enumerate}

Imposing the supplementary condition of \emph{metric compatibility} $\nabla\ind{_{a}} g\ind{_{b c}} = 0$, defines a unique covariant derivative operator $\nabla\ind{_{a}}$. Using the Levi-Civita connection, we may express the action of the covariant derivative operator $\nabla\ind{_{a}}$ on a type-$(r, s)$ tensor $T$ in terms of the ordinary partial derivative operator $\partial\ind{_{a}}$ in a coordinate basis $\{ x\ind{^{a}} \}$ as
\begin{equation}
\label{eqn:covariant_derivative_tensor}
\nabla\ind{_{a}} T\ind{^{b_{1} \ldots b_{r}}_{c_{1} \ldots c_{s}}} = \partial\ind{_{a}} T\ind{^{b_{1} \ldots b_{r}}_{c_{1} \ldots c_{s}}} + \sum_{i} \Gamma\ind{^{b_{i}}_{a d}} T\ind{^{b_{1} \ldots d \ldots b_{r}}_{c_{1} \ldots c_{s}}} - \sum_{i} \Gamma\ind{^{d}_{a c_{i}}} T\ind{^{b_{1} \ldots b_{r}}_{c_{1} \ldots d \ldots c_{s}}},
\end{equation}
where there is one term for each contravariant (covariant) index of $T$ which comes with a coefficient of $+1$ ($-1$).\footnote{One may, of course, define a covariant derivative using connection other than the Levi-Civita connection; for a more complete discussion, see e.g.~Wald \cite{Wald1984}.} The \emph{connection coefficients} $\Gamma\ind{^{a}_{bc}}$ are known as the \emph{Christoffel symbols}, which are defined by
\begin{equation}
\label{eqn:christoffel_symbols}
\Gamma\ind{^{a}_{b c}} = \frac{1}{2} g\ind{^{a d}} \left( \partial\ind{_{b}} g\ind{_{d c}} + \partial\ind{_{c}} g\ind{_{b d}} - \partial\ind{_{d}} g\ind{_{b c}} \right).
\end{equation}
The Christoffel symbols are symmetric in their lower indices (i.e., $\Gamma\ind{^{a}_{b c}} = \Gamma\ind{^{a}_{c b}}$), due to the torsion-free property of the covariant derivative. We caution here that the connection coefficients $\Gamma\ind{^{a}_{b c}}$ are \emph{not} tensors as they do not obey the tensor transformation law \eqref{eqn:tensor_transformation_law} under changes of coordinates. However, the covariant derivative of a tensor field \emph{does} transform covariantly.

Special cases of \eqref{eqn:covariant_derivative_tensor} are the action of the covariant derivative on functions $f$, vectors $v\ind{^{a}}$, and one-forms $\sigma\ind{_{a}}$; these are, respectively, given by
\begin{equation}
\nabla\ind{_{a}} f = \partial\ind{_{a}} f , \qquad \nabla\ind{_{a}} v\ind{^{b}} = \partial\ind{_{a}} v\ind{^{b}} + \Gamma\ind{^{b}_{a c}} v\ind{^{c}}, \qquad \nabla\ind{_{a}} \sigma\ind{_{b}} = \partial\ind{_{a}} \sigma\ind{_{b}} - \Gamma\ind{^{c}_{a b}} \sigma\ind{_{c}}.
\end{equation}
In general, we will denote the covariant derivative using a semi-colon (and the partial derivative using a comma), e.g.~$\nabla\ind{_{a}} v\ind{^{b}} = v\ind{^{b}_{; a}} = v\ind{^{b}_{, a}} + \Gamma\ind{^{b}_{a c}} v\ind{^{c}}$.

Having defined the covariant derivative $\nabla\ind{_{a}}$, we may now describe the notion of parallel transport of a vector (or, more generally, a tensor) along a curve $\gamma$ with tangent vector $u\ind{^{a}}$. A vector $v\ind{^{a}}$ at each point on $\gamma$ is said to be \emph{parallel-transported} (or \emph{parallel-propagated}) along the curve $\gamma$ if its covariant derivative vanishes along the curve, i.e.,
\begin{equation}
\label{eqn:parallel_transport_vector}
u\ind{^{b}} \nabla\ind{_{b}} v\ind{^{a}} = 0.
\end{equation}
In general, a tensor field of type $(r, s)$ with components $T\ind{^{a_{1} \ldots a_{r}}_{b_{1} \ldots b_{s}}}$ is parallel-transported along $\gamma$ if
\begin{equation}
u\ind{^{c}} \nabla\ind{_{c}} T\ind{^{a_{1} \ldots a_{r}}_{b_{1} \ldots b_{s}}} = 0
\end{equation}
along the curve. In a coordinate basis $\{x\ind{^{a}}\}$, one may write the equation of parallel transport of a vector field \eqref{eqn:parallel_transport_vector} as $u\ind{^{b}} \partial\ind{_{b}} v\ind{^{a}} + \Gamma\ind{^{a}_{b c}} u\ind{^{b}} v\ind{^{c}} = 0$. If the curve is parametrised by a parameter $\lambda \in \mathbb{R}$, then one may use the chain rule to rewrite this as
\begin{equation}
\label{eqn:parallel_transport_coordinates}
\dot{v}\ind{^{a}} + \Gamma\ind{^{a}_{b c}} u\ind{^{b}} v\ind{^{c}} = 0,
\end{equation}
where an overdot denotes differentiation with respect to $\lambda$.


\subsection{Lie differentiation}
\label{sec:lie_differentiation}

The \emph{commutator} of two vector fields $u$ and $v$ is itself a vector field, denoted $[u, v]$. It is defined by its action on smooth functions $f$:
\begin{equation}
[u, v] f = u (v ( f ) ) - v ( u ( f ) ).
\end{equation}
For all vector fields $u$, $v$ and $w$, the commutator satisfies the anti-symmetry property and the Jacobi identity:
\begin{enumerate}[(i)]
\item $[u, v] = - [v , u]$;
\item $[u, [v, w]] + [v, [w, u]] + [w, [u, v]] = 0$.
\end{enumerate}

On an $n$-dimensional manifold $\mathcal{M}$, one may use the commutator of vector fields to define the \emph{Lie derivative} of a type-$(r, s)$ tensor field $T$ along the vector field $v$, which is itself a type-$(r, s)$ tensor field, denoted $\mathcal{L}_{v} T$. The Lie derivative evaluates the change of a tensor field along the flow induced by a vector field.

For smooth scalar functions $f$, the Lie derivative of $f$ along the vector field $u$ is given by
\begin{equation}
\mathcal{L}_{u} f = u(f) = u\ind{^{a}} \partial\ind{_{a}} f.
\end{equation}
The Lie derivative of a vector field $v$ along the vector field $u$ is defined by
\begin{equation}
\mathcal{L}_{u} v = [u , v].
\end{equation}
This definition may be extended to tensors of arbitrary type, by demanding that the Lie derivative satisfies the following properties.
\begin{enumerate}[(i)]
\item Linearity: $\mathcal{L}_{u} \left( \alpha X + \beta Y \right) = \alpha \mathcal{L}_{u} X + \beta \mathcal{L}_{u} Y$, for all type-$(r, s)$ tensors $X, Y$, and all $\alpha, \beta \in \mathbb{R}$.
\item Leibniz rule on the tensor product of two tensors: $\mathcal{L}_{u} \left( X \otimes Y \right) = ( \mathcal{L}_{u} X ) \otimes Y + X \otimes \left( \mathcal{L}_{u}  Y \right)$ for all tensors $X, Y$.
\item ``Product rule'' on the contraction of a vector and a one-form: $\mathcal{L}_{u} \langle \sigma, v \rangle = \langle \mathcal{L}_{u} \sigma, v \rangle + \langle \sigma, \mathcal{L}_{u}  v \rangle$, for all one-forms $\sigma$ and vectors $v$.
\end{enumerate}
The above properties may be used to deduce the components of the Lie derivative of an arbitrary type-$(r, s)$ tensor field:
\begin{equation}
\label{eqn:lie_derivative_tensor}
(\mathcal{L}_{u} T )\ind{^{a_{1} \ldots a_{r}}_{b_{1} \ldots b_{s}}} = u\ind{^{c}} T\ind{^{a_{1} \ldots a_{r}}_{b_{1} \ldots b_{s} , c}} - \sum_{i} T\ind{^{a_{1} \ldots c \ldots a_{r}}_{b_{1} \ldots b_{s}}} u\ind{^{a_{i}}_{, c}} + \sum_{i} T\ind{^{a_{1} \ldots a_{r}}_{b_{1} \ldots c \ldots b_{s}}} u\ind{^{c}_{, b_{i}}} ,
\end{equation}
where there is one term for each index, and the covariant (contravariant) indices come with a coefficient of $+1$ ($-1$).

For the metric tensor $g\ind{_{a b}}$, one may replace the partial derivatives on the right-hand side of \eqref{eqn:lie_derivative_tensor} with covariant derivatives using the Levi-Civita connection. This gives the Lie derivative of the metric tensor along the vector field $u$:
\begin{equation}
\label{eqn:lie_derivative_metric}
(\mathcal{L}_{u} g )\ind{_{a b}} = u\ind{^{c}} g\ind{_{a b ; c}} + g\ind{_{c b}} u\ind{^{c}_{; a}} + g\ind{_{a c}} u\ind{^{c}_{; b}} = u\ind{_{a ; b}} + u\ind{_{b ; a}},
\end{equation}
where we have used the metric compatibility of the covariant derivative, i.e., $g\ind{_{a b ; c}} = 0$.

A tensor field $T$ of type $(r, s)$ is said to be \emph{Lie-transported} along the curve $\gamma$ with tangent vector field $u$ if its Lie derivative along the curve vanishes, i.e., $\mathcal{L}_{u} T = 0$.

We shall see later that the Lie derivative plays an important role in describing the symmetries of the gravitational field in general relativity.
%

\subsection{Killing vectors}
\label{sec:killing_vectors}

Symmetries are ubiquitous in physics and play a fundamental role in general relativity. Here, we briefly describe explicit continuous symmetries of spacetime. A spacetime with metric $g$ admits a \emph{continuous symmetry} (\emph{isometry}) if there exists a continuous transformation of the spacetime into itself which preserves the metric.

Such transformations are encoded by \emph{Killing vectors} $\xi$; the isometry condition states that the Lie derivative of the metric with respect to $\xi$ vanishes, i.e.,
\begin{equation}
\mathcal{L}_{\xi} g = 0.
\end{equation}
Geometrically, the condition $\mathcal{L}_{\xi} g = 0$ says that the metric is invariant under the flow of $\xi$.

Using the metric connection $\nabla\ind{_{a}}$ in local coordinates, we see from \eqref{eqn:lie_derivative_metric} that the isometry condition is equivalent to the \emph{Killing vector equation},
\begin{equation}
\label{eqn:killing_vector_equation}
\nabla\ind{_{a}}\xi\ind{_{b}} + \nabla\ind{_{b}}\xi\ind{_{a}} = 0.
\end{equation}
This equation is often written in the form $\xi\ind{_{(a ; b)}} = 0$.

If the metric coefficients $g\ind{_{a b}}$ in some basis $\left\{ \ed x\ind{^{a}} \right\}$ are independent of one of the coordinate $x\ind{^{\alpha}}$ (for some fixed $\alpha$), then $\xi\ind{^{a}} = \delta\ind{^{a}_{\alpha}}$ is a Killing vector field in the basis $\{ \partial\ind{_{a}} \}$. To see this, we first note that $g\ind{_{a b , \alpha}} = 0$. We have $\xi\ind{^{a}} = \delta\ind{^{a}_{\alpha}}$, so $\xi\ind{_{a}} = g\ind{_{a b}} \xi\ind{^{b}} = g\ind{_{a b}} \delta\ind{^{b}_{\alpha}} = g\ind{_{a \alpha}}$. The left-hand side of the Killing vector equation \eqref{eqn:killing_vector_equation} is then
\begin{equation}
\label{eqn:killing_equation_lhs}
\xi\ind{_{a ; b}} + \xi\ind{_{b ; a}} = \xi\ind{_{a , b}} + \xi\ind{_{b , a}} - 2 \Gamma\ind{^{c}_{a b}} \xi\ind{_{c}} = g\ind{_{a \alpha , b}} + g\ind{_{b \alpha , a}} - g\ind{^{c d}} \left( g\ind{_{b d , a}} + g\ind{_{a d , a}} - g\ind{_{a b , d}} \right) g\ind{_{c \alpha}}.
\end{equation}
Now, using the fact that $g\ind{^{c d}} g\ind{_{c \alpha}} = \delta\ind{^{d}_{\alpha}}$, the right-hand side of \eqref{eqn:killing_equation_lhs} becomes
\begin{equation}
g\ind{_{a \alpha , b}} + g\ind{_{b \alpha , a}} - \left( g\ind{_{b \alpha , a}} + g\ind{_{a \alpha , a}} - g\ind{_{a b , \alpha}} \right) = - g\ind{_{a b , \alpha}} = 0 ,
\end{equation}
by virtue of the fact that $g\ind{_{a b}}$ is independent of $x\ind{^{\alpha}}$. Hence, $\xi\ind{^{a}}$ is a Killing vector.
%

\subsection{Stationary and static spacetimes}
\label{sec:static_and_stationary_spacetimes}

Consider a spacetime $(\mathcal{M}, g\ind{_{a b}})$ with local coordinates $x\ind{^{a}}$. A non-vanishing one-form field $u\ind{_{a}}$ is said to be \emph{hypersurface-orthogonal} if it is orthogonal to a family of $\Phi = \text{constant}$ hypersurfaces for some function $\Phi$ with $\partial\ind{_{a}} \Phi \neq 0$. In other words, $u\ind{_{a}} = \alpha \partial\ind{_{a}} \Phi$ for some non-zero function of the spacetime coordinates $\alpha$. The Frobenius theorem says that, locally, the definition of hypersurface-orthogonality is equivalent to $u\ind{_{[ a}} \partial\ind{_{b}} u\ind{_{c ]}} = u\ind{_{[ a}} \nabla\ind{_{b}} u\ind{_{c ]}} = 0$. A vector field $u\ind{^{a}}$ is hypersurface-orthogonal if and only if the corresponding one-form field $u\ind{_{a}} = g\ind{_{a b}} u\ind{^{b}}$ is hypersurface-orthogonal.

A spacetime is \emph{stationary} if it admits a timelike Killing vector $\xi$. If this is the case, there exist local coordinates $(t, x\ind{^{i}})$ in which $\xi = \partial\ind{_{t}}$, $g\ind{_{t t}} < 0$, and $g\ind{_{a b, t}} = 0$ (i.e., the metric is independent of $t$). The off-diagonal terms $g\ind{_{t i}}$ need not vanish.

A spacetime is \emph{static} if (i) it admits a timelike hypersurface-orthogonal Killing vector $\xi$, and (ii) it can be globally foliated by hypersurfaces that are orthogonal to $\xi$. The first of these criteria ensures that such hypersurfaces exist locally, but not necessarily globally. In this case, there exist local coordinates $(t, x\ind{^{i}})$ in which $\xi = \partial\ind{_{t}}$, $g\ind{_{t t}} < 0$, $g\ind{_{a b, t}} = 0$, and $g\ind{_{t i}} = 0$.

\subsection{Curvature}
\label{sec:curvature}

The \emph{Riemann curvature tensor}, which has components $R\ind{^{a}_{b c d}}$, is a type-$(1, 3)$ tensor which maps a one-form $\sigma$ and three vectors $w$, $u$ and $v$ into a real number:
\begin{align}
R\ind{^{a}_{b c d}} \sigma\ind{_{a}} w\ind{^{b}} u\ind{^{c}} v\ind{^{d}}
&= \langle \sigma , ( \nabla\ind{_{u}} \nabla\ind{_{v}} - \nabla\ind{_{v}} \nabla\ind{_{u}} - \nabla\ind{_{[u, v]}} ) w \rangle \\
&= \sigma\ind{_{a}} \left[ u\ind{^{c}} ( v\ind{^{d}} w\ind{^{a}_{; d}} )\ind{_{; c}} - v\ind{^{c}} ( u\ind{^{d}} w\ind{^{a}_{; d}} )\ind{_{; c}} - ( u\ind{^{c}} v\ind{^{d}_{; c}} - v\ind{^{c}} u\ind{^{d}_{; c}} ) w\ind{^{a}_{; d}} \right] \\
&= \sigma\ind{_{a}} ( w\ind{^{a}_{; d c}} - w\ind{^{a}_{; c d}} ) u\ind{^{c}} v\ind{^{d}} .
\end{align}
Since $\sigma$, $u$ and $v$ are arbitrary, we arrive at the condition
\begin{equation}
\label{eqn:ricci_identity}
R\ind{^{a}_{b c d}} w\ind{^{b}} = w\ind{^{a}_{; d c}} - w\ind{^{a}_{; c d}},
\end{equation}
which is known as the \emph{Ricci identity}.

One may use the expression for the covariant derivative of a tensor \eqref{eqn:covariant_derivative_tensor} to write the components of the Riemann tensor in terms of a coordinate basis as
\begin{equation}
\label{eqn:riemann_tensor_components}
R\ind{^{a}_{b c d}} = \partial\ind{_{c}} \Gamma\ind{^{a}_{b d}} - \partial\ind{_{d}} \Gamma\ind{^{a}_{b c}} + \Gamma\ind{^{e}_{b d}} \Gamma\ind{^{a}_{e c}} - \Gamma\ind{^{e}_{b c}} \Gamma\ind{^{a}_{e d}} .
\end{equation}
The components $R\ind{_{a b c d}} = g\ind{_{a e}} R\ind{^{e}_{b c d}}$ obey the symmetry relations
\begin{equation}
\label{eqn:riemann_tensor_identities}
R\ind{_{a b c d}} = - R\ind{_{a b d c}} = - R\ind{_{b a c d}}, \qquad R\ind{_{a b c d}} = R\ind{_{c d a b}}, \qquad R\ind{_{a [b c d]}} = 0.
\end{equation}
The last of these is known as the \emph{first (algebraic) Bianchi identity}. The components of the covariant derivative of the Riemann tensor obey the \emph{second (differential) Bianchi identity},
\begin{equation}
\label{eqn:second_bianchi_identity}
R\ind{_{ a b [ c d ; e ] }} = 0.
\end{equation}

The \emph{Ricci curvature tensor} $R\ind{_{a b}}$ is defined as the trace of the Riemann tensor over its first and third indices:
\begin{equation}
R\ind{_{a b}} = R\ind{^{c}_{a c b}} .
\end{equation}
The Ricci tensor satisfies the symmetry property $R\ind{_{a b}} = R\ind{_{b a}}$. The \emph{Ricci scalar} $R$ is defined as the trace of the Ricci tensor:
\begin{equation}
R = R\ind{^{a}_{a}}.
\end{equation}

The trace-free part of the Riemann tensor is the \emph{Weyl tensor}. For $n$-dimensional manifolds with $n \geq 3$, the Weyl tensor has components
\begin{equation}
\label{eqn:weyl_tensor_definition}
C\ind{_{a b c d}} = R\ind{_{a b c d}} - \frac{2}{n - 2} ( g\ind{_{ a [ c }} R\ind{_{ d ] b }} - g\ind{_{ b [ c }} R\ind{_{ d ] a}} ) + \frac{2}{(n - 1)(n - 2)} R g\ind{_{a [ c }} g\ind{_{ b ] d }} .
\end{equation}
The Weyl tensor satisfies the same symmetry properties as the Riemann tensor, and is trace-free on all of its indices.

The \emph{Einstein tensor} is defined as
\begin{equation}
\label{eqn:einstein_tensor}
G\ind{_{a b}} = R\ind{_{a b}} - \frac{1}{2} R g\ind{_{a b}}.
\end{equation}
Clearly, the Einstein tensor is symmetric, i.e., $G\ind{_{a b}} = G\ind{_{b a}}$. Taking the trace of both sides of \eqref{eqn:einstein_tensor}, we see that $G\ind{^{a}_{a}} = \frac{2 - n}{2} R$. In four dimensions, this is $G\ind{^{a}_{a}} = - R$. The Einstein tensor is therefore referred to as the \emph{trace-reversed Ricci tensor}. By contracting the second Bianchi identity \eqref{eqn:second_bianchi_identity}, one may show that the Einstein tensor is divergence-free, i.e.,
\begin{equation}
\nabla\ind{^{a}} G\ind{_{a b}} = 0.
\end{equation}


\section{Geodesics}
\label{sec:geodesics}

We now turn to a topic of central importance to this thesis -- the notion of geodesic motion. Intuitively, a geodesic is a ``straight line'' on a curved manifold. Here, we present a formal definition of a geodesic, derive the geodesic equation, discuss causal geodesics, and present the Lagrangian and Hamiltonian formulations of geodesic motion on curved manifolds. We restrict our attention to the special case of four-dimensional spacetime manifolds $\mathcal{M}$, endowed with a metric $g\ind{_{a b}}$ of Lorentzian signature, although the concepts reviewed here can be generalised to arbitrary $n$-dimensional pseudo-Riemannian manifolds.

\subsection{Geodesic equation}
\label{sec:geodesic_equation}

\subsubsection{Geodesic curves, parallel transport and affine parameters}

On spacetime, a \emph{geodesic} is a curve $\gamma$ whose tangent vector $u\ind{^{a}}$ is parallel-transported along itself, i.e.,
\begin{equation}
\label{eqn:affine_geodesic}
u\ind{^{b}} \nabla\ind{_{b}} u\ind{^{a}} = 0.
\end{equation}
In fact, for a curve on a manifold to be a geodesic, we only require that the parallel-transport of the tangent vector points in the same direction as the tangent vector at each point along the curve; the tangent vector does not necessarily have to be of the same length. This yields the weaker condition
\begin{equation}
\label{eqn:non_affine_geodesic}
u\ind{^{b}} \nabla\ind{_{b}} u\ind{^{a}} = \alpha u\ind{^{a}},
\end{equation}
where $\alpha$ is some arbitrary (smooth) function along the curve. One may show that, given a curve which satisfies \eqref{eqn:non_affine_geodesic}, it is always possible to reparametrise so that it satisfies the stronger condition \eqref{eqn:affine_geodesic}. The latter is referred to as an \emph{affine parametrisation}, and can be used without loss of generality. Throughout this work, we shall define a geodesic to be a curve which satisfies \eqref{eqn:affine_geodesic}.

In some coordinate system, an affinely parametrised geodesic $\gamma(\lambda)$ is a curve $x\ind{^{a}}(\lambda)$. Using \eqref{eqn:parallel_transport_coordinates}, the tangent vector $u\ind{^{a}} = \dot{x}\ind{^{a}}$ to the geodesic satisfies
\begin{equation}
\dot{u}\ind{^{a}} + \Gamma\ind{^{a}_{b c}} u\ind{^{b}} u\ind{^{c}} = 0,
\end{equation}
where an overdot denotes differentiation with respect to the affine parameter $\lambda$. The components of the tangent vector are related to the coordinates by $u\ind{^{a}} = \dot{x}\ind{^{a}}$, so the geodesic equation can be written
\begin{equation}
\label{eqn:geodesic_equation}
\ddot{x}\ind{^{a}} + \Gamma\ind{^{a}_{b c}} \dot{x}\ind{^{b}} \dot{x}\ind{^{c}} = 0.
\end{equation}
This is actually a coupled system of four second-order ordinary differential equations for each of the spacetime coordinates $x\ind{^{a}}(\lambda)$. For every initial value $x\ind{^{a}}(0)$ and $\dot{x}\ind{^{a}}(0)$, a unique solution to this system is guaranteed to exist. In other words, given a point $p \in \mathcal{M}$ and a tangent vector $u\ind{^{a}} \in T_{p} \mathcal{M}$, there always exists a unique geodesic through $p$ with tangent vector $u\ind{^{a}}$.

\subsubsection{Timelike, null and spacelike geodesics}

A geodesic $\gamma$ with tangent $u\ind{^{a}}$ is said to be \emph{timelike} if $g\ind{_{a b}} u\ind{^{a}} u\ind{^{b}} < 0$ everywhere along $\gamma$; it is said to be \emph{null} if $g\ind{_{a b}} u\ind{^{a}} u\ind{^{b}} = 0$ everywhere along $\gamma$; or it is said to be \emph{spacelike} if $g\ind{_{a b}} u\ind{^{a}} u\ind{^{b}} > 0$ everywhere along $\gamma$. The first two types of geodesic are referred to as \emph{causal} geodesics. For timelike geodesics, the \emph{proper time} $\tau$ measured along the curve is defined to be
\begin{equation}
\tau = \int_{\gamma} \sqrt{ - g\ind{_{a b}} u\ind{^{a}} u\ind{^{b}} } \, \ed \lambda.
\end{equation}
Similarly, along a spacelike geodesic, the \emph{proper length} $\ell$ is defined as
\begin{equation}
\ell = \int_{\gamma} \sqrt{g\ind{_{a b}} u\ind{^{a}} u\ind{^{b}} } \, \ed \lambda.
\end{equation}
It is straightforward to show that the proper time and proper length are independent of the choice of parametrisation. For null geodesics, there is no invariant notion of proper time or proper length: a null geodesic can be said to have zero proper time and proper length.

Along an affinely parametrised geodesic with tangent vector $u\ind{^{a}}(\lambda)$, the norm $g\ind{_{a b}} u\ind{^{a}} u\ind{^{b}} = u\ind{^{a}} u\ind{_{a}}$ is constant:
\begin{equation}
\frac{\ed}{\ed \lambda} \left( u\ind{^{a}} u\ind{_{a}} \right) = u\ind{^{b}} \left( u\ind{^{a}} u\ind{_{a}} \right)\ind{_{; b}} = u\ind{^{b}} u\ind{^{a}_{; b}} u\ind{_{a}} + u\ind{^{a}} u\ind{^{b}} u\ind{_{a ; b}} = 0.
\end{equation}
For null geodesics, the norm is always zero. For timelike (spacelike) geodesics which are parametrised by proper time (proper length), the norm is $- 1$ ($+ 1$).

In general relativity, a freely falling \emph{massive} particle follows a timelike geodesic, whereas a freely propagating \emph{massless} particle (e.g.~a photon) follows a null geodesic. The terms null geodesic, lightlike geodesic and light ray are used interchangeably.

\subsubsection{Geodesic completeness}

A geodesic $x\ind{^{a}}(\lambda)$ for which the affine parameter $\lambda$ ranges over all of $\mathbb{R}$ is said to be \emph{complete}. A geodesic that is not complete is called \emph{incomplete}. A spacetime $(\mathcal{M}, g)$ is said to be \emph{geodesically complete} if every geodesic can be extended into a complete geodesic. A spacetime $(\mathcal{M}, g)$ is said to be \emph{timelike} (respectively \emph{null} or \emph{spacelike}) \emph{geodesically complete} if every \emph{timelike} (respectively \emph{null} or \emph{spacelike}) geodesic can be extended into a complete one.

The notion of geodesic (in)completeness is central to the definition of singularities in general relativity. If a spacetime $(\mathcal{M}, g)$ is geodesically incomplete, and it is not possible to extend $(\mathcal{M}, g)$ into a larger spacetime that is geodesically complete, we say that $(\mathcal{M}, g)$ is \emph{singular}. If any scalar quantity which is polynomial in $R\ind{^{a}_{b c d}}$ or its covariant derivatives diverges along an incomplete geodesic, the spacetime is singular. In this case, the spacetime is said to have a \emph{scalar curvature singularity}. Alternatively, a component of $R\ind{^{a}_{b c d}}$ or its covariant derivative(s) in a parallel-transported tetrad (see Section \ref{sec:newman_penrose_tetrads}) may blow up; this is referred to as a parallel-propagated curvature singularity \cite{Wald1984}.

\subsubsection{Conformal transformations}

Consider a conformal transformation of the metric $g\ind{_{a b}} \mapsto \tilde{g}\ind{_{a b}} = \Omega^{2} g\ind{_{a b}}$, where $\Omega > 0$ is a smooth function of the spacetime coordinates. Clearly, a conformal transformation preserves causal structure; that is, if a vector is a timelike, null or spacelike with respect to the metric $g\ind{_{a b}}$, then it has the same property with respect to the conformally related metric $\tilde{g}\ind{_{a b}}$.

Let us consider the geodesics of the conformally related spacetimes $(\mathcal{M}, g\ind{_{a b}})$ and $(\mathcal{M}, \tilde{g}\ind{_{a b}})$. Using the properties of the covariant derivative (Section \ref{sec:covariant_differentiation_parallel_transport}), one can show that, for an arbitrary vector field $v\ind{^{a}}$, the covariant derivatives $\nabla\ind{_{a}}$ and $\tilde{\nabla}\ind{_{a}}$ are related by \cite{Wald1984}
\begin{equation}
\label{eqn:cov_der_vector_conformally_related}
\tilde{\nabla}\ind{_{a}} v\ind{^{b}} = \nabla\ind{_{a}} v\ind{^{b}} + D\ind{^{b}_{a c}} v\ind{^{c}},
\end{equation}
where the \emph{connection coefficients} are given by
\begin{equation}
\label{eqn:general_connection_coefficients}
D\ind{^{a}_{b c}} = \frac{1}{2} \tilde{g}\ind{^{a d}} \left( \nabla\ind{_{b}} \tilde{g}\ind{_{d c}} + \nabla\ind{_{c}} \tilde{g}\ind{_{b d}} - \nabla\ind{_{d}} \tilde{g}\ind{_{b c}} \right).
\end{equation}
This is a generalisation of the relationship between the covariant derivative operator and the partial derivative operator, where the connection coefficients are simply the Christoffel symbols \eqref{eqn:christoffel_symbols}. Using the Leibniz rule and metric compatibility of $\nabla\ind{_{a}}$, we find
\begin{equation}
\nabla\ind{_{a}} \tilde{g}\ind{_{b c}} = \nabla\ind{_{a}} ( \Omega^{2} g\ind{_{b c}} ) = 2 \Omega g\ind{_{b c}} \nabla\ind{_{a}} \Omega + \Omega^{2} \nabla\ind{_{a}} g\ind{_{b c}} = 2 \Omega g\ind{_{b c}} \nabla\ind{_{a}} \Omega .
\end{equation}
The connection coefficients \eqref{eqn:general_connection_coefficients} can therefore be expressed in terms of $g\ind{_{a b}}$ and $\Omega$ as
\begin{equation}
\label{eqn:connection_coefficients_conformally_related}
D\ind{^{a}_{b c}} = 2 \delta\ind{^{a}_{ ( b }} \nabla\ind{_{c )}} \ln{\Omega} - g\ind{_{b c}} g\ind{^{a d}} \nabla\ind{_{d}} \ln{\Omega}.
\end{equation}

Let $\gamma(\lambda)$ be a geodesic of the spacetime $(\mathcal{M}, g\ind{_{a b}})$ with tangent vector $u\ind{^{a}}$. Using \eqref{eqn:cov_der_vector_conformally_related} and \eqref{eqn:connection_coefficients_conformally_related}, we obtain
\begin{equation}
\label{eqn:geodesics_conformally_related}
u\ind{^{b}} \tilde{\nabla}\ind{_{b}} u\ind{^{a}}
= u\ind{^{b}} \nabla\ind{_{b}} u\ind{^{a}} + D\ind{^{a}_{b c}} u\ind{^{b}} u\ind{^{c}}
= 2 \left( u\ind{^{b}} \nabla\ind{_{b}} \ln{\Omega} \right) u\ind{^{a}} - \left( g\ind{_{b c}} u\ind{^{b}} u\ind{^{c}} \right) g\ind{^{a d}} \nabla\ind{_{d}} \ln{\Omega} .
\end{equation}
In general, we see that $\gamma$ is \emph{not} a geodesic with respect to the metric $\tilde{g}\ind{_{a b}}$; however, if $\gamma$ is a \emph{null} geodesic, then the term $g\ind{_{b c}} u\ind{^{b}} u\ind{^{c}}$ on the right-hand side of \eqref{eqn:geodesics_conformally_related} will vanish. This leaves us with $u\ind{^{b}} \tilde{\nabla}\ind{_{b}} u\ind{^{a}} = 2 \left( u\ind{^{b}} \nabla\ind{_{b}} \ln{\Omega} \right) u\ind{^{a}}$, which is nothing more than the equation for a non-affinely-parametrised geodesic \eqref{eqn:non_affine_geodesic}. This shows that null geodesics are conformally invariant.
%

\subsection{Lagrangian formulation}
\label{sec:lagrangian_formulation_geodesics}

One may derive the geodesic equation \eqref{eqn:geodesic_equation} using the calculus of variations and Hamilton's principle of least action \cite{Arnold1989}. Our starting point is the \emph{action functional}, which is taken to be the spacetime distance between two fixed endpoints,
\begin{equation}
\label{eqn:action_functional_lagrangian}
S[x\ind{^{a}}(\lambda)] = \int L^{\prime} \, \ed \lambda , \qquad L^{\prime} = \sqrt{ \left| g\ind{_{a b}} \dot{x}\ind{^{a}} \dot{x}\ind{^{b}} \right| } ,
\end{equation}
where $L^{\prime}$ is the \emph{Lagrangian function}, and an overdot denotes differentiation with respect to the parameter $\lambda$. This definition is somewhat degenerate, as the action is invariant under arbitrary reparametrisations of the curve. The requirement that the action functional \eqref{eqn:action_functional_lagrangian} is stationary (i.e., $\delta S = 0$) yields the Euler--Lagrange equations of motion
\begin{equation}
\label{eqn:euler_lagrange_equations_geodesic}
\frac{\ed}{\ed \lambda} \left( \frac{\partial L^{\prime}}{\partial \dot{x}\ind{^{a}}} \right) - \frac{\partial L^{\prime}}{\partial x\ind{^{a}}} = 0.
\end{equation}
Inserting the Lagrangian $L^{\prime} = \sqrt{ \left| g\ind{_{a b}} \dot{x}\ind{^{a}} \dot{x}\ind{^{b}} \right| }$ into the Euler--Lagrange equations gives the equation for non-affinely parametrised geodesics \eqref{eqn:non_affine_geodesic}. Demanding that the curve be affinely parametrised, we arrive at the geodesic equation \eqref{eqn:geodesic_equation}. Geodesics are thus spacetime paths $x\ind{^{a}}(\lambda)$ which extremise the action \eqref{eqn:action_functional_lagrangian}.

The Lagrangian function \eqref{eqn:action_functional_lagrangian} is not unique. In fact, any function which yields the same Euler--Lagrange equations of motion is valid. For example, the geodesic equation \eqref{eqn:geodesic_equation} also follows from extremising the action
\begin{equation}
\label{eqn:action_functional_lagrangian_square}
S[x\ind{^{a}}(\lambda)] = \int L \, \ed \lambda , \qquad L = \frac{1}{2} g\ind{_{a b}} \dot{x}\ind{^{a}} \dot{x}\ind{^{b}} .
\end{equation}
Unlike \eqref{eqn:action_functional_lagrangian}, the action \eqref{eqn:action_functional_lagrangian_square} is extremal \emph{only} for affinely parametrised curves. We favour the action \eqref{eqn:action_functional_lagrangian_square} for two reasons: (i) there is no degeneracy associated with reparametrisations of the curve; (ii) the Lagrangian is more tractable, as it does not feature a square root. We note here that the Lagrangian itself is constant along geodesics, with $L = 0$ in the null case, which will be of particular interest in this work.

The flow determined by the Euler--Lagrange equations \eqref{eqn:euler_lagrange_equations_geodesic} with Lagrangian \eqref{eqn:action_functional_lagrangian} or \eqref{eqn:action_functional_lagrangian_square} is called the \emph{geodesic flow}. The geodesics are then the projections of the integral curves of the geodesic flow onto the manifold $\mathcal{M}$. Physically speaking, the geodesic flow governs the motion of a particle that is not subject to any external forces, therefore moving freely on the manifold $\mathcal{M}$.

In the Lagrangian picture, the geodesic flow is a dynamical system on the tangent bundle $T \mathcal{M}$; for each value of the parameter $\lambda$, the system is completely determined in local coordinates by the position vector $x\ind{^{a}}(\lambda)$ and the tangent vector (or \emph{velocity}) $\dot{x}\ind{^{a}}(\lambda)$. The cotangent bundle is interpreted as the \emph{configuration space} of Lagrangian mechanics.
%

\subsection{Hamiltonian formulation}
\label{sec:hamiltonian_formulation_geodesics}

With a change of perspective, the geodesic flow can be viewed as a dynamical system on the cotangent bundle $T^{\ast} \mathcal{M}$, where the state of the system is determined in local coordinates by the position vector $x\ind{^{a}}(\lambda)$ and the momentum one-form $p\ind{_{a}}(\lambda)$; this is sometimes referred to as the \emph{cogeodesic flow}. In this picture, the geodesic flow becomes a Hamiltonian flow on \emph{phase space}, where the latter is the cotangent bundle. One may pass from the Lagrangian formulation to the Hamiltonian formulation using the following standard procedure.

First, one may write down the \emph{Hamiltonian function} $H$ (which is the Legendre transform of the Lagrangian function $L$), and the \emph{conjugate momenta} $p\ind{_{a}}$; these take the form
\begin{equation}
H = \dot{x}\ind{^{a}} p\ind{_{a}} - L, \qquad p\ind{_{a}} = \frac{\partial L}{\partial \dot{x}\ind{^{a}}},
\end{equation}
where it is understood that the velocities $\dot{x}\ind{^{a}}$ are to be replaced by the momenta $p\ind{_{a}}$ in the definition of the Hamiltonian $H$. Using the Lagrangian \eqref{eqn:action_functional_lagrangian_square}, the Hamiltonian for geodesics is given by
\begin{equation}
\label{eqn:geoesic_hamiltonian}
H = \frac{1}{2} g\ind{^{a b}} p\ind{_{a}} p\ind{_{b}} ,
\end{equation}
where the conjugate momenta are
\begin{equation}
\label{eqn:geodesics_conjugate_momenta}
p\ind{_{a}} = g\ind{_{a b}} \dot{x}\ind{^{a}} .
\end{equation}
Clearly the momenta $p\ind{_{a}}$ are the covariant components of the tangent vector to the geodesic. Moreover, the Hamiltonian \eqref{eqn:geoesic_hamiltonian} is manifestly covariant.

A standard variational approach yields \emph{Hamilton's equations}, a system of $2 n$ coupled first-order ordinary differential equations for each of the $n$ spacetime coordinates $x\ind{^{a}}$ and their conjugate momenta $p\ind{_{a}}$, given by
\begin{align}
\label{eqn:geoesic_hamiltons_equations}
\dot{x}\ind{^{a}} = \frac{\partial H}{\partial p\ind{_{a}}}, \qquad \dot{p}\ind{_{a}} = - \frac{\partial H}{\partial x\ind{^{a}}}.
\end{align}
Using the Hamiltonian \eqref{eqn:geoesic_hamiltonian}, Hamilton's equations \eqref{eqn:geoesic_hamiltons_equations} become
\begin{equation}
\label{eqn:geoesic_hamiltons_equations_2}
\dot{x}\ind{^{a}} = g\ind{^{a b}} p\ind{_{b}}, \qquad \dot{p}\ind{_{a}} = - \frac{1}{2} g\ind{^{b c}_{, a}} p\ind{_{b}} p\ind{_{c}}.
\end{equation}
The first of these is clearly equivalent to the definition of the conjugate momenta \eqref{eqn:geodesics_conjugate_momenta}. The equations \eqref{eqn:geoesic_hamiltons_equations_2} may be combined to give the familiar geodesic equation $p\ind{^{b}} \nabla\ind{_{b}} p\ind{_{a}} = 0$.

We remark here that the Hamiltonian function \eqref{eqn:geoesic_hamiltonian} is conserved along geodesics. In the case of timelike geodesics, i.e., the paths of freely falling massive particles, we have $H = - \frac{1}{2} m^{2}$, where $m$ is the mass of the particle. In the null case, we see that the Hamiltonian vanishes along null rays. In general, the constancy of the Hamiltonian is referred to as the \emph{Hamiltonian constraint}; for null geodesics, the constraint $H = 0$ is referred to as the \emph{null condition}.

In Section \ref{sec:killing_vectors}, we saw that continuous spacetime symmetries are encoded by Killing vectors. These spacetime symmetries can always be ``lifted up'' to symmetries on phase space. In Section \ref{sec:killing_objects_symmetries_integrability}, we will discuss the relationship between Killing vectors (or, more generally, higher-rank Killing tensors) and conserved quantities on phase space.
%
%

\subsection{Geodesic deviation equation}
\label{sec:geodesic_deviation_equation}

Our final task for this section is to consider the behaviour of a \emph{congruence} (or \emph{family}) of non-intersecting geodesics. Let $x\ind{^{a}}(\lambda, \sigma)$ be a congruence of geodesics, where $\lambda$ is an affine parameter along each curve, and $\sigma$ parametrises the congruence, i.e., each geodesic is distinguished by the value of the parameter $\sigma$. The tangent vector field is given by
\begin{equation}
u\ind{^{a}} = \frac{\partial x\ind{^{a}}}{\partial \lambda},
\end{equation}
which satisfies the geodesic equation $u\ind{^{b}} \nabla\ind{_{b}} u\ind{^{a}} = 0$. We define the \emph{separation vector} (or \emph{deviation vector}) between neighbouring rays in the congruence to be
\begin{equation}
\xi\ind{^{a}} = \frac{\partial x\ind{^{a}}}{\partial \sigma} .
\end{equation}

It is straightforward to show, using the symmetry of mixed partial derivatives, that $\xi$ is Lie-transported along each geodesic in the congruence:
\begin{equation}
\label{eqn:lie_derivative_separation_tangent}
\mathcal{L}_{u} \xi\ind{^{a}} = [ u, \xi ]\ind{^{a}} = u\ind{^{b}} \partial\ind{_{b}} \xi\ind{^{a}} - \xi\ind{^{b}} \partial\ind{_{b}} u\ind{^{a}} = \frac{\partial^{2} x\ind{^{a}}}{\partial \lambda \partial \sigma} - \frac{\partial^{2} x\ind{^{a}}}{\partial \sigma \partial \lambda} = 0.
\end{equation}
It follows from the torsion-free property of the Christoffel symbols ($\Gamma\ind{^{a}_{[b c]}} = 0$) that the partial derivatives in \eqref{eqn:lie_derivative_separation_tangent} may be replaced by covariant derivatives. We then find
\begin{equation}
\label{eqn:parallel_transport_deviation_tangent}
u\ind{^{b}} \nabla\ind{_{b}} \xi\ind{^{a}} = \xi\ind{^{b}} \nabla\ind{_{b}} u\ind{^{a}},
\end{equation}
which says that the parallel transport of $\xi$ along $u$ is equivalent to the parallel transport of $u$ along $\xi$. A consequence of this result is that the quantity $\xi\ind{^{a}} u\ind{_{a}}$ is conserved along each geodesic, i.e., $\frac{\ed}{\ed \lambda} ( \xi\ind{^{a}} u\ind{_{a}} ) = 0$.

Recall that the commutator of covariant derivatives acting on a vector field satisfies
\begin{equation}
\nabla\ind{_{a}} \nabla\ind{_{b}} u\ind{^{c}} - \nabla\ind{_{b}} \nabla\ind{_{a}} u\ind{^{c}} = - R\ind{^{c}_{d a b}} u\ind{^{d}},
\end{equation}
where $R\ind{^{a}_{b c d}}$ are the components of the Riemann curvature tensor. Applying the directional derivative operator $u\ind{^{c}} \nabla\ind{_{c}}$ to both sides of \eqref{eqn:parallel_transport_deviation_tangent} and using the above results, we see that the acceleration of the deviation vector $\xi\ind{^{a}}$ is given by
\begin{equation}
\frac{\eD^{2} \xi\ind{^{a}}}{\ed \lambda^{2}} = R\ind{^{a}_{b c d}} u\ind{^{b}} u\ind{^{c}} \xi\ind{^{d}},
\end{equation}
where $\frac{\eD}{\ed \lambda} = u\ind{^{a}} \nabla\ind{_{a}}$ denotes the covariant derivative along $u\ind{^{a}}$. This equation is known as the \emph{geodesic deviation equation} (or \emph{Jacobi equation}). It describes how spacetime curvature (i.e., the Riemann curvature term on the right-hand side) is responsible for the relative acceleration of neighbouring geodesics in a congruence, even if the rays are initially parallel.
%

\section{Field equations and solutions}

\subsection{Einstein field equations}

The cornerstone of Einstein's theory of general relativity is Einstein's field equation, which couples the (dynamical) spacetime background to the matter fields present in spacetime. The field equations may be derived using a principle of least action. Our starting point is the action functional
\begin{equation}
\label{eqn:einstein_hilbert_action}
S =\int \left( \mathscr{L}_{\text{EH}} + \mathscr{L}_{\text{M}} \right) \sqrt{- \det{g}} \, \ed^{4} x, \qquad \mathscr{L}_{\text{EH}} = \frac{1}{16 \pi} R,
\end{equation}
where $\mathscr{L}_{\text{EH}}$ is the \emph{Einstein--Hilbert Lagrangian} for the gravitational field; $\mathscr{L}_\text{M}$ is the \emph{matter Lagrangian}, which contains information about the matter fields present in spacetime; $\det{g}$ is the metric determinant; and $R$ is the Ricci curvature scalar. Variation of the action \eqref{eqn:einstein_hilbert_action} yields the \emph{Einstein field equation},
\begin{equation}
\label{eqn:einstein_field_equations}
G\ind{_{a b}} = 8 \pi T \ind{_{a b}},
\end{equation}
where $G\ind{_{a b}} = R\ind{_{a b}} - \frac{1}{2} R g\ind{_{a b}}$ is the Einstein tensor (see Section \ref{sec:curvature}), and $T\ind{_{a b}}$ is the \emph{stress--energy tensor}. Here, we outline the key properties of the field equation \eqref{eqn:einstein_field_equations}. Firstly, the field equation is tensorial; in fact, there are ten independent components, due to the fact that the Einstein tensor and the stress--energy tensor are symmetric. We will often refer to \eqref{eqn:einstein_field_equations} as the Einstein field equations (using the plural) to reflect this. Secondly, the field equations constitute a system of coupled second-order differential equations for the components of the metric tensor $g\ind{_{a b}}$. Thirdly, the fact that the Einstein tensor is divergence-free (see Section \ref{sec:curvature}) means that the stress--energy tensor also satisfies the same property, i.e., $\nabla\ind{^{a}} T\ind{_{a b}} = 0$. Finally, the constant of proportionality between the Einstein tensor and stress--energy tensor in \eqref{eqn:einstein_field_equations} -- which in non-geometrised units is $\frac{8 \pi G}{c^{2}}$, where $G$ is Newton's gravitational constant and $c$ is the speed of light \emph{in vacuo} -- is chosen so that we recover Newton's theory of gravitation in the limit of slowly varying and weak gravitational fields.

Taking the trace of \eqref{eqn:einstein_field_equations} with respect to the metric gives $-R = 8 \pi T$, where $T = T\ind{^{a}_{a}}$ denotes the trace of the stress--energy tensor. Replacing the term involving $R$ in \eqref{eqn:einstein_field_equations}, one obtains the \emph{trace-reversed} form of the Einstein field equations,
\begin{equation}
\label{eqn:trace_reversed_einstein_field_equations}
R\ind{_{a b}} = 8 \pi \left( T \ind{_{a b}} - \frac{1}{2} T g\ind{_{a b}} \right).
\end{equation}

In pure vacuum, the stress--energy tensor vanishes, i.e., $T\ind{_{a b}} = 0$. It is straightforward to see from \eqref{eqn:einstein_field_equations} that the vacuum field equations may be expressed in the form $G\ind{_{a b}} = 0$. Equivalently, by setting $T\ind{_{a b}} = 0$ in the trace-reversed field equations \eqref{eqn:trace_reversed_einstein_field_equations}, the vacuum field equations can be written in the form $R\ind{_{a b}} = 0$. Solutions to the vacuum field equations of general relativity are referred to as \emph{vacuum solutions}. The simplest example of such a solution is Minkowski spacetime (see Section \ref{sec:minkowski_spacetime}). The Schwarzschild and Kerr black hole spacetimes are also examples of vacuum solutions (see Section \ref{sec:black_hole_spacetimes}). In general, manifolds with a vanishing Ricci tensor ($R\ind{_{a b}} = 0$) are referred to as \emph{Ricci-flat} manifolds.

\subsection{Einstein--Maxwell field equations}
\label{sec:einstein_maxwell_field_equations}

The Einstein--Maxwell field equations of gravitation and electromagnetism may be obtained with the inclusion of the free-field Lagrangian $\mathscr{L}_{\text{EM}} = - \frac{1}{4} F\ind{_{a b}} F\ind{^{a b}}$ in the action \eqref{eqn:einstein_hilbert_action}, where $F\ind{_{a b}}$ is the \emph{Faraday tensor}, an antisymmetric rank-two tensor (i.e., a two-form) which encodes the electromagnetic field. The Euler--Lagrange equations for the new action yield the \emph{Einstein--Maxwell equations}
\begin{equation}
\label{eqn:einstein_maxwell_equations_1}
G\ind{_{a b}} = 8 \pi T \ind{_{a b}} , \qquad T\ind{_{a b}} = \frac{1}{4 \pi} \left( F\ind{_{a c}} F\ind{_{b}^{c}} - \frac{1}{4} g\ind{_{a b}} F\ind{^{c d}} F\ind{_{c d}}  \right) ,
\end{equation}
\begin{equation}
\label{eqn:einstein_maxwell_equations_2}
\nabla\ind{_{b}} F\ind{^{a b}} = \mu_{0} J\ind{^{a}} , \qquad \nabla\ind{_{b}} \vp{F}^{\star} F\ind{^{a b}} = 0 ,
\end{equation}
where $\vp{F}^{\star} F\ind{^{a b}}$ is the Hodge dual of the Faraday tensor; and $J\ind{^{a}}$ is the four-current, which is divergence-free ($\nabla\ind{_{a}} J\ind{^{a}} = 0$). In the absence of sources ($J\ind{^{a}} = 0$), the first of Maxwell's equations \eqref{eqn:einstein_maxwell_equations_2} is simply $\nabla\ind{_{b}} F\ind{^{a b}} = 0$.

In the language of differential forms, the Faraday two-form is \emph{closed}, i.e., $\ed F = 0$, so $F$ must be locally exact ($F = \ed A$ for some one-form potential $A$) by Poincar\'{e}'s lemma. Hence, the components of the Faraday tensor can be expressed in terms of the one-form potential $A\ind{_{a}}$ as
\begin{equation}
F\ind{_{a b}} = \nabla\ind{_{a}} A\ind{_{b}} - \nabla\ind{_{b}} A\ind{_{a}}.
\end{equation}

Taking the trace of the electromagnetic stress--energy tensor \eqref{eqn:einstein_maxwell_equations_1}, it is clear that $T = T\ind{^{a}_{a}} = 0$. Using the trace-reversed field equations \eqref{eqn:trace_reversed_einstein_field_equations}, we see that
\begin{equation}
R\ind{_{a b}} = 8 \pi T\ind{_{a b}} = 2 \left( F\ind{_{a c}} F\ind{_{b}^{c}} - \frac{1}{4} g\ind{_{a b}} F\ind{^{c d}} F\ind{_{c d}} \right)
\end{equation}
for pure electromagnetism.


\subsection{Stationary axisymmetric solutions}
\label{sec:stationary_axisymmetric_solutions}

A solution to the Einstein--Maxwell equations which will be of central importance to this thesis is that of a rotating source with a stationary, axially symmetric gravitational field. The most general such solution is the \emph{Weyl--Lewis--Papapetrou geometry}, which is described in Weyl--Lewis--Papapetrou coordinates $\{ t, \rho, z, \phi \}$ by the line element
\begin{equation}
\label{eqn:weyl_solution_line_element}
\ed s^{2} = g\ind{_{a b}} \ed x\ind{^{a}} \ed x\ind{^{b}} = - f \left(\ed t + w \, \ed \phi \right)^{2} + \frac{1}{f} \left[ e^{2 \gamma} \left( \ed \rho^{2} + \ed z^{2} \right) + \rho^{2} \, \ed \phi^{2} \right],
\end{equation}
and electromagnetic one-form potential
\begin{equation}
\label{eqn:weyl_solution_four_potential}
A\ind{_{a}} \ed x\ind{^{a}} = A\ind{_{t}} \ed t + A\ind{_{\phi}} \ed \phi,
\end{equation}
where $f$, $\gamma$, $w$, $A\ind{_{t}}$ and $A\ind{_{\phi}}$ are functions of the coordinates $\rho$ and $z$ only.

The metric represents a stationary, axisymmetric solution as the metric \eqref{eqn:weyl_solution_line_element} admits a timelike Killing vector $\partial\ind{_{t}}$ and a spacelike Killing vector $\partial\ind{_{\phi}}$, and the coordinate $\phi$ is taken to be periodic with $\phi \in [ 0, 2\pi )$. The function $w$ which appears in \eqref{eqn:weyl_solution_line_element} represents a rotation about the symmetry axis $\rho = 0$. In the case $w = 0$, we recover the so-called \emph{Weyl solution} for a \emph{static} axisymmetric gravitational field. In pure vacuum, the one-form potential \eqref{eqn:weyl_solution_four_potential} vanishes.

For the Weyl--Lewis--Papapetrou metric \eqref{eqn:weyl_solution_line_element} with one-form potential \eqref{eqn:weyl_solution_four_potential}, the electrovacuum field equations are non-linear, but are known to be integrable nevertheless. One may use the Ernst formulation in terms of complex potentials to solve the field equations \cite{Ernst1968, StephaniKramerMacCallumEtAl2003, GriffithsPodolsky2009}.
%

\section{Black holes}
\label{sec:black_holes}

\subsection{Minkowski spacetime and asymptotic flatness}
\label{sec:minkowski_spacetime}

The simplest solution to Einstein's vacuum field equations is \emph{Minkowski spacetime} (or \emph{flat spacetime}), which is described in Cartesian coordinates $\{ t, x, y, z \}$ by the line element
\begin{equation}
\label{eqn:minkowski_line_element}
\ed s^{2} = \eta\ind{_{a b}} \ed x\ind{^{a}} \ed x\ind{^{b}} = - \ed t^{2} + \ed x^{2} + \ed y^{2} + \ed z^{2},
\end{equation}
where we use the standard notation $\eta\ind{_{a b}}$ to denote the components of the Minkowski metric.

We may use the Minkowski metric to provide a coordinate-dependent definition of asymptotic flatness. A spacetime $(\mathcal{M}, g\ind{_{a b}})$ is \emph{asymptotically flat} if there exist some coordinates $\{ t, x, y, z \}$ such that the metric can be expressed in the form $g\ind{_{a b}} = \eta\ind{_{a b}} + h\ind{_{a b}}$, with $\lim_{r \rightarrow \infty} h\ind{_{a b}} = O \left(\frac{1}{r} \right)$, where $r^{2} = x^{2} + y^{2} + z^{z}$. For a more rigorous, coordinate-invariant definition of asymptotic flatness, we refer the reader to Wald \cite{Wald1984}.

\subsection{Event horizon}
\label{sec:event_horizon}

Consider a spacetime $(\mathcal{M}, g\ind{_{a b}})$. The infinities of this spacetime are identified as follows. \emph{Future} (\emph{past}) \emph{null infinity}, denoted $\mathscr{I}^{+}$ ($\mathscr{I}^{-}$), is the future (past) infinity of all null curves. \emph{Future} (\emph{past}) \emph{timelike infinity}, denoted $i^{+}$ ($i^{-}$), is the future (past) infinity of all timelike curves. \emph{Spacelike infinity}, denoted $i^{0}$, is the future (past) infinity of all spacelike curves. For a more detailed description of these infinities, and the notion of a \emph{conformal boundary}, see \cite{Wald1984}.

To define a black hole, we consider a spacetime which has a connected future null infinity $\mathscr{I}^{+}$. If $\mathscr{I}^{+}$ is not connected, it is sufficient to consider its connected components separately. The \emph{causal past} of $\mathscr{I}^{+}$, denoted $I^{-} (\mathscr{I}^{+})$, is the region of $\mathcal{M}$ from which there exist causal curves which reach $\mathscr{I}^{+}$. A spacetime $(\mathcal{M}, g)$ contains a \emph{black hole} if $I^{-} (\mathscr{I}^{+}) \neq \mathcal{M}$. The black hole region is then defined as $\mathcal{B} = \mathcal{M} \setminus I^{-} (\mathscr{I}^{+})$. The \emph{event horizon} of the black hole is the boundary of the black hole: $\mathcal{H} = \partial \mathcal{B}$. (Analogous definitions for a white hole and a white hole horizon can be formulated by time inversion, i.e., replacing future null infinity with past null infinity, and the causal future with the causal past.)

\subsection{Black hole spacetimes}
\label{sec:black_hole_spacetimes}

\subsubsection{Schwarzschild solution}

The first exact solution to Einstein's vacuum field equations was uncovered by Schwarzschild in 1916 \cite{Schwarzschild1916}. The solution describes the exterior spacetime of a spherically symmetric non-rotating gravitational source.

In spherical coordinates $\{ t, r, \theta, \phi \}$, the Schwarzschild solution is described by the line element
\begin{equation}
\label{eqn:schwarzschild_line_element}
\ed s^{2} = g\ind{_{a b}} \ed x\ind{^{a}} \ed x\ind{^{b}} = - f(r) \ed t^{2} + \frac{\ed r^{2}}{f(r)} + r^{2} \left( \ed \theta^{2} + \sin^{2}{\theta} \, \ed \phi^{2} \right), \qquad f(r) = 1 - \frac{2 M}{r}.
\end{equation}
The parameter $M$, which is interpreted as the mass of the source, completely determines the solution. Far from the source ($r \rightarrow \infty$) the line element \eqref{eqn:schwarzschild_line_element} approaches the Minkowski line element expressed in standard spherical polar coordinates. Note that, in the case $M = 0$, the Schwarzschild spacetime reduces to that of Minkowski.

The domain $r > 2 M$ is referred to as the \emph{exterior region}. The \emph{black hole region} of Schwarzschild spacetime is given by $0 < r \leq 2M$; the \emph{event horizon} of the black hole is the null surface $r = 2M$. From the form of the metric \eqref{eqn:schwarzschild_line_element}, it would appear that $r = 2 M$ is a singularity of the spacetime; however, this can be removed by transforming to an alternative coordinate system (e.g.~Kruskal--Szekeres or Eddington--Finkelstein). Thus, $r = 2 M$ is only a coordinate singularity. The Schwarzschild solution does, however, admit a physical singularity at $r = 0$. To see that this is the case, one may compute the Kretschmann invariant, which is given by $R\ind{_{a b c d}} R\ind{^{a b c d}} = \frac{48 M^{2}}{r^{6}}$ in spherical coordinates. Clearly, this quantity diverges as $r \rightarrow 0$.

The metric \eqref{eqn:schwarzschild_line_element} is independent of $t$, so $\partial\ind{_{t}}$ is a Killing vector. This is timelike for $r > 2 M$, so the exterior Schwarzschild spacetime is (at least) stationary. In fact, $\partial\ind{_{t}}$ is hypersurface-orthogonal, so the exterior Schwarzschild spacetime is \emph{static}. The metric \eqref{eqn:schwarzschild_line_element} is also independent of $\phi$, so $\partial\ind{_{\phi}}$ is a Killing vector. This is spacelike in the exterior region. There are two further spacelike Killing vectors, given by $\sin{\phi} \, \partial\ind{_{\theta}} + \cot{\theta} \cos{\phi} \, \partial\ind{_{\phi}}$ and $\cos{\phi} \, \partial\ind{_{\theta}} - \cot{\theta} \sin{\phi} \, \partial\ind{_{\phi}}$. The three spacelike Killing vectors comprise the $\mathrm{SO}(3)$ rotations; they are precisely the Killing vectors admitted by the unit two-sphere $S^{2}$. The Schwarzschild spacetime is therefore \emph{spherically symmetric}.

Birkhoff \cite{BirkhoffLanger1923} demonstrated that the Schwarzschild solution \eqref{eqn:schwarzschild_line_element} is the \emph{unique} spherically symmetric solution to Einstein's field equations in vacuum; this statement is known as \emph{Birkhoff's theorem}.

\subsubsection{Kerr solution}

In 1963, almost half a century after the development of Einstein's field equations, Kerr discovered a solution to the vacuum field equations which describes a rotating black hole \cite{Kerr1963}. The exterior spacetime geometry may be described in Boyer--Lindquist coordinates $\{ t, r, \theta, \phi \}$ \cite{BoyerLindquist1967} by the line element
\begin{align}
\ed s^{2} &= g\ind{_{a b}} \ed x\ind{^{a}} \ed x\ind{^{b}} \\
\begin{split}
&= - \left( 1 - \frac{2 M r}{\Sigma} \right) \ed t^{2} + \frac{\Sigma}{\Delta} \, \ed r^{2} + \Sigma \, \ed \theta^{2} + \left(r^{2} + a^{2} + \frac{2 M a^{2} r}{\Sigma} \sin^{2} \theta\right) \sin^{2} \theta \, \ed \phi^{2} \\ & \qquad - \frac{4 M a r \sin^{2} \theta}{\Sigma} \, \ed t \, \ed \phi,
\label{eqn:kerr_line_element_boyer_lindquist}
\end{split}
\end{align}
where $\Sigma(r, \theta) = r^{2} + a^{2} \cos^{2}{\theta}$ and $\Delta(r) = r^{2} - 2 M r + a^{2}$. Here, $M$ and $J = a M$ are the mass and angular momentum of the black hole, respectively. The interpretation of these parameters is best understood by considering the asymptotic form of the metric \eqref{eqn:kerr_line_element_boyer_lindquist}, as shown in \cite{WiltshireVisserScott2009}.

The Kerr spacetime has two event horizons at $r_{\pm} = M \pm \sqrt{M^{2} - a^{2} \cos^{2}{\theta}}$. There exist two \emph{ergosurfaces} at $r^{\textrm{ergo}}_{\pm} = M \pm \sqrt{M^{2} - a^{2} \cos^{2} \theta}$. There is a physical curvature singularity (at which $R\ind{_{a b c d}} R\ind{^{a b c d}} \rightarrow \infty$) at $r = 0$ and $\theta = \frac{\pi}{2}$; this is in fact a ring singularity, which is best seen by transforming to Kerr--Schild coordinates \cite{Kerr1965}. For a comprehensive discussion of the key features of the Kerr spacetime, see e.g.~\cite{Chandrasekhar1989, WiltshireVisserScott2009}.

Clearly, the metric \eqref{eqn:kerr_line_element_boyer_lindquist} is independent of the coordinates $t$ and $\phi$. In Boyer--Lindquist coordinates, the vectors $\xi_{(t)} = \partial\ind{_{t}}$ and $\xi_{(\phi)} = \partial\ind{_{\phi}}$ are therefore Killing vectors, which satisfy $\xi\ind{_{(a ; b)}} = 0$. These vectors correspond to time translation and rotation about the symmetry axis, respectively. In addition, any constant coefficient linear combination of these two Killing vectors is also a Killing vector. This exhausts the set of all Killing vectors of the Kerr spacetime. The spacetime is stationary (but not static), and axisymmetric. We note also that the metric \eqref{eqn:kerr_line_element_boyer_lindquist} is invariant under simultaneous inversion of $t$ and $\phi$ ($t \mapsto - t$ and $\phi \mapsto - \phi$).

%

\subsubsection{Kerr--Newman solution}

In 1965, Newman \cite{NewmanCouchChinnaparedEtAl1965, NewmanJanis1965} discovered the solution to the Einstein--Maxwell field equations in \emph{electrovacuum} which describes a rotating black hole endowed with an electric charge; this can be viewed as a generalisation of the Kerr metric. In Boyer--Lindquist coordinates, the line element is given by \eqref{eqn:kerr_line_element_boyer_lindquist} with $\Sigma(r, \theta) = r^{2} + a^{2} \cos^{2}{\theta}$ and $\Delta(r) = r^{2} - 2 M r + a^{2} + Q^{2}$. Here, $M$ and $J = a M$ are again interpreted as the mass and angular momentum of the black hole, respectively; the parameter $Q$ is interpreted as the black hole's electric charge. Again, there exist a pair of event horizons, located at Boyer--Lindquist radii $r_{\pm} = M \pm \sqrt{M^{2} - (a^{2} + Q^{2})}$. For $M^{2} > a^{2} + Q^{2}$, there are two distinct event horizons at $r_{\pm}$. In the case $M^{2} = a^{2} + Q^{2}$, the black hole is referred to as \emph{extremal}, and the inner and outer horizons coincide at $r = M$. If $M^{2} < a^{2} + Q^{2}$, the Kerr--Newman solution describes a \emph{naked singularity} \cite{Penrose1973}.

As with the uncharged Kerr black hole, the Kerr--Newman solution is stationary and axisymmetric; these symmetries are encoded by the Killing vector fields $\xi_{(t)} = \partial\ind{_{t}}$ and $\xi_{(\phi)} = \partial\ind{_{\phi}}$. In Section \ref{sec:killing_objects_symmetries_integrability}, we will discuss the importance of these (and other) symmetries in the context of geodesic motion on the Kerr--Newman spacetime.

The Kerr--Newman family of solutions is characterised by three parameters: mass $M$; spin $a$; and electric charge $Q$. The uncharged ($Q = 0$) non-rotating ($a = 0$) case is simply the Schwarzschild solution, described above. The charged ($Q \neq 0$) non-rotating ($a = 0$) case is the Reissner--Nordstr\"{o}m solution, which is a static spherically symmetric solution to the electrovacuum field equations, discovered by Reissner \cite{Reissner1916}, Nordstr\"{o}m \cite{Nordstroem1918} and others \cite{Weyl1917, Jeffery1921} between 1916 and 1921. The uncharged ($Q = 0$) rotating ($a \neq 0$) case is the Kerr solution, described above.

It should be emphasised that, in the case of rotating spacetimes, there is no analogue of Birkhoff's theorem: it is \emph{not} true in general that the vacuum geometry in the exterior region of a rotating object is given by (part of) the Kerr geometry. However, there are a number of powerful uniqueness theorems in situations with less symmetry. The so-called ``no-hair theorem'' states that isolated (mathematical) black holes in stationary axisymmetric electrovacuum are remarkably simple objects, and are uniquely characterised by three (externally observable) parameters: mass $M$, angular momentum $J$, and electric charge $Q$ \cite{Heusler1996}. (The latter is thought to be negligible for astrophysical black holes.) In other words, black holes are described by the Kerr(--Newman) solution to the Einstein(--Maxwell) equations.
%

\section{Newman--Penrose formalism}
\label{sec:newman_penrose}

The Newman--Penrose formalism \cite{NewmanPenrose1962} is a special case of the tetrad formalism \cite{Chandrasekhar1989}, in which tensorial quantities in general relativity are projected onto a complex null tetrad. This formalism is well-adapted to spacetime symmetries, and is a useful tool in the treatment of radiation on curved spacetime. In this section, we present a brief review of the tetrad formalism in general relativity, before outlining the key features of the (tensorial) Newman--Penrose formalism, which are relevant for the purposes of this thesis. A more exhaustive review of the Newman--Penrose formalism is presented by Stephani \emph{et al.} \cite{StephaniKramerMacCallumEtAl2003}, using both tensor and spinor notation.
%

\subsection{Tetrad formalism}
\label{sec:newman_penrose_tetrads}

\subsubsection{Tetrad representation}

At each point in spacetime, one may construct a local basis of four contravariant vectors, i.e., a \emph{tetrad} $\{ e\ind{_{(\alpha)}^{a}} \}$ \cite{Chandrasekhar1989}. Here, lowercase Greek letters enclosed in parentheses denote \emph{tetrad indices}, whereas lowercase Latin letters are reserved for \emph{tensor indices}. To each contravariant vector, we associate a covariant vector
\begin{equation}
e\ind{_{(\alpha) a}} = g\ind{_{a b}} e\ind{_{(\alpha)}^{b}},
\end{equation}
so that spacetime indices are lowered (raised) with the spacetime metric (inverse metric). The inverse of $e\ind{_{(\alpha)}^{a}}$, denoted $e\ind{^{(\alpha)}_{a}}$, is defined such that $e\ind{_{(\alpha)}^{a}} \, e\ind{^{(\beta)}_{a}} = \delta\ind{_{(\alpha)}^{(\beta)}}$ and $e\ind{_{(\alpha)}^{a}} \, e\ind{^{(\alpha)}_{b}} = \delta\ind{^{a}_{b}}$, where the summation over tetrad/spacetime indices is assumed. Furthermore, we assume that
\begin{equation}
\label{eqn:tetrad_inner_product_matrix}
e\ind{_{(\alpha)}^{a}} \, e\ind{_{(\beta) a}} = S\ind{_{(\alpha) (\beta)}} ,
\end{equation}
where $S\ind{_{(\alpha) (\beta)}}$ is a symmetric matrix, often referred to as the \emph{inner product matrix}. When all of the inner products between the tetrad legs (i.e., all components of $S\ind{_{(\alpha) (\beta)}}$) are constant, the tetrad is referred to as a \emph{rigid frame}. Let $S\ind{^{(\alpha) (\beta)}}$ be the inverse of $S\ind{_{(\alpha) (\beta)}}$, so that $S\ind{^{(\alpha) (\gamma)}} S\ind{_{(\gamma) (\beta)}} = \delta\ind{^{(\alpha)}_{(\beta)}}$. Combining the above definitions, we see that
\begin{equation}
S\ind{_{(\alpha) (\beta)}} \, e\ind{^{(\alpha)}_{a}} = e\ind{_{(\beta) a}}, \qquad S\ind{^{(\alpha) (\beta)}} \, e\ind{_{(\alpha) a}} = e\ind{^{(\beta)}_{a}},
\end{equation}
so that tetrad indices are lowered (raised) with the inner product matrix (inverse inner product matrix). Importantly, we also find
\begin{equation}
e\ind{_{(\alpha) a}} e\ind{^{(\alpha)}_{b}} = g\ind{_{a b}}, \qquad e\ind{_{(\alpha)}^{a}} e\ind{^{(\alpha) b}} = g\ind{^{a b}},
\end{equation}
so the metric and its inverse can be expressed as an inner product of the tetrad legs.

Given any type-$(r,s)$ tensor field with components $T\ind{^{a_{1} \ldots a_{r}}_{b_{1} \ldots b_{s}}}$, we can project it onto the tetrad to obtain the \emph{tetrad components}, viz.~
\begin{equation}
T\ind{^{(\alpha_{1}) \ldots (\alpha_{r})}_{(\beta_{1}) \ldots (\beta_{s})}} = e\ind{^{(\alpha_{1})}_{a_{1}}} \, \ldots \, e\ind{^{(\alpha_{r})}_{a_{r}}} \, e\ind{_{(\beta_{1})}^{b_{1}}} \, \ldots \, e\ind{_{(\beta_{s})}^{b_{s}}} \, T\ind{^{a_{1} \ldots a_{r}}_{b_{1} \ldots b_{s}}} .
\end{equation}
Similarly, we may pass from tetrad indices to spacetime indices according to
\begin{equation}
T\ind{^{a_{1} \ldots a_{r}}_{b_{1} \ldots b_{s}}} = e\ind{_{(\alpha_{1})}^{a_{1}}} \, \ldots \, e\ind{_{(\alpha_{r})}^{a_{r}}} \, e\ind{^{(\beta_{1})}_{b_{1}}} \, \ldots \, e\ind{^{(\beta_{s})}_{b_{s}}} \, T\ind{^{(\alpha_{1}) \ldots (\alpha_{r})}_{(\beta_{1}) \ldots (\beta_{s})}} .
\end{equation}

\subsubsection{Ricci rotation coefficients}

The covariant derivative of a tetrad leg $e\ind{_{(\beta)}}$ in the direction of another tetrad leg $e\ind{_{(\alpha)}}$ can be expanded in the tetrad in terms of connection coefficients (see Section \ref{sec:covariant_differentiation_parallel_transport}) as follows \cite{StephaniKramerMacCallumEtAl2003, Chandrasekhar1989}:
\begin{equation}
e\ind{_{(\alpha)}^{b}} \nabla\ind{_{b}} e\ind{_{(\beta)}^{a}} = \mathit{\Gamma}\ind{^{(\gamma)}_{(\alpha) (\beta)}} e\ind{_{(\gamma)}^{a}} .
\end{equation}
Contracting this equation with the basis vector $e\ind{^{(\delta)}_{a}}$, using the properties of the tetrad listed above, and relabelling the indices, we arrive at the expression
\begin{equation}
\label{eqn:ricci_rotation_coefficients}
\mathit{\Gamma}\ind{_{(\alpha) (\beta) (\gamma)}} = e\ind{_{(\alpha)}^{a}} \, e\ind{_{(\beta) a ; b}} \, e\ind{_{(\gamma)}^{b}} .
\end{equation}
The quantities $\mathit{\Gamma}\ind{_{(\alpha) (\beta) (\gamma)}}$ are called the \emph{Ricci rotation coefficients}.

If $\{ e\ind{_{(\alpha)}} \}$ is a rigid frame, such that the inner products $g\ind{_{a b}} \, e\ind{_{(\alpha)}^{a}} \, e\ind{_{(\beta)}^{b}} = S\ind{_{(\alpha) (\beta)}}$ are all constant, then the Ricci rotation coefficients are antisymmetric in the first pair of indices, i.e., $\mathit{\Gamma}\ind{_{(\alpha) (\beta) (\gamma)}} = - \mathit{\Gamma}\ind{_{(\beta) (\alpha) (\gamma)}}$.

\subsubsection{Ricci and Bianchi identities}

The Ricci identity \eqref{eqn:ricci_identity} for the tetrad $\{ e\ind{_{(\alpha)}} \}$ is
\begin{equation}
\label{eqn:ricci_identity_tetrad_legs}
e\ind{_{ (\alpha) a ; b c }} - e\ind{_{ (\alpha) a ; c b }} = R\ind{_{ d a b c }} \, e\ind{_{(\alpha)}^{d}} .
\end{equation}
Projecting \eqref{eqn:ricci_identity_tetrad_legs} onto the tetrad, and replacing the covariant derivatives of the tetrad legs using \eqref{eqn:ricci_rotation_coefficients}, we may express the projection of the Riemann tensor onto the tetrad in terms of the Ricci rotation coefficients as
\begin{equation}
\label{eqn:riemann_tensor_tetrad_projection}
\begin{split}
R\ind{_{(\alpha)(\beta)(\gamma)(\delta)}} &= \mathit{\Gamma}\ind{_{(\alpha)(\beta)(\delta) ; (\gamma)}} - \mathit{\Gamma}\ind{_{(\alpha)(\beta)(\gamma) ; (\delta)}}
+ \mathit{\Gamma}\ind{_{(\alpha)(\beta)(\eta)}}  \mathit{\Gamma}\ind{_{(\delta)}^{(\eta)}_{(\gamma)}} \\
& \qquad -
\mathit{\Gamma}\ind{_{(\alpha)(\beta)(\eta)}} \mathit{\Gamma}\ind{_{(\gamma)}^{(\eta)}_{(\delta)}} + \mathit{\Gamma}\ind{_{(\eta)(\alpha)(\gamma)}} \mathit{\Gamma}\ind{_{(\beta)}^{(\eta)}_{(\delta)}} - \mathit{\Gamma}\ind{_{(\eta)(\alpha)(\delta)}} \mathit{\Gamma}\ind{_{(\beta)}^{(\eta)}_{(\gamma)}} ,
\end{split}
\end{equation}
where we have used the notation $\mathit{\Gamma}\ind{_{(\alpha)(\beta)(\delta) ; (\gamma)}} = e\ind{_{(\gamma)}^{c}} \, \mathit{\Gamma}\ind{_{(\alpha)(\beta)(\delta) ; c}}$.

Similarly, one may project the second (differential) Bianchi identity \eqref{eqn:second_bianchi_identity} onto the tetrad $\{ e\ind{_{(\alpha)}} \}$. The tetrad components of the covariant derivative of the Riemann tensor are
\begin{equation}
\label{eqn:covariant_derivative_riemann_tetrad_projection}
\begin{split}
R\ind{_{(\alpha)(\beta)(\gamma)(\delta); (\eta)}} &= R\ind{_{(\alpha)(\beta)(\gamma)(\delta) , (\eta)}} -
\mathit{\Gamma}\ind{_{(\mu)(\alpha)(\eta)}} \, R\ind{^{(\mu)}_{(\beta)(\gamma)(\delta)}} - \mathit{\Gamma}\ind{_{(\mu)(\beta)(\eta)}} \, R\ind{_{(\alpha)}^{(\mu)}_{(\gamma)(\delta)}} \\
& \qquad -
\mathit{\Gamma}\ind{_{(\mu)(\gamma)(\eta)}} \, R\ind{_{(\alpha)(\beta)}^{(\mu)}_{(\delta)}} - \mathit{\Gamma}\ind{_{(\mu)(\delta)(\eta)}} \, R\ind{_{(\alpha)(\beta)(\gamma)}^{(\mu)}} ,
\end{split}
\end{equation}
where $R\ind{_{(\alpha)(\beta)(\gamma)(\delta); (\eta)}} = e\ind{_{(\eta)}^{a}} R\ind{_{(\alpha)(\beta)(\gamma)(\delta); a}}$ and $R\ind{_{(\alpha)(\beta)(\gamma)(\delta), (\eta)}} = e\ind{_{(\eta)}^{a}} R\ind{_{(\alpha)(\beta)(\gamma)(\delta), a}}$. The projection of the second Bianchi identity \eqref{eqn:second_bianchi_identity} onto the tetrad is then given by anti-symmetrising \eqref{eqn:covariant_derivative_riemann_tetrad_projection} over its last three indices.

\subsubsection{Orthonormal tetrads}

A special case of the tetrad formalism involves choosing an orthonormal tetrad at each spacetime point. An \emph{orthonormal tetrad} consists of one unit timelike vector and three unit spacelike vectors whose inner products satisfy
\begin{equation}
g\ind{_{a b}} \, e\ind{_{(\alpha)}^{a}} \, e\ind{_{(\beta)}^{b}} = \eta\ind{_{(\alpha) (\beta)}},
\end{equation}
where $\eta\ind{_{(\alpha) (\beta)}}$ are the components of the Minkowski metric; this is a special case of \eqref{eqn:tetrad_inner_product_matrix}. The spacetime metric and its inverse may be expressed in terms of the orthonormal tetrad, respectively, as
\begin{align}
g\ind{_{a b}} &= - e\ind{^{(0)}_{a}} e\ind{^{(0)}_{b}} + e\ind{^{(1)}_{a}} e\ind{^{(1)}_{b}} + e\ind{^{(2)}_{a}} e\ind{^{(2)}_{b}} + e\ind{^{(3)}_{a}} e\ind{^{(3)}_{b}} , \\
g\ind{^{a b}} &= - e\ind{_{(0)}^{a}} e\ind{_{(0)}^{b}} + e\ind{_{(1)}^{a}} e\ind{_{(1)}^{b}} + e\ind{_{(2)}^{a}} e\ind{_{(2)}^{b}} + e\ind{_{(3)}^{a}} e\ind{_{(3)}^{b}} .
\end{align}

\subsubsection{Complex null tetrads}

In general relativity, complex null tetrads also play an important role. A \emph{complex null tetrad} $\{ k, n, m, \overline{m} \}$ consists of two real null vectors ($k$ and $n$) and a pair of complex conjugate null vectors ($m$ and $\overline{m}$), which satisfy
\begin{equation}
k\ind{^{a}} n\ind{_{a}} = -1, \qquad m\ind{^{a}} \overline{m}\ind{_{a}} = 1,
\end{equation}
with all other inner products zero. The metric and inverse metric can be written, respectively, in the form
\begin{align}
g\ind{_{a b}} &= - k\ind{_{a}} n\ind{_{b}} - n\ind{_{a}} k\ind{_{b}} + m\ind{_{a}} \overline{m}\ind{_{b}} + \overline{m}\ind{_{a}} m\ind{_{b}}, \\
g\ind{^{a b}} &= - k\ind{^{a}} n\ind{^{b}} - n\ind{^{a}} k\ind{^{b}} + m\ind{^{a}} \overline{m}\ind{^{b}} + \overline{m}\ind{^{a}} m\ind{^{b}}.
\end{align}

One may always construct a complex null tetrad $\{ k, n, m, \ol{m} \}$ from a real orthonormal tetrad $\{ e_{(0)}, e_{(1)}, e_{(2)}, e_{(3)} \}$ as follows:
\begin{align}
k &= \frac{1}{\sqrt{2}} \left( e_{(0)} + e_{(3)} \right), & n &= \frac{1}{\sqrt{2}} \left( e_{(0)} - e_{(3)} \right), \\
m &= \frac{1}{\sqrt{2}} \left( e_{(1)} + i e_{(2)} \right), & \ol{m} &= \frac{1}{\sqrt{2}} \left( e_{(1)} - i e_{(2)} \right) .
\end{align}

It is standard to introduce four directional derivatives along each of the legs of the complex null tetrad,
\begin{equation}
\label{eqn:np_directional_derivatives}
D = k\ind{^{a}} \nabla\ind{_{a}}, \qquad \mathit{\Delta} = n\ind{^{a}} \nabla\ind{_{a}}, \qquad \delta = m\ind{^{a}} \nabla\ind{_{a}}, \qquad \ol{\delta} = \ol{m}\ind{^{a}} \nabla\ind{_{a}}.
\end{equation}

\subsection{Transformation laws for complex null tetrads}
\label{sec:np_lorentz_transformations}

A complex null tetrad $\{k, n, m, \ol{m} \}$ may be transformed in the following ways \cite{StephaniKramerMacCallumEtAl2003}.
\begin{enumerate}[(i)]
	\item Null rotations which keep $k$ fixed:
	\begin{equation}
	\label{eqn:lt_k}
	k^{\prime} = k, \qquad m^{\prime} = m + B k, \qquad n^{\prime} = n + B \overline{m} + \overline{B} m + B \overline{B} k,
	\end{equation}
	where $B$ is a complex function.
	\item Null rotations which keep $n$ fixed:
	\begin{equation}
	\label{eqn:lt_n}
	n^{\prime} = n, \qquad m^{\prime} = m + E n, \qquad k^{\prime} = k + E \overline{m} + \overline{E} m + E \overline{E} n,
	\end{equation}
	where $E$ is a complex function.
	\item Spatial rotations in the $(m, \overline{m})$-plane:
	\begin{equation}
	\label{eqn:lt_m_mbar}
	m^{\prime} = e^{i \Theta} m,
	\end{equation}
	where $\Theta$ is a real function.
	\item Boosts in the $(k, n)$-plane:
	\begin{equation}
	\label{eqn:lt_k_n}
	k^{\prime} = A k, \qquad n^{\prime} = \frac{1}{A} n,
	\end{equation}
	where $A > 0$ is a real function.
\end{enumerate}
Together, the transformations \eqref{eqn:lt_k}--\eqref{eqn:lt_k_n}, which contain six real parameters, represent the six-parameter group of Lorentz transformations.

\subsection{Spin coefficients}
\label{sec:newman_penrose_spin_coefficients}

Newman and Penrose \cite{NewmanPenrose1962} introduce a set of twelve independent complex scalars, which are defined as linear combinations of the Ricci rotation coefficients \eqref{eqn:ricci_rotation_coefficients} for the complex null tetrad $\{ k, n, m, \ol{m} \}$. The \emph{spin coefficients} (or \emph{Newman--Penrose scalars}) are \cite{StephaniKramerMacCallumEtAl2003, Chandrasekhar1989}
\begin{align}
\kappa &= - \mathit{\Gamma}\ind{_{a b c}} m\ind{^{a}} k\ind{^{b}} k\ind{^{c}} = - \nps{m}{k}{k}, &
\rho &= - \mathit{\Gamma}\ind{_{a b c}} m\ind{^{a}} k\ind{^{b}} \ol{m}\ind{^{c}} = - \nps{m}{k}{\ol{m}},  \label{eqn:np_scalar_kappa_rho} \\
\sigma &= - \mathit{\Gamma}\ind{_{a b c}} m\ind{^{a}} k\ind{^{b}} m\ind{^{c}} = - \nps{m}{k}{m}, &
\tau &= - \mathit{\Gamma}\ind{_{a b c}} m\ind{^{a}} k\ind{^{b}} n\ind{^{c}} = - \nps{m}{k}{n}, \\
\pi &= \mathit{\Gamma}\ind{_{a b c}} \ol{m}\ind{^{a}} n\ind{^{b}} k\ind{^{c}} = \nps{\ol{m}}{n}{k}, &
\nu &= \mathit{\Gamma}\ind{_{a b c}} \ol{m}\ind{^{a}} n\ind{^{b}} n\ind{^{c}} = \nps{\ol{m}}{n}{n}, \\
\mu &= \mathit{\Gamma}\ind{_{a b c}} \ol{m}\ind{^{a}} n\ind{^{b}} m\ind{^{c}} = \nps{\ol{m}}{n}{m}, &
\lambda &= \mathit{\Gamma}\ind{_{a b c}} \ol{m}\ind{^{a}} n\ind{^{b}} \ol{m}\ind{^{c}} = \nps{\ol{m}}{n}{\ol{m}}, \\
\begin{split}
\varepsilon &=
\frac{1}{2} \left( \mathit{\Gamma}\ind{_{a b c}} \ol{m}\ind{^{a}} m\ind{^{b}} k\ind{^{c}} - \mathit{\Gamma}\ind{_{a b c}} n\ind{^{a}} k\ind{^{b}} k\ind{^{c}} \right) \\
&= \frac{1}{2} \left( \nps{\ol{m}}{m}{k} - \nps{n}{k}{k} \right),
\end{split}
&
\begin{split}
\gamma &= \frac{1}{2} \left( \mathit{\Gamma}\ind{_{a b c}} k\ind{^{a}} n\ind{^{b}} k\ind{^{c}} - \mathit{\Gamma}\ind{_{a b c}} m\ind{^{a}} \ol{m}\ind{^{b}} n\ind{^{c}} \right) \\
&= \frac{1}{2} \left( \nps{k}{n}{k} - \nps{m}{\ol{m}}{n} \right),
\end{split}
\\
\begin{split}
\alpha &= \frac{1}{2} \left( \mathit{\Gamma}\ind{_{a b c}} k\ind{^{a}} n\ind{^{b}} \ol{m}\ind{^{c}} - \mathit{\Gamma}\ind{_{a b c}} \ol{m}\ind{^{a}} \ol{m}\ind{^{b}} \ol{m}\ind{^{c}} \right) \\
&= \frac{1}{2} \left( \nps{k}{n}{\ol{m}} - \nps{\ol{m}}{\ol{m}}{\ol{m}} \right),
\end{split}
&
\begin{split}
\beta &= \frac{1}{2} \left( \mathit{\Gamma}\ind{_{a b c}} \ol{m}\ind{^{a}} m\ind{^{b}} m\ind{^{c}} - \mathit{\Gamma}\ind{_{a b c}} n\ind{^{a}} k\ind{^{b}} m\ind{^{c}} \right) \\
&= \frac{1}{2} \left( \nps{\ol{m}}{m}{m} - \nps{n}{k}{m} \right) . \label{eqn:np_scalar_alpha_beta}
\end{split}
\end{align}
The twelve spin coefficients are the primary quantities in the Newman--Penrose formalism. For this reason, the formalism is sometimes referred to in the literature as the \emph{spin coefficient formalism}.

In many situations, the complex null tetrad will be transformed using one of the Lorentz transformations (i)--(iv) from Section \ref{sec:np_lorentz_transformations}. The corresponding transformation laws for the Newman--Penrose scalars can be derived using the definitions involving projections of the tetrad legs \eqref{eqn:np_scalar_kappa_rho}--\eqref{eqn:np_scalar_alpha_beta}; alternatively, these transformation laws can be found in Chapter 7 of \cite{StephaniKramerMacCallumEtAl2003}.

The directional derivatives \eqref{eqn:np_directional_derivatives} satisfy a set of commutation relations involving the Newman--Penrose scalars, which arise from the metric compatibility and torsion-free property of the covariant derivative; the commutator identities are
\begin{align}
\mathit{\Delta} D - D \mathit{\Delta} &= (\gamma + \ol{\gamma}) D + (\varepsilon + \ol{\varepsilon}) \mathit{\Delta} - (\ol{\tau} + \pi) \delta - (\tau + \ol{\pi}) \ol{\delta} , \\
\delta D - D \delta &= (\ol{\alpha} + \beta - \ol{\pi}) D + \kappa \mathit{\Delta} - (\ol{\rho} + \varepsilon - \ol{\varepsilon}) \delta - \sigma \ol{\delta} , \\
\delta \mathit{\Delta} - \mathit{\Delta} \delta &= - \ol{\nu} D + (\tau - \ol{\alpha} - \beta) \mathit{\Delta} + (\mu - \gamma + \ol{\gamma}) \delta + \ol{\lambda} \, \ol{\delta} , \\
\ol{\delta} \delta - \delta \ol{\delta} &= (\ol{\mu} - \mu) D + (\ol{\rho} - \rho) \mathit{\Delta} + (\alpha - \ol{\beta}) \delta - (\ol{\alpha} - \beta) \ol{\delta} .
\end{align}


\subsection{Weyl scalars}
\label{sec:newman_penrose_weyl_scalars}

In Section \ref{sec:curvature}, we introduced the Weyl tensor $C\ind{_{a b c d}}$, the trace-free part of the Riemann curvature tensor; see \eqref{eqn:weyl_tensor_definition}. In four spacetime dimensions, the Weyl tensor has ten independent components, which are encoded in five complex scalars in the Newman--Penrose formalism. The \emph{Weyl scalars} are
\begin{align}
\Psi_{0} &= C\ind{_{a b c d}} k\ind{^{a}} m\ind{^{b}} k\ind{^{c}} m\ind{^{d}} , \label{eqn:np_weyl_scalar_0} \\
\Psi_{1} &= C\ind{_{a b c d}} k\ind{^{a}} n\ind{^{b}} k\ind{^{c}} m\ind{^{d}} , \\
\Psi_{2} &= C\ind{_{a b c d}} k\ind{^{a}} m\ind{^{b}} \ol{m}\ind{^{c}} n\ind{^{d}} , \\
\Psi_{3} &= C\ind{_{a b c d}} k\ind{^{a}} n\ind{^{b}} n\ind{^{c}} \ol{m}\ind{^{d}} , \\
\Psi_{4} &= C\ind{_{a b c d}} \ol{m}\ind{^{a}} n\ind{^{b}} \ol{m}\ind{^{c}} n\ind{^{d}} . \label{eqn:np_weyl_scalar_4}
\end{align}

The Ricci tensor $R\ind{_{a b}}$ has ten independent components. These are encoded in the Ricci curvature scalar $R$, and the six complex \emph{Ricci scalars} and their complex conjugates:
\begin{align}
\Phi_{0 0} &= \frac{1}{2} R\ind{_{a b}} k\ind{^{a}} k\ind{^{b}} = \ol{\Phi}_{0 0} , & \Phi_{0 1} &= \frac{1}{2} R\ind{_{a b}} k\ind{^{a}} m\ind{^{b}} = \ol{\Phi}_{1 0} , \label{eqn:np_ricci_scalar_00_01} \\
\Phi_{0 2} &= \frac{1}{2} R\ind{_{a b}} m\ind{^{a}} m\ind{^{b}} = \ol{\Phi}_{2 0} , & \Phi_{1 1} &= \frac{1}{2} R\ind{_{a b}} \left( k\ind{^{a}} n\ind{^{b}} + m\ind{^{a}} \ol{m}\ind{^{b}} \right) = \ol{\Phi}_{1 1} , \\
\Phi_{1 2} &= \frac{1}{2} R\ind{_{a b}} n\ind{^{a}} m\ind{^{b}} = \ol{\Phi}_{2 1} , & \Phi_{2 2} &= \frac{1}{2} R\ind{_{a b}} n\ind{^{a}} n\ind{^{b}} = \ol{\Phi}_{2 2} . \label{eqn:np_ricci_scalar_12_22}
\end{align}
We note that there are nine independent components of the complex Ricci scalars \eqref{eqn:np_ricci_scalar_00_01}--\eqref{eqn:np_ricci_scalar_12_22}. In these definitions, one may replace the Ricci tensor $R\ind{_{a b}}$ with its trace-free counterpart $S\ind{_{a b}} = R\ind{_{a b}} - \frac{1}{4} R g\ind{_{a b}}$ if so desired.

\subsection{Newman--Penrose field equations}
\label{sec:np_field_equations}

When projected onto the complex null tetrad $\{ k, n, m, \ol{m} \}$, the Ricci identities \eqref{eqn:riemann_tensor_tetrad_projection} give rise to a system of $36$ equations involving the Newman--Penrose scalars \eqref{eqn:np_scalar_kappa_rho}--\eqref{eqn:np_scalar_alpha_beta}, their directional derivatives along tetrad legs \eqref{eqn:np_directional_derivatives}, and the Weyl and Ricci scalars \eqref{eqn:np_weyl_scalar_0}--\eqref{eqn:np_ricci_scalar_12_22}. The full set of \emph{Ricci equations} (or \emph{Newman--Penrose field equations}) are written out explicitly in Chapter 7 of \cite{StephaniKramerMacCallumEtAl2003}.

In addition, one may project the Bianchi identities onto the complex null tetrad $\{ k, n, m, \ol{m} \}$, which can be achieved by anti-symmetrising the expression \eqref{eqn:covariant_derivative_riemann_tetrad_projection} over its last three indices. This yield a system of equations involving the Newman--Penrose scalars, the Weyl and Ricci scalars and their directional derivatives. The full set of such equations, referred to as the \emph{Bianchi equations}, can again be found in Chapter 7 of \cite{StephaniKramerMacCallumEtAl2003}.

\subsection{Geroch--Held--Penrose formalism}
\label{sec:ghp_formalism}

Geroch \emph{et al.} \cite{GerochHeldPenrose1973} developed a formalism for the treatment of general relativity that lies between a fully covariant formulation and the Newman--Penrose spin-coefficient formalism; this approach is referred to as the \emph{Geroch--Held--Penrose formalism}. In this approach, only a \emph{pair} of null directions -- as opposed to an entire null tetrad -- is considered at each spacetime point. The formulae in this Geroch--Held--Penrose formalism are simpler than those of the spin-coefficient formalism of Newman and Penrose.

The most general transformation which preserves the pair of null directions is a two-parameter subgroup of the full group of Lorentz transformations, corresponding to a rotation in the $(m, \ol{m})$-plane and a boost in the $(k, n)$-plane; see transformations (iii) and (iv) of Section \ref{sec:np_lorentz_transformations}. Under this transformation, the tetrad legs transform according to
\begin{align}
\label{eqn:ghp_tetrad_transformation}
k^{\prime} &= A k, & n^{\prime} &= \frac{1}{A} n, & m^{\prime} &= e^{i \Theta} m, &
\ol{m}^{\prime} &= e^{- i \Theta} \ol{m},
\end{align}
where $A = C \ol{C}$ and $e^{i \Theta} = C \ol{C}\vp{C}^{-1}$, with $C$ complex.

A scalar quantity $z$ which transforms according to $z^{\prime} = C^{p} \ol{C}\vp{C}^{q} z$ under \eqref{eqn:ghp_tetrad_transformation} is called a \emph{scalar of type $(p, q)$}. The scalar is said to have \emph{spin weight} $\mathfrak{s} = \frac{1}{2} (p - q)$, and \emph{boost weight} $\mathfrak{b} = \frac{1}{2}(p + q)$. The spin (boost) weight can be obtained easily by counting the number of times the legs $m$ and $\ol{m}$ ($k$ and $n$) appear in the definition of the quantity: we add $1$ for the number of times $m$ ($k$) occurs, and subtract $1$ for the number of times $\ol{m}$ ($n$) occurs. For example, the Newman--Penrose scalar $\rho = - \nps{m}{k}{\ol{m}}$ has spin weight $\mathfrak{s} = 0$, since there is one $m$ and one $\ol{m}$ in the definition of $\rho$; and it has boost weight $\mathfrak{b} = 1$, since $k$ appears once and $n$ does not appear. We note here that certain quantities (e.g.~$\alpha$) do not have a well-defined spin/boost weight.

\section{Integrability and chaos in general relativity}
\label{sec:integrability_and_chaos_in_gr}

The main theme of this thesis is the study of light propagation on black hole spacetimes. As described in Section \ref{sec:geodesics}, light follows null geodesics on curved spacetime; this can be modelled using a Hamiltonian formalism, where the geodesics are integral curves of Hamilton's equations with Hamiltonian function $H = \frac{1}{2} g\ind{^{a b}} p\ind{_{a}} p\ind{_{b}}$. The themes of integrability and chaos in Hamiltonian dynamical systems will be of central importance to our understanding of light propagation on black hole spacetimes in general relativity.

In this section, we review some key features of the theory of (non-linear) dynamical systems. In particular, we will discuss the relationship between spacetime symmetries, conserved quantities along geodesics, and integrability in Section \ref{sec:killing_objects_symmetries_integrability}. We will also present a review of some important features of the theory of deterministic chaos in Section \ref{sec:chaotic_dynamical_systems}.

\subsection{Dynamical systems}
\label{sec:dynamical_systems_review}

Dynamical systems theory is concerned with the long-term behaviour of evolving systems. In general, a \emph{dynamical system} consists of a \emph{state space} (or \emph{phase space}), a space whose points describe the state of the system at any instant of time; and a rule which governs the time-evolution of points in the state space. For a comprehensive overview of this topic, see e.g.~\cite{Devaney1989, Glendinning1994, AlligoodSauerYorke1996, BrinStuck2002, Ott2002, Gaspard2005}. Here, we review some key concepts and definitions.

Mathematically, a \emph{discrete-time dynamical system} consists of a non-empty set $X$ (the state space) and a map $f \colon X \to X$ (the time-evolution rule). For $n \in \mathbb{N}$, the $n$th iterate of $f$ is given by $f^{n} = f \circ \ldots \circ f$, the $n$-fold composition of $f$, where $f^{0}$ is the identity map. If $f$ is invertible, then $f^{-n} = f^{-1} \circ \ldots \circ f^{-1}$, i.e., the inverse of $f$ composed with itself $n$ times. The iterates satisfy $f^{n + m} = f^{n} \circ f^{m}$, so they form a group if $f$ is invertible. In this context, a discrete-time dynamical system is often referred to as a \emph{map}.

Typically, the set $X$ will be endowed with some additional structure which is preserved by the map $f$. For example, $(X, f)$ could be a smooth manifold and a smooth map (or a diffeomorphism in the case of invertible maps).

A \emph{continuous time dynamical system} consists of a space $X$ and a one-parameter family of maps $\left\{ f^{\lambda} \colon X \to X \, | \, \lambda \in \mathbb{R} \right\}$ which form a one-parameter group, i.e., $f^{\lambda + \tau} = f^{\lambda} \circ f^{\tau}$ and $f^{0}$ is the identity map. Here, $\lambda$ is the continuous time parameter. In this context, the dynamical system is called a \emph{flow}. The map is invertible, since $\left( f^{\lambda} \right)^{-1} = f^{-\lambda}$. For a fixed value $\lambda = \lambda_{0}$, the iterates $\left( f^{\lambda_{0}} \right)^{n}$ form a discrete-time dynamical system.

The time-evolution of many systems of interest is governed by systems of ordinary differential equations. If there is no time-dependent forcing, then the system will be \emph{autonomous}. We consider a continuous-time dynamical system described by a set of first-order ordinary differential equations
\begin{equation}
\label{eqn:dynamical_system_odes}
\dot{z}(\lambda) = X(z),
\end{equation}
where $X$ is a vector field, $z \in \mathcal{P}$ represents a point in phase space $\mathcal{P}$, and an overdot denotes differentiation with respect to the time parameter $\lambda$. The system \eqref{eqn:dynamical_system_odes} induces a flow
\begin{equation}
z = \varphi_{X}(z_{0}, \lambda),
\end{equation}
which is a non-linear function of $\lambda$ and the initial data $z_{0}$. Systems of higher-order ordinary differential equations can, in general, be reduced to the form \eqref{eqn:dynamical_system_odes}. Phase space volumes may be conserved if the vector field $F$ is divergence-free; otherwise, phase space volumes may expand or contract. Dynamical systems are divided into two main groups: Hamiltonian (conservative) systems and dissipative systems. In this work we focus on the former.

\subsection{Hamiltonian systems}

A particularly special class of dynamical system is a Hamiltonian system. In the Hamiltonian approach, a dynamical system is described in terms of phase space, which can be represented geometrically using symplectic geometry. For a more detailed description of Hamiltonian systems, see the textbooks by Arnold \cite{Arnold1989}, Ozorio de Almeida \cite{OzoriodeAlmeida1990}, and Lowenstein \cite{Lowenstein2012}, for example.

Let $\mathcal{P}$ be a $2 N$-dimensional manifold. A \emph{symplectic structure} on $\mathcal{P}$ is a non-degenerate two-form $\Omega$ which is closed ($\ed \Omega = 0$). The pair $(\mathcal{P}, \Omega)$ is called a \emph{symplectic manifold}; it describes a dynamical system with $N$ degrees of freedom. The existence of a non-degenerate two-form $\Omega$ implies that the manifold $\mathcal{P}$ must be even-dimensional.

Let $z\ind{^{A}}$ ($A \in \left\{ 1, \ldots, 2 N \right\}$) denote coordinates on $\mathcal{P}$. The components of the symplectic form $\Omega\ind{_{A B}}$ are given by an anti-symmetric non-degenerate matrix. The inverse symplectic form, whose components are written $\Omega\ind{^{A B}}$, is defined via the relation $\Omega\ind{_{A C}} \Omega\ind{^{B C}} = \delta\ind{_{A}^{B}}$.

An \emph{observable} is a scalar function on $\mathcal{P}$. In autonomous Hamiltonian systems, the observables do not explicitly depend on time. For an observable $F$, the symplectic form defines the corresponding \emph{Hamiltonian vector field} $X_{F}$, which is given in components by
\begin{equation}
{X_{F}}\ind{^{A}} = \Omega\ind{^{A B}} F\ind{_{, B}} ,
\end{equation}
where a comma denotes the partial derivative with respect to the phase space coordinate $z\ind{^{A}}$, i.e., $F\ind{_{, A}} = \frac{\partial F}{\partial z\ind{^{A}}}$. The integral curves of the vector field $X_{F}$ determine a map from $\mathcal{P}$ to itself; this is called a \emph{Hamiltonian flow}. Introducing a parameter $\lambda$ and local coordinates $z\ind{^{A}}(\lambda)$, the integral curves of the Hamiltonian flow are given by
\begin{equation}
X_{F} = \frac{\ed z\ind{^{A}}}{\ed \lambda} \partial\ind{_{A}} = \dot{z}\ind{^{A}} \partial\ind{_{A}}.
\end{equation}

For a pair of observables $F$ and $G$, the symplectic structure $\Omega$ gives another observable $K = \left\{ F, G \right\}$, which is the \emph{Poisson bracket}, given by
\begin{equation}
\left\{ F, G \right\} = \Omega\ind{^{A B}} F\ind{_{, A}} G\ind{_{, B}} .
\end{equation}
The Poisson bracket satisfies the following properties.
\begin{enumerate}[(i)]
\item Antisymmetry: $\left\{ F, G \right\} = - \left\{ G, F \right\}$ for all observables $F$ and $G$ on $\mathcal{P}$. This follows from the fact that $\Omega$ is a two-form
\item Jacobi identity: $ \left\{ F, \left\{ G , K \right\} \right\} + \left\{ G, \left\{ K , F \right\} \right\} + \left\{ K, \left\{ F , G \right\} \right\} = 0$, for all observables $F$, $G$ and $K$ on $\mathcal{P}$. This follows from the fact that $\Omega$ is closed.
\end{enumerate}
A pair of observables $F$ and $G$ are said to \emph{Poisson commute} if their Poisson bracket vanishes, i.e., $\left\{ F, G \right\} = 0$. Such observables are said to be \emph{in involution}. These properties ensure that observables on phase space form a Lie algebra with respect to the Poisson bracket. This is related to the Lie algebra of (Hamiltonian) vector fields via
\begin{equation}
\label{eqn:lie_bracket_hamiltonian_vector_fields}
[ X_{F}, X_{G} ] = - X_{ \left\{ F, G \right\} } ,
\end{equation}
where $[X , Y]$ denotes the commutator of the vector fields $X$ and $Y$ (see Section \ref{sec:lie_differentiation}). If a pair of observables are in involution, then their corresponding Hamiltonian vector fields commute.

Hamiltonian vector fields preserve the symplectic two-form $\Omega$, that is, $\mathcal{L}_{X_{F}} \Omega = 0$. Using the Leibniz rule for the Lie derivative of the exterior product, one may demonstrate that the volume form $\Omega^{\wedge N} = \Omega \wedge \ldots \wedge \Omega$ ($N$ times) induced by the symplectic structure is preserved by the Hamiltonian vector field $X_{F}$; in other words, $\mathcal{L}_{X_{F}} \Omega^{\wedge N} = 0$. This is the famous \emph{Liouville's theorem}, which says that the natural volume form on a symplectic manifold is invariant under the Hamiltonian flow.

The symplectic structure ensures the existence of a special class of coordinates, which simplify the expressions discussed above. The existence of such coordinates follows from the \emph{Darboux theorem}, which states that, in the vicinity of a point of $\mathcal{P}$, it is possible to choose \emph{canonical coordinates} $z\ind{^{A}} = (q\ind{^{a}}, p\ind{_{b}}) =  ( q\ind{^{1}}, \ldots, q\ind{^{N}}, p\ind{_{1}}, \ldots, p\ind{_{N}} )$ (with $a, b \in \left\{ 1, \ldots, N \right\}$), in which the symplectic form $\Omega$ can be written in canonical form as
\begin{equation}
\Omega = \ed q\ind{^{a}} \wedge \ed p\ind{_{a}} .
\end{equation}
In canonical coordinates, the Hamiltonian vector field takes the form
\begin{equation}
\label{eqn:hamiltonian_vector_field}
X_{F} = \frac{\partial F}{\partial p\ind{_{a}}} \frac{\partial}{\partial q\ind{^{a}}} - \frac{\partial F}{\partial q\ind{^{a}}} \frac{\partial}{\partial p\ind{_{a}}} ,
\end{equation}
and the Poisson bracket is
\begin{equation}
\left\{ F, G \right\} = \frac{\partial F}{\partial q\ind{^{a}}} \frac{\partial G}{\partial p\ind{_{a}}} - \frac{\partial F}{\partial p\ind{_{a}}} \frac{\partial G}{\partial q\ind{^{a}}} .
\end{equation}

The dynamics on phase space is determined by the \emph{Hamiltonian function} $H$, a given scalar function on $\mathcal{P}$. (For autonomous systems, $H$ does not explicitly depend on time.) The time-evolution is determined by the Hamiltonian flow corresponding to the Hamiltonian function: phase space trajectories are the integral curves of the Hamiltonian vector field corresponding to $H$, given in canonical coordinates by
\begin{equation}
\label{eqn:hamiltonian_vector_field_hamiltonian_function}
X_{H} = \dot{z}\ind{^{A}} \partial\ind{_{A}} = \dot{q}\ind{^{a}} \frac{\partial}{\partial q\ind{^{a}}} + \dot{p}\ind{_{a}} \frac{\partial}{\partial p\ind{_{a}}} .
\end{equation}
Comparison of \eqref{eqn:hamiltonian_vector_field} with \eqref{eqn:hamiltonian_vector_field_hamiltonian_function} yields \emph{Hamilton's equations},
\begin{equation}
\label{eqn:hamiltons_canonical_equations}
\dot{q}\ind{^{a}} = \frac{\partial H}{\partial p\ind{_{a}}}, \qquad \dot{p}\ind{_{a}} = - \frac{\partial H}{\partial q\ind{^{a}}} .
\end{equation}
The time-evolution of an observable $F$ on $\mathcal{P}$ is given by
\begin{equation}
\dot{F} = \left\{ F, H \right\} .
\end{equation}

In the case of geodesic motion on a spacetime $(\mathcal{M}, g)$, the phase space is the cotangent bundle $\mathcal{P} = T^{\ast} \mathcal{M}$, and the canonical coordinates on $\mathcal{P}$ are the spacetime coordinates $\{ x\ind{^{a}} \}$ and their conjugate momenta $\{ p\ind{_{a}} \}$ (see Section \ref{sec:hamiltonian_formulation_geodesics}). In this work, we are concerned only with the study of null geodesic motion on four-dimensional spacetime $(\mathcal{M}, g\ind{_{a b}})$, so the phase space $\mathcal{P}$ is eight-dimensional and there are four degrees of freedom. We will use either $q\ind{^{a}}$ or $x\ind{^{a}}$ to denote spacetime coordinates, depending on the context.

\subsection{Integrability in Hamiltonian systems}
\label{sec:integrability}

A \emph{conserved quantity} -- also referred to as a \emph{constant} or \emph{integral of motion} -- is an observable which remains constant along phase space trajectories. Clearly, an observable $F$ is a conserved quantity if it Poisson commutes with the Hamiltonian, i.e., $\left\{ F, H \right\} = 0$.

Each observable $F$ on phase space induces a Hamiltonian flow, which can be viewed as a transformation on phase space. If the observable is a constant of motion, then the corresponding Hamiltonian vector field commutes with the time-evolution $X_{H}$: $[X_{F}, X_{H}] = 0$, by \eqref{eqn:lie_bracket_hamiltonian_vector_fields}. Any trajectory which satisfies Hamilton's equations \eqref{eqn:hamiltons_canonical_equations} can be mapped to another trajectory by the transformation induced by $F$. This says that conserved quantities generate symmetries  of the time-evolution of a dynamical system. Frolov \emph{et al.} \cite{FrolovKrtousKubiznak2017} note this in the following theorem, which can be viewed as \emph{Noether's theorem} on phase space. Let $Y$ be a vector field which preserves both the symplectic structure and the Hamiltonian, i.e., $\mathcal{L}_{Y} \Omega = 0$ and $\mathcal{L}_{Y} H = 0$. Then there exists a constant of motion $I$, such that $Y = X_{I}$.

Of particular interest are dynamical systems which admit multiple observables which mutually Poisson commute (i.e., they are in involution). A dynamical system which admits the maximum possible number of independent commuting observables is said to be \emph{completely integrable}. The time-evolution of such systems is ``regular'': trajectories remain in well-defined submanifolds of phase space, and can be obtained using a systematic procedure. When the system is non completely integrable, it is called \emph{non-integrable}. In such systems, we generically observe rich chaotic motion in phase space; see Section \ref{sec:chaotic_dynamical_systems}.

A Hamiltonian system with $N$ degrees of freedom is said to be \emph{Liouville integrable} if it admits $N$ independent constants of motion $F\ind{_{a}}$ which are in involution with one another, $\left\{ F\ind{_{a}}, F\ind{_{b}} \right\} = 0$ (for all $a, b \in \left\{1, 2, \ldots, N \right\}$); one of the $F\ind{_{a}}$ is the Hamiltonian function itself, say $H = F_{1}$. The fact that the $F\ind{_{a}}$ are independent means that no one of these integrals of motion can be expressed as a function of the other $N - 1$ conserved quantities. The requirement that an integrable system has $N$ independent integrals of motion implies that the dynamical trajectories lie on $N$-dimensional surface defined by the equations
\begin{equation}
\label{eqn:conserved_quantities_torus}
F\ind{_{a}} \left( p_{1}, \ldots, p\ind{_{N}}, q_{1}, \ldots, q\ind{_{N}} \right) = C_{a} ,
\end{equation}
where $C_{a}$ are constants given by the values of the phase space functions $F\ind{_{a}}$ along the dynamical trajectories. The requirement that the constants of motion be in involution restricts the topology of the $N$-dimensional surface \eqref{eqn:conserved_quantities_torus}: it must be an $N$-dimensional torus \cite{Arnold1989, OzoriodeAlmeida1990}.

The $N$-tori in a completely integrable system with $N$ degrees of freedom in the $2 N$-dimensional phase space are often referred to as \emph{invariant tori} \cite{Berry1978}. This is due to the fact that any dynamical trajectory which starts on one of the $N$-tori remains on it forever. Different initial conditions lead to nested $N$-tori in phase space which do not intersect. The motion on these $N$-tori is characterised by a set of $N$ fundamental frequencies $\omega_{i}$, $i \in \left\{ 1, 2, \ldots, N \right\}$. The trajectory then has $n$ associated fundamental periods, given by $\frac{2 \pi}{\omega_{i}}$. If the trajectory is \emph{closed} on the $N$-torus, then it is \emph{exactly periodic}. In order for this to be the case, the ratio of every pair of frequencies must be a rational number, i.e., $\frac{\omega_{i}}{\omega_{j}} \in \mathbb{Q}$ for all $i, j \in \left\{ 1, 2, \ldots, N \right\}$. In such circumstances, the $N$-torus is called \emph{resonant}. Conversely, if the frequency ratio $\frac{\omega_{i}}{\omega_{j}}$ is irrational, then the orbit will not be closed on the $N$-torus. In this case, we have \emph{quasi-periodicity}, and the orbit is said to be \emph{ergodic} on the $N$-torus \cite{Berry1978}.

Many completely integrable systems may be solved analytically by quadrature (i.e., using a finite number of algebraic operations and integrations). This prescription, which involves transforming to \emph{action--angle variables}, is codified in \emph{Liouville's procedure}, which is summarised in Appendix B of \cite{FrolovKrtousKubiznak2017}.
%

\subsection{Killing objects, symmetries and integrability}
\label{sec:killing_objects_symmetries_integrability}

Symmetries play a key role in general relativity, both in the search for cherished exact solutions to Einstein's field equations, and in the study of the integrability properties of wave equations and geodesic motion on a fixed spacetime background. Here, we briefly review the relationship between Killing vectors and Killing tensors (referred to collectively as Killing objects), symmetries on spacetime and phase space, and complete integrability of geodesic motion. A highly comprehensive account of this topic can be found in the review by Frolov \emph{et al.} \cite{FrolovKrtousKubiznak2017}, and the review by Cariglia \cite{Cariglia2014}.

As discussed in Section \ref{sec:geodesics}, the motion of a free particle on a spacetime $(\mathcal{M}, g\ind{_{a b}})$ follows a causal geodesic $x\ind{^{a}}(\lambda)$, which is completely determined by the spacetime geometry, i.e., by the Hamiltonian $H = \frac{1}{2} g\ind{^{a b}} p\ind{_{a}} p\ind{_{b}}$, where $p\ind{_{a}} = g\ind{_{a b}} \dot{x}\ind{^{a}}$ are the canonical momenta, and an overdot denotes differentiation with respect to an affine parameter $\lambda$.

\subsubsection{Killing vectors and explicit symmetries}

We saw in Section \ref{sec:killing_vectors} that Killing vector fields generate isometries of spacetime. According to Noether's theorem, there is a one-to-one correspondence between these isometries and constants of motion along geodesics. The latter can be expressed in terms of the Killing vector $\xi\ind{^{a}}$ and the momentum $p\ind{_{a}}$ as $I = \xi\ind{^{a}} p\ind{_{a}}$. It is quick to demonstrate that $I$ is indeed an integral of motion:
\begin{equation}
\dot{I} = p\ind{^{b}} I\ind{_{; b}} = p\ind{^{b}} \left( \xi\ind{^{a}} p\ind{_{a}} \right)\ind{_{; b}} = p\ind{^{b}} \xi\ind{_{a ; b}} p\ind{^{a}} +  p\ind{^{b}} p\ind{_{a ; b}} \xi\ind{^{a}} = \xi\ind{_{( a ; b )}} p\ind{^{a}} p\ind{^{b}} = 0,
\end{equation}
where we have used the Killing vector equation $\xi\ind{_{(a ; b)}} = 0$, and the geodesic equation $p\ind{^{b}} p\ind{_{a ; b}} = 0$ in the final step. On phase space $\mathcal{P}$, the Hamiltonian vector field corresponding to the constant of motion $I$ is given by \cite{FrolovKrtousKubiznak2017}
\begin{equation}
X_{I} = \xi\ind{^{a}} \frac{\partial}{\partial x\ind{^{a}}} - \xi\ind{^{b}_{, a}} p\ind{_{b}} \frac{\partial}{\partial p\ind{_{a}}} .
\end{equation}
When projected back onto the spacetime manifold $\mathcal{M}$, the Hamiltonian vector field reduces to the Killing vector field $\xi$. If the canonical projection of a phase space symmetry yields a well-defined spacetime symmetry, the symmetry is said to be \emph{explicit}. Killing vectors therefore correspond to explicit symmetries. The projection of the phase space symmetry must depend only on the spacetime coordinates. It follows that the conserved quantity $I$ must be linear in the momentum $p\ind{_{a}}$. Thus, the constants of motion along a particle's trajectories in spacetime which correspond to explicit symmetries must be linear in the particle's momentum \cite{FrolovKrtousKubiznak2017}.

\subsubsection{Killing tensors and hidden symmetries}

In addition to explicit symmetries on spacetime induced by Killing vectors, there may exist symmetries which do not have a simple description on spacetime. Such symmetries are referred to as \emph{hidden} symmetries, and are generated by higher-rank Killing tensors. A \emph{Killing tensor} $K\ind{^{a_{1} \ldots a_{r}}}$ of type $(r, 0)$ is a totally symmetric tensor which satisfies the \emph{Killing tensor equation},
\begin{equation}
\label{eqn:killing_tensor_equation}
\nabla\ind{^{( b}} K\ind{^{a_{1} \ldots a_{r} )}} = 0 .
\end{equation}
A type-$(1, 0)$ Killing tensor is a Killing vector. A trivial example of a type-$(2, 0)$ Killing tensor which exists on every spacetime is the (inverse) metric tensor $g\ind{^{a b}}$; this fact follows from the metric compatibility of the covariant derivative, i.e., $g\ind{_{a b ; c}} = 0$.

Using the Killing tensor equation \eqref{eqn:killing_tensor_equation} and the geodesic equation, one may show that, if $K\ind{^{a_{1} \ldots a_{r}}}$ is a type-$(r, 0)$ Killing tensor, the scalar quantity
\begin{equation}
K = K\ind{^{a_{1} \ldots a_{r}}} p\ind{_{a_{1}}} \ldots p\ind{_{a_{r}}}
\end{equation}
is conserved along geodesics. A special case of this result is the so-called Hamiltonian constraint (see Section \ref{sec:geodesics}): the Hamiltonian function $H = \frac{1}{2} g\ind{^{a b}} p\ind{_{a}} p\ind{_{b}}$ is constant along dynamical trajectories (i.e., the geodesics), due to the fact that $g\ind{^{a b}}$ is a (trivial) Killing tensor.

\subsubsection{Conformal Killing vectors, conformal Killing tensors and null geodesics}

Our primary focus in this thesis is the propagation of electromagnetic radiation -- in particular, light rays -- on curved spacetime. We shall therefore now turn our attention to generalisations of Killing vectors and Killing tensors which provide conserved quantities only along \emph{null} geodesics. A \emph{conformal Killing vector} is a vector $\xi$ which satisfies
\begin{equation}
\label{eqn:conformal_killing_vector_equation}
\nabla\ind{^{( a }} \xi\ind{^{ b )}} = \alpha g\ind{^{a b}},
\end{equation}
for some function $\alpha$. This is known as the \emph{conformal Killing vector equation}. Clearly, if $\alpha = 0$, \eqref{eqn:conformal_killing_vector_equation} reduces to the Killing vector equation \eqref{eqn:killing_vector_equation}. Now, given a conformal Killing vector $\xi$, it is quick to show that the observable $I = \xi\ind{^{a}} p\ind{_{a}}$ is conserved along \emph{null} geodesics (which have $g\ind{^{a b}} p\ind{_{a}} p\ind{_{b}} = 0$):
\begin{equation}
\dot{I} = p\ind{^{b}} I\ind{_{; b}} = p\ind{^{b}} \left( \xi\ind{^{a}} p\ind{_{a}} \right)\ind{_{; b}} = p\ind{^{b}} \xi\ind{_{a ; b}} p\ind{^{a}} +  p\ind{^{b}} p\ind{_{a ; b}} \xi\ind{^{a}} = \xi\ind{_{( a ; b )}} p\ind{^{a}} p\ind{^{b}} = \alpha g\ind{^{a b}} p\ind{_{a}} p\ind{_{b}} = 0.
\end{equation}
It is also possible to show that both Killing vectors and conformal Killing vectors yield conserved quantities for any matter fields whose stress--energy tensors $T\ind{^{a b}}$ are (i) divergence-free ($\nabla\ind{_{a}} T\ind{^{a b}} = 0$), and (ii) traceless ($T = T\ind{^{a}_{a}} = 0$). In particular, the current $J\ind{^{a}} = T\ind{^{a b}} \xi\ind{_{b}}$ is conserved: $\nabla\ind{_{a}} J\ind{^{a}} = 0$ \cite{FrolovKrtousKubiznak2017}.

A symmetric tensor $K\ind{^{a_{1} \ldots a_{r}}}$ of type $(r, 0)$ is a \emph{conformal Killing tensor} if it satisfies the \emph{conformal Killing tensor equation}
\begin{equation}
\label{eqn:conformal_killing_tensor_equation}
\nabla\ind{^{( b}} K\ind{^{a_{1} \ldots a_{r} )}} = g\ind{^{( b a_{1} }} \alpha\ind{^{ a_{2} \ldots a_{r} )}} ,
\end{equation}
where $\alpha$ is a symmetric tensor of type $(r - 1, 0)$. Of course, in the case $\alpha = 0$, \eqref{eqn:conformal_killing_tensor_equation} reduces to the Killing tensor equation \eqref{eqn:killing_tensor_equation}. The observable $K = K\ind{^{a_{1} \ldots a_{r}}} p\ind{_{a_{1}}} \ldots p\ind{_{a_{r}}}$ is conserved along \emph{null} geodesics if $K\ind{^{a_{1} \ldots a_{r}}}$ is a conformal Killing tensor.

\subsubsection{The principal tensor}

We now turn our attention to a highly important geometrical object -- the \emph{principal tensor} -- which is responsible for a complete set of explicit and hidden symmetries, and whose existence uniquely determines the spacetime geometry, which is given by the Kerr--NUT--(anti-)de Sitter metric \cite{FrolovKrtousKubiznak2017}. The \emph{principal tensor} $h\ind{_{a b}}$ is a non-degenerate closed conformal Killing--Yano two-form, which obeys
\begin{equation}
\label{eqn:ccky_tensor_governing_equation}
\nabla\ind{_{c}} h\ind{_{a b}} = g\ind{_{c a}} \xi\ind{_{b}} - g\ind{_{c b}} \xi\ind{_{a}} ,
\end{equation}
where $\xi\ind{^{a}}$ is an associated vector field. Since $h$ is closed, there exists (locally) a potential one-form, say $b$, such that $h = \ed b$. By contracting \eqref{eqn:ccky_tensor_governing_equation} with $g\ind{^{c b}}$ and using the antisymmetry of $h\ind{_{ab}}$, one may show that, in four spacetime dimensions, the vector $\xi\ind{^{a}}$ takes the form
\begin{equation}
\label{eqn:killing_vector_principal_tensor}
\xi\ind{_{a}} = - \frac{1}{3} \nabla\ind{^{b}} h\ind{_{a b}}.
\end{equation}
It is possible to show that \cite{Krtous2007, KrtousFrolovKubiznak2008, HouriOotaYasui2009, YasuiHouri2011}
\begin{equation}
\mathcal{L}_{\xi} g = 0;
\end{equation}
in other words, $\xi\ind{^{a}}$ defined in \eqref{eqn:killing_vector_principal_tensor} is a Killing vector; this is referred to as the \emph{primary Killing vector}.

The existence of a principal tensor is responsible for some remarkable properties. The algebraic properties of the principal tensor may be exploited to construct a \emph{(special) Darboux frame} \cite{FrolovKrtousKubiznak2017}. Moreover, the principal tensor is responsible for the existence of a rich symmetry structure, referred to as the \emph{Killing tower}. In particular, the principal tensor can be used to generate (i) a family of closed conformal Killing--Yano forms, generated from exterior powers of the principal tensor; (ii) a family of Killing--Yano forms, which are the Hodge duals of the closed conformal Killing--Yano forms; (iii) rank-two Killing tensors, which are the ``squares'' of the Killing--Yano forms; (iv) rank-two conformal Killing tensors, which are the squares of the closed conformal Killing--Yano forms; and (v) Killing vectors, which are given by contracting the Killing tensors with the primary Killing vector. For a more detailed description of this procedure, see Chapter 2 of the review by Frolov \emph{et al.} \cite{FrolovKrtousKubiznak2017}.

The most general higher-dimensional black hole solution to Einstein's field equations -- the (on-shell) Kerr--NUT--(anti-)de Sitter metric -- admits a principal tensor. This result was first proved for Myers--Perry black holes \cite{FrolovKubiznak2007}, and later generalised to the more general family of Kerr--NUT--(anti-)de Sitter spacetimes \cite{KubiznakFrolov2007}. In fact, the most general vacuum solution with a cosmological constant to the Einstein field equations in any number of spacetime dimensions ($n \geq 4$) that admits a principal tensor is precisely the Kerr--NUT--(anti-)de Sitter metric \cite{HouriOotaYasui2007, KrtousFrolovKubiznak2008, HouriOotaYasui2009}.

Clearly, the very existence of a principal restricts the allowed geometry. In fact, the existence of this special geometrical object determines the geometry entirely: the most general geometry that is consistent with the existence of the principal tensor is the so-called \emph{off-shell} Kerr--NUT--(anti-)de Sitter geometry; see \cite{FrolovKrtousKubiznak2017} and references therein.

\subsubsection{The role of the principal tensor on Kerr spacetime}

Let us now illustrate some of the key concepts described above by using the Kerr metric as an example. As described in Section \ref{sec:black_hole_spacetimes}, the exterior of the Kerr spacetime (i.e., outside the outer event horizon) is usually expressed in terms of Boyer--Lindquist coordinates $\{ t, r, \theta, \phi \}$. One may instead express the metric in \emph{canonical coordinates} $\{ \tau = t - a \phi, r, y = a \cos{\theta}, \psi = \frac{\phi}{a} \}$, where $a$ is the spin of the black hole. The coordinates $r$ and $y$ are related to the eigenvalues of the principal tensor, and the coordinates $\tau$ and $\psi$ are related to the Killing vectors which correspond to stationarity and axisymmetry, respectively. In canonical coordinates, the line element reads
\begin{equation}
\ed s^{2} = g\ind{_{a b}} \ed x\ind{^{a}} \ed x\ind{^{b}} = - \frac{\Delta_{r}}{\Sigma} \left( \ed \tau + y^{2} \ed \psi \right)^{2} + \frac{\Delta_{y}}{\Sigma} \left( \ed \tau - r^{2} \ed \psi \right)^{2} + \frac{\Sigma}{\Delta_{r}} \, \ed r^{2} + \frac{\Sigma}{\Delta_{y}} \, \ed y^{2} ,
\end{equation}
where $\Sigma = r^{2} + y^{2}$, $\Delta_{r} = r^{2} - 2 M r + a^{2}$ and $\Delta_{y} = a^{2} - y^{2}$. Replacing the canonical coordinates by the Boyer--Lindquist coordinates, we arrive at the familiar form of the Kerr metric \eqref{eqn:kerr_line_element_boyer_lindquist}.

The Kerr metric admits a pair of Killing vectors, namely $\xi_{(\tau)} = \partial\ind{_{\tau}} = \partial\ind{_{t}}$ and $\xi_{(\psi)} = \partial\ind{_{\psi}} = a^{2} \, \partial\ind{_{t}} + a \, \partial\ind{_{\phi}}$. As discussed above, the Kerr spacetime admits a principal tensor (i.e., a closed conformal Killing--Yano two-form) which satisfies $h\ind{_{a b ; c}} = g\ind{_{c a}} {\xi_{(\tau)}}\ind{_{b}} - g\ind{_{c b}} {\xi_{(\tau)}}\ind{_{a}}$. The primary Killing vector can be derived from the principal tensor by noting that ${\xi_{(\tau)}}\ind{^{a}} = \frac{1}{3} h\ind{^{b a}_{; b}}$. In canonical coordinates, the principal tensor takes the form
\begin{equation}
h = y \, \ed y \wedge \left( \ed \tau - r^{2} \ed \psi \right) - r \, \ed r \wedge  \left( \ed \tau + y^{2} \ed \psi \right) .
\end{equation}
The fact that $h$ is closed means that it can be expressed in terms of a one-form potential $b$ ($h = \ed b$) which in canonical coordinates takes the form
\begin{equation}
b = - \frac{1}{2} \left[ \left( r^{2} - y^{2} \right) \ed \tau + r^{2} y^{2} \ed \psi \right] .
\end{equation}
The Hodge dual of the principal tensor is the Killing--Yano tensor $f = \vp{h}^{\star} h$; in component form this is $f\ind{_{a b}} = \frac{1}{2} \epsilon\ind{_{a b c d}} h\ind{^{c d}}$, where $\epsilon\ind{_{a b c d}}$ is the Levi-Civita tensor. In canonical coordinates, the Killing--Yano tensor is
\begin{equation}
f = r \, \ed r \wedge \left( \ed \tau - r^{2} \ed \psi \right) + y \, \ed y \wedge  \left( \ed \tau + y^{2} \ed \psi \right) .
\end{equation}
The Killing--Yano tensor has a totally antisymmetric derivative: $f\ind{_{a b ; c}} = f\ind{_{[a b ; c]}}$. The ``square'' of the Killing--Yano tensor is the Killing tensor $K\ind{_{a b}} = f\ind{_{a c}} f\ind{_{b}^{c}}$, which satisfies the Killing tensor equation $K\ind{_{(a b ; c)}} = 0$. The ``square'' of the conformal Killing--Yano tensor is the conformal Killing tensor $Q\ind{_{a b}} = h\ind{_{a c}} h\ind{_{b}^{c}}$, which satisfies the conformal Killing tensor equation $Q\ind{_{(a b ; c)}} = g\ind{_{( a b}} q\ind{_{c )}}$, where the rank-one tensor $q\ind{_{a}}$ can be expressed in terms of the principal tensor and the primary Killing vector as $q\ind{_{a}} = h\ind{_{a b}} {\xi_{(\tau)}}\ind{^{b}}$. The secondary Killing vector is related to the primary Killing vector and the Killing tensor via ${\xi_{(\psi)}} = - K\ind{^{a}_{b}} {\xi_{(\tau)}}\ind{^{b}}$.

The conserved quantities on phase space which are induced by the Killing vectors ${\xi_{(\tau)}}\ind{^{a}}$ and ${\xi_{(\psi)}}\ind{^{a}}$, the Killing tensor $K\ind{_{a b}}$, and the metric tensor $g\ind{_{a b}}$ (which is itself a trivial Killing tensor) are all independent and in involution (i.e., they mutually Poisson commute). This fact ensures that the geodesic motion on Kerr spacetime is completely integrable (see Section \ref{sec:integrability}). The separability of geodesic motion on Kerr spacetime was first demonstrated by Carter \cite{Carter1968b}. Subsequently, in 1973, Floyd \cite{Floyd1973} and Penrose \cite{Penrose1973} demonstrated that the Kerr spacetime admits a rank-two Killing tensor $K\ind{^{a b}}$ which can be constructed from the dual of the closed conformal Killing--Yano tensor, or more directly from the Killing--Yano tensor. The Killing tensor is responsible for the existence of a fourth ``hidden'' constant of the motion, which can be expressed in the form $K = K\ind{^{a b}} p\ind{_{a}} p\ind{_{b}}$; this is referred to as the \emph{Carter constant}. Thus, the principal tensor reveals ``hidden'' symmetries of Kerr spacetime which guarantee that the geodesic dynamics are integrable in the sense of Liouville.

There are a number of further important consequences for geodesic motion on Kerr spacetime, which are related to the existence of the principal tensor $f\ind{_{a b}}$. Firstly, the vector $L\ind{^{a}} = f\ind{^{a b}} p\ind{_{b}}$ satisfies the equation of parallel transport $p\ind{^{b}} \nabla\ind{_{b}} L\ind{^{a}} = 0$, so gyroscopic precession is straightforward \cite{Marck1983b}; this fact will be important in Chapter \ref{chap:geometric_optics_kerr}. Secondly, gravitational Faraday rotation of the polarisation plane of light is trivial: the complex quantity $z = \left( f\ind{^{a b}} + i h\ind{^{a b}} \right) p\ind{_{a}} v\ind{_{b}}$ is constant along a null ray with momentum $p\ind{_{a}}$ for any parallel-transported vector $v\ind{^{a}}$ \cite{Penrose1973}.

The hidden symmetry brought about by the principal tensor is also implicated in a number of other important results relating to the separability of wave equations on Kerr spacetime. Firstly, the Klein--Gordon equation $\Box \Phi = g\ind{^{a b}} \nabla\ind{_{a}} \nabla\ind{_{b}} \Phi = 0$ for a scalar field $\Phi$ admits separable solutions of the form $\Phi = e^{- i \omega t} e^{i m \phi} R(r) S(\theta)$ in Boyer--Lindquist coordinates, where $R(r)$ and $S(\theta)$ obey second-order ordinary differential equations \cite{BrillChrzanowskiPereiraEtAl1972}. Secondly, the solutions to the Dirac equation are separable, and the governing equations reduce to a set of two coupled ordinary differential equations \cite{Unruh1973, Chandrasekhar1976}. Thirdly, Maxwell's equations admit separable solutions for the Maxwell scalars of extreme spin weight $\pm 1$, given by $\phi_{0} = F\ind{_{a b}} k\ind{^{a}} m\ind{^{b}}$ and $\phi_{2} = F\ind{_{a b}} n\ind{^{a}} \ol{m}\ind{^{b}}$, where $\{ k, n, m, \ol{m} \}$ is a certain complex null tetrad (see Section \ref{sec:newman_penrose_tetrads}), and $F\ind{_{a b}}$ is the Faraday two-form (see Sections \ref{sec:einstein_maxwell_field_equations} and \ref{sec:electromagnetism_in_curved_spacetime}) \cite{Teukolsky1972}. Finally, the equations governing gravitational perturbations admit separable solutions for the Weyl scalars of extreme spin weight $\pm 2$, given by $\Psi_{0} = C\ind{_{a b c d}} k\ind{^{a}} m\ind{^{b}} k\ind{^{c}} m\ind{^{d}}$ and $\Psi_{4} = C\ind{_{a b c d}} \ol{m}\ind{^{a}} n\ind{^{b}} \ol{m}\ind{^{c}} n\ind{^{d}}$ (see Section \ref{sec:newman_penrose_weyl_scalars}) \cite{Teukolsky1973}.

\subsection{Non-integrable systems and chaos}
\label{sec:chaotic_dynamical_systems}

In Section \ref{sec:integrability}, we saw that, in Hamiltonian systems, the existence of a maximal number of independent conserved quantities which are in involution guarantees the Liouville integrability of motion in phase space. In such circumstances, the equations of motion can be solved by quadrature, either analytically or using numerical techniques. In general, however, this is almost never possible.

When a Hamiltonian system does not admit a maximal set of independent Poisson-commuting integrals of motion, i.e., the system is non-integrable, the trajectories have much less restriction on their motion in phase space: they are not required to lie on invariant tori. In these situations, which are generic, one observes \emph{sensitive dependence on initial conditions} in (regions of) phase space. Inevitably, \emph{chaotic} behaviour will occur in non-integrable systems.

In this section, we review some fundamental aspects of deterministic chaos, with a particular emphasis on chaotic scattering and open Hamiltonian systems; we also discuss some of the hallmarks of chaotic motion, such as fractal structures in phase space.

\subsubsection{Deterministic chaos and fractals in dynamical systems}

The solutions of simple non-linear dynamical systems can behave in an extremely complicated fashion \cite{Devaney1989, Glendinning1994}. Heuristically, a chaotic solution is aperiodic, and initially neighbouring solutions diverge exponentially in time. The former property indicates that the solution is \emph{irregular} or \emph{disordered} in some sense, and the latter property, known as \emph{sensitive dependence on initial conditions}, suggests that even the smallest changes in the initial state of the system can yield wildly different outcomes. Sensitive dependence on initial conditions makes long-term predictions of the system very difficult, despite the deterministic nature of equations which govern the dynamics. This type of behaviour is known as \emph{deterministic chaos} (or simply \emph{chaos}). A precise mathematical definitions of chaos in the context of maps (i.e., discrete-time dynamical systems) is given by Devaney \cite{Devaney1989}. Often, the term ``chaos'' is used to refer only to motion on compact manifolds. Compact means that systems with escapes are excluded; in such cases, the term ``chaotic scattering'' is used, rather than ``chaos''. Extensive reviews of deterministic chaos can be found in a range of textbooks and articles, e.g.~\cite{Devaney1989, Glendinning1994, AlligoodSauerYorke1996, BrinStuck2002, Ott2002, Gaspard2005}.

Chaotic behaviour is typically characterised using a variety of mathematical methods, which can be either qualitative or quantitative in nature. For example, a common diagnostic of chaos is the calculation of \emph{Lyapunov exponents} \cite{AlligoodSauerYorke1996}, which quantify the rate of separation of initially neighbouring trajectories.

Another hallmark of chaos is the presence of fractal structures in phase space \cite{AguirreVianaSanjuan2009}. Loosely speaking, a fractal is a complicated geometrical object which exhibits irregular structure at arbitrarily small scales \cite{AlligoodSauerYorke1996, Falconer1997, Falconer2004}. Much like with the notion of chaos, there have been a variety of attempts to give a precise mathematical definition of a fractal; however, many of these definitions are not general enough to capture the rich structure of many sets which are intuitively fractals. Falconer \cite{Falconer1997, Falconer2004} considers a fractal to be a set $F$ in Euclidean space which exhibits all or most of the following characteristic features.
\begin{enumerate}[(i)]
\item The set $F$ has a fine structure, i.e., it exhibits irregular detail at arbitrarily small scales.
\item Due to its ``irregularity'', $F$ cannot be described using the language of calculus or traditional Euclidean geometry, either locally or globally.
\item In many cases, $F$ will exhibit self-similarity or self-affinity, perhaps in a statistical or approximate manner.
\item The fractal dimension of $F$ is strictly greater than its topological dimension (for some suitable definition of fractal dimension, e.g.~the Hausdorff dimension).
\item A simple, perhaps recursive, procedure (e.g.~an iterated function system) can be used to generate $F$.
\item The set $F$ will often have a ``natural'' appearance.
\end{enumerate}

There exist a multitude of fractals which possess some or all of the properties listed above. A particularly interesting class of fractals are those which are invariant under simple families of transformations, including self-similar, self-affine, approximately self-similar, and statistically self-similar fractals. Some of the best-known examples of fractals are self-similar, e.g.~the middle-thirds Cantor set, the von Koch curve, the Sierpinski triangle and the Sierpinski carpet. Further popular examples of fractals are those which occur as the attractors of (mathematical) dynamical systems, such as the Mandelbrot set or Julia sets, both of which are the result of complex iterated function systems.

Fractals, however, are not only a mathematical curiosity: they arise naturally in non-linear dynamical systems, in such a way that the two concepts are intimately related. For a detailed account of the relationship between fractal structures and non-linear dynamics, we refer the reader to the review by Aguirre \emph{et al.} \cite{AguirreVianaSanjuan2009}.
%

\subsubsection{Poincar\'{e} sections}

We now turn to an important technique in chaotic dynamics (in the case of bounded motion), originally suggested by Poincar\'{e} \cite{Poincare1899}, which permits us to study the breakdown in integrability, the destruction of invariant tori, and the nature of motion in the gaps between the destroyed tori \cite{Berry1978}. This method is particularly well suited to Hamiltonian systems in two dimensions (although it can be employed in higher-dimensional systems). In such situations, the phase space is spanned by four canonical coordinates, say $\{ q\ind{^{1}}, q\ind{^{2}}, p\ind{_{1}}, p\ind{_{2}} \}$, and the motion is confined to an \emph{energy hypersurface} $\mathcal{E}$, a three-dimensional hypersurface described by the Hamiltonian constraint $H( q\ind{^{1}}, q\ind{^{2}}, p\ind{_{1}}, p\ind{_{2}} ) = h$, where $h$ is the (conserved) total energy. (We note that, in the case of null geodesic motion in general relativity, the Hamiltonian constraint is $H = 0$ for \emph{all} trajectories in phase space.)

Rather than trace out the orbits which wander through the three-dimensional energy hypersurface, it is advantageous to study a two-dimensional slice through the surface; this is known as a \emph{Poincar\'{e} section} (or \emph{surface of section}). A typical choice for the section is the surface $\mathcal{S}_{1}$, given by the intersection of the hypersurface $\mathcal{E}$ with $q\ind{^{2}} = 0$; this surface of section has coordinates $(q\ind{^{1}}, p\ind{_{1}})$. (Of course, one may define the surface of section $\mathcal{S}_{2}$ analogously, or choose a more general two-dimensional surface which intersects $\mathcal{E}$.) Specifying a point on $\mathcal{S}_{1}$ then completely determines a dynamical trajectory (modulo a direction): $q\ind{^{1}}$, $p\ind{_{1}}$ and $q\ind{^{2}} = 0$ are all specified, and a rearrangement of the Hamiltonian constraint $H( q\ind{^{1}}, q\ind{^{2}}, p\ind{_{1}}, p\ind{_{2}} ) = h$ allows us to determine $p\ind{_{2}}$ up to a sign (since $H$ is assumed to be quadratic in the momenta). By convention, this sign is taken to be positive.

A trajectory which begins on $\mathcal{S}_{1}$ will subsequently cross it repeatedly as it moves on $\mathcal{E}$. The \emph{Poincar\'{e} map} $P \colon \mathcal{S}_{1} \to \mathcal{S}_{1}$ maps a trajectory that begins on $\mathcal{S}_{1}$ to its next point of intersection with $\mathcal{S}_{1}$. The Poincar\'{e} map reduces a continuous-time dynamical system (a flow) on a three-dimensional energy surface to a discrete-time dynamical system (a map) on a two-dimensional surface of section. An important property of the map $P$ is that it is \emph{area preserving} \cite{Berry1978}.

The intersections of an orbit with the Poincar\'{e} section reveal whether the motion is integrable. If this is the case, the system will not explore the whole of the three-dimensional hypersurface $\mathcal{E}$ ergodically; rather, the motion will be restricted to a two-torus (see Section \ref{sec:integrability}), which intersects the surface of section in a closed curve. All iterates of this orbit under the action of the Poincar\'{e} map must lie on this curve. For \emph{periodic} orbits, some iterate will coincide with the initial position on the surface of section, with the order of the iterate depending on the commensurability of the system's characteristic frequencies; such orbits are fixed points of the map $P^{k}$, where $k$ is the order of the iterate. Most curves, however, are formed by the intersection of irrational (non-resonant) tori with the Poincar\'{e} section. All of these curves are \emph{invariant curves} of the Poincar\'{e} map, as they are mapped to themselves by $P$.

Consider a perturbed system with Hamiltonian $H = H_{0} + \varepsilon H_{1}$, where $H_{0}$ is the Hamiltonian of an integrable system. When $\varepsilon \ll 1$, the system is said to be \emph{near-integrable}. As one increases $\varepsilon$, the system undergoes a transition from integrability to non-integrability, which obeys the Kolmogorov--Arnold--Moser (KAM) theorem \cite{Tabor1989}. This says that for sufficiently small perturbations $\varepsilon$, almost all non-resonant tori of the integrable system survive, but are slightly deformed. Such tori are known as \emph{KAM tori}. Here, ``almost all'' means that the measure of the complement of the KAM tori is small, and tends to zero with the magnitude of the perturbation $\varepsilon$. The surviving tori are those which are ``sufficiently irrational'' \cite{OzoriodeAlmeida1990, Lowenstein2012}.

In the case of non-integrable systems, the invariant tori do not exist, and orbits explore a three-dimensional region of the energy hypersurface $\mathcal{E}$. The intersection of such orbits with the Poincar\'{e} section do not lie on a curve, but instead cover a two-dimensional region of $\mathcal{S}_{1}$.

A typical Poincar\'{e} section will exhibit invariant curves, resonant ``islands'' (i.e., intersections of the nested tori with the surface of section), and chaotic regions; the latter may border the islands, or fill more extensive regions of the surface of section. The ordered phase space of the integrable system becomes highly intricate under perturbations. As the perturbation is increased, the system moves further away from integrability; more and more invariant tori are destroyed, and the chaotic regions grow and merge with one another, filling up the surface of section.

The usefulness of Poincar\'{e} sections is best demonstrated by considering an example. In Appendix \ref{chap:appendix_b}, we illustrate the method by applying it to the H\'{e}non--Heiles Hamiltonian system \cite{HenonHeiles1964} as a pedagogical example. This will be of particular importance in Chapter \ref{chap:stable_photon_orbits}, in which we use Poincar\'{e} sections to analyse stable bounded null geodesics around a pair of black holes.
%

\subsubsection{Chaotic scattering}

We now turn our attention to the case of \emph{unbounded} chaotic motion, in which trajectories are permitted to reach infinity. Motion of this type is referred to as \emph{chaotic scattering}. In its most general sense, a \emph{scattering system} is the problem of obtaining a relationship between an ``input variable'' (e.g.~the impact parameter, or the initial angle), which characterises an \emph{initial state} of a dynamical system (usually a Hamiltonian system), and an ``output variable'' (e.g.~the scattering angle, or some discrete label), which characterises the \emph{final state} of the dynamical system \cite{OttTel1993}. Typically, a scattering problem deals with the motion of a particle described by the Hamiltonian
\begin{equation}
H(q, p) = H_{0}(q, p) + V(q) ,
\end{equation}
where $V$ is a potential which is zero (or sufficiently small) outside of some bounded region of space which is referred to as the \emph{scattering region} \cite{Ott2002}. The particle can effectively be treated as a free particle, governed by the Hamiltonian $H_{0}$, far from the scattering region.

\emph{Chaotic scattering} is the study of scattering systems which exhibit extreme sensitive dependence on initial conditions. Eckhardt \cite{Eckhardt1987} defines scattering in a Hamiltonian system as \emph{irregular} (or \emph{chaotic}) if there exists, on some manifold of initial data, an infinity of distinct scattering singularities of zero (Lebesgue) measure, which are typically arranged into a fractal set. In this context, a \emph{scattering singularity} is an initial value (input) for which the scattering process (output) is undefined. In regions of chaotic scattering (i.e., in neighbourhoods of scattering singularities), small perturbations in the initial conditions lead to completely different outcomes (final states) in the scattering process.

In a general scattering system, the \emph{classical repellor}, denoted $\Omega_{\text{R}}$, is the set of all scattering singularities, which correspond to non-escaping orbits \cite{GaspardRice1989, Gaspard2005}. In general, $\Omega_{\text{R}}$ may be a simple set which contains a finite or countable number of scattering singularities, corresponding to periodic orbits. On the other hand, $\Omega_{\text{R}}$ may be a fractal set (of measure zero) which comprises a countable set of periodic orbits, and an uncountable set of aperiodic orbits. The former case corresponds to regular scattering, whilst the latter represents chaotic scattering. If the repellor is a fractal, it is referred to as a \emph{strange repellor}.

Chaotic scattering occurs in wide variety of contexts, including the three-disc model \cite{Eckhardt1987, GaspardRice1989}, and the H\'{e}non--Heiles Hamiltonian system of celestial mechanics \cite{HenonHeiles1964}. The phenomenology of chaotic scattering is reviewed in a range of references, e.g.~\cite{Eckhardt1988, Gaspard2005, SeoaneSanjuan2012}.

\subsubsection{Open Hamiltonian systems and exit basins}

An \emph{open Hamiltonian system} is a Hamiltonian dynamical system in which the particle is permitted to escape from the scattering region to infinity, via one ore more \emph{escapes} (or \emph{exits}) in phase space \cite{AguirreVianaSanjuan2009}.

The dynamics of open Hamiltonian systems can be understood in more detail by constructing exit basins in phase space, in a fashion analogous to the basins of attraction for dissipative dynamical systems \cite{Ott1981, Milnor1985, EckmannRuelle1985}. Dissipative dynamical systems often possess more than one \emph{attractor}. A \emph{basin of attraction} is defined as a subset of state space such that all initial conditions chosen from this set asymptote towards a given attractor in forward time. In an open Hamiltonian system with $N \geq 2$ exits in phase space, the \emph{exit basin} corresponding to the $i$th exit in phase space is defined to be the set of all initial conditions which escape the scattering region through exit $i$; see Section \ref{sec:chaotic_dynamical_systems} and references therein.

When more than one basin is present, the basins will be separated by a \emph{basin boundary}. A point $p$ is a \emph{boundary point} of the basin $B$ if $b_{\varepsilon}(p) \cap B \neq \emptyset$ and $b_{\varepsilon}(p) \cap B^{\text{c}} \neq \emptyset$ for all $\varepsilon > 0$, where $B^{\text{c}}$ is the complement of the set $B$, and $b_{\varepsilon}(p)$ denotes the open ball of radius $\varepsilon$ centred on $p$. The basin boundary, denoted $\partial B$, is then the set of all boundary points of $B$. 

The exit basins of an open Hamiltonian system may be visualised by integrating the equations of motion for a fine grid of initial conditions, then colour-coding the initial data according the escape through which the corresponding trajectories leave the scattering region. The result of this procedure is a so-called \emph{exit basin diagram} \cite{AguirreVianaSanjuan2009}, which provides a useful tool with which one can analyse the global dynamics of open Hamiltonian systems. In Hamiltonian systems with more than two degrees of freedom (i.e., the phase space has dimension four or higher), one must use a Poincar\'{e} section to realise the exit basin diagram. This involves choosing an one- or two-dimensional surface in phase space, from which the initial conditions are chosen. The trajectories are then integrated forwards in time, and the initial data on the surface are colour-coded according to the final state of the corresponding trajectory. (Note that, in this case, the motion is unbounded, so the Poincar\'{e} map is not well-defined.)

If the system under consideration is chaotic, one typically observes \emph{fractal} structures in the exit basin diagrams; in particular, the boundaries between any two exit basins may be fractal. In this case, there exist a variety of methods which can be used to characterise (both qualitatively and quantitatively) the nature of these fractal structures and the chaotic dynamics; see the review by Aguirre \emph{et al.} \cite{AguirreVianaSanjuan2009} for further details.
%

\subsubsection{Chaos in general relativity}

General relativity is an inherently non-linear theory. The Einstein field equations \eqref{eqn:einstein_field_equations}, which govern the geometry of spacetime, are a set of coupled non-linear partial differential equations; this means that the spacetime dynamics can, in principle, exhibit chaotic behaviour. This has been the subject of a range of studies over the last four decades, with a particular emphasis on the study of chaos in the spacetime dynamics of cosmological Bianchi IX (Mixmaster) models \cite{ChernoffBarrow1983, HobillBurdColey1994, CornishLevin1997}.

During the last thirty years, investigations of chaos in general relativity have followed an alternative route: the analysis of geodesic motion on a \emph{fixed} spacetime background. It is common to start with a background on which the geodesic motion is integrable (e.g. Schwarzschild or Kerr), and then consider the effect of various types of ``perturbations''. These perturbations typically reduce the symmetry required for complete integrability, and rich chaotic behaviour emerges naturally. Such behaviour can be characterised using a variety of methods. The fact that geodesic motion in general relativity is amenable to Hamiltonian methods means that existing techniques can be borrowed from the theory of non-linear dynamical systems. However, care must be taken to ensure that these descriptors of chaos are coordinate-invariant. Here, we review some key studies of chaos in geodesic motion in general relativity. (Further details can be found in the comprehensive summary presented by Semer\'{a}k and Sukov\'{a} in \cite{SemerakSukova2010}.)

A range of authors have considered the effect of endowing a test particle with spin, looking at the dynamics of the Mathisson--Papapetrou--Dixon equations on Schwarzschild spacetime \cite{SuzukiMaeda1997}, and Kerr spacetime \cite{Hartl2003a, Hartl2003b}. Others have focussed on chaos in gravitational waves. In 1992, Bombelli and Calzetta \cite{BombelliCalzetta1992} studied the influence of small periodic perturbations (which serve as a model for weak gravitational waves) on Schwarzschild spacetime. Further studies of chaos in gravitational waves were carried out by Levin \cite{Levin2000} and others \cite{LetelierVieira1997, Cornish2001}.

Various authors have looked at perturbing the background itself by endowing it with higher multipoles. For example, Vieira and Letelier \cite{VieiraLetelier1996} analysed the effect of perturbations by quadrupolar and octupolar moments on the timelike geodesics of Schwarzschild spacetime. Later, by considering exit basins in phase space, de Moura and Letelier \cite{DeMouraLetelier2000} considered the escape dynamics of timelike and null geodesics from a static black hole endowed with dipolar, quadrupolar and octupolar moments.

Rather than artificially superposing a regular background with higher order multipoles, one may also analyse exact solutions which incorporate these multipoles. Such solutions are typically stationary and axisymmetric. In 1996, Sota \emph{et al.} \cite{SotaSuzukiMaeda1996} studied the Zipoy--Voorhees family of solutions, which includes Minkowski, Schwarzschild and Chazy--Curzon spacetimes as special cases. The authors used techniques such as Poincar\'{e} sections and Lyapunov exponents to diagnose chaos in the motion of free test particles.

Motivated by astrophysical scenarios, a range of studies have looked at relativistic core--shell models, such as the superposition of a black hole and a self-gravitating disc or ring \cite{DeMouraLetelier2000}; see \cite{SemerakSukova2010} and references therein for a review.

Finally, there has been much interest in the analysis of regular and chaotic dynamics in the motion of free particles around two (or more) fixed centres. This began with the work of Contopoulos \cite{Contopoulos1990, Contopoulos1991, Contopoulos2002}, who demonstrated that the motion of free particles around two fixed black holes in the Majumdar--Papapetrou spacetime is chaotic. This topic will be of central importance in Chapters \ref{chap:binary_black_hole_shadows}--\ref{chap:stable_photon_orbits} of this thesis; a review of key contributions to this field will be provided there.

\chapter{Gravitational lensing} \label{chap:gravitational_lensing}

\section{Introduction}

\emph{Gravitational lensing} is a general term which accounts for all effects of a gravitational field on the propagation of electromagnetic radiation \cite{Perlick2004}. Typically, this electromagnetic radiation is described by light rays, using a high-frequency (geometrical optics) approximation. The gravitational field which affects these light rays can itself be described using a variety of approaches.

In situations where the gravitational field is ``weak'', quasi-Newtonian approximations may be employed to reduce the formalism of general relativity to a Newtonian setting; this is often referred to as \emph{perturbative lensing}. The review by Wambsganss \cite{Wambsganss1998} provides a comprehensive overview of observational aspects and key methods of perturbative lensing, with an emphasis on gravitational lensing as an astrophysical tool. Due to the fact that the quasi-Newtonian formalism applies in situations where the magnitude of the Newtonian gravitational potential is small (i.e., in the weak-field limit), perturbative lensing is sometimes referred to as \emph{weak-field gravitational lensing}.

On the other hand, one may analyse \emph{lensing from a spacetime perspective} (also known as \emph{non-perturbative lensing}). In this approach, the propagation of light rays is described by null geodesics on a (four-dimensional) Lorentzian spacetime, which is a solution to the Einstein field equations of general relativity. Of course, the latter condition may be dropped in favour of an alternative theory of gravity, if so desired. For an in-depth account of gravitational lensing from a spacetime perspective, see the review by Perlick \cite{Perlick2004}, and references therein. Non-perturbative lensing can be used to describe gravitational phenomena in both weak- and strong-field regimes. In situations where we are only concerned with the latter, we use the terminology \emph{strong-field gravitational lensing}.


In this chapter, we present a brief review of gravitational lensing. Section \ref{sec:weak_field_gravitational_lensing} concerns perturbative lensing. We begin with a short historical account of the origins of the theory, illustrated by important theoretical and observational milestones; we then review the list of observed gravitational lensing phenomena; and we present a brief account of the mathematical formalism of perturbative lensing. In Section \ref{sec:strong_field_gravitational_lensing}, we discuss key elements of the theory of non-perturbative gravitational lensing, within the framework of four-dimensional general relativity. First, we review Maxwell's theory of electromagnetism in curved spacetime. We then describe the (leading-order) geometrical optics approximation, which allows the electromagnetic field to be described in terms of null rays on curved spacetime. We conclude with an introduction to some key features of gravitational lensing by black holes, including light-rings (i.e., unstable photon orbits) and black hole shadows.

\section{Perturbative lensing}
\label{sec:weak_field_gravitational_lensing}

\subsection{History of gravitational lensing}
\label{sec:history_gravitational_lensing}

The idea of light rays being affected by a gravitational field can be traced back to Isaac Newton. In his 1704 treatise \emph{Opticks}, Newton queries whether (gravitational) bodies act upon light at a distance, and by their action bend its rays \cite{Newton1979}. It was not until 1784 that Henry Cavendish calculated the deflection angle of light due to a point mass, using Newton's gravitational theory and Newtonian optics, in which light is assumed to be composed of corpuscles that are affected by a gravitational field in the same way as material particles.

Interestingly, it was Cavendish's correspondence with his friend and colleague John Michell which sparked the former's interest in the effect of gravity on light. In 1783, Michell wrote a letter to Cavendish \cite{Michell1784} which proposed a method to measure the mass of stars by measuring the reduction in the speed of the light corpuscles which propagate from the star to the Earth. It was this paper and the correspondence with Michell that prompted Cavendish to consider the deflection of light due to a gravitational field. Cavendish, however, never published his calculation of light deflection; evidence of this work was only found after Cavendish's death \cite{Cavendish2011}.

In his 1783 letter, Michell also proposed that there might exist ``dark stars'', objects which are so dense that not even light can escape from their surface; such a body would therefore be invisible, and only detectable by its gravitational influence on nearby stars \cite{Schaffer1979}. This proposal is now widely quoted as the origin of the concept of a black hole -- an object so dense that not even light can escape its gravitational pull. Over a decade later, Pierre-Simon Laplace, independently of Michell, calculated the conditions required for the escape velocity from the surface of a body to be greater than the velocity of light \cite{Laplace2009}. Curiously, it is the work of Laplace that inspired Johann Georg von Soldner to provide a detailed calculation of the deflection angle suffered by a light ray due to the gravitational attraction of a massive body \cite{Soldner1801, Jaki1978}: von Soldner found that a ray passing close to a body of mass $M$ is bent through the angle
\begin{equation}
\alpha_{\text{N}} = \frac{2 G M}{c^{2} b} ,
\end{equation}
where $b$ is the ray's perpendicular impact parameter, $G$ is Newton's gravitational constant, and $c$ is the speed of light. For a ray which grazes the limb of the Sun, the deflection angle is $\alpha_{\odot} \approx 0.87~\text{arcsec}$. An account of the contributions of Cavendish and von Soldner to the field of gravitational lensing is provided by Will \cite{Will1988}.

In 1911, Albert Einstein calculated the deflection angle of light by the Sun using the equivalence principle of his special theory of relativity \cite{Einstein1911}. The result of Einstein's calculation matched that of von Soldner, of whom he knew nothing about. A few years later, in 1914, at the request of Einstein, the German astronomer Erwin Freundlich led an expedition to the Crimean peninsula, with the aim of observing light deflection by the Sun during a total solar eclipse. Unfortunately, the outbreak of the First World War prevented Freundlich's team from completing the expedition \cite{Dodelson2017}.

Equipped with his recently developed general theory of relativity, Einstein revisited the calculation of light deflection in 1915. Using a linearised version of the field equations, Einstein found that the deflection angle was actually twice the Newtonian prediction of Cavendish and von Soldner \cite{Einstein1915, Will2006}:
\begin{equation}
\label{eqn:einstein_deflection_angle}
\alpha_{\text{E}} = \frac{4 G M}{c^{2} b} ,
\end{equation}
with $\alpha_{\odot} \approx 1.75~\text{arcsec}$ for a ray which grazes the Sun's surface. Using the Schwarzschild solution to Einstein's field equations, which describes a spherically symmetric non-rotating source (see Section \ref{sec:black_holes}), the general-relativistic deflection angle can be calculated exactly in terms of an elliptic integral \cite{Darwin1959, BisnovatyiKoganTsupko2008}.

The theoretical prediction of light deflection by the Sun due to general relativity was verified experimentally during a total solar eclipse in May 1919, by a team of astronomers led by Sir Frank Watson Dyson and Sir Arthur Eddington \cite{DysonEddingtonDavidson1920}. The accuracy of the 1919 observations was sufficient to distinguish between the predictions of Newtonian gravity and Einstein's theory of general relativity. This observation helped to establish the general relativity as a fundamental physical theory of gravitation.

In 2004, the deflection by the Sun of radio waves emitted by distant compact radio sources was measured using very-long-baseline interferometry; this confirmed the general-relativistic prediction of $1.75~\text{arcsec}$ to within $0.2\%$ \cite{ShapiroDavisLebachEtAl2004}. More recently, during the total solar eclipse of August 2017, Einstein's value was observed for visible light, with an accuracy of $3\%$ \cite{Bruns2018}.
%

\subsection{Gravitational lensing phenomena}
\label{sec:gravitational_lensing_phenomena}

The success of the 1919 observations of light deflection by the Sun marked the birth of gravitational lensing. In the decades that followed, gravitational lensing was developed as a theoretical tool.

Chwolson \cite{Chwolson1924} proposed the idea of a ``fictitious double star'' in 1924, and described the symmetric case in which the source, lens and observer are perfectly aligned, resulting in a ring-like image of the source. In 1936, Einstein described the result of the latter lens configuration as a ``luminous circle'' \cite{Einstein1936}; today this phenomenon is known as an \emph{Einstein ring}. Curiously, Einstein disregarded his results, claiming that there was no hope of observing such an effect directly.

One year later, influenced by the work of Einstein, Zwicky \cite{Zwicky1937} posited that distant galaxies could act as sources and as gravitational lenses. Zwicky noted that galaxy--galaxy lensing could be used as an observational tool to test general relativity, and that these lensing effects could be used as a natural telescope to observe distant galaxies, due to the magnification effect of gravitational lensing.

The field of gravitational lensing remained dormant until the early 1960s, when interest in the subject was reignited by the theoretical developments of Klimov \cite{Klimov1963}, Liebes \cite{Liebes1964} and Refsdal \cite{Refsdal1964a}. In particular, Refsdal  established the quasi-Newtonian formalism for gravitational lensing, based on the assumptions of weak gravitational fields and small deflection angles; and demonstrated that gravitational lensing could be used as a tool to measure the Hubble constant by measuring the time delay between two images of a single source \cite{Refsdal1964b}. Interestingly, earlier that decade, Sachs \cite{Sachs1961} provided a fully covariant formulation of how a bundle of light rays is distorted by a gravitational field in the context of gravitational radiation; this will be discussed in Section \ref{sec:strong_field_gravitational_lensing}.

The first detection of a quasar -- a luminous, compact and very distant source -- came in 1963 \cite{Schmidt1963}. Such an object would be an ideal source for Zwicky's galaxy--galaxy lensing, and it was hoped that the observation of gravitational lensing would soon follow. Fifteen years later, Walsh \emph{et al.} \cite{WalshCarswellWeymann1979} discovered the ``twin quasar'' Q0957+561, and confirmed that this was in fact one quasar, of which two images are produced by gravitational lensing. This was the first detection of a gravitational lensing effect since the observation of light deflection by the Sun in 1919, and signalled the birth of gravitational lensing as an observational field \cite{Wambsganss1998}.

To date, the list of observed lensing phenomena includes multiply imaged quasars; rings; giant luminous arcs; image distortion by weak (or statistical) lensing; and galactic microlensing. Detailed reviews of these lensing effects can be found in a range of sources; see for example \cite{Wambsganss1998, SchneiderEhlersFalco1992, SchneiderKochanekWambsganss2006, Dodelson2017} and references therein.
%

\subsection{Perturbative lensing formalism}

In this section, we review some basic aspects of the mathematical formalism of perturbative gravitational lensing. In its most general sense, solving for the motion of light rays on an arbitrary curved spacetime is a non-trivial theoretical task. However, in many cases, one may assume that the spacetime geometry is well-described by the Friedmann--Lema\^{i}tre--Robertson--Walker metric, and that the matter inhomogeneities which are responsible for gravitational lensing can be treated as (small) local perturbations. The mathematical formalism of linearised gravity and perturbative gravitational lensing is reviewed in a range of sources, e.g.~\cite{Wambsganss1998, SchneiderEhlersFalco1992, NarayanBartelmann1996, SchneiderKochanekWambsganss2006, Dodelson2017}. Here, we follow the discussion presented by Carroll \cite{Carroll2019}.

For simplicity, we restrict our attention to cases where the background spacetime is well described by the flat Minkowski metric $\ed s^{2} = \eta\ind{_{a b}} \ed x\ind{^{a}} \ed x\ind{^{b}} = - \ed t^{2} + \ed x^{2} + \ed y^{2} + \ed z^{2}$. The fact that the gravitational field is weak means that one may express the spacetime metric as
\begin{equation}
\label{eqn:perturbed_metric_linearised_gravity}
g\ind{_{a b}} = \eta\ind{_{a b}} + h\ind{_{a b}}, \qquad \left| h\ind{_{a b}} \right| \ll 1 ,
\end{equation}
where $h\ind{_{a b}}$ is regarded as a small perturbation, and $\left| h\ind{_{a b}} \right|$ denotes the magnitude of a typical non-zero component of the perturbation. (Note that we returned to units in which $c = 1 = G$ for simplicity.) In linearised gravity, we keep only terms which are linear in the perturbation $h\ind{_{a b}}$; higher-order terms are discarded. It follows that $g\ind{^{a b}} = \eta\ind{^{a b}} - h\ind{^{a b}}$, where $h\ind{^{a b}} = \eta\ind{^{a c}} \eta\ind{^{b d}} h\ind{_{c d}}$. A key consequence of this is that spacetime indices may be lowered and raised using the Minkowski metric and its inverse, as any corrections will be higher than first order in the perturbation $h\ind{_{a b}}$.

Typically, one assumes that the source of the gravitational field is static and Newtonian. Moreover, the gravitational source is modelled as dust, a perfect fluid with vanishing pressure. In this approximation, the perturbed metric reads
\begin{equation}
\label{eqn:static_newtonian_line_element}
\ed s^{2} = g\ind{_{a b}} \ed x\ind{^{a}} \ed x\ind{^{b}} = - \left( 1 + 2 \Phi \right) \ed t^{2} + \left( 1 - 2 \Phi \right) \left( \ed x^{2} + \ed y^{2} + \ed z^{2} \right) ,
\end{equation}
in Cartesian coordinates $\left\{ t, x, y, z \right\}$. Comparing \eqref{eqn:perturbed_metric_linearised_gravity} and \eqref{eqn:static_newtonian_line_element}, we see that $h\ind{_{a b}} = - 2 \Phi \, \delta\ind{_{a b}}$ and $\left| \Phi \right| \ll 1$. The Newtonian gravitational potential $\Phi$ satisfies Poisson's equation $\nabla^{2} \Phi = 4 \pi \varrho$, where $\varrho$ is the mass density of the gravitational source in the rest frame of the dust.

We are interested in the propagation of light rays on the geometry \eqref{eqn:static_newtonian_line_element}. We therefore wish to solve the perturbed null geodesic equation for a ray $x\ind{^{a}}(\lambda)$, where $\lambda$ is an affine parameter. As in the case of the metric tensor \eqref{eqn:perturbed_metric_linearised_gravity}, we decompose the ray into a background path (in Minkowski spacetime) plus a perturbation:
\begin{equation}
\label{eqn:perturbed_null_geodesic}
x\ind{^{a}}(\lambda) = {x_{(0)}}\ind{^{a}}(\lambda) + {x_{(1)}}\ind{^{a}}(\lambda) .
\end{equation}
Here, ${x_{(0)}}\ind{^{a}}(\lambda)$ solves the geodesic equation on Minkowski spacetime; the path is therefore a straight line. To solve for the perturbation ${x_{(1)}}\ind{^{a}}(\lambda)$, we adopt the \emph{Born approximation} \cite{Dodelson2017}, i.e., we evaluate all quantities along the background path rather than the actual path. This procedure is valid because we assume that $\left| \Phi \right| \ll 1$, so we expect the deviation to be small.

We introduce the notation
\begin{equation}
u\ind{^{a}} = (u^{0}, \mathbf{u}) = {\dot{x}_{(0)}} \vp{x}\ind{^{a}} , \qquad
v\ind{^{a}} = (v^{0}, \mathbf{v}) = {\dot{x}_{(1)}} \vp{x}\ind{^{a}} ,
\end{equation}
where an overdot denotes differentiation with respect to the affine parameter $\lambda$. The path \eqref{eqn:perturbed_null_geodesic} must be null, i.e., $g\ind{_{a b}} \dot{x}\ind{^{a}} \dot{x}\ind{^{b}} = 0$. This equation must be solved order-by-order in the perturbation. At zeroth order, this is nothing more than $\eta\ind{_{a b}} u\ind{^{a}} u\ind{^{b}} = 0$, which says that $u\ind{^{a}}$ a null vector with respect to the background (Minkowski) spacetime; this condition may be written
\begin{equation}
- \left( u\ind{^{0}} \right)^{2} + \left| \mathbf{u} \right|^{2} = 0 .
\end{equation}
We define $u^{2} = \left( u\ind{^{0}} \right)^{2} = \left| \mathbf{u} \right|^{2}$. At first order in the perturbation, the null condition is $2 \eta\ind{_{a b}} u\ind{^{a}} v\ind{^{b}} + h\ind{_{a b}} u\ind{^{a}} u\ind{^{b}} = 0$, which may be written in the form
\begin{equation}
\label{eqn:first_order_perturbation_null_condition}
- u\ind{^{0}} v\ind{^{0}} + \mathbf{u} \cdot \mathbf{v} = 2 u^{2} \Phi .
\end{equation}

The perturbed geodesic equation is
\begin{equation}
\label{eqn:perturbed_geodesic_equation}
\ddot{x}\ind{^{a}} + \Gamma\ind{^{a}_{b c}} \dot{x}\ind{^{b}} \dot{x}\ind{^{c}} = 0 ,
\end{equation}
where, to first order in the perturbation, the Christoffel symbols are given by
\begin{equation}
\Gamma\ind{^{a}_{b c}}
= \frac{1}{2} g\ind{^{a d}} \left( \partial\ind{_{b}} g\ind{_{d c}} + \partial\ind{_{c}} g\ind{_{b d}} - \partial\ind{_{d}} g\ind{_{b c}} \right)
= \frac{1}{2} \eta\ind{^{a d}} \left( \partial\ind{_{b}} h\ind{_{d c}} + \partial\ind{_{c}} h\ind{_{b d}} - \partial\ind{_{d}} h\ind{_{b c}} \right) .
\end{equation}
For the static Newtonian gravitational field with line element \eqref{eqn:static_newtonian_line_element}, the only non-zero Christoffel symbols are
\begin{align}
\Gamma\ind{^{0}_{i 0}} &= \partial\ind{_{i}} \Phi , &
\Gamma\ind{^{i}_{0 0}} &= \partial\ind{_{i}} \Phi , &
\Gamma\ind{^{i}_{j k}} &= \delta\ind{_{j k}} \partial\ind{_{i}} \Phi - \delta\ind{_{i k}} \partial\ind{_{j}} \Phi - \delta\ind{_{i j}} \partial\ind{_{k}} \Phi .
\end{align}
At zeroth order, \eqref{eqn:perturbed_geodesic_equation} reduces to $\dot{u}\ind{^{a}} = 0$. This tells us that ${x_{(0)}}\ind{^{a}}(\lambda)$ is a straight line (i.e., a geodesic in Minkowski spacetime), as anticipated. At first order in the perturbation, \eqref{eqn:perturbed_geodesic_equation} yields $\dot{v}\ind{^{a}} + \Gamma\ind{^{a}_{b c}} u\ind{^{b}} u\ind{^{c}} = 0$. The temporal and spatial components of this equation are
\begin{equation}
\label{eqn:components_perturbation_wave_vector}
\dot{v}\ind{^{0}} = - 2 u \left( \mathbf{u} \cdot \bnab \Phi \right), \qquad
\dot{\mathbf{v}} = - 2 u^{2} \bnab_{\bot} \Phi ,
\end{equation}
where the \emph{transverse gradient} is defined as
\begin{equation}
\bnab_{\bot} \Phi = \bnab \Phi - \bnab_{\parallel} \Phi = \bnab \Phi - \frac{1}{u^{2}} \left( \mathbf{u} \cdot \bnab \Phi \right) \mathbf{u} .
\end{equation}

To first order in the perturbation, $\mathbf{u}$ is orthogonal to $\mathbf{v}$. To demonstrate this, one may obtain an expression for $v\ind{^{0}}$ by integrating the first equation of \eqref{eqn:components_perturbation_wave_vector} along the background path:
\begin{equation}
\label{eqn:temporal_component_perturbation_wave_vector}
v\ind{^{0}}
= - 2 u \int \left( \mathbf{u} \cdot \bnab \Phi \right) \ed \lambda
= - 2 u \int \left( \bnab \Phi \cdot \dot{\mathbf{x}}_{(0)} \right) \ed \lambda
= - 2 u \int \bnab \Phi \cdot \ed \mathbf{x}_{(0)}
= - 2 u \Phi ,
\end{equation}
where the integration constant is chosen to ensure that $v\ind{^{0}} = 0$ when $\Phi = 0$, which comes from the fact that the deviation vanishes in the absence of gravitational sources. Inserting \eqref{eqn:temporal_component_perturbation_wave_vector} into \eqref{eqn:first_order_perturbation_null_condition} shows that $\mathbf{u} \cdot \mathbf{v} = u\ind{^{0}} v\ind{^{0}} + 2 u^{2} \Phi = 0$, as claimed. Alternatively, one may note that
\begin{equation}
\label{eqn:conserved_quantity_background_ray}
\frac{\ed}{\ed \lambda} \left( \mathbf{u} \cdot \mathbf{v} \right)
= \dot{\mathbf{u}} \cdot \mathbf{v} + \mathbf{u} \cdot \dot{\mathbf{v}}
= - 2 u^{2} \left( \mathbf{u} \cdot \bnab_{\bot} \Phi \right)
= 0 ,
\end{equation}
where we have used \eqref{eqn:components_perturbation_wave_vector}. The quantity $\mathbf{u} \cdot \mathbf{v}$ is therefore conserved along the background ray. Of course, there is some freedom in the choice of initial direction for the perturbation. Choosing this to point along the background ray is equivalent to rescaling the affine parameter $\lambda$. Our choice is to take $\mathbf{u} \cdot \mathbf{v} = 0$ initially; \eqref{eqn:conserved_quantity_background_ray} ensures that this remains so along the background ray.

The deflection angle $\boldsymbol{\alpha}$ (discussed in Section \ref{sec:history_gravitational_lensing}) is the amount by which the spatial wave vector $\dot{\mathbf{x}}$ is deflected as it travels from the light source to the observer in the gravitational field \eqref{eqn:static_newtonian_line_element}. The deflection angle is a two-dimensional vector which lies in the plane perpendicular to $\mathbf{u}$; this may be expressed in the form
\begin{equation}
\label{eqn:deflection_angle_change_in_wave_vector}
\boldsymbol{\alpha} = - \frac{\Delta \mathbf{v}}{u} ,
\end{equation}
where the minus sign is chosen to account for the fact that the deflection is measured by an observer looking backwards along the null ray. The quantity $\Delta \mathbf{v}$ can be calculated from the second equation of \eqref{eqn:components_perturbation_wave_vector}:
\begin{equation}
\Delta \mathbf{v} = \int \dot{\mathbf{v}} \, \ed \lambda = - 2 u^{2} \int \bnab_{\bot} \Phi \, \ed \lambda .
\end{equation}
One can then express deflection angle \eqref{eqn:deflection_angle_change_in_wave_vector} as an integral over the background path as
\begin{equation}
\label{eqn:deflection_angle_integral}
\boldsymbol{\alpha} = 2 \int \bnab_{\bot} \Phi \, \ed \ell ,
\end{equation}
where $\ell = u \lambda$ is the spatial distance travelled along the unperturbed ray in the background spacetime.

To illustrate the application of \eqref{eqn:deflection_angle_integral}, let us consider the gravitational deflection of light due to a point mass. We set up the Cartesian coordinates so that the background path in Minkowski spacetime is along the $x$-direction. The impact parameter $b$ is measured along the  $y$-direction, which is the transverse direction from the path to the mass at the point of closest approach. The problem is effectively two-dimensional: one can always rotate the Cartesian axes $\{ x, y, z \}$ so that the point mass, the ray, the light source and the observer all lie in the $(x, y)$-plane; we may therefore neglect the $z$-direction.

The Newtonian gravitational potential for a point mass is
\begin{equation}
\Phi = - \frac{M}{r} = - \frac{M}{\sqrt{x^{2} + y^{2}}} ,
\end{equation}
where $M$ is the mass of the gravitational source. The transverse gradient of the potential, evaluated on the background ray, is
\begin{equation}
\bnab_{\bot} \Phi = \frac{M b}{\left(x^{2} + b^{2}\right)^{3/2}} \, \mathbf{e}_{y} ,
\end{equation}
where $\mathbf{e}_{y}$ is a unit vector in the $y$-direction.
Assuming that the observer and source are located at $x \rightarrow \pm \infty$, the deflection angle \eqref{eqn:deflection_angle_integral} is
\begin{equation}
\label{eqn:einstein_deflection_angle_point_mass}
\boldsymbol{\alpha} = 2 M b \int_{- \infty}^{\infty} \frac{\ed x}{\left(x^{2} + b^{2}\right)^{3/2}} \, \mathbf{e}_{y} = \frac{4 M}{b} \, \mathbf{e}_{y} .
\end{equation}
Transforming \eqref{eqn:einstein_deflection_angle_point_mass} to units in which the speed of light and Newton's gravitational constant are equal to $c$ and $G$, respectively, we recover the expression \eqref{eqn:einstein_deflection_angle} for the Einstein deflection angle.

An alternative approach to perturbative lensing involves considering the propagation of photons in a medium with effective refractive index $n = 1 - 2 \Phi = 1 + 2 \left| \Phi \right|$. The equations of motion for light rays can then be derived using Fermat's principle of least time \cite{SchneiderEhlersFalco1992, Carroll2019}. As in the case of standard ray optics, light travelling through a medium with a refractive index $n > 1$ travels slower than in vacuum.
%

\section{Non-perturbative lensing}
\label{sec:strong_field_gravitational_lensing}

The observation of a range of gravitational lensing effects over the last forty years has played a crucial role in demonstrating the superiority of general relativity over the Newtonian theory. However, the mathematical formalism relies on quasi-Newtonian approximations, in which the gravitational field is assumed to be weak, and deflection angles are assumed to be small \cite{Wambsganss1998}; indeed, the largest of the observed lensing effects described in the previous section is only of the order of a few tens of arcseconds (see \cite{InadaOguriPindorEtAl2003}, for example).

Einstein's theory of general relativity leads to extreme gravitational objects, including neutron stars and black holes. Electromagnetic radiation propagates through the strong gravitational fields in the vicinity of such compact objects. It is therefore natural to want to build up an understanding of the signatures of \emph{strong-field gravitational lensing}, associated with the near-field regions of extreme compact objects, where quasi-Newtonian approximations break down. In order to describe this strong-field gravitational lensing, we employ \emph{non-perturbative lensing} \cite{Perlick2004}, in which light rays follow null geodesics on a four-dimensional spacetime $(\mathcal{M}, g\ind{_{a b}})$, which is a solution to the field equations of general relativity.

\subsection{Electromagnetism in curved spacetime}
\label{sec:electromagnetism_in_curved_spacetime}

In its most general sense, gravitational lensing is concerned with all effects of a gravitational field (i.e., spacetime curvature) on the propagation of electromagnetic radiation. Here, we briefly review some fundamental aspects of Maxwell's theory of electromagnetism on curved spacetime.

Electromagnetism is governed by the \emph{Faraday tensor} $F\ind{_{a b}}$, which is antisymmetric type-$(0, 2)$ tensor field (i.e., a two-form). At a particular spacetime point, the electric and magnetic fields are dependent on the choice of Lorentz frame. An observer moving on a timelike worldline with unit tangent vector $u\ind{^{a}}$ and orthonormal spatial frame $\{ e\ind{_{(i)}^{a}} \}$ would measure electric and magnetic fields with components $E\ind{_{(i)}} = F\ind{_{a b}} e\ind{_{(i)}^{a}} u\ind{^{b}}$ and $B\ind{_{(i)}} = \vp{F}^{\star} F\ind{_{a b}} e\ind{_{(i)}^{a}} u\ind{^{b}}$, respectively. Here,
\begin{equation}
\label{eqn:hodge_dual_faraday}
\vp{F}^{\star} F\ind{_{a b}} = \frac{1}{2} \epsilon\ind{_{a b c d}} F\ind{^{c d}}
\end{equation}
is the Hodge dual of the Faraday two-form, where $\epsilon\ind{_{a b c d}}$ is the Levi-Civita tensor (see Section \ref{sec:metric_tensor}).

When describing electromagnetism in curved spacetime, it is convenient to introduce a complexified version of the Faraday tensor,
\begin{equation}
\label{eqn:complexified_faraday_tensor}
\mathcal{F}\ind{_{a b}} = F\ind{_{a b}} + i \vp{F}^{\star} F\ind{_{a b}}.
\end{equation}
Since $\vp{\mathcal{F}}^{\star} \mathcal{F}\ind{_{a b}} = - i \mathcal{F}\ind{_{a b}}$, the complex bivector $\mathcal{F}\ind{_{a b}}$ is self-dual. We also note that $\mathcal{F}\ind{_{a b}} \ol{\mathcal{F}}\ind{^{a b}} = 0$, where an overbar denotes complex conjugation. From the complex bivector \eqref{eqn:complexified_faraday_tensor}, one may define a complex three-vector $\boldsymbol{\mathcal{F}}$ whose components are given by $\mathcal{F}\ind{_{(i)}} = \mathcal{F}\ind{_{a b}} e\ind{_{(i)}^{a}} u\ind{^{b}}$; the real (imaginary) part of $\boldsymbol{\mathcal{F}} = \mathbf{E} + i \mathbf{B}$ gives the electric (magnetic) field measured by an observer moving along the timelike worldline with tangent vector $u\ind{^{a}}$ with orthonormal spatial frame $\{ e\ind{_{(i)}^{a}} \}$.

In charge-free regions of spacetime, the Faraday two-form $F\ind{_{a b}}$ satisfies Maxwell's equations,
\begin{equation}
\label{eqn:maxwell_equations_faraday}
\nabla\ind{_{b}} F\ind{^{a b}} = 0, \qquad \nabla\ind{_{b}} \vp{F}^{\star} F\ind{^{a b}} = 0.
\end{equation}
The second of these equations is equivalent to the Bianchi identity for the Faraday tensor: $\nabla\ind{_{ [ a }} F\ind{_{ c d ] }} = 0$. Hence, $F$ is closed ($\ed F = 0$), so $F$ is (locally) exact ($F = \ed A$, for some one-form potential $A$) by the Poincar\'{e} lemma; see Section \ref{sec:vectors_one_forms_tensors}. The Faraday tensor may be expressed in terms of the potential $A\ind{_{a}}$ as
\begin{equation}
\label{eqn:faraday_tensor_potential}
F\ind{_{a b}} = 2 \nabla\ind{_{[ a}} A\ind{_{b ]}}.
\end{equation}
Expanding out the covariant derivatives in \eqref{eqn:faraday_tensor_potential}, we see that the connection coefficients will cancel by antisymmetry; this permits us to write $F\ind{_{a b}} = 2 \partial\ind{_{[ a}} A\ind{_{b ]}}$. The fact that the Faraday tensor is locally exact means that it is invariant under gauge transformations of the form $A \mapsto A^{\prime} = A + \ed G$, where $G$ is any scalar field.

In the absence of charges, one may express Maxwell's equations in terms of the complexified Faraday tensor \eqref{eqn:complexified_faraday_tensor} as
\begin{equation}
\label{eqn:maxwells_equations_complex}
\nabla\ind{_{b}} \mathcal{F}\ind{^{a b}} = 0, \qquad \nabla\ind{_{ [ a }} \mathcal{F}\ind{_{ b c ] }} = 0 .
\end{equation}

It is possible to derive a wave equation for the Faraday tensor, by taking a derivative of the first of Maxwell's equations \eqref{eqn:maxwell_equations_faraday}, applying the commutator identity for covariant derivatives, and the Bianchi identity. This yields
\begin{equation}
\label{eqn:faraday_wave_equation}
\Box F\ind{_{a b}} + 2 R\ind{_{a c b d}} F\ind{^{c d}} + R\ind{_{a}^{c}} F\ind{_{b c}} - R\ind{_{b}^{c}} F\ind{_{a c}} = 0,
\end{equation}
where $\Box = \nabla\ind{^{a}} \nabla\ind{_{a}}$ is the d'Alembertian operator. In \eqref{eqn:faraday_wave_equation}, one may replace the Faraday tensor $F\ind{_{a b}}$ with its complexified counterpart $\mathcal{F}\ind{_{a b}}$, as we only consider the source-free Maxwell equations \eqref{eqn:maxwell_equations_faraday}.

One may instead derive a wave equation for the vector potential: in the absence of charges, this reads
\begin{equation}
\label{eqn:wave_equation_vector_potential}
\Box A\ind{^{a}} - R\ind{^{a}_{b}} A\ind{^{b}} - \nabla\ind{^{a}} \left( \nabla\ind{_{b}} A\ind{^{b}} \right) = 0 .
\end{equation}
Adopting the Lorenz gauge condition $\nabla\ind{_{a}} A\ind{^{a}} = 0$ eliminates the final term on the left-hand side of \eqref{eqn:wave_equation_vector_potential}.

When expressed in terms of the complexified Faraday tensor \eqref{eqn:complexified_faraday_tensor}, the electromagnetic stress--energy tensor (see Section \ref{sec:einstein_maxwell_field_equations}) takes a particularly simple form:
\begin{equation}
\label{eqn:stress_energy_complexified_faraday}
T\ind{_{a b}} = \frac{1}{4 \pi} \left( F\ind{_{a c}} F\ind{_{b}^{c}} - \frac{1}{4} g\ind{_{a b}} F\ind{_{c d}} F\ind{^{c d}} \right)
= \frac{1}{8 \pi} \on{Re}(\mathcal{F}\ind{_{a}^{c}} \ol{\mathcal{F}}\ind{_{b c}}).
\end{equation}
Clearly, the stress--energy tensor is trace-free: $T\ind{^{a}_{a}} = \frac{1}{2} \mathcal{F}\ind{_{a b}} \ol{\mathcal{F}}\ind{^{a b}} = 0$. Furthermore, in the absence of electromagnetic sources, the stress--energy tensor \eqref{eqn:stress_energy_complexified_faraday} satisfies the conservation equation $\nabla\ind{_{b}} T\ind{^{a b}} = 0$.
%

\subsection{Geometric optics}
\label{sec:geometric_optics_review}

\subsubsection{Leading-order geometric optics approximation}

Consider an electromagnetic field propagating in curved spacetime. In particular, let us assume that the wavelength is short in comparison with other physically relevant length scales, and that the inverse frequency is short in comparison with relevant time scales. In such situations, one may introduce a geometric optics ansatz for the vector potential $A\ind{_{a}}$, and insert this into the wave equation \eqref{eqn:wave_equation_vector_potential}, restricting to Lorenz gauge, $A\ind{^{a}_{; a}} = 0$ \cite{MisnerThorneWheeler1973, PoissonWill2014}.

Alternatively, one may introduce a geometric optics ansatz for the Faraday tensor itself (or its complexified counterpart) \cite{KristianSachs1966, Ehlers1967, Anile1976}. This is particularly useful when calculating the stress--energy tensor \eqref{eqn:stress_energy_complexified_faraday}; moreover, results obtained by working with the Faraday tensor are manifestly gauge-invariant, in contrast to those obtained from the vector potential $A\ind{^{a}}$. Here, we summarise the key points of the formalism presented by Dolan \cite{Dolan2017, Dolan2018} for the electromagnetic field; similar approaches can be taken for scalar and gravitational fields \cite{Dolan2017}.

In this approach, we introduce the geometric optics ansatz for \eqref{eqn:complexified_faraday_tensor} of the form
\begin{equation}
\label{eqn:complex_faraday_go_ansatz}
\mathcal{F}\ind{_{a b}} = \mathcal{A} \mathfrak{f}\ind{_{a b}} \exp{(i \omega \Phi)},
\end{equation}
where $\omega$ is an order-counting parameter which is related to the frequency; $\mathcal{A}$ is the (real) amplitude; $\Phi$ is the (real) phase; and $\mathfrak{f}\ind{_{a b}}$ is the polarisation bivector, a complex self-dual bivector field. Substitution of the ansatz \eqref{eqn:complex_faraday_go_ansatz} into the wave equation for $\mathcal{F}\ind{_{a b}}$ gives
\begin{equation}
\label{eqn:go_expansion_omega}
- \omega^{2} k\ind{^{c}} k\ind{_{c}} \mathfrak{f}\ind{_{a b}} + i \omega \left[ \left( 2 k\ind{^{c}} \mathcal{A}\ind{_{; c}} + k\ind{^{c}_{; c}} \mathcal{A} \right) \mathfrak{f}\ind{_{a b}} + \mathcal{A} k\ind{^{c}} \mathfrak{f}\ind{_{a b ; c}} \right] + O( \omega^{0} ) = 0,
\end{equation}
where $k\ind{^{a}} = g\ind{^{a b}} \nabla{_{b}} \Phi$ is the gradient of the phase.

At leading order, we see from \eqref{eqn:go_expansion_omega} that $k\ind{^{a}}$ is a null vector field ($k\ind{^{a}} k\ind{_{a }} = 0$). Since this vector field is both null and a gradient, it is quick to show that $k\ind{^{a}}$ satisfies the geodesic equation, $k\ind{^{b}} k\ind{_{a ; b}} = 0$. In other words, the integral curves of $k\ind{^{a}}$ -- spacetime paths\footnote{Here, an overdot denotes differentiation with respect to the affine parameter $s$.} $x\ind{^{a}}(s)$ with $k\ind{^{a}} = \dot{x}\ind{^{a}}$ -- are null geodesics which lie in constant-phase hypersurfaces ($\Phi = \textrm{constant}$); these are known as \emph{null generators}, and may be obtained from the Hamiltonian $H = \frac{1}{2} g\ind{^{a b}} k\ind{_{a}} k\ind{_{b}}$, where the ``conjugate momenta'' are $k\ind{_{a}} = g\ind{_{a b}} \dot{x}\ind{^{b}}$, and the Hamiltonian function $H$ vanishes along null geodesics. Furthermore, it follows directly from the fact that $k\ind{_{a}} = \nabla\ind{_{a}} \Phi$, and the equality of mixed partial derivatives, that $\nabla\ind{_{ [ a }} k\ind{_{ b]}} = \nabla\ind{_{ [ a }} \nabla \ind{_{ b]}} \Phi = 0$. Hence, the integral curves of $k\ind{^{a}}$ form a \emph{twist-free} null congruence.

Proceeding to $O(\omega)$, we see from \eqref{eqn:go_expansion_omega} that we require $\left( 2 k\ind{^{c}} \mathcal{A}\ind{_{; c}} + k\ind{^{c}_{; c}} \mathcal{A} \right) \mathfrak{f}\ind{_{a b}} + \mathcal{A} k\ind{^{c}} \mathfrak{f}\ind{_{a b ; c}} = 0$. Exploiting the ambiguity in the definition of the amplitude and polarisation bivector in \eqref{eqn:complex_faraday_go_ansatz}, we split the equations such that
\begin{align}
k\ind{^{a}} \mathcal{A}\ind{_{;a}} &= - \frac{1}{2} \vartheta \mathcal{A}, \label{eqn:transport_equation_amplitude} \\
k\ind{^{c}} \mathfrak{f}\ind{_{a b ; c}} &= 0, \label{eqn:transport_equation_polarisation_bivector}
\end{align}
where $\vartheta = k\ind{^{a}_{; a}}$ is the \emph{expansion scalar} \cite{Poisson2004}. By \eqref{eqn:transport_equation_amplitude}, the flux $\mathcal{A}^{2} k\ind{^{a}}$ is conserved: $\nabla\ind{_{a}} \left( \mathcal{A}^{2} k\ind{^{a}} \right) = 0$. The transport equation \eqref{eqn:transport_equation_polarisation_bivector} says that the polarisation bivector $\mathfrak{f}\ind{_{a b}}$ is parallel-transported along the null generators. Moreover, at leading order, Maxwell's equation $\nabla\ind{_{b}} \mathcal{F}\ind{^{a b}} = 0$ implies that the polarisation bivector is transverse: $\mathfrak{f}\ind{_{a b}} k\ind{^{b}} = 0$.

A circularly polarised wave which satisfies $\mathfrak{f}\ind{_{a b}} k\ind{^{b}} = 0$ and $k\ind{^{c}} \mathfrak{f}\ind{_{a b ; c}} = 0$ can be constructed by choosing
\begin{equation}
\label{eqn:circular_polarisation_bivector}
\mathfrak{f}\ind{_{a b}} = 2 k\ind{_{[ a }} m\ind{_{ b ]}} ,
\end{equation}
where $k\ind{^{a}}$ is the gradient of the phase, and $m\ind{^{a}}$ is a complex null ($m\ind{^{a}} m\ind{_{a}} = 0$) vector field which is transverse ($k\ind{^{a}} m\ind{_{a}} = 0$) and satisfies $m\ind{^{a}} \ol{m}\ind{_{a}} = 1$. In addition, $m\ind{^{a}}$ is required to satisfy the transport equation $k\ind{^{b}} m\ind{^{a}_{; b}} = \alpha(s) k\ind{^{a}}$, where $\alpha(s)$ is a scalar function along the null generator. We have $\alpha = 0$ in the case where $m\ind{^{a}}$ is parallel-transported along the null generators. (We note that a linearly polarised wave may be constructed by superposing two circularly polarised waves \cite{Dolan2017}.)

The complex null vector $m\ind{^{a}}$ may be constructed from the legs of an orthonormal triad $\{ e\ind{_{(i)}^{a}} \}$, viz.~$m\ind{^{a}} = \frac{1}{\sqrt{2}} ( e\ind{_{(1)}^{a}} + i e\ind{_{(2)}^{a}} )$; see Section \ref{sec:newman_penrose_tetrads}. The handedness of the circularly polarised wave depends on the sign of $\omega$ and the handedness of $m\ind{^{a}}$. We hereafter assume that $m\ind{^{a}}$ is constructed such that $i \epsilon\ind{_{a b c d}} u\ind{^{a}} k\ind{^{b}} m\ind{^{c}} \ol{m}\ind{^{d}} > 0$ for any future-pointing timelike vector $u\ind{^{a}}$. The wave is then right-hand (left-hand) polarised if $\omega > 0$ ($\omega < 0$). We note that the Lorentz transformation $m\ind{^{a}} \mapsto e^{i \Theta} m\ind{^{a}} + B k\ind{^{a}}$, where $\Theta$ is real and $B$ is complex, preserves the handedness, transversality property, and the parallel-transport of $m\ind{^{a}}$ (see Section \ref{sec:np_lorentz_transformations}).

One may introduce an auxiliary null vector $n\ind{^{a}}$, which is future-pointing, and satisfies $k\ind{^{a}} n\ind{_{a}} = -1$ with all other inner products zero. The tetrad $\{ k\ind{^{a}}, n\ind{^{a}}, m\ind{^{a}}, \ol{m}\ind{^{a}} \}$ is then a complex null tetrad (see Section \ref{sec:newman_penrose_tetrads}): the Newman--Penrose formalism is therefore well-adapted to describe the geometric optics approximation for the electromagnetic field on curved spacetime. It is convenient to choose this tetrad to be parallel-propagated along null geodesics with tangent vector field $k\ind{^{a}}$.

Inserting \eqref{eqn:complex_faraday_go_ansatz} and \eqref{eqn:circular_polarisation_bivector} into \eqref{eqn:stress_energy_complexified_faraday}, we find that the electromagnetic stress--energy tensor takes the form of a null fluid at leading order:
\begin{equation}
T\ind{_{a b}} = \frac{1}{8 \pi} \mathcal{A}^{2} k\ind{_{a}} k\ind{_{b}} .
\end{equation}
Conservation of the stress--energy tensor implies that the flux is conserved at leading order in $\omega$, i.e., $\nabla\ind{_{a}} \left( \mathcal{A}^{2} k\ind{^{a}} \right) = 0$.
%

\subsubsection{Gravitational lensing from geometric optics}

In order to calculate many quantities of interest to quantify gravitational lensing phenomena (e.g.~image distortion and brightness), one must solve the geodesic deviation equation (or Jacobi equation), which describes how spacetime curvature leads to a relative acceleration between neighbouring geodesics in a congruence. The geodesic deviation equation was derived for a general geodesic congruence in Section \ref{sec:geodesic_deviation_equation}. We recall here that, for a \emph{null} geodesic congruence $x\ind{^{a}}(s, v)$ with tangent vector field $k\ind{^{a}} = \frac{\partial x\ind{^{a}}}{\partial s}$ and separation vector $\xi\ind{^{a}} = \frac{\partial x\ind{^{a}}}{\partial v}$, the geodesic deviation equation is
\begin{equation}
\label{eqn:null_geodesic_deviation_equation}
D^{2} \xi\ind{^{a}} = - R\ind{^{a}_{b c d}} k\ind{^{b}} \xi\ind{^{c}} k\ind{^{d}} ,
\end{equation}
where $D = k\ind{^{a}} \nabla\ind{_{a}}$ denotes the directional derivative along the tangent vector field. Recall from Section \ref{sec:geodesic_deviation_equation} that $\mathcal{L}_{k} \xi = 0$ implies that the quantity $\xi\ind{^{a}} k\ind{_{a}}$ is conserved along each null ray. If we choose $\xi\ind{^{a}}$ to be initially transverse to $k\ind{^{a}}$ ($\xi\ind{^{a}} k\ind{_{a}} = 0$), then it will remain so along the ray.

The null geodesic deviation equation \eqref{eqn:null_geodesic_deviation_equation} describes the evolution of an infinitesimal bundle of rays with elliptical cross-section along the central null geodesic $x\ind{^{a}}(s, 0)$. In this context, an \emph{infinitesimal bundle of rays} is the set \cite{Perlick2004}
\begin{equation}
\mathcal{B} = \left\{ \alpha_{1} {\xi_{1}}\ind{^{a}} + \alpha_{2} {\xi_{2}}\ind{^{a}} \, | \, \alpha_{1}, \alpha_{2} \in \mathbb{R}, \, \alpha_{1}^{2} + \alpha_{2}^{2} \leq 1 \right\} ,
\end{equation}
where $\xi\ind{_{1}}$ and $\xi\ind{_{2}}$ are two vector fields which satisfy \eqref{eqn:null_geodesic_deviation_equation} and $g\ind{_{a b}} k\ind{^{a}} \xi\ind{^{b}} = 0$, such that $\xi\ind{_{1}}(s)$, $\xi\ind{_{2}}(s)$ and $k\ind{^{a}}(s)$ are linearly independent for almost all values of the affine parameter $s$ \cite{Perlick2004}.

A standard approach \cite{Perlick2004} to solving \eqref{eqn:null_geodesic_deviation_equation} is to introduce a Sachs basis, then decompose \eqref{eqn:null_geodesic_deviation_equation} into the Sachs equations for the optical scalars\cite{Sachs1961, JordanEhlersSachs2013}. Here, we favour a slightly different approach, which involves working with the (parallel-propagated) complex null tetrad $\{ k\ind{^{a}}, n\ind{^{a}}, m\ind{^{a}}, \ol{m}\ind{^{a}} \}$ introduced above. The deviation vector $\xi\ind{^{a}}$, restricted to the central null geodesic, may be decomposed in terms of the complex null tetrad along that geodesic:
\begin{equation}
\label{eqn:deviation_vector_decompose_tetrad}
\xi\ind{^{a}} = a(s) k\ind{^{a}} + b(s) n\ind{^{a}} + \ol{c}(s) m\ind{^{a}} + c(s) \ol{m}\ind{^{a}} ,
\end{equation}
where $a$ and $b$ are real functions along the ray, and $c$ is complex. Substitution of \eqref{eqn:deviation_vector_decompose_tetrad} into \eqref{eqn:null_geodesic_deviation_equation} yields a hierarchical system of coupled second-order ordinary differential equations \cite{Dolan2017, Dolan2018}, namely
\begin{align}
\ddot{b} &= 0 , \\
\ddot{a} &= b R\ind{_{a b c d}} k\ind{^{a}} n\ind{^{b}} k\ind{^{c}} n\ind{^{d}} + c R\ind{_{a b c d}} k\ind{^{a}} n\ind{^{b}} k\ind{^{c}} \ol{m}\ind{^{d}} + \ol{c} R\ind{_{a b c d}} k\ind{^{a}} n\ind{^{b}} k\ind{^{c}} m\ind{^{d}}, \\
\ddot{c} &= - b R\ind{_{a b c d}} k\ind{^{a}} m\ind{^{b}} k\ind{^{c}} n\ind{^{d}} - c R\ind{_{a b c d}} k\ind{^{a}} m\ind{^{b}} k\ind{^{c}} \ol{m}\ind{^{d}} - \ol{c} R\ind{_{a b c d}} k\ind{^{a}} m\ind{^{b}} k\ind{^{c}} m\ind{^{d}} .
\end{align}
The first of these equations is consistent with the fact that $\xi\ind{^{a}} k\ind{_{a}} = - b$ is conserved. Setting $b = 0$ ensures that $\xi\ind{^{a}}$ is transverse to $k\ind{^{a}}$. The equations then read
\begin{align}
\ddot{a} &= \left( \ol{\Phi}_{00} + \ol{\Psi}_{1} \right) c + \left( \Phi_{00} + \Psi_{1} \right) \ol{c}, \label{eqn:a_double_dot} \\
\ddot{c} &= - \Phi_{00} c - \Psi_{0} \ol{c} , \label{eqn:c_double_dot}
\end{align}
where $\Phi_{0 0} = \frac{1}{2} R\ind{_{a b}} k\ind{^{a}} k\ind{^{b}}$ is a complex Ricci scalar, and $\Psi_{0} = C\ind{_{a b c d}} k\ind{^{a}} m\ind{^{b}} k\ind{^{b}} m\ind{^{d}}$ and $\Psi_{1} = C\ind{_{a b c d}} k\ind{^{a}} n\ind{^{b}} k\ind{^{b}} m\ind{^{d}}$ are complex Weyl scalars; see Section \ref{sec:newman_penrose} for a review of the Newman--Penrose formalism.

Applying the directional derivative $D$ to $\xi\ind{^{a}} m\ind{_{a}}$, and using the definition \eqref{eqn:deviation_vector_decompose_tetrad}, as well as the Lie-transport ($\mathcal{L}_{k} \xi\ind{^{a}} = 0$) and transversality ($k\ind{^{a}} \xi\ind{_{a}} = - b = 0$) properties of the deviation vector, we find $\dot{c} = - \rho c - \sigma \ol{c}$, where $\rho = - \nps{m}{k}{\ol{m}}$, and $\sigma = - \nps{m}{k}{m}$ are Newman--Penrose scalars. Inserting this into \eqref{eqn:c_double_dot} and equating coefficients of $c$ and $\ol{c}$ yields the first-order transport equations
\begin{align}
\dot{\rho} &= \rho^{2} + \sigma \ol{\sigma} + \Phi_{0 0} , \label{eqn:sachs_equations_1} \\
\dot{\sigma} &= \sigma \left( \rho + \ol{\rho} \right) + \Psi_{0} , \label{eqn:sachs_equations_2}
\end{align}
which are known as the \emph{Sachs equations} \cite{Sachs1961, JordanEhlersSachs2013}. The real and imaginary parts of $- \rho = \theta + i \varpi$ and $- \sigma = \varsigma_{1} + i \varsigma_{2}$ are known collectively as the \emph{optical scalars} \cite{JordanEhlersSachs2013, Kantowski1968, FrolovNovikov1998}; in particular, $\theta = \frac{1}{2} \vartheta = \frac{1}{2} k\ind{^{a}_{; a}}$ is the \emph{expansion}, $\varpi$ is the \emph{twist}, and $(\varsigma_{1}, \varsigma_{2})$ is the \emph{shear}. In the case of a hypersurface-orthogonal null congruence (e.g.~the geometric optics congruence where $k\ind{_{a}} = \nabla\ind{_{a}} \Phi$ is orthogonal to constant-phase hypersurfaces $\Phi = \text{constant}$), we have $\varpi = 0$. Such a congruence is called \emph{twist-free}.

One limitation of this approach is that the Sachs equations \eqref{eqn:sachs_equations_1}--\eqref{eqn:sachs_equations_2} suffer from divergences at \emph{caustic points}, where neighbouring rays in the congruence cross \cite{Perlick2004, Dolan2017, Dolan2018}; we revisit this issue in Chapter \ref{chap:geometric_optics_kerr} in the context of (higher-order) geometric optics on Kerr spacetime. The second-order equation \eqref{eqn:c_double_dot} does not exhibit the same pathological behaviour at caustic points. One may determine the optical scalars $(\rho, \sigma)$ by solving \eqref{eqn:c_double_dot}, then inverting the matrix equation
\begin{equation}
\label{eqn:optical_scalars_from_c}
\left[
\begin{array}{c}
\dot{c}_{1} \\
\dot{c}_{2}
\end{array}
\right]
=
-
\left[
\begin{array}{c c}
c_{1} & \ol{c}_{1} \\
c_{2} & \ol{c}_{2}
\end{array}
\right]
\left[
\begin{array}{c}
\rho \\
\sigma
\end{array}
\right] ,
\end{equation}
where $c_{1}$ and $c_{2}$ are any linearly independent pair of solutions to the second-order equation \eqref{eqn:c_double_dot}. We note that the matrix on the right-hand side of \eqref{eqn:optical_scalars_from_c} is non-invertible at points where $\on{Im}{\left( c_{1} \ol{c}_{2} \right)} = 0$, which are precisely the caustic points.

Using a similar argument to that presented above, one may show that $\dot{a} = \ol{\tau} c + \tau \ol{c}$, where $\tau = - \nps{m}{k}{n}$ is a Newman--Penrose scalar. One may find $\tau$ by inverting
\begin{equation}
\label{eqn:tau_from_a_c}
\left[
\begin{array}{c}
\dot{a}_{1} \\
\dot{a}_{2}
\end{array}
\right]
=
\left[
\begin{array}{c c}
c_{1} & \ol{c}_{1} \\
c_{2} & \ol{c}_{2}
\end{array}
\right]
\left[
\begin{array}{c}
\ol{\tau} \\
\tau
\end{array}
\right] ,
\end{equation}
for any linearly independent pair of solutions $(a_{1}, c_{1})$ and $(a_{2}, c_{2})$ to the second-order system of ordinary differential equations \eqref{eqn:a_double_dot}--\eqref{eqn:c_double_dot}. Again, this breaks down at caustic points, where neighbouring rays in the bundle cross.

Let ${e_{(1)}}\ind{^{a}}$ and ${e_{(2)}}\ind{^{a}}$ be orthogonal unit vectors which lie in the instantaneous electromagnetic wavefront, i.e., they are orthogonal to $k\ind{^{a}}$. The complex number $c = \frac{1}{\sqrt{2}} \left( x + i y \right)$ corresponds to a point in the instantaneous wavefront with position vector $\hat{\xi}\ind{^{a}} = \ol{c} m\ind{^{a}} + c \ol{m}\ind{^{a}} = x {e_{(1)}}\ind{^{a}} + y {e_{(2)}}\ind{^{a}}$, where we recall the definition of the complex null vector $m\ind{^{a}} = \frac{1}{\sqrt{2}} ({e_{(1)}}\ind{^{a}} + i {e_{(2)}}\ind{^{a}})$; and we note that ${e_{(1)}}\ind{^{a}} = \frac{1}{\sqrt{2}} \left( m\ind{^{a}} + \ol{m}\ind{^{a}} \right)$ and ${e_{(2)}}\ind{^{a}} = -\frac{i}{\sqrt{2}} \left( m\ind{^{a}} - \ol{m}\ind{^{a}} \right)$. If $c_{1}$ and $c_{2}$ are any pair of linearly independent solutions to \eqref{eqn:c_double_dot}, then the function $c(\phi) = \cos{\phi} \, c_{1} + \sin{\phi} \, c_{2}$ parametrises an infinitesimal bundle of rays with elliptical cross-section which lies in the wavefront. One can demonstrate that the principal axes of the ellipse are given by $c_{+} = \cos{\phi_{0}} \, c_{1} + \sin{\phi_{0}} \, c_{2}$ and $c_{-} = - \sin{\phi_{0}} \, c_{1} + \cos{\phi_{0}} \, c_{2}$, with $\tan{\left( 2 \phi_{0} \right)} = \frac{2 \on{Re}{\left( c_{1} \ol{c}_{2} \right)}}{\left| c_{1} \right|^{2} - \left| c_{2} \right|^{2}}$. The semi-major and semi-minor axes are $d_{\pm} = \sqrt{2} \left| c_{\pm} \right|$, with $d_{+} d_{-} = 2 \left| \on{Im}{\left( c_{1} \ol{c}_{2} \right)} \right|$ and $d_{+}^{2} + d_{-}^{2} = 2 ( \left| c_{1} \right|^{2} + \left| c_{2} \right|^{2} )$. The parameters $d_{+}$ and $d_{-}$ are sometimes referred to as the \emph{shape parameters} of the bundle \cite{Perlick2004}. The cross-sectional area of the infinitesimal bundle of rays is $A = \pi d_{+} d_{-}$; this quantity satisfies the transport equation
\begin{equation}
\label{eqn:cross_sectional_area_transport_equation}
D A = - \left( \rho + \ol{\rho} \right) A = \vartheta A,
\end{equation}
where $\rho = - \nps{m}{k}{\ol{m}}$ is a Newman--Penrose scalar and $\vartheta$ is the expansion scalar, cf.~\eqref{eqn:transport_equation_amplitude}.

Using a derivation analogous to that of the geodesic deviation equation (Section \ref{sec:geodesic_deviation_equation}), Dolan \cite{Dolan2017, Dolan2018} also derives the \emph{differential precession equation},
\begin{equation}
\label{eqn:differential_precession_equation}
D \zeta\ind{^{a}} = - R\ind{^{a}_{b c d}} m\ind{^{b}} \xi\ind{^{c}} k\ind{^{d}} ,
\end{equation}
where $\zeta\ind{^{a}} = \left. \xi\ind{^{b}} \nabla\ind{_{b}} m\ind{^{a}} \right|_{C}$ is the \emph{precession vector}, defined on an observer's worldline $C$. Projecting on the null tetrad, $\zeta\ind{^{a}} = v(s) k\ind{^{a}} + \dot{c}(s) n\ind{^{a}} + w(s) m\ind{^{a}}$, yields a system of first-order ordinary differential equations for the coefficients $v$ and $w$ which read
\begin{align}
\dot{v} &= c R\ind{_{a b c d}} n\ind{^{a}} m\ind{^{b}} \ol{m}\ind{^{c}} k\ind{^{d}} + \ol{c} R\ind{_{a b c d}} n\ind{^{a}} m\ind{^{b}} m\ind{^{c}} k\ind{^{d}} , \\
\dot{w} &= - c R\ind{_{a b c d}} \ol{m}\ind{^{a}} m\ind{^{b}} \ol{m}\ind{^{c}} k\ind{^{d}} - \ol{c} R\ind{_{a b c d}} \ol{m}\ind{^{a}} m\ind{^{b}} m\ind{^{c}} k\ind{^{d}} .
\end{align}
In a Ricci-flat spacetime, this system of equations is simply
\begin{align}
\dot{v} &= \ol{\Psi}_{2} c , \label{eqn:v_dot} \\
\dot{w} &= - \ol{\Psi}_{1} c - \Psi_{1} \ol{c} , \label{eqn:w_dot}
\end{align}
where $\Psi_{2} = C\ind{_{a b c d}} k\ind{^{a}} m\ind{^{b}} \ol{m}\ind{^{c}} n\ind{^{d}}$ is a complex Weyl scalar. Taking a derivative of $\zeta\ind{^{a}}$ contracted with the legs of the complex null tetrad, one may show that $v = \ol{\mu} c + \ol{\lambda} \ol{c}$ and $w = - \ol{\chi} c + \chi \ol{c}$, where $\mu$, $\lambda$ and $\chi = \beta - \ol{\alpha}$ are Newman--Penrose scalars (Section \ref{sec:newman_penrose_spin_coefficients}). The Newman--Penrose quantities $\mu$, $\lambda$ and $\chi$ can therefore be obtained by inverting
\begin{equation}
\left[
\begin{array}{c}
v_{1} \\
v_{2}
\end{array}
\right]
=
\left[
\begin{array}{c c}
\ol{c}_{1} & c_{1} \\
\ol{c}_{2} & c_{2}
\end{array}
\right]
\left[
\begin{array}{c}
\ol{\mu} \\
\ol{\lambda}
\end{array}
\right] ,
\qquad
\left[
\begin{array}{c}
w_{1} \\
w_{2}
\end{array}
\right]
=
\left[
\begin{array}{c c}
\ol{c}_{1} & c_{1} \\
\ol{c}_{2} & c_{2}
\end{array}
\right]
\left[
\begin{array}{c}
\chi \\
- \ol{\chi}
\end{array}
\right] ,
\end{equation}
for any pair of linearly independent solutions $(c_{1}, v_{1}, w_{1})$ and $(c_{2}, v_{2}, w_{2})$ to the equations \eqref{eqn:c_double_dot}, \eqref{eqn:v_dot} and \eqref{eqn:w_dot}. This inversion breaks down at caustic points; however, the first-order ordinary differential equations \eqref{eqn:v_dot}--\eqref{eqn:w_dot} can be evolved through caustic points without issue, as in the case of \eqref{eqn:a_double_dot}--\eqref{eqn:c_double_dot}.
%

\subsection{Photon orbits}
\label{sec:photon_orbits_review}

In general relativity, ultra-compact objects -- such as black holes -- are responsible for spacetime curvature, which can cause extreme local light deflection. In fact, such objects (by definition) possess light-rings (i.e., circular photon orbits), and are able to bend light through an arbitrarily large angle \cite{CunhaHerdeiro2018}. We consider null geodesic motion on Kerr spacetime as an illustrative example. (This is considered in more detail in Chapters \ref{chap:stable_photon_orbits} and \ref{chap:geometric_optics_kerr}, so we only outline the key details here.)

The Kerr solution, given in Boyer--Lindquist coordinates $\{ t, r, \theta, \phi\}$ in \eqref{eqn:kerr_line_element_boyer_lindquist}, describes the exterior spacetime of a rotating black hole of mass $M$ with spin $a$ (see Section \ref{sec:black_holes} for a review). The spacetime admits a pair of commuting Killing vectors $\xi_{(t)} = \partial\ind{_{t}}$ and $\xi_{(\phi)} = \partial\ind{_{\phi}}$, which correspond to the stationarity and axisymmetry of the geometry. In Boyer--Lindquist coordinates, the geodesic motion is Liouville integrable thanks to the existence of the principal tensor $h\ind{_{a b}}$, from which one can generate the Killing tensor $K\ind{_{a b}}$ (see Section \ref{sec:killing_objects_symmetries_integrability}). Along a null geodesic with tangent vector field $k\ind{^{a}}$, the Hamiltonian $H = \frac{1}{2} g\ind{^{a b}} k\ind{_{a}} k\ind{_{b}}$, total energy $E = - {\xi_{(t)}}\ind{^{a}} k\ind{_{a}}$, azimuthal angular momentum $L\ind{_{z}} = {\xi_{(\phi)}}\ind{^{a}} k\ind{_{a}}$, and Carter constant $K = K\ind{^{a b}} k\ind{_{a}} k\ind{_{b}}$ are all conserved, with $H = 0$ in the null case.

Null geodesics in the equatorial plane ($\theta = \frac{\pi}{2}$), for which $K = (a E - L\ind{_{z}})^{2}$, are governed by the radial equation \cite{Chandrasekhar1983}
\begin{equation}
\dot{r} = \frac{{L\ind{_{z}}}^{2}}{b^{2}} \left[ 1 + \frac{(a^{2} - b^{2})}{r^{2}} + \frac{2 M \left( b - a \right)^{2}}{r^{3}} \right] ,
\end{equation}
where an overdot denotes differentiation with respect to an affine parameter, and $b$ is an \emph{impact parameter}, defined as the ratio of the azimuthal angular momentum to the energy:
\begin{equation}
b = \frac{L\ind{_{z}}}{E}.
\end{equation}
Unstable circular photon orbits, which neither plunge into the black hole nor escape to spatial infinity, are possible if the pair of conditions $\dot{r} = 0 = \ddot{r}$ are met. Prograde ($+$) and retrograde ($-$) perpetual orbits exist at Boyer--Lindquist radii \cite{Chandrasekhar1983}
\begin{equation}
\hat{r}_{\pm} = 2 M \left[ 1 + \cos{\left( \frac{2}{3} \arccos{\left( \mp \frac{a}{M} \right)} \right)} \right].
\end{equation}
The corresponding critical impact parameters are given by
\begin{equation}
b_{\pm} = \pm 3 \sqrt{M \hat{r}_{\pm}} - a,
\end{equation}
where $b_{-} < 0$ by definition for the retrograde orbit. In the literature, the circular photon orbits of constant Boyer--Lindquist radii are referred to as \emph{unstable light-rings}. In the non-rotating case ($a = 0$), the Kerr spacetime reduces to the Schwarzschild solution, in which there exists a unique unstable light-ring at $r = 3 M$, with critical impact parameter $b = 3 \sqrt{3} M$.

The null orbits around a Schwarzschild black hole are necessarily confined to a plane passing through the black hole's centre (which can be taken to be the equatorial plane after a suitable change of coordinates), due to the spherical symmetry of the spacetime geometry. In Kerr spacetime, however, the lack of spherical symmetry means that non-planar orbits are permitted. In fact, there exists a set of so-called \emph{spherical photon orbits} -- unstable perpetual null geodesics of constant Boyer--Lindquist radii -- which are not necessarily confined to the plane $\theta = \frac{\pi}{2}$ \cite{Teo2003, Hod2013}. The subfamily of spherical photon orbits which are restricted to the equatorial plane are precisely the two light-rings located at $r = \hat{r}_{\pm}$.

In stationary axisymmetric spacetimes whose motion is not (necessarily) Liouville integrable, there may exist generalisations of the spherical photon orbits which are present in Kerr spacetime. In recent work, Cunha \emph{et al.} \cite{CunhaHerdeiroRadu2017} have generalised the notion of a spherical photon orbit to a generic stationary axisymmetric spacetime $(\mathcal{M}, g\ind{_{a b}})$. Let $\gamma(s)$ be an affinely parametrised null geodesic. Then $\gamma(s)$ is a \emph{fundamental photon orbit} if it is restricted to a compact spatial region (i.e., it is a bound state), and there exists a value $T > 0$ such that $\gamma(s) = \gamma(s + T)$ for all $s \in \mathbb{R}$ (i.e., it is periodic with period $T$).

Fundamental photon orbits on Kerr spacetime are precisely the spherical photon orbits, which include the equatorial light-rings as a subset. In the exterior region (i.e., outside the outer event horizon), all spherical photon orbits are unstable. More generally, however, fundamental photon orbits can be \emph{stable}, leading to interesting phenomenological features, such as spacetime instabilities. This will be central to the work presented in Chapter \ref{chap:stable_photon_orbits}.

\subsection{Black hole shadows}
\label{sec:black_hole_shadows_review}

Black holes are characterised by their event horizon, a one-way causal boundary in spacetime beyond which nothing -- not even light -- can escape once it has fallen inside (see Section \ref{sec:event_horizon}). For an observer in spacetime, the \emph{black hole shadow} is the apparent optical image of the black hole. In a given observation frame, the black hole shadow corresponds to the set of all null directions in the local sky which originate (asymptotically) from the black hole's event horizon: the shadow thus corresponds to a lack of radiation received by the observer. In the high-frequency (geometric optics) approximation, the shadow is associated with the black hole's light absorption cross-section. Moreover, the outline of the shadow is intimately related to the gravitational lensing of electromagnetic radiation in the strong-field region; the shadow therefore encodes important information about the geometry of spacetime close to the black hole.

To illustrate the concept of a black hole shadow, consider the case of null geodesics on Schwarzschild spacetime. The unstable light-ring at $r = 3 M$ plays an important role: this orbit defines the boundary between the null rays which plunge into the event horizon of the black hole, and those which escape to infinity. The shadow of the Schwarzschild black hole is simply a circular disc in the sky, whose radius is determined by the unstable light-ring. For a static observer in Schwarzschild spacetime, the angular radius of the black hole shadow was calculated by Synge \cite{Synge1966}. (In fact, Synge calculated the ``escape cone'' of light, which is simply the complement of the shadow on the observer's local sky.)

In 1979, Luminet \cite{Luminet1979} determined the image of a ``bare'' Schwarzschild black hole, illuminated by a distant light source from directly behind. In the same paper, Luminet also analysed the more astrophysically relevant case of a Schwarzschild black hole surrounded by an emitting accretion disc. In situations where one is only concerned with the outline of the shadow, the system (i.e., the black hole and the observer) is assumed to be surrounded by a \emph{celestial sphere} at infinity, which acts as a light source.
%

\subsubsection{Kerr black hole shadow}

Here we consider the shadow of a Kerr black hole of mass $M$ and spin $a$ (see Section \ref{sec:black_holes}). For a null geodesic with energy $E$, angular momentum $L\ind{_{z}}$, and Carter constant $K$, the motion is determined by the independent \emph{impact parameters} \cite{Bardeen1973}
\begin{equation}
\label{eqn:kerr_impact_parameters}
b = \frac{L\ind{_{z}}}{E}, \qquad \chi = \frac{K - \left( a E - {L\ind{_{z}}}^{2} \right)}{E^{2}} .
\end{equation}
The shadow of the Kerr black hole is determined by the spherical photon orbits of constant Boyer--Lindquist radius $r$ (satisfying $\dot{r} = 0 = \ddot{r}$). These spherical photon orbits separate the rays which are captured by the black hole from those which escape to infinity. The former correspond to the black hole shadow; the latter correspond to regions of the observer's local sky which are illuminated by light from the celestial sphere. For a spherical photon orbit with Boyer--Lindquist radius $r$, the impact parameters \eqref{eqn:kerr_impact_parameters} are \cite{Teo2003}
\begin{align}
b &= \frac{r^{3} - 3 M r^{2} + a^{2} r + M a^{2}}{a (M - r)} , \label{eqn:kerr_impact_parameter_spo_1} \\
\chi &= \frac{r^{2}}{r^{2} - a^{2}} \left( 3 r^{2} + a^{2} - b^{2} \right) . \label{eqn:kerr_impact_parameter_spo_2}
\end{align}

Consider a static observer at infinity with a \emph{viewing angle} of $\theta = \theta_{0}$ in Boyer--Lindquist coordinates. The boundary of the shadow is the locus of points in the observer's local sky which correspond to null rays that asymptote to the spherical photon orbits; the orbital parameters are therefore determined by \eqref{eqn:kerr_impact_parameter_spo_1} and \eqref{eqn:kerr_impact_parameter_spo_2}. The observer will set up an \emph{image plane} which has local Cartesian-type coordinates $(X, Y)$, which are related to the impact parameters $(b, \chi)$ by \cite{Bardeen1973, Chandrasekhar1976, CunhaHerdeiro2018}
\begin{equation}
\label{eqn:kerr_shadow_parametric_form}
X = - b \on{cosec}{\theta_{0}}, \qquad Y = \pm \sqrt{\chi + a^{2}\cos^{2}{\theta_{0}} - b^{2} \cot^{2}{\theta_{0}}} .
\end{equation}
The shadow boundary is then defined as a parametric curve in the $(X, Y)$-plane, where $X$ and $Y$ depend on the Boyer--Lindquist radius $r$ of a spherical photon orbit through the parameters $b$ and $\chi$. The shadow boundary demarcates all of the rays which reach the observer from those which plunge into the black hole. Recently, Cunha and Herdeiro \cite{CunhaHerdeiro2018} provided an expression for the shadow boundary as a function $Y(X)$, rather than in the parametric form \eqref{eqn:kerr_shadow_parametric_form}.

\begin{figure}
\begin{center}
\subfigure[$a = 0$]{
\includegraphics[height=0.31\textwidth]{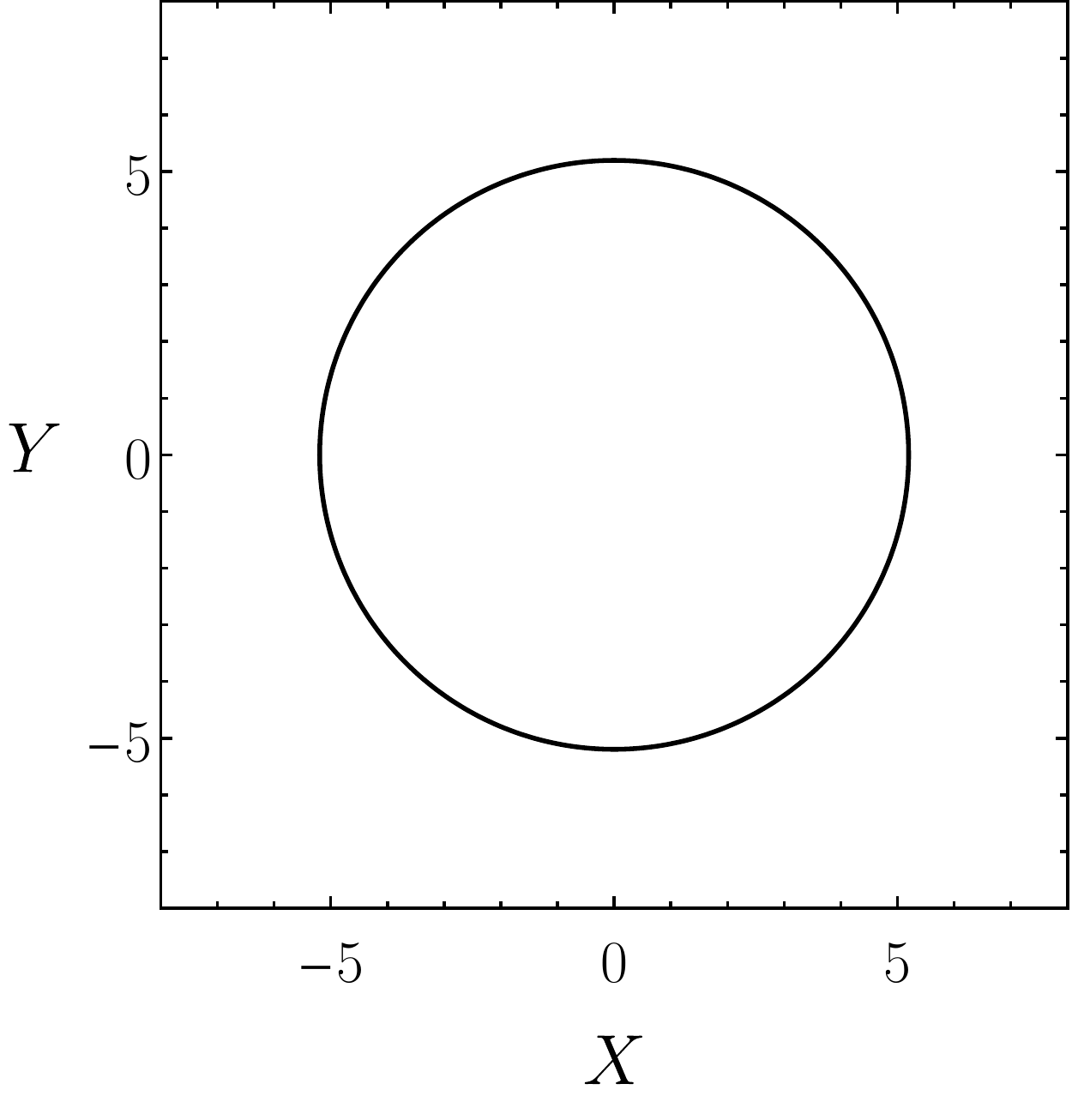} \label{fig:kerr_shadow_a_0000}} \hfill
\subfigure[$a = 0.8 M$]{
\includegraphics[height=0.31\textwidth]{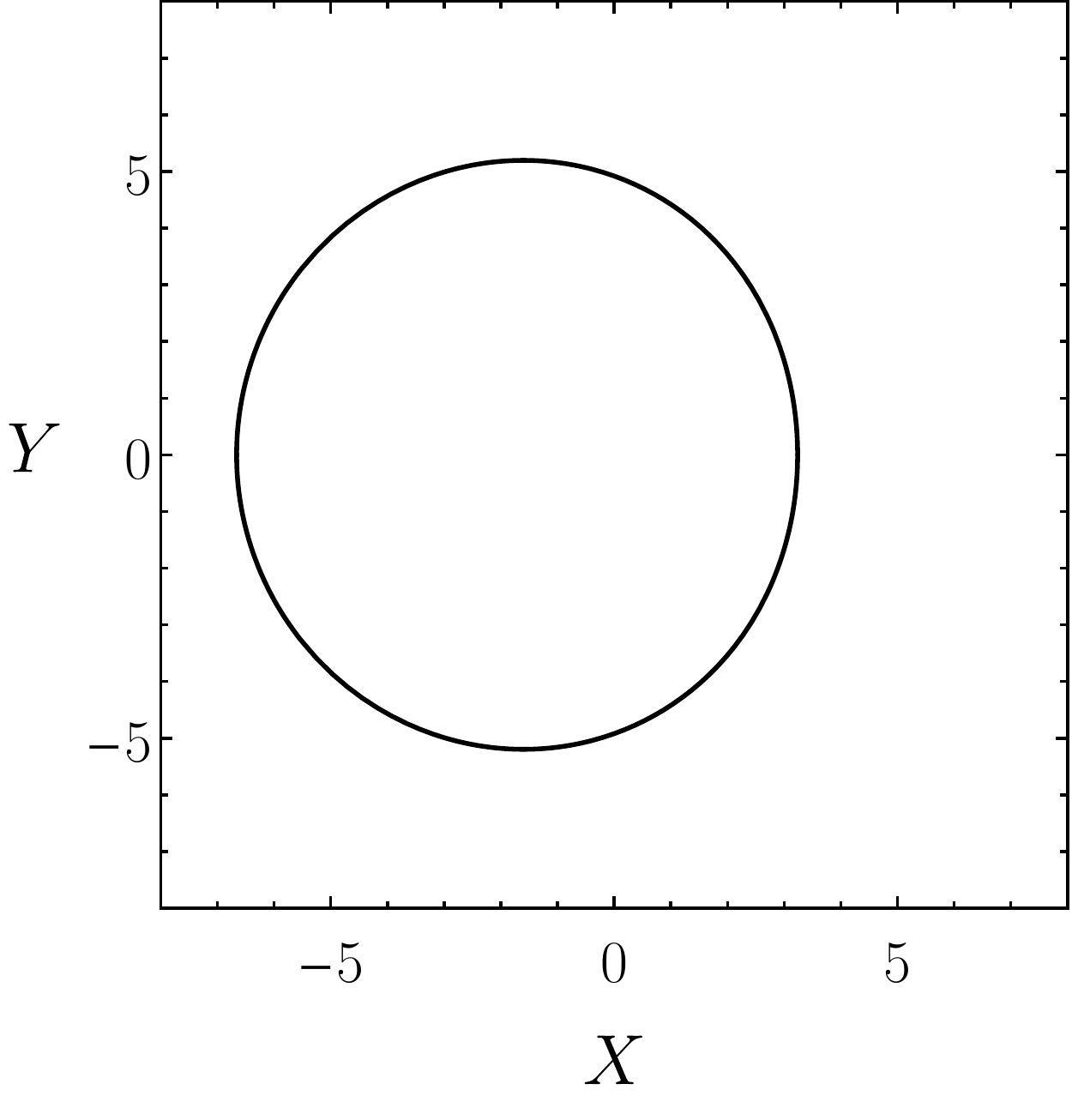} \label{fig:kerr_shadow_a_0800}} \hfill
\subfigure[$a = M$]{
\includegraphics[height=0.31\textwidth]{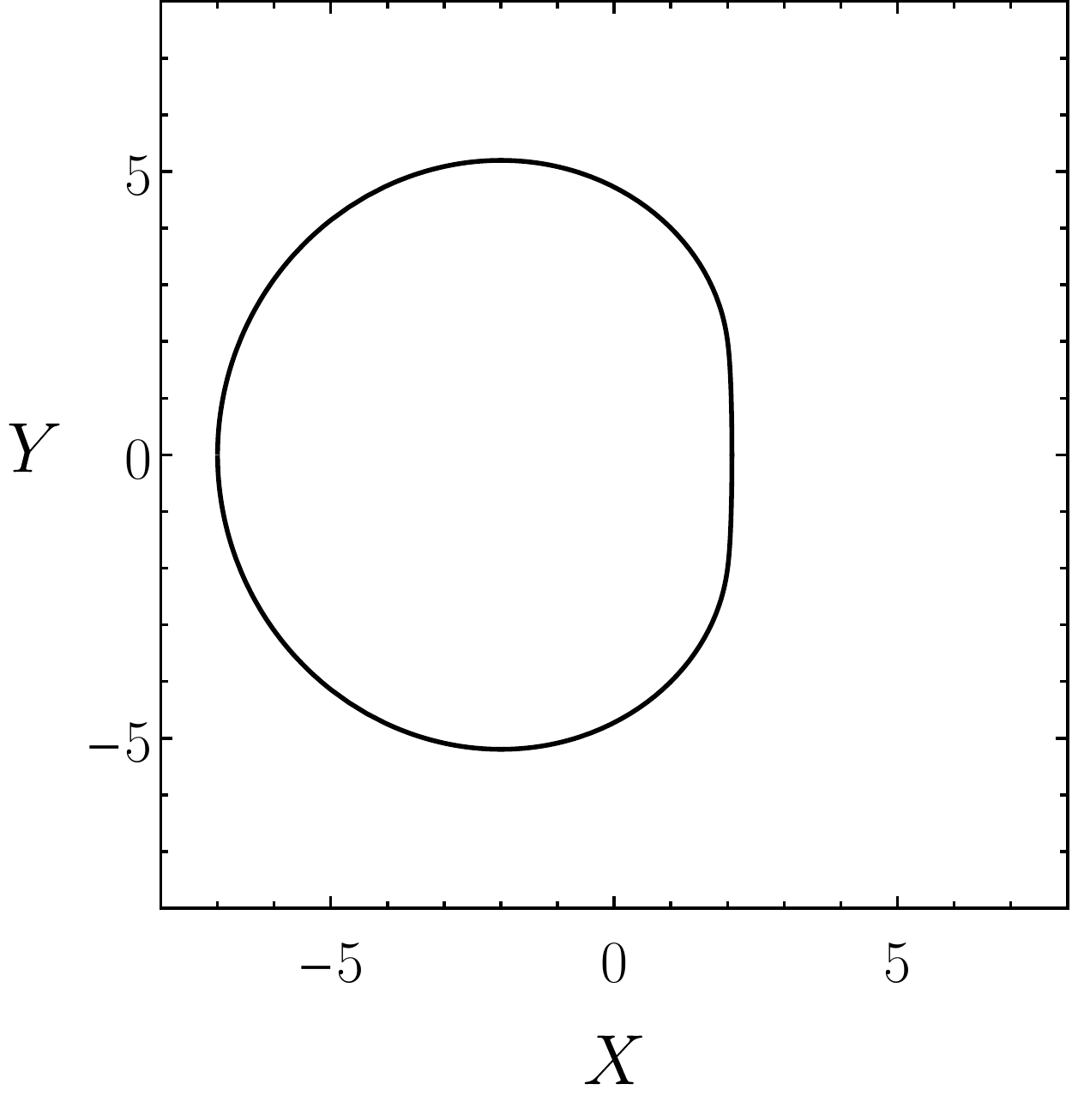} \label{fig:kerr_shadow_a_0999}} \hfill
\caption{Boundary of a Kerr black hole shadow for a selection of values of the spin parameter $a$, as seen by a distant observer with a viewing angle of $\theta = \frac{\pi}{2}$ and image plane coordinates $(X, Y)$. In the Schwarzschild case ($a = 0$), the shadow boundary is a circle of radius $3 \sqrt{3} M$; the shadow is entirely determined by the unstable photon orbit at $r = 3 M$ with impact parameter $b = 3 \sqrt{3} M$, due to the spherical symmetry of the spacetime. As one increases the value of $a$, the shadow boundary becomes deformed and asymmetric, due to the frame-dragging effects caused by the rotation of the black hole.}
\label{fig:kerr_shadow_a}
\end{center}
\end{figure}

In Figure \ref{fig:kerr_shadow_a}, we show the shadow boundary of a Kerr black hole of mass $M$ and spin parameter $a$, for an observer located in the equatorial plane ($\theta = \frac{\pi}{2}$). Figure \ref{fig:kerr_shadow_a_0000} shows the Schwarzschild case ($a = 0$); the shadow boundary is a circle whose radius in the image plane is determined by the equatorial unstable photon orbit (light-ring) at $r = 3 M$ with impact parameter $b = 3 \sqrt{3} M$. As the value of $a$ is increased from zero, as shown in Figures \ref{fig:kerr_shadow_a_0800} and \ref{fig:kerr_shadow_a_0999}, the frame-dragging effects are clearly visible: the shadow is asymmetric, due to the fact that $\left| b_{-} \right| > \left| b_{+} \right|$ (i.e., there is more absorption of retrograde rays than prograde rays); and it is deformed from circularity, exhibiting a characteristic ``D''-shape in the extremal ($a = M$) case.

\subsubsection{Ray-tracing}

Often it is impractical (or impossible) to calculate the shadow boundary analytically in terms of non-escaping photon orbits, e.g.~in situations where the geodesic motion is non-integrable. In such cases, the null geodesic equations must be solved using numerical methods. Rather than evolve all light rays which begin on the celestial sphere and determine which of these rays reach the observer, it is computationally more effective to reverse this process, tracing rays away from the observer, backwards in time. In this approach, known as \emph{backwards ray-tracing}, the black hole shadow corresponds to the set of all null geodesics which asymptote to the black hole's event horizon when traced backwards in time from the observer.

The use of ray-tracing in general relativity dates back to 1970s, when Cunningham and Bardeen \cite{CunninghamBardeen1973} determined the optical appearance of a star orbiting an extremal Kerr black hole, Cunningham \cite{Cunningham1975} calculated the emission spectrum of an accretion disc around a Kerr black hole, and Luminet \cite{Luminet1979} determined the image of an accretion disc around a Schwarzschild black hole.

Since then, a variety of numerical ray-tracing codes have been developed, including GeoKerr \cite{DexterAgol2009}, GeodesicViewer \cite{MuellerGrave2010}, GYOTO \cite{VincentPaumardGourgoulhonEtAl2011}, GeoViS \cite{Mueller2014}, and PyHole \cite{CunhaGroverHerdeiroEtAl2016}. Ray-tracing has been employed in a range of situations to understand gravitational lensing phenomena, such as the optical appearance of the black hole (i.e., the shadow) \cite{FalckeMeliaAgol1999, JohannsenPsaltis2010}; images of accretion structures around black holes \cite{Marck1996, DexterAgol2009}; the apparent appearance of stars orbiting compact objects \cite{LevinPerez-Giz2008, Mueller2009}; magnetohydrodynamical processes occurring around black holes \cite{NobleLeungGammieEtAl2007}; the polarisation of radiation propagating in the vicinity of compact objects \cite{ShcherbakovHuang2010}; the shadow of a dynamical binary black hole system \cite{BohnThroweHebertEtAl2015}; and the lensing dynamics of hairy black holes \cite{CunhaHerdeiroRaduEtAl2015}.
%

\chapter{Binary black hole shadows and chaotic scattering} \label{chap:binary_black_hole_shadows}

\section{Introduction}

Pairs of rotating black holes that orbit one another -- known as binary black holes -- have recently been detected. In 2015, the first direct detection of gravitational waves was made by the Advanced LIGO experiment, which confirmed the existence of binary black holes in nature \cite{Abbottothers2016a}. Since then, a host of gravitational-wave signals from merging binary black holes have been observed by the LIGO--Virgo collaboration \cite{Abbottothers2016b, AbbottAbbottAbbottEtAl2016, Abbottothers2017a, Abbottothers2017b, Abbottothers2017c}. More recently, the era of multi-messenger astronomy began with the observation of a gravitational-wave signal from binary neutron star merger \cite{Abbottothers2017d}, which was accompanied, approximately 1.7 seconds later, by a gamma-ray burst \cite{Abbottothers2017e}. A catalogue of compact binary coalescences observed by LIGO and Virgo during the first and second observing runs can be found in \cite{AbbottAbbottAbbottEtAl2016, Abbottothers2018}. 

The Event Horizon Telescope (EHT) -- a global array of radio telescopes, which employs very-long-baseline interferometry (VLBI) to create an Earth-scale virtual telescope -- began to observe nearby galactic centres, including the Milky Way and Messier 87 (M87), in 2017 \cite{FalckeMeliaAgol1999, DoelemanAgolBackerEtAl2009, Falcke2017, Psaltis2018}. The EHT's principal aim is to image the environment of supermassive black hole candidates which reside at the centre of these galaxies, with the hope of observing gravitational lensing phenomena associated with the black hole, such as multiple images, Einstein rings, and the black hole shadow itself. On 10 April 2019, the EHT collaboration presented the first event-horizon-scale image of the supermassive black hole candidate M87$^{\ast}$ (at the centre of the galaxy M87) from VLBI observations conducted in April 2017 \cite{EHTC2019a, EHTC2019b, EHTC2019c, EHTC2019d, EHTC2019e, EHTC2019f}. The image captured by the EHT is consistent with expectations for the shadow of a Kerr black hole, as predicted by general relativity (see Section \ref{sec:black_hole_shadows_review}).

Inspired by the recent detection of gravitational waves from merging binary black holes, and the first direct image of a black hole's event horizon, a strand of recent work has focussed on the strong-field gravitational lensing effects of binary black holes. The geometry of a dynamical binary, such as those observed by the LIGO--Virgo collaboration, is not known analytically as a solution to Einstein's field equations. In order to analyse the lensing effects of such systems, one must perform ray-tracing on top of fully non-linear numerical simulations. Bohn \emph{et al.} \cite{BohnThroweHebertEtAl2015} consider what an observer in the vicinity of a binary black hole system would actually \emph{see} as the two black holes orbit, inspiral and merge. The authors find that a binary black hole shadow is not simply the superposition of two singleton black hole shadows. In addition, the binary shadow exhibits intricate structures, including ``eyebrow'' features -- arc-shaped shadows which surround the primary globular shadow. Such features were anticipated by Nitta \emph{et al.} \cite{NittaChibaSugiyama2011} and Yumoto \emph{et al.} \cite{YumotoNittaChibaEtAl2012} in their studies of colliding black holes using the cosmological Kastor--Traschen solution \cite{KastorTraschen1993}.

Analysing the lensing effects of dynamical binary black holes is a formidable task, because it requires computationally expensive numerical simulations of the field equations of general relativity. In this chapter, based on the work presented in \cite{ShipleyDolan2016}, our principal aim is to understand the \emph{qualitative} features of binary black hole shadows. We therefore choose to focus on an imitative closed-form model: the Majumdar--Papapetrou binary black hole solution \cite{Majumdar1947, Papapetrou1947}, a static solution to the Einstein--Maxwell equations which describes a pair of extremally charged (Reissner--Nordstr\"{o}m) black holes whose gravitational attraction and electrostatic repulsion are in perfect balance. The Majumdar--Papapetrou solution has been studied in detail by Hartle and Hawking \cite{HartleHawking1972} and Chandrasekhar \cite{Chandrasekhar1989}.

In Section \ref{sec:killing_objects_symmetries_integrability}, we saw that a singleton Kerr--Newman black hole admits a Killing tensor, which gives rise to a fourth ``hidden'' constant of geodesic motion (Carter's constant), permitting the separability and Liouville integrability of the geodesic equations. By contrast, the presence of a pair of black holes in the Majumdar--Papapetrou binary black hole system reduces the symmetry, eliminating the Killing tensor associated with the Carter constant. As a result, the null geodesic equations, which govern the motion of photons on the Majumdar--Papapetrou geometry, are non-integrable; we should therefore anticipate rich gravitational lensing phenomena around binary black holes.

Here, building on the work of Contopoulos \cite{Contopoulos1990, Contopoulos1991, ContopoulosPapadaki1993, ContopoulosVoglisEfthymiopoulos1999, Contopoulos2002, ContopoulosHarsoula2004, ContopoulosHarsoula2005} and others
\cite{Yurtsever1995, DettmannFrankelCornish1994, DettmannFrankelCornish1995, CornishFrankel1997, CornishGibbons1997, Levin1999, Levin2000, LevinOReillyCopeland2000, HobillBurdColey1994, SotaSuzukiMaeda1996, SzydlowskiKrawiec1996, DeMouraLetelier2000, HananRadu2007, AlonsoRuizSanchez-Hernandez2008}, we advance the view that binary black hole systems are natural examples of \emph{chaotic scattering}: the system admits (we shall show) a fractal set of scattering singularities of measure zero. (See Section \ref{sec:chaotic_dynamical_systems} for a review of chaotic scattering in Hamiltonian systems.) In regions of chaotic scattering, a small uncertainty in fixing the initial conditions can lead to a drastically different outcome in the final state of the system. For the Majumdar--Papapetrou binary black hole system, the scattering singularities are associated with \emph{perpetual orbits}, i.e., unstable null geodesics which are neither scattered nor absorbed by the black holes.

This chapter is based on \cite{ShipleyDolan2016}. Section \ref{sec:majumdar_papapetrou_solution} contains a review of the two-centre Majumdar--Papapetrou solution. In Section \ref{sec:majumdar_papapetrou_null_geodesic_structure}, we discuss the null geodesic structure of the Majumdar--Papapetrou spacetime, with a focus on the Lagrangian and Hamiltonian formalisms, and a review of integrability and chaotic motion of null geodesics. Next, in Section \ref{sec:planar_geodesics_one_dim_shadows}, we analyse a planar scattering problem and its associated one-dimensional binary black hole shadows. In Section \ref{sec:symbolic_dynamics}, we introduce symbolic dynamics, considering the Gaspard--Rice three-disc scatterer as an illustrative example; and use the technique to describe the null geodesics of the Majumdar--Papapetrou binary black hole. In Section \ref{sec:cantor_like_structure}, we use symbolic dynamics to understand the ordering of the scattering singularities in the initial data; we construct the one-dimensional binary black hole shadows using an iterative procedure akin to that of the Cantor set, based on symbolic dynamics; we demonstrate self-similarity of the one-dimensional shadow explicitly; and we discuss quantitative measures of chaos which can be employed to understand the fractal structure of the strange repellor (see Section \ref{sec:chaotic_dynamical_systems}). In Section \ref{sec:non_planar_rays_two_dim_shadows}, we analyse non-planar null geodesics, which can be understood using an effective null geodesic potential; we highlight the existence of stable bounded photon orbits; and we present ray-traced images of two-dimensional binary black hole shadows for the Majumdar--Papapetrou system. In Section \ref{sec:through_the_event_horizons}, we consider the consequences of following rays through the event horizons in the maximally extended Majumdar--Papapetrou binary black hole spacetime. Finally, we conclude with a discussion of the work presented in this chapter in Section \ref{sec:discussion_bbh_shadows}.


\section{Majumdar--Papapetrou solution}
\label{sec:majumdar_papapetrou_solution}

The Majumdar--Papapetrou spacetime is a solution to the Einstein--Maxwell equations of gravity and electromagnetism in electrovacuum. The geometry is described in isotropic coordinates $\{t, x, y, z \}$ by the line element
\begin{equation}
\label{eqn:mp_general_line_element}
\ed s^{2} = g\ind{_{a b}} \ed x\ind{^{a}} \ed x\ind{^{b}} = - \frac{\ed t^{2}}{U^{2}} + U^{2} \left( \ed x^{2} + \ed y^{2} + \ed z^{2} \right),
\end{equation}
and electromagnetic one-form potential
\begin{equation}
\label{eqn:mp_electromagnetic_potential}
A\ind{_{a}} \ed x\ind{^{a}} = \frac{1}{U} \ed t,
\end{equation}
where $U(x, y, z)$ is any function of the \emph{spatial} coordinates which satisfies Laplace's equation $\nabla^{2} U = 0$ on three-dimensional auxiliary Euclidean space. The linearity of Laplace's equation means that solutions to \eqref{eqn:mp_general_line_element} may be generated by linear superposition, a feature which is rare in general relativity due to the inherent non-linearity of Einstein's field equations.

The Majumdar--Papapetrou solution to the Einstein--Maxwell equations was discovered independently by Majumdar \cite{Majumdar1947} and Papapetrou \cite{Papapetrou1947} in 1947. The Majumdar--Papapetrou spacetime was studied by Hartle and Hawking \cite{HartleHawking1972}, who found that for a particular solution $U$, the Majumdar--Papapetrou geometry describes a static configuration of $N$ extremally charged Reissner--Nordstr\"{o}m black holes.

\subsection{Majumdar--Papapetrou di-hole}
\label{sec:mp_dihole_review}

The Majumdar--Papapetrou solution that describes an assemblage of $N$ extremal Reissner--Nordstr\"{o}m black holes in static equilibrium is given by the line element \eqref{eqn:mp_general_line_element} and electromagnetic one-form potential \eqref{eqn:mp_electromagnetic_potential} with
\begin{equation}
\label{eqn:function_in_mp_metric}
U = 1 + \sum_{j = 1}^{N} \frac{M_{j}}{\sqrt{(x - x_{j})^{2} + (y - y_{j})^{2} + (z - z_{j})^{2}}},
\end{equation}
where $M_{j}$ is the mass of the $j$th black hole, and its location in the background spatial coordinates is $(x, y, z) = (x_{j}, y_{j}, z_{j})$. Each black hole is extremally charged ($Q_{j} = M_{j}$); the gravitational attraction and electrostatic repulsion between the black holes are in perfect balance, ensuring a static spacetime.

We consider the Majumdar--Papapetrou binary black hole (or \emph{di-hole}) solution, where \eqref{eqn:function_in_mp_metric} is given by
\begin{equation}
\label{eqn:mp_metric_function}
U = 1 + \frac{M_{+}}{\sqrt{x^{2} + y^{2} + (z - z_{+})^{2}}} + \frac{M_{-}}{\sqrt{x^{2} + y^{2} + (z - z_{-})^{2}}}.
\end{equation}
The two black holes of mass $M_{\pm}$ are located on the $z$-axis at $z_{\pm} = \pm \frac{d M_{\mp}}{M_{+} + M_{-}}$, where $d$ is the coordinate separation between the black holes in the background coordinates, and the centre of mass is at the origin of the coordinate system. The Majumdar--Papapetrou di-hole spacetime is symmetric about the axis connecting the black holes -- here the $z$-axis. The black holes' event horizons are located at the ``points'' $(x, y, z) = (0, 0, z_{\pm})$; these are in fact null surfaces of topology $\mathbb{R} \times S^{2}$, with non-zero area \cite{HartleHawking1972}. The geometry \eqref{eqn:mp_general_line_element} with \eqref{eqn:mp_metric_function} is an example of a Weyl spacetime (see Section \ref{sec:stationary_axisymmetric_solutions}).


\section{Null geodesic structure}
\label{sec:majumdar_papapetrou_null_geodesic_structure}

To analyse the lensing effects of the Majumdar--Papapetrou di-hole, it is necessary to study the null geodesic structure of the spacetime. In this section, we briefly review the Lagrangian and Hamiltonian formalism for null geodesics, before analysing the null geodesic structure of the Majumdar--Papapetrou di-hole geometry. For a more detailed review of geodesic motion in general relativity, see Section \ref{sec:geodesics}.
%

\subsection{Lagrangian formalism}

As outlined in Section \ref{sec:lagrangian_formulation_geodesics}, a standard approach to solving for the geodesic motion on curved spacetime is to adopt the Lagrangian formalism \cite{MisnerThorneWheeler1973}. In this scheme, one begins with the action, which is a functional of the spacetime coordinates, given by $S = \int L \, \ed \lambda$, with Lagrangian $L = \frac{1}{2} g\ind{_{a b}} \dot{q}\ind{^{a}} \dot{q}\ind{^{b}}$. Here, $q\ind{^{a}}(\lambda)$ is a spacetime path, and an overdot denotes the derivative with respect to the affine parameter $\lambda$. Along geodesics, the Lagrangian is conserved, with $L = 0$ in the case of null geodesics. The geodesics are given by solutions to the Euler--Lagrange equations,
\begin{equation}
\label{eqn:mp_euler_lagranage}
\frac{\ed}{\ed \lambda} \left( \frac{\partial L}{\partial \dot{q}\ind{^{a}}} \right) = \frac{\partial L}{\partial q\ind{^{a}}}.
\end{equation}

For null geodesics on the Majumdar--Papapetrou geometry \eqref{eqn:mp_general_line_element}, the Lagrangian takes the form
\begin{equation}
\label{eqn:mp_lagrangian}
L = \frac{1}{2} \left[ -\frac{1}{U^{2}} \dot{t}^{2} + U^{2} \left( \dot{x}^{2} + \dot{y}^{2} + \dot{z}^{2} \right) \right] = 0,
\end{equation}
where $U$ is given by \eqref{eqn:mp_metric_function} for the Majumdar--Papapetrou di-hole. Inserting the Lagrangian \eqref{eqn:mp_lagrangian} into the Euler--Lagrange equations \eqref{eqn:mp_euler_lagranage} yields a system of four second-order coupled non-linear ordinary differential equations for the path in spacetime $q\ind{^{a}}(\lambda) = \left( t(\lambda), x(\lambda), y(\lambda), z(\lambda) \right)$.

We recall that the Lagrangian \eqref{eqn:mp_lagrangian} is not unique: any Lagrangian which yields the same equations of motion is valid (see Section \ref{sec:lagrangian_formulation_geodesics}). Some alternative choices of Lagrangian for the Majumdar--Papapetrou di-hole are outlined in \cite{ShipleyDolan2016}.

\subsection{Hamiltonian formalism}
\label{sec:mp_dihole_hamiltonian_formalism}

In the Hamiltonian approach, one uses the Lagrangian function to define the canonical momenta as $p\ind{_{a}} = \frac{\partial L}{\partial \dot{q}\ind{^{a}}} = g\ind{_{a b}} \dot{q}\ind{^{b}}$. Performing a Legendre transformation $H = \dot{q}\ind{^{a}} p\ind{_{a}} - L$, the corresponding Hamiltonian is then given in coordinates $\{ q\ind{^{a}}, p\ind{_{b}} \}$ by the Hamiltonian function $H(q\ind{^{a}} , p\ind{_{a}}) = \frac{1}{2} g\ind{^{a b}} p\ind{_{a}} p\ind{_{b}}$, where $g\ind{^{a b}}$ are the contravariant components of the metric tensor. Geodesics of the spacetime geometry are then the integral curves of Hamilton's equations, a system of first-order differential equations given by
\begin{equation}
\label{eqn:mp_dihole_hamiltons_equations}
\dot{q}\ind{^{a}} = \frac{\partial H}{\partial p\ind{_{a}}}, \qquad
\dot{p}\ind{_{a}} = -\frac{\partial H}{\partial q\ind{^{a}}}.
\end{equation}
Along geodesics, the Hamiltonian is conserved, with $H = 0$ in the case of null rays. (See Section \ref{sec:hamiltonian_formulation_geodesics} for a review.)

In isotropic Cartesian coordinates $\{t, x, y, z\}$, the Hamiltonian for null geodesics on the Majumdar--Papapetrou di-hole spacetime takes the form
\begin{equation}
\label{eqn:mp_hamiltonian_cartesian}
H = \frac{1}{2} \left[ - U^{2} {p\ind{_{t}}}^{2} + \frac{1}{U^{2}} \left( {p\ind{_{x}}}^{2} + {p\ind{_{y}}}^{2} + {p\ind{_{z}}}^{2} \right) \right] .
\end{equation}
Hamilton's equations \eqref{eqn:mp_dihole_hamiltons_equations} with Hamiltonian function \eqref{eqn:mp_hamiltonian_cartesian} are given by
\begin{align}
\label{eqn:mp_hamilton_equations_cartesian}
\dot{t} &= -U^{2} p\ind{_{t}}, &
\dot{p}\ind{_{t}} &= 0, &
\dot{x} &= \frac{p\ind{_{x}}}{U^{2}}, &
\dot{p}\ind{_{x}} &= - \frac{\partial U}{\partial x} \left[ U {p\ind{_{t}}}^{2} + \frac{1}{U^{3}} \left( {p\ind{_{x}}}^{2} + {p\ind{_{y}}}^{2} + {p\ind{_{z}}}^{2} \right) \right],
\end{align}
with similar expressions for $y$ and $z$ by symmetry of \eqref{eqn:mp_hamiltonian_cartesian}. The metric is independent of coordinate time $t$, so the momentum $p\ind{_{t}} = - E$ is a constant of motion. Rescaling $p\ind{_{t}}$ is equivalent to rescaling the affine parameter $\lambda$; we may therefore choose to fix $p\ind{_{t}} = - 1$ in the following analysis without loss of generality.

Rearranging the Hamiltonian constraint $H = 0$ for the Hamiltonian function \eqref{eqn:mp_hamiltonian_cartesian}, and inserting the result into the equations for $\dot{p}\ind{_{i}}$ given in \eqref{eqn:mp_hamilton_equations_cartesian}, we see that the geodesic equations for the \emph{spatial} components of the four-momentum can be expressed in the form $\dot{p}\ind{_{i}} = - \left( U^{2} \right)\ind{_{, i}}$. This resembles Newton's second law of motion, in which the force is expressed as the gradient of a scalar potential.

The diffeomorphism invariance of general relativity allows us to perform a change of coordinates without altering the physics of the problem at hand. We therefore choose to recast the problem in a coordinate system which is well-adapted to the symmetry of the configuration. The Majumdar--Papapetrou di-hole geometry is symmetric about the $z$-axis; it is therefore natural to choose cylindrical polar coordinates $\{t, \rho, \phi, z\}$, related to Cartesian coordinates in the standard way: $\rho^{2} = x^{2} + y^{2}$, $\phi = \arctan{\left( \frac{y}{x} \right)}$. Having set $p\ind{_{t}} = -1$, the Hamiltonian \eqref{eqn:mp_hamiltonian_cartesian} now reads
\begin{equation}
\label{eqn:mp_hamiltonian_cylindrical}
H = \frac{1}{2} \left[ \frac{1}{U^{2}} \left( {p\ind{_{\rho}}}^{2} + {p\ind{_{z}}}^{2} + \rho^{2} {p\ind{_{\phi}}}^{2} \right) - U^{2} \right].
\end{equation}

In cylindrical coordinates, it is straightforward to see that $\partial\ind{_{t}}$ ($\partial\ind{_{\phi}}$) is a Killing vector which generates time-translational (axial) symmetry. Hence, the coordinates $t$ and $\phi$ are ignorable, and their conjugate momenta are conserved along geodesics. The physical interpretations of $-p\ind{_{t}}$ and $p\ind{_{\phi}}$ are, respectively, the total energy and axial angular momentum of the photon at infinity.

For the Majumdar--Papapetrou geometry in four spacetime dimensions, the full phase space is eight-dimensional, spanned by the four spacetime coordinates and their conjugate momenta $\{q\ind{^{a}}, p\ind{_{a}}\}$. The conserved momenta allow us to focus on a reduced four-dimensional phase space on which the dynamics is governed by two pairs of conjugate variables $\{\rho, z, p\ind{_{\rho}}, p\ind{_{z}} \}$, and the null condition $H = 0$. This constraint allows us to express one of the phase space coordinates (e.g. $p\ind{_{z}}$) in terms of the other three.

Recall from Section \ref{sec:geodesic_equation} that the null geodesics of the conformally related metrics $g\ind{_{a b}}$ and $\tilde{g}\ind{_{a b}} = \Omega^{2} g\ind{_{a b}}$ coincide, where $\Omega > 0$ is a (smooth) function of the spacetime coordinates. Performing a conformal transformation with $\Omega = \frac{1}{U}$ allows us to recast the Hamiltonian \eqref{eqn:mp_hamiltonian_cylindrical} in canonical form as
\begin{equation}
\label{eqn:mp_hamiltonian_canonical}
\tilde{H} = \frac{1}{2} \left( {p\ind{_{\rho}}}^{2} + {p\ind{_{z}}}^{2} \right) + V, \qquad
V(\rho, z) = - \frac{1}{2 \rho^{2}} (h - p\ind{_{\phi}})(h + p\ind{_{\phi}}),
\end{equation}
where we have factorised the potential $V(\rho, z)$ by introducing the \emph{effective potential} (or \emph{height function})
\begin{equation}
\label{eqn:mp_effective_potential}
h(\rho, z) = \rho \, U^{2}.
\end{equation}
The Hamiltonian \eqref{eqn:mp_hamiltonian_canonical} will be used throughout our analysis of null geodesics on the Majumdar--Papapetrou di-hole spacetime, with the effective potential \eqref{eqn:mp_effective_potential} playing a key role. (We hereafter omit the tilde from the Hamiltonian \eqref{eqn:mp_hamiltonian_canonical}, simply writing $H$.) Using the Hamiltonian \eqref{eqn:mp_hamiltonian_canonical}, the geodesic equations (i.e., Hamilton's equations) are
\begin{align}
\dot{\rho} &= p\ind{_{\rho}}, & \dot{z} &= p\ind{_{z}}, & \dot{\phi} &= \frac{p\ind{_{\phi}}}{\rho^{2}}, \label{eqn:hamiltons_equations_cylndrical_1} \\
\dot{p}\ind{_{\rho}} &= U^{2} \frac{\partial \left( U^{2} \right)}{\partial \rho} + \frac{{p\ind{_{\phi}}}^{2}}{\rho^{3}}, & \dot{p}\ind{_{z}} &= U^{2} \frac{\partial \left( U^{2} \right)}{\partial z}, & \dot{p}\ind{_{\phi}} &= 0. \label{eqn:hamiltons_equations_cylndrical_2}
\end{align}

\subsection{Integrability and chaotic scattering}
\label{sec:integrability_chaos}

The Newtonian analogue of geodesic motion on the Majumdar--Papapetrou spacetime is the problem of two fixed centres (also known as Euler's three-body problem or the two-centre Kepler problem) \cite{MurrayDermott1999, Contopoulos2002}. This is a special case of the classical three-body problem of three point masses described by Newton's laws of motion and Newton's law of universal gravitation. The general three-body problem has no closed-form solution.

Remarkably, Euler's restricted three-body problem is soluble. In the case of two fixed centres, the equations governing the motion of a test particle are \emph{separable} in spheroidal coordinates. This is due to the fact that, besides the total energy and one component of the angular momentum, there exists an additional ``hidden'' constant of motion -- Whittaker's constant \cite{CoulsonJoseph1967, Lynden-Bell2003} (cf.~Carter's constant \cite{Carter1968b} for geodesic motion on Kerr spacetime). The problem is therefore Liouville integrable; see Section \ref{sec:integrability_and_chaos_in_gr}.

In contrast to motion in Euler's three-body problem, geodesic motion on the Majumdar--Papapetrou di-hole spacetime -- as well as other relativistic generalisations, such as the Weyl--Bach di-hole \cite{BachWeyl2012, CoelhoHerdeiro2009} -- is not integrable, as there is no known analogue of Whittaker's constant \cite{Will2009}.\footnote{In \cite{CoelhoHerdeiro2009}, the authors demonstrate that, for the special case of equal-mass black holes, the $\mathbb{Z}_{2}$-symmetric Weyl--Bach di-hole admits a $(2+1)$-dimensional totally geodesic submanifold -- the equatorial plane -- on which the geodesic motion is Liouville integrable. This is also explored by Assump\c{c}ao \emph{at al.} \cite{AssumpcaoCardosoIshibashiEtAl2018}.} The lack of Liouville integrability on the Majumdar--Papapetrou di-hole spacetime results in rich phenomena. Contopoulos revealed that the Majumdar--Papapetrou di-hole exhibits chaotic dynamics and self-similarity \cite{Contopoulos1990, Contopoulos1991, Contopoulos2002}. This paved the way for deeper insights into chaos in binary systems in general relativity from Contopoulos and collaborators \cite{ContopoulosPapadaki1993, ContopoulosVoglisEfthymiopoulos1999, ContopoulosHarsoula2004, ContopoulosHarsoula2005}; Yurtsever \cite{Yurtsever1995}; Cornish, Dettmann, Frankel, Gibbons, Levin and collaborators \cite{DettmannFrankelCornish1994, DettmannFrankelCornish1995, CornishFrankel1997, CornishGibbons1997, Levin1999, Levin2000, LevinOReillyCopeland2000}; and others \cite{HobillBurdColey1994, SotaSuzukiMaeda1996, SzydlowskiKrawiec1996, DeMouraLetelier2000, HananRadu2007, AlonsoRuizSanchez-Hernandez2008}. In this chapter, we extend on these influential perspectives, advancing the view that the scattering of photons by a pair of black holes in the Majumdar--Papapetrou di-hole spacetime is a natural example of \emph{chaotic scattering} (see Section \ref{sec:chaotic_dynamical_systems} for a review).

For the Majumdar--Papapetrou di-hole, a scattering problem is set up as follows. We consider the evolution of null geodesics (light rays) for different initial conditions (impact parameters). The final state of the photon is characterised by three possible outcomes. A photon may (i) cross the event horizon of the upper black hole; (ii) cross the event horizon of the lower black hole; (iii) escape to spatial infinity. In this system,  the scattering singularities correspond to the null geodesics which are non-escaping (``perpetual'') orbits in forward time ($\lambda \rightarrow \infty$). The non-escaping orbits are neither scattered nor absorbed by the black holes. The scattering singularities, which comprise the \emph{repellor} (Section \ref{sec:chaotic_dynamical_systems}), can be divided into two classes: the subset of \emph{periodic} orbits; and the subset of \emph{aperiodic} orbits. In the following sections, we shall explore the structure of these subsets of the initial data, and analyse some examples of non-escaping orbits.

\section{Planar geodesics and one-dimensional shadows}
\label{sec:planar_geodesics_one_dim_shadows}

%
In this section, we consider null geodesics on the Majumdar--Papapetrou di-hole spacetime in standard coordinates $\{t, x, y, z \}$, using the Hamiltonian formalism outlined in Section \ref{sec:mp_dihole_hamiltonian_formalism}. The line element is given by \eqref{eqn:mp_general_line_element} and the electromagnetic one-form potential by \eqref{eqn:mp_electromagnetic_potential}, where the function $U$ appearing in both quantities is given by \eqref{eqn:mp_metric_function}. For simplicity, we consider the symmetric case of equal-mass black holes ($M_{+} = M_{-} = M$) located at the points $z = \pm \frac{d}{2}$, where $d$ is the coordinate separation between the black holes. We employ units in which $M = 1$.

We shall examine a highly symmetric scattering problem, in which null rays are confined to the meridian plane ($y = 0$). The conservation of $p_{\phi} = x p\ind{_{y}} - y p\ind{_{x}}$ ensures that rays which start in the meridian plane (with $y = 0 = p\ind{_{y}}$) remain in the plane. We consider rays which start at the centre of mass $(x, z) = (0, 0)$ between the black holes. The scattering scenario is one-dimensional: the initial state of the system is characterised by the angle $\alpha$ made by the momentum two-vector $(p\ind{_{x}}, p\ind{_{z}})$ and the positive $x$-axis. The final state of the scattering process is described by three distinct possibilities: (i) the photon crosses the event horizon of the upper black hole; (ii) the photon crosses the event horizon of the lower black hole; (iii) the photon escapes to spatial infinity.

The initial data (with $\lambda = 0$) for this scattering problem are taken to be
\begin{align}
\label{eqn:mp_one_dim_initial_data}
t &= x = y = z = 0, & p\ind{_{t}} &= - 1 & p\ind{_{y}} &= 0, & p\ind{_{x}} &= U_{0}^{2} \cos{\alpha}, & p\ind{_{z}} &= U_{0}^{2} \sin{\alpha},
\end{align}
where $\alpha$ is the initial angle and $U_{0} = U(0, 0, 0) = 1 + \frac{4}{d}$. Taking $d = 2$ as a default value, we numerically evolve the null geodesic equations (Hamilton's equations) with initial data given by \eqref{eqn:mp_one_dim_initial_data} for $0 \leq \alpha \leq \frac{\pi}{2}$; we need only consider rays whose initial momentum two-vector $(p\ind{_{x}}, p\ind{_{z}})$ points into the first quadrant ($x \geq 0, z \geq 0$) due to the symmetry of the scattering problem.

\begin{figure}
\begin{center}
\begin{tabular}{c c}
\subfigure[Scattering set-up]{
\begin{tabular}{c}
\includegraphics[width=0.43\textwidth]{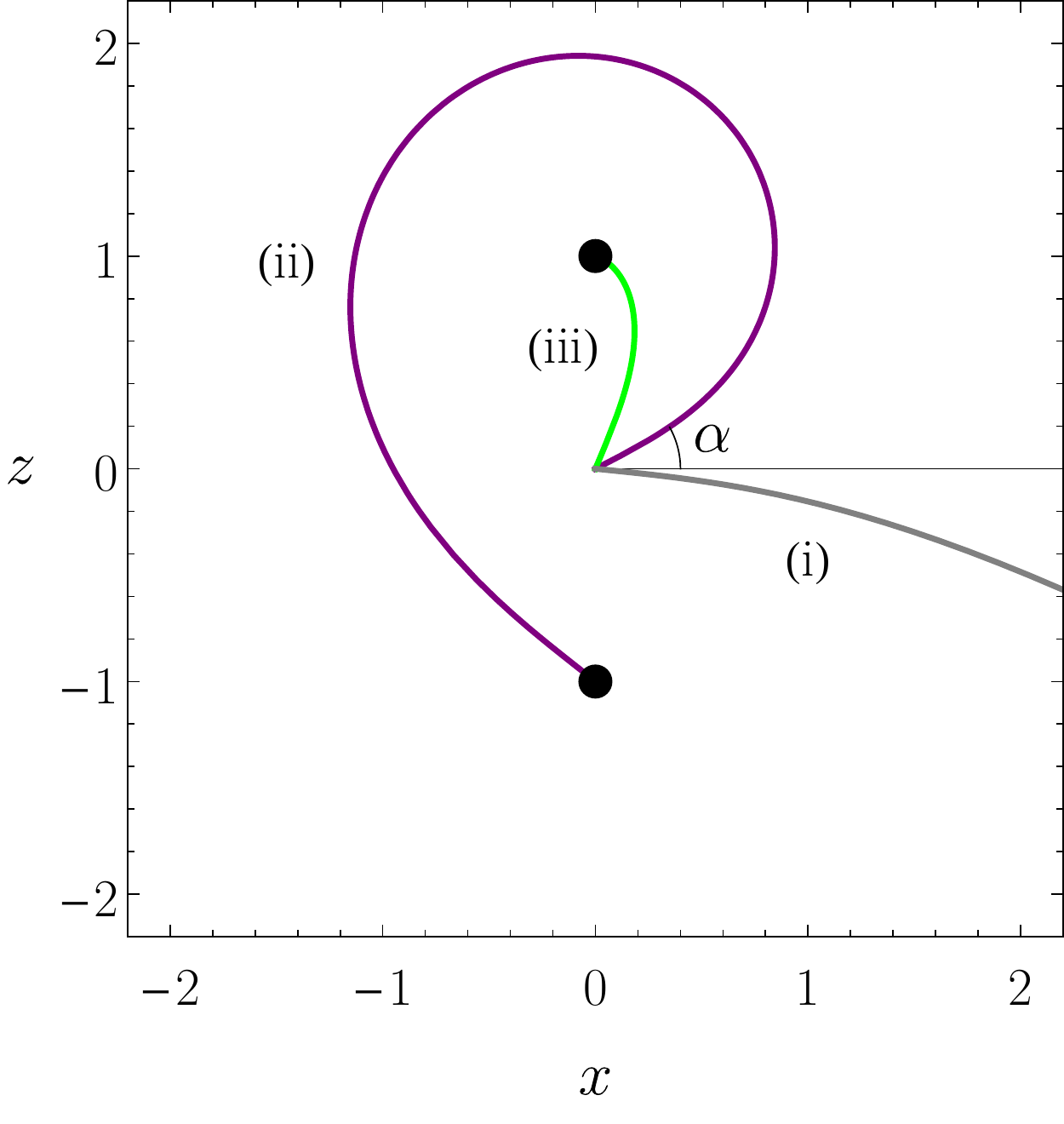}
\label{fig:mp_one_dim_scattering_set_up}
\end{tabular}
}
&
\begin{tabular}{c}
{\vspace{-0.5em}} \\
{\subfigure[Basins for $0 \leq \alpha \leq \frac{\pi}{2}$]{
\includegraphics[width=0.43\textwidth]{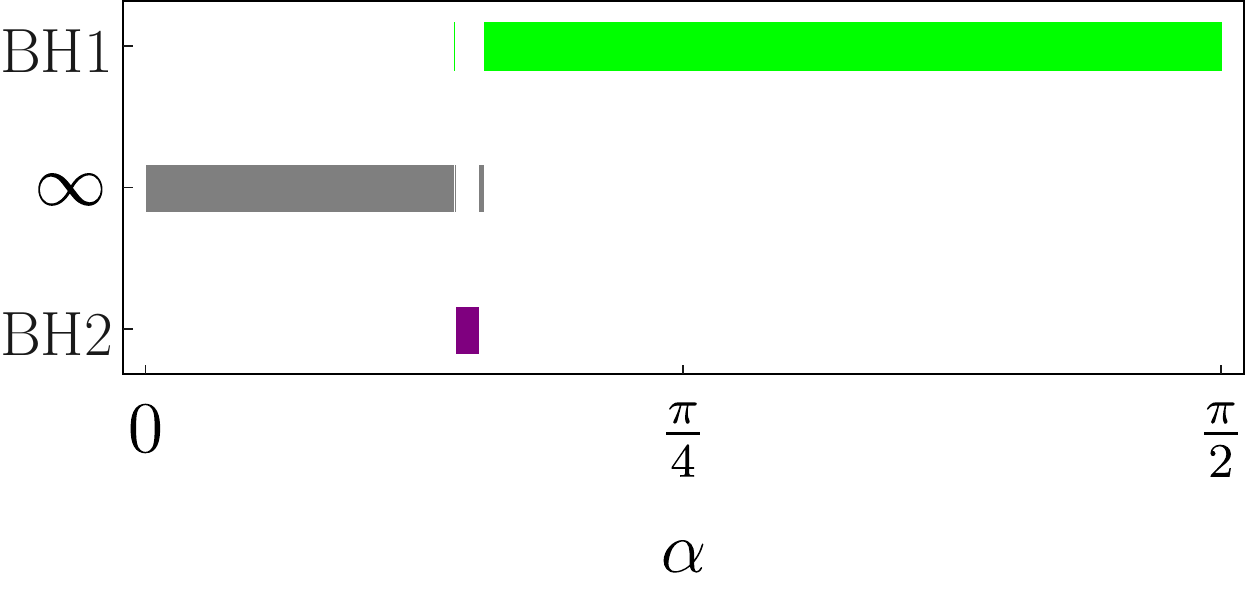}
\label{fig:mp_one_dim_basins_full}} } \\
{\subfigure[Basins for $0.43 \leq \alpha \leq 0.53$]{
\includegraphics[width=0.43\textwidth]{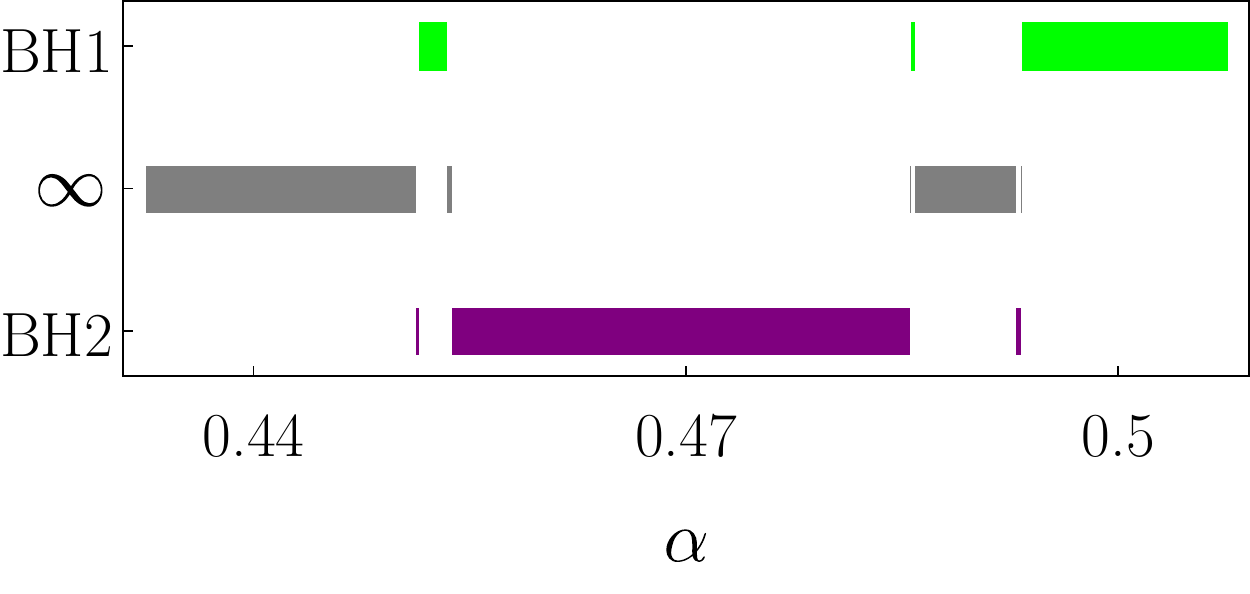}
\label{fig:mp_one_dim_basins_zoom}}}
\end{tabular}
\end{tabular}
\end{center}
\caption{Set-up of the one-parameter scattering problem and corresponding one-dimensional exit basins. (a) Rays which emanate from the centre of mass between the black holes with initial scattering angle $\alpha$ may: (i) escape to spatial infinity; (ii) fall into the lower black hole; (iii) fall into the upper black hole. (b) One-dimensional exit basin diagram for $\alpha \in [0, \frac{\pi}{2}]$. (c) Magnification of the exit basin diagram from (b), with $\alpha \in [0.43, 0.53]$.}
\end{figure}

Figure \ref{fig:mp_one_dim_scattering_set_up} shows examples of null geodesics emanating from the origin for different values of the initial angle $\alpha$. We see that, for $\alpha \sim 0$, the photon will escape to spatial infinity; if $\alpha \sim \frac{\pi}{2}$, the ray will fall directly into the upper black hole; for intermediate values of $\alpha$, the ray may orbit the upper black hole before plunging into the lower black hole, or it may pass between the black holes before escaping to spatial infinity. Figure \ref{fig:mp_one_dim_basins_full} shows the final fate of the photon as a function of $\alpha$. This is an example of a one-dimensional \emph{exit basin diagram} (see Section \ref{sec:chaotic_dynamical_systems}), which allows us to gain insight into the nature of scattering by a pair of fixed black holes.

%
The exit basin diagram presented in Figure \ref{fig:mp_one_dim_basins_full} can be thought of as a one-dimensional binary black hole shadow for an observer located at the centre of mass. For the one-parameter scattering problem outlined above, the shadow (or exit basin) of the upper/lower black hole can be defined formally as
\begin{equation}
\label{eqn:basin_upper_lower_definition}
\mathcal{B}_{\pm} = \lim_{\varepsilon \rightarrow 0} \left\{ \alpha \in [0, 2 \pi) \, \left| \, x^{2}(\lambda; \alpha) + \left( z(\lambda; \alpha) - z_{\pm} \right)^{2} \leq \varepsilon , \, \lambda \in (- \infty, 0 ) \right. \right\},
\end{equation}
where $(x, z) = (0, z_{\pm})$ is the location of the upper/lower black hole. Here, the notation $x(\lambda; \alpha)$ means $x(\lambda)$ with initial data $\alpha$ specified by \eqref{eqn:mp_one_dim_initial_data}. The binary black hole shadow is then the union of the two exit basins: $\mathcal{B}_{\textrm{S}} = \mathcal{B}_{+} \cup \mathcal{B}_{-}$. Similarly, the basin corresponding to spatial infinity can be defined as
\begin{equation}
\label{eqn:basin_infinity_definition}
\mathcal{B}_{\infty} = \lim_{R \rightarrow \infty} \left\{ \alpha \in [0, 2 \pi) \, \left| \, \lim_{\lambda \rightarrow - \infty} r(\lambda; \alpha) \geq R \right. \right\},
\end{equation}
where $r = \sqrt{x^{2} + z^{2}}$ is the radial distance from the origin in the meridian plane. The set of scattering singularities in the initial data is given by $\left[ 0, 2 \pi \right) \setminus ( \mathcal{B}_{\textrm{S}} \cup \mathcal{B}_{\infty} )$.

In Figure \ref{fig:mp_one_dim_basins_zoom}, we magnify an interesting region of the exit basin diagram from Figure \ref{fig:mp_one_dim_basins_full} which demonstrates rich structure for intermediate scattering angles $\alpha \in [0.43, 0.53]$. Inspecting this figure, it appears that the one-dimensional shadows exhibit intricate detail on smaller and smaller scales, hinting at self-similarity. In Section \ref{sec:symbolic_dynamics}, we will analyse the trajectories of the planar one-parameter scattering problem in more detail. Then, in Section \ref{sec:ordering_perpetual_orbits}, we will then demonstrate that this analysis can be used to understand the rich structure hinted at in Figure \ref{fig:mp_one_dim_basins_zoom} in greater detail.

\section{Symbolic dynamics}
\label{sec:symbolic_dynamics}

\subsection{Chaotic scattering and symbolic dynamics}

One way to demonstrate that a system is chaotic is through the use of \emph{symbolic dynamics}, which provides a succinct description of the dynamics of chaotic systems. Symbolic dynamics describes the topology of trajectories in phase space, encoding trajectories as a string of abstract symbols. This method provides a coordinate-invariant method of characterising chaos, which is of particular importance when studying the dynamics of general-relativistic systems, due to the general covariance of the theory. Furthermore, symbolic dynamics can be studied \emph{analytically}, despite the fact that the equations of motion are themselves non-integrable.

In the study of chaotic scattering, the subset of the initial data corresponding to trajectories which asymptote to unstable \emph{periodic} orbits, denoted by $\Omega_{\textrm{P}}$, is particularly important. Moreover, the subset of unstable \emph{aperiodic} orbits $\Omega_{\textrm{A}}$ also plays in important role. The development of a suitable symbolic code allows one to develop a greater understanding of the rich structure of the strange repellor $\Omega_{\textrm{R}} = \Omega_{\textrm{P}} \cup \Omega_{\textrm{A}}$ (see Section \ref{sec:chaotic_dynamical_systems}).
%

The aim of this section is to develop a symbolic dynamics for the Majumdar--Papapetrou di-hole. Rather than give a formal overview of symbolic dynamics, we will instead review two symbolic codes for the Gaspard--Rice three-disc model \cite{GaspardRice1989}. The first symbolic coding, referred to here as \emph{collision dynamics} was introduced in \cite{Eckhardt1987}; the second, which we refer to as \emph{decision dynamics} is an alternative symbolic code which allows a greater understanding of the ordering of non-escaping orbits in the initial data, i.e., the structure of the repellor $\Omega_{\textrm{R}}$. We will then turn our attention to the development of collision dynamics and decision dynamics for the Majumdar--Papapetrou di-hole.


\subsection{Symbolic dynamics for the Gaspard--Rice system}
\label{sec:symbolic_dynamics_gaspard_rice}

\begin{figure}
\begin{center}
\includegraphics[width=0.45\textwidth]{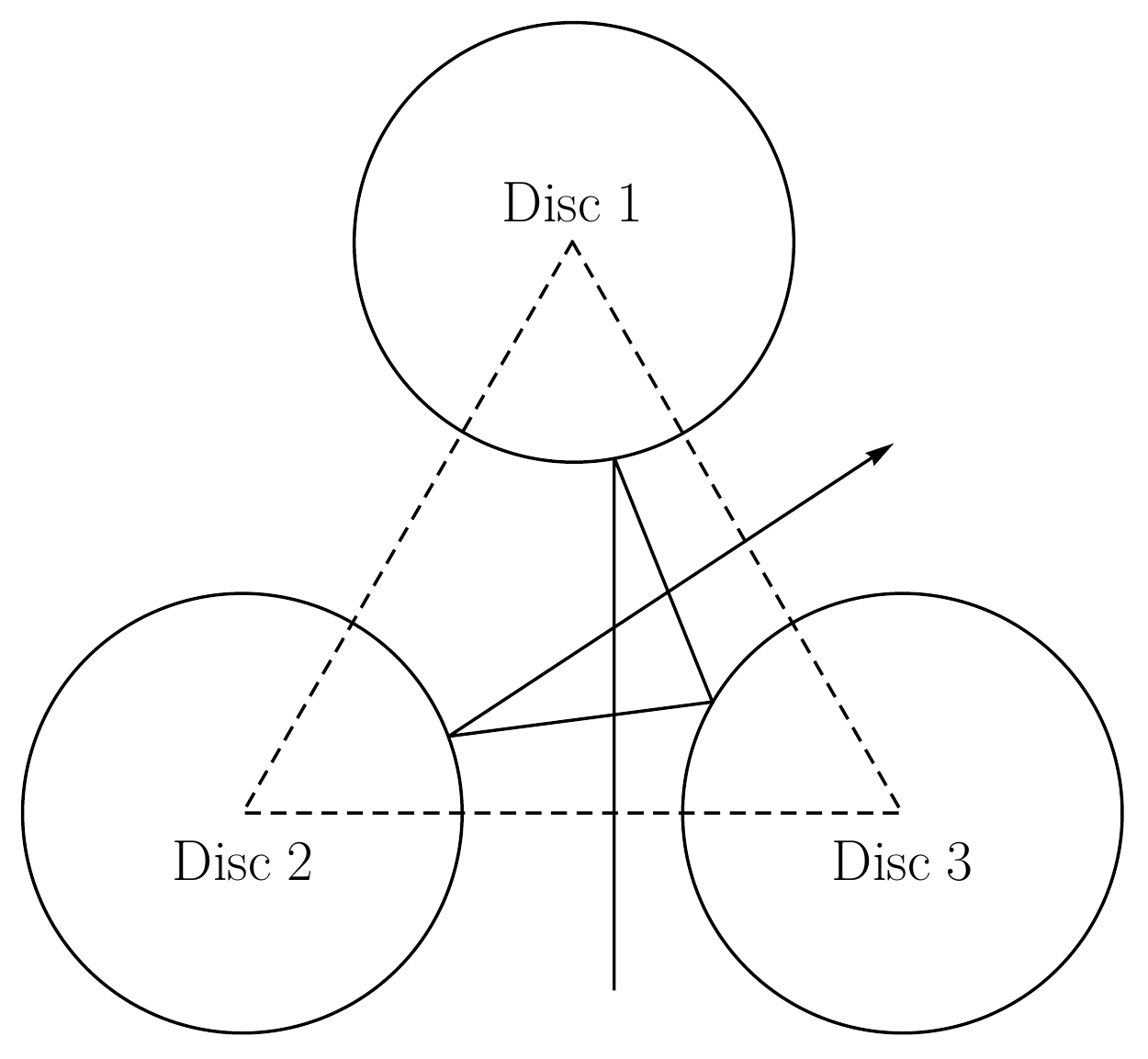}
\caption{Scattering geometry for the Gaspard--Rice three-disc model. A particle undergoes perfectly elastic scattering with three hard discs, centred on the vertices of an equilateral triangle. Using collision dynamics, the reference trajectory -- which first hits Disc 1, then Disc 3, then Disc 2, before escaping to infinity -- can be encoded as the sequence of symbols $132$.}
\label{fig:three_disc_set_up}
\end{center}
\end{figure}

The canonical example of a chaotic scatterer is the three-disc model proposed by Eckhardt \cite{Eckhardt1987}, which has been studied extensively by Gaspard and Rice \cite{GaspardRice1989}, and others \cite{PoonCamposOttEtAl1996, AguirreSanjuan2003}. In this two-dimensional system, an incoming particle, whose initial state is characterised by a pair of initial conditions (i.e., position and momentum), undergoes perfectly elastic collisions with three fixed hard discs (typically of equal radius and situated at the vertices of an equilateral triangle) until it escapes to infinity. The set-up is shown in Figure \ref{fig:three_disc_set_up}.

The qualitative features of the system can be understood through the use of symbolic dynamics. The standard approach is to assign a label from the alphabet $\mathcal{A} = \{1, 2, 3 \}$ to each of the discs. A trajectory is then labelled by a sequence of digits from this alphabet, each of which records the collision of the particle with the disc of the same label. One must also impose the requirement that no digit follows itself (i.e., the particle may not collide with the same disc twice in a row). For example, the sequence $132$ labels a trajectory which first hits Disc 1, then Disc 3, then Disc 2, before escaping from the scattering region (see Figure \ref{fig:three_disc_set_up}).

In principle, there exist sequences of arbitrary length in this symbolic dynamics representation. The set of perpetual non-escaping orbits are encoded by (bi-)infinite sequences of digits from $\mathcal{A}$. Formally, these are sequences $\mathcal{X}$ which belong to the set $\Sigma = \left\{ \mathcal{X} = (\mathcal{X}_{i})_{i \in \mathbb{F}} \,| \, \mathcal{X}_{i} \in \mathcal{A} \right\}$, where $\mathbb{F} = \mathbb{N}$ ($\mathbb{F} = \mathbb{Z}$) for infinite (bi-infinite) sequences. Here, we are mostly concerned with trajectories which start far from the scattering region. We therefore only consider finite sequences (escaping trajectories) or infinite sequences (non-escaping trajectories), rather than bi-infinite sequences. An important subset of $\Sigma$ is the \emph{countable} set of (eventually) recurring sequences describing (asymptotically) periodic orbits; the embedding of this set in $\Sigma$ is akin to the embedding of $\mathbb{Q}$ in $\mathbb{R}$. There also exists an \emph{uncountable} set of non-recurring infinite sequences, which describe aperiodic orbits.

There is an alternative approach to labelling the trajectories of the three-disc system. After each collision with a disc, a trajectory can be continued in three possible ways: the particle may hit the disc on the left of the current disc (0); it may collide with the disc on the right (2); or it may escape the scattering region (1). In this scheme, the perpetual orbits are described by (bi-)infinite sequences which do not contain the digit $1$. There is a natural mapping between the symbolic representation of non-escaping orbits in this picture and the ternary representation of the middle-thirds Cantor set; this will be discussed in more detail in Section \ref{sec:construction_cantor_like_set}.

We distinguish between these two types of symbolic dynamics as follows: the latter approach is referred to as \emph{decision dynamics}, whilst the former is termed \emph{collision dynamics}. In decision dynamics, repeated neighbouring digits are permitted; this is not the case in collision dynamics. Moreover, the representation of orbits in decision dynamics provides natural insight into the structure of the strange repellor $\Omega_{\textrm{R}}$ -- the subset of initial data corresponding to the non-escaping perpetual orbits.

The decision dynamics representation using the alphabet $\{0, 1, 2 \}$ is useful for understanding the structure of $\Omega_{\textrm{R}}$. However, in open Hamiltonian systems, one is typically concerned with the \emph{escaping} trajectories, as well as the non-escaping ones. In the three-disc model, one may wish to record the exit through which the particle left the scattering region. We may therefore reformulate our symbolic code to account for this as follows. A particle which has just collided with a disc may: escape through the exit on the left (0); collide with the disc on the left (1); escape through the exit opposite (2); collide with the disc on the right (3); escape through the exit on the right (4).

In this symbolic code, one can describe non-escaping orbits as infinite sequences which do not contain the digits $0$, $2$, or $4$. A key point is that finite sequences encode \emph{families} of trajectories, which correspond to open sets in the initial data. Conversely, infinite sequences describe \emph{unique} non-escaping orbits, which correspond to scattering singularities in the initial data.
%

\subsection{Symbolic dynamics for the Majumdar--Papapetrou di-hole}
\label{sec:symbolic_dynamics_mp}

\begin{figure}
\begin{center}
\includegraphics[width=0.45\textwidth]{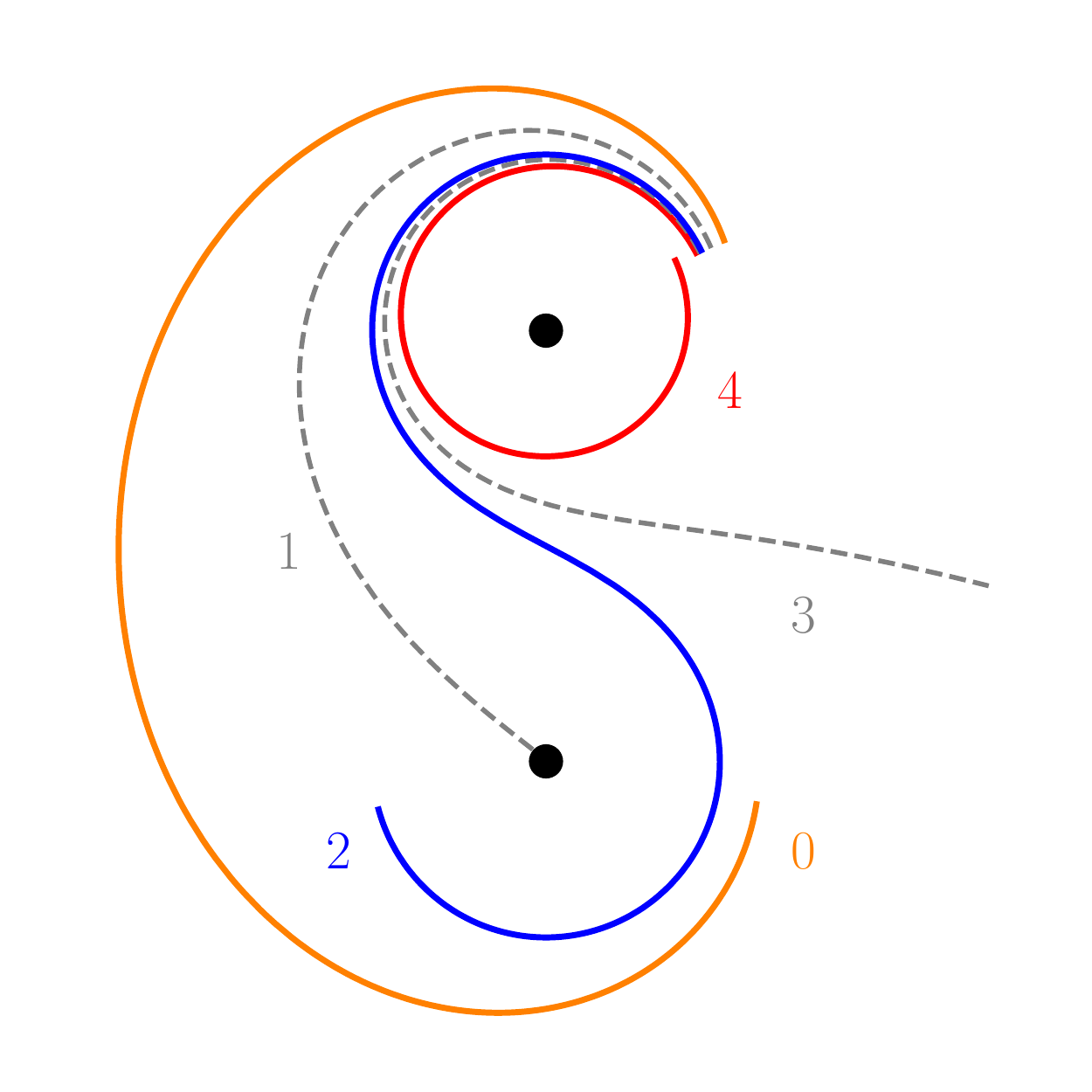}
\end{center}
\caption{Schematic diagram of ``decision dynamics'' -- a version of symbolic dynamics -- for the Majumdar--Papapetrou di-hole.}
\label{fig:mp_decision_dynamics}
\end{figure}

We now turn our attention to a symbolic dynamics representation of null geodesics on the Majumdar--Papapetrou di-hole spacetime. We develop the symbolic coding by considering a null geodesic in a congruence which has reached a ``decision point''. This is illustrated in Figure \ref{fig:mp_decision_dynamics}. The geodesic may follow a path around the opposite black hole in the same sense ($0$); around the other black hole in the opposite sense ($2$); or around the same black hole in the same sense ($4$). Moreover, the photon could fall into a black hole ($1$); or escape to infinity ($3$). In order to avoid double-counting, we do not enumerate the possibility of falling into the ``other'' black hole; this is accounted for at a previous or subsequent decision point. The rays which plunge directly into a black hole, or directly escape to infinity do not generate interesting structure in the exit basins (shadows).

Null rays confined to the meridian plane in the Majumdar--Papapetrou di-hole spacetime can be described using the base-$5$ symbolic alphabet $\mathcal{A} = \{0, 1, 2, 3, 4 \}$. The non-escaping orbits -- which remain in the strong-field region without plunging into a black hole or escaping to infinity -- are described by infinite sequences of \emph{even} digits from the alphabet $\mathcal{A}$, i.e., sequences from the set
\begin{equation}
\Sigma_{\textrm{R}} = \left\{ \mathcal{X} = (\mathcal{X}_{i})_{i \in \mathbb{N}} \,| \, \mathcal{X}_{i} \in \left\{ 0, 2, 4 \right\} \right\} ,
\end{equation}
The rays which escape the system (by crossing an event horizon or reaching spatial infinity) are represented by sequences of finite length $k \in \mathbb{N}$, whose first $k - 1$ digits are even, and which terminate in the digit $1$ (absorbed) or $3$ (scattered); formally, these are sequences from the set
\begin{equation}
\Sigma_{\textrm{E}} = \left\{ \mathcal{X} = (\mathcal{X}_{i})_{i = 1}^{k} \,| \, \mathcal{X}_{i} \in \left\{ 0, 2, 4 \right\}, 1 \leq i \leq k - 1, \mathcal{X}_{k} \in \left\{ 1, 3 \right\}, \forall k \in \mathbb{N} \right\} .
\end{equation}
Sequences in $\Sigma_{\textrm{E}}$ (which are of finite length) describe families of orbits, which are given by open sets in the initial data. (In fact, any sequence of finite length describes a family of orbits; such a sequence need not belong to $\Sigma_{\textrm{E}}$. For example, the sequence $024$ represents the family of trajectories which first make decision $0$, then decision $2$, then decision $4$. The corresponding initial conditions form a set of finite measure.) Sequences which belong to $\Sigma_{\textrm{R}}$ encode unique perpetual orbits, each of which generates a scattering singularity in the initial data. These constitute the repellor $\Omega_{\textrm{R}}$.

\begin{figure}[t]
\begin{center}
\subfigure[$\overline{0}$]{
{\includegraphics[width=0.23\textwidth]{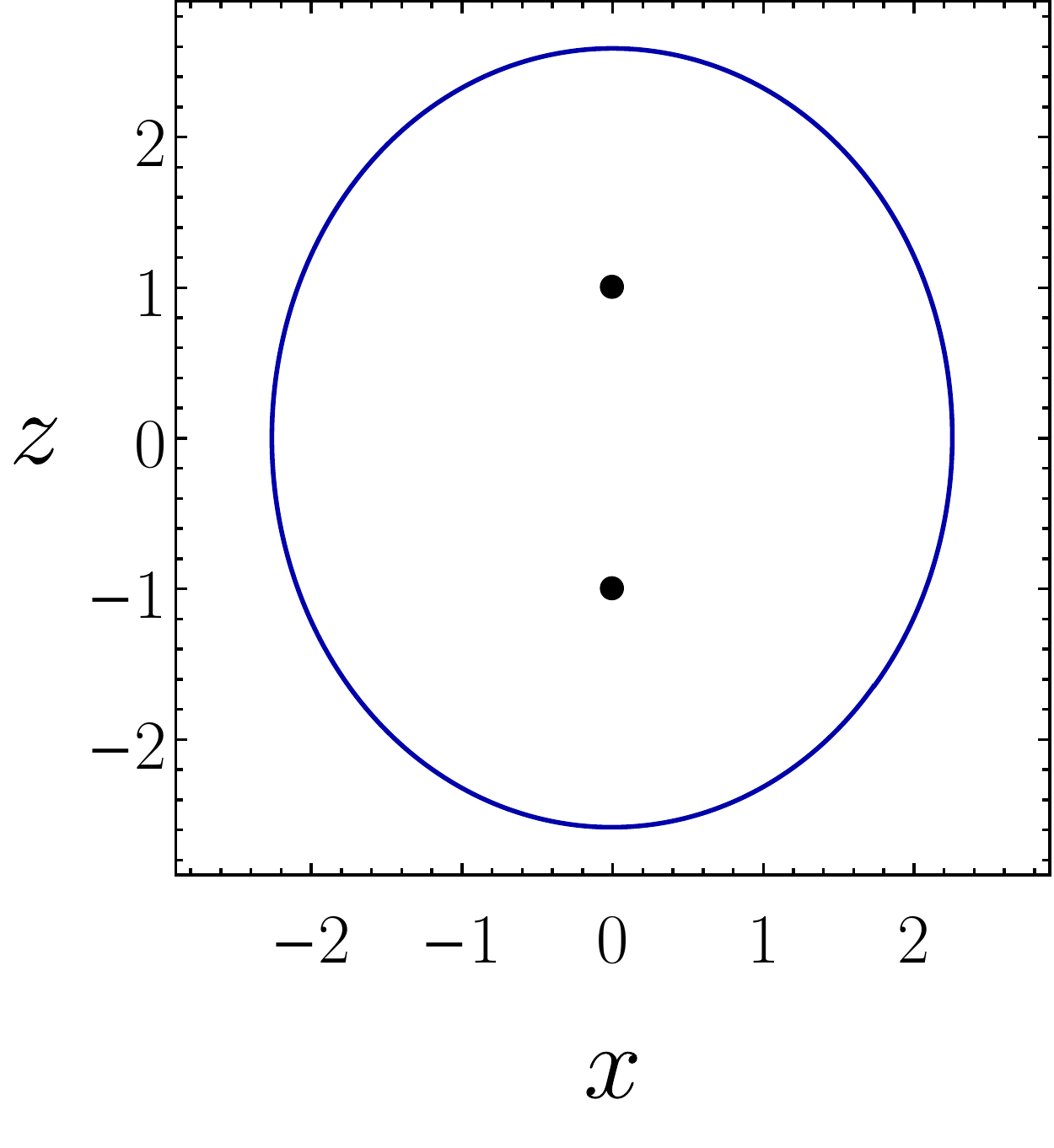}}\label{fig:mp_planar_ray_0}}
\subfigure[$\overline{2}$]{
{\includegraphics[width=0.23\textwidth]{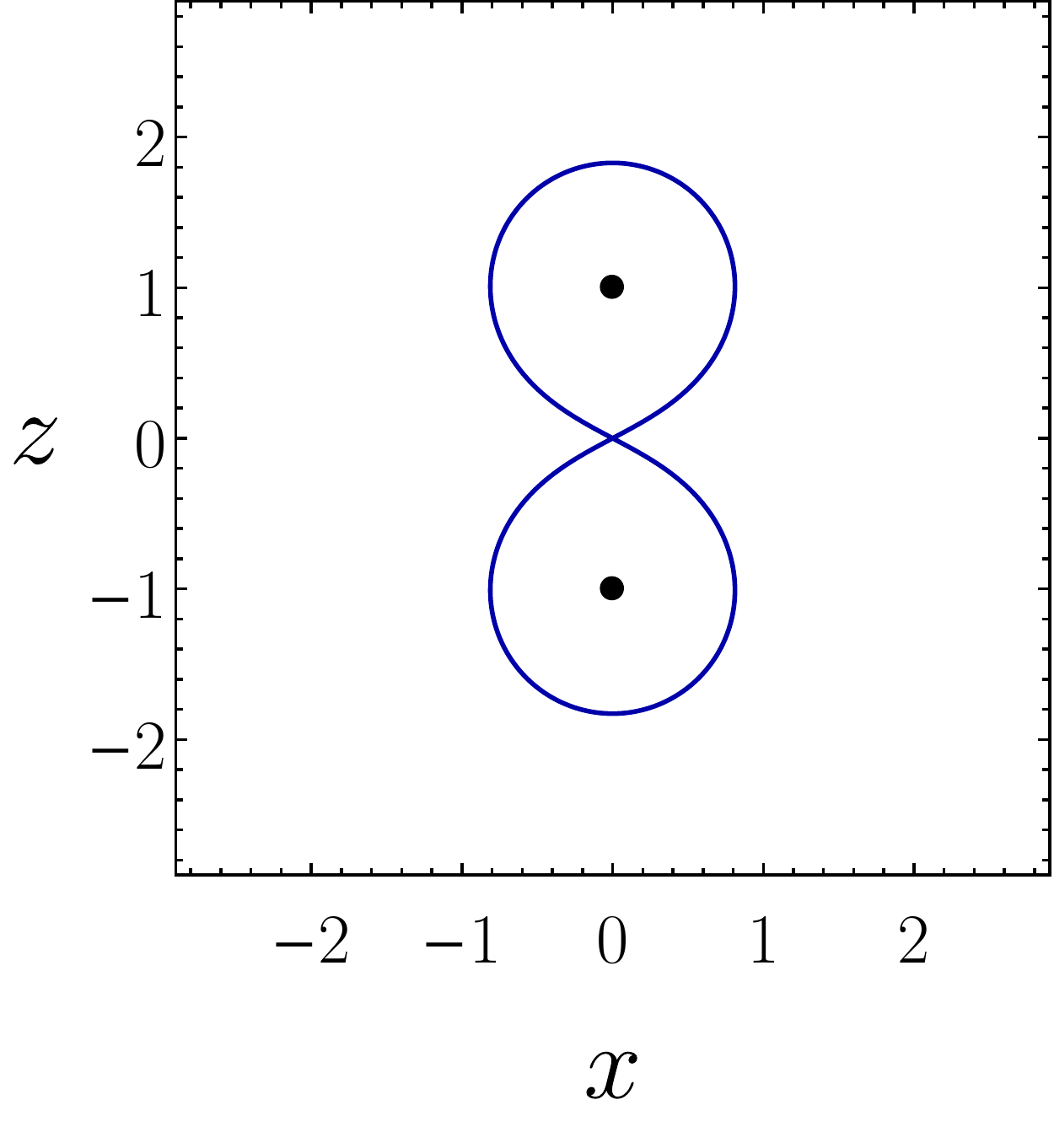}}\label{fig:mp_planar_ray_2}}
\subfigure[$\overline{4}$]{
{\includegraphics[width=0.23\textwidth]{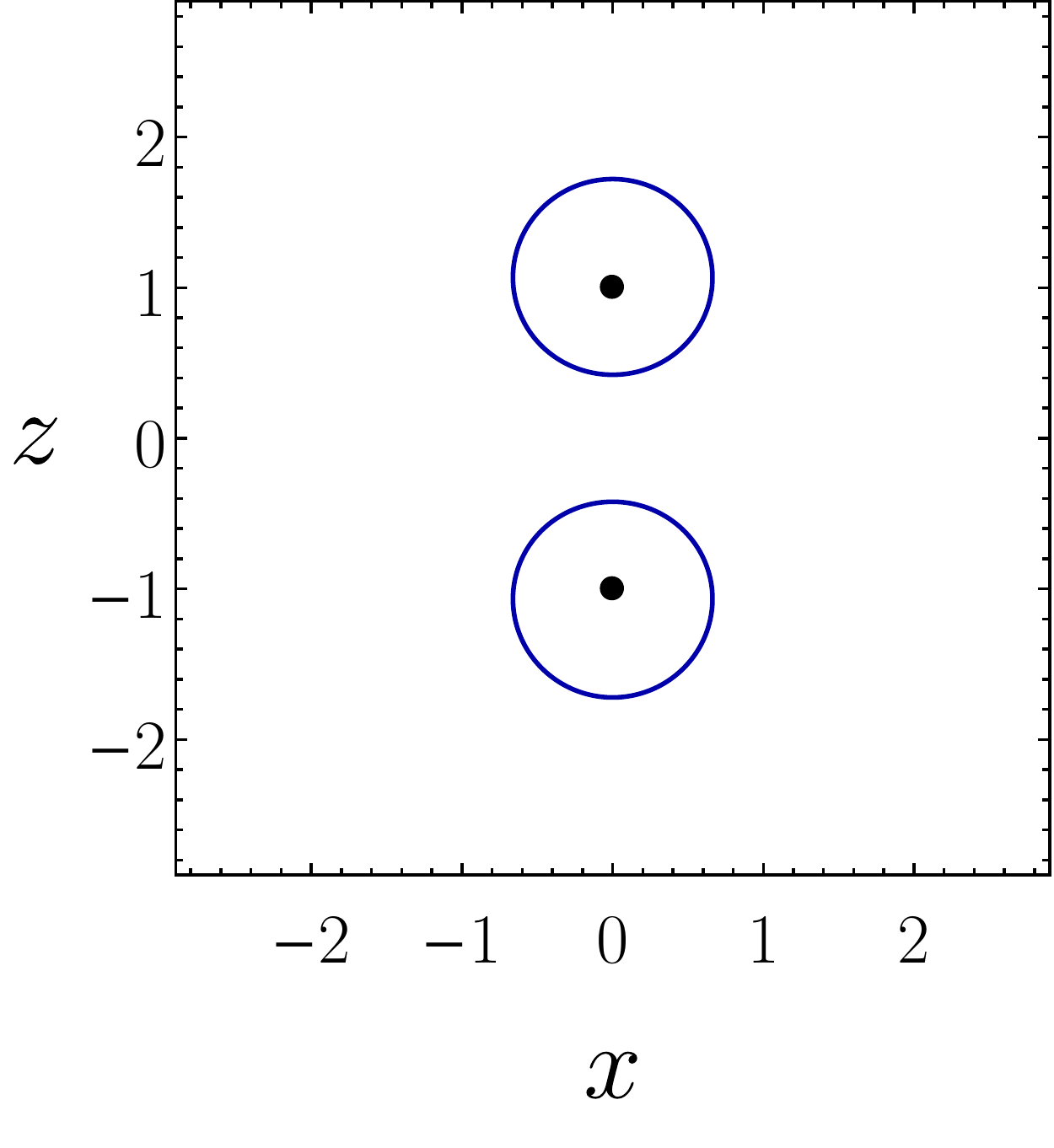}}\label{fig:mp_planar_ray_4}}
\subfigure[$\overline{02}$]{
{\includegraphics[width=0.23\textwidth]{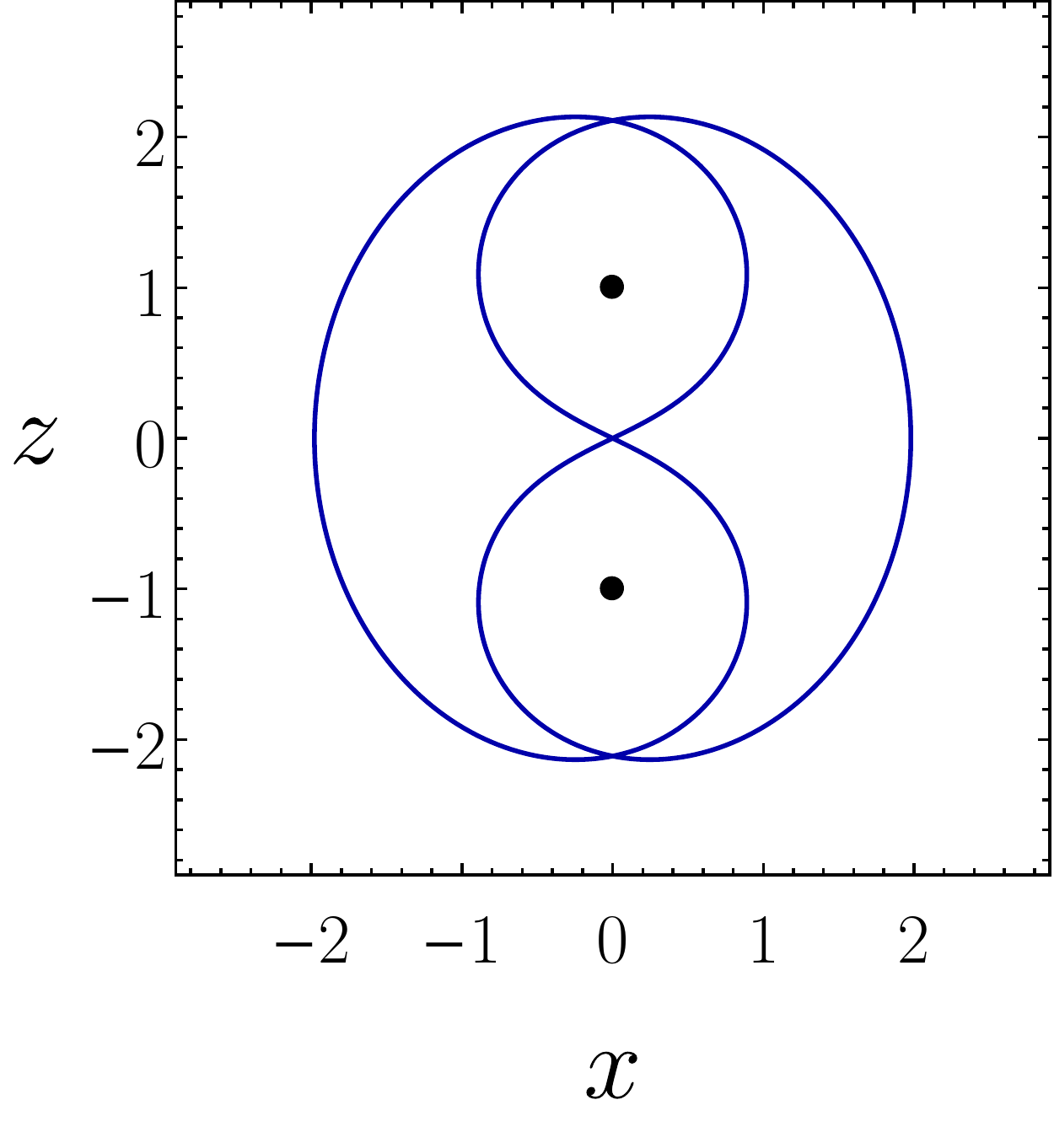}}\label{fig:mp_planar_ray_02}}
\subfigure[One-dimensional shadow]{
{\includegraphics[width=0.45\textwidth]{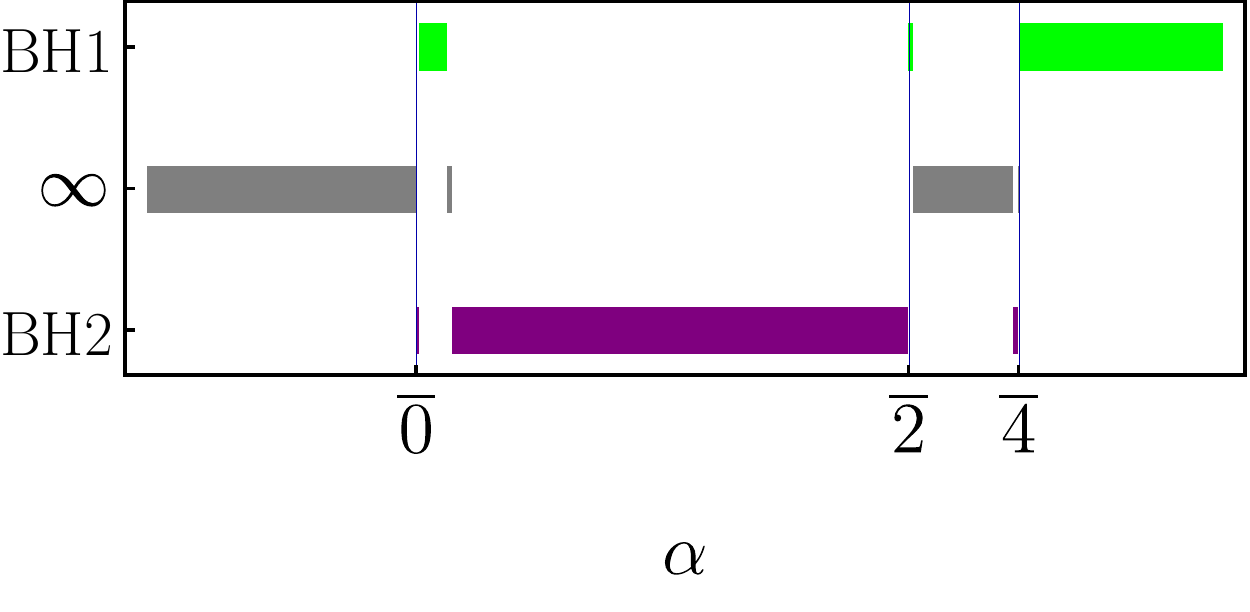}}\label{fig:mp_one_dim_shadow_fundamental_orbit_labels}}
\caption{Examples of planar periodic null orbits with zero angular momentum $p\ind{_{\phi}}$. We choose $M_{\pm} = 1$ and $d = 2$ as default parameter values. In our symbolic code, these orbits are represented by the recurring sequences (a) $\overline{0}$, (b) $\overline{2}$, (c) $\overline{0}$, and (d) $\overline{02}$. The orbits (a)--(c) are referred to as the \emph{fundamental} periodic orbits. Orbit (d) undergoes ``dynamical transitions'' between orbits (a) and (b); this is reflected in its symbolic representation. (e) One-dimensional binary black hole shadow. The vertical lines show the location of the fundamental orbits (a)--(c) in the initial data. \label{fig:mp_planar_rays}}
\end{center}
\end{figure}

An important subset of the perpetual orbits is the set of periodic orbits. In the decision dynamics scheme, periodic orbits are represented by repeating sequences from $\Sigma_{\textrm{R}}$. For example, the recurring sequence $\overline{0} = 000\cdots$ corresponds to the periodic orbit which loops around both black holes; the sequence $\overline{2}$ corresponds to a figure-of-eight orbit; and $\overline{4}$ describes the null geodesic which orbits an individual black hole. These three periodic orbits, and an additional example, are shown in Figure \ref{fig:mp_planar_rays}. A generic non-escaping orbit need not be periodic, however. The representation of aperiodic orbits is given by non-recurring sequences from $\Sigma_{\textrm{R}}$.

The encoding of orbits using a base-$5$ symbolic alphabet provides a natural mapping between the initial data corresponding to the perpetual orbits and the elements of the $5$-adic Cantor set. (This will be discussed in more detail in Section \ref{sec:cantor_like_structure}.)

Cornish and Gibbons \cite{CornishGibbons1997} provide an alternative symbolic coding for planar unstable photon orbits in the Majumdar--Papapetrou di-hole spacetime.\footnote{In fact, Cornish and Gibbons \cite{CornishGibbons1997} analyse a one-parameter family of solutions to the Einstein--Maxwell--dilaton theories, of which the Majumdar--Papapetrou di-hole geometries are a subfamily. The authors explore transitions between regular and chaotic motion as the dilaton coupling is varied, and the integrable Kaluza--Klein limit. The article also contains a discussion of the topological entropy of the symbolic coding.} In their approach, a null geodesic is described by a sequence of digits from the alphabet $\left\{ +1, 0, -1 \right\}$ which record the passage of the geodesic through three ``windows'' placed on the symmetry axis. The digits $+1$, $0$, and $-1$ correspond, respectively, to the intersection of the geodesic with the open intervals $z \in (z_{+}, + \infty)$, $z \in (z_{-}, z_{+})$, and $z \in (- \infty, z_{-})$. We classify this symbolic code as an example of collision dynamics (cf.~the three-disc model in Section \ref{sec:symbolic_dynamics_gaspard_rice}). In this work, we favour decision dynamics for the Majumdar--Papapetrou di-hole, as it provides a better description of the structure of the strange repellor $\Omega_{\textrm{R}}$, and it may be used to describe both the escaping and non-escaping trajectories.

Alonso \emph{et al.} \cite{AlonsoRuizSanchez-Hernandez2008} employ a yet another symbolic code to describe non-escaping orbits in the meridian plane. The authors use a type of collision dynamics, describing orbits with symbols from the alphabet $\{ +, \circ, - \}$, depending on where the ray intersects the symmetry axis. This choice of symbolic coding is similar to that of Cornish and Gibbons \cite{CornishGibbons1997}.

\section{Cantor-like structure of the one-dimensional shadow}
\label{sec:cantor_like_structure}

\subsection{Ordering of non-escaping orbits in the initial data}
\label{sec:ordering_perpetual_orbits}

The non-escaping orbits are of importance for a number of reasons. Firstly, in a general open Hamiltonian dynamical system, the non-escaping trajectories correspond to scattering singularities in the initial data: they form the repellor $\Omega_{\textrm{R}}$ (see Section \ref{sec:chaotic_dynamical_systems}). The structure of the strange repellor provides useful information about the dynamics of chaotic scattering systems. Secondly, in the context of scattering and absorption of null geodesics by black holes, the perpetual orbits lie on the boundary between the absorbed and scattered rays; that is, they correspond to the \emph{boundary} of the black hole shadow. In order to understand the structure of the shadows of a pair of fixed black holes, one must therefore understand the structure of the perpetual orbits in the associated scattering problem.

We now apply the symbolic code developed in Section \ref{sec:symbolic_dynamics_mp} to gain insight into the ordering and organisation of the non-escaping orbits in the initial data, for the one-dimensional scattering problem described in Section \ref{sec:planar_geodesics_one_dim_shadows}. Consider the one-dimensional shadow (or exit basin diagram) for $\alpha \in \left[ 0, \frac{\pi}{2} \right]$, presented in Figure \ref{fig:mp_one_dim_basins_full}. Note that we need only consider this restricted domain due to the symmetry of the equal-mass scattering problem. As described in Section \ref{sec:planar_geodesics_one_dim_shadows}, all rays with $\alpha \sim 0$ will escape to infinity, and rays with $\alpha \sim \frac{\pi}{2}$ will plunge directly into the upper black hole. For intermediate values of $\alpha$, we observe interesting structure in the exit basins. In particular, we highlight three regions of interest; see Figure \ref{fig:mp_one_dim_basins_zoom_0}. Outside of these open intervals, there are no scattering singularities: all trajectories are either absorbed or scattered by the system. However, each of these open intervals contains infinitely many scattering singularities.

To the left of the left-hand interval, all trajectories escape to infinity. Between the left and middle intervals, all rays cross the event horizon of the lower black hole. All initial conditions which lie between the middle and right-hand intervals correspond to rays which escape to infinity. Finally, to the right of the right-hand interval, all trajectories fall into the upper black hole.

Let us consider what happens as one increases the scattering parameter $\alpha$ continuously, starting from $\alpha = 0$. It is clear from the exit basin diagram that, for $\alpha$ sufficiently small, all rays escape to infinity. This is the case until we reach some critical value of $\alpha$ which corresponds to the \emph{first} perpetual orbit. This orbit asymptotes to the perpetual orbit which has symbolic representation $\overline{0}$ in decision dynamics; see Figure \ref{fig:mp_planar_ray_0}.


In a similar manner, let us consider the effect of continuously decreasing $\alpha$, starting from $\alpha = \frac{\pi}{2}$. Initially, all trajectories fall into the upper black hole, until we reach a critical value of $\alpha$, corresponding to the \emph{last} perpetual orbit (i.e., the one with the largest value of $\alpha \in \left[ 0 , \frac{\pi}{2} \right]$). This is a periodic orbit, which loops only around the upper black hole. In decision dynamics, this orbit is represented by the recurring sequence $\overline{4}$; see Figure \ref{fig:mp_planar_ray_4}.

At first sight, it appears from Figure \ref{fig:mp_one_dim_shadow_fundamental_orbit_labels} that the ordering of the perpetual orbits in the initial data exactly matches the natural ordering of their symbolic representations, when the latter are interpreted as elements of the unit interval in base-$5$; however, this is not the case.

The set of all possible symbolic sequences for planar trajectories is $\Sigma = \Sigma_{\textrm{R}} \cup \Sigma_{\textrm{E}}$, where sequences in $\Sigma_{\textrm{R}}$ describe non-escaping trajectories, and sequences in $\Sigma_{\textrm{E}}$ describe trajectories which fall into a black hole or escape to infinity; see Section \ref{sec:symbolic_dynamics_mp}. We now define an \emph{ordering function} $F \colon \Sigma \rightarrow [0, 1]$, which maps a decision dynamics sequence to an element of the unit interval, represented in base-$5$. We demand that the ordering of elements in $\operatorname{im}{(F)} \subseteq [0, 1]$ exactly matches the ordering of the scattering singularities in the initial data; that is, $F(\mathcal{X}^{(1)}) < F(\mathcal{X}^{(2)})$ if and only if $\alpha^{(1)} < \alpha^{(2)}$, where $\mathcal{X}^{(i)} \in \Sigma$ is any sequence and $\alpha^{(i)}$ is its corresponding initial condition.

The ordering function is defined as follows. We first define a \emph{parity-reordering operation} $\mathcal{P} \colon \Sigma \rightarrow \Sigma$, which maps each digit $\mathcal{X}_{i}$ of a sequence $\mathcal{X} = \mathcal{X}_{1} \mathcal{X}_{2} \mathcal{X}_{3} \cdots \in \Sigma$ to a digit $\widetilde{\mathcal{X}}_{i}$ of the sequence $\mathcal{P}(\mathcal{X})$, according to the following procedure which keeps track of parity $P$. We begin with $P = +1$ and consider the sequence $\mathcal{X}$ from left to right, examining each digit $\mathcal{X}_{i}$ in turn, as follows.
\begin{enumerate}
\item[(1)] Set
\begin{equation}
\widetilde{\mathcal{X}}_{i} =
\begin{cases}
\mathcal{X}_{i}, & \text{if~} P = + 1; \\
4 - \mathcal{X}_{i}, & \text{if~} P = -1.
\end{cases}
\end{equation}
\item[(2)] Reverse the parity ($P \mapsto - P$) if $\mathcal{X}_{i} = 2$.
\item[(3)] Iterate $i \mapsto i + 1$.
\end{enumerate}
We note that $\mathcal{P}^{2}(\mathcal{X}) = \mathcal{X}$, so $\mathcal{P}$ is an involution. The parity-reordering operation can be understood in terms of the geometry of rays: the digit $2$ describes a null geodesic which passes between the black holes, changing the sense of the orbit from clockwise to anti-clockwise and vice versa. Hence, the ordering of digits in our symbolic code must be reversed. (It will be convenient to write $\mathcal{P}(\mathcal{X}) = \widetilde{\mathcal{X}}$ to denote parity-reordered sequences.)

Next, we define $f \colon \Sigma \rightarrow [0, 1]$ to be the function which maps a (parity-reordered) sequence to a real number in the unit interval represented in base-$5$. We take $f(\mathcal{Y}) = 0.\mathcal{Y}_{1}\mathcal{Y}_{2}\mathcal{Y}_{3} \cdots$, where $\mathcal{Y} = \mathcal{Y}_{1}\mathcal{Y}_{2}\mathcal{Y}_{3} \cdots \in \Sigma$ is any (parity-reordered) sequence. Finally, we take the ordering function to be $F(\mathcal{X}) = (f \circ \mathcal{P})(\mathcal{X})$ for all sequences $\mathcal{X} \in \Sigma$.

To illustrate how $F$ acts on sequences in $\Sigma$, consider the pair of recurring sequences $\mathcal{X}^{(1)} = \overline{20}$, $\mathcal{X}^{(2)} = \overline{24}$. The parity-reordered sequences are $\mathcal{P}(\mathcal{X}^{(1)}) = \overline{2420}$ and $\mathcal{P}(\mathcal{X}^{(2)}) = \overline{2024}$. We therefore find $F(\mathcal{X}^{(1)}) = 0.\overline{2420}$ and $F(\mathcal{X}^{(2)}) = 0.\overline{2024}$. Since $F(\mathcal{X}^{(2)}) < F(\mathcal{X}^{(1)})$, the perpetual orbit described by the sequence $\mathcal{X}^{(2)}$ will precede that represented by $\mathcal{X}^{(1)}$ in the initial data; in other words, $\alpha^{(2)} < \alpha^{(1)}$.

\subsection{Constructing a Cantor-like set on the initial data}
\label{sec:construction_cantor_like_set}

\subsubsection{Cantor sets}

The \emph{middle-thirds Cantor set} $\mathcal{C}$ is a paradigm of fractal geometry; its complex structure can be arrived at using a remarkably simple iterative procedure \cite{Falconer2004}. Starting with the closed unit interval $I = [0, 1]$, one first removes the open middle-third interval $\left(\frac{1}{3}, \frac{2}{3} \right)$, which leaves two closed intervals $\left[0, \frac{1}{3} \right]$ and $\left[\frac{2}{3}, 1\right]$. At the next step, one removes the open middle-third of the remaining closed intervals. This is repeated \emph{ad infinitum}, until one is left with a set of distinct points of zero Lebesgue measure which were not removed at any step.

In general, a set $\Lambda$ is a \emph{Cantor set} if it is a closed, totally disconnected, and perfect subset of $I$ \cite{Devaney1989}. Recall that a set is \emph{closed} if it contains all of its boundary points; a set is \emph{totally disconnected} if it contains no intervals; and a set is \emph{perfect} if every point in it is a limit point of other points in the set. Let $A$ be a subset of a topological space $X$. A point $p \in X$ is a \emph{limit point} of $A$ if every open neighbourhood of $p$ contains at least one point of $A$ which is different from $p$.

Related Cantor-like sets may be generated by removing multiple intervals at each step, or by varying the proportionate width of the interval(s) removed. A straightforward generalisation of the middle-thirds Cantor set is the \emph{middle-$\gamma$ Cantor set} (or \emph{$\gamma$-Cantor set}), denoted $\mathcal{C}_{\gamma}$ \cite{Falconer1997, Falconer2004, Devaney1989}. This can again be constructed iteratively on the unit interval: at each stage one removes the open middle interval of length $\gamma \in (0, 1)$. Clearly the case $\gamma = \frac{1}{3}$ gives rise to the standard middle-thirds Cantor set $\mathcal{C}$.

The middle-thirds Cantor set is an example of a \emph{fractal} (see Section \ref{sec:chaotic_dynamical_systems}), a set which is self-similar under magnification \cite{Devaney1989}. Consider only those points of the Cantor set which lie in the left-hand interval $\left[ 0, \frac{1}{3} \right]$. Magnifying this region by a factor of three, the portion of the Cantor set in $\left[ 0, \frac{1}{3} \right]$ looks exactly the same as the original set in $I$. (More formally, the linear map $L(x) = 3 x$ is a homeomorphism from the piece of the Cantor set in $\left[ 0, \frac{1}{3} \right]$ to the whole Cantor set.) This self-similarity is not restricted to one level of magnification: enlarging any part of the Cantor set at the $k$th iteration in its construction by a factor of $3^{k}$ yields the original set. The $\gamma$-Cantor set also exhibits self-similarity.

The fractal structure of the Cantor set (and its complement in the unit interval) can be understood in more depth by considering the ternary representation of numbers in the unit interval $I$. It is well-known that the middle-thirds Cantor set consists of all elements of $I$ which can be represented in base-$3$ using only the digits $0$ and $2$; see Theorem 4.2 of \cite{AlligoodSauerYorke1996}, for example. The action of deleting the middle-third interval at the $k$th iteration is equivalent to removing from $I$ all elements with a ternary representation whose $k$th digit is $1$, and whose first $k - 1$ digits consist only of the digits $0$ and $2$.

\subsubsection{Cantor basins}

Consider the iterative process used to generate the $\gamma$-Cantor set. One can construct a pair of basins $\left\{ B_{i} \, | \, i \in \{1, 2\} \right\}$, which we refer to as \emph{Cantor basins}, from the complement of $\mathcal{C}_{\gamma}$ in $I$ as follows. Denote by $J_{k}$ the union of the $2^{k - 1}$ open intervals which are \emph{removed} from $I$ at the $k$th iteration in the construction of $\mathcal{C}_{\gamma}$. Now, add the open set $J_{k}$ to the basin $B_{i}$, where $i = 1$ ($i = 2$) if $k$ is odd (even). The basins can then be expressed as an infinite union of disjoint open sets:
\begin{equation}
B_{1} = \bigcup_{l = 1}^{\infty} J_{2 l - 1}, \qquad
B_{2} = \bigcup_{l = 1}^{\infty} J_{2 l}.
\end{equation}
Clearly, $\partial B_{1} = \partial B_{2} = \mathcal{C}_{\gamma}$, i.e., the basins $B_{i}$ share a common boundary, namely the $\gamma$-Cantor set.

It is possible to generalise the procedure outlined above in order to construct an $N$-basin set $\left\{ B_{i} \, | \, i \in \{1, 2, \ldots, N\}, N \in \mathbb{N} \right\}$, where $\partial B_{1} = \partial B_{2} = \ldots = \partial B_{N} = \mathcal{C}_{\gamma}$. Again, we denote by $J_{k}$ the union of the open intervals removed at the $k$th step, and we define the $N$ disjoint basins as
\begin{equation}
B_{i} = \bigcup_{k = i \bmod{N}} J_{k}, \qquad i \in \left\{1, 2, \ldots, N \right\}.
\end{equation}

The construction of the Cantor basins associated with the middle-thirds Cantor set can be understood by considering the ternary representation of elements in the unit interval, which can be viewed as a ``symbolic dynamics'' for the Cantor set; see Figure 4.1 of \cite{AlligoodSauerYorke1996}. For simplicity, consider the case of two basins $\{ B_{1}, B_{2} \}$. All points $p \in I$ with a ternary expansion whose first $1$ appears in the $k$th slot will belong to $B_{1}$ ($B_{2}$) if $k$ is odd (even). For example, all points with ternary expansion $0.021\cdots$ are removed at the third iteration, so they will belong to $B_{1}$ (in the case of two basins). The boundary points (contained in neither basin) are numbers in $I$ whose ternary expansion does not contain the digit $1$. (These are precisely the elements of the Cantor set.) In the language of dynamical systems, the Cantor set (i.e., the boundary of the Cantor basins) is a strange repellor.

In the case of Cantor basins with $\gamma \neq \frac{1}{3}$, the symbolic alphabet $\{ 0, 1, 2 \}$ does not correspond to the digits in the ternary representation of element of the unit interval; rather, the digits $0$, $1$ and $2$ represent the left, middle, and right intervals, respectively, at each step in the construction of the set. For example, the symbolic code $021$ (represented in base-$3$ by the number $0.201$) corresponds to the interval which is arrived at by choosing first the closed left-hand interval ($0$); choosing the closed right-hand sub-interval at the next iteration ($2$); and finally choosing the open middle sub-interval ($1$). It is straightforward to see that, in the case of two basins $\{ B_{1}, B_{2} \}$, the interval represented by the symbolic code $021$ would belong to basin $B_{1}$. We emphasise that it is \emph{not} the case that each element of the open interval represented by $201$ will have a ternary expansion which begins $0.201$. This is only true for $\gamma = \frac{1}{3}$.

\begin{figure}
\begin{center}
\subfigure[Cantor basins in ${\left[0,1\right]}$]{\includegraphics[width=0.45\textwidth]{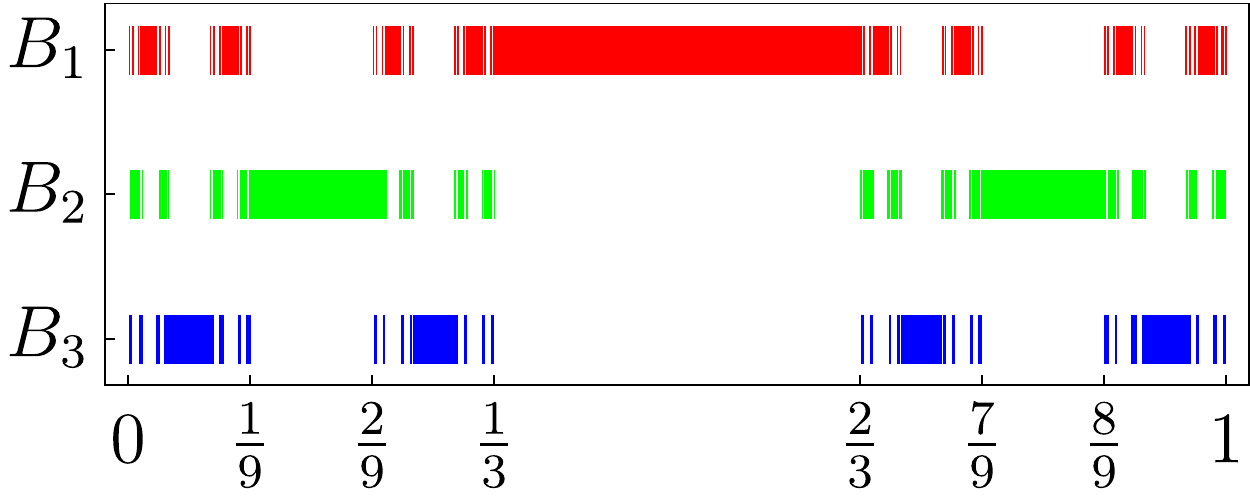} \label{fig:cantor_basin_middle_third}}
\hspace{1em}
\subfigure[Magnified Cantor basins in ${\left[0, \frac{1}{3}\right]}$]{\includegraphics[width=0.45\textwidth]{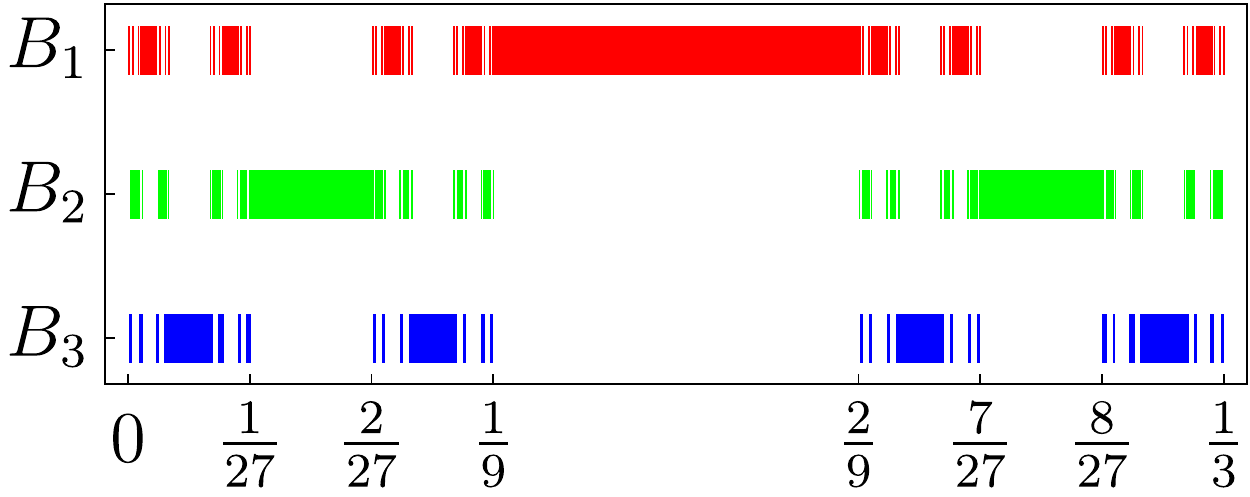} \label{fig:cantor_basin_middle_third_zoom}}
\caption{Three-colour Cantor basins constructed using the iterative procedure used to generate the middle-thirds Cantor set. (a) Three basins $\{ B_{1}, B_{2}, B_{3} \}$ on the unit interval, coloured red, green and blue, respectively. The boundary of the basins is precisely the Cantor set. (b) Magnification of the basins of (a), showing the interval $\left[ 0, \frac{1}{3} \right]$. One can clearly see that the basins are self-similar under this magnification. \label{fig:three_colour_cantor_basins}}
\end{center}
\end{figure}

In Figure \ref{fig:three_colour_cantor_basins}, we plot $N = 3$ Cantor basins which are generated from the construction of the canonical middle-thirds Cantor set $\mathcal{C}$ on the unit interval $\left[ 0, 1 \right]$. The basins $B_{1}$, $B_{2}$ and $B_{3}$ are coloured red, green and blue, respectively. From Figure \ref{fig:cantor_basin_middle_third}, one can see that the open middle-third interval $\left(\frac{1}{3}, \frac{2}{3}\right)$, which is removed at the first step in the construction of the middle-thirds Cantor set, belongs to be basin $B_{1}$; the union of the open middle-third intervals of the two remaining intervals, given by $\left(\frac{1}{9}, \frac{2}{9}\right) \cup  \left(\frac{7}{9}, \frac{8}{9}\right)$, belong to the basin $B_{2}$, and so on. The boundary of the three basins is the Cantor set. In Figure \ref{fig:cantor_basin_middle_third_zoom}, we magnify the Cantor basins of Figure \ref{fig:cantor_basin_middle_third} by a factor of $3$, considering the left-hand interval $\left[ 0, \frac{1}{3} \right]$. In this magnified figure, the self-similarity of the Cantor basins is manifest. This magnification process (i.e., zooming in on the closed left-hand interval) is equivalent to making the ``decision'' represented by the digit $0$: all elements of the interval $\left[0, \frac{1}{3} \right]$ are represented in symbolic dynamics by sequences whose first digit is $0$.
%

\subsubsection{Constructing one-dimensional binary black hole shadows}

We may use decision dynamics as a guide to develop a similar iterative process to construct the one-dimensional shadow on the initial data, defined as $\mathcal{B}_{\textrm{S}} = \mathcal{B}_{+} \cup \mathcal{B}_{-}$ in \eqref{eqn:basin_upper_lower_definition}. Denote by $\alpha_{\widetilde{\mathcal{X}}}$ an initial value corresponding to a perpetual orbit, with representation $\mathcal{X}$ in symbolic dynamics, and parity-reordered sequence $\mathcal{P}(\mathcal{X}) = \widetilde{\mathcal{X}}$ (see Section \ref{sec:ordering_perpetual_orbits}). We need only consider the ``interesting'' interval $C = [\alpha_{\overline{0}}, \alpha_{\overline{4}}]$, in which all of the perpetual orbits lie. (Recall that $\alpha_{\overline{0}}$ and $\alpha_{\overline{4}}$ denote the initial values corresponding to the first and last non-escaping orbits, respectively.) From $C$ one may remove two open intervals $O_{1} = (\alpha_{0 \overline{4}}, \alpha_{2 \overline{0}})$ and $O_{3} = (\alpha_{2 \overline{4}}, \alpha_{4 \overline{0}})$, which correspond to null geodesics which immediately fall into the lower black hole or escape to spatial infinity, respectively. The open interval $O_{1}$ forms part of the one-dimensional black hole shadow ($O_{1} \subset \mathcal{B}_{\textrm{S}}$). Similarly, $O_{3}$ forms part of the exit basin corresponding to spatial infinity ($O_{3} \subset \mathcal{B}_{\infty}$).

\begin{figure}[t]
\begin{center}
\includegraphics[width=0.7\textwidth]{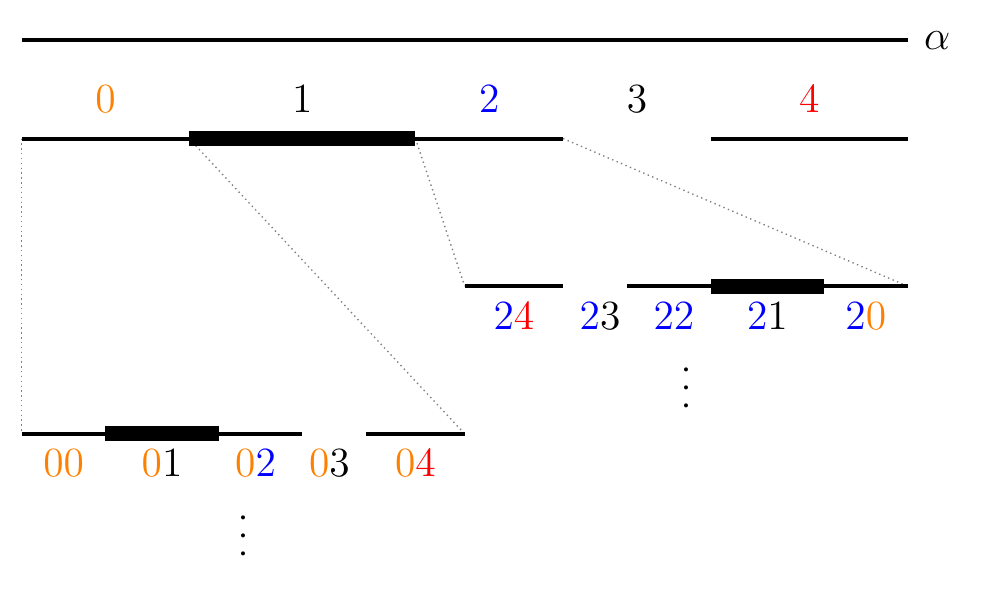}
\caption{Schematic diagram which demonstrates the iterative procedure based on ``decision dynamics'' used to construct a one-dimensional binary black hole shadow. Beginning with a domain of the initial data $\alpha$, one first removes two open intervals, corresponding to decision $1$ (capture by a black hole) and decision $3$ (escape to infinity). The open interval corresponding to decision $1$ (depicted as a thick black line) constitutes part of the black hole shadow; the interval corresponding to decision $3$ (white) does not lie in the shadow. At the next stage, one iterates on the remaining closed intervals, which comprise the initial data corresponding to trajectories whose symbolic representations start with the digit $0$, $2$ or $4$, removing two open intervals at each stage. Iterating \emph{ad infinitum}, we are left with an uncountable infinity of distinct points of measure zero, which correspond to the perpetual orbits.}
\label{fig:mp_shadow_construction}
\end{center}
\end{figure}

We now iterate this procedure on the remaining closed intervals. The process of iterating is equivalent to following geodesics which linger in the strong-field region until they reach the next ``decision point''. This iterative procedure is demonstrated using a schematic diagram, presented in Figure \ref{fig:mp_shadow_construction}.

Suppose we have iterated $k$ times, and we are considering the closed interval $C_{\widetilde{\mathcal{X}}}$, where $\widetilde{\mathcal{X}} = \mathcal{P}(\mathcal{X})$ is a parity-reordered sequence in decision dynamics of length $k$, which only contains \emph{even} digits from the symbolic alphabet. Iterating involves removing the pair of open intervals $O_{\widetilde{\mathcal{X}} 1} = \left( \alpha_{\widetilde{\mathcal{X}} 0 \overline{4}}, \alpha_{\widetilde{\mathcal{X}} 2 \overline{0}} \right)$ and $O_{\widetilde{\mathcal{X}} 3} = \left( \alpha_{\widetilde{\mathcal{X}} 2 \overline{4}}, \alpha_{\widetilde{\mathcal{X}} 4 \overline{0}} \right)$. (Here, $\widetilde{\mathcal{X}} 0 \overline{4}$ represents the concatenation of the sequences $\widetilde{\mathcal{X}}$ and $0 \overline{4}$, for example.) The subset $O_{\widetilde{\mathcal{X}} 1} \subset \mathcal{B}_{\textrm{S}}$ is added to the one-dimensional shadow at each stage; the subset $O_{\widetilde{\mathcal{X}} 3} \subset \mathcal{B}_{\infty}$ is added to the exit basin corresponding to spatial infinity. We note at this stage that $O_{\widetilde{\mathcal{X}} 1}$ and $O_{\widetilde{\mathcal{X}} 3}$ are the open intervals in initial data $\alpha$ which correspond to the set of null geodesics which go through the decision dynamics sequence $\mathcal{X}$, before falling into a black hole, or escaping to infinity, respectively. In the former case, one can determine the black hole into which the photon plunges as follows: count the number of digits $n$ in the sequence $\mathcal{X}$ that are not equal to $4$; if $n$ is even (odd), the geodesic falls into the lower (upper) black hole.

Having iterated $k$ times, we have partitioned $C$ into $3^{k} - 1$ open intervals
\begin{equation}
\left\{ \{ O_{1}, O_{3} \}, \{ O_{01}, O_{03}, O_{21}, O_{23}, O_{41}, O_{43} \}, \ldots \right\},
\end{equation}
which correspond to geodesics which make at most $k - 1$ decisions (represented by even digits in decision dynamics), before falling into a black hole or escaping to spatial infinity; and $3^{k}$ closed intervals
\begin{equation}
\left\{ C, \{ C_{0}, C_{2}, C_{4} \}, \{ C_{00}, C_{02}, C_{04}, C_{20}, C_{22}, C_{24}, C_{40}, C_{42}, C_{44} \}, \ldots \right\},
\end{equation}
which are made up of initial conditions corresponding to rays which linger in the strong-field region long enough to make $k$ decisions, but whose ultimate fate has not yet been determined at this level of precision. (Here, $C$ can be thought of as the set of all initial data corresponding to geodesics which are yet to go through a decision sequence, i.e., those which have been through an \emph{empty} sequence.)

In the limit $k \rightarrow \infty$, one is left with an infinite number of closed sets of zero measure whose union is isomorphic to the $5$-adic Cantor set; and an infinite number of open sets whose union is isomorphic to the complement of the $5$-adic Cantor set in the unit interval. The binary black hole shadow is then the disjoint union of open subsets $O_{\widetilde{\mathcal{X}} 1}$, i.e., the set
\begin{equation}
\mathcal{B}_{\textrm{S}} = \bigcup_{\mathcal{X} \in \hat{\Sigma}} O_{\widetilde{\mathcal{X}} 1},
\end{equation}
where $\hat{\Sigma} $ is the set of all decision dynamics sequences $\mathcal{X}$ of arbitrary length, which do not contain the digits $1$ or $3$.

At each stage of the iterative procedure outlined above, the length of the open intervals $O_{\widetilde{\mathcal{X}} 1}$ and $O_{\widetilde{\mathcal{X}} 3}$ as a proportion of the closed interval $C_{\widetilde{\mathcal{X}}}$ from which they are removed will depend on the entire history of the geodesic; and, thus, on all preceding digits (decisions) in the trajectory's symbolic representation $\mathcal{X}$. In practice, the length will depend most strongly on the most recent decision taken; the dependence on previous decisions will be exponentially suppressed. We anticipate that the structure of the one-dimensional shadow will be self-similar; this will be demonstrated in Section \ref{sec:demonstrating_self_similarity}.
%

\subsection{Demonstrating self-similarity in the one-dimensional shadow}
\label{sec:demonstrating_self_similarity}

\begin{figure}
\begin{center}
\subfigure[$0.43 \leq \alpha \leq 0.51$]{\includegraphics[width=0.45\textwidth]{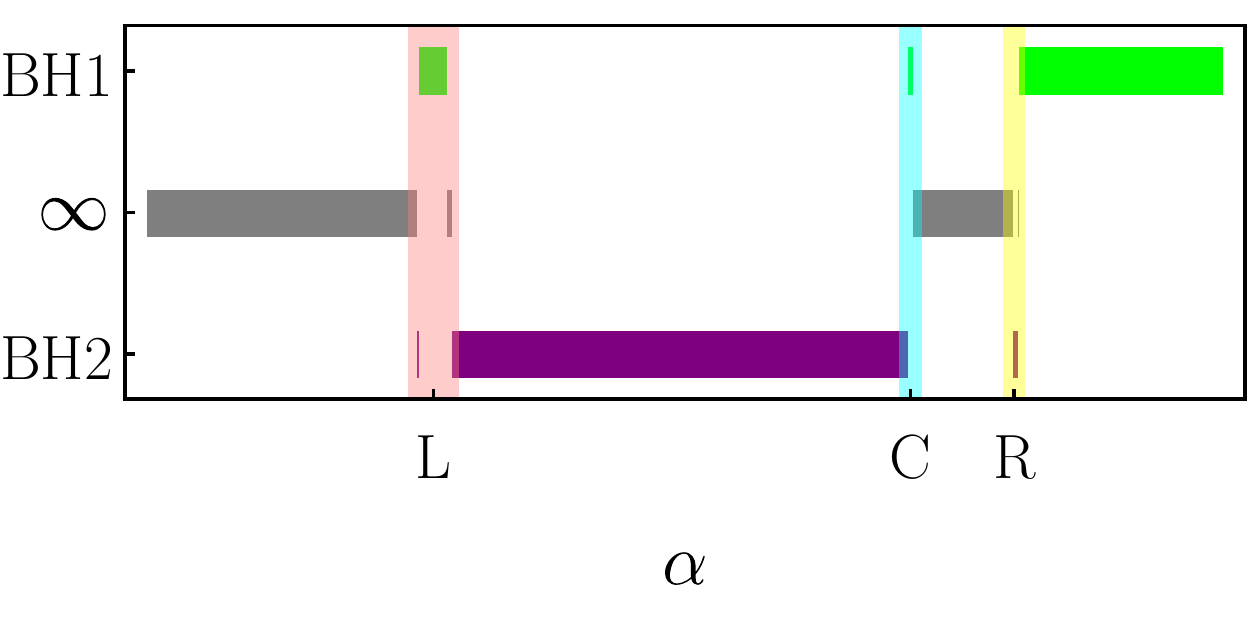} \hspace{1em}\label{fig:mp_one_dim_basins_zoom_0}}
\subfigure[$0.44865 \leq \alpha \leq 0.45375$ (L)]{\includegraphics[width=0.45\textwidth]{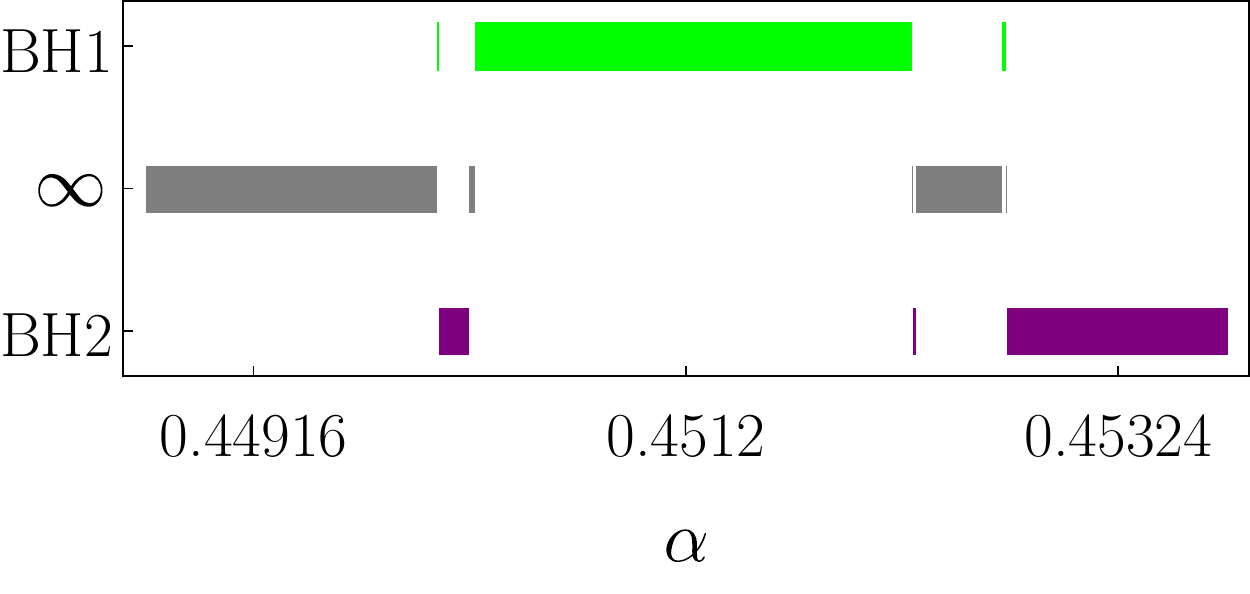}\label{fig:mp_one_dim_basins_zoom_1l}}
\subfigure[$0.48646 \leq \alpha \leq 0.48715$ (C)]{\includegraphics[width=0.45\textwidth]{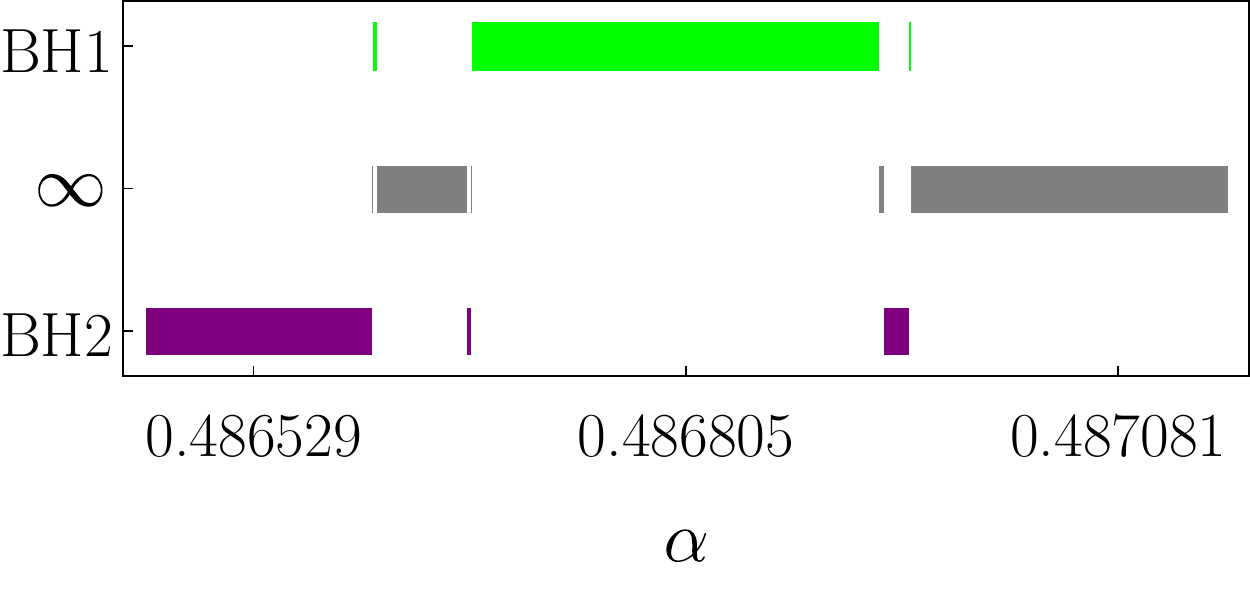} \hspace{1em}\label{fig:mp_one_dim_basins_zoom_1m}}
\subfigure[$0.49419 \leq \alpha \leq 0.49505$ (R)]{\includegraphics[width=0.45\textwidth]{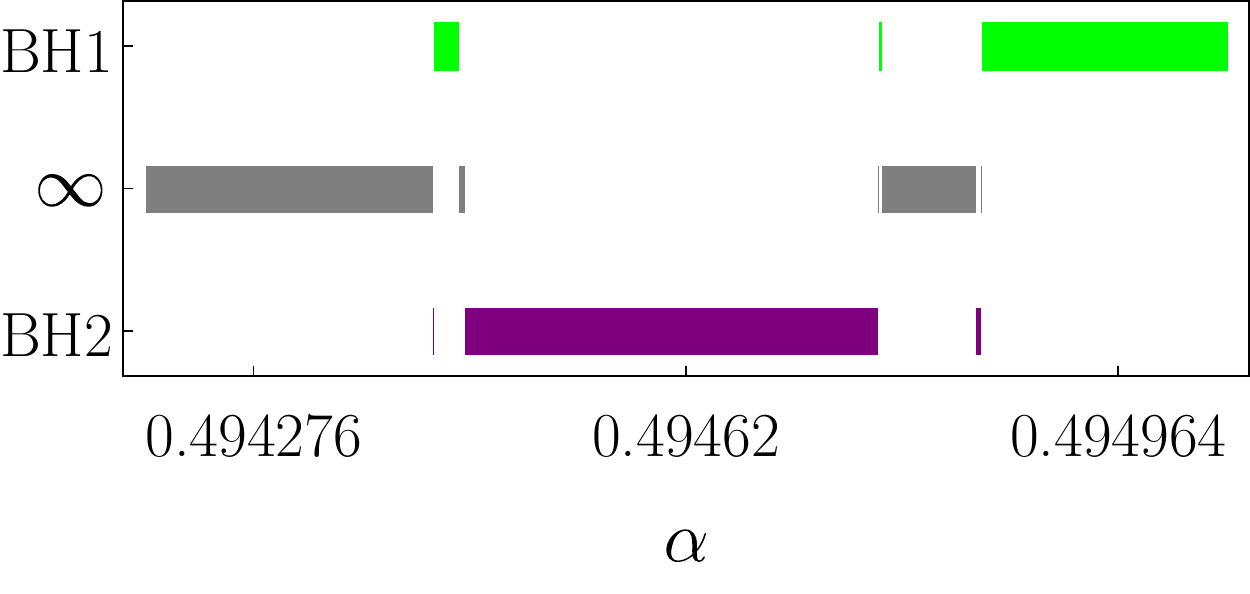}\label{fig:mp_one_dim_basins_zoom_1r}}
\subfigure[$0.44992 \leq \alpha \leq 0.4503$ (LL)]{\includegraphics[width=0.45\textwidth]{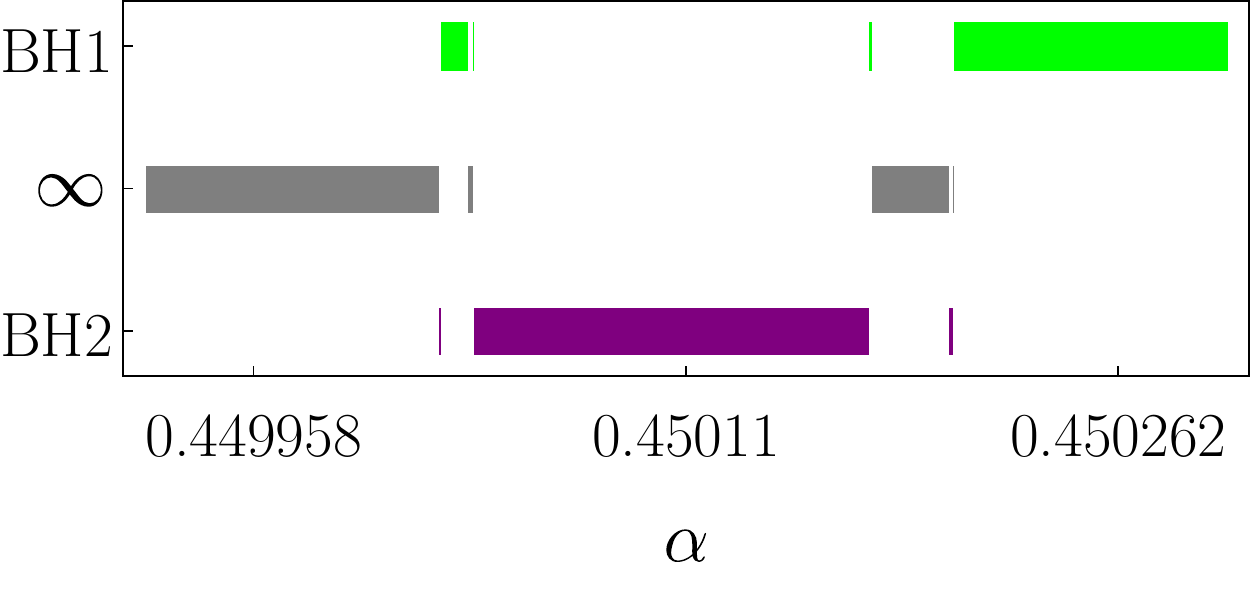} \hspace{1em}\label{fig:mp_one_dim_basins_zoom_2l}}
\subfigure[$0.4869264 \leq \alpha \leq 0.4869332$ (CC)]{\includegraphics[width=0.45\textwidth]{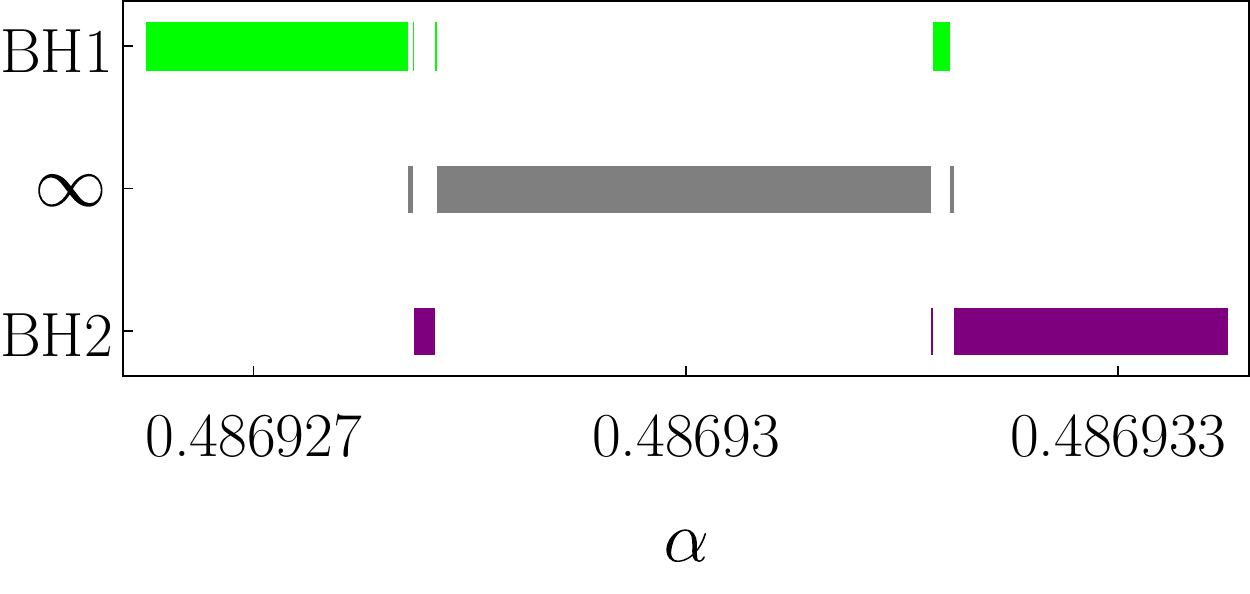}\label{fig:mp_one_dim_basins_zoom_2m}}
\subfigure[$0.4948485 \leq \alpha \leq 0.494857$ (RR)]{\includegraphics[width=0.45\textwidth]{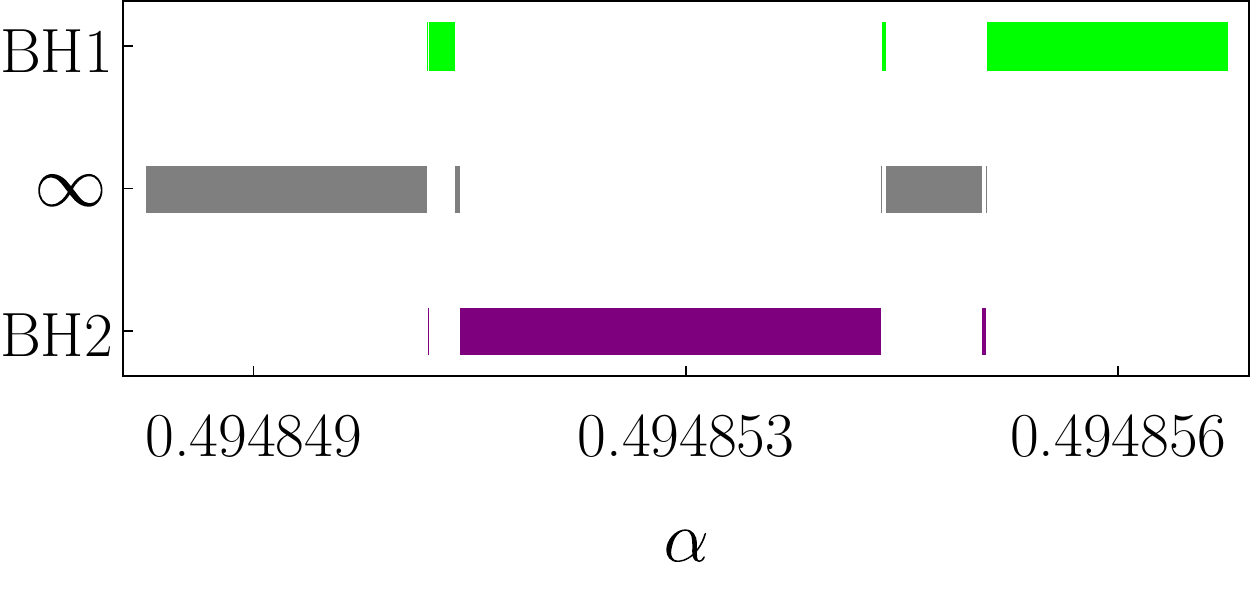} \hspace{1em}\label{fig:mp_one_dim_basins_zoom_2r}}
\subfigure[$0.450016 \leq \alpha \leq 0.450042$ (LLL)]{\includegraphics[width=0.45\textwidth]{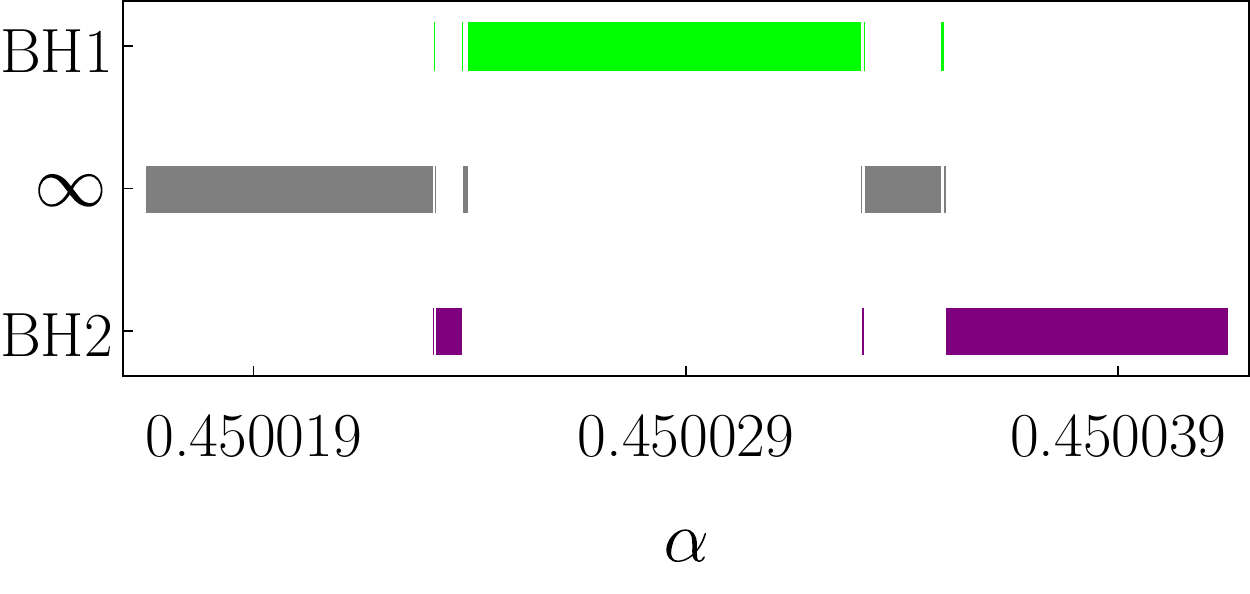}\label{fig:mp_one_dim_basins_zoom_3l}}
\subfigure[$0.486931336 \leq \alpha \leq 0.486931354$ (CCC)]{\includegraphics[width=0.45\textwidth]{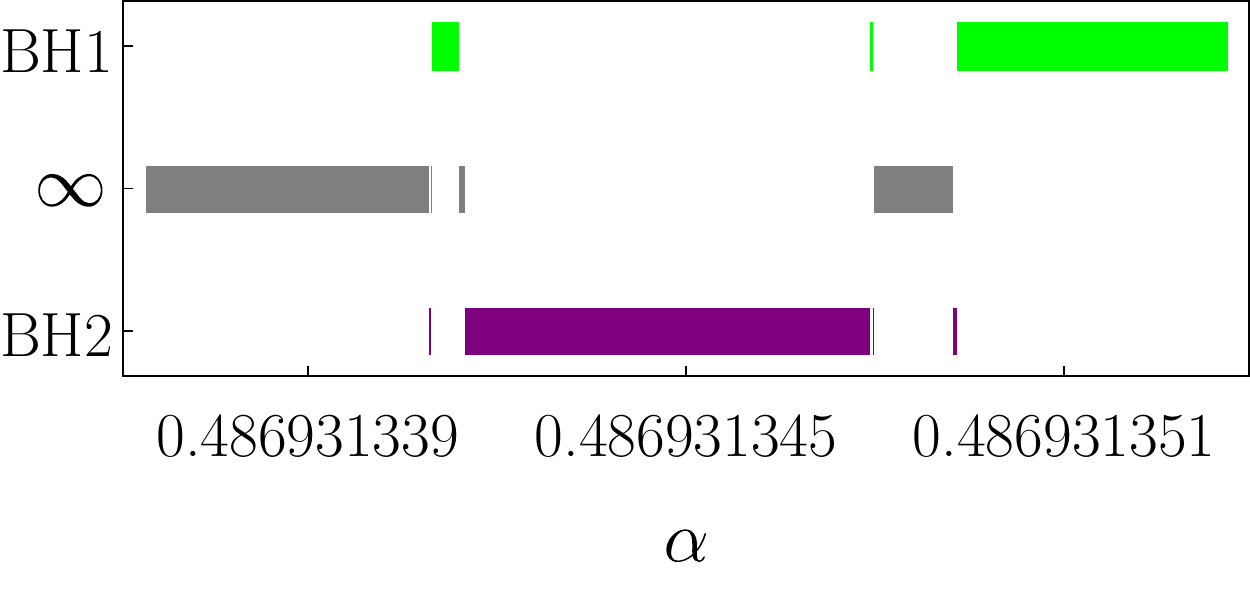} \hspace{1em}\label{fig:mp_one_dim_basins_zoom_3m}}
\subfigure[$0.49485505 \leq \alpha \leq 0.49485514$ (RRR)]{\includegraphics[width=0.45\textwidth]{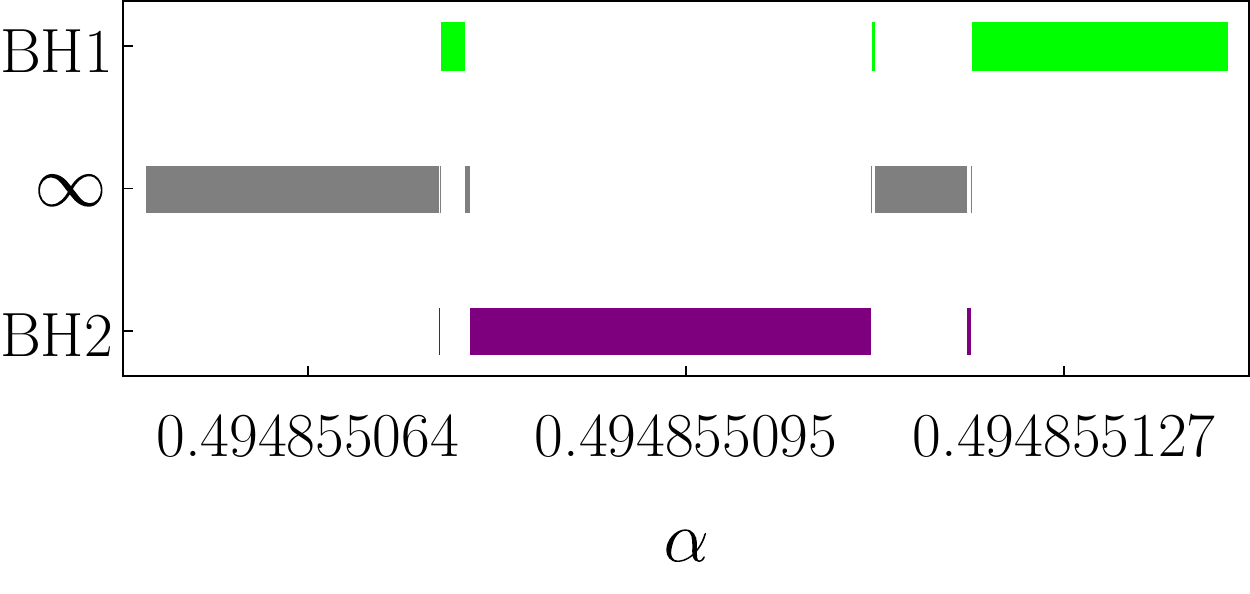}\label{fig:mp_one_dim_basins_zoom_3r}}
\caption{One-dimensional exit basins (shadows) for the equal-mass Majumdar--Papapetrou di-hole, with coordinate separation parameter $d = 2 M$. (a) There are three regions of interest on the left (L), in the centre (C), and on the right (R), highlighted in pink, cyan, and yellow, respectively. The exit basin diagrams (b)--(j) are obtained by successively magnifying the left (L), central (C), or right (R) intervals of the ``parent'' exit basin diagram (a). \label{fig:mp_one_dim_basins_all}}
\end{center}
\end{figure}

We now use our symbolic code as a guide to understand the fractal properties -- namely self-similarity -- of the one-dimensional shadow. In Figure \ref{fig:mp_one_dim_basins_zoom_0}, we present the one-dimensional binary black hole shadow (i.e., exit basin diagram) for the scattering problem introduced in Section \ref{sec:planar_geodesics_one_dim_shadows}; see Figure \ref{fig:mp_one_dim_scattering_set_up} for the set-up of the problem. It is clear from Figure \ref{fig:mp_one_dim_basins_zoom_0} that there are three open intervals of interest on the left (L), in the centre (C) and on the right (R). To the left of the first open interval (L), all rays escape to infinity. In between the first and second of these open intervals (L and C), all rays plunge directly into the lower black hole. All rays with initial data taken from the closed interval between the second and third open intervals (C and R) escape to infinity. Finally, all initial conditions to the right of the third open interval (R) plunge directly into the upper black hole. The highlighted intervals possess rich fractal structure. Here, we shall demonstrate that the shadow is self-similar, by magnifying each of these open intervals successively.

First, consider the leftmost open interval (L) highlighted in Figure \ref{fig:mp_one_dim_basins_zoom_0}. Magnifying this open interval yields Figure \ref{fig:mp_one_dim_basins_zoom_1l}. This magnification procedure is equivalent to making the decision $0$ in symbolic dynamics: all rays with initial data in this open set travel around the opposite black hole in the same sense. We see that the magnified exit basin diagram is qualitatively similar to its ``parent'', shown in Figure \ref{fig:mp_one_dim_basins_zoom_0}. In fact, this image is identical, but the rays which plunge into the upper and lower black holes have been interchanged. We may continue in this fashion, magnifying the leftmost open interval (i.e., repeatedly making the decision $0$). This yields Figures \ref{fig:mp_one_dim_basins_zoom_2l} and \ref{fig:mp_one_dim_basins_zoom_3l}. Each time we magnify the left-hand interval, we obtain a new self-similar exit basin diagram but with the basins corresponding to the upper/lower black holes exchanged.

Next, we magnify the central interval (C) of the parent exit basin diagram, shown in Figure \ref{fig:mp_one_dim_basins_zoom_0}. This is equivalent to repeatedly making the decision $2$ in symbolic dynamics. The resulting exit basin diagrams are shown in Figures \ref{fig:mp_one_dim_basins_zoom_1m}, \ref{fig:mp_one_dim_basins_zoom_2m} and \ref{fig:mp_one_dim_basins_zoom_3m}. We observe that, with each magnification, the trajectories which plunge into the upper and lower black holes are reversed (as with decision $0$). Moreover, the exit basin diagram is reflected in $\alpha$ about the centre of the open interval, because making the decision $2$ reverses the sense of the orbit. This is equivalent to reversing the parity ($P \mapsto - P$) in the parity-reordering operation $\mathcal{P}$ (see Section \ref{sec:construction_cantor_like_set}).

Finally, we magnify the right-hand interval (R) of Figure \ref{fig:mp_one_dim_basins_zoom_0}, which is equivalent to making the decision $4$. This yields Figures \ref{fig:mp_one_dim_basins_zoom_1r}, \ref{fig:mp_one_dim_basins_zoom_2r} and \ref{fig:mp_one_dim_basins_zoom_3r}. In this case, each magnified exit basin diagram looks the same as the parent exit basin diagram.

In Figure \ref{fig:mp_one_dim_basins_all}, we have repeatedly made the \emph{same} decision (either $0$, $2$ or $4$) at each stage. This process yields a particular class of self-similar exit basin diagrams, in which the proportion of the intervals corresponding to each decision are similar at each level of magnification. One could instead make \emph{different} decisions at each stage. In this case, the resulting exit basins would again be self-similar, but the relative proportions of the intervals would depend principally on the previous digit in the decision dynamics representation of the trajectory.

We remark that the one-dimensional shadows of Figure \ref{fig:mp_one_dim_basins_all} are qualitatively similar to the three-colour Cantor basins of Figure \ref{fig:three_colour_cantor_basins}. In the construction of the Cantor basins, we remove one open interval from the remaining closed intervals at each step; in the construction of the one-dimensional binary black hole shadows, we remove two open intervals each time we iterate. As we have seen, in both cases, the iterative procedure and the resulting self-similar basins can be understood through the use of decision dynamics.
%

\subsection{The strange repellor and chaotic scattering}
\label{sec:strange_repellor}

For a general dynamical system which evolves in time $t$, the set of unstable (kinematically unbounded) trajectories which remain confined to the scattering region as $t \rightarrow \infty$ constitutes the repellor $\Omega_{\textrm{R}}$ \cite{GaspardRice1989}; see Section \ref{sec:chaotic_dynamical_systems}. For example, consider the two-disc scatterer. In this model, $\Omega_{\textrm{R}}$ consists of a unique trajectory, which forever bounces between the two discs, along the straight line connecting their centres. All other trajectories scatter off the discs before reaching spatial infinity. The two-disc model exhibits \emph{regular} dynamics, and the Kolmogorov--Sinai entropy is zero. The repellor $\Omega_{\textrm{R}}$ consists of a unique scattering singularity: the repellor is therefore \emph{regular}.

We recall Eckhardt's definition \cite{Eckhardt1987} of chaotic scattering (as stated in Section \ref{sec:chaotic_dynamical_systems}): scattering in a Hamiltonian system is irregular or chaotic if there exists, on some manifold of initial data, an infinity of distinct scattering singularities of measure zero which are typically arranged into a fractal set.

In contrast to the (regular) two-disc system, the repellor for the Gaspard--Rice three-disc system (see Section \ref{sec:symbolic_dynamics_gaspard_rice}) forms a Cantor-like set: an uncountably infinite fractal set \cite{Eckhardt1988}. We recall that, in such cases, the repellor is called \emph{irregular} (or \emph{strange}). According to Eckhardt's definition, the three-disc system therefore exhibits chaotic scattering.

In this chapter, we have demonstrated that a strange repellor $\Omega_{\textrm{R}}$ exists for planar null geodesics on the Majumdar--Papapetrou di-hole spacetime. Moreover, the arguments presented Sections \ref{sec:ordering_perpetual_orbits}--\ref{sec:demonstrating_self_similarity} demonstrate that $\Omega_{\textrm{R}}$ is a Cantor-like set. We conclude therefore that the scattering of null rays on the Majumdar--Papapetrou di-hole spacetime is a natural exemplar of chaotic scattering: there exists a Cantor-like set of scattering singularities, corresponding to the perpetual orbits.

\subsection{Measures of chaos}
\label{sec:measures_of_chaos}

Gaspard and Rice \cite{GaspardRice1989} characterise the strange repellor $\Omega_{\textrm{R}}$, and the natural measure it supports, using quantities such as Lyapunov exponents, the Kolmogorov--Sinai entropy per unit time, the Hausdorff dimension, the information dimension, the escape rate and the time-delay function. For scattering processes in general relativity, one must use such measures of chaos with caution: many of the quantities listed here are coordinate-dependent.

\begin{figure}
\begin{center}
\subfigure[$0.43 \leq \alpha \leq 0.51$]{
\includegraphics[height=0.45\textwidth]{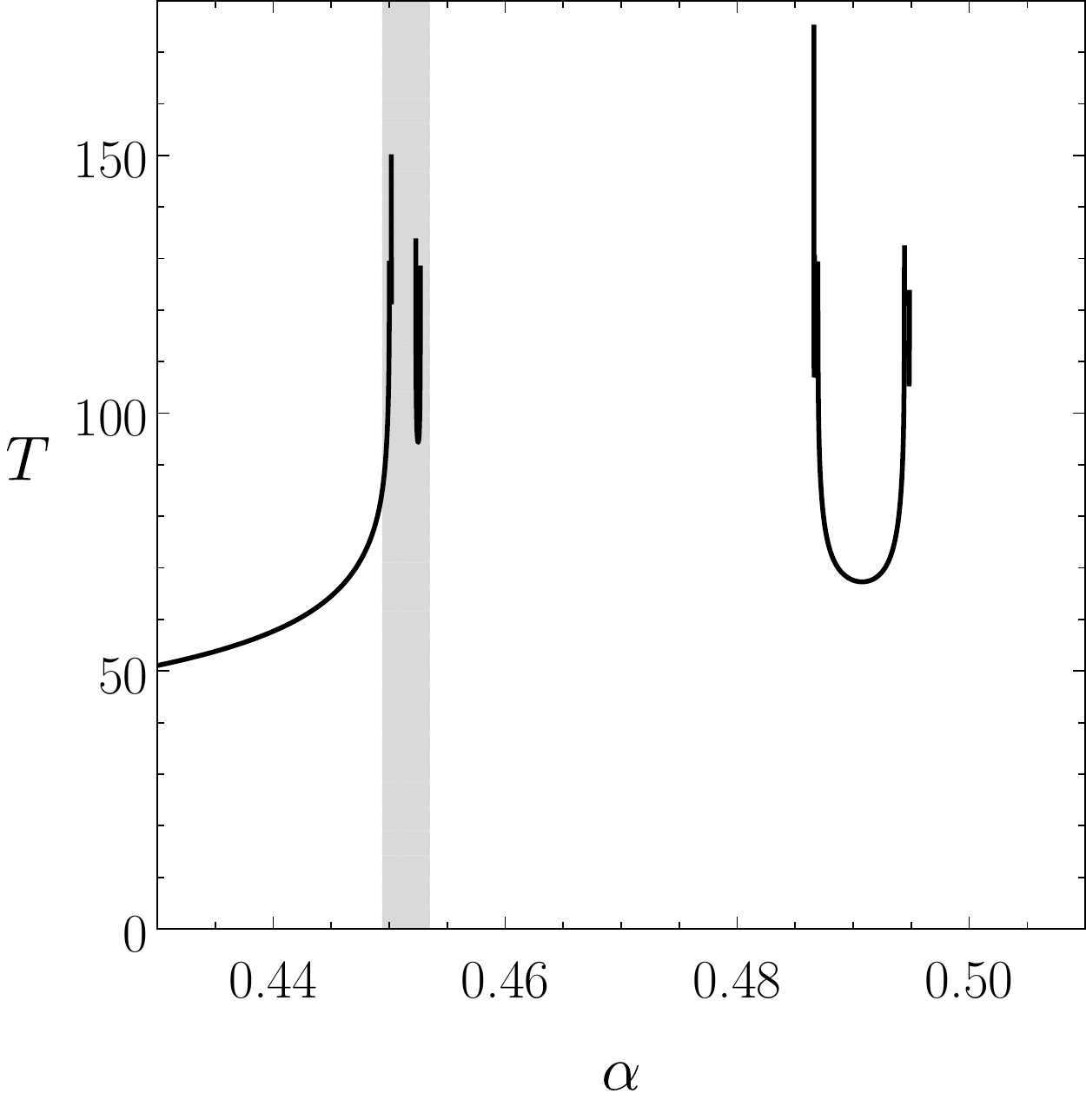} \label{fig:mp_time_delay_main} \hspace{1em}}
\hspace{0.5em}
\subfigure[$0.4494 \leq \alpha \leq 0.4535$]{
\includegraphics[height=0.45\textwidth]{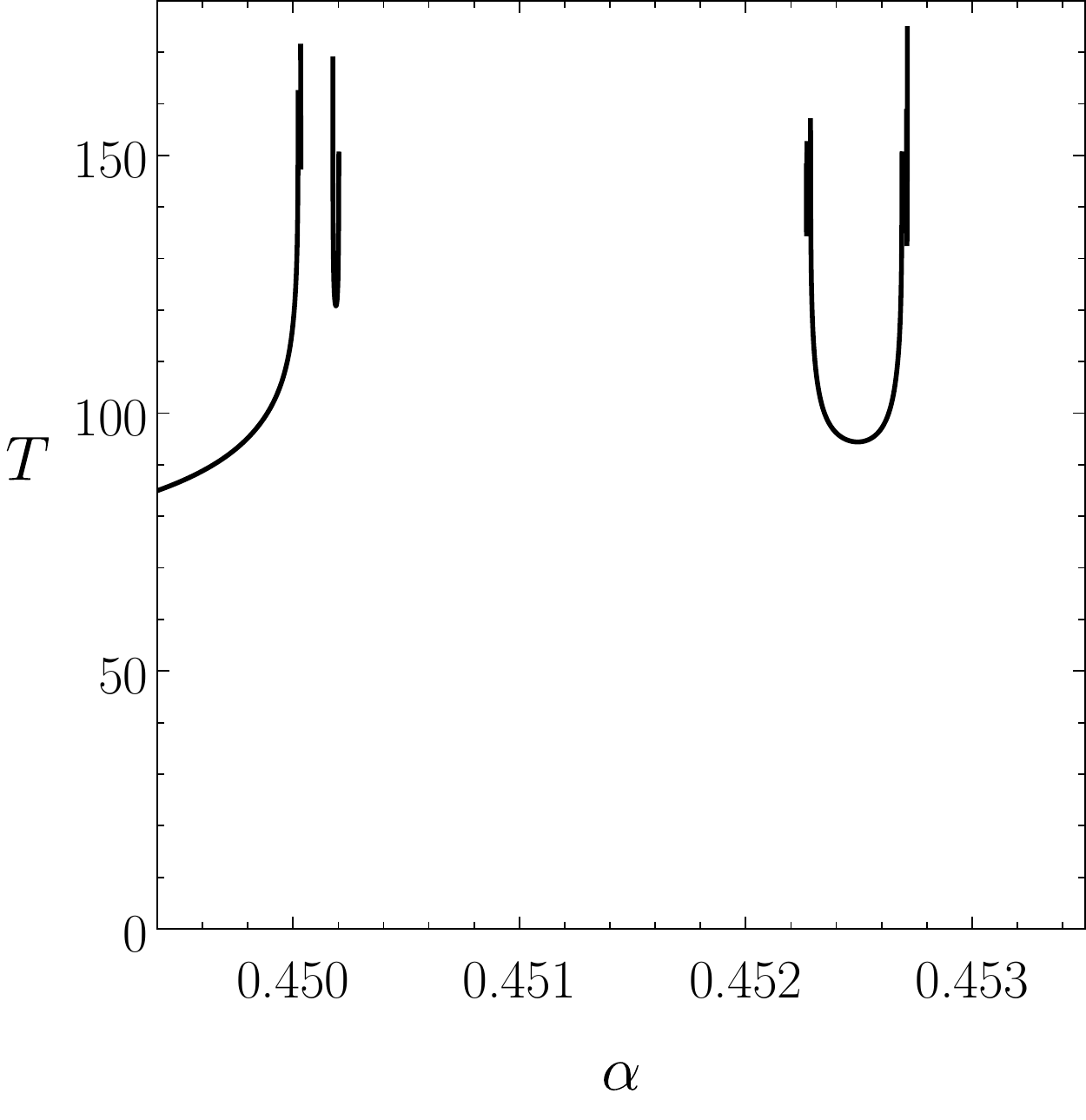} \label{fig:mp_time_delay_zoom} \hspace{1em}}
\end{center}
\caption{(a) Time-delay function $T(\alpha)$ for scattering by the equal-mass Majumdar--Papapetrou di-hole with initial data $\alpha$. (The parameters are taken to be $M_{\pm} = M = 1$, $d = 2 M$.) The time-delay is defined to be the Killing time $t$ taken for a null ray starting at the centre of mass to escape to a large radius. The function is undefined for rays which approach a black hole horizon; $T(\alpha) \rightarrow \infty$ in the approach to a scattering singularity. The time-delay function inherits the self-similar Cantor-like structure of the strange repellor. (b) Magnification of $T(\alpha)$ which highlights the self-similarity.}
\label{fig:mp_time_delay}
\end{figure}

Here, we use the time-delay function to characterise the strange repellor $\Omega_{\textrm{R}}$ in the Majumdar--Papapetrou di-hole scattering problem. The time-delay $T(\alpha)$, shown in Figure \ref{fig:mp_time_delay}, is defined as the coordinate time $t$ taken by a null geodesic with initial data given by \eqref{eqn:mp_one_dim_initial_data} to reach some large fixed radius $r_{\textrm{max}} \gg 1$. It should be noted that $T(\alpha)$ is undefined for rays which fall into a black hole horizon, because Killing time $t$ diverges for orbits which approach an event horizon. The time-delay $T(\alpha)$ also diverges as one approaches a scattering singularity in the initial data. This is not because it approaches a horizon, but because it asymptotes towards a perpetual orbit which lingers in the scattering region as $t \rightarrow \infty$. The time-delay function inherits the Cantor-like structure of the one-dimensional shadow, due to the distribution of the scattering singularities in the initial data; this self-similarity can be seen in Figure \ref{fig:mp_time_delay_zoom}, in which we magnify a domain of the time-delay function presented in Figure \ref{fig:mp_time_delay_main}. The time-delay function is thus qualitatively similar to that used by Gaspard and Rice to characterise the strange repellor of the three-disc model (see Figure 2 of \cite{GaspardRice1989}).

In addition to the time-delay function, which we use here as a diagnostic for chaotic scattering, one could employ other measures to characterise the chaotic motion of (timelike and null) geodesics in the Majumdar--Papapetrou di-hole system, and related spacetimes. Here, we present a brief account of some important contributions to this area.

Dettmann \emph{et al.} \cite{DettmannFrankelCornish1994, DettmannFrankelCornish1995} investigate the phase space for trajectories of charged test particles with arbitrary values of the energy and angular momentum, extending on the work of Contopoulos \cite{Contopoulos1990, Contopoulos1991}. In particular, the authors analyse exit basins for a scattering problem in the meridian plane similar to that set up in Section \ref{sec:planar_geodesics_one_dim_shadows}. They demonstrate that the boundary of the exit basins corresponding to the two black holes scales as a fractal in a diffeomorphism-invariant manner, by calculating the box-counting dimension of the boundary numerically. Moreover, the authors find that chaotic motion of \emph{complete} timelike geodesics (i.e., those which do not plunge into a black hole) is well-described by Lyapunov exponents.

Cornish and Gibbons \cite{CornishGibbons1997} also study geodesic motion in the field of two fixed centres (black holes) described by a family of Einstein--Maxwell--dilaton theories, which includes the Majumdar--Papapetrou di-hole (Einstein--Maxwell two-centre problem) as a special case. The authors favour invariant methods, such as exit basins, fractal dimensions and topological entropies, over non-invariant methods, such as Lyapunov exponents and metric entropy. It is argued that geodesics will be chaotic if there exists a \emph{chaotic invariant set} of orbits, i.e., a strange repellor (see Section \ref{sec:strange_repellor}). This set is understood as the boundary between different outcomes of the scattering problem, and the authors compute the exit basins of a scattering problem in the meridian plane, similar to this work (Section \ref{sec:planar_geodesics_one_dim_shadows}) and Dettmann \emph{et al.} \cite{DettmannFrankelCornish1994, DettmannFrankelCornish1995}. Intriguingly, the results of \cite{CornishGibbons1997} suggest that there is a transition between regular and chaotic geodesic motion as the dilaton coupling is varied. As described in Section \ref{sec:symbolic_dynamics_mp}, Cornish and Gibbons develop a symbolic code (``collision dynamics'') to describe the chaotic motion of null rays in the meridian plane of the two-centre Einstein--Maxwell--dilaton problem.

The numerical analysis performed by Contopoulos and others (described above) demonstrates that the motion of null geodesics on the Majumdar--Papapetrou di-hole spacetime is chaotic. Yurtsever \cite{Yurtsever1995} performs a geometrical analysis of the null geodesic flow on the Majumdar--Papapetrou spacetime, by reducing the problem to that of geodesic motion on a negatively curved Riemannian manifold. (Heuristically, the negative curvature is responsible for exponential divergence of trajectories in the geodesic flow -- a key ingredient for chaotic motion.) The author then presents a precise formulation of chaotic motion in terms of the chaotic invariant set, before proving that the null geodesic flow of the two-centre Majumdar--Papapetrou problem satisfies this definition of chaos.

More recently, a study of the escape dynamics of photons from the field of two fixed extremal Reissner--Nordstr\"{o}m black holes has been carried out by Alonso \emph{et al.} \cite{AlonsoRuizSanchez-Hernandez2008}. Considering a scattering problem in the meridian plane (i.e., photons with $p\ind{_{\phi}} = 0$), the authors identify the non-escaping periodic orbits which play a key role in the escape dynamics of photons from the two centres. In addition, Alonso \emph{et al.} describe these orbits using a symbolic code similar to that of Cornish and Gibbons, and calculate characteristic quantities of chaos including the mean Lyapunov exponent of the repellor, the escape rate, the Kolmogorov--Sinai entropy, the topological entropy, the Hausdorff dimension, and the partial information dimension.

%

\section{Non-planar geodesics and two-dimensional shadows}
\label{sec:non_planar_rays_two_dim_shadows}

\subsection{Non-planar rays}
\label{sec:non_planar_rays}

We now turn to non-planar motion around the two black holes, which is governed by the Hamiltonian \eqref{eqn:mp_hamiltonian_canonical}, with a non-zero conserved azimuthal angular momentum $p\ind{_{\phi}}$. As in the case of scattering in the meridian plane (Section \ref{sec:planar_geodesics_one_dim_shadows}), it is beneficial to consider the fundamental perpetual null orbits (see Figure \ref{fig:mp_planar_rays}). Where two or more distinct but ``dynamically connected'' fundamental orbits exist, we anticipate chaotic scattering, and thus a Cantor-like set of scattering singularities in initial data. In this section, we shall demonstrate that this is not the only possibility when considering rays with $p\ind{_{\phi}} \neq 0$.

\begin{figure}
\begin{center}
\hfill
\subfigure[]{
{\includegraphics[width=0.45\textwidth]{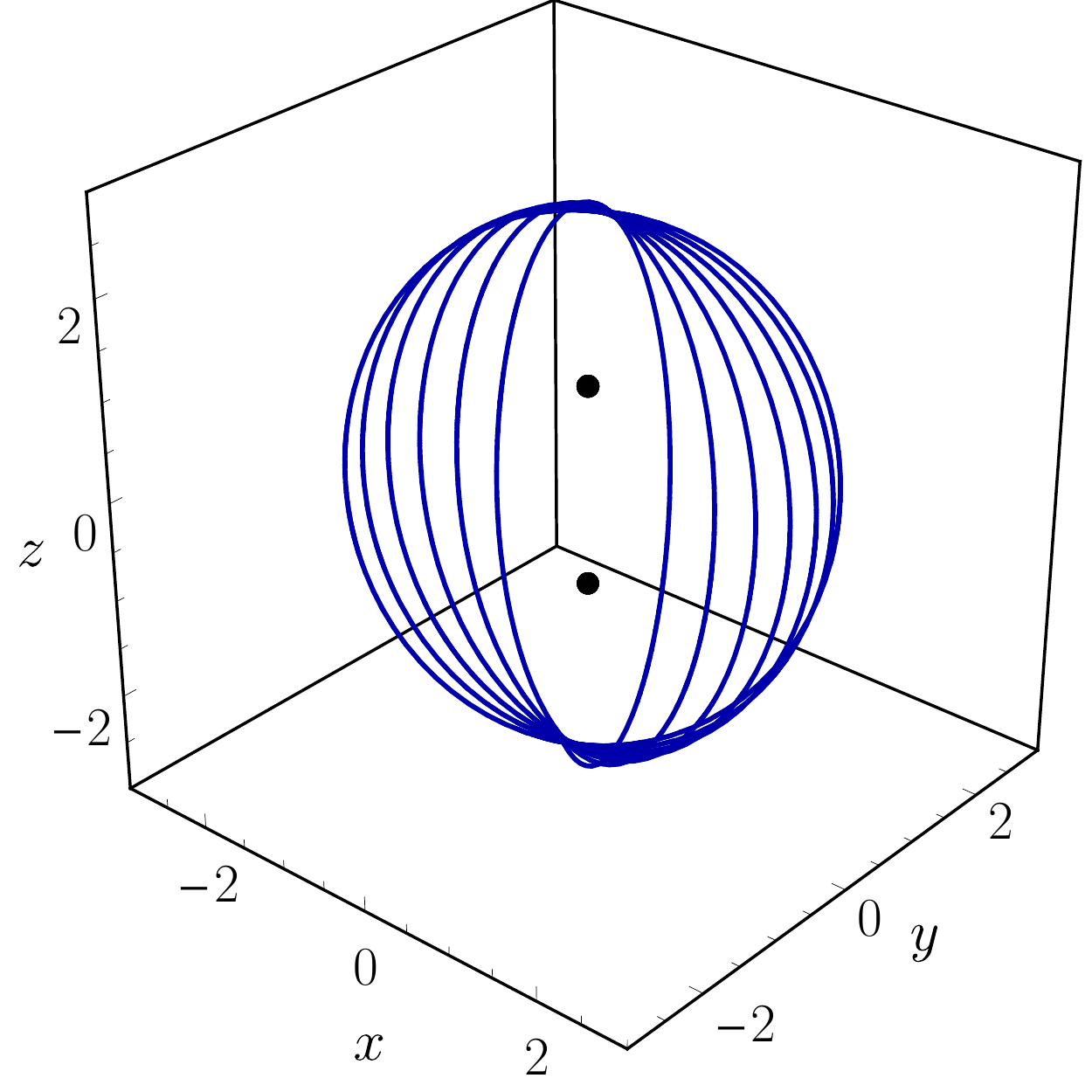}}
\label{fig:mp_non_planar_ray_0}}
\hfill
\subfigure[]{
{\includegraphics[width=0.45\textwidth]{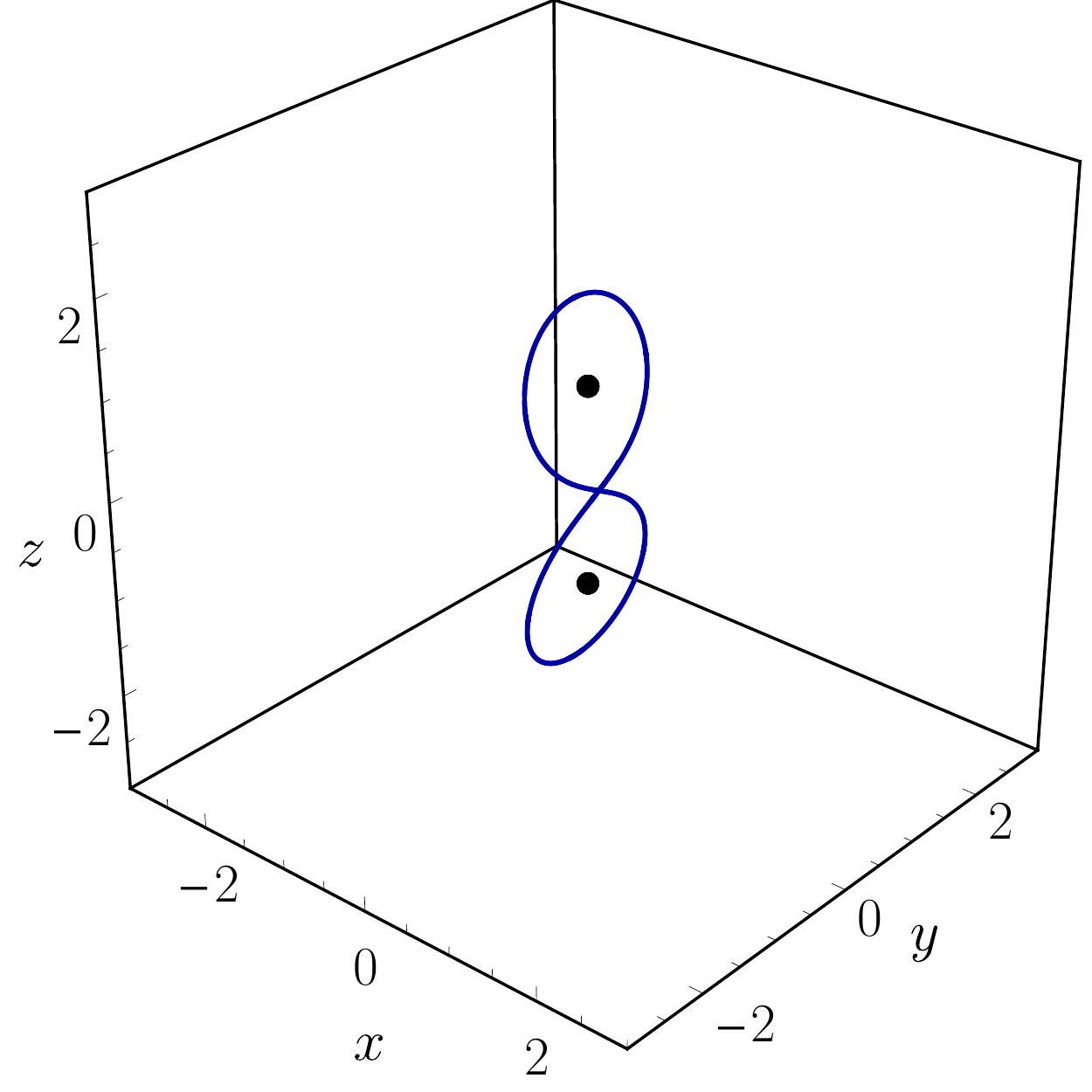}}
\label{fig:mp_non_planar_ray_2}}
\hfill
\subfigure[]{
{\includegraphics[width=0.45\textwidth]{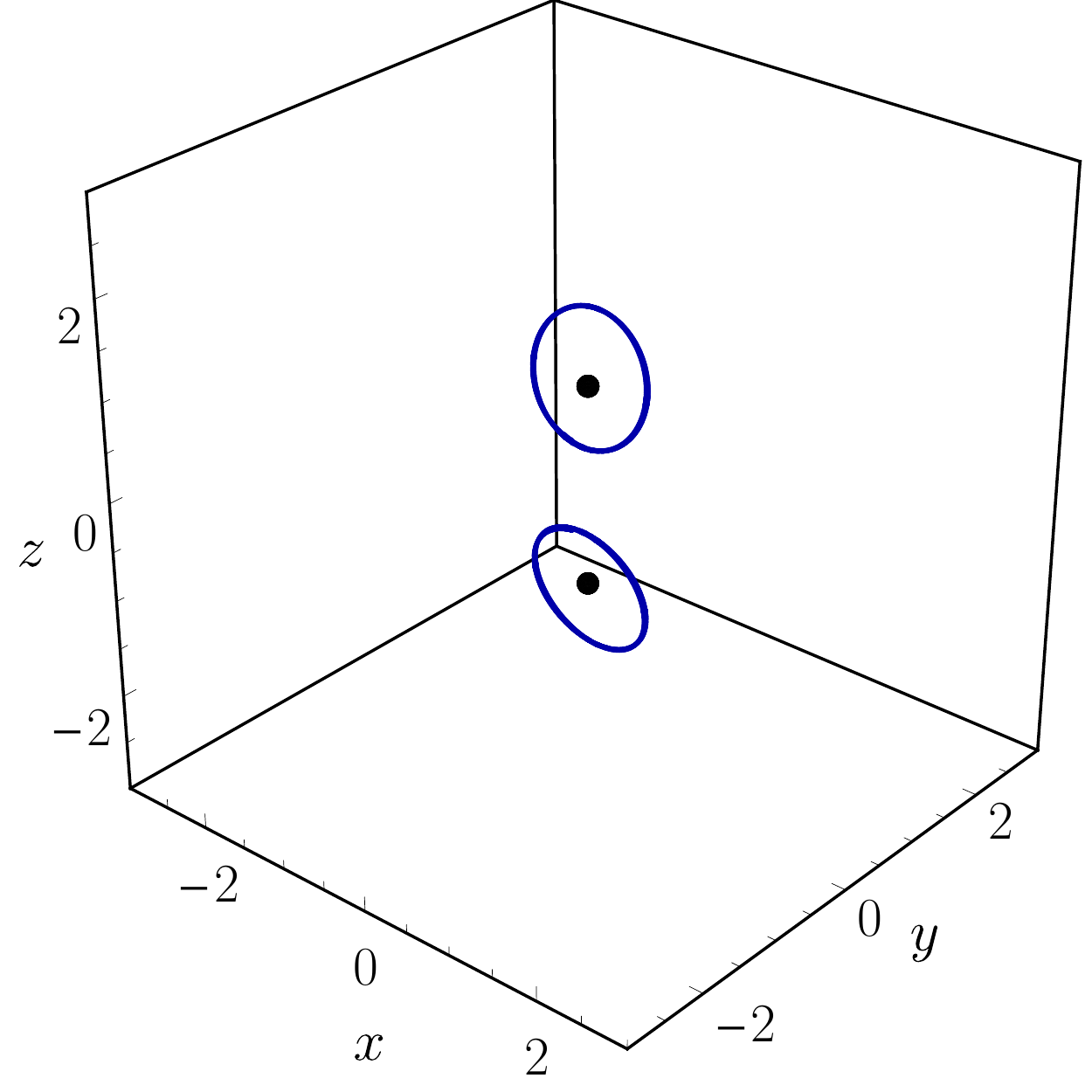}}
\label{fig:mp_non_planar_ray_4}}
\hfill
\subfigure[]{
{\includegraphics[width=0.45\textwidth]{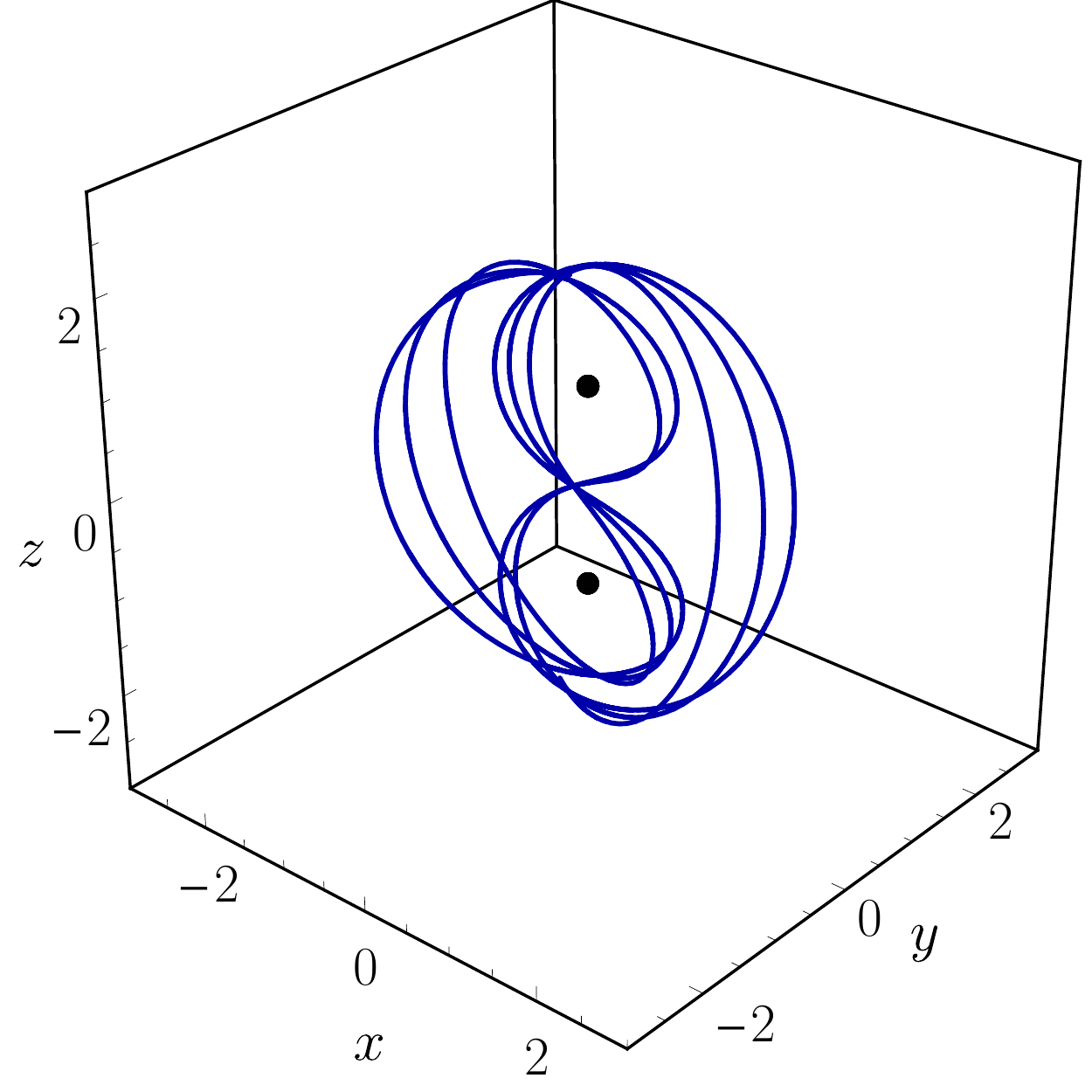}}
\label{fig:mp_non_planar_ray_02}}
\hfill
\caption{Examples of non-planar periodic null orbits with non-zero angular momentum $p\ind{_{\phi}}$, shown in $\mathbb{R}^{3}$. We choose $M_{\pm} = 1$, $d = 2$ and $p\ind{_{\phi}} = 1$ as default values. (a)--(c) Fundamental periodic orbits with symbolic representations (a) $\overline{0}$, (b) $\overline{2}$, and (c) $\overline{4}$. (d) A periodic orbit which transitions between two fundamental orbits. This orbit has symbolic representation $\overline{02}$ in decision dynamics. These are the non-planar versions of the fundamental orbits of Figure \ref{fig:mp_planar_rays}. \label{fig:mp_non_planar_rays}}
\end{center}
\end{figure}

\subsubsection{Fundamental orbits with non-zero angular momentum}

Figure \ref{fig:mp_non_planar_rays} shows a selection of non-escaping orbits on the Cartesian $(x, y, z)$-axes in $\mathbb{R}^{3}$ for a non-zero value of the azimuthal angular momentum $p\ind{_{\phi}}$. (The parameters are chosen to be $M_{\pm} = M = 1$, $d = 2$ and $p\ind{_{\phi}} = 1$.) Figures \ref{fig:mp_non_planar_ray_0}--\ref{fig:mp_non_planar_ray_4} show the non-planar versions of the fundamental orbits of Figure \ref{fig:mp_planar_rays}, which are described by decision dynamics sequences (a) $\overline{0}$, (b) $\overline{2}$ and (c) $\overline{4}$. Figure \ref{fig:mp_non_planar_ray_02} shows an example of an orbit which transitions between the fundamental orbits $\overline{0}$ and $\overline{2}$; its representation in decision dynamics is $\overline{02}$; see Figure \ref{fig:mp_planar_ray_02}. This case indicates that transitions between the three fundamental null orbits are still possible, despite the motion being non-planar.

\begin{figure}
\begin{center}
\subfigure[$\overline{0}$]{
\includegraphics[width=0.22\textwidth]{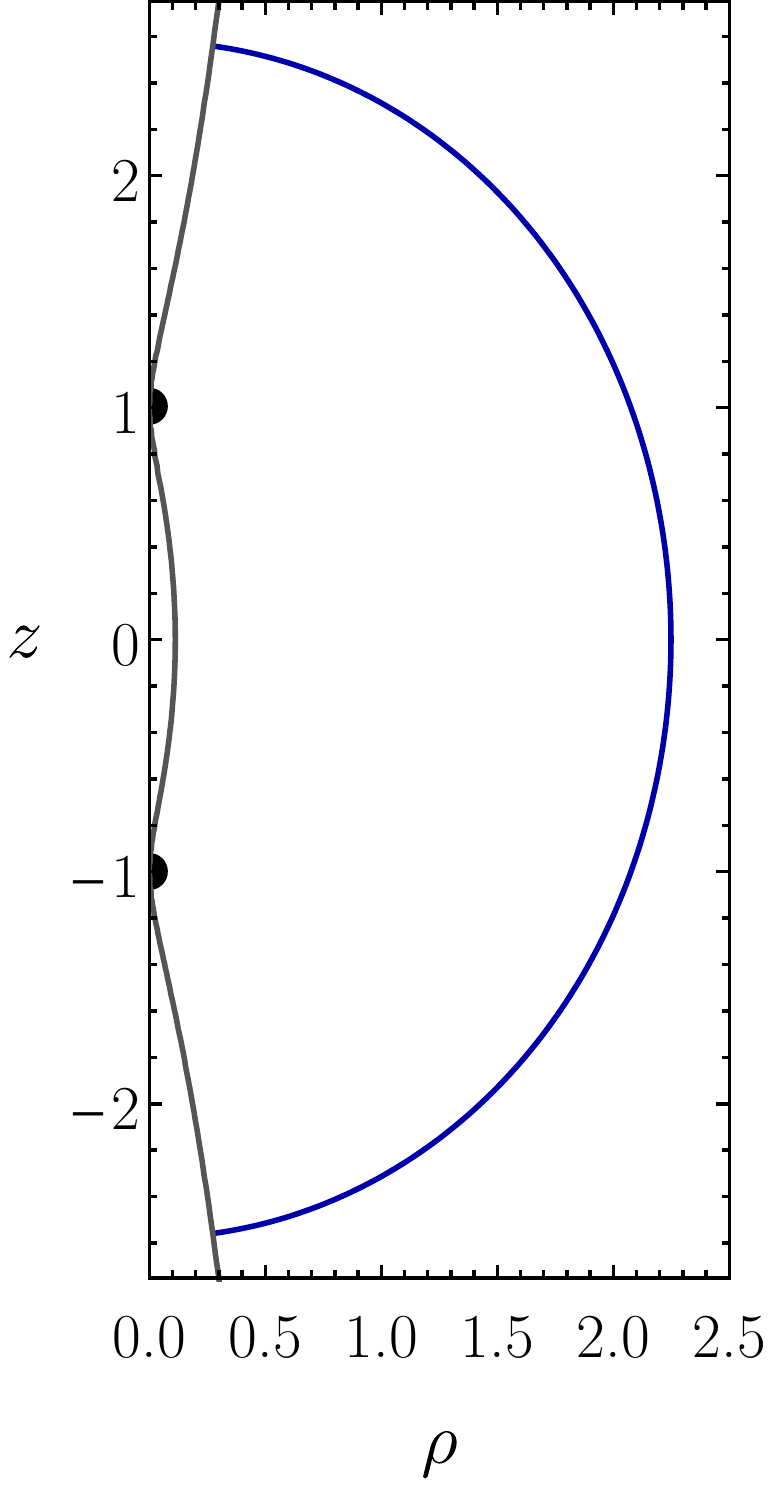} \label{fig:mp_non_planar_ray_rho_z_0}}
\subfigure[$\overline{2}$]{
\includegraphics[width=0.22\textwidth]{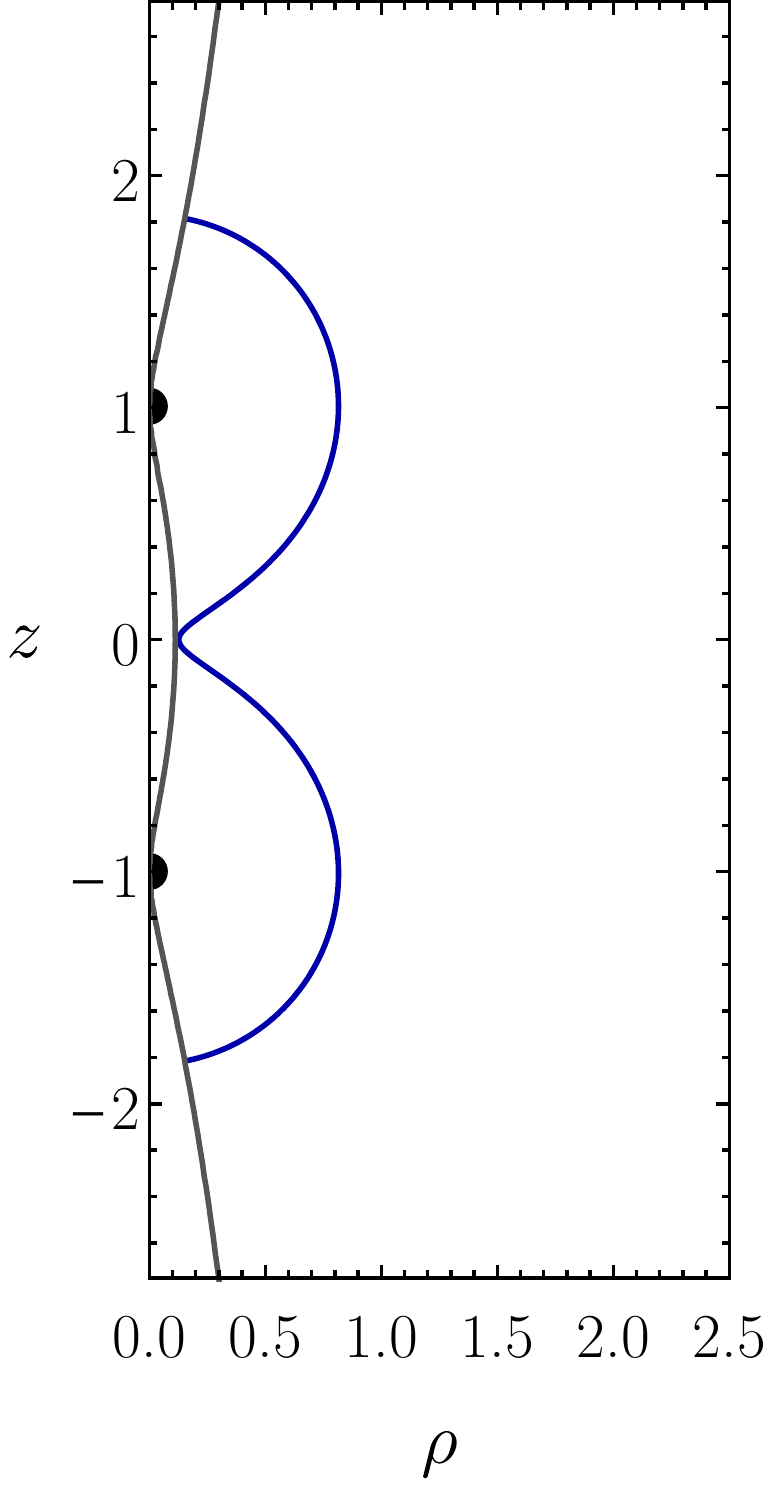} \label{fig:mp_non_planar_ray_rho_z_2}}
\subfigure[$\overline{4}$]{
\includegraphics[width=0.22\textwidth]{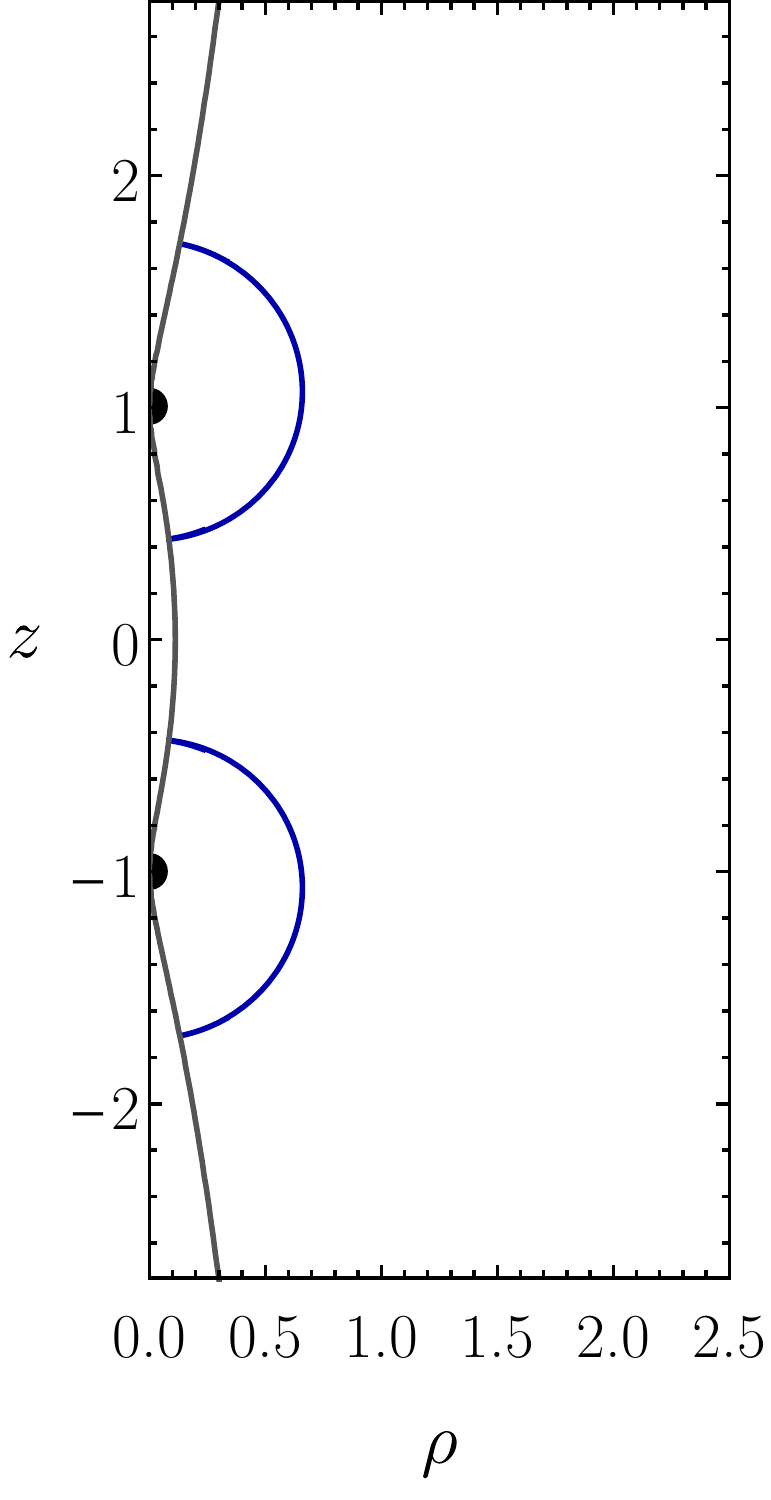} \label{fig:mp_non_planar_ray_rho_z_4}}
\subfigure[$\overline{02}$]{
\includegraphics[width=0.22\textwidth]{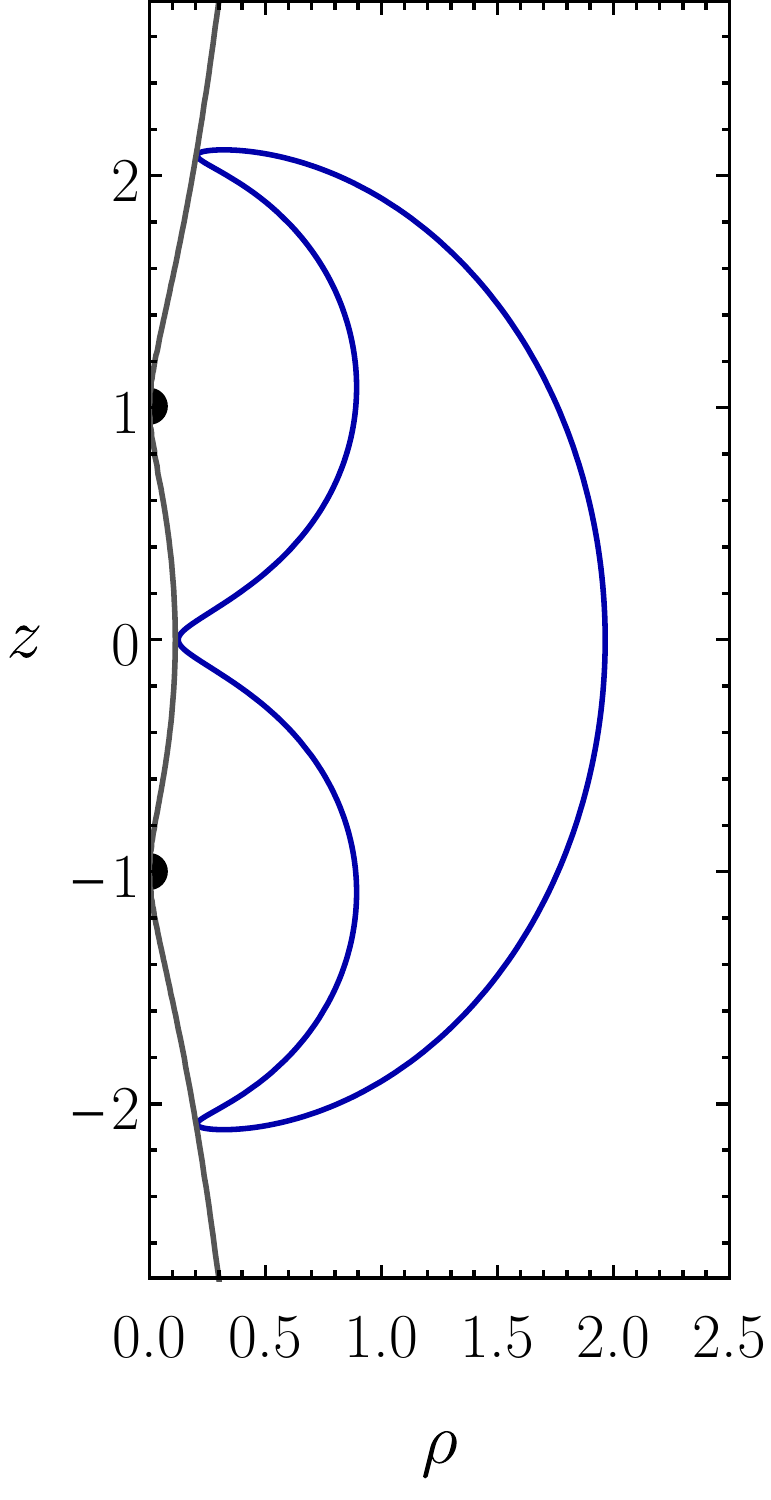} \label{fig:mp_non_planar_ray_rho_z_02}}
\caption{The non-planar periodic null orbits with non-zero angular momentum $p\ind{_{\phi}}$ of Figure \ref{fig:mp_non_planar_rays}, projected onto the $(\rho, z)$-plane. (a)--(c) Fundamental orbits described by the decision dynamics sequences (a) $\overline{0}$, (b) $\overline{2}$, and (c) $\overline{4}$. (d) A ray which transitions between the $\overline{0}$- and $\overline{2}$-orbits, represented by the decision dynamics sequence $\overline{02}$. The grey curve is the contour $h(\rho, z) = p\ind{_{\phi}}$. Orbits (a)--(c) touch the contour in such a way that the two-momentum is orthogonal to the contour; orbit (d) does not touch the contour. \label{fig:mp_non_planar_rays_rho_z}}
\end{center}
\end{figure}

In Figure \ref{fig:mp_non_planar_rays_rho_z}, we show the orbits of Figure \ref{fig:mp_non_planar_rays}, projected onto the $(\rho,z)$-plane. Clearly, the orbits are periodic in the $(\rho, z)$-plane; however, it is clear from Figure \ref{fig:mp_non_planar_rays} that the non-commensurate motion in $\phi$ means that the geodesics are not closed (in general) in three dimensions: the rays of Figures \ref{fig:mp_non_planar_ray_0} and \ref{fig:mp_non_planar_ray_02} trace out two-surfaces in $\mathbb{R}^{3}$. (We only show part of the orbit for clarity.)

\subsubsection{Analysis of the effective potential}

In Section \ref{sec:mp_dihole_hamiltonian_formalism}, we introduced the effective potential (or height function) $h(\rho, z) = \rho \, U^{2}$, which allowed us to express the geodesic potential $V(\rho, z ; p\ind{_{\phi}})$ in a convenient factorised form. A straightforward rearrangement of the null condition $H = 0$, with Hamiltonian \eqref{eqn:mp_hamiltonian_canonical}, yields the \emph{energy equation},
\begin{equation}
\label{eqn:mp_energy_equation}
{p\ind{_{\rho}}}^{2} + {p\ind{_{z}}}^{2} = \frac{1}{\rho^{2}} \left(h + p\ind{_{\phi}} \right) \left(h - p\ind{_{\phi}} \right).
\end{equation}
Restricting our attention to $p\ind{_{\phi}} \geq 0$, the effective potential $h$ (which is clearly non-negative) determines the sign of the right-hand side of \eqref{eqn:mp_energy_equation} via the factor $( h - p\ind{_{\phi}} )$. The solutions to the equation $h(\rho, z) = p\ind{_{\phi}}$ (i.e., the \emph{contours} of $h$) are curves in the $(\rho, z)$-plane on which a ray may be instantaneously at rest, due to the fact that $p\ind{_{\rho}} = 0 = p\ind{_{z}}$ (and thus $\dot{\rho} = 0 = \dot{z}$) by the energy equation \eqref{eqn:mp_energy_equation}. For a given value of the azimuthal angular momentum $p\ind{_{\phi}}$, the positivity of the left-hand side of \eqref{eqn:mp_energy_equation} indicates that the contour $h = p\ind{_{\phi}}$ demarcates a \emph{forbidden region} (corresponding to $h < p\ind{_{\phi}}$) which the ray cannot access. The subset of the $(\rho, z)$-plane defined by the inequality $h \geq p\ind{_{\phi}}$ is referred to as the \emph{allowed region}. For a ray with $p\ind{_{\phi}} > 0$, the forbidden region typically includes the whole symmetry axis ($z$-axis), with the possible exception of the black hole horizons, which are located at $z = z_{\pm}$.

\begin{figure}
\begin{center}
\subfigure[$p\ind{_{\phi}} = 4$]{
\includegraphics[width=0.22\textwidth]{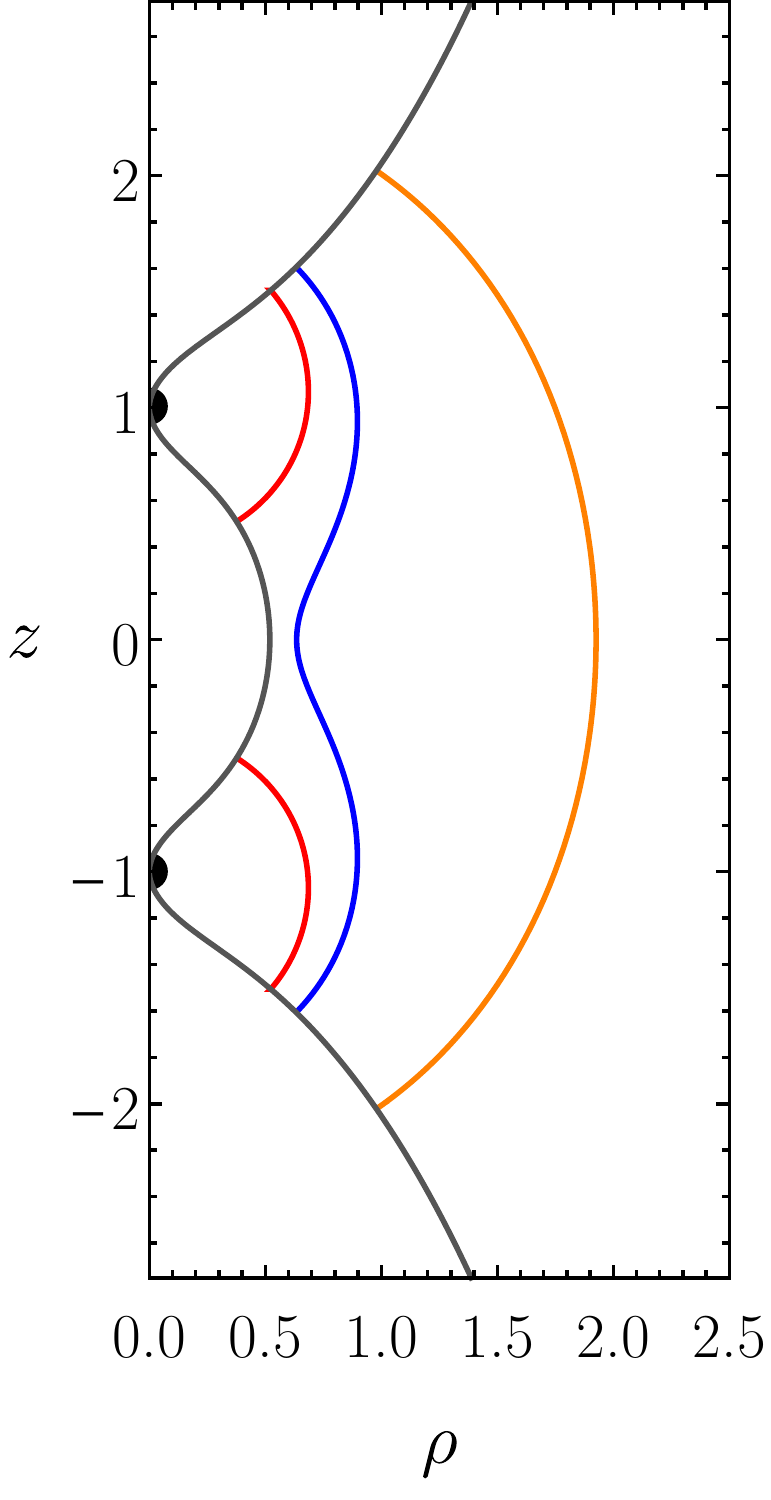} \label{fig:mp_fundamental_orbits_p_phi_400}}
\subfigure[$p\ind{_{\phi}} = 5.08$]{
\includegraphics[width=0.22\textwidth]{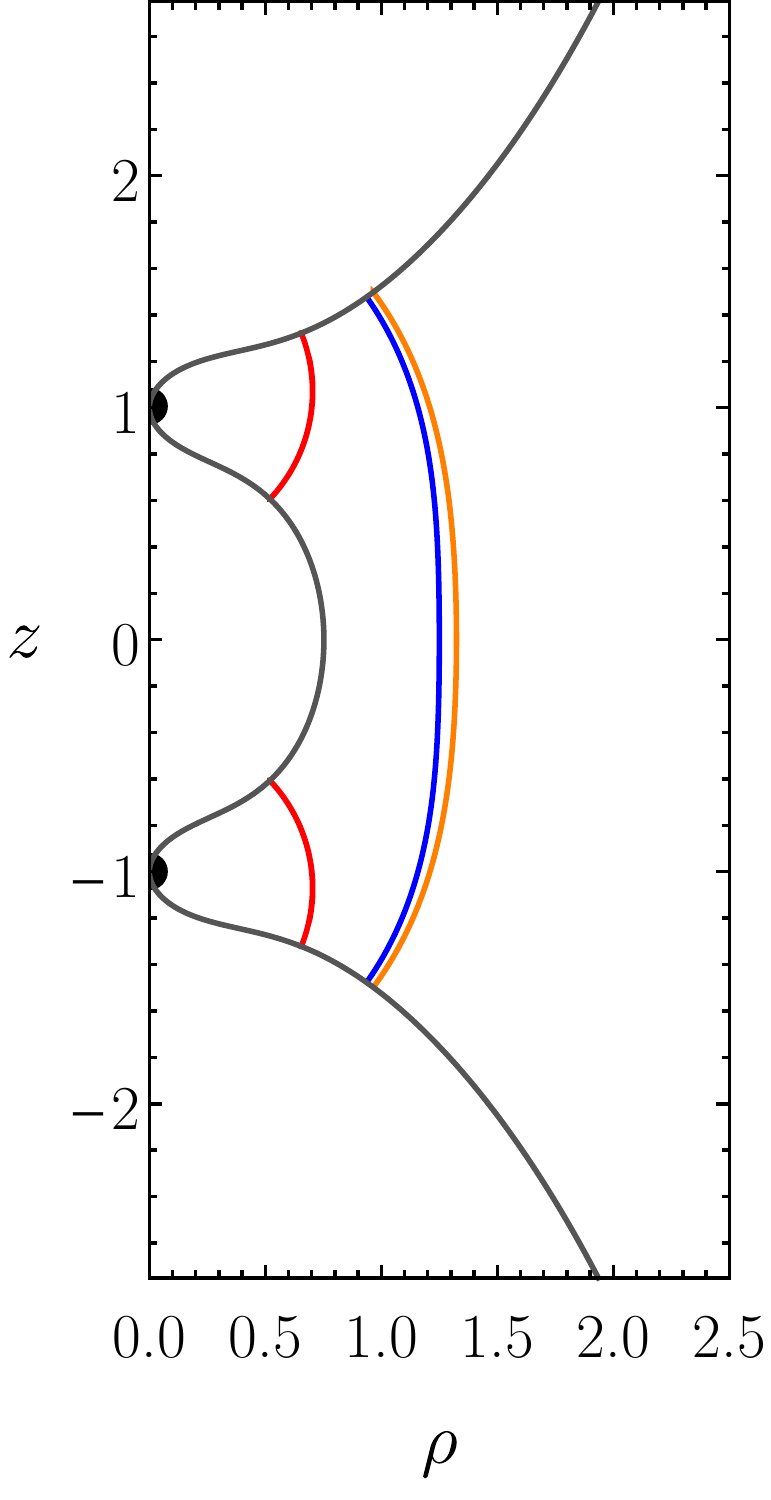} \label{fig:mp_fundamental_orbits_p_phi_508}}
\subfigure[$p\ind{_{\phi}} = 5.9$]{
\includegraphics[width=0.22\textwidth]{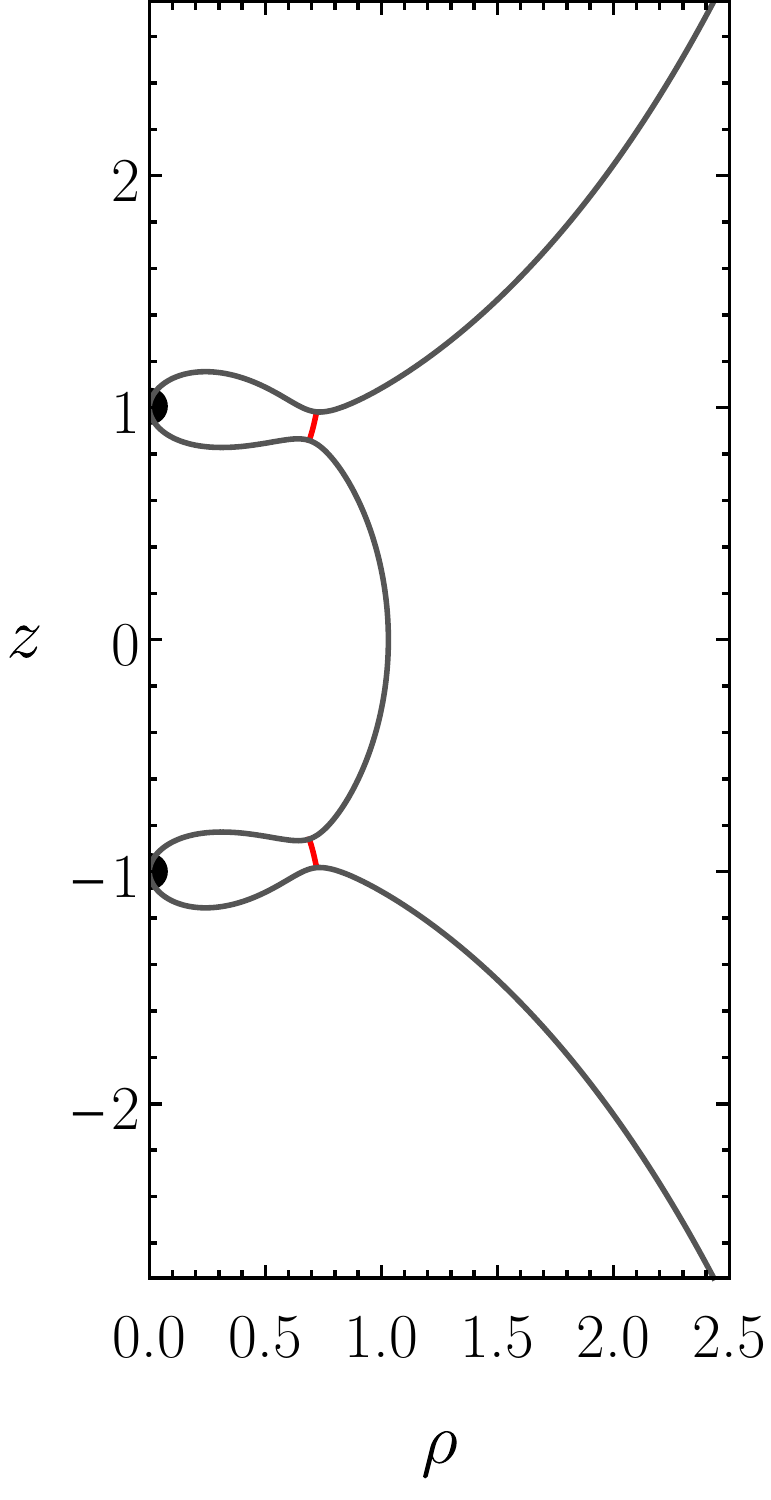} \label{fig:mp_fundamental_orbits_p_phi_590}}
\subfigure[$p\ind{_{\phi}} = 5.92214$]{
\includegraphics[width=0.22\textwidth]{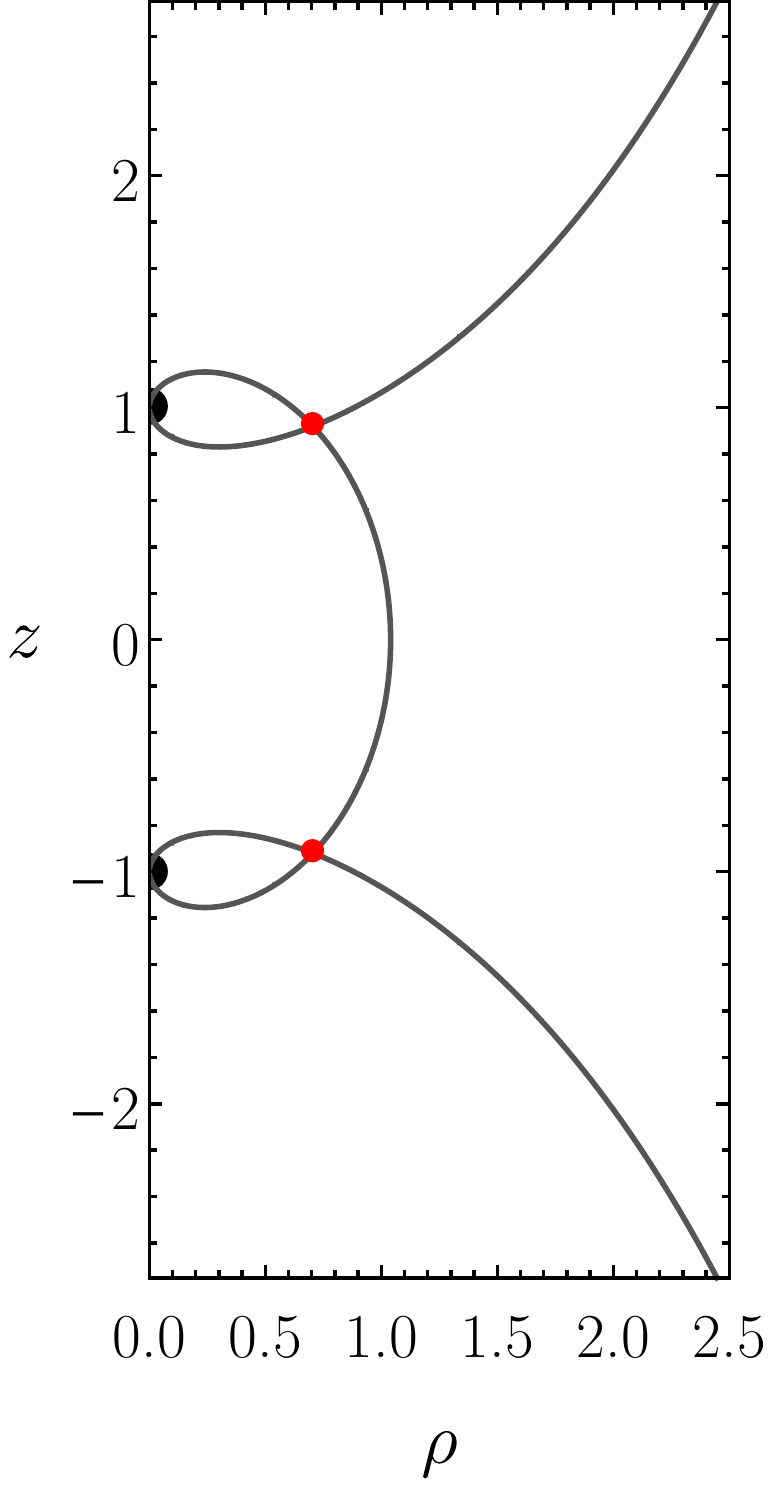} \label{fig:mp_fundamental_orbits_p_phi_592}}
\caption{Fundamental photon orbits for the Majumdar--Papapetrou binary di-hole with coordinate separation $d = 2 M$, projected onto the $(\rho, z)$-plane. (a) All three fundamental orbits exist, and are dynamically connected. (b) The outer orbits move closer together until they merge. (c) Only the inner pair of fundamental orbits exist. (d) The inner fundamental orbits exist up to $p\ind{_{\phi}} = {p\ind{_{\phi}}}^{\ast}$, the value of the azimuthal angular momentum for the (unstable) non-planar circular photon orbits of constant $\rho$ and $z$. In each case, the contour $h(\rho, z) = p\ind{_{\phi}}$ is shown as a grey curve. \label{fig:mp_fundamental_orbits_p_phi}}
\end{center}
\end{figure}

Consider a ray with angular momentum $p\ind{_{\phi}}$. Let $\Gamma$ denote the contour $h = p\ind{_{\phi}}$ in the $(\rho, z)$-plane. If the ray ``touches'' the contour, then $\dot{\rho} = 0 = \dot{z}$ (and thus $p\ind{_{\rho}} = 0 = p\ind{_{z}}$), instantaneously. Hamilton's equations for the momenta conjugate to $\rho$ and $z$ are
\begin{equation}
\left. \dot{p}\ind{_{\rho}} \right|_{\Gamma} = \left. \frac{p\ind{_{\phi}}}{\rho^{2}} h\ind{_{, \rho}} \right|_{\Gamma}, \qquad
\left. \dot{p}\ind{_{z}} \right|_{\Gamma} = \left. \frac{p\ind{_{\phi}}}{\rho^{2}} h\ind{_{, z}} \right|_{\Gamma},
\end{equation}
where we have used the fact that $\left. h \right|_{\Gamma} = p\ind{_{\phi}}$. As a result, rays which touch the contour $\Gamma$ must have a two-momentum $(\dot{p}\ind{_{\rho}}, \dot{p}\ind{_{z}})$ which is parallel to the two-vector $\bnab h$, where $\bnab = \left( \partial\ind{_{\rho}}, \partial\ind{_{z}} \right)$ denotes a two-gradient operator. The vector $\bnab h$ is clearly normal to the curve $h = p\ind{_{\phi}}$ in the $(\rho, z)$-plane: rays which touch the contour $\Gamma$ do so in a way which is orthogonal to $\Gamma$. This can be seen in Figures \ref{fig:mp_non_planar_ray_rho_z_0}--\ref{fig:mp_non_planar_ray_rho_z_4}, in which the three fundamental orbits touch the contour $h = p\ind{_{\phi}}$ orthogonally. Note that the ray in Figure \ref{fig:mp_non_planar_ray_rho_z_02} does not quite touch the contour.

In Figure \ref{fig:mp_fundamental_orbits_p_phi}, we show the effect of varying $p\ind{_{\phi}}$ on the contours of $h$ and the fundamental orbits, for an equal-mass di-hole with separation parameter $d = 2$. As $p\ind{_{\phi}}$ is increased, the fundamental orbits with symbolic representations $\overline{0}$ and $\overline{2}$ move closer together. When a critical value $p\ind{_{\phi}} = \hat{p}\ind{_{\phi}}$ is reached ($\hat{p}\ind{_{\phi}} \approx 5.09$ for $d = 2$), the $\overline{0}$- and $\overline{2}$-orbits coincide (as shown in Figure \ref{fig:mp_fundamental_orbits_p_phi_508} for $d = 2$). For $p\ind{_{\phi}} > \hat{p}\ind{_{\phi}}$, these two fundamental orbits cease to exist (as in Figure \ref{fig:mp_fundamental_orbits_p_phi_590}, for example). The fundamental orbits with decision dynamics representation $\overline{4}$ (light-rings around the individual black holes) persist until the contour ``pinches off'' at $p\ind{_{\phi}} = {p\ind{_{\phi}}}^{\ast}$ (${p\ind{_{\phi}}}^{\ast} \approx 5.92214$ for $d = 2$), as shown in Figure \ref{fig:mp_fundamental_orbits_p_phi_592}. The black holes are inaccessible to a ray incident from infinity with $p\ind{_{\phi}} > {p\ind{_{\phi}}}^{\ast}$. (Note that $\hat{p}\ind{_{\phi}} < {p\ind{_{\phi}}}^{\ast}$ for $d = 2$.)

\subsubsection{Fixed points of the effective potential}

\begin{figure}
\begin{center}
\subfigure[$d = \frac{1}{2} M$]{
\includegraphics[width=0.45\textwidth]{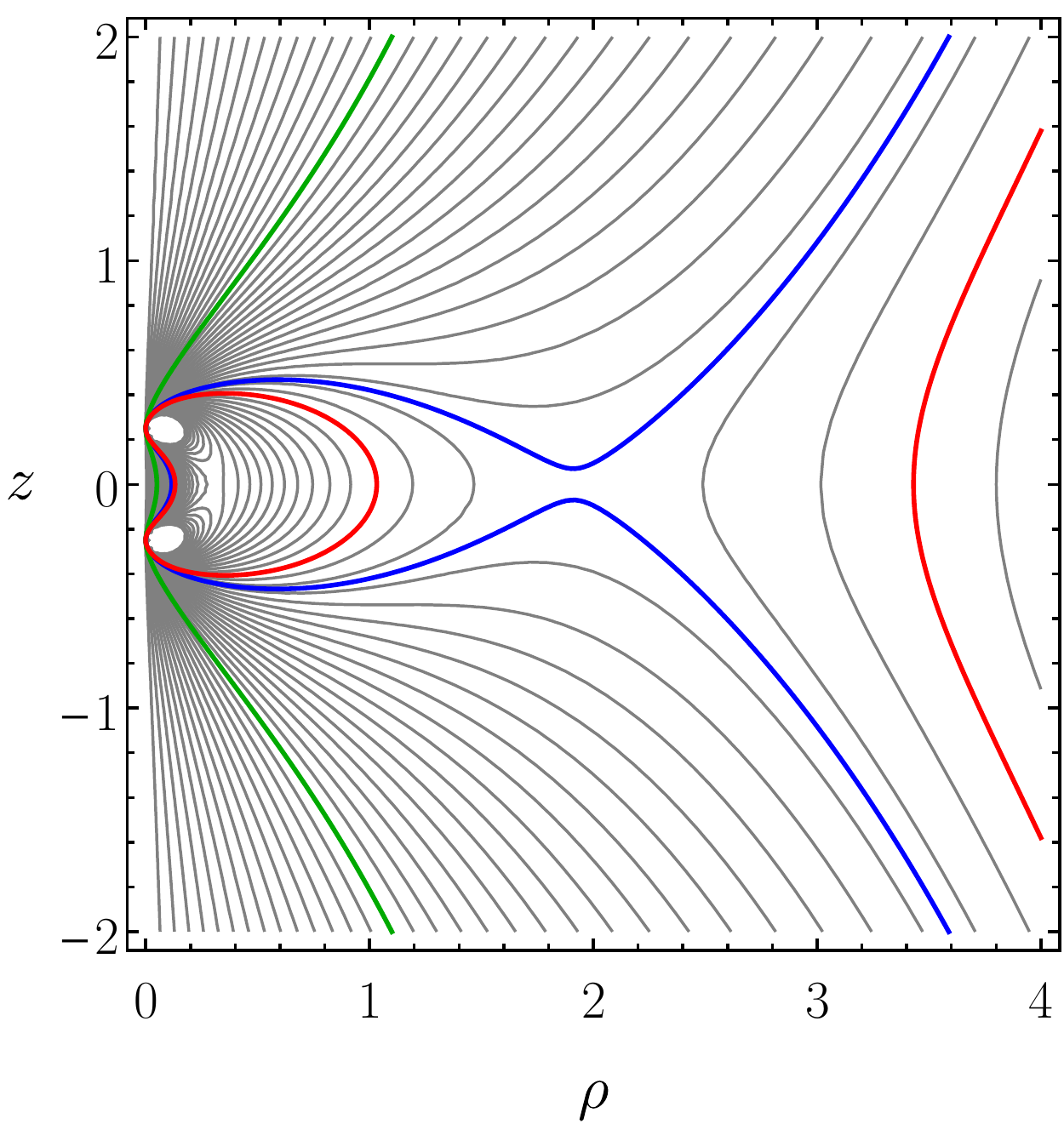} \label{fig:mp_height_function_contours_d05}}
\hspace{1em}
\subfigure[$d = 2 M$]{
\includegraphics[width=0.45\textwidth]{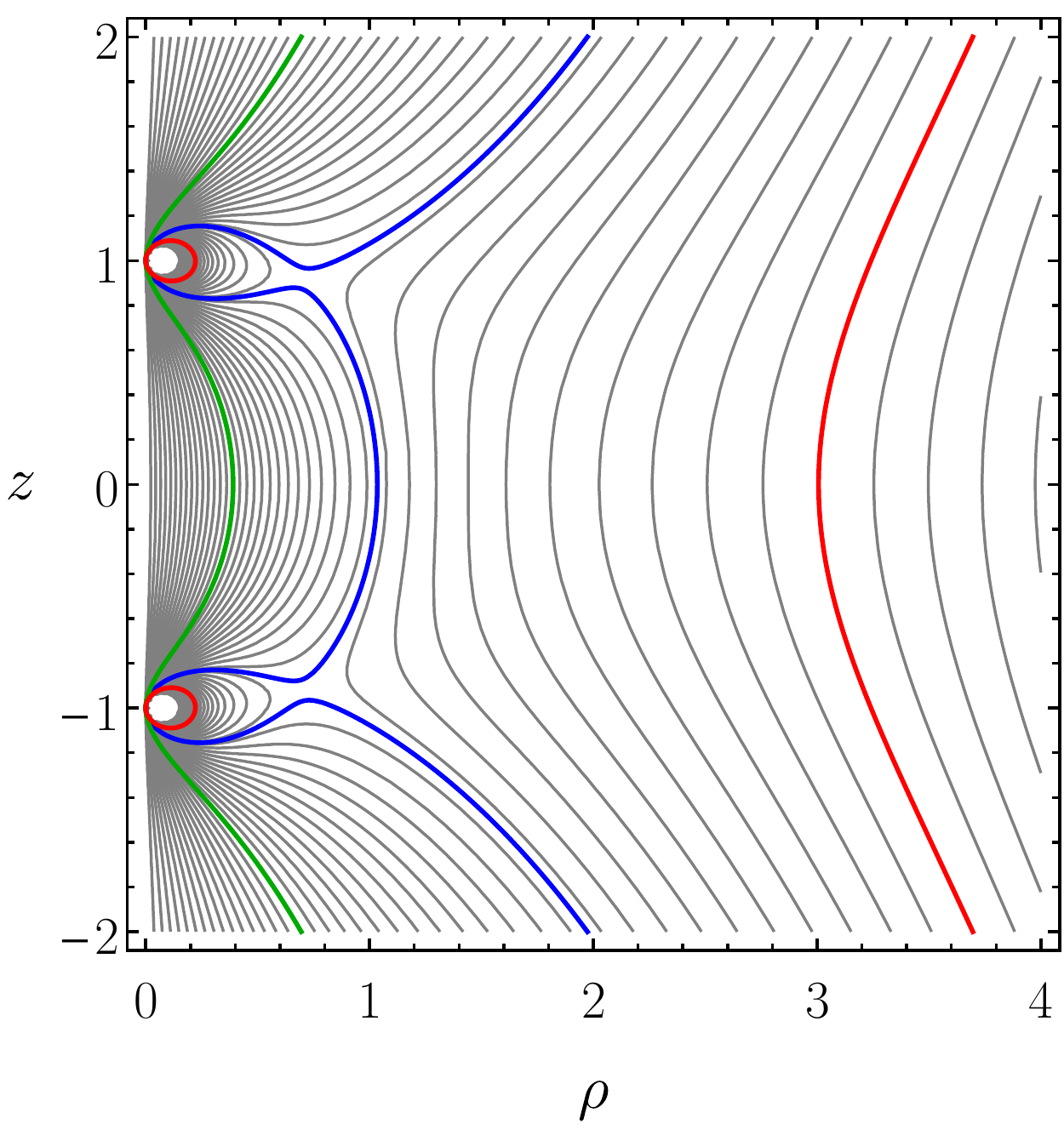} \label{fig:mp_height_function_contours_d2}}
\caption{Contours of the effective potential $h(\rho, z) = \rho \, U^{2}$ for (a) $d = \frac{1}{2} M$, and (b) $d = 2 M$. We highlight three typical cases: ${p\ind{_{\phi}}}^{A}$ [red], ${p\ind{_{\phi}}}^{B}$ [blue], ${p\ind{_{\phi}}}^{C}$ [green], where ${p\ind{_{\phi}}}^{A} > {p\ind{_{\phi}}}^{B} > {p\ind{_{\phi}}}^{C}$. Rays incident from infinity with $p\ind{_{\phi}} = {p\ind{_{\phi}}}^{A}$ are forbidden from accessing the black holes by angular momentum; rays with $p\ind{_{\phi}} = {p\ind{_{\phi}}}^{B}$ or $p\ind{_{\phi}} = {p\ind{_{\phi}}}^{C}$ are permitted to plunge into either black hole. For $p\ind{_{\phi}} = {p\ind{_{\phi}}}^{B}$, the ray would have to pass through a narrow channel demarcated by the contour.
\label{fig:mp_height_function_contours_d}}
\end{center}
\end{figure}

It is clear from Figure \ref{fig:mp_height_function_contours_d} that the morphology of the contours of $h$ depends on the separation between the centres $d$. For widely separated black holes, the Majumdar--Papapetrou di-hole system effectively behaves like a pair of isolated extremal Reissner--Nordstr\"{o}m black holes, as shown in Figure \ref{fig:mp_height_function_contours_d05}. On the other hand, for a tightly bound di-hole, the system effectively resembles a single distorted black hole, as can be inferred from Figure \ref{fig:mp_height_function_contours_d2}. In order to better understand the system's dependence on $d$, let us consider the fixed points of $h$ and its critical contours, i.e., those which pass through the fixed points.

In Appendix \ref{chap:appendix_a}, we present a classification of the stationary points of the effective potential $h$ both in and out of the equatorial plane. Here, we aim to deepen our understanding of the system by considering the equatorial fixed points of $h$, i.e., those which lie in the plane $z = 0$. (This can be achieved by considering the one-dimensional effective potential $\hat{h} = h(\rho, 0) = 1 + \frac{2}{R}$, where $R^{2} = \rho^{2} + \frac{d^{2}}{4}$.) First, define the critical values of the separation parameter
\begin{equation}
d_{1} = \sqrt{\frac{16}{27}} M, \qquad
d_{2} = \sqrt{\frac{32}{27}} M.
\end{equation}
Using the method presented in Appendix \ref{chap:appendix_a}, we find that, for $d > d_{2}$, $h$ has no stationary points in the equatorial plane. In the case $d = d_{2}$, there is a ``cusp'' in the equatorial plane at $\rho = \frac{\sqrt{5}}{2} d_{2}$, which corresponds to a point of inflection in the $\rho$-direction and a maximum in the $z$-direction. For separations in the range $d_{1} < d < d_{2}$, $h$ admits two equatorial fixed points: a saddle point at $\rho = \rho_{+}$, and a maximum at $\rho = \rho_{-}$, where $\rho_{+} > \rho_{-} > \frac{d}{\sqrt{2}}$. For $d \leq d_{1}$, $h$ has a pair of equatorial saddle points at $\rho = \rho_{\pm}$, where $\rho_{-} < \frac{d}{\sqrt{2}} < \rho_{+}$. In addition, we find that non-planar saddle points of $h$ are permitted, provided that $d > d_{1}$.

\begin{figure}
\begin{center}
\subfigure[$\frac{d}{M} = \sqrt{\frac{16}{27}}$]{
\includegraphics[width=0.22\textwidth]{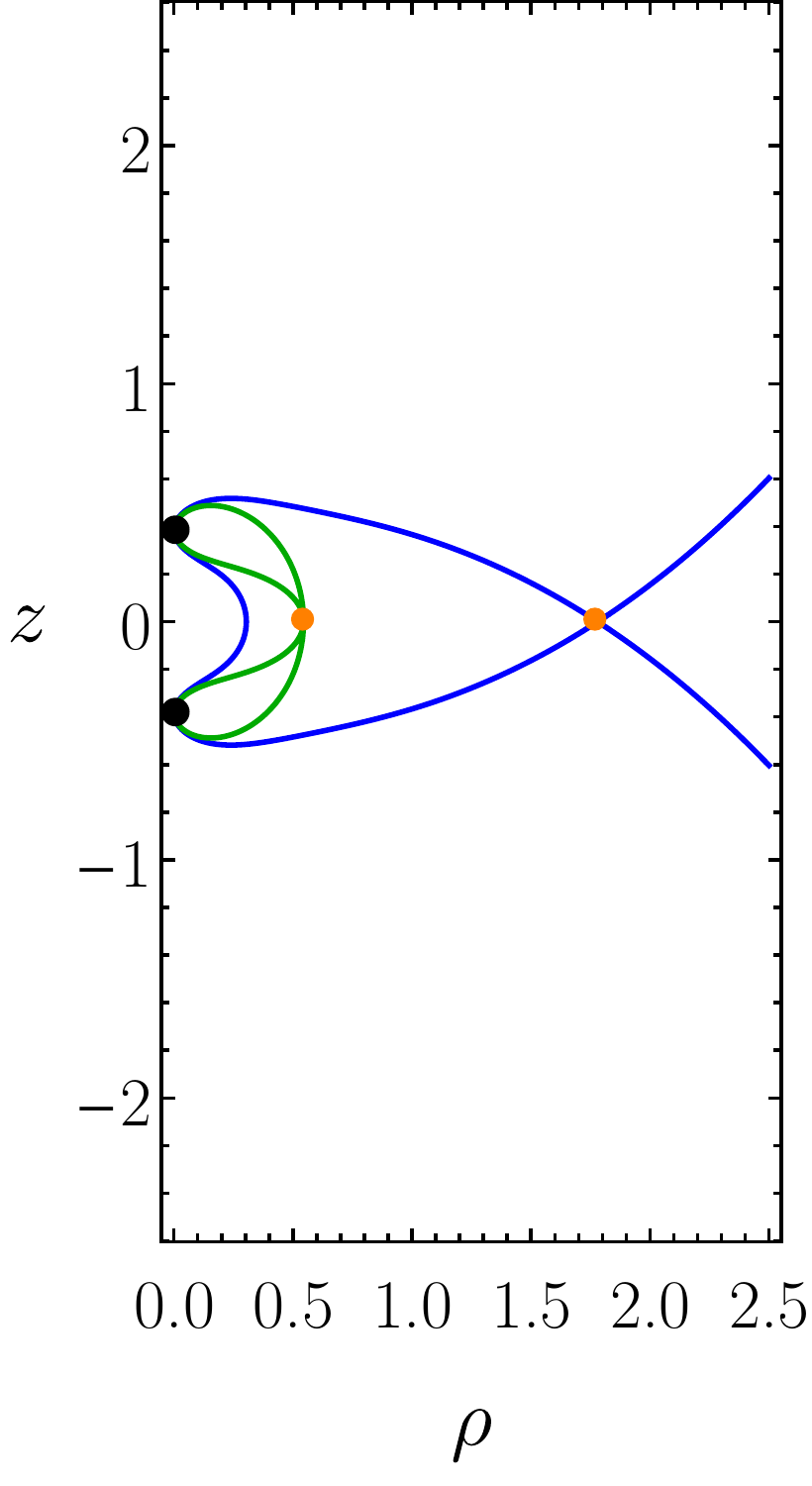} \label{fig:mp_critical_contours_d_2}}
\subfigure[$\frac{d}{M} = 1$]{
\includegraphics[width=0.22\textwidth]{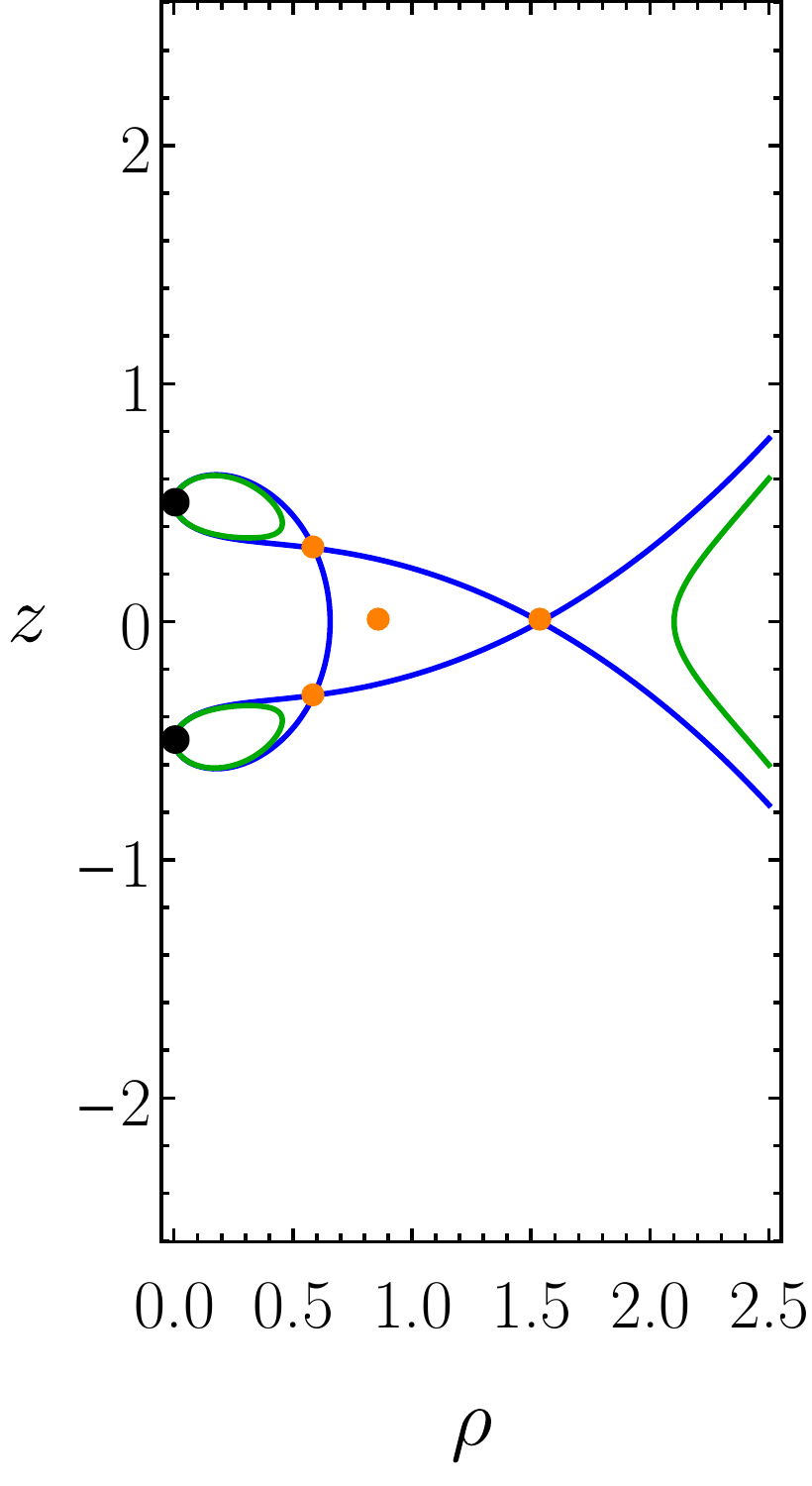} \label{fig:mp_critical_contours_d_3}}
\subfigure[$\frac{d}{M} = \sqrt{\frac{32}{27}}$]{
\includegraphics[width=0.22\textwidth]{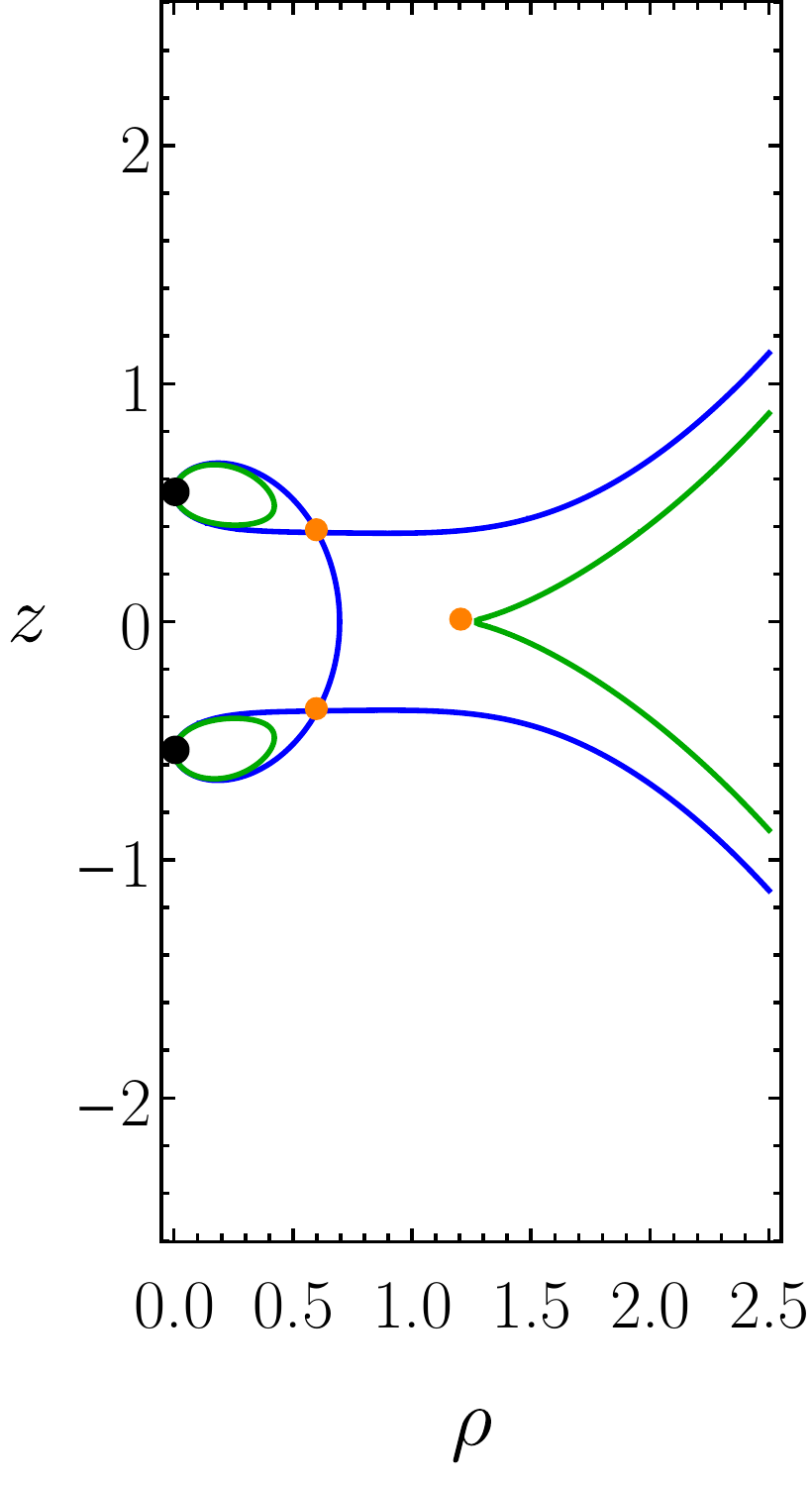} \label{fig:mp_critical_contours_d_4}}
\subfigure[$\frac{d}{M} = 2$]{
\includegraphics[width=0.22\textwidth]{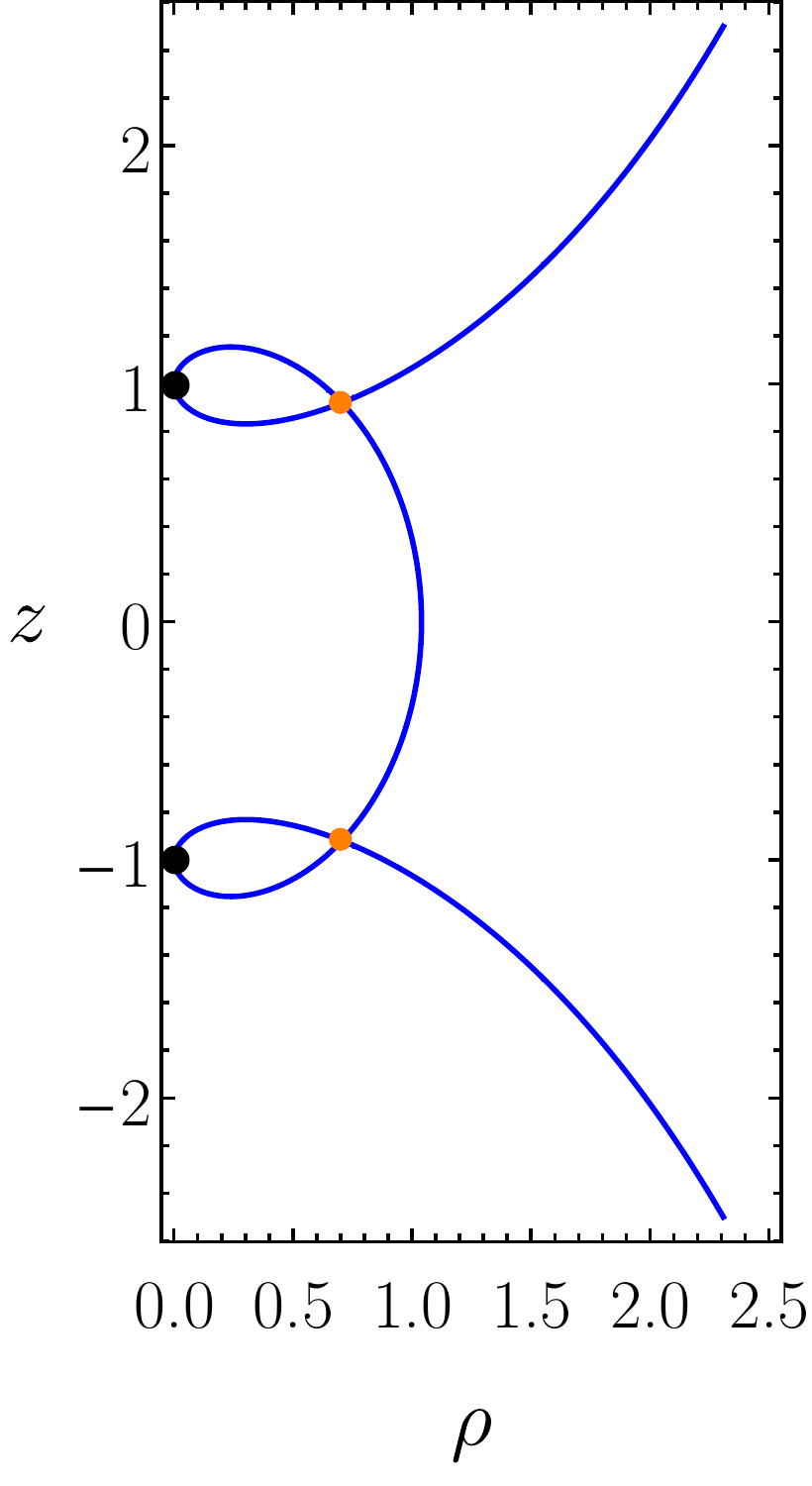} \label{fig:mp_critical_contours_d_5}}
\caption{Critical contours and fixed points of the effective potential $h(\rho, z) = \rho \, U^{2}$ for a range of values of the coordinate separation $d$, in the equal-mass Majumdar--Papapetrou di-hole spacetime. (a) For $\frac{d}{M} \leq \sqrt{\frac{16}{27}}$, $h$ has two saddle points in the equatorial ($z = 0$) plane. (b) For $\sqrt{\frac{16}{27}} < \frac{d}{M} < \sqrt{\frac{32}{27}}$, $h$ has one equatorial saddle point and a pair of non-equatorial saddles; $h$ also possesses a maximum in the equatorial plane. Stable bounded photon orbits are permitted in this regime. (c) For $\frac{d}{M} = \sqrt{\frac{32}{27}}$, there are two non-planar saddles and a ``cusp'' in the equatorial plane (marginal case). (d) For $\frac{d}{M} > \sqrt{\frac{32}{27}}$, $h$ admits two saddle points out of the equatorial plane. \label{fig:mp_critical_contours}}
\end{center}
\end{figure}

In Figure \ref{fig:mp_critical_contours}, we show the fixed points and critical contours of the height function $h(\rho, z) = \rho \, U^{2}$ for a selection of values of the separation parameter $d$. The case $d = M$, shown in Figure \ref{fig:mp_critical_contours_d_3}, is special: $h$ has three saddle points (one of which is in the equatorial plane), which are all connected by the \emph{same} contour. This contour encloses a \emph{maximum} of $h$. Remarkably, it is possible to find closed-form expressions for the stationary points and critical contours in terms of the golden ratio $\varphi = \frac{1}{2} \left( 1 + \sqrt{5} \right)$. The maximum is located at $\rho_{(1)} = \frac{\sqrt{3}}{2} M$, with ${p\ind{_{\phi}}}^{(1)} = \frac{9 \sqrt{3}}{2} M$. There are three saddles at $\rho_{(2)} = \frac{1}{2} 5^{1/4} \varphi^{3/2} M$, $z = 0$ and $\rho_{(3)} = \frac{1}{2} 5^{1/4} \varphi^{- 1/2} M$, $\pm z_{(3)} = \pm \frac{M}{2 \varphi}$. It is straightforward to check that all three saddle points are connected by the contour $h = {p\ind{_{\phi}}}^{(2)} = {p\ind{_{\phi}}}^{(3)} = \frac{1}{2} 5^{5/4} \varphi^{3/2} M$. (See Appendix \ref{chap:appendix_a} for the derivation of these results.)

The saddle points of $h$, which correspond to unstable circular photon orbits of constant $\rho$ and $z$, can be viewed as unstable ``Lagrange points'' for null geodesics in the Majumdar--Papapetrou di-hole system. Perturbing $p\ind{_{\phi}}$ slightly from the value corresponding to a saddle point has the effect of opening a narrow channel, demarcated by neighbouring contours, in the vicinity of the saddle point. Hence, where there exist saddle points, there also exist neighbouring contours on either side of the saddle which are almost parallel to one another. Because null geodesics ``touch'' the contour orthogonally, we find that generic periodic unstable null orbits which ``bounce'' between the neighbouring parts of the contour will occur in these cases. These unstable orbits are born from perturbing the unstable ``Lagrange points''. Comparing Figures \ref{fig:mp_fundamental_orbits_p_phi_590} and \ref{fig:mp_fundamental_orbits_p_phi_592}, we see that the $\overline{4}$ fundamental orbits are born from the non-equatorial saddle points in the case $d = 2 M$.
%

\subsubsection{Bounded null geodesics}

The existence of a maximum of the effective potential $h$ in the regime $d_{1} < d < d_{2}$ implies the existence of \emph{bounded} null geodesics. These are confined to a compact subregion of the $(\rho, z)$-plane. Using the relationship between the geodesic potential $V(\rho, z)$ and the effective potential $h(\rho, z)$ given in \eqref{eqn:mp_hamiltonian_canonical}, one can show that a local maximum of $h$ corresponds to a local minimum of the potential $V$; hence, the circular orbit of constant $\rho$ and $z$ corresponding to the maximum of $h$ is \emph{stable}. The existence and phenomenology of stable null geodesics will be explored in the context of stationary axisymmetric electrovacuum spacetimes in Chapter \ref{chap:stable_photon_orbits}.

\begin{figure}
\begin{center}
\subfigure[]{\includegraphics[height=0.31\textwidth]{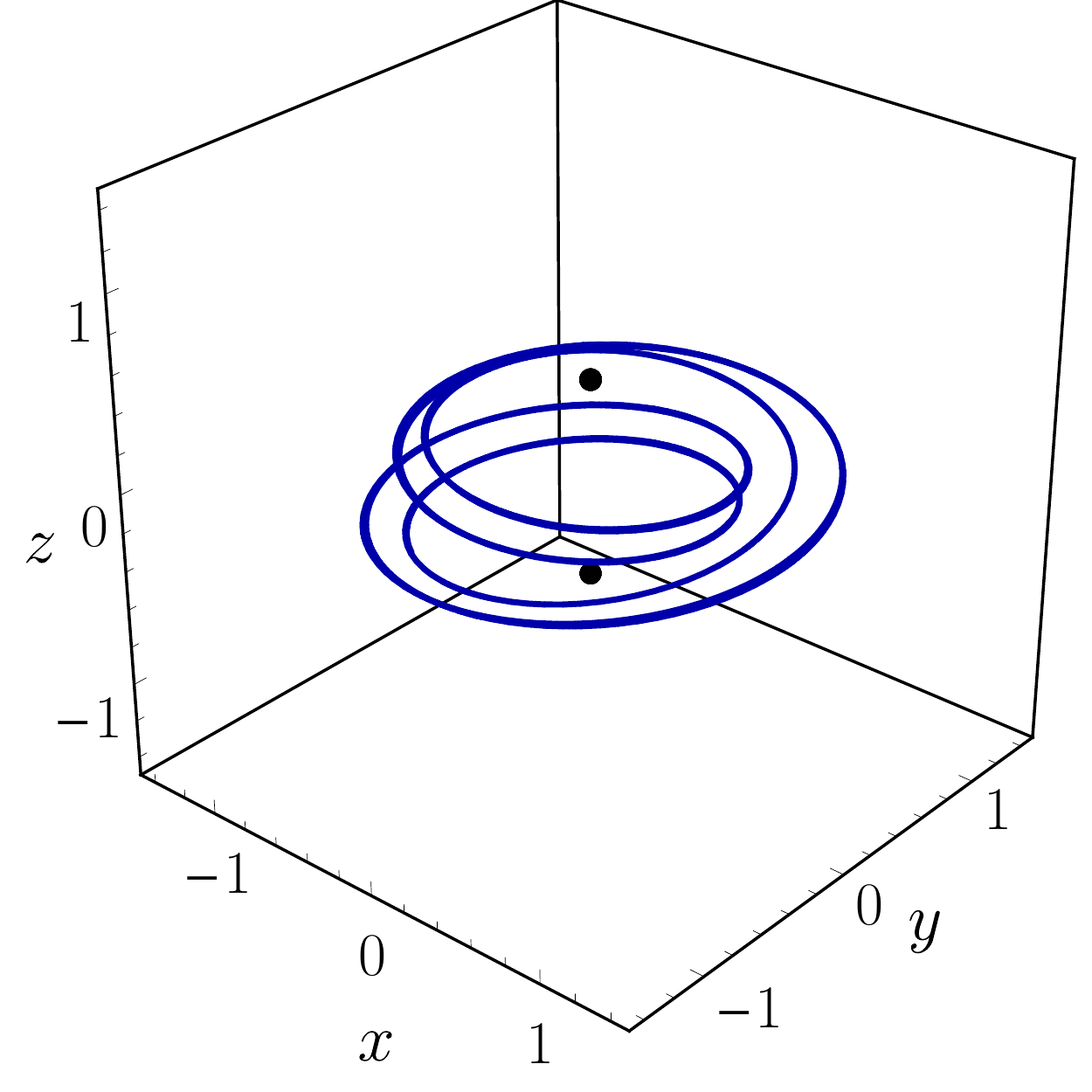} \label{fig:mp_bounded_null_geodesic_d1_3d}}
\hspace{0.5em}
\subfigure[]{\includegraphics[height=0.31\textwidth]{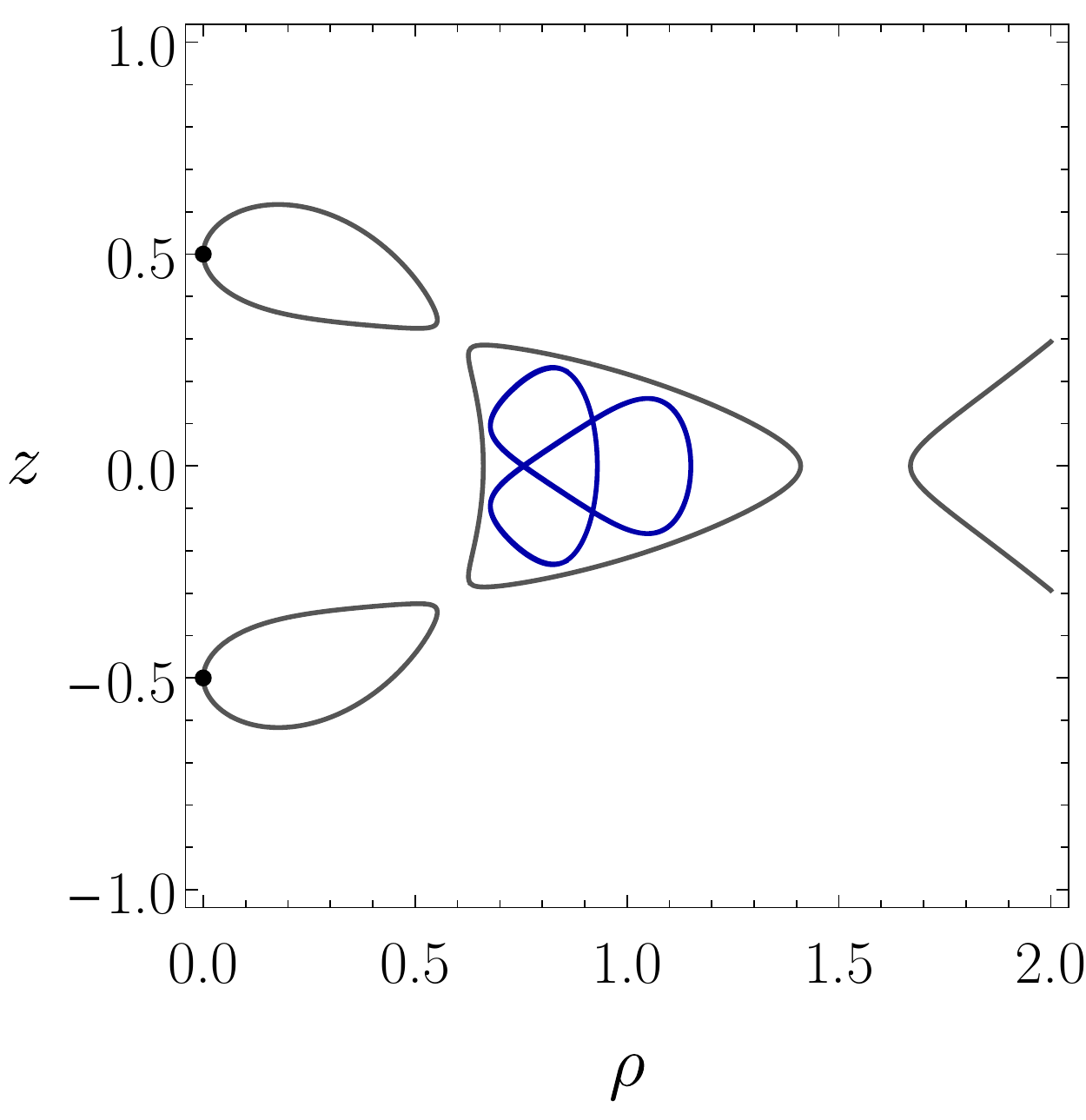} \label{fig:mp_bounded_null_geodesic_d1_2d}}
\subfigure[]{\includegraphics[height=0.31\textwidth]{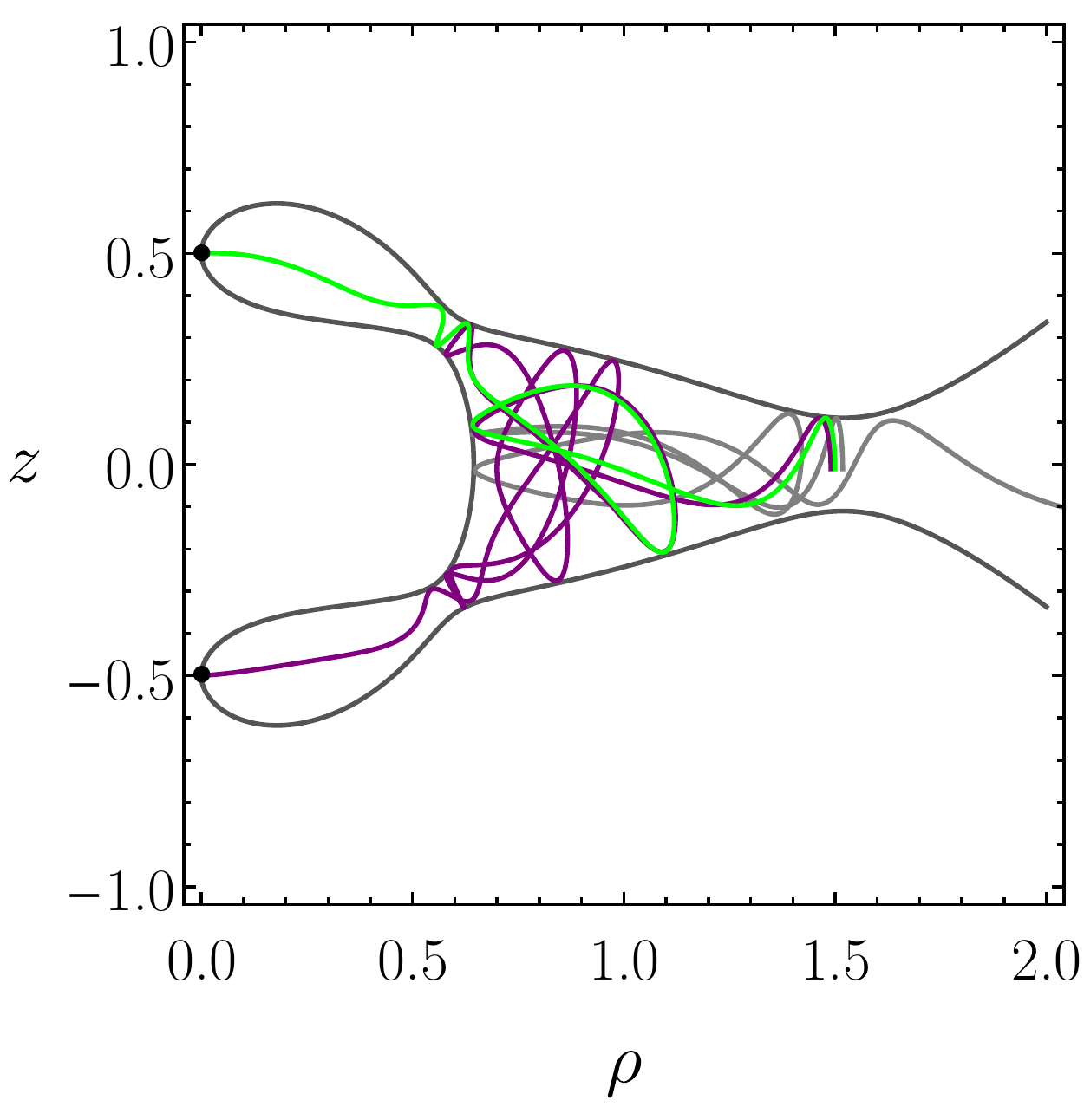} \label{fig:mp_narrow_escapes_trajectories}}
\caption{Examples of rays in the ``pocket'' for and equal-mass Majumdar--Papapetrou di-hole with $d = M$. (a) A bounded null geodesic with $p\ind{_{\phi}} = 7.7 > {p\ind{_{\phi}}}^{\ast}$, shown on the $(x, y, z)$-axes in $\mathbb{R}^{3}$. (b) Projection of the orbit in (a) onto the $(\rho, z)$-plane [dark blue], with the energy contour $h(\rho, z) = p\ind{_{\phi}}$ [dark grey]. (c) Null geodesics with $p\ind{_{\phi}} = 7.68 < {p\ind{_{\phi}}}^{\ast}$ in the ``pocket'' with three narrow escape channels leading to the black holes and to infinity. These initially neighbouring rays end up in different final states. \label{fig:mp_bounded_null_geodesic_d1}}
\end{center}
\end{figure}

Figure \ref{fig:mp_bounded_null_geodesic_d1} shows an example of a bounded null geodesic for a di-hole with $d = M$. In Figure \ref{fig:mp_bounded_null_geodesic_d1_3d}, we show the orbit in $\mathbb{R}^{3}$ on a set of Cartesian axes; the orbit is clearly confined to a toroidal region around the two black holes. Figure \ref{fig:mp_bounded_null_geodesic_d1_2d} shows the orbit's projection onto the $(\rho, z)$-plane and the corresponding contour $h = p\ind{_{\phi}}$; the periodicity of the orbit is manifest.

Decreasing $p\ind{_{\phi}}$ from ${p\ind{_{\phi}}}^{\ast}$, it is possible to construct a ``pocket'' in the $(\rho, z)$-plane which connects the scattering region to the two black holes and to spatial infinity via three narrow channels; see Figure \ref{fig:mp_narrow_escapes_trajectories}. The pocket and the escape channels are demarcated by the contour $h = {p\ind{_{\phi}}}^{\ast} - \Delta p\ind{_{\phi}}$, where $\Delta p\ind{_{\phi}} \gtrsim 0$ is a small perturbation. As $\Delta p\ind{_{\phi}} \rightarrow 0$, the width of the escapes goes to zero. In Figure \ref{fig:mp_narrow_escapes_trajectories}, we show three initially neighbouring rays which begin in the ``pocket'' (inside the scattering region), but which end up in different final states.
%

\subsection{Binary black hole shadows}
\label{sec:binary_black_hole_shadows}

\subsubsection{Ray-tracing set-up and numerical method}

In the section, we revisit the definition of a black hole shadow (see Section \ref{sec:black_hole_shadows_review}). Bohn \emph{et al.} \cite{BohnThroweHebertEtAl2015} employ a ray-tracing approach, and define a shadow as ``a region of the [observer's] image where geodesics are traced backwards in time from the camera to the black hole''. The camera provides a natural one-to-one correspondence between a ``pixel'' on a two-dimensional black hole shadow image and a null geodesic.

Our primary aim is to study the structure of binary black hole shadows for the Majumdar--Papapetrou di-hole geometry. To realise the black hole shadows, we trace rays which pass orthogonally through a planar surface (the image plane) with centre $(x, y, z) = (x_{0}, 0, z_{0})$ in Cartesian coordinates, where $r_{0} = \sqrt{x_{0}^{2} + z_{0}^{2}}$ is the distance from the centre of the surface to the origin (taken to be the centre of mass). In our computations, we typically take $r_{0} = 50 M$. The angle of incidence $\theta$ is defined via the equations $\sin{\theta} = \frac{x_{0}}{r_{0}}$, $\cos{\theta} = -\frac{z_{0}}{r_{0}}$; that is, $\theta$ is measured with respect to the negative $z$-axis in the anti-clockwise sense.

The observer's two-dimensional image plane is labelled by a local system of Cartesian-type coordinates $(X, Y)$. We determine whether the point $(X, Y)$ is in the shadow by integrating Hamilton's equations backwards in time, taking as initial data
\begin{align}
x(0) &= x_{0} + X \cos{\theta}, & y(0) &= Y, & z(0) &= z_{0} - X \sin{\theta}, \label{eqn:ray_tracing_ics_1} \\
p\ind{_{x}}(0) &= - U_{0}^{2} \sin{\theta}, & p\ind{_{y}}(0) &= 0, & p\ind{_{z}}(0) &= U_{0}^{2} \cos{\theta}, \label{eqn:ray_tracing_ics_2}
\end{align}
where $U_{0} = \left. U(x, y, z) \right|_{\lambda = 0}$. If the ray approaches a black hole horizon, then the point $(X, Y)$ is assumed to be in the shadow. If the ray approaches spatial infinity, then the point $(X, Y)$ is in the non-shadow region.

As described in Section \ref{sec:integrability_chaos}, geodesic motion on the Majumdar--Papapetrou di-hole spacetime is non-integrable. We therefore resort to numerical methods to evolve the equations of motion. In particular, we perform ray-tracing by numerically evolving Hamilton's equations subject to the initial data \eqref{eqn:ray_tracing_ics_1}--\eqref{eqn:ray_tracing_ics_2}. This is achieved by employing Mathematica's \texttt{NDSolve} function.

In practice, the shadow is realised using a finite-resolution grid of pixels on the image plane. A \emph{pixel} is defined to be a square region of the image plane with side length $L$ and centre $(X, Y)$. The pixel with midpoint $(X, Y)$ is in the black hole shadow if and only if the null geodesic with initial data given by \eqref{eqn:ray_tracing_ics_1}--\eqref{eqn:ray_tracing_ics_2} approaches a black hole horizon. Practically, this involves implementing one of two halting conditions. Our numerical integration is stopped when either
\begin{equation}
x^{2}(\lambda) + y^{2}(\lambda) + \left( z(\lambda) - z_{\pm} \right)^{2} \leq \varepsilon, \qquad 0 < \varepsilon \ll 1;
\end{equation}
or when
\begin{equation}
x^{2}(\lambda) + y^{2}(\lambda) + z^{2}(\lambda) > r_{0}.
\end{equation}
The former case indicates absorption by the black hole with horizon at $z = z_{\pm}$; the latter case indicates scattering.

In the approach outline above, the image plane defines a set of local observers at each point $(X, Y)$. With a change of emphasis, one can instead define a black hole shadow with respect to a single observer, by performing ray-tracing from a single point in spacetime, varying the initial elevation and azimuth of the ray's two-momentum. The two definitions are essentially equivalent as $r_{0} \rightarrow \infty$.


\subsubsection{Gallery of two-dimensional shadows}

\begin{figure}
\begin{center}
\subfigure[$\theta = 0^{\circ}$]{\includegraphics[width=0.19\textwidth]{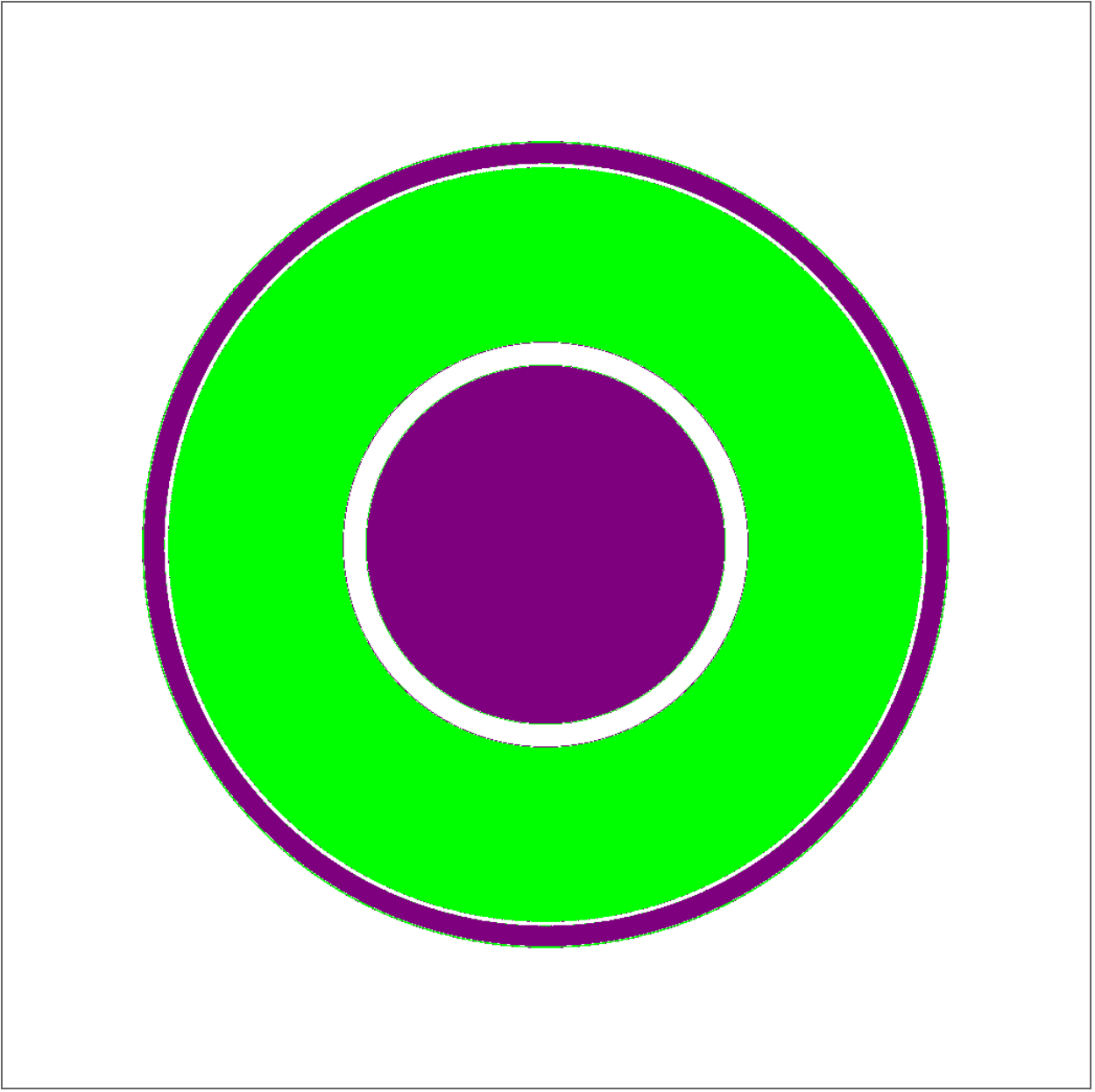}\label{fig:mp_shadows_gallery_00}}
\subfigure[$\theta = 10^{\circ}$]{\includegraphics[width=0.19\textwidth]{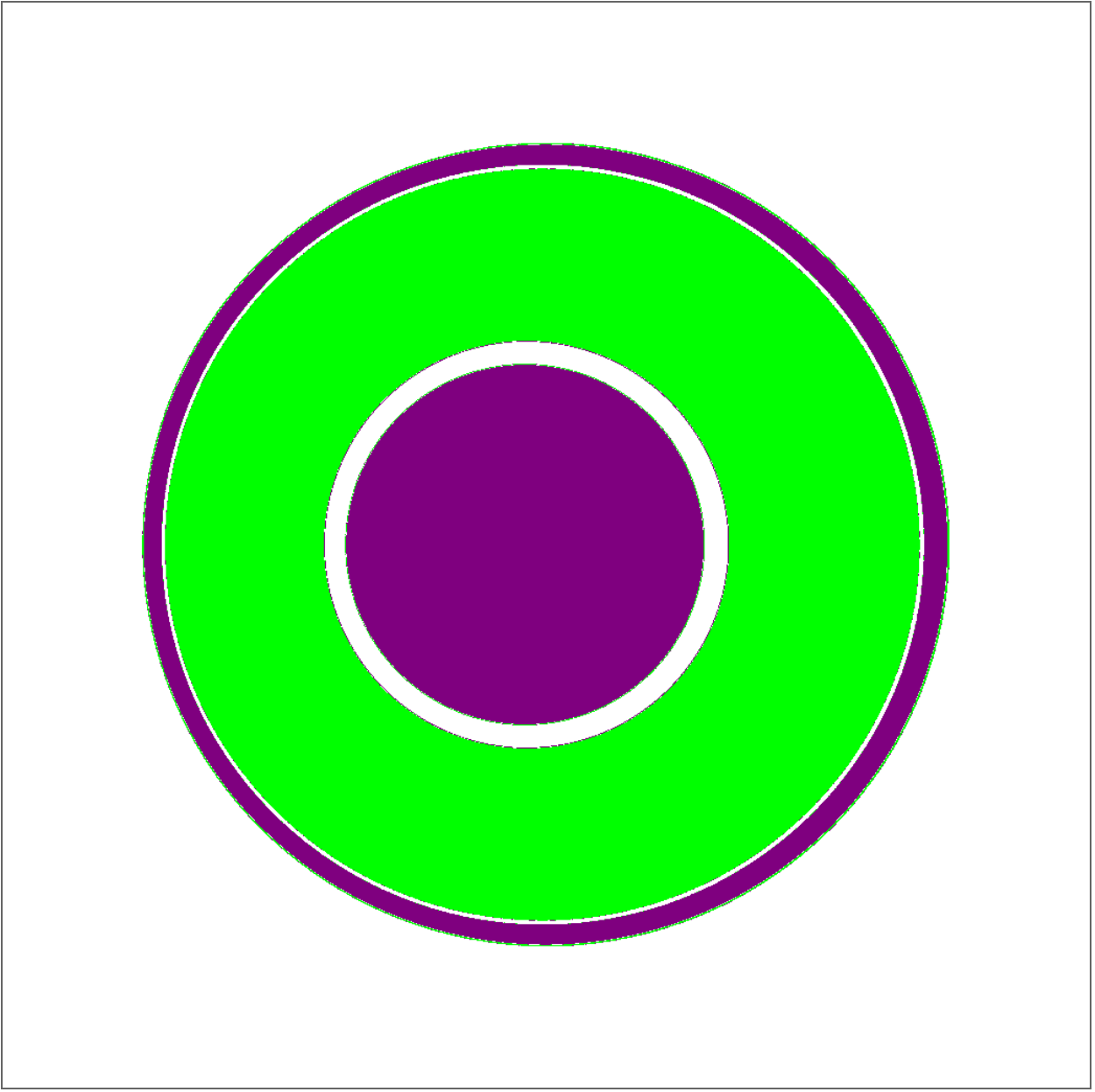}\label{fig:mp_shadows_gallery_10}}
\subfigure[$\theta = 20^{\circ}$]{\includegraphics[width=0.19\textwidth]{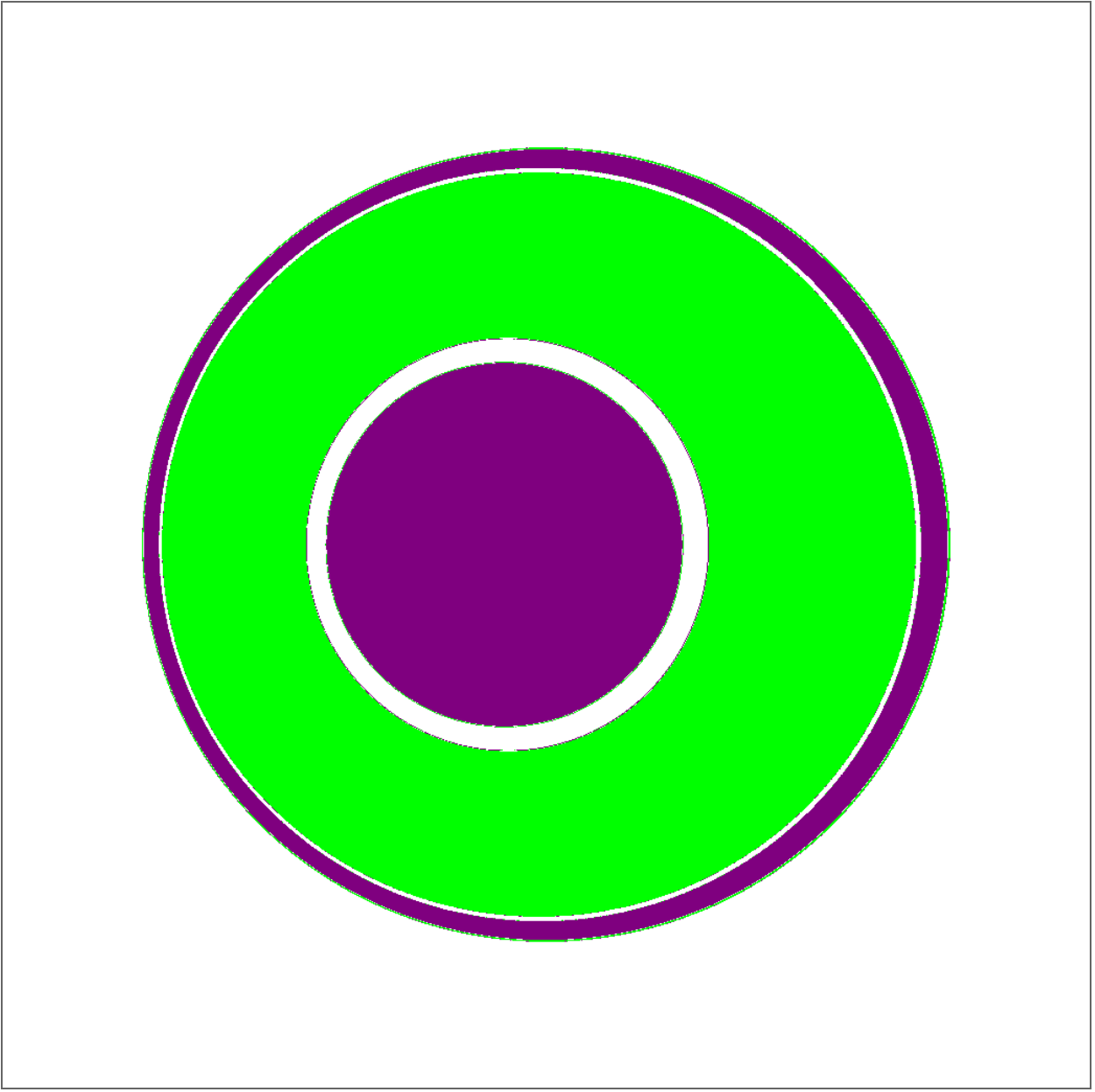}\label{fig:mp_shadows_gallery_20}}
\subfigure[$\theta = 30^{\circ}$]{\includegraphics[width=0.19\textwidth]{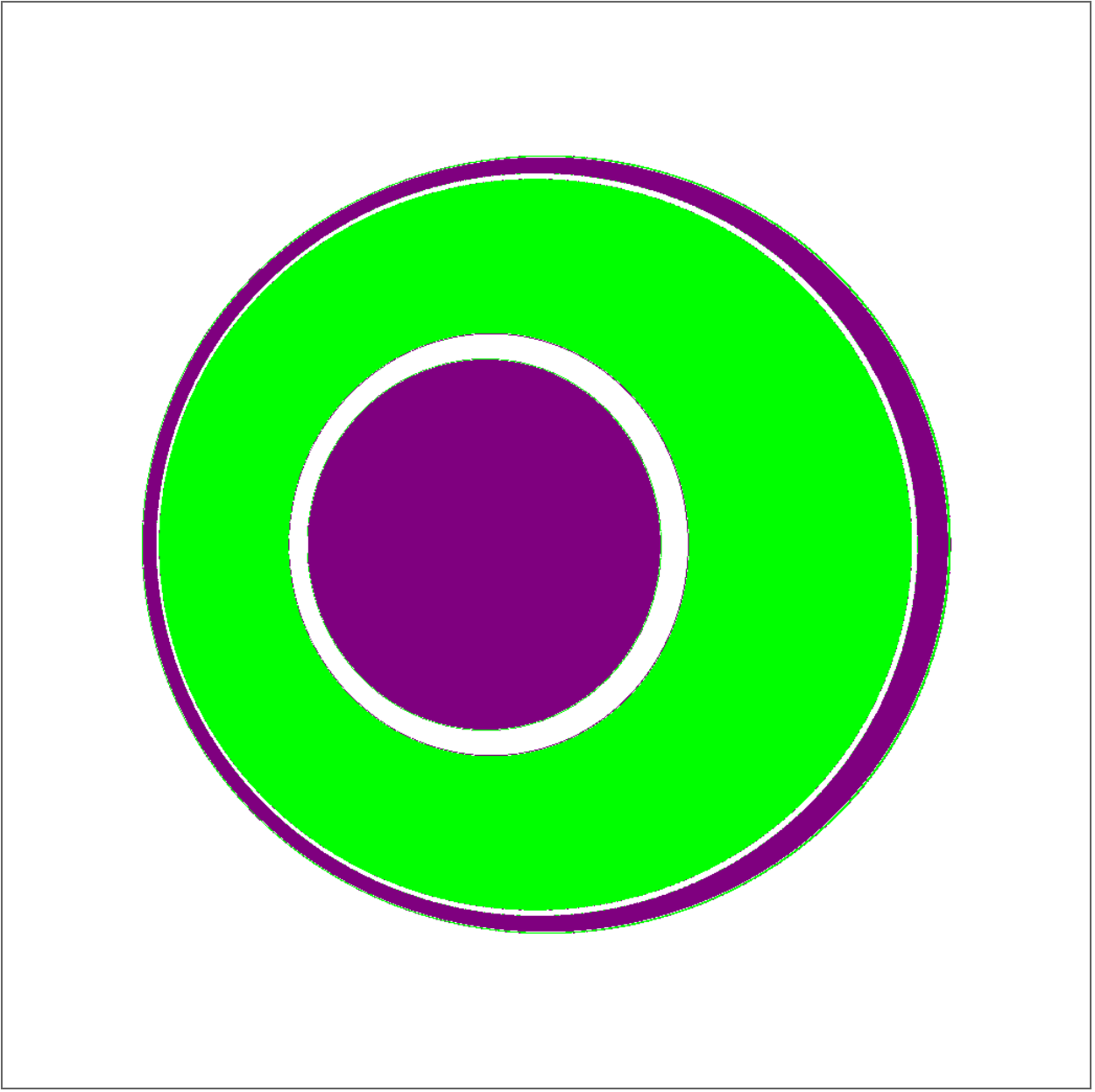}\label{fig:mp_shadows_gallery_30}}
\subfigure[$\theta = 40^{\circ}$]{\includegraphics[width=0.19\textwidth]{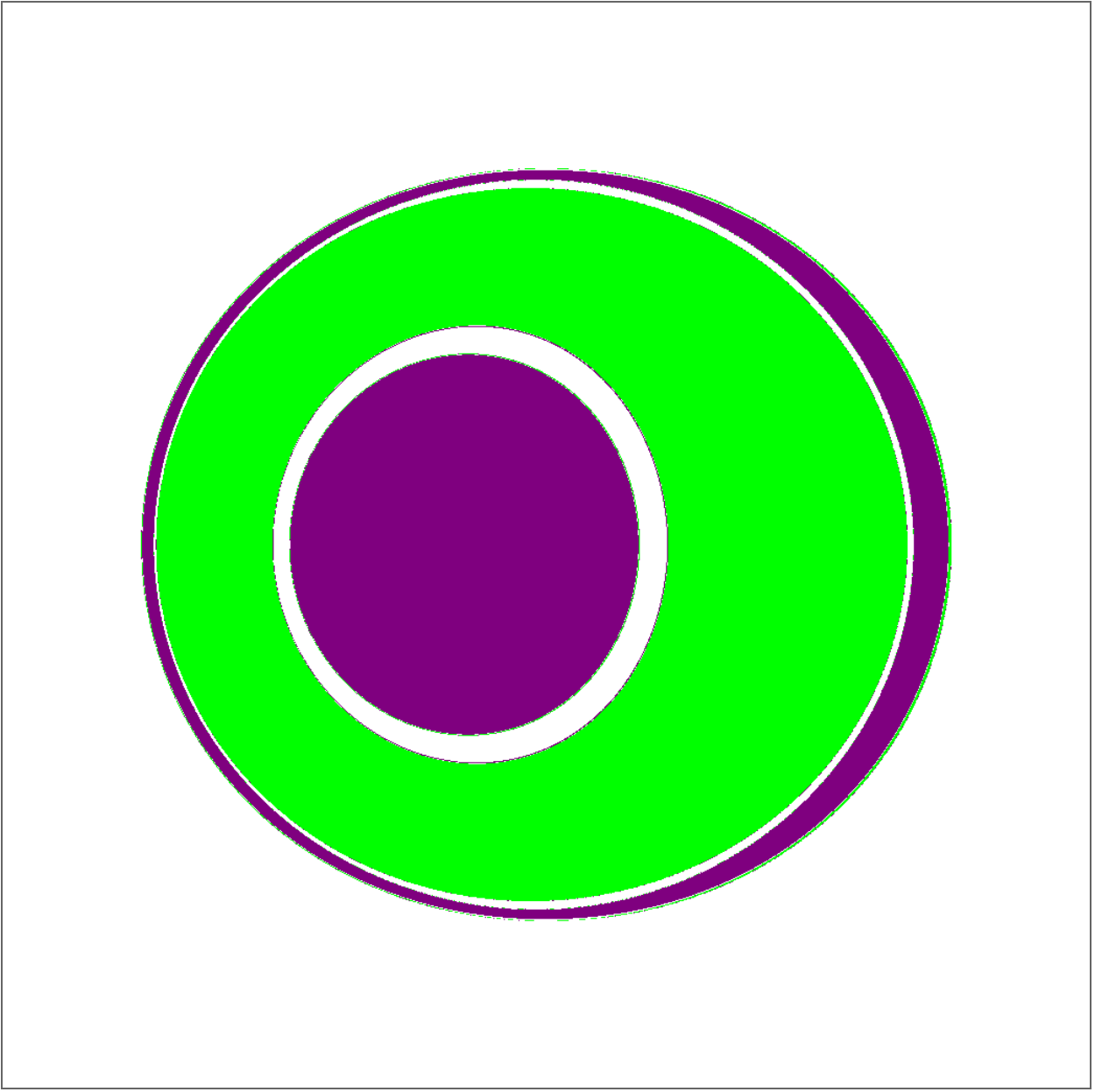}\label{fig:mp_shadows_gallery_40}}
\subfigure[$\theta = 50^{\circ}$]{\includegraphics[width=0.19\textwidth]{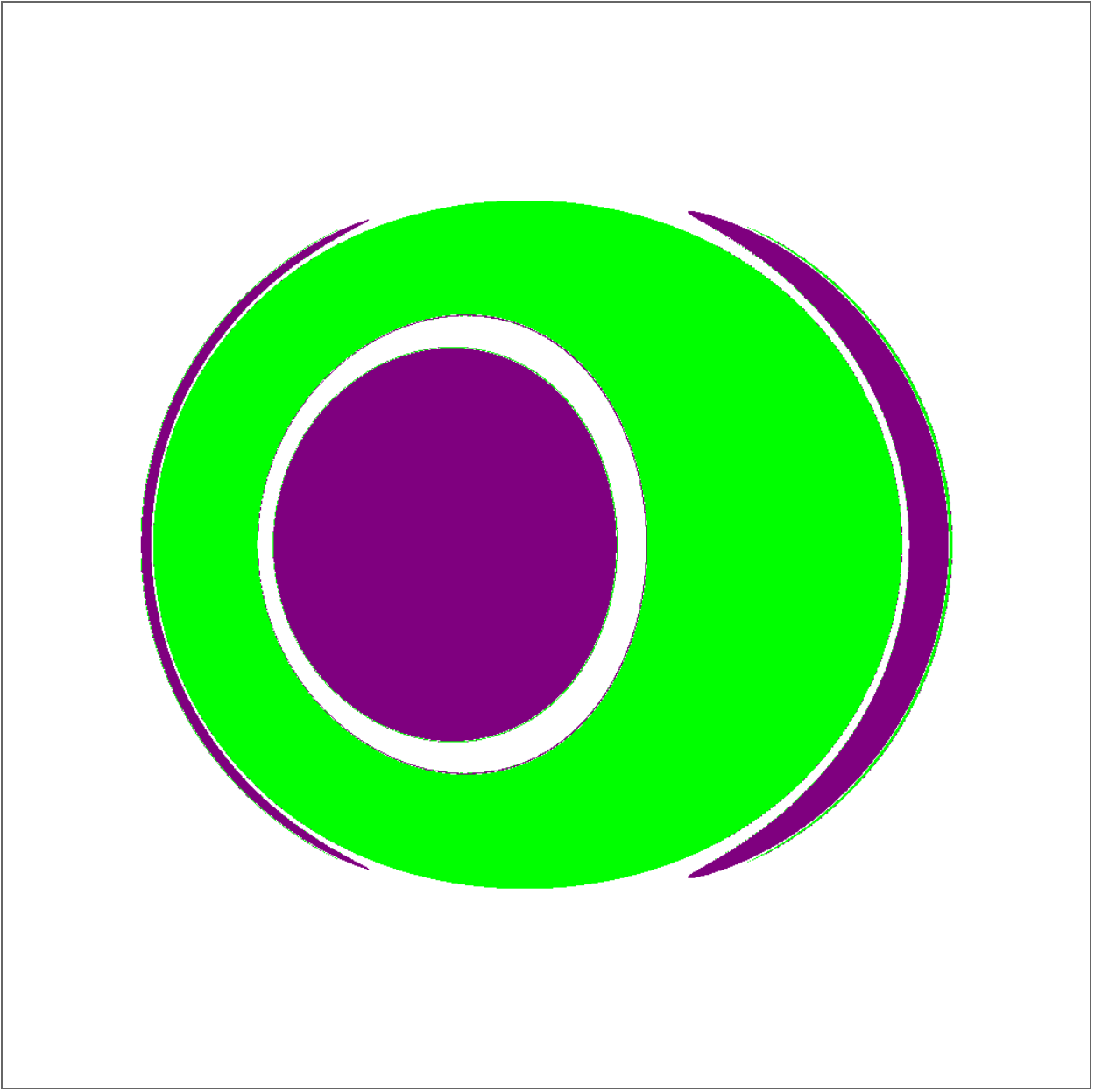}\label{fig:mp_shadows_gallery_50}}
\subfigure[$\theta = 60^{\circ}$]{\includegraphics[width=0.19\textwidth]{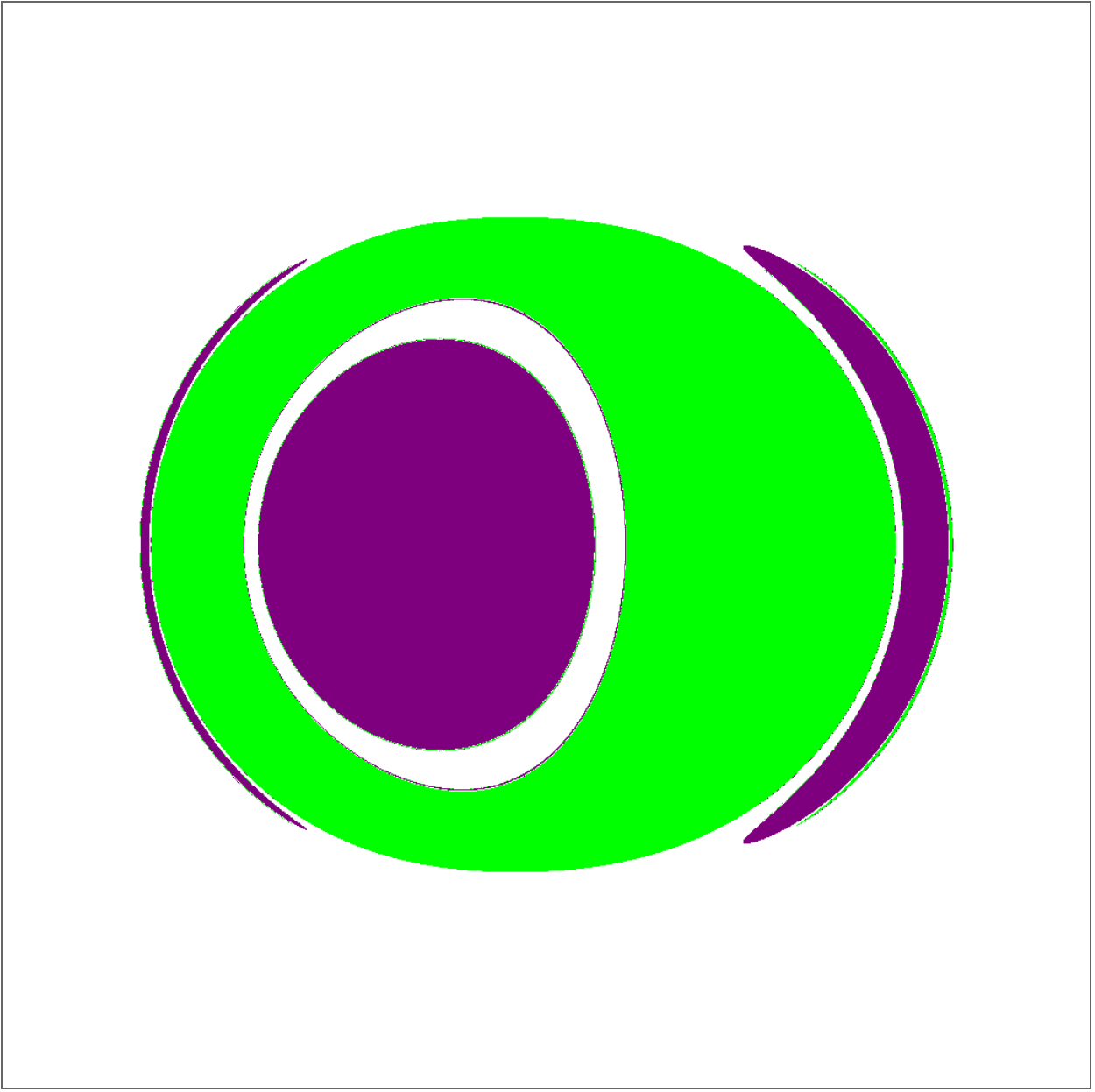}\label{fig:mp_shadows_gallery_60}}
\subfigure[$\theta = 70^{\circ}$]{\includegraphics[width=0.19\textwidth]{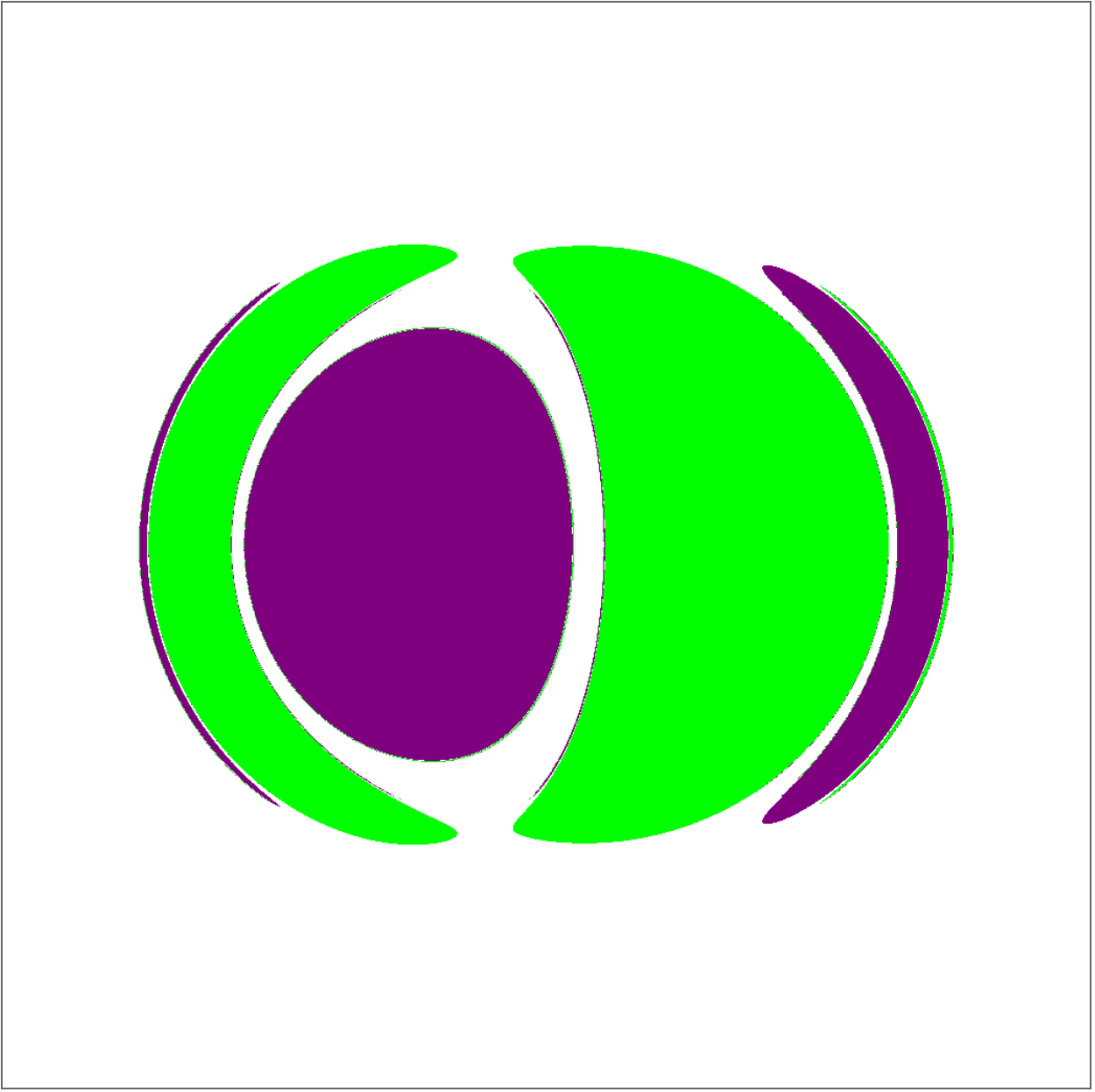}\label{fig:mp_shadows_gallery_70}}
\subfigure[$\theta = 80^{\circ}$]{\includegraphics[width=0.19\textwidth]{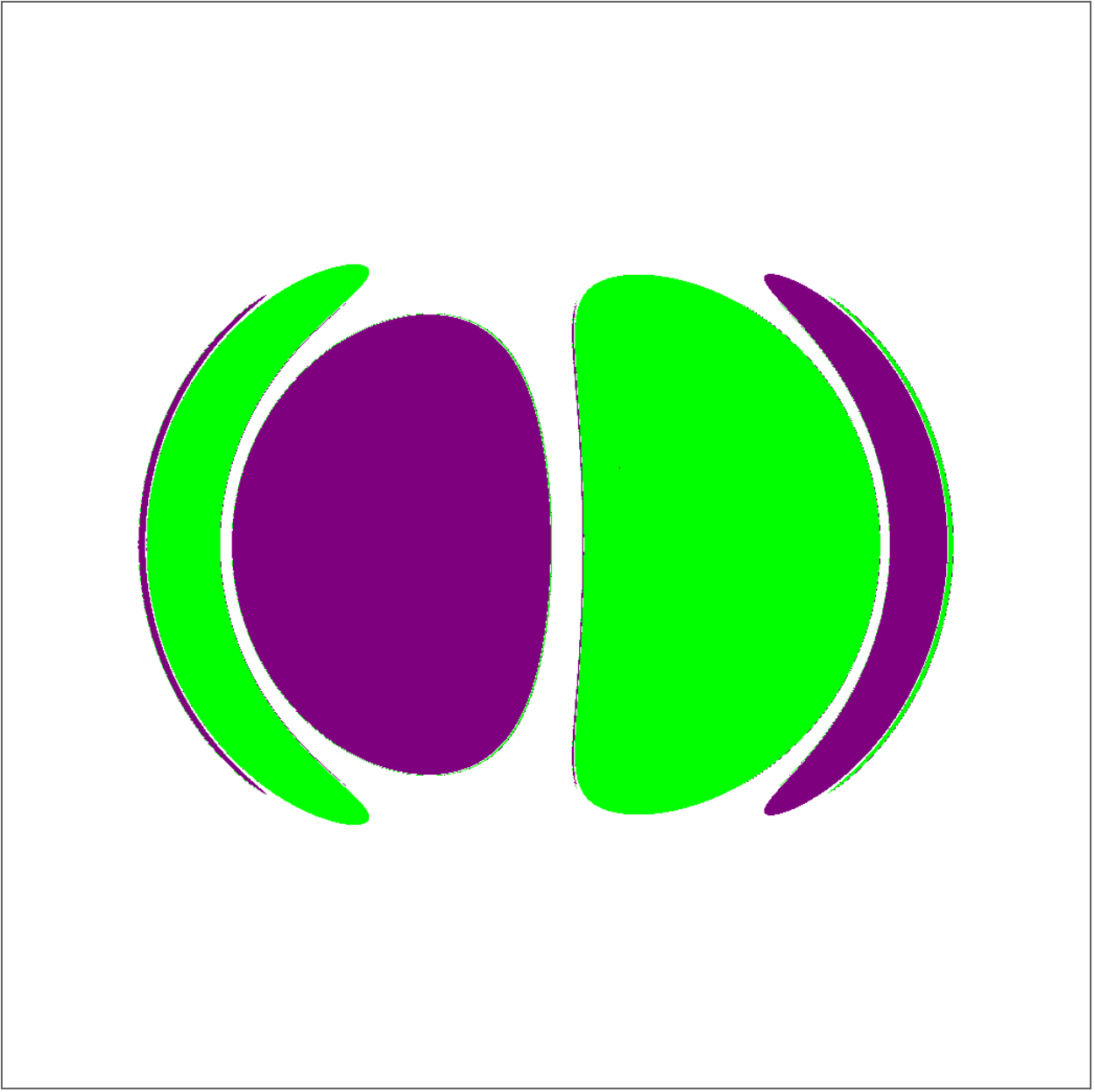}\label{fig:mp_shadows_gallery_80}}
\subfigure[$\theta = 90^{\circ}$]{\includegraphics[width=0.19\textwidth]{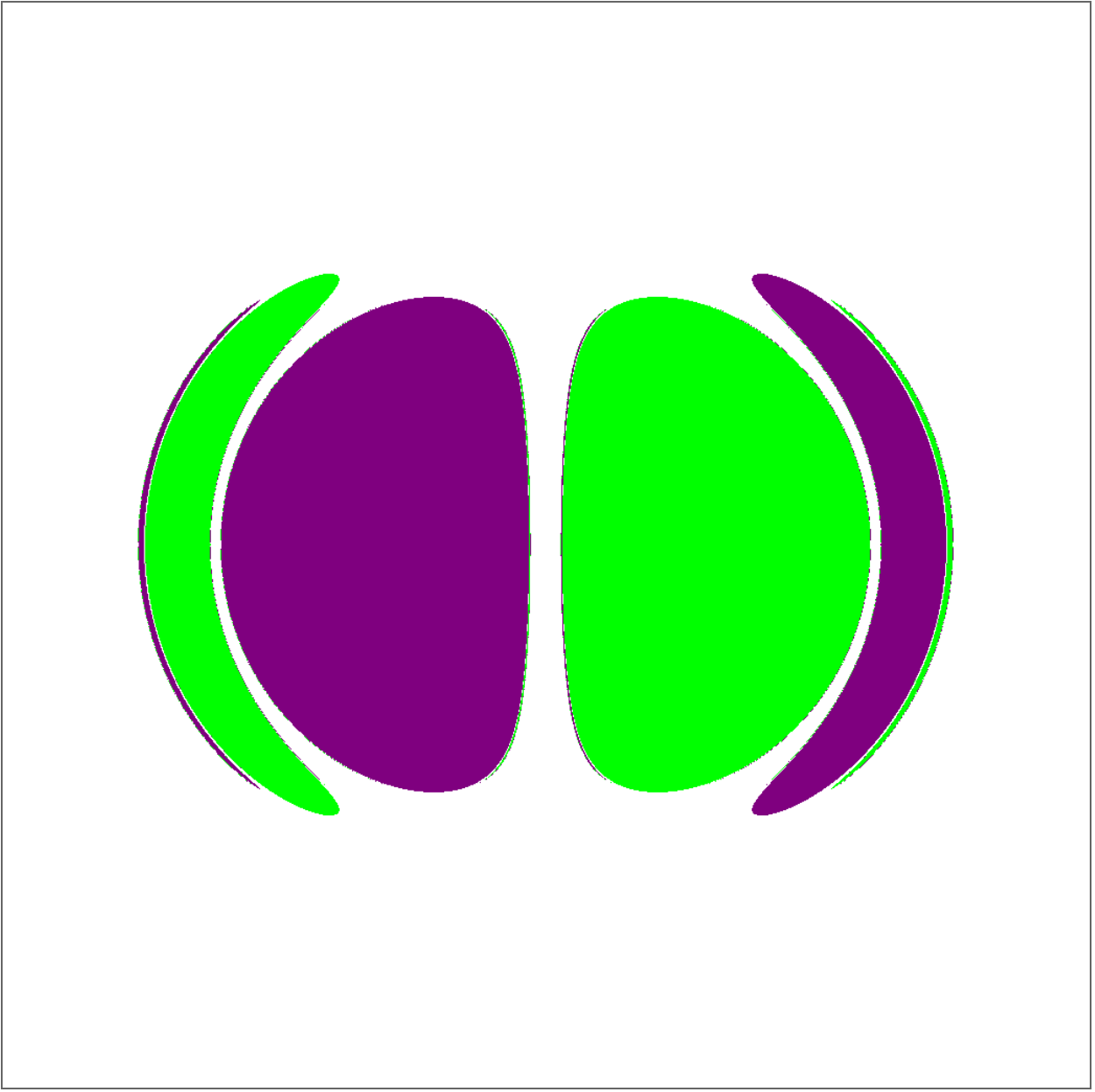}\label{fig:mp_shadows_gallery_90}}
\end{center}
\caption{Gallery of two-dimensional binary black hole shadows for the equal-mass ($M_{\pm} = M$) Majumdar--Papapetrou di-hole with coordinate separation $d = 2 M$. Initial conditions corresponding to rays which plunge into the upper (lower) black hole are presented in green (purple). \label{fig:mp_shadows_gallery}}
\end{figure}

In Figure \ref{fig:mp_shadows_gallery}, we present a gallery of two-dimensional binary black hole shadows cast by the equal-mass ($M_{\pm} = M = 1$) Majumdar--Papapetrou di-hole system with coordinate separation $d = 2$, for a selection of values of the angle of incidence $\theta$. The shadows are realised by using the ray-tracing approach described above. Pixels which belong to the shadow of the upper (lower) black hole are coloured green (purple); those which correspond to spatial infinity are coloured white. We observe that, as anticipated from the simulations by Bohn \emph{et al.} \cite{BohnThroweHebertEtAl2015}, the binary shadow is \emph{not} simply the superposition of two singleton black hole shadows. Rather, each black hole has a primary shadow -- either topologically equivalent to a disc or an annulus -- surrounded by a hierarchy of disconnected subsidiary components.

Let us consider the effect of increasing $\theta$ in Figure \ref{fig:mp_shadows_gallery}. For $0^{\circ} \leq \theta \leq 40^{\circ}$, the shadow is distorted but remains qualitatively similar to the $\theta = 0^{\circ}$ case. When $\theta \sim 50^{\circ}$, we clearly see eyebrow-like features \cite{BohnThroweHebertEtAl2015} for the first time: the secondary annular shadow corresponding to the lower black hole (depicted in purple) has fragmented into two disconnected arc-like components, as shown in Figure \ref{fig:mp_shadows_gallery_50}. When we reach $\theta \sim 70^{\circ}$, the primary annular shadow corresponding to the upper black hole (plotted in green) splits into two disconnected components -- one of which forms the primary globular shadow; the other forms a secondary eyebrow -- as shown in Figure \ref{fig:mp_shadows_gallery_70}.

When $\theta = 90^{\circ}$, the observer has a ``side-on'' view of the black holes. In this case, we observe two primary shadows of equal size and shape. The primary shadow is surrounded by a hierarchy of self-similar eyebrows, which have a Cantor-like structure. This case was shown in Figure 3(b) of \cite{HananRadu2007} and Figure 2 of \cite{YumotoNittaChibaEtAl2012}.

One could form a ``flip book'' animation using Figures \ref{fig:mp_shadows_gallery_00}--\ref{fig:mp_shadows_gallery_90}. This would mimic the effect of an observer passing by a pair of fixed black holes; or a static observer viewing a pair of rigidly rotating black holes.

%

\subsubsection{On-axis case}

\begin{figure}
\begin{center}
\begin{tabular}{c c}
{
\begin{tabular}{c}
\subfigure[Rays]{\includegraphics[width=0.42\textwidth]{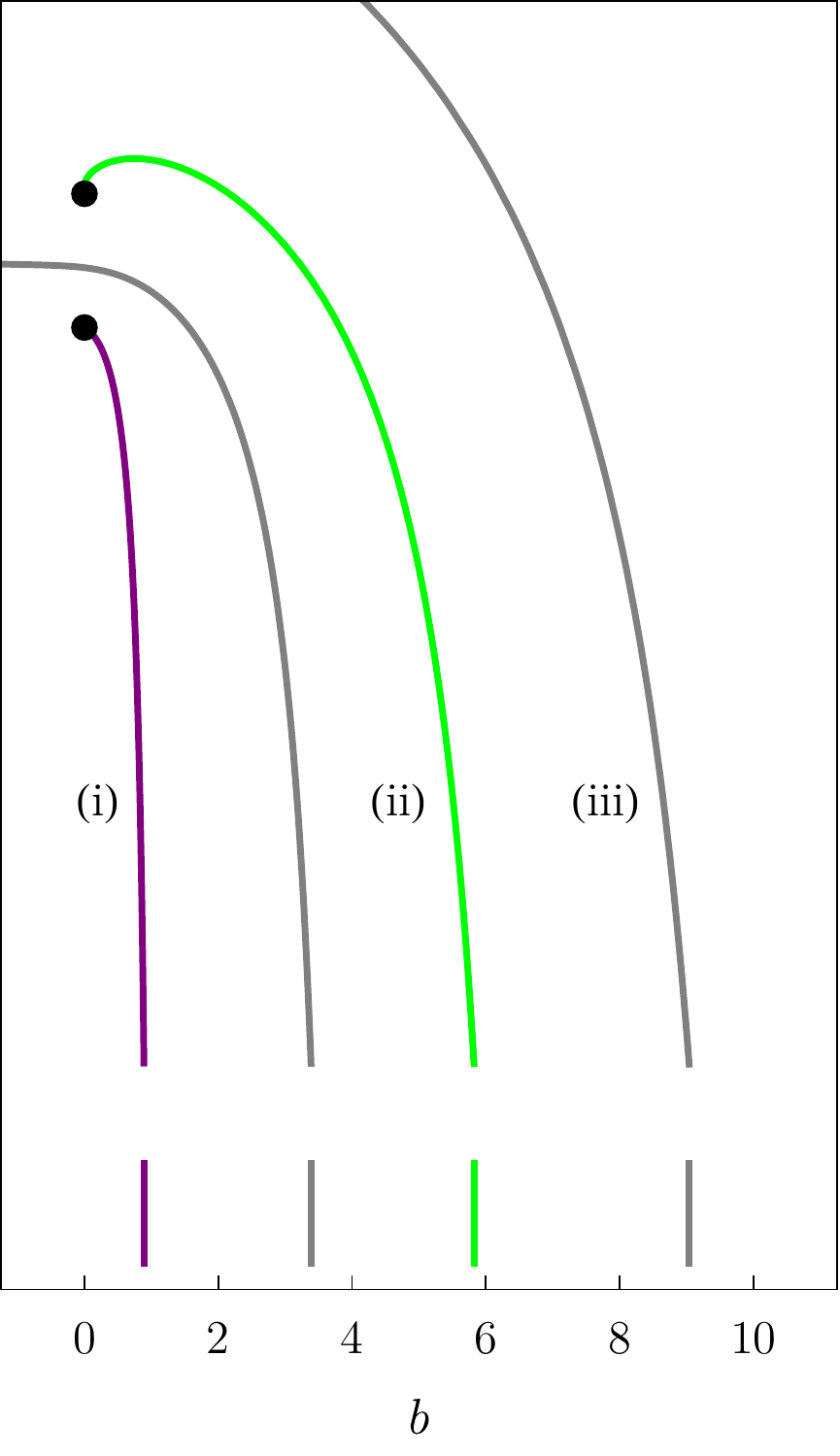} \label{fig:mp_impact_parameter_rays}}
\end{tabular}
}
&
{
\begin{tabular}{c}
{
\subfigure[Two-dimensional shadow]{\includegraphics[width=0.45\textwidth]{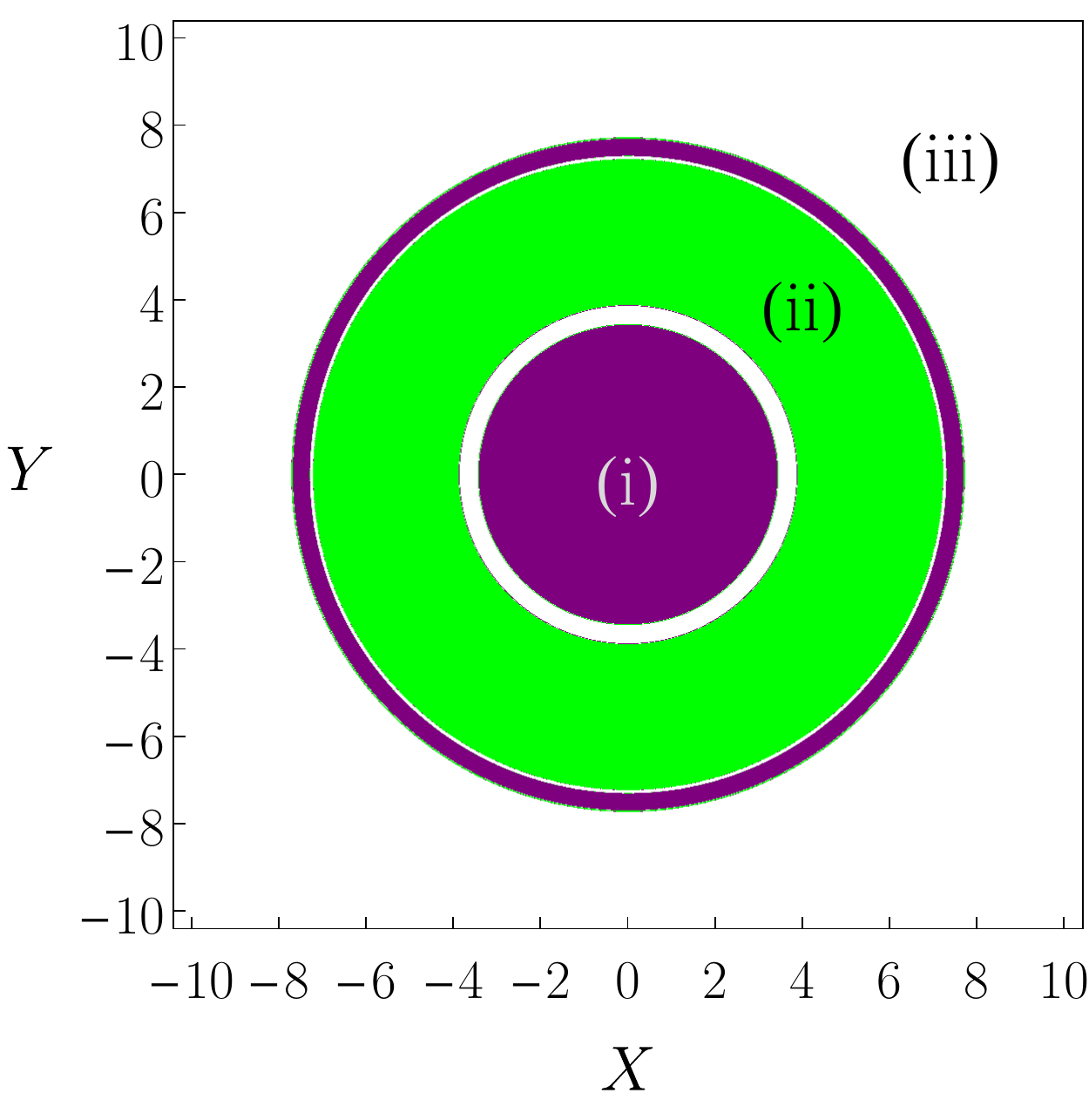} \label{fig:mp_shadow_d2_theta_00_labelled}}
}
\\
{\vspace{-1em}}
\\
{
\subfigure[One-dimensional shadow]{\includegraphics[width=0.45\textwidth]{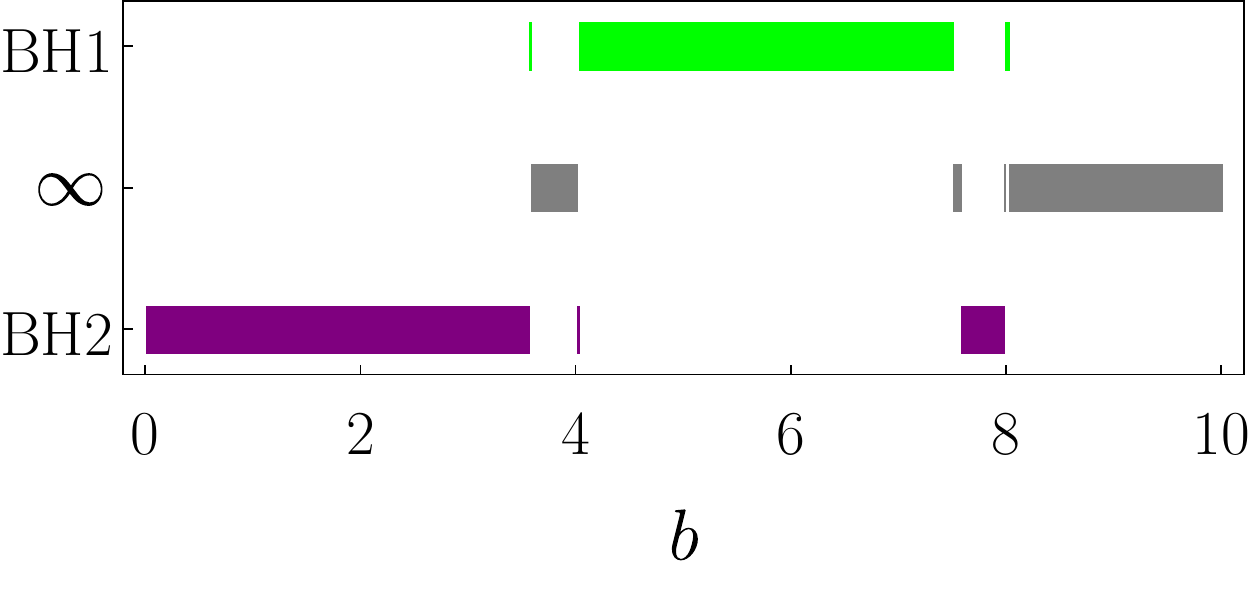} \label{fig:mp_impact_parameter_shadow}}
}
\end{tabular}
}
\end{tabular}
\end{center}
\caption{On-axis shadow and the impact parameter. (a) Null rays on an initial data surface labelled by the impact parameter $b$. (b) Two-dimensional shadow viewed by an observer on the symmetry axis. (c) One-dimensional shadow for rays with impact parameter $b$. Rays can (i) fall directly into the lower black hole [purple]; (ii) plunge directly into the upper black hole [green]; (iii) escape to infinity [grey/white]. In between these possibilities, we observe rich dynamics which results in self-similar shadows. \label{fig:mp_impact_parameter}}
\end{figure}

Let us consider the $\theta = 0^{\circ}$ case, shown in Figure \ref{fig:mp_shadows_gallery_00}, in more detail. This depicts the Majumdar--Papapetrou di-hole shadow as seen by an observer located on the symmetry axis. Figure \ref{fig:mp_impact_parameter_rays} shows a one-parameter scattering problem, in which rays are fired towards the black holes from infinity, initially parallel to the symmetry axis, with impact parameter $b$. In Figure \ref{fig:mp_shadow_d2_theta_00_labelled}, we show the two-dimensional shadow in the image plane. By symmetry, this shadow may be constructed from the ``area of revolution'' of the one-dimensional shadow of Figure \ref{fig:mp_impact_parameter_shadow}, which depicts the fate of the rays as a function of the \emph{impact parameter} $b$, which is the perpendicular distance from the initial position of the ray to the symmetry axis.

Let us analyse the crude features of the shadow as $b$ is increased from zero. For $b \sim 0$, all rays plunge directly into the lower black hole. Such rays are responsible for the primary disc-shaped shadow of the lower black hole, depicted in purple. Then, there is an interval in initial data corresponding to rays which pass between the two black holes and escape to infinity. These rays correspond to the ``gap'' between the primary shadows. Then, there is an open interval in $b$ corresponding to rays which plunge directly into the upper black hole. This open interval in initial data corresponds to the primary annular shadow of the upper black hole, shown in green. Finally, for sufficiently large values of $b$, all rays escape to infinity.

This crude analysis overlooks the self-similar features of the on-axis shadow. A more careful analysis may be performed with the aid of symbolic dynamics, introduced in Section \ref{sec:symbolic_dynamics_mp}. The outer edge of the primary shadow of the lower black hole (i.e., the purple disc in the centre of the image) corresponds to the fundamental orbit with decision dynamics representation $\overline{4}$. Similarly, the outer edge of the shadow corresponds to the fundamental orbit which is represented in decision dynamics by the sequence $\overline{0}$. As described in Section \ref{sec:cantor_like_structure}, there exists an uncountable infinity of impact parameters $b_{\mathcal{X}}$ which correspond to the non-escaping perpetual orbits. Here, $\mathcal{X}$ is a sequence in decision dynamics of infinite length which does not contain the digits $1$ or $3$. The ordering of $b_{\mathcal{X}}$ in initial data is determined by the ordering function $F$, defined in Section \ref{sec:ordering_perpetual_orbits}; however, the ordering is now reversed, so we begin with parity $P = - 1$ in the parity-reordering operation $\mathcal{P}$. The one-dimensional shadow on initial data $b$ may be constructed iteratively using symbolic dynamics, as in Section \ref{sec:construction_cantor_like_set}. Hence, the one-dimensional shadow has a Cantor-like structure which is inherited by the two-dimensional shadow due to the symmetry of the on-axis scattering problem.

\begin{figure}
\begin{center}
\begin{tabular}{c c}
\subfigure[Two-dimensional shadow ($d = 2$)]{
\begin{tabular}{c}
\includegraphics[width=0.44\textwidth]{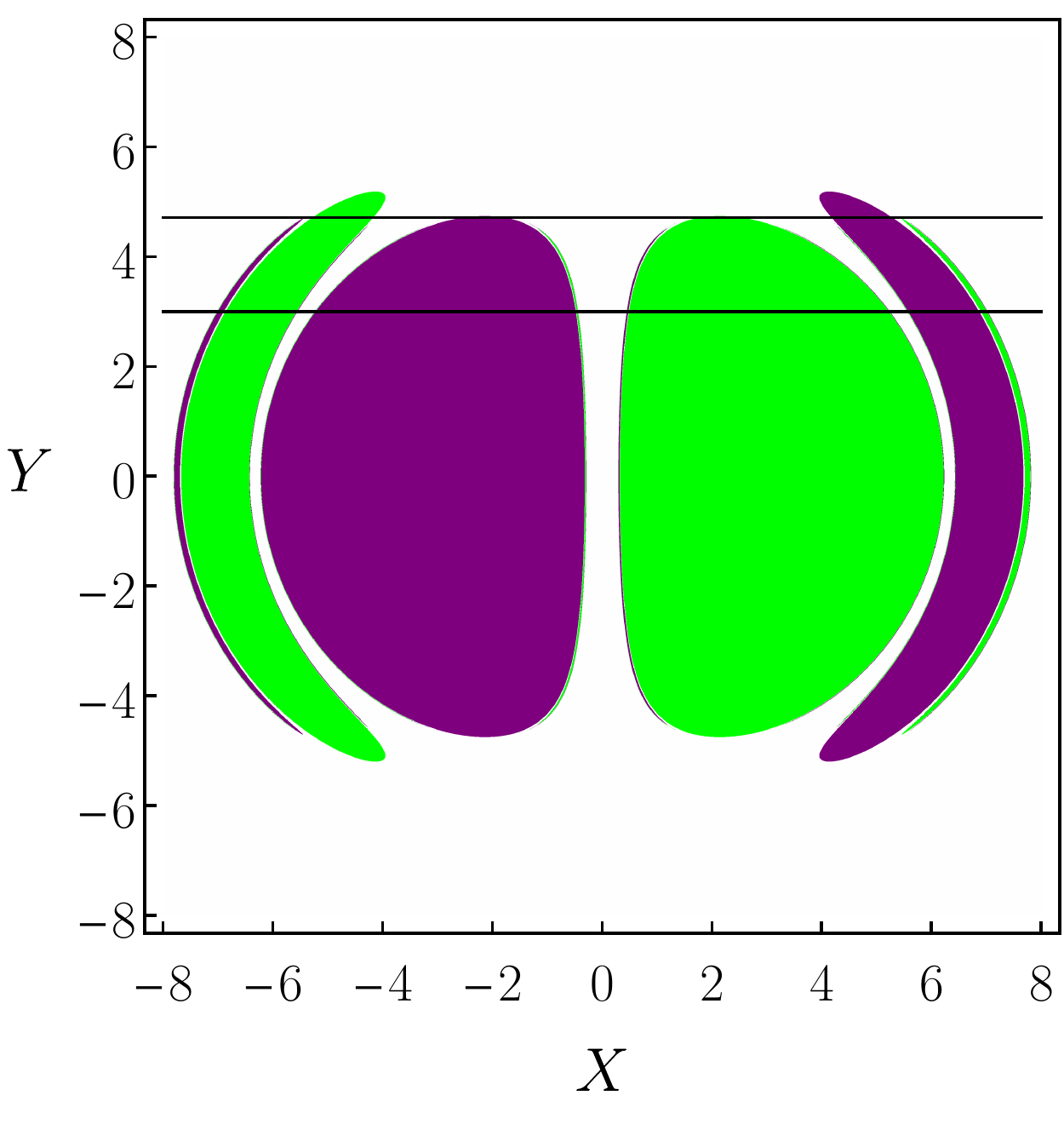}
\label{fig:mp_shadow_structure_d2}
\end{tabular}
}
&
\begin{tabular}{c}
{\vspace{-0.5em}}
\\
{\subfigure[Regular shadow ($d = 2$)]{
\includegraphics[width=0.44\textwidth]{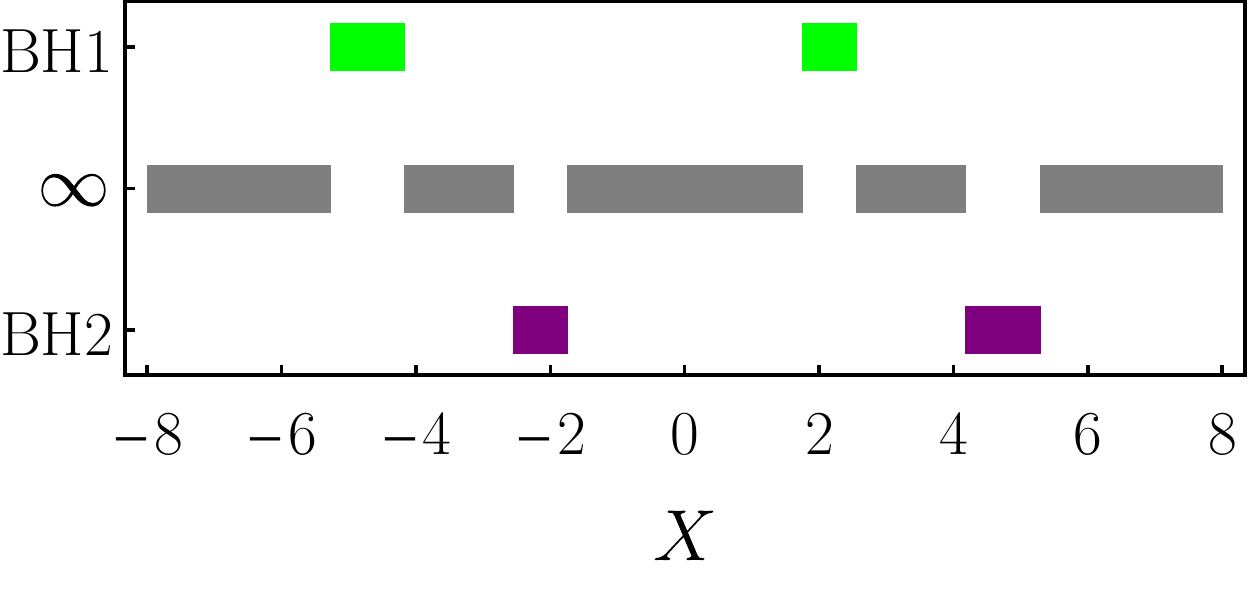}
\label{fig:mp_one_dim_shadow_d2_regular}}}
\\
{\subfigure[Cantor-like shadow ($d = 2$)]{
\includegraphics[width=0.44\textwidth]{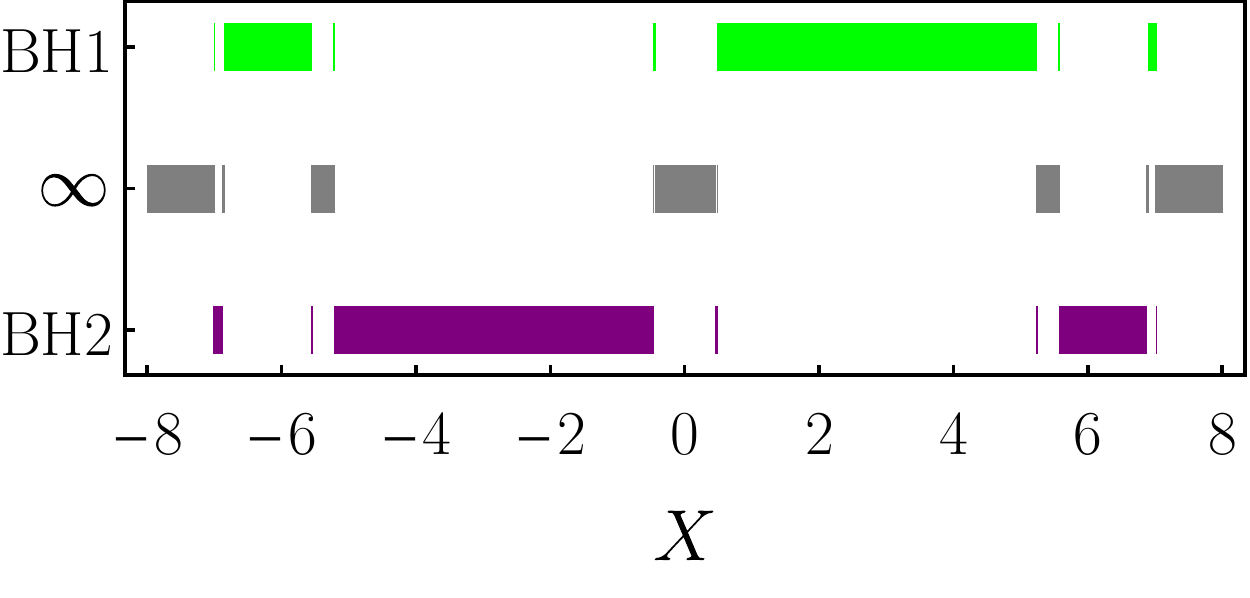}
\label{fig:mp_one_dim_shadow_d2_cantor}}}
\end{tabular}
\\
\subfigure[Two-dimensional shadow ($d = 1$)]{
\begin{tabular}{c}
\includegraphics[width=0.44\textwidth]{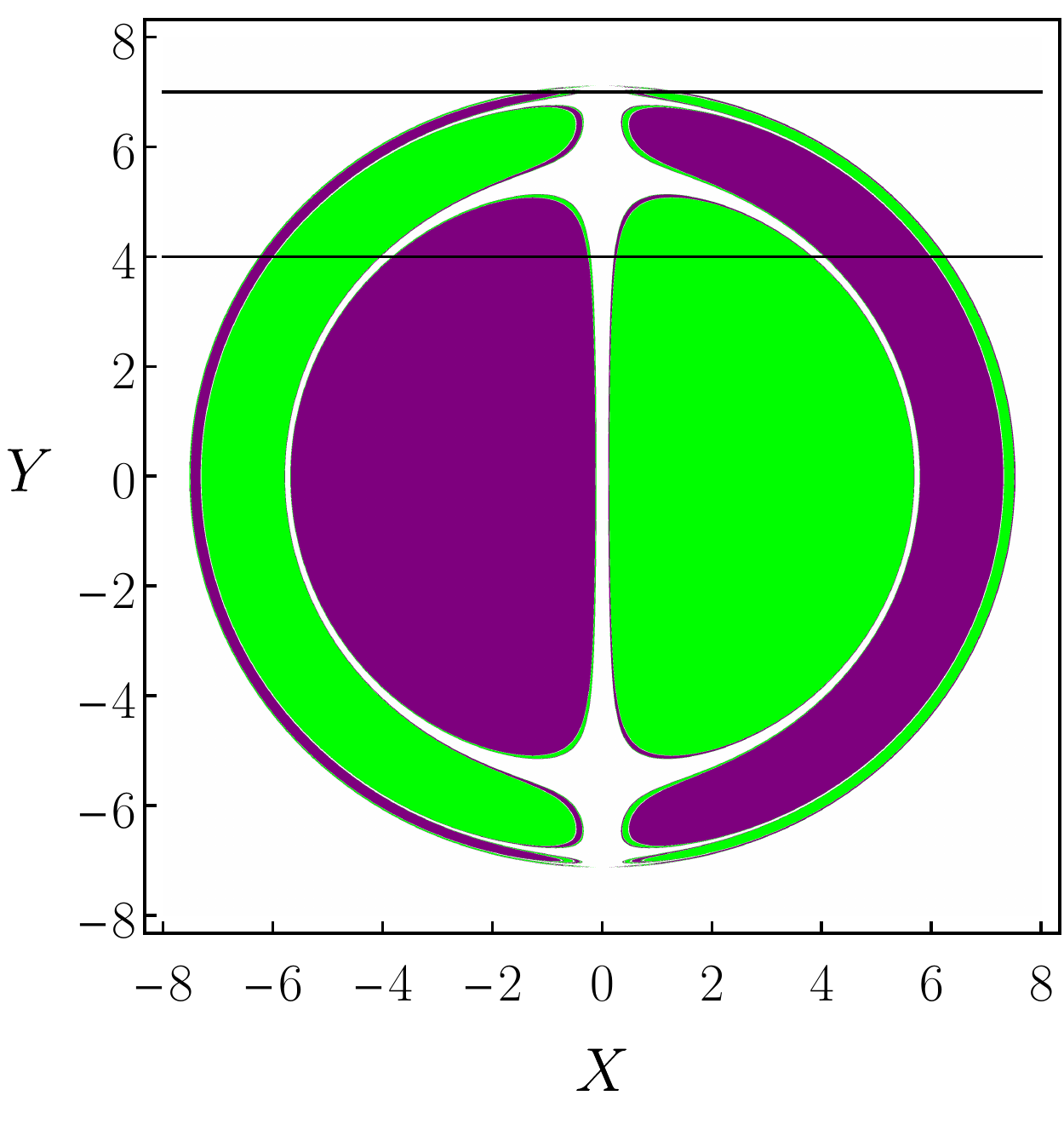}
\label{fig:mp_shadow_structure_d1}
\end{tabular}
}
&
\begin{tabular}{c}
{\vspace{-0.5em}}
\\
{\subfigure[``Highly fractal'' shadow ($d = 1$)]{
\includegraphics[width=0.44\textwidth]{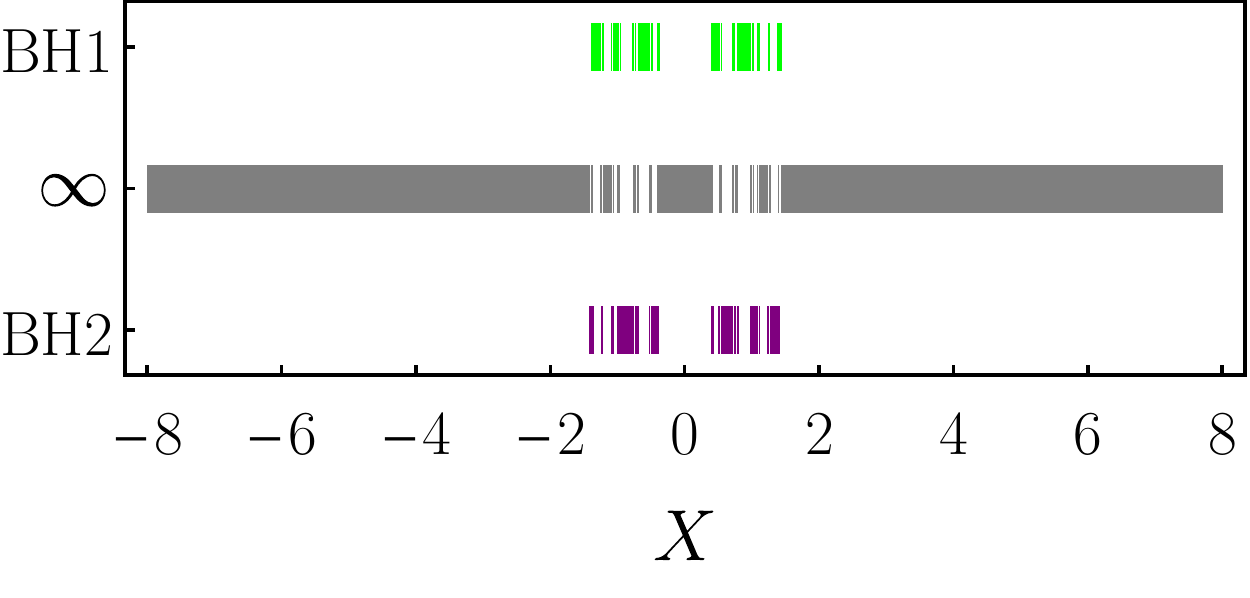}
\label{fig:mp_one_dim_shadow_d1_chaotic}}}
\\
{\subfigure[Cantor-like shadow ($d = 1$)]{
\includegraphics[width=0.44\textwidth]{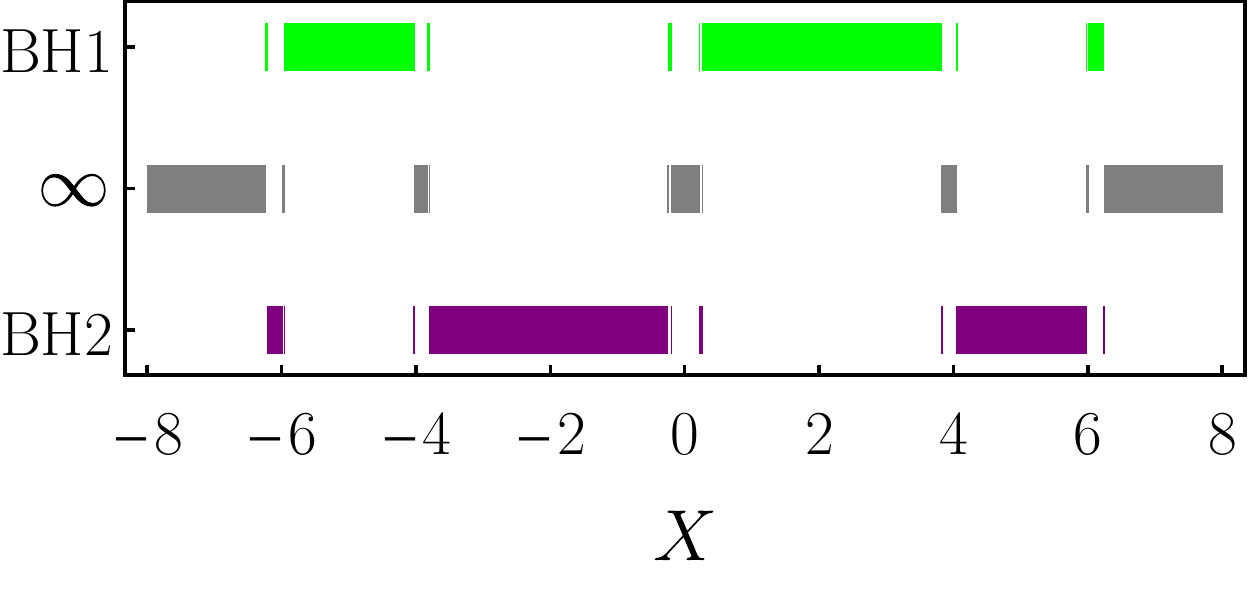}
\label{fig:mp_one_dim_shadow_d1_cantor}}}
\end{tabular}
\end{tabular}
\end{center}
\caption{Fractal structure of the two-dimensional shadows of the equal-mass ($M_{\pm} = M = 1$) Majumdar--Papapetrou di-hole, shown in the $(X, Y)$-plane. The two-dimensional shadow can be viewed as the union of one-dimensional shadows for a fixed value of $p\ind{_{\phi}}$ (which corresponds to a fixed value of $Y$). (a) Two-dimensional shadow image for $d = 2$. (b) One-dimensional slice with $d = 2$ at $Y = 4.7$, which exhibits regular structure due to the non-existence of the $\overline{0}$ and $\overline{2}$ fundamental orbits. (c) Cantor-like one-dimensional slice at $Y = 3$ for $d = 2$. (d) Two-dimensional shadow for $d = 1$. (e) One-dimensional shadow for $d = 1$ with $Y = 7$ which exhibits ``highly fractalised'' structure, due to the existence of a scattering region with three narrow ``throats''; see Figure \ref{fig:mp_narrow_escapes_trajectories}. (f) Cantor-like one-dimensional slice at $Y = 4$ for $d = 1$. \label{fig:mp_shadow_decomposition}}
\end{figure}
%

\subsubsection{Side-on case}

We now turn our attention to the case $\theta = 90^{\circ}$, in which an observer views the system from ``side-on'', and the centre of the image plane is in the equatorial ($z = 0$) plane. In this case, the two-dimensional black hole shadow image, with image plane coordinates $(X, Y)$, can be viewed as a set of one-dimensional shadows, each given by a fixed value of the conserved azimuthal angular momentum $p\ind{_{\phi}}$. To see this, consider the change of spacetime coordinates $x\ind{^{a}} \mapsto x\ind{^{a^{\prime}}}$. Under such a transformation, the momenta transform according to $p\ind{_{a}} \mapsto p\ind{_{a^{\prime}}} = \frac{\partial x\ind{^{a}}}{\partial x\ind{^{a^{\prime}}}} p \ind{_{a}}$. The standard relationship between Cartesian coordinates $\{x, y, z\}$ and cylindrical polar coordinates $\{\rho, z, \phi \}$ then yields the relationship $p\ind{_{\phi}} = x p\ind{_{y}} - y p\ind{_{x}}$. A photon with initial momentum normal to the image plane has $p\ind{_{y}}(0) = 0 = p\ind{_{z}}(0)$ and $p\ind{_{x}}(0) = -U_{0}^{2}$, by the Hamiltonian constraint. Hence, the azimuthal angular momentum can be related to the image plane coordinates via $p\ind{_{\phi}} = Y U_{0}^{2}$. Moreover, for an observer in the far-field ($r_{0} \rightarrow \infty$), we have $U_{0} \rightarrow 1$. Hence, a scattering problem with $p\ind{_{\phi}} = \text{constant}$ admits a one-dimensional shadow, which corresponds to a curve $Y = \text{constant}$ in the two-dimensional shadow image. This is shown in Figure \ref{fig:mp_shadow_decomposition}.

Consider first the case $d = 2$, shown in Figures \ref{fig:mp_shadow_structure_d2}--\ref{fig:mp_one_dim_shadow_d2_cantor}. One-dimensional slices corresponding to sufficiently small values of $p\ind{_{\phi}}$ exhibit self-similar Cantor-like structure, as shown in Figure \ref{fig:mp_one_dim_shadow_d2_cantor}. However, for large values of $p\ind{_{\phi}}$, the one-dimensional slices (e.g.~close to the top of the shadow) do \emph{not} exhibit fractal structure. This qualitative difference in shadow structure was anticipated in the analysis of the non-planar fundamental photon orbits in Section \ref{sec:non_planar_rays}. In Figure \ref{fig:mp_fundamental_orbits_p_phi_590}, we show a value of the angular momentum $p\ind{_{\phi}}$ for which the fundamental orbits $\overline{0}$ and $\overline{2}$ are forbidden, but for which the pair of fundamental orbits with representation $\overline{4}$ exist, allowing absorption by the black holes. In this regime, dynamical transitions between fundamental orbits are not possible, and chaotic scattering does not occur; however, the shadow is non-empty because absorption is permitted. As a result, the one-dimensional shadows will be regular (not Cantor-like), as illustrated in Figure \ref{fig:mp_one_dim_shadow_d2_regular}.

In the case $d = 1$, the one-dimensional scattering problem for values of $p\ind{_{\phi}}$ close to ${p\ind{_{\phi}}}^{\ast}$ is associated with the existence of a ``pocket'' feature with three narrow escape channels; see Figure \ref{fig:mp_narrow_escapes_trajectories}. The pocket acts as a ``randomising'' region, and the orbits which enter are highly chaotic. The resulting one-dimensional shadows are therefore ``highly fractalised'', as can be seen in Figure \ref{fig:mp_one_dim_shadow_d1_chaotic}.

The analysis carried out in this section demonstrates that there is clearly a qualitative difference between the shadows for $d = 1$ and $d = 2$. This could be anticipated from the analysis of the effective potential $h$; see Section \ref{sec:non_planar_rays}, in particular Figures \ref{fig:mp_height_function_contours_d} and \ref{fig:mp_critical_contours}. In particular, we observed a ``phase change'' in the behaviour of null geodesics as the separation was varied. This will be explored in more depth using quantitative methods in Chapter \ref{chap:fractal_structures}.

\section{Following rays through the event horizons}
\label{sec:through_the_event_horizons}

In this section, we consider the consequences of following null geodesics \emph{through} the event horizons in the maximally extended Majumdar--Papapetrou di-hole spacetime \cite{HartleHawking1972}. As we shall show, following geodesics in this way results in richer chaotic phenomena.

We recall that the ``points'' $(x, y, z) = (0, 0, z_{\pm})$ are \emph{not} curvature singularities but coordinate singularities at which $U \rightarrow \infty$ \cite{HartleHawking1972}. At these points, the coordinate time $t$ diverges ($t \rightarrow \infty$); however, the affine parameter for each geodesic remains finite, as do curvature invariants such as the Kretschmann scalar $R\ind{_{a b c d}} R\ind{^{a b c d}}$. (The ``points'' at which $U \rightarrow \infty$ are actually null \emph{surfaces} of finite area with topology $\mathbb{R} \times S^{2}$, which correspond to the \emph{event horizons} of the two black holes.)

\subsection{Extremal Reissner--Nordstr\"{o}m case}

Before considering the behaviour of null geodesics which pass through the black hole event horizons in the Majumdar--Papapetrou di-hole spacetime, it will be beneficial to review the singleton black hole case, i.e., the (maximally extended) extremal Reissner--Nordstr\"{o}m spacetime. The spacetime metric in isotropic spherical polar coordinates $\{ t, r, \theta, \phi \}$ takes the form
\begin{equation}
\label{eqn:extremal_rn_metric_isotropic}
\ed s^{2} = - \frac{\ed t^{2}}{U^{2}} + U^{2} \left( \ed r^{2} + r^{2} \, \ed \Omega^{2} \right), \qquad
U(r) = 1 + \frac{M}{r},
\end{equation}
where $\ed \Omega^{2} = \ed \theta^{2} + \sin^{2}{\theta} \, \ed \phi^{2}$ is the line element on the unit two-sphere. The point $r = 0$ in isotropic coordinates corresponds to a null surface of finite area with topology $\mathbb{R} \times S^{2}$ -- the event horizon of the black hole.

The metric \eqref{eqn:extremal_rn_metric_isotropic} may be transformed into a more familiar form via the change of coordinates $\hat{r} = r + M$, so that the line element in standard (Schwarzschild-type) coordinates $\{ t, \hat{r}, \theta, \phi \}$ reads
\begin{equation}
\label{eqn:extremal_rn_metric_schw_type}
\ed s^{2} = - f(\hat{r}) \, \ed t^{2} + \frac{\ed r^{2}}{f(\hat{r})} + \hat{r}^{2} \, \ed \Omega^{2}, \qquad
f(\hat{r}) = \left(1 - \frac{M}{\hat{r}} \right)^{2}.
\end{equation}
One may instead choose to replace the temporal coordinate $t$ with an ingoing ($-$) or outgoing ($+$) null coordinate $w_{\pm} = t \pm F(r)$, where $F^{\prime}(r) = U(r)$. Performing this coordinate transformation leads to the line element
\begin{equation}
\ed s^{2} = g\ind{_{a b}} \ed x\ind{^{a}} \ed x\ind{^{b}} = - \frac{\ed w_{\pm}^{2}}{U^{2}} \pm \ed w_{\pm} \ed r + U^{2} r^{2} \, \ed \Omega^{2}.
\end{equation}
These coordinates are similar to the familiar Eddington--Finkelstein coordinates \cite{Eddington1924, Finkelstein1958} for the Schwarzschild black hole. In the $\{ w_{\pm}, r, \theta, \phi \}$ coordinate system, the metric components $g\ind{^{a b}}$ are \emph{regular} at the horizon $r = 0$. One may therefore employ the Hamiltonian $H = \frac{1}{2} g\ind{^{a b}} p\ind{_{a}} p\ind{_{b}}$ with the ingoing (outgoing) null coordinate $w_{-}$ ($w_{+}$) to follow null rays into (out of) the black hole horizon.

\begin{figure}[h]
\begin{center}
\includegraphics[width=0.3\textwidth]{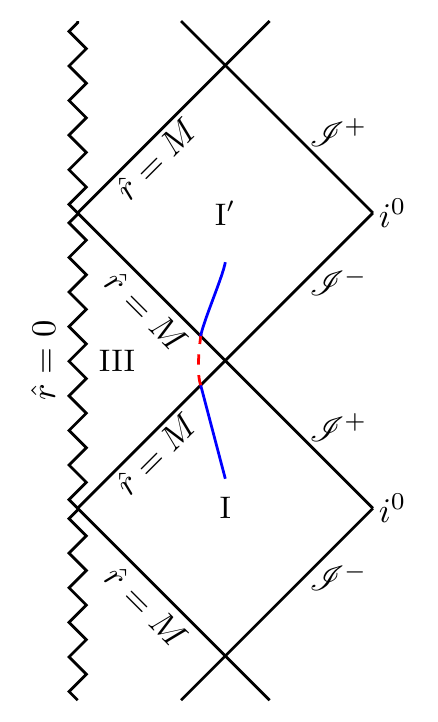}
\caption{Penrose--Carter diagram for the extremal Reissner--Nordstr\"{o}m black hole. The diagram includes an example of a null geodesic which is emitted by a source in Region I, then crosses the black hole horizon from Region I to Region III, before emerging from a white hole horizon in a new asymptotically flat spacetime, Region I$^{\prime}$. The part of the trajectory which is evolved using Eddington--Finkelstein-type coordinates is shown as a red dashed curve. The transition from ingoing to outgoing null coordinates occurs at the turning point, where $\frac{\ed}{\ed \lambda} \hat{r} = 0$.}
\label{fig:rn_penrose_carter}
\end{center}
\end{figure}

In Figure \ref{fig:rn_penrose_carter}, we present a Penrose--Carter diagram \cite{Penrose2011, Carter1966} which illustrates the causal structure of a (singleton) extremal Reissner--Nordstr\"{o}m black hole of mass $M$. In Schwarzschild-type coordinates \eqref{eqn:extremal_rn_metric_schw_type}, the black hole horizon is located at $\hat{r} = M$, and the curvature singularity is located at $\hat{r} = 0$. In isotropic coordinates \eqref{eqn:extremal_rn_metric_isotropic}, the event horizon is located at $r = 0$. Null geodesics may be evolved through the black hole horizon from Region I ($r > 0$) to Region III ($r < 0$) by switching to \emph{ingoing} Eddington--Finkelstein-type coordinates. All rays inside Region III which have non-zero angular momentum will avoid the timelike singularity, and hence will emerge through thew white hole horizon into a new asymptotically flat spacetime (Region I$^{\prime}$). An example of a ray which has been extended through the black hole horizon, and then through the white hole horizon into a new asymptotically flat spacetime is shown in Figure \ref{fig:rn_penrose_carter}.

\subsection{Majumdar--Papapetrou di-hole case}

In the case of the Majumdar--Papapetrou di-hole, one may track ingoing/outgoing null rays as follows. Suppose that we wish to follow a null geodesic through the event horizon of the lower black hole. We transform from standard isotropic coordinates to a spherical coordinate system $\{ t, r, \theta, \phi \}$ centred on the event horizon of the lower black hole $(x, y, z) = (0, 0, z_{-})$. In this coordinate system, the line element is given by
\begin{align}
\ed s^{2} &= - \frac{\ed t^{2}}{U^{2}} + U^{2} \left( \ed r^{2} + r^{2} \, \ed \Omega^{2} \right), &
U(r, \theta) &= 1 + \frac{M_{-}}{r} + \frac{M_{+}}{\sqrt{r^{2} - 2 d r \cos{\theta} + d^{2}}}.
\end{align}
We introduce the radial function $V(r) = 1 + \frac{M_{-}}{r} + \frac{M_{+}}{d}$ and the outgoing/ingoing null coordinate $w_{\pm} = t \pm F(r)$, such that $F^{\prime}(r) = V(r)$. In coordinates $\{ w_\pm, r, \theta, \phi \}$ the line element reads
\begin{equation}
\ed s^{2} = g\ind{_{a b}} \ed x\ind{^{a}} \ed x\ind{^{b}} = - \frac{\ed w_{\pm}^{2}}{U^{2}} \pm \frac{2 V^{2}}{U^{2}} \, \ed w_{\pm} \ed r + \frac{U^{4} - V^{4}}{U^{2}} \ed r^{2} + U^{2} r^{2} \, \ed \Omega^{2},
\end{equation}
and the corresponding Hamiltonian is given by
\begin{equation}
H = \frac{1}{2 U^{2}} \left[ - \left( U^{4} - V^{4} \right) {p\ind{_{w_{\pm}}}}^{2} \pm 2 V^{2} p\ind{_{w_{\pm}}} p\ind{_{r}} + {p\ind{_{r}}}^{2} + \frac{1}{r^{2}} {p\ind{_{\theta}}}^{2} + \frac{1}{ r^{2} \sin^{2}{\theta} } {p\ind{_{\phi}}}^{2}  \right]. \label{eqn:mp_hamiltonian_cross_horizon}
\end{equation}
The momenta are related by $p\ind{_{w_{\pm}}} = p\ind{_{t}} \pm V(r) p\ind{_{r}}$. We note here that
\begin{equation}
\lim_{r \rightarrow 0} \frac{U^{4} - V^{4}}{U^{2}} = \frac{4 M_{+} M_{-} \cos{\theta}}{d^{2}}, \qquad
\lim_{r \rightarrow 0} \frac{V^{2}}{U^{2}} = 1, \qquad
\lim_{r \rightarrow 0} \frac{1}{U^{2} r^{2}} = \frac{1}{M_{-}},
\end{equation}
so the Hamiltonian \eqref{eqn:mp_hamiltonian_cross_horizon} is non-singular as we approach the event horizon of the lower black hole, $r = 0$. We may therefore evolve Hamilton's equations through the coordinate singularity at $r = 0$, using the outgoing/ingoing null coordinate $w_{\pm}$.

In practice, we may follow a ray which begins in the exterior asymptotically flat region of the Majumdar--Papapetrou di-hole spacetime, until it reaches the vicinity of a black hole horizon. The null geodesic may then be evolved through the horizon (from $r > 0$ to $r < 0$) using a coordinate patch which is centred on the relevant black hole and a null ingoing coordinate $w_{-}$. All rays with non-zero angular momentum will avoid the timelike singularity; they will therefore reach a turning point at which $\dot{r} = 0$. At this point, one may switch from the ingoing to the outgoing null coordinate $w_{-} \mapsto w_{+}$, switching also the radial momenta ${p\ind{_{r}}}^{-} \mapsto {p\ind{_{r}}}^{+} = {p\ind{_{r}}}^{-} - 2 V^{2} p\ind{_{t}}$. One may track the null ray from the interior region to a new asymptotically flat exterior region (through a white hole horizon) from $r < 0$ to $r > 0$, before switching back to isotropic coordinates. This process may be repeated as necessary, for all rays which plunge into a black hole horizon and emerge through a white hole horizon.

\begin{figure}
\begin{center}
\begin{tabular}{c c}
\hfill
{\includegraphics[width=0.45\textwidth]{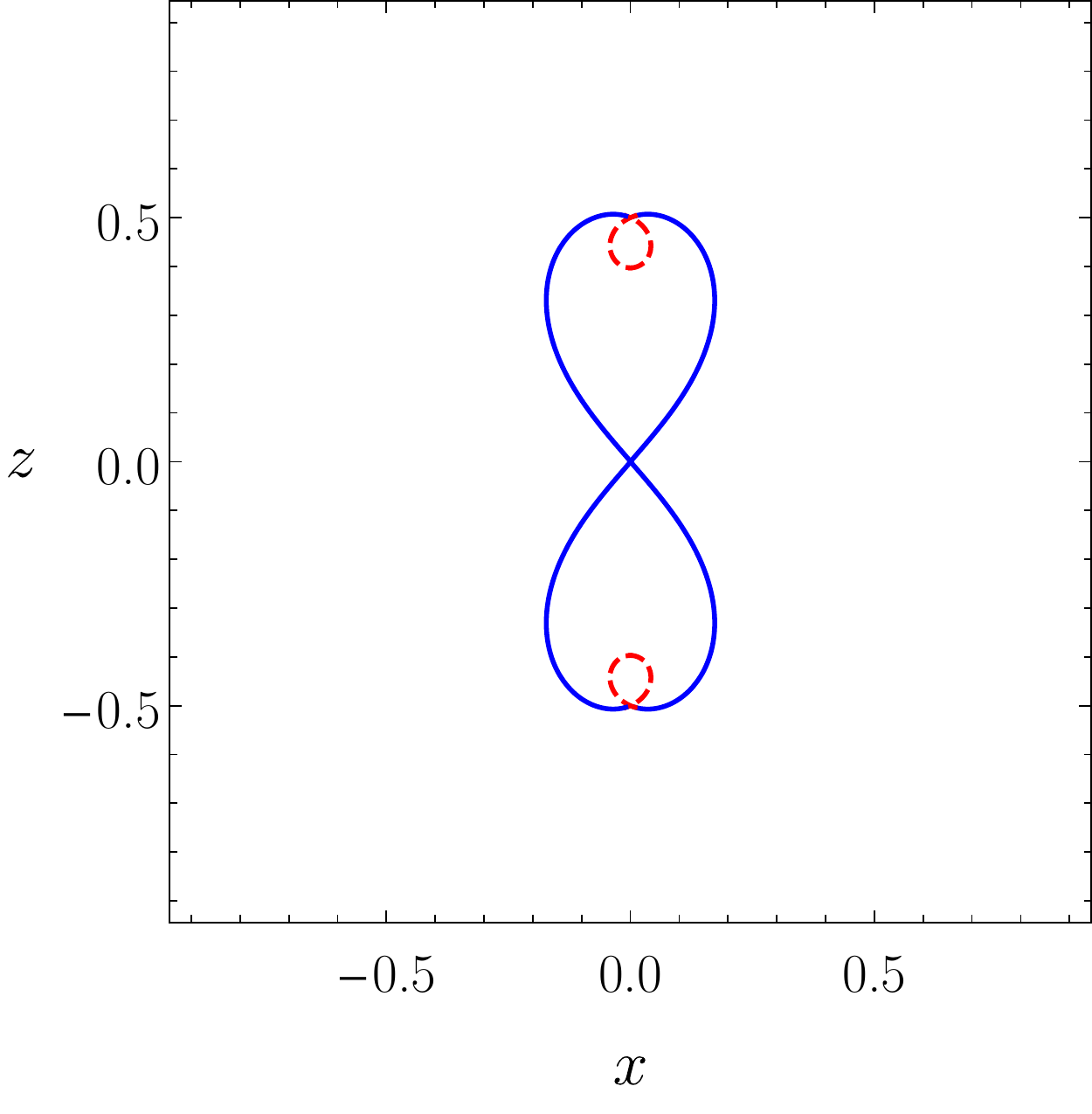}}
\hfill
&
\hfill
{\includegraphics[width=0.45\textwidth]{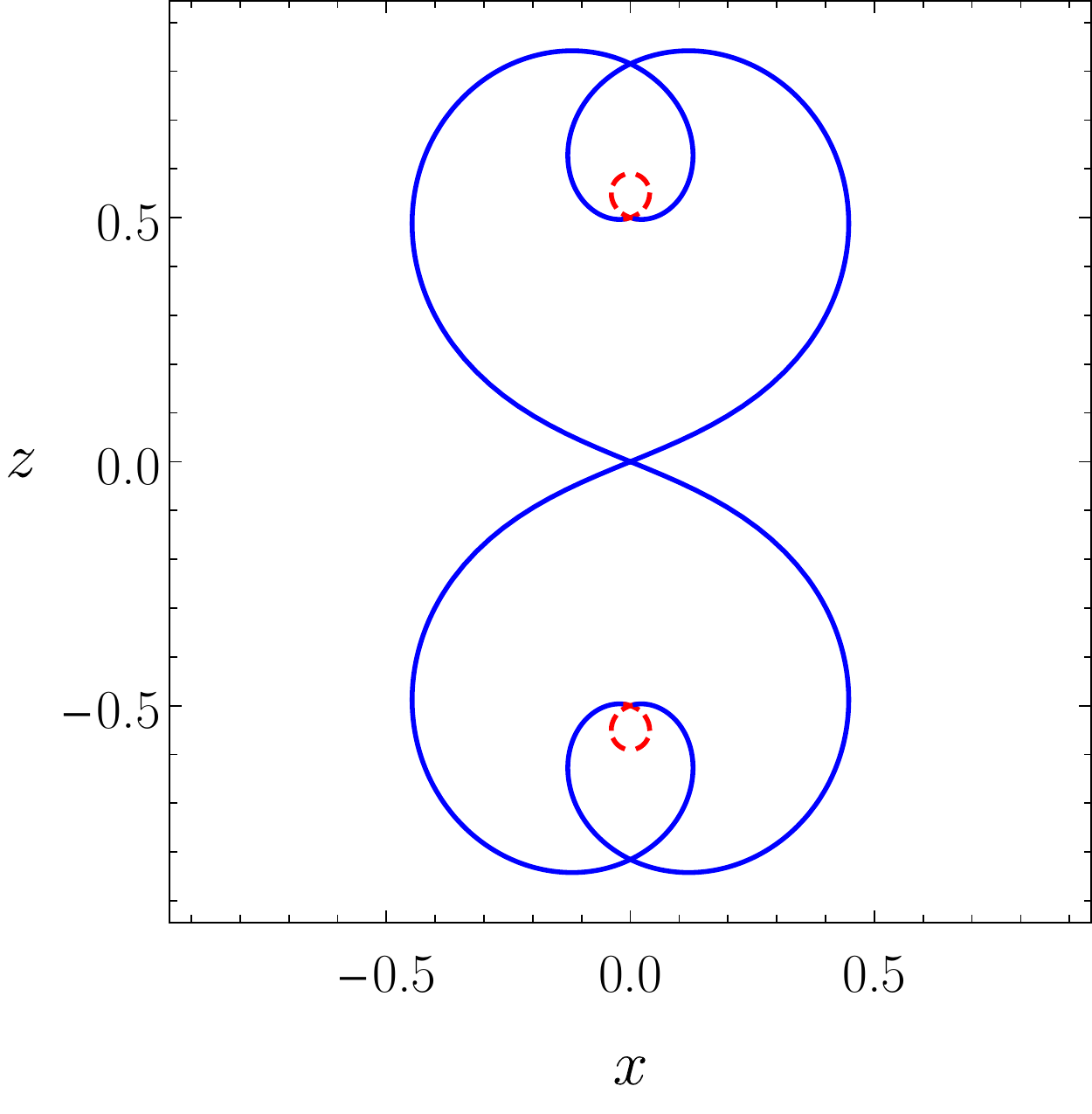}}
\hfill
\end{tabular}
\caption{Examples of periodic null geodesics which pass through the event horizons of both black holes in the Majumdar--Papapetrou di-hole spacetime. The sources are separated by coordinate distance $d = 1$. Rays which emanate from the centre of mass between the black holes fall into the upper black hole. The geodesics are then tracked using the ingoing ($-$) and outgoing ($+$) null coordinates $w_{\pm}$; the switch from ingoing to outgoing coordinates occurs at the symmetry point (i.e., where the ray crosses the $z$-axis).}
\label{fig:mp_periodic_through_horizon}
\end{center}
\end{figure}

Figure \ref{fig:mp_periodic_through_horizon} shows examples of \emph{periodic} null geodesics which pass through the black holes' event horizons in the Majumdar--Papapetrou di-hole spacetime. (We choose $d = 1$ as the coordinate separation between the holes.) Parts of the trajectory which are evolved in isotropic coordinates are shown in blue; the part of the geodesic which was evolved using the null coordinates $w_{\pm}$ is depicted as a red dashed line. The switch from ingoing ($-$) to outgoing ($+$) coordinates occurs at the symmetry point. The exterior ($r > 0$) and interior ($r < 0$) are presented on the same plot by only plotting the magnitude of $r$ in each case, i.e., we plot the coordinates $(x, z) = (|r|\cos{\theta}, |r|\sin{\theta})$. The figure is slightly misleading: each time the photon passes through a white hole horizon, it emerges into a \emph{new} asymptotically flat exterior region. All of these regions are overlaid in Figure \ref{fig:mp_periodic_through_horizon}.
%

\section{Discussion}
\label{sec:discussion_bbh_shadows}

\subsection{Extensions}
%

\subsubsection{Chaotic lensing in other spacetimes}

At the beginning of this chapter, the Majumdar--Papapetrou di-hole spacetime was introduced as a surrogate model for a dynamical binary black hole system in the late stages of its inspiral, before merger. We caution that the Majumdar--Papapetrou di-hole fails to be a ``physically realistic'' surrogate in a number of ways. Firstly, the components of a realistic binary (and indeed all astrophysical black holes) are thought to have negligible charge. (Recall that each black hole in the Majumdar--Papapetrou system is extremally charged to ensure the system is static.) Secondly, the black holes may have non-zero spin $a$, which could be a significant fraction of the Kerr bound $a = M$. Thirdly, a dynamical binary has orbital angular momentum, i.e., the two black holes orbit one another about the centre of mass, so $p\ind{_{\phi}}$ will not be a constant of the motion. Lastly, the black holes will spiral inwards as the emit energy in the form of gravitational waves; the system will therefore not be static (or even stationary), so $p\ind{_{t}}$ will not be conserved.

Recall that Bohn \emph{et al.} \cite{BohnThroweHebertEtAl2015} have successfully performed ray-tracing to study strong-field gravitational lensing on a dynamical binary black hole spacetime. However, a range of other authors have adopted methods similar to those presented here, choosing instead to analyse closed-form models and surrogates. Let us briefly review some key studies here.

The Majumdar--Papapetrou di-hole spacetime is a subfamily of more general classes of exact solutions to the Einstein--Maxwell equations, including (i) the Israel--Wilson spacetimes \cite{IsraelWilson1972}; (ii) the higher-dimensional Majumdar--Papapetrou solutions \cite{HananRadu2007}; (iii) the Kastor--Traschen class of cosmological multi-black-hole spacetimes \cite{KastorTraschen1993}; (iv) the Bret\'{o}n--Manko--Aguilar solutions \cite{BretonMankoAguilarSanchez1998}; and (v) the double-Kerr--Newman solution \cite{MankoMartinRuiz1994}.

In \cite{NittaChibaSugiyama2011, YumotoNittaChibaEtAl2012} the authors study the Kastor--Traschen di-hole solution with a positive cosmological constant. This has the effect of pushing the black holes together, which mimics a head-on collision between the black holes. The shadows exhibit the same eyebrow-like features as those of the Majumdar--Papapetrou di-hole discussed in the present chapter.

A static solution to the Einstein field equations comprising a pair of uncharged black holes was first studied by Bach and Weyl \cite{BachWeyl2012}. In the absence of charges, a ``Weyl strut'' (i.e., a conical singularity along the symmetry axis connecting the sources) is required to maintain equilibrium between the black holes. This solution, occasionally referred to as the double-Schwarzschild black hole, was studied in the context of geodesic motion by Coelho and Herdeiro in 2009 \cite{CoelhoHerdeiro2009}.

More recently, Cunha \emph{et al.} \cite{CunhaHerdeiroRodriguez2018} have employed a novel method to reproduce the lensing effects of dynamical binaries (e.g.~those studied by Bohn \emph{et al.} \cite{BohnThroweHebertEtAl2015}), by considering ``quasi-static'' ray-tracing on spacetimes given by closed-form solutions to the Einstein field equations, including the static double-Schwarzschild binary \cite{BachWeyl2012} and the stationary double-Kerr binary \cite{KramerNeugebauer1980}. In the latter case, the authors produce images of the shadows of the exact stationary co-rotating and counter-rotating stationary double-Kerr binary black hole configurations, which allows one to assess the impact of the intrinsic spin of the black holes on the binary shadows, contrasting it with the effect of the orbital angular momentum of the system. In related work \cite{CunhaHerdeiroRodriguez2018a}, Cunha \emph{et al.} use the double-Schwarzschild binary black hole to analyse whether black hole shadows can be used as a probe of the geometry of the black hole's event horizon. The authors conclude that the conical deficit along the symmetry axis leaves no imprint on the black hole shadow.

Wang \emph{et al.} \cite{WangChenJing2018a} have studied the shadows cast by the Bonnor di-hole. They find that the presence of a magnetic dipole renders the null geodesic motion non-integrable, which results in chaotic patterns in the black hole shadows similar to those of the Majumdar--Papapetrou di-hole.

In this chapter, we have focussed on the physically interesting case of the shadow of a pair of black holes. Chaotic scattering of photons, however, is not unique to this scenario. In fact, assemblages of $N \geq 3$ black holes will exhibit even richer chaotic phenomena. Intriguingly, there are a number of singleton compact objects which have been shown to exhibit chaotic signatures, including perturbed or tidally distorted black holes \cite{BombelliCalzetta1992, AbdolrahimiMannTzounis2015, WangChenJing2018b, WangChenJing2019}, Kerr black holes with scalar hair \cite{CunhaHerdeiroRaduEtAl2015, CunhaHerdeiroRaduEtAl2016} and boson stars \cite{CunhaGroverHerdeiroEtAl2016, CunhaFontHerdeiroEtAl2017}.
%

\subsubsection{Wave propagation on binary black hole spacetimes}

In the case of isolated black holes, (unstable) light-rings are related to important physical phenomena, such as strong gravitational lensing (e.g.~the black hole shadow), the absorption cross-section of radiation incident on the black hole, and the relaxation of the black hole through quasinormal ringing. If two black holes are present (e.g.~during the inspiral phase of a binary black hole merger), the non-escaping photon orbits exist, but their relationship to other aspects of binary black hole physics has not been well studied.

In this work, we have explored the relationship between non-escaping perpetual orbits of the Majumdar--Papapetrou binary black hole system and its black hole shadows. Recently, Assump\c{c}ao \emph{et al.} \cite{AssumpcaoCardosoIshibashiEtAl2018} have provided the first steps towards an understanding of wave dynamics around black hole binaries by considering the Majumdar--Papapetrou di-hole and the double-sink solution, a fluid mechanical analogue which describes a pair of ``sonic holes''. In particular, the authors evolve the Klein--Gordon equation for a massless scalar field on the $(2 + 1)$-dimensional double-sink geometry and explore the connection between the perturbations and the fundamental photon orbits (considered in this work). Further aspects of the physics of black hole binaries -- including null geodesics, ringdown modes and energy extraction processes -- have been studied by Bernard \emph{et al.} \cite{BernardCardosoIkedaEtAl2019}.
%
%

\subsection{Conclusions}

Here, we outline the main conclusions of the work presented in this chapter, and discuss the possible implications of our results for gravitational lensing by dynamical binaries.

Firstly, we have demonstrated that chaotic scattering arises generically in the context of null geodesics motion around black holes when the spacetime admits more than one fundamental photon orbit, provided that such orbits are distinct and dynamically connected (i.e., a null geodesic may travel between the asymptotic neighbourhoods of the fundamental null orbits). The existence of more than one fundamental null orbit generates an uncountable infinity of non-escaping null orbits, which are neither absorbed nor scattered by the black holes. These orbits correspond to a fractal set of scattering singularities in the initial data (i.e., the strange repellor), for which the scattering process is undefined.

One key insight of this work is that black hole shadows can be viewed as the exit basins of an open Hamiltonian system. In the case of Majumdar--Papapetrou di-hole shadows, the (strange) repellor corresponds to the (fractal) boundary of the black hole shadows. One can therefore employ techniques from the theory of chaotic dynamical systems to understand chaotic scattering by binary black holes and the associated fractal structure of the black hole shadows.

Using decision dynamics for the Majumdar--Papapetrou di-hole (Section \ref{sec:symbolic_dynamics_mp}), we have been able to successfully describe the ordering of the non-escaping orbits in initial data in the case of null geodesic scattering in the meridian plane (Section \ref{sec:ordering_perpetual_orbits}). Moreover, we demonstrated that the one-dimensional binary black hole shadow may be constructed using an iterative procedure which involves deleting open intervals of initial data corresponding to scattered or absorbed rays; one may understand the procedure with the aid of symbolic dynamics (Section \ref{sec:construction_cantor_like_set}). This iterative process is similar to that which is used to construct the Cantor set. By considering a particular one-parameter scattering problem, we demonstrated that the one-dimensional shadow is manifestly self-similar (Section \ref{sec:demonstrating_self_similarity}).

When motion is restricted to the meridian plane ($p\ind{_{\phi}} = 0$), there exist three distinct fundamental periodic orbits (Figure \ref{fig:mp_planar_rays}). We find that these orbits are dynamically connected: a null ray can transition between the asymptotic neighbourhoods of these fundamental obits by making a sequence of ``decisions'', which are described by our choice of symbolic dynamics.

For $p\ind{_{\phi}} \neq 0$, motion is not confined to the meridian plane. In this case, the existence of the fundamental photon orbits depends on the value of the angular momentum $p\ind{_{\phi}}$. We find that a two-dimensional shadow image can be decomposed into one-dimensional shadows, each of which corresponds to a value of $p\ind{_{\phi}} = \textrm{constant}$. When transitions between fundamental orbits are not possible, the corresponding one-dimensional shadow is regular, rather than fractal.

Non-planar motion and the effect of angular momentum can be understood through the introduction of an effective potential (or height function) for null geodesics. It is favourable to work with this effective potential (rather than the geodesic potential) as it is independent of the photon's orbital parameters, i.e., energy and angular momentum. The circular null orbits can be understood by classifying the stationary points of the effective potential.

Using the equal-mass Majumdar--Papapetrou di-hole as a case study, we analysed the effect of varying the coordinate separation parameter $d$. As the black holes are brought together, the morphology of the contours of the effective potential is affected. This changes the character of the fundamental periodic photon orbits. Once the value of $d$ is sufficiently small, light-rings about the individual components of the binary are forbidden; such orbits are supplanted by a fundamental periodic orbit which is symmetrical about the equatorial plane. Of course, a more detailed analysis would be required to understand the case of an unequal-mass binary.

Intriguingly, we uncover the possibility of \emph{stable bounded photon orbits} around the two black holes in the Majumdar--Papapetrou di-hole spacetime for coordinate separations in the range $\sqrt{\frac{16}{27}} < \frac{d}{M} < \sqrt{\frac{32}{27}}$. The existence and phenomenology of stable photon orbits will be investigated in the context of stationary axisymmetric spacetimes in Chapter \ref{chap:stable_photon_orbits} of this thesis. In the case of quasi-bounded orbits, there exists a ``pocket'' feature in phase space with three narrow escapes which lead to the two black holes and spatial infinity. This acts as a ``randomising region'', and yields qualitatively different chaotic behaviour which is associated with ``highly fractalised'' regions of the binary black hole shadow (see Figure \ref{fig:mp_shadow_decomposition}). This will be explored further in Chapter \ref{chap:fractal_structures}.

We presented a practical method to track rays through event horizons in the maximally extended Majumdar--Papapetrou di-hole spacetime, based on the construction of ingoing/outgoing null coordinates centred on the event horizon of each black hole. When allowing for this possibility, we see that there is the potential for richer behaviour in the periodic orbits and chaotic orbits in the maximally extended spacetime.

Let us conclude with some comments on the physical implications of this work for real dynamical binary black holes. The Majumdar--Papapetrou di-hole, considered here as a surrogate model, is highly symmetric. The staticity and axisymmetry of the model mean that the momenta $p\ind{_{t}}$ and $p\ind{_{\phi}}$ are conserved along null geodesics. Moreover, the two-dimensional shadow images can be decomposed into one-dimensional shadows, each of which corresponds to a fixed value of $p\ind{_{\phi}}$. In these one-dimensional shadows, we observed Cantor-like, regular, and ``highly fractalised'' structures, which could be understood by considering the existence of the fundamental photon orbits and the contours of the effective null geodesic potential. Of course, a dynamical binary will be neither static nor axially symmetric. It is therefore an open question whether all three types of structure are permitted in realistic binary black hole shadows. To investigate this possibility further, one may wish to analyse a sample of the ray-traced simulations generated by Bohn \emph{et al.} \cite{BohnThroweHebertEtAl2015}, and investigate the existence of fundamental periodic null orbits in this case. 

\chapter{Fractal structures in binary black hole shadows} \label{chap:fractal_structures}

\section{Introduction}

A key feature of general relativity is the gravitational lensing of light by massive bodies (e.g.~stars or black holes). Such objects generate spacetime curvature, which results in the deviation of null geodesics on the curved background; see Chapter \ref{chap:gravitational_lensing} for a review.

Black holes, which possess an event horizon, will cast a shadow -- a two-dimensional region of an observer's local sky which cannot be illuminated by distant light sources due to the blockage of the black hole. A shadow can equivalently be conceived as the set of all photon initial conditions on the observer's ``image plane'' which, when traced backwards in time from the observer, asymptote towards the event horizon of the black hole. In the language of non-linear dynamics, black hole shadows can be viewed as exit basins (see Section \ref{sec:chaotic_dynamical_systems}) in an open Hamiltonian system with two escapes -- one of which corresponds to the event horizon of the black hole; the other to spatial infinity.

%
%
For the scattering of null geodesics by a binary black hole system, such as the static Majumdar--Papapetrou di-hole (see Chapter \ref{chap:binary_black_hole_shadows}), there exist three escapes: a photon may plunge into either of the black holes, or it may escape to spatial infinity. It is therefore natural to associate three exit basins with these escapes. In the case of the Majumdar--Papapetrou di-hole, both the exit basins in phase space and the black hole shadows exhibit a rich variety of structure: they may possess both fractal and regular (i.e., non-fractal) regions; moreover, we shall show that, in certain parameter regimes, the exit basins may possess the stronger \emph{Wada property}, in which all three basins share a common fractal boundary.

In 1917, Yoneyama \cite{Yoneyama1917} proposed the \emph{lakes of Wada} as a curious example of three open sets in $\mathbb{R}^{2}$ which all share the \emph{same} fractal boundary; Yoneyama attributed the construction of the lakes of Wada to his supervisor Takeo Wada, after whom they are named. Close to the end of the century, in 1991, Kennedy and Yorke \cite{KennedyYorke1991} demonstrated that the Wada property is not only a topological curiosity, but that three or more open sets which share a common fractal boundary may arise (generically) in non-linear dynamical systems. In the years that followed, the Wada property has been shown to exist in the basins of a range of chaotic systems, including the H\'{e}non--Heiles Hamiltonian system \cite{HenonHeiles1964, AguirreVallejoSanjuan2001}, the Gaspard--Rice three-disc system \cite{GaspardRice1989, PoonCamposOttEtAl1996}, and the Duffing oscillator \cite{AguirreSanjuan2002}. For a comprehensive review, see \cite{AguirreVianaSanjuan2009}.

A key consequence of the existence of Wada basins in the phase space of scattering problems is the difficulty which arises when predicting the final state of the system. If there is some small uncertainty in fixing the initial conditions close to a Wada boundary, this leads to a high level of indeterminacy and an extreme sensitive dependence on initial conditions, despite the system being completely deterministic. In the context of scattering by a pair of black holes, a photon whose initial conditions are chosen close to a Wada boundary in phase space could end up in one of three final states: the photon could plunge into either black hole or escape to spatial infinity.

In this chapter, we employ a recently developed numerical method \cite{DazaWagemakersSanjuan2018} to test for the existence of Wada basins. This method relies on a simple observation: the boundaries of Wada basins are invariant under the pairwise merging of the basins. In \cite{DazaWagemakersSanjuan2018}, Daza \emph{et al.} develop the ``merging method'', and apply it to three canonical systems which are known to exhibit the Wada property: the forced damped pendulum; the Newton fractal; and the H\'{e}non--Heiles Hamiltonian. The merging method takes as its input only the exit basin diagram at finite resolution (e.g.~an image of the basins in phase space or the black hole shadows). Practically, the merging method determines whether an exit basin diagram possesses the Wada property up to a certain resolution.

The majority of the work presented in this chapter is based on \cite{DazaShipleyDolanEtAl2018}. In Section \ref{sec:mp_hamiltonian_review_wada}, we review the Majumdar--Papapetrou di-hole system -- used here as a toy model for a more realistic dynamical black hole binary -- from the perspective of Hamiltonian dynamics. We review the spacetime geometry and its geodesics (Section \ref{sec:mp_geometry_and_rays_recap}); describe the exit basins in phase space (Section \ref{sec:exit_basins_phase_space}); and describe the structure of the binary black hole shadows (Section \ref{sec:mp_black_hole_shadows_wada}). Section \ref{sec:uncertainty_definitions_numerical} contains a review of two quantitative measures of fractality in exit basins: the fractal dimension and the uncertainty exponent. In Section \ref{sec:uncertaint_exponent_cantor_basins}, we apply these methods to the Cantor basins as a pedagogical example, before using them to distinguish between regular and fractal parts of binary black hole shadows in Section \ref{sec:uncertainty_exponent_shadows}. We then turn our attention to the more restrictive property of Wada. We review the Wada property in Section \ref{sec:wada_property}, with an emphasis on the exit basins of open Hamiltonian systems. In Section \ref{sec:test_wada_basins}, we outline how one can verify that exit basins exhibit the Wada property using numerical algorithms. Section \ref{sec:wada_merging_method} contains a review of the merging method, which was first presented in \cite{DazaWagemakersSanjuan2018}. We apply this method to the exit basins in phase space (Section \ref{sec:results_wada_exit_basins}), and to the binary black hole shadows of the Majumdar--Papapetrou di-hole (Section \ref{sec:results_wada_shadows}). We discuss our results and potential extensions to this work in Section \ref{sec:discussion_fractal_structures}.

\section{Majumdar--Papapetrou open Hamiltonian system}
\label{sec:mp_hamiltonian_review_wada}

The Majumdar--Papapetrou di-hole is a static axisymmetric solution to the Einstein--Maxwell equations, which describes the exterior spacetime of a pair of extremally charged Reissner--Nordstr\"{o}m black holes, each with its mass parameter equal to its charge parameter ($M = Q$). This model was introduced and discussed in detail in Chapter \ref{chap:binary_black_hole_shadows}. For ease of reference, we review the key features of the model here, from the perspective of Hamiltonian dynamics.

\subsection{Spacetime geometry and Hamiltonian formalism for null rays}
\label{sec:mp_geometry_and_rays_recap}

In Weyl--Lewis--Papapetrou coordinates $\left\{ t, \rho, z, \phi \right\}$, the line element for the Majumdar--Papapetrou di-hole is given by \eqref{eqn:mp_general_line_element} with one-form potential \eqref{eqn:mp_electromagnetic_potential}. Choosing $M_{\pm} = M$ (equal-mass case), the function appearing in the metric and electromagnetic potential takes the form
\begin{equation}
U(\rho, z) = 1 + \frac{M}{\sqrt{\rho^{2} + \left(z - \frac{d}{2} \right)^{2}}} + \frac{M}{\sqrt{\rho^{2} + \left(z + \frac{d}{2} \right)^{2}}} .
\end{equation}
We recall that $M$ is the mass of each black hole, and $d$ is the separation between the centres. We hereafter employ units in which $M = 1$. In the chosen coordinate system, the black holes' event horizons appear as points, located on the symmetry axis at $z = \pm \frac{d}{2}$; we recall that these are actually null surfaces of topology $\mathbb{R} \times S^{2}$.

The geodesics of the Majumdar--Papapetrou di-hole $q\ind{^{a}}(\lambda)$ are the integral curves of Hamilton's equations, with Hamiltonian $H(q, p) = \frac{1}{2} g\ind{^{a b}} p \ind{_{a}} p\ind{_{b}}$, where $g\ind{^{a b}}$ are the contravariant components of the metric tensor, $p\ind{_{a}} = g\ind{_{a b}} \dot{q}\ind{^{a}}$ are the canonical momenta, and an overdot denotes differentiation with respect to the affine parameter $\lambda$. The time independence and axial symmetry of the spacetime mean that $t$ and $\phi$ are ignorable coordinates; thus, their conjugate momenta $p\ind{_{t}}$ and $p\ind{_{\phi}}$ are constants of motion. Moreover, the Hamiltonian $H$ is conserved along geodesics. In the case of null geodesics, we have $H = 0$, and we may set $p\ind{_{t}} = - 1$ without loss of generality, by availing the freedom to rescale the affine parameter.

We recall that, after performing a conformal transformation of the metric, we are able to express the Hamiltonian as
\begin{equation}
\label{eqn:mp_hamiltonian_canonical_recall}
H = \frac{1}{2} \left( {p\indices{_{\rho}}}^{2} + {p\indices{_{z}}}^{2} \right) + V, \qquad V(\rho, z) = - \frac{1}{2 \rho^{2}} (h - p\indices{_{\phi}})(h + p\indices{_{\phi}}),
\end{equation}
where the \emph{effective potential} (or \emph{height function}) is
\begin{equation}
\label{eqn:mp_effective_potential_recall}
h(\rho, z) = \rho \, U^{2},
\end{equation}
and $p\ind{_{\phi}}$ is a free parameter which is conserved along rays (see Section \ref{sec:mp_dihole_hamiltonian_formalism}).

The null condition $H = 0$ and the positivity of the kinetic term in the Hamiltonian \eqref{eqn:mp_hamiltonian_canonical_recall} together imply that the geodesic potential must satisfy $V(\rho, z) \leq 0$; this inequality defines the \emph{allowed regions} in configuration space. The contours $h = \pm p\ind{_{\phi}}$ (which are equivalent to $V = 0$) demarcate the boundary of the allowed regions in the $(\rho, z)$-plane.

The full phase space, spanned by the four spacetime coordinates and their conjugate momenta, $\left\{ q\ind{^{a}}, p\ind{_{b}} \right\}$, is eight-dimensional. The symmetries of the spacetime give rise to conserved quantities (associated with the ignorable coordinates $t$ and $\phi$). This permits us to focus on a four-dimensional \emph{reduced} phase space, with two pairs of conjugate variables $\left\{ \rho, z, p\ind{_{\rho}}, p\ind{_{z}} \right\}$, and an additional constraint $H = 0$. The null condition $H = 0$ allows one of the coordinates in the reduced phase space (e.g.~$p\ind{_{z}}$) to be expressed as a function of the other three coordinates: the motion is restricted to a three-dimensional energy hypersurface which lies in the reduced phase space, described by $H = 0$.

In the reduced phase space, the Majumdar--Papapetrou di-hole shares qualitative features with the H\'{e}non--Heiles Hamiltonian system \cite{HenonHeiles1964}, which has become a paradigm for two-dimensional time-independent Hamiltonian scattering since its introduction in the 1960s. We review the H\'{e}non--Heiles system in Appendix \ref{chap:appendix_b}.
%

\subsection{Exit basins in phase space}
\label{sec:exit_basins_phase_space}

\subsubsection{Exit basins of the Majumdar--Papapetrou di-hole}

Here, we construct exit basins in a two-dimensional subspace of the reduced phase space for the Majumdar--Papapetrou di-hole with $d = 1$. The initial conditions used to construct the exit basins are chosen in close analogy with those of a study of the exit basins of the H\'{e}non--Heiles Hamiltonian system \cite{AguirreVallejoSanjuan2001}.

\begin{figure}
\begin{center}
\subfigure[Contours of $h(\rho, z)$]{
\includegraphics[width=0.31\textwidth]{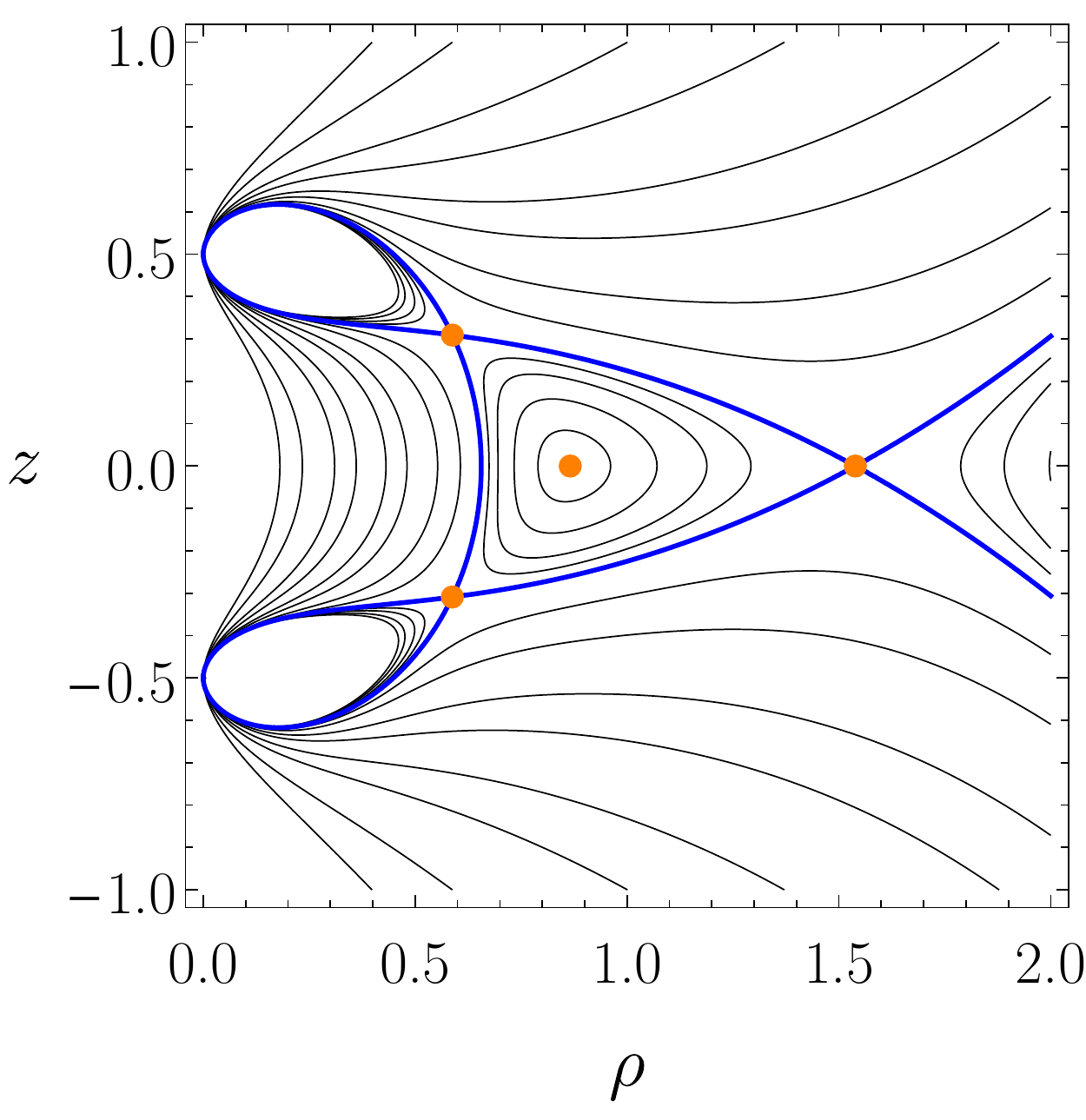} \label{fig:mp_effective_potential_contours}} \hfill
\subfigure[$(\rho, z)$-space]{
\includegraphics[width=0.31\textwidth]{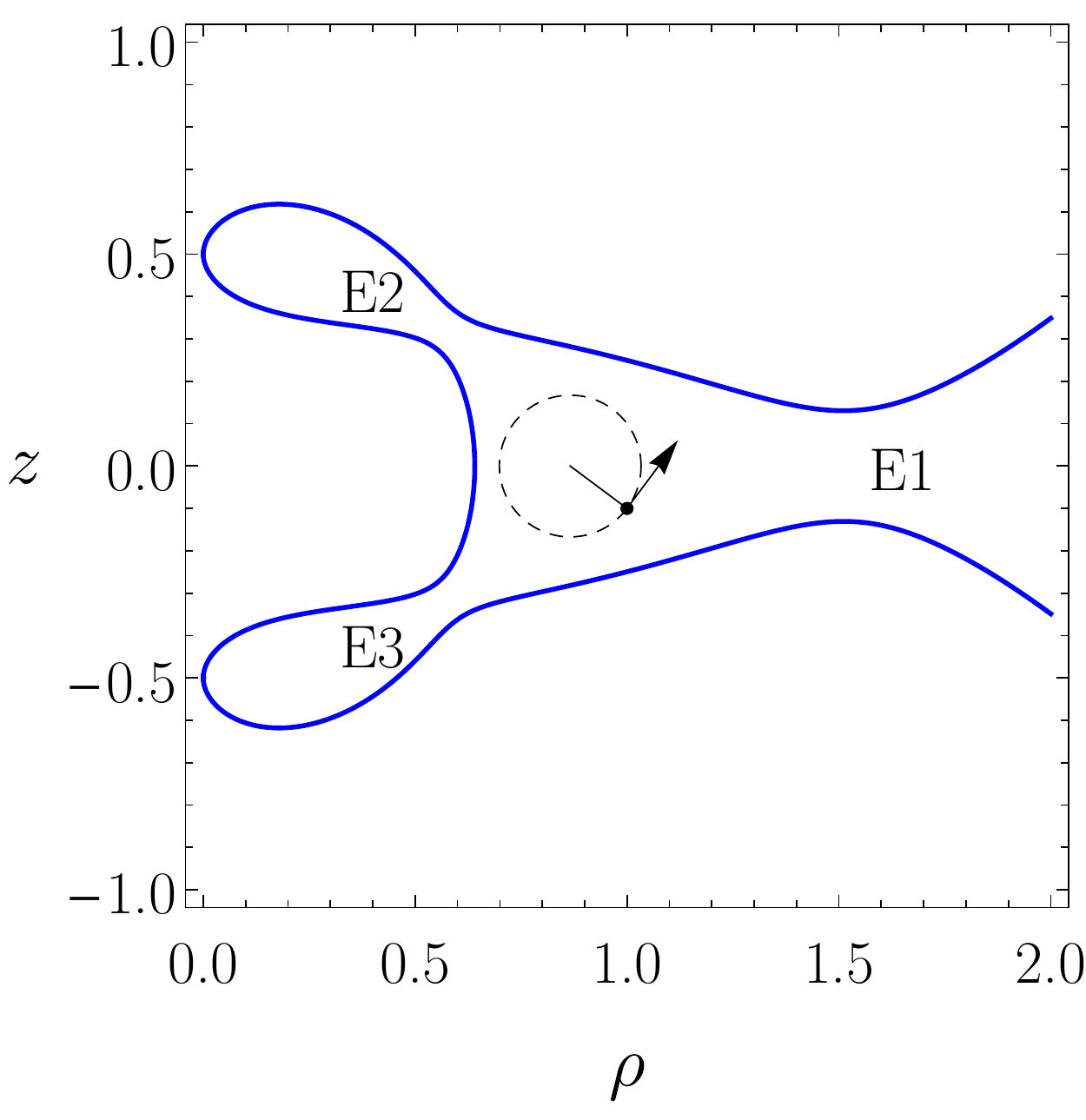} \label{fig:mp_rho_z_basin_set_up}} \hfill
\subfigure[$(\rho, p\ind{_{\rho}})$-space]{
\includegraphics[width=0.31\textwidth]{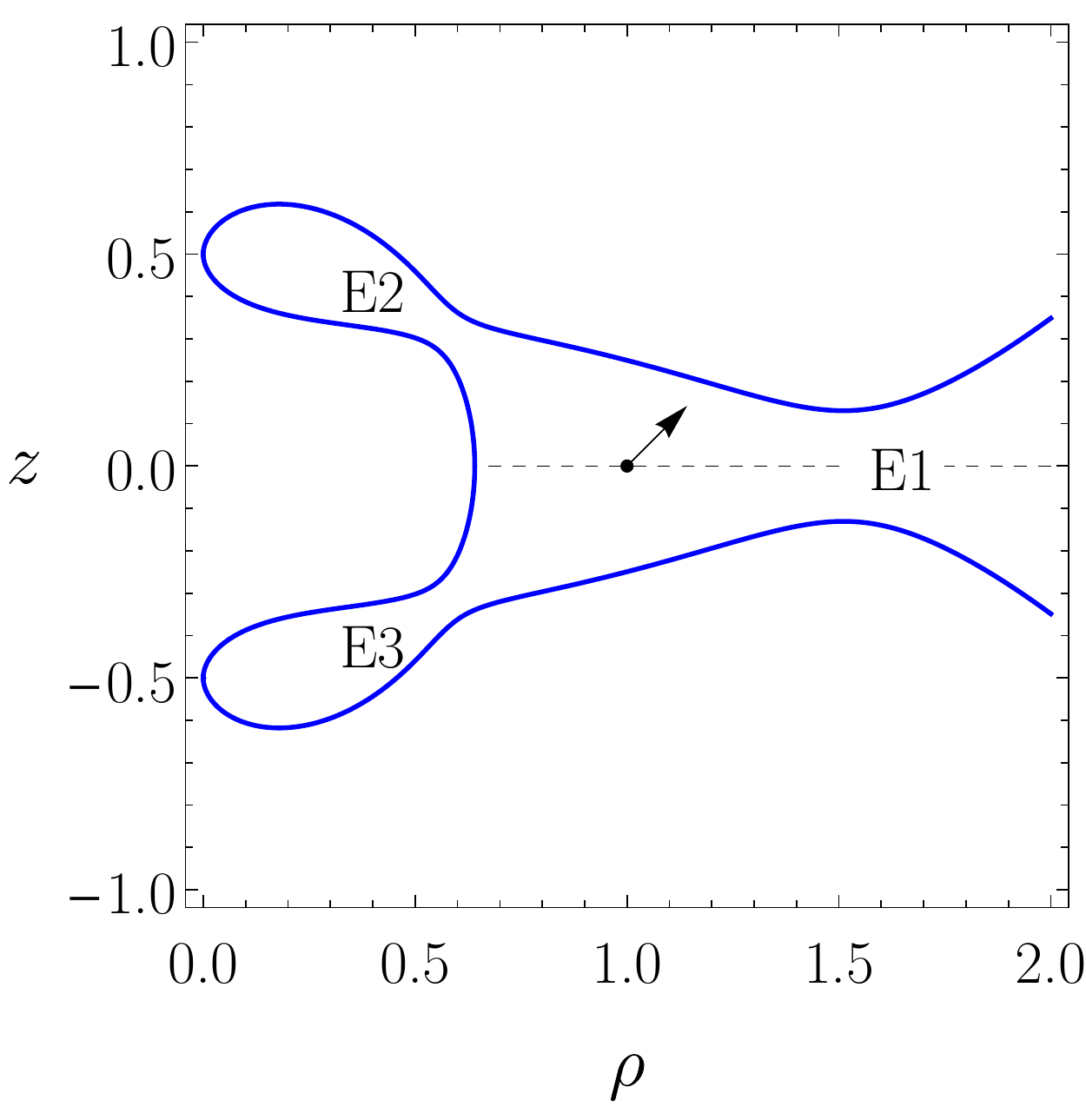} \label{fig:mp_rho_p_rho_basin_set_up}}
\caption{(a) Contours of the effective potential $h(\rho, z) = \rho \, U^{2}$ for the equal mass ($M_{\pm} = M = 1$) Majumdar--Papapetrou di-hole with separation parameter $d = M = 1$. A critical contour passes through three saddle points, and encloses a local maximum. (b)  Initial conditions for exit basins in the $(\rho, z)$-plane. The initial two-momentum $(p\ind{_{\rho}}, p\ind{_{z}})$ is tangent to the circle centred on the local maximum of $h$ that passes through the initial condition. (c) Initial conditions for exit basins in the $(\rho, p\ind{_{\rho}})$-plane. The photon is fired from the $\rho$-axis ($z = 0$) and the initial values of $\rho$ and $p\ind{_{\rho}}$ are varied.}
\label{fig:mp_exit_basin_set_up}
\end{center}
\end{figure}

To construct the exit basins, we focus on the case of equal-mass black holes separated by coordinate distance $d = 1$, due to the high symmetry of this scenario. The effective potential $h$ possesses three saddle points, one of which is in the equatorial plane at $(\rho, z) = \left( \frac{1}{2} 5^{1/4} \varphi^{3/2}, 0 \right)$, and the other two are out of the plane at $(\rho, z) = \left( \frac{1}{2} 5^{1/4} \varphi^{-1/2}, \pm \frac{1}{2 \varphi} \right)$, where $\varphi = \frac{1}{2} \left( 1 + \sqrt{5} \right)$ is the golden ratio. All three of the saddle points lie on a single critical contour $h = {p\ind{_{\phi}}}^{\ast} = \frac{1}{2} 5^{5/4} \varphi^{3/2}$. The critical contour encloses a local maximum of $h$, which is located at $(\rho, z) = \left( \rho_{(1)}, z_{(1)} \right) = \left( \frac{\sqrt{3}}{2}, 0 \right)$, with $h = {p\ind{_{\phi}}}^{(1)} = \frac{9 \sqrt{3}}{2}$. A derivation of the closed-form expressions for the location of the fixed points and critical contours of $h$ is given in Appendix \ref{chap:appendix_a}. The stationary points and contours of $h$ are presented in Figure \ref{fig:mp_effective_potential_contours}. Contrasting Figures \ref{fig:mp_effective_potential_contours} and \ref{fig:hh_contours}, one can clearly see a qualitative resemblance between the contours of the Majumdar--Papapetrou effective potential \eqref{eqn:mp_effective_potential_recall} and those of the H\'{e}non--Heiles potential \eqref{eqn:hh_potential}.

For $p\ind{_{\phi}} \geq {p\ind{_{\phi}}}^{\ast}$, the equipotential curves $h = p\ind{_{\phi}}$ form closed curves on a subregion of the $(\rho, z)$-plane, close to the local maximum of $h$: there is a closed region of phase space, in which orbits are kinematically bounded. Conversely, for $p\ind{_{\phi}} < {p\ind{_{\phi}}}^{\ast}$, the contours $h = p\ind{_{\phi}}$ are open: null rays are permitted to plunge into either of the black holes or escape to infinity. In this chapter, we focus on the latter case, treating the Majumdar--Papapetrou di-hole as a novel example of a two-dimensional time-independent open Hamiltonian system with three escapes in phase space. We express the azimuthal angular momentum as $p\ind{_{\phi}} = {p\ind{_{\phi}}}^{\ast} - \Delta p\ind{_{\phi}}$, where $0 < \Delta p\ind{_{\phi}} \leq {p\ind{_{\phi}}}^{\ast}$.

\begin{figure}[h]
\begin{center}
\subfigure[$\Delta p\ind{_{\phi}} = 0.05$]{
\includegraphics[width=0.31\textwidth]{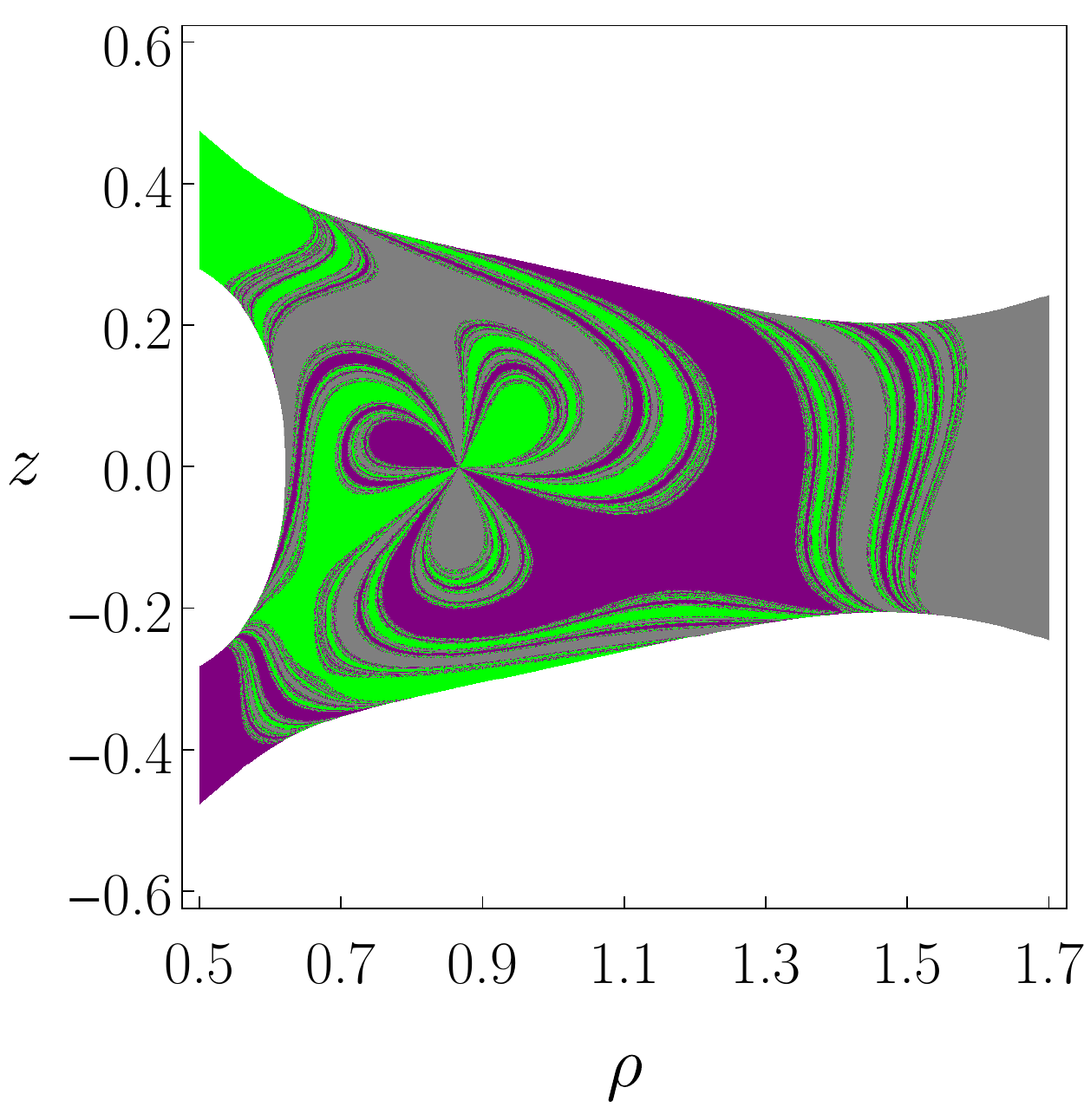} \label{fig:mp_rho_z_basin_005}} \hfill
\subfigure[$\Delta p\ind{_{\phi}} = 0.03$]{
\includegraphics[width=0.31\textwidth]{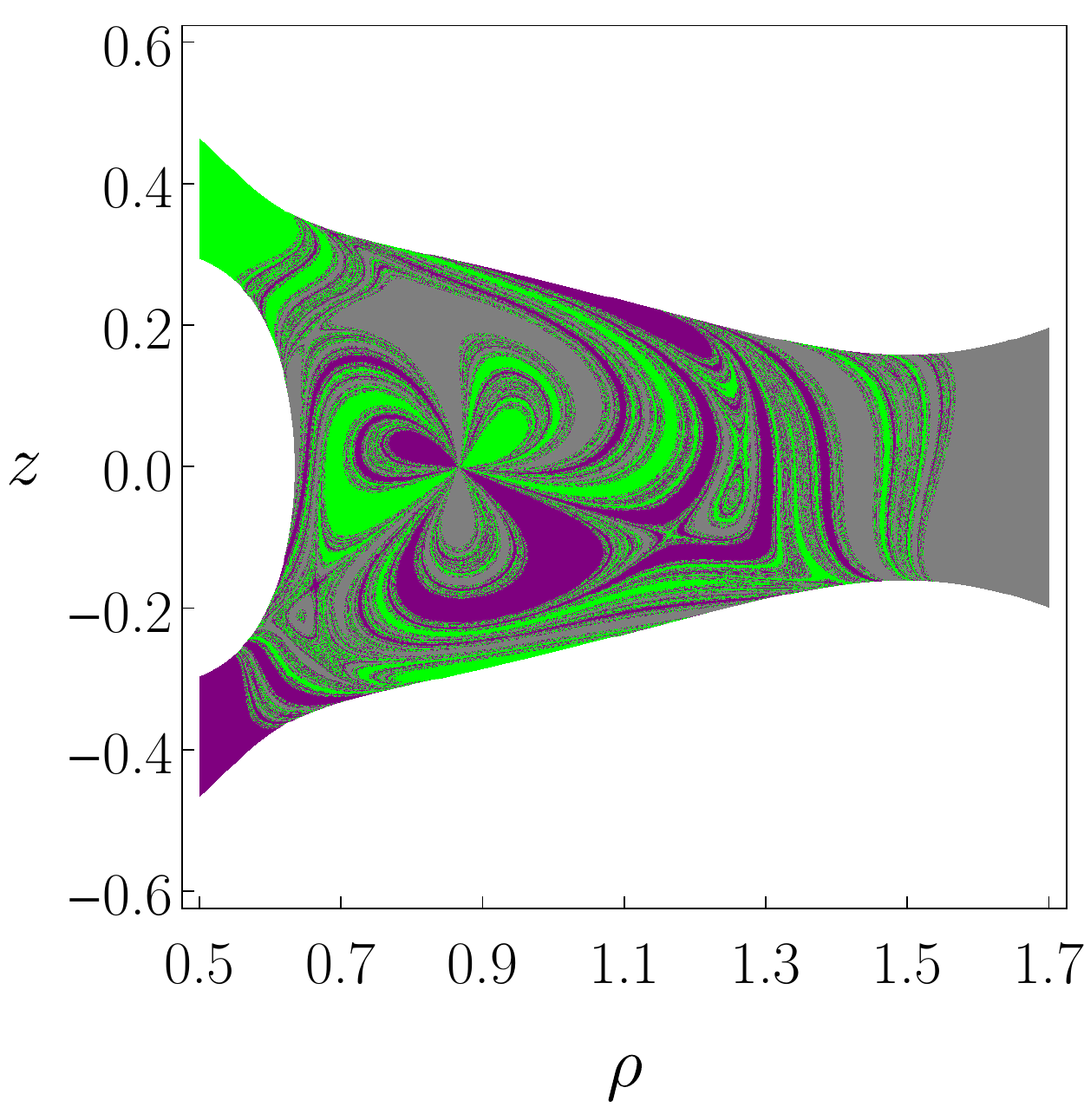} \label{fig:mp_rho_z_basin_003}} \hfill
\subfigure[$\Delta p\ind{_{\phi}} = 0.01$]{
\includegraphics[width=0.31\textwidth]{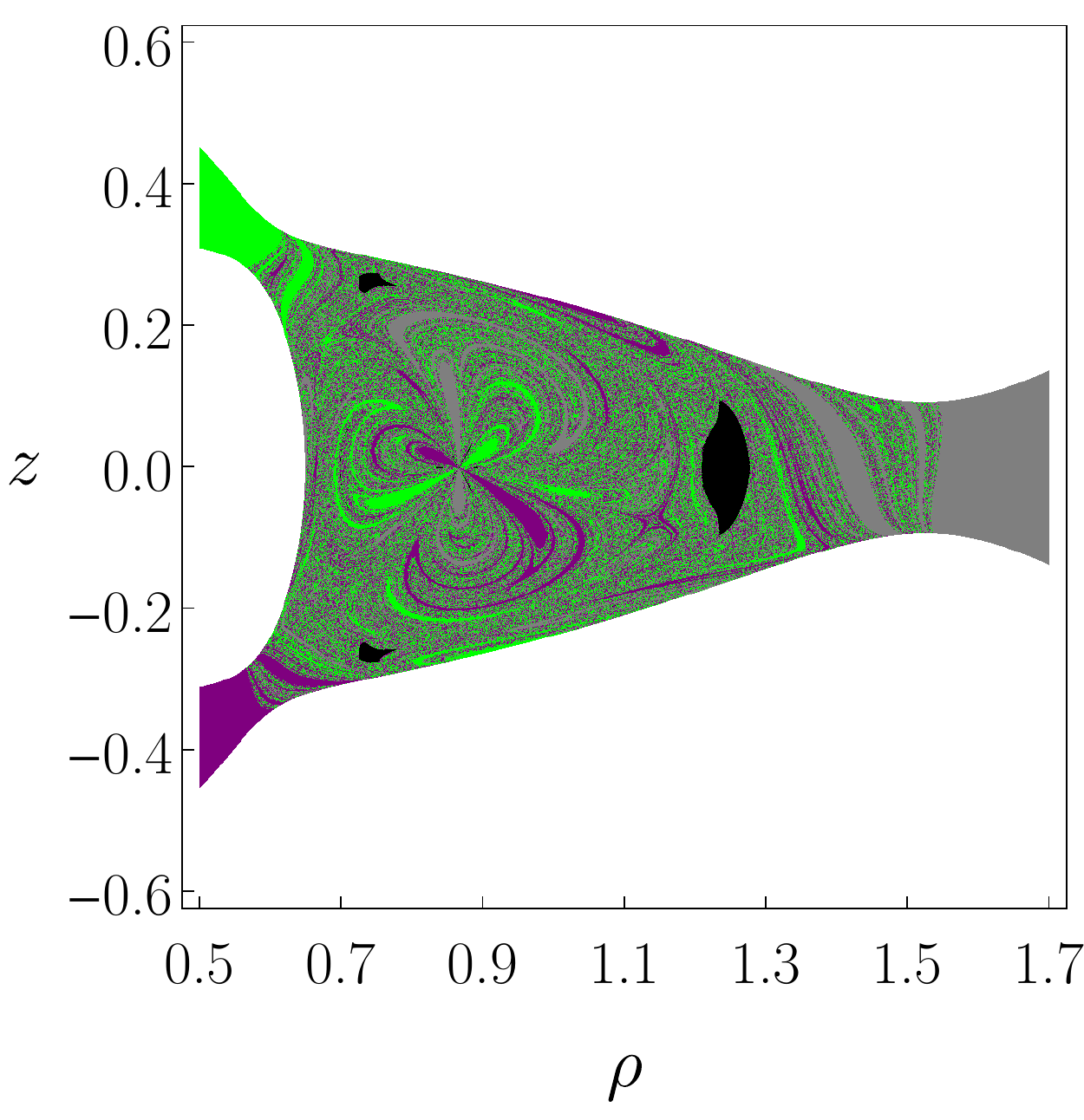} \label{fig:mp_rho_z_basin_001}}
\subfigure[$\Delta p\ind{_{\phi}} = 0.05$]{
\includegraphics[width=0.31\textwidth]{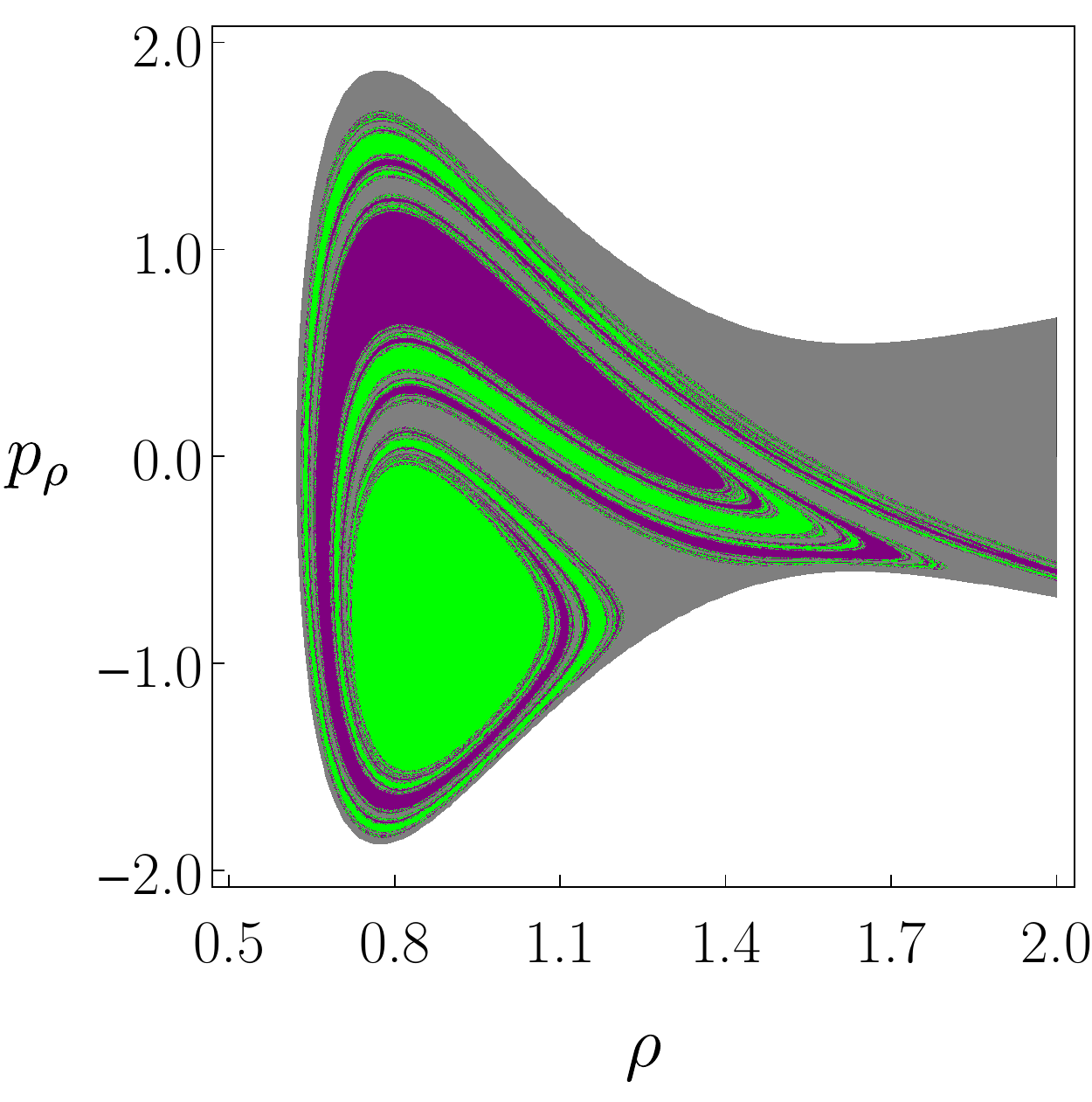} \label{fig:mp_rho_p_rho_basin_005}} \hfill
\subfigure[$\Delta p\ind{_{\phi}} = 0.03$]{
\includegraphics[width=0.31\textwidth]{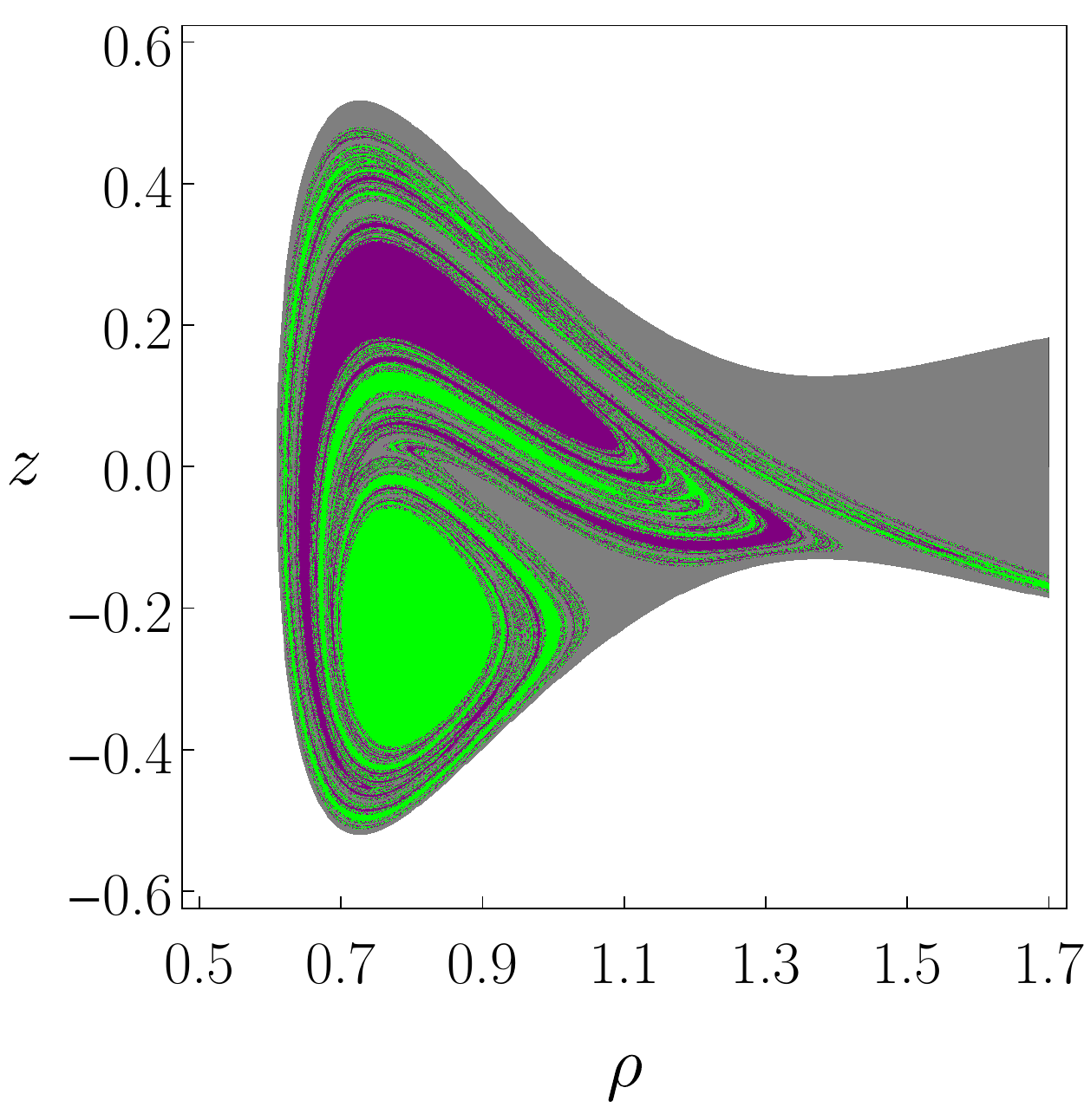} \label{fig:mp_rho_p_rho_basin_003}} \hfill
\subfigure[$\Delta p\ind{_{\phi}} = 0.01$]{
\includegraphics[width=0.31\textwidth]{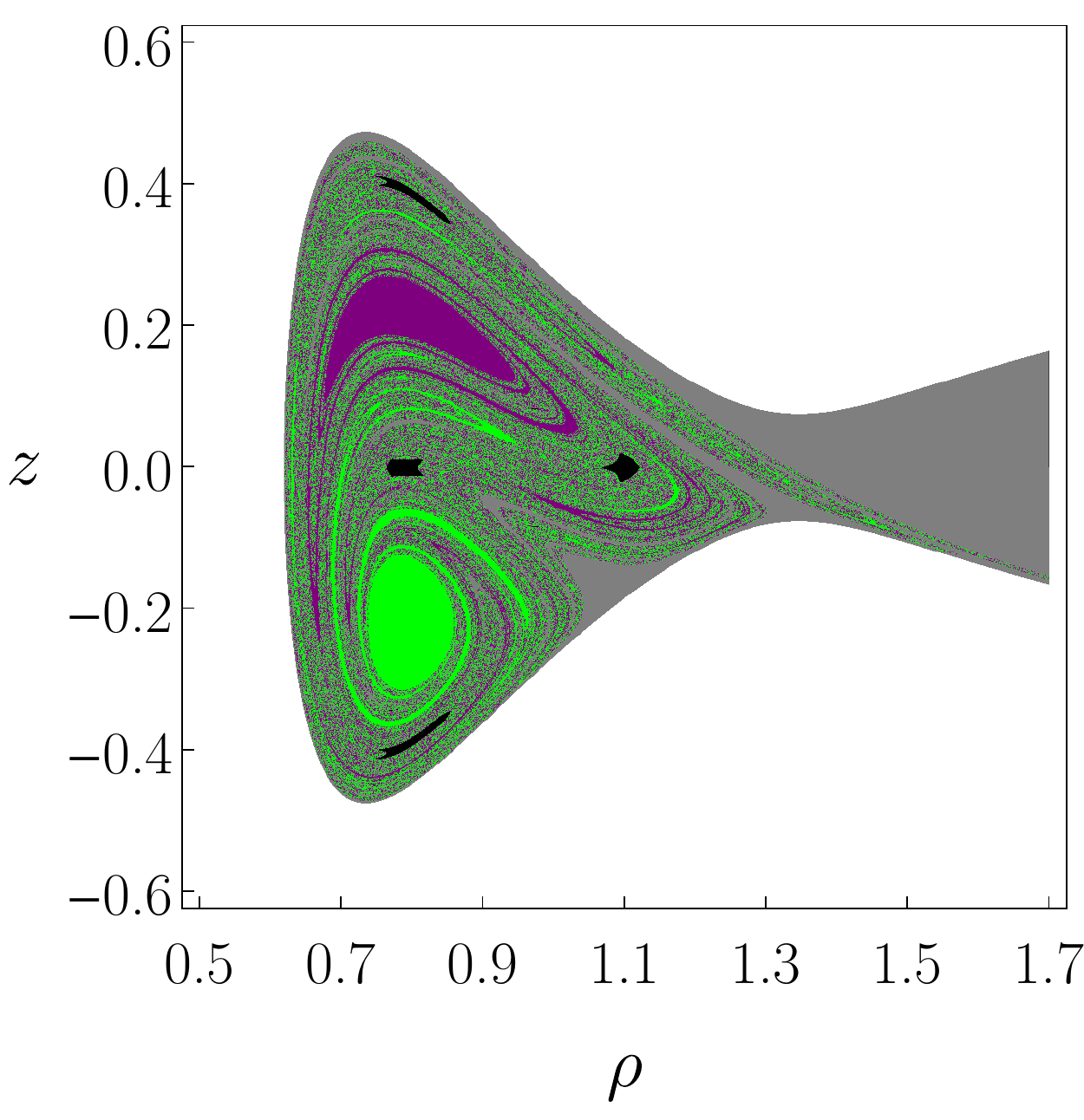} \label{fig:mp_rho_p_rho_basin_001}}
\caption{Exit basins in the Majumdar--Papapetrou open Hamiltonian system for (a)--(c) initial conditions in $(\rho, z)$-space; and (d)--(f) initial conditions in $(\rho, p\ind{_{\rho}})$-space. (Both choices of initial conditions are inspired by studies of the H\'{e}non--Heiles system, and are described in the text.) The initial data which lead to spatial infinity are plotted in grey, and those which asymptote to the event horizon of the upper (lower) black hole are shown in green (purple). As one decreases the parameter $\Delta p\ind{_{\phi}} = {p\ind{_{\phi}}}^{\ast} - p\ind{_{\phi}}$, the widths of the three escapes decrease, and KAM islands of stability (depicted in black) appear.}
\label{fig:mp_exit_basins}
\end{center}
\end{figure}

Our first choice of initial conditions is to fix the coordinates $\rho$ and $z$, and choose the initial two-momentum $\left( p\ind{_{\rho}}, p\ind{_{z}} \right)$ to be tangent (in the anti-clockwise sense) to the circle of radius $\sqrt{\left(\rho - \rho_{(1)}\right)^{2} + \left(z - z_{(1)}\right)^{2}}$, centred on the maximum of $h$, which is located at $\left( \rho_{(1)}, z_{(1)} \right) = \left( \frac{\sqrt{3}}{2}, 0 \right)$; see Figure \ref{fig:mp_rho_z_basin_set_up} for the set-up. The corresponding exit basins are plotted in the $(\rho, z)$-plane; see Figures \ref{fig:mp_rho_z_basin_005}--\ref{fig:mp_rho_z_basin_001} for the exit basin diagrams. The second choice of initial conditions is to fix $z = 0$, and vary the values of $\rho$ and $p\ind{_{\rho}}$, as shown in Figure \ref{fig:mp_rho_p_rho_basin_set_up}. The exit basin diagrams are plotted in the $(\rho, p\ind{_{\rho}})$-plane; see Figures \ref{fig:mp_rho_p_rho_basin_005}--\ref{fig:mp_rho_p_rho_basin_001}.


In Figure \ref{fig:mp_exit_basins}, we present a selection of exit basins for various values of the angular momentum parameter $\Delta p\ind{_{\phi}} = {p\ind{_{\phi}}}^{\ast} - p\ind{_{\phi}}$, employing both choices of initial conditions described above. To realise the exit basins in phase space, we consider a fine grid of initial conditions in either the $(\rho, z)$-plane or the $(\rho, p\ind{_{\rho}})$-plane. We then integrate Hamilton's equations subject to these initial conditions, until a halting condition is satisfied: the numerical integrator is stopped when either $\rho^{2}(\lambda) + \left( z(\lambda) - z_{\pm} \right)^{2} \leq \varepsilon$, where $0 < \varepsilon \ll 1$; or when $\rho^{2}(\lambda) + z^{2}(\lambda) \geq R$, where $R \gg 1$ is some sufficiently large radius. The former case corresponds to absorption by the black hole located at $z = z_{\pm}$, and the latter case indicates scattering.

In order to visualise the exit basins, we colour the initial conditions grey if they lead to infinity, green for the upper black hole, and purple for the lower black hole (cf.~the shadows of Chapter \ref{chap:binary_black_hole_shadows}). In each of the exit basins diagrams of Figure \ref{fig:mp_exit_basins}, we observe an intricate fractal-like structure. For larger values of $\Delta p\ind{_{\phi}}$, the basins are more regular with ``sharper'' boundaries. As $\Delta p\ind{_{\phi}} \rightarrow 0$, the three basins become more intertwined and exhibit ``irregular'' structure. Moreover, in this limit, Kolmogorov--Arnold--Moser (KAM) islands of stability \cite{AguirreVianaSanjuan2009} (depicted in black) appear. The KAM islands consist of the initial conditions corresponding to trajectories which never escape the scattering region as $\lambda \rightarrow \infty$, despite the system being open (i.e., $\Delta p\ind{_{\phi}} > 0$). Trajectories inside KAM islands never escape to infinity nor end up in either of the black holes; rather, they wander forever through the scattering region in a quasi-periodic fashion. The KAM islands are organised in a fractal hierarchy, and have non-zero (Lebesgue) measure, as can be inferred from Figures \ref{fig:mp_rho_z_basin_001} and \ref{fig:mp_rho_p_rho_basin_001}. As the value of $\Delta p\ind{_{\phi}}$ is increased (i.e., the width of the escape channels increases), all trajectories escape the scattering region through one of the three escapes, and the KAM islands disappear. A similar effect is seen in the H\'{e}non--Heiles Hamiltonian system as one increases the total energy $E$. A general discussion of the limit of small escapes in open Hamiltonian systems is presented in \cite{AguirreSanjuan2003}, where the authors employ the Gaspard--Rice three-disc model and the H\'{e}non--Heiles Hamiltonian as illustrative examples.

\subsubsection{Comparison with the H\'{e}non--Heiles Hamiltonian system}

The H\'{e}non--Heiles system was investigated in the context of chaotic scattering in \cite{AguirreVallejoSanjuan2001}, for conserved energies in the range where trajectories are permitted to escape from the scattering region (see Appendix \ref{chap:appendix_b} for a summary). Exit basins were employed to develop a qualitative understanding of the dynamics in the (open) H\'{e}non--Heiles Hamiltonian system. The authors demonstrate, using various computational methods, that (in a certain parameter regime) the exit basins are not only fractal, but possess the more restrictive Wada property; that is, each point on a basin boundary belongs to the boundary of all three basins.

By comparing the exit basin diagrams, one can clearly see a resemblance between the Majumdar--Papapetrou di-hole system (Figure \ref{fig:mp_exit_basins}) and the H\'{e}non--Heiles system (Figure \ref{fig:hh_exit_basins}). The basins appear to be ``topologically equivalent''; however, the H\'{e}non--Heiles basins are more symmetric due to the rotational symmetry of the potential. Given the striking similarities between the exit basins of two systems, and that the basins of the H\'{e}non--Heiles system are known to exhibit the Wada property, we anticipate that the basins of the Majumdar--Papapetrou system will satisfy the same property. This will be explored in more depth in Section \ref{sec:test_wada_basins}.
%

\section{Transition in shadow structure}
\label{sec:mp_black_hole_shadows_wada}

\subsection{Structure of Majumdar--Papapetrou di-hole shadows}
\label{sec:mp_shadow_structure_wada}

Here, we consider shadows of the equal-mass ($M_{\pm} = M = 1$) Majumdar--Papapetrou di-hole, where the separation between the centres is $d$. The shadows were constructed using ray-tracing in Section \ref{sec:binary_black_hole_shadows}. In Figures \ref{fig:mp_shadows_gallery}--\ref{fig:mp_shadow_decomposition}, we present Majumdar--Papapetrou di-hole shadows on the observer's image plane -- spanned by the coordinates $(X, Y)$ -- for a range of viewing angles $\theta$, and separations $d$.

In the equatorial case ($\theta = \frac{\pi}{2}$), the two-dimensional shadow images can be viewed as the union of one-dimensional shadows of constant $p\ind{_{\phi}}$ (equivalently $Y$); see Figure \ref{fig:mp_shadow_decomposition}. In Section \ref{sec:binary_black_hole_shadows}, we saw that the one-dimensional shadows can either exhibit Cantor-like (fractal) structure, or they can be regular. As a consequence, the two-dimensional binary black hole shadow image can either be ``homogeneous'' or ``heterogeneous''. In the former case, each one-dimensional shadow is Cantor-like; in the latter, the two-dimensional shadow is the union of Cantor-like shadows and regular shadows. In particular, we saw that the shadow for $d = 1$ is homogeneous (Figures \ref{fig:mp_shadow_structure_d1}--\ref{fig:mp_one_dim_shadow_d1_cantor}), whereas the shadow for $d = 2$ is heterogeneous (Figures \ref{fig:mp_shadow_structure_d2}--\ref{fig:mp_one_dim_shadow_d2_cantor}).

We anticipate that there will be some critical value of the separation, say $d = \hat{d}$, such that two-dimensional Majumdar--Papapetrou di-hole shadows with $d < \hat{d}$ will be homogeneous, whereas shadows with $d > \hat{d}$ will be heterogeneous. In Section \ref{sec:mp_wada_photon_orbits}, we describe the role of the so-called fundamental photon orbits in the structure of one-dimensional shadows. Then, in Section \ref{sec:critical_separation}, we provide a practical method which can be used to calculate the critical value of the separation parameter $\hat{d}$, which marks the ``phase transition'' in the shadow structure from homogeneous to heterogeneous.

\subsection{Role of photon orbits}
\label{sec:mp_wada_photon_orbits}

Expanding on the discussion in Sections \ref{sec:binary_black_hole_shadows} and \ref{sec:mp_shadow_structure_wada}, we now turn our attention to the so-called fundamental photon orbits, a special class of null rays which play an important role in the Majumdar--Papapetrou di-hole spacetime. We recall from Section \ref{sec:mp_hamiltonian_review_wada} that the allowed regions of phase space, accessible to a null geodesic with azimuthal angular momentum $p\ind{_{\phi}}$, are demarcated by the contours of the effective potential \eqref{eqn:mp_effective_potential_recall}. For an equal-mass Majumdar--Papapetrou di-hole, a \emph{fundamental photon orbit} is a null geodesic $q\ind{^{a}}(\lambda)$ which satisfies the following properties: (i) it is restricted to a compact subset of the $(\rho, z)$-plane; (ii) it is periodic, i.e., there exists some $T > 0$ such that $q\ind{^{a}}(\lambda) = q\ind{^{a}}(\lambda + T)$ for all $\lambda \in \mathbb{R}$; (iii) it is unstable; (iv) it touches the contour $h(\rho, z) = p\ind{_{\phi}}$ in such a way that, locally, it is orthogonal to the contour; and (v) the radial momentum $p\ind{_{\rho}}$ vanishes if the orbit passes through the equatorial plane, by symmetry.

More general null geodesics are permitted in the Majumdar--Papapetrou di-hole spacetime which satisfy some (but not all) of properties (i)--(v). For example, null geodesics which satisfy (i) and (ii) but which are \emph{stable} are permitted; these orbits are explored in Chapter \ref{chap:stable_photon_orbits}. In \cite{CunhaHerdeiroRadu2017}, Cunha \emph{et al.} present a classification scheme for generic fundamental photon orbits in stationary axisymmetric spacetimes, which need only satisfy properties (i) and (ii). The role of light-rings and generic fundamental photon orbits in the analysis of strong-field gravitational lensing is discussed in \cite{CunhaHerdeiro2018}. (In this work, we reserve the label ``fundamental'' for a photon orbit which satisfies all of the properties (i)--(v) from the above list.)
%

Figure \ref{fig:mp_fundamental_orbits_p_phi} shows examples of the three types of fundamental photon orbit around the equal-mass Majumdar--Papapetrou di-hole with $d = 2$, projected onto the $(\rho, z)$-plane. These fundamental orbits were discussed in the case of motion restricted to the meridian plane ($p\ind{_{\phi}} = 0$) and non-planar motion ($p\ind{_{\phi}} \neq 0$) in Chapter \ref{chap:binary_black_hole_shadows}. Moreover, these orbits can be understood using symbolic dynamics (see Section \ref{sec:symbolic_dynamics_mp}). For the remainder of this chapter, we label the three types of fundamental photon orbits as follows: (I) a one-component light-ring around only one of the black holes; (II) a figure-of-eight orbit around both black holes; (III) a two-component light-ring around both black holes. Recall that, in decision dynamics (Section \ref{sec:symbolic_dynamics_mp}), the fundamental orbits have representations (I) $\ol{0}$; (II) $\ol{2}$; (III) $\ol{4}$.

In Figures \ref{fig:mp_fundamental_orbits_p_phi_400}--\ref{fig:mp_fundamental_orbits_p_phi_592}, we demonstrate the effect of increasing $p\ind{_{\phi}}$ on the fundamental orbits, for the case of equal-mass black holes separated by coordinate distance $d = 2$. As one increases $p\ind{_{\phi}}$ from zero, the contour $h(\rho, z) = p\ind{_{\phi}}$ moves away from the symmetry axis and orbits II and III move closer together. These orbits merge together at $p\ind{_{\phi}} = \hat{p}\ind{_{\phi}} \approx 5.09$; beyond this value ($p\ind{_{\phi}} > \hat{p}\ind{_{\phi}}$), the type II and III orbits no longer exist. We see that the type I orbits persist until $p\ind{_{\phi}} = {p\ind{_{\phi}}}^{\ast}$, which is the value of the azimuthal angular momentum for the pair of non-planar unstable circular photon orbits of constant $\rho$ and $z$. These unstable circular orbits correspond to the non-planar saddle points of the effective potential $h(\rho, z)$; see Figure \ref{fig:mp_fundamental_orbits_p_phi_592}. We note that a ray with $p\ind{_{\phi}} > {p\ind{_{\phi}}}^{\ast}$ which starts far from the scattering region is forbidden from accessing the black holes by the contour $h = p\ind{_{\phi}}$; the value ${p\ind{_{\phi}}}^{\ast}$ corresponds to the top of the binary black hole shadow image (i.e., the last value of $Y$ for which absorption is permitted); see Figure \ref{fig:mp_shadow_decomposition}.

As demonstrated in Chapter \ref{chap:binary_black_hole_shadows}, when all three types of orbit exist and are ``dynamically connected'' -- i.e., if rays are permitted to transition between the asymptotic neighbourhoods of two or more fundamental orbits -- then chaotic scattering arises naturally. Moreover, we find that this chaotic scattering is responsible for Cantor-like fractal structures in one-dimensional binary black hole shadows with $p\ind{_{\phi}} = \text{constant}$. For a given value of $d$, the fundamental photon orbits are dynamically connected for $p\ind{_{\phi}} < \hat{p}\ind{_{\phi}}$. Indeed, in Chapter \ref{chap:binary_black_hole_shadows}, we demonstrated that, for a given value of $p\ind{_{\phi}} < \hat{p}\ind{_{\phi}}$, the one-dimensional shadow has a Cantor-like structure; see Figure \ref{fig:mp_shadow_decomposition}. We note that it is not sufficient for two separate orbits of type I to exist: these are not typically dynamically connected in the absence of type II and III orbits.

In Chapter \ref{chap:binary_black_hole_shadows}, we describe how the structure of the one-dimensional shadows for coordinate separation $d = 2$ changes as $p\ind{_{\phi}}$ varies. For $p\ind{_{\phi}} > {p\ind{_{\phi}}}^{\ast}$, the ray is forbidden from accessing the black holes by the contour $h = p\ind{_{\phi}}$: the corresponding one-dimensional shadow is the empty set. As describe above, when $\hat{p}\ind{_{\phi}} < p\ind{_{\phi}} < {p\ind{_{\phi}}}^{\ast}$, the orbits of type II and III do not exist and the type I orbits are not dynamically connected: the shadow is therefore regular in this regime. Finally, for $0 < p\ind{_{\phi}} < \hat{p}\ind{_{\phi}}$, all orbits of types I--III exist and are dynamically connected: the one-dimensional shadow exhibits a Cantor-like fractal structure.

Heterogeneous shadows are admitted by equal-mass di-holes with coordinate separation $d$, such that the coexistence condition $\hat{p}\ind{_{\phi}} < {p\ind{_{\phi}}}^{\ast}$ is satisfied. In the following section, we demonstrate that this condition is only met for sufficiently separated black holes with $d > \hat{d}$; for $d < \hat{d}$, the coexistence condition is not met, and the two-dimensional shadow will be homogeneous (i.e., each one-dimensional shadow will be Cantor like).
%

\subsection{Calculating the critical separation}
\label{sec:critical_separation}

Here, we briefly outline a method to calculate the critical value of the separation $\hat{d}$, introduced in Section \ref{sec:mp_wada_photon_orbits}. The critical separation $d = \hat{d}$ marks a transition from homogeneous ($d < \hat{d}$) to heterogeneous ($d > \hat{d}$) binary black hole shadows.

Recall from Section \ref{sec:mp_wada_photon_orbits} that, for a fixed value of $d$, the type II and III fundamental photon orbits merge at $p\ind{_{\phi}} = \hat{p}\ind{_{\phi}}$. We seek the value of the coordinate separation between a pair of equal-mass black holes in the Majumdar--Papapetrou di-hole which gives rise to a single outer fundamental photon orbit when $\hat{p}\ind{_{\phi}} = {p\ind{_{\phi}}}^{\ast}$; that is, the value of $d$ for which the type II and III fundamental photon orbits merge at exactly the value of $p\ind{_{\phi}}$ for which the two black holes become inaccessible to a ray incident from infinity.

We first choose a value of the coordinate separation $d$. The location of the stationary points of $h(\rho, z)$, and thus the corresponding value of ${p\ind{_{\phi}}}^{\ast}$, are found using the method presented in Appendix \ref{chap:appendix_a}. We then consider rays which start on the contour $h = {p\ind{_{\phi}}}^{\ast}$, with $\rho(0) = \rho_{0}$. The initial value $z(0) = z_{0} > 0$ is found by numerically solving the equation $h(\rho_{0}, z_{0}) = {p\ind{_{\phi}}}^{\ast}$ for $z_{0}$. A ray which starts on the contour has $p\ind{_{\rho}}(0) = 0 = p\ind{_{z}}(0)$ from the Hamiltonian constraint $H = 0$.

Having fixed the initial conditions $(\rho(0), z(0), p\ind{_{\rho}}(0), p\ind{_{z}}(0)) = (\rho_{0}, z_{0}, 0, 0)$, we evolve the geodesic equations (i.e., Hamilton's equations) until the ray passes through the equatorial plane ($z = 0$). At this point, we record the value of $\vartheta = \frac{\pi}{2} + \arctan{\left( \frac{p\ind{_{z}}}{p\ind{_{\rho}}} \right)}$, where $\arctan{\left( \frac{p\ind{_{z}}}{p\ind{_{\rho}}} \right)}$ is the angle made by the tangent vector (i.e., the two-momentum) and the equatorial plane. By symmetry, the type II and III fundamental orbits must have $\vartheta = 0$. Hence, the zeros of the function $\vartheta(\rho_{0})$ determine the location of the initial conditions $\rho(0) = \rho_{0}$ (on the contour $h = {p\ind{_{\phi}}}^{\ast}$) of the type II and III fundamental orbits.

\begin{figure}
\begin{center}
\includegraphics[width=0.45\textwidth]{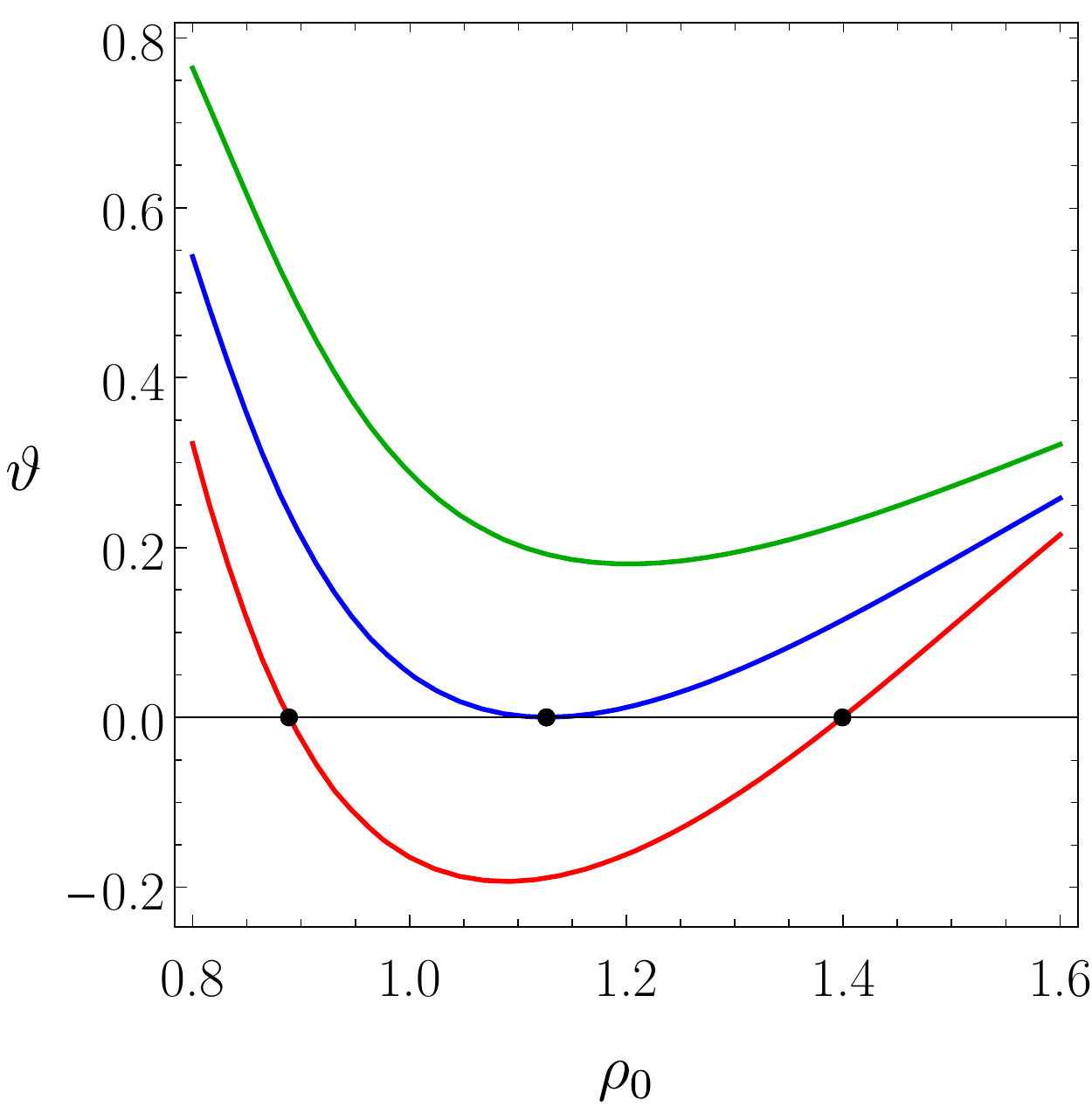}
\caption{Results of the algorithm used to search for initial data $\rho(0) = \rho_{0}$ corresponding to the type II and III fundamental orbits, for which $\vartheta = 0$. We show three representative examples for varying coordinate separations $d$: (i) $d < \hat{d}$ [red]; (ii) $d = \hat{d}$ [blue]; (iii) $d > \hat{d}$ [green]. The function $\vartheta(\rho_{0})$ has (i) two zeros, so both type II and III orbits exist for all $p\ind{_{\phi}} \in [0, {p\ind{_{\phi}}}^{\ast} ]$; (ii) a single zero, so there is a single outer (type II/III) orbit; (iii) no zeros, so there exists some $\hat{p}\ind{_{\phi}} < {p\ind{_{\phi}}}^{\ast}$ beyond which the type II and III orbits no longer exist.}
\label{fig:mp_critical_separation}
\end{center}
\end{figure}

In Figure \ref{fig:mp_critical_separation}, we show the function $\vartheta(\rho_{0})$ for three representative values of the separation $d$. In each case, the curve has been determined numerically using the method described above. We are interested in the value $d = \hat{d}$ for which $\vartheta(\rho_{0})$ admits a single zero, i.e., there exists a single outer (type II/III) fundamental orbit. We find that
\begin{equation}
\hat{d} \approx 1.2085 M
\end{equation}
for the equal-mass Majumdar--Papapetrou di-hole. This is shown as a blue curve in Figure \ref{fig:mp_critical_separation}. In the case of unequal masses, the $\mathbb{Z}_{2}$ reflection symmetry is broken; a more detailed analysis of the structure of the fundamental orbits would be required to determine the value of $\hat{d}$ in this case.

For di-holes with a sufficiently small separation ($d < \hat{d}$), there exist three types of fundamental orbits for all $p\ind{_{\phi}} \in [0, {p\ind{_{\phi}}}^{\ast} ]$. In this regime, we anticipate that the boundary of the Majumdar--Papapetrou binary black hole shadow will be entirely fractal (homogeneous). On the other hand, for sufficiently separated black holes ($d > \hat{d}$), there exists some range of values $p\ind{_{\phi}} \in ( \hat{p}\ind{_{\phi}}, {p\ind{_{\phi}}}^{\ast} )$ in which the type II and III fundamental orbits no longer exist, but absorption by the black holes is still possible. (This phenomenon is demonstrated in Figure \ref{fig:mp_fundamental_orbits_p_phi}, where we consider the case $d = 2M > \hat{d}$.) In such situations, we anticipate that the regions of the binary black hole shadow boundary corresponding to $p\ind{_{\phi}} \in ( \hat{p}\ind{_{\phi}}, {p\ind{_{\phi}}}^{\ast} )$ will no longer be fractal. The two-dimensional shadow image will therefore be heterogeneous.

\section{Uncertainty exponent and fractal dimension}
\label{sec:uncertainty_exponent}

\subsection{Definitions and numerical method}
\label{sec:uncertainty_definitions_numerical}

\subsubsection{Hausdorff dimension of fractal sets}

Informally, a geometrical figure which contains $m$ copies of itself which are scaled by a factor $\frac{1}{s}$ is said to have \emph{Hausdorff dimension} $D_{\textrm{H}}$, where $m = s^{D_{\textrm{H}}}$ \cite{Ott2002, Falconer2004}. (Note that the Hausdorff dimension is often referred to as the \emph{Hausdorff--Besicovitch dimension}.) Inverting this scaling law yields
\begin{equation}
\label{eqn:hausdorff_dimension}
D_{\textrm{H}} = \frac{\ln{m}}{\ln{s}} .
\end{equation}
A more formal definition of the Hausdorff dimension, which relies on the notion of Hausdorff measure, is given in e.g.~\cite{Falconer2004}. The definition \eqref{eqn:hausdorff_dimension} will be sufficient for the purpose of this work.

To illustrate the (heuristic) calculation of the Hausdorff dimension, let us consider some simple geometric objects. A line segment contains $m$ equal-sized copies of itself, each of which is scaled by a factor $\frac{1}{m}$, where $m \in \mathbb{N}$. The Hausdorff dimension of a line segment is therefore $D_{\textrm{H}} = \frac{\ln{m}}{\ln{m}} = 1$. Similarly, a square can be divided into $k^{2}$ smaller squares, which are scaled by a factor $\frac{1}{k}$, where $k \in \mathbb{N}$. The Hausdorff dimension of a square is therefore $D_{\textrm{H}} = \frac{\ln{k^{2}}}{\ln{k}} = 2$. In both of these cases, the Hausdorff dimension takes an integer value which agrees with our usual notion of dimension (i.e., the topological dimension) \cite{Falconer2004}.

The Hausdorff dimension is intimately related to the notions of scaling and self-similarity, and provides a natural way to characterise fractals. The Hausdorff dimension is therefore often referred to as the \emph{fractal dimension}. In fact, some attempts to provide a rigorous mathematical definition of a fractal are dependent on the concept of Hausdorff dimension. For example, Mandelbrot \cite{Mandelbrot1982} conceived that a fractal is a set whose Hausdorff dimension is strictly greater than its topological dimension. This definition proved to be unsatisfactory, as it excludes a large number of sets which clearly ought to be defined as fractals; for example, certain space-filling curves (e.g.~the Hilbert curve) do not meet Mandelbrot's criterion.

Consider the Cantor set, which is an example of a self-similar fractal constructed on the unit interval (see Section \ref{sec:construction_cantor_like_set}). Heuristically, we may calculate the Hausdorff dimension $D_{\textrm{H}}$ of the Cantor set $\mathcal{C}$ as follows. The Cantor set contains two copies of itself, $\mathcal{C} \cap \left[ 0, \frac{1}{3} \right]$ and $\mathcal{C} \cap \left[ \frac{2}{3}, 1 \right]$, each of which has been scaled by a factor $\frac{1}{3}$. The Hausdorff dimension of $\mathcal{C}$ is thus $D_{\textrm{H}} = \frac{\ln{2}}{\ln{3}} \approx 0.631$. The Hausdorff dimension of the Cantor set is between zero (the Hausdorff dimension of a countable set) and one (the Hausdorff dimension of the unit interval).

Similarly, we may consider the $\gamma$-Cantor set $\mathcal{C}_{\gamma}$, defined in Section \ref{sec:construction_cantor_like_set}. This set contains two copies of itself which have been scaled by a factor $\frac{1 - \gamma}{2}$. The Hausdorff dimension of $\mathcal{C}_{\gamma}$ is therefore
\begin{equation}
\label{eqn:hausdorff_gamma_cantor_set}
D_{\text{H}}(\gamma) = \frac{\ln{2}}{\ln{\left( \frac{2}{1 - \gamma}\right)}} .
\end{equation}
Varying $\gamma \in \left( 0, 1 \right)$ gives any Hausdorff dimension $D_{\text{H}}(\gamma) \in \left( 0, 1 \right) $. Clearly, the case $\gamma = \frac{1}{3}$ gives the Hausdorff dimension of the canonical middle-thirds Cantor set.
%

\subsubsection{Uncertainty exponent}

Grebogi \emph{et al.} \cite{GrebogiMcDonaldOttEtAl1983} considered the \emph{uncertainty} (or \emph{unpredictability}) associated with basin boundaries in dynamical systems, introducing the \emph{uncertainty  exponent} as a quantitative measure of the indeterminacy in final-state prediction when fractal boundaries are present. Here, we review the definition of the uncertainty exponent, and discuss its usefulness in distinguishing between regular and fractal boundaries.

Consider an initial uncertainty $\varepsilon$ in the determination of the initial conditions, which can, in general, be due to numerical or experimental error, or a small perturbation of the initial data. Given this uncertainty, we compute the fraction of initial conditions $f$ which are \emph{uncertain}, i.e., the proportion of initial conditions that are within a distance $\varepsilon$ of the basin boundary. Computing $f$ for different values of $\varepsilon$, we see that $f \sim \varepsilon$ for smooth boundaries. However, for fractal boundaries, the fraction of uncertain initial conditions obeys a power-law scaling of the form $f \sim \varepsilon^{\alpha}$. Here, $\alpha$ is the \emph{uncertainty exponent}, which takes values between zero (total fractality) and one (regularity).

The uncertainty exponent is related to the dimension $D$ of the space in which the object is embedded, and the Hausdorff dimension of the fractal boundary $D_{\text{H}}$, via \cite{GrebogiMcDonaldOttEtAl1983}
\begin{equation}
\label{eqn:uncertainty_exponent_hausdorff_dim}
\alpha = D - D_{\text{H}}.
\end{equation}
(A proof of this statement is given in \cite{McDonaldGrebogiOttEtAl1985}.) The uncertainty exponent of the Cantor basins, whose boundary is the middle-thirds Cantor set, is $\alpha = 1 - \frac{\ln{2}}{\ln{3}} \approx 0.369$. This indicates that the Cantor basins are indeed fractal.

\subsubsection{Numerical method to calculate the uncertainty exponent}

In practice, the uncertainty exponent is typically computed numerically by analysing how the fraction of uncertain initial conditions $f$ varies with the uncertainty $\varepsilon$. For a given initial condition $x$, one first finds the basin to which $x$ belongs. Next, one determines the basin for (a sample of) initial conditions $y$ which satisfy $|x - y| \leq \varepsilon$, for some suitable distance measure $\left| \cdot \right|$. (Here, we use the Euclidean distance on $\mathbb{R}^{n}$.) If all of these initial conditions $y$ belong to the \emph{same} basin as $x$, then the point $x$ is labelled \emph{certain}, otherwise it is labelled \emph{uncertain}.

For a fixed value of $\varepsilon$, a large number $N$ of initial conditions $x$ are chosen -- either randomly or from a uniform grid. These are labelled as certain or uncertain according to the method described above. The total number of uncertain initial conditions is denoted $N_{\text{u}}$. Then $f(\varepsilon) = \frac{N_{\text{u}}}{N}$, the fraction of initial conditions which are uncertain. This procedure is repeated for a range of values of the tolerance parameter $\varepsilon$, in order to determine the power-law relationship $f \sim \varepsilon^{\alpha}$. The slope of the graph of $\log{f}$ versus $\log{\varepsilon}$ is precisely the uncertainty exponent $\alpha$.

\subsection{Uncertainty exponent for Cantor basins}
\label{sec:uncertaint_exponent_cantor_basins}

To test and calibrate the method outlined in Section \ref{sec:uncertainty_definitions_numerical}, we compute $\alpha$ numerically for the two-colour $\gamma$-Cantor basins $\left\{ B_{1}, B_{2} \right\}$ (defined in Section \ref{sec:construction_cantor_like_set}), and we compare this with an exact result.

The numerical calculation of the uncertainty exponent is performed using a fixed set of $N = 10^{8}$ equally spaced ``initial conditions'' in the unit interval $I = \left[ 0, 1 \right]$. For a fixed value of $\varepsilon$, we consider each of the $N$ initial conditions $x \in I$ in turn. The initial condition $x$ is classified as certain or uncertain using the procedure outlined above. That is, by looking at the subset of the $N$ initial conditions which are within a distance $\varepsilon$, and determining to which basin each one belongs. If all such points belong to the same basin as the initial condition under consideration, then the initial condition $x$ is labelled as certain; otherwise, $x$ is labelled uncertain. The uncertainty exponent $\alpha$ is then obtained from the slope of $\log{f}$ versus $\log{\varepsilon}$, where $f$ is the fraction of uncertain initial conditions.

For the $\gamma$-Cantor basins, the uncertainty exponent $\alpha(\gamma)$ may be obtained in closed form by virtue of \eqref{eqn:hausdorff_gamma_cantor_set} and \eqref{eqn:uncertainty_exponent_hausdorff_dim}. That is, by subtracting the Hausdorff dimension of the $\gamma$-Cantor set -- the boundary of the $\gamma$-Cantor basins -- from unity (the topological dimension of the unit interval). This yields
\begin{equation}
\label{eqn:uncertainty_gamma_cantor_set}
\alpha(\gamma) = 1 + \frac{\ln{2}}{\ln{\left( \frac{1 - \gamma}{2}\right)}}, \qquad \gamma \in \left( 0, 1 \right).
\end{equation}
The function $\alpha(\gamma)$ is monotonically increasing on $\gamma \in \left( 0 , 1 \right)$ with $\lim_{\gamma \rightarrow 0^{+}} \alpha(\gamma) = 0$ and $\lim_{\gamma \rightarrow 1^{-}} \alpha(\gamma) = 1$. This indicates that the basins become \emph{more uncertain} (i.e., $\alpha$ increases) as we decrease the parameter $\gamma$.

\begin{figure}
\begin{center}
\begin{tabular}{c}
{
\subfigure[$\gamma = \frac{1}{2}$]{\includegraphics[width=0.42\textwidth]{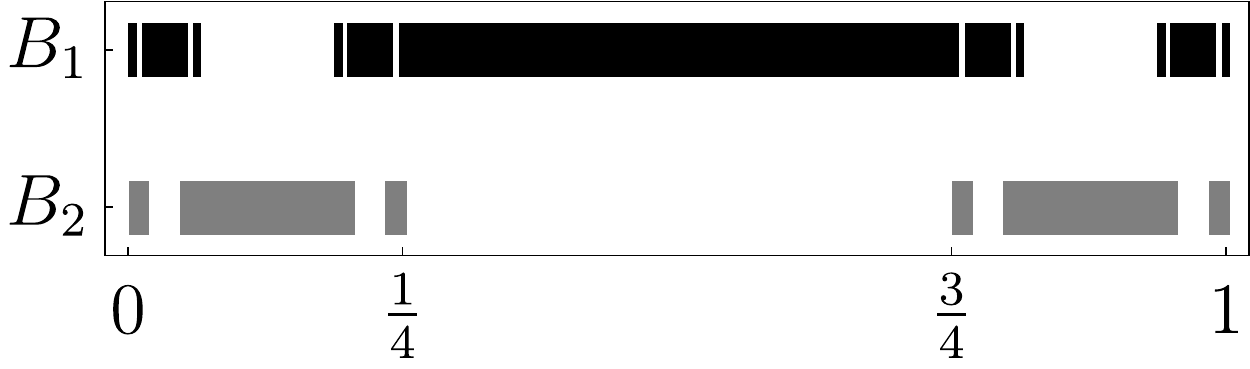} \label{fig:cantor_basins_2_middle_half}}
}
\\
{
\subfigure[$\gamma = \frac{1}{3}$]{\includegraphics[width=0.42\textwidth]{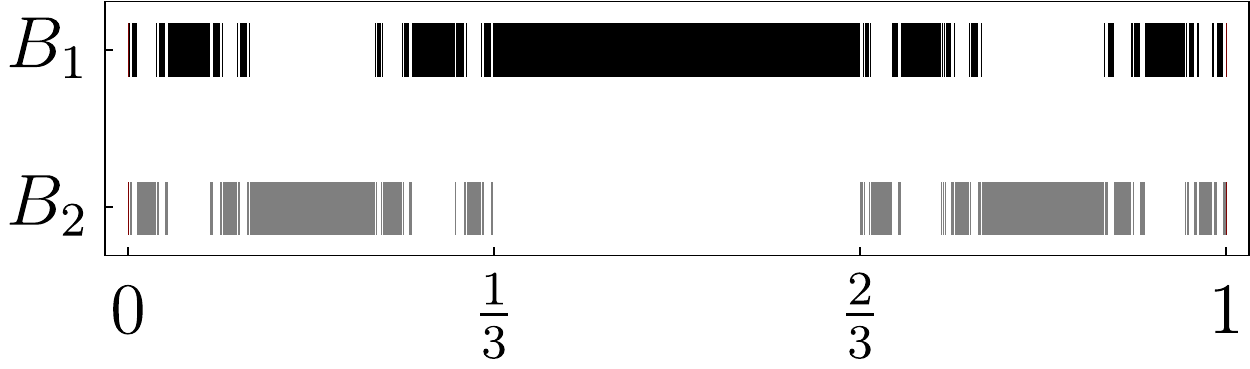} \label{fig:cantor_basins_2_middle_third}}
}
\\
{
\subfigure[$\gamma = \frac{1}{9}$]{\includegraphics[width=0.42\textwidth]{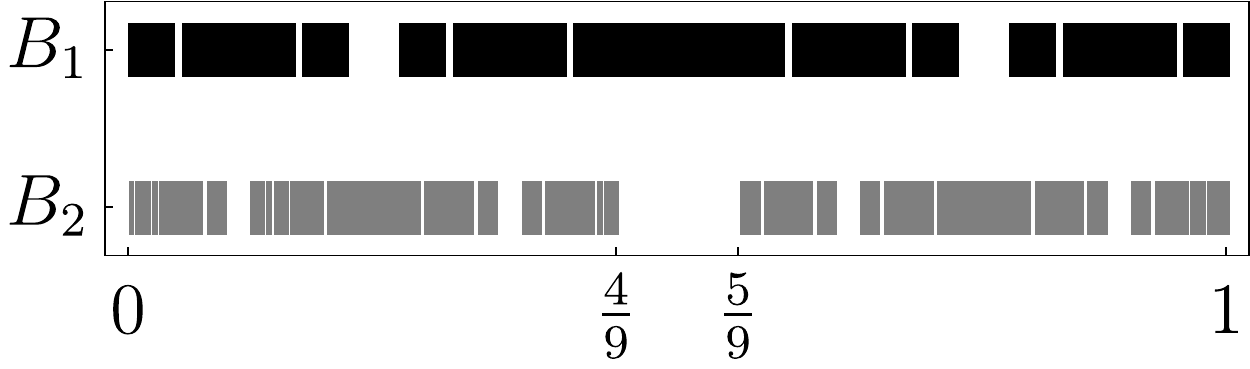} \label{fig:cantor_basins_2_middle_ninth}}
}
\end{tabular}
\begin{tabular}{c}
{
\subfigure[Uncertainty exponent]{\includegraphics[width=0.44\textwidth]{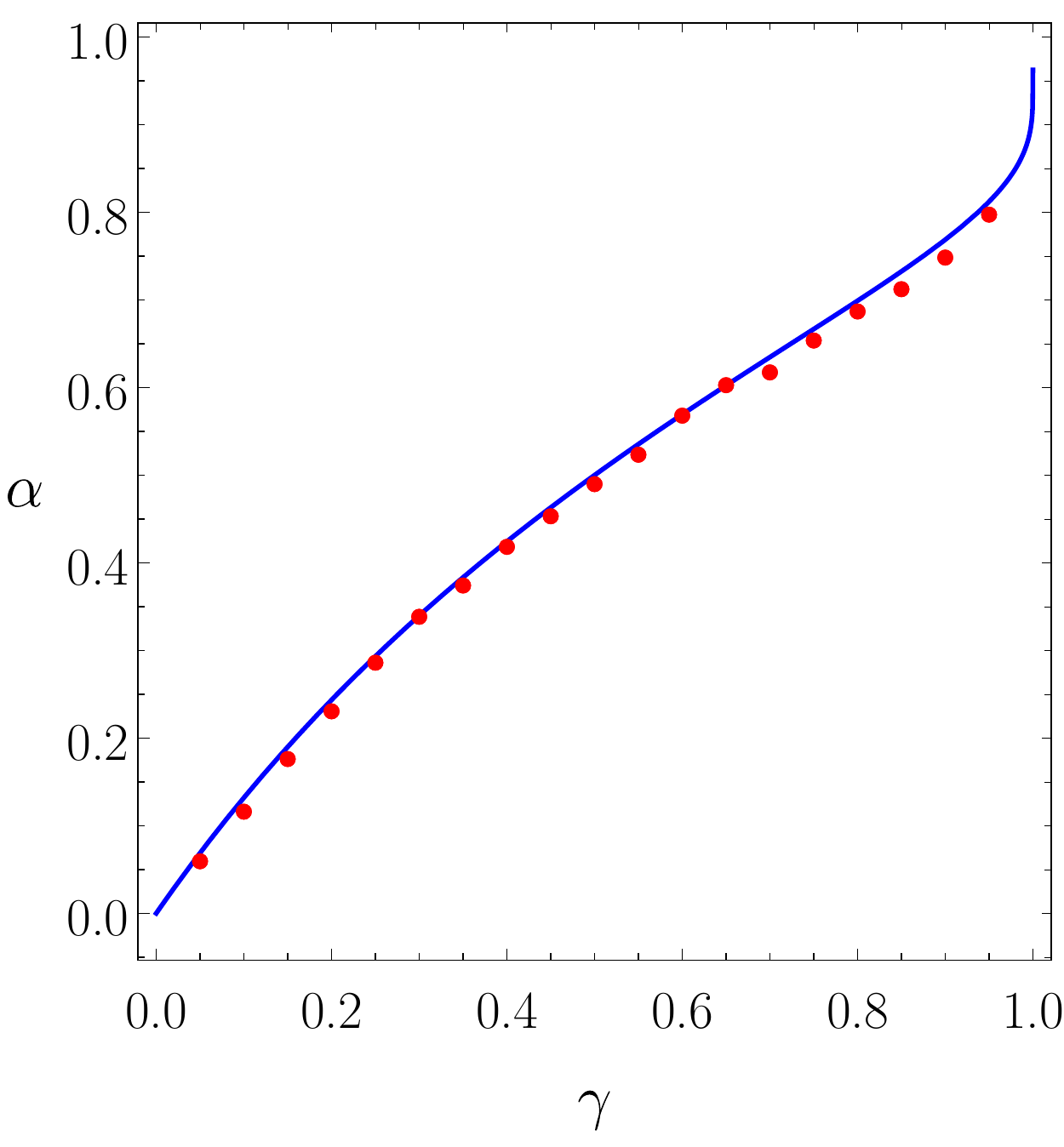}
\label{fig:cantor_basin_uncertainty_exponent}}
}
\end{tabular}
\caption{Uncertainty exponent for Cantor basins. (a)--(c) Two-colour $\gamma$-Cantor basins $\{ B_{1}, B_{2} \}$ for a selection of values of $\gamma$. (d) Uncertainty exponent $\alpha$ for the $\gamma$-Cantor basins. The graph of the analytical expression is shown in blue; the numerical values, computed for $\gamma \in \left\{ 0.05, 0.10, \ldots 0.95 \right\}$ using the method described in the text, are shown in red. The uncertainty exponent increases with $\gamma$. This indicates that the smaller $\gamma$ is, the more uncertain (i.e., the more ``fractalised'') the $\gamma$-Cantor basins are. \label{fig:uncertainty_exponent_for_cantor_basin}}
\end{center}
\end{figure}

In Figures \ref{fig:cantor_basins_2_middle_half}--\ref{fig:cantor_basins_2_middle_ninth} we show examples of two-colour $\gamma$-Cantor basins for $\gamma \in \left\{ \frac{1}{2}, \frac{1}{3}, \frac{1}{9} \right\}$. We consider the case of two basins because the uncertainty exponent is concerned only with the \emph{boundary} of the basins, not the number of basins. We recall that the boundary of the $\gamma$-Cantor basins is the middle-$\gamma$ Cantor set, no matter how many basins we choose to construct; see Section \ref{sec:construction_cantor_like_set}. One can see by inspection of Figures \ref{fig:cantor_basins_2_middle_half}--\ref{fig:cantor_basins_2_middle_ninth} that the basins become more ``fractalised'' (i.e., more uncertain) as one increases $\gamma$.

Figure \ref{fig:cantor_basin_uncertainty_exponent} shows the uncertainty exponent for the $\gamma$-Cantor basins, calculated using (i) the closed-form expression for the uncertainty exponent \eqref{eqn:uncertainty_gamma_cantor_set} [blue curve]; and (ii) the numerical method outlined above with $\gamma \in \left\{ 0.05, 0.10, \ldots 0.95 \right\}$ and a resolution of $N = 10^{8}$ initial conditions [red points]. We see that there is good agreement between the analytical and numerical results. Better agreement could be obtained for the more uncertain basins (which have $\gamma \approx 1$) by calculating $\alpha$ numerically using a finer grid of initial conditions; however, this would be more computationally expensive.
%
%

\subsection{Uncertainty exponent for binary black hole shadows}
\label{sec:uncertainty_exponent_shadows}

In this section, we compute the uncertainty exponent of the boundary of one-dimensional equal-mass Majumdar--Papapetrou di-hole shadows of constant $Y$, in the cases $d = 1$ and $d = 2$, for a viewing angle of $\theta = \frac{\pi}{2}$. Recall that each slice of constant $Y$ is related to a fixed value of the azimuthal angular momentum $p\ind{_\phi}$, as described in Section \ref{sec:binary_black_hole_shadows}. Here, we compute the uncertainty exponent using the numerical method outlined in Section \ref{sec:uncertainty_definitions_numerical}, for one-dimensional shadows with $Y \in \left\{ -8.0, 7.9, \ldots, 8.0 \right\}$, using $N = 10^{7}$ initial conditions in each slice. Once the uncertainty exponent has been calculated, one can obtain the Hausdorff dimension of the boundary using \eqref{eqn:uncertainty_exponent_hausdorff_dim} with $D = 1$. In the following analysis, we discuss only $Y \geq 0$, because the equatorial ($\theta = \frac{\pi}{2}$) shadows are symmetric about $Y = 0$.

\begin{figure}
\begin{center}
\subfigure[$d = 2$]{
\includegraphics[height=0.45\textwidth]{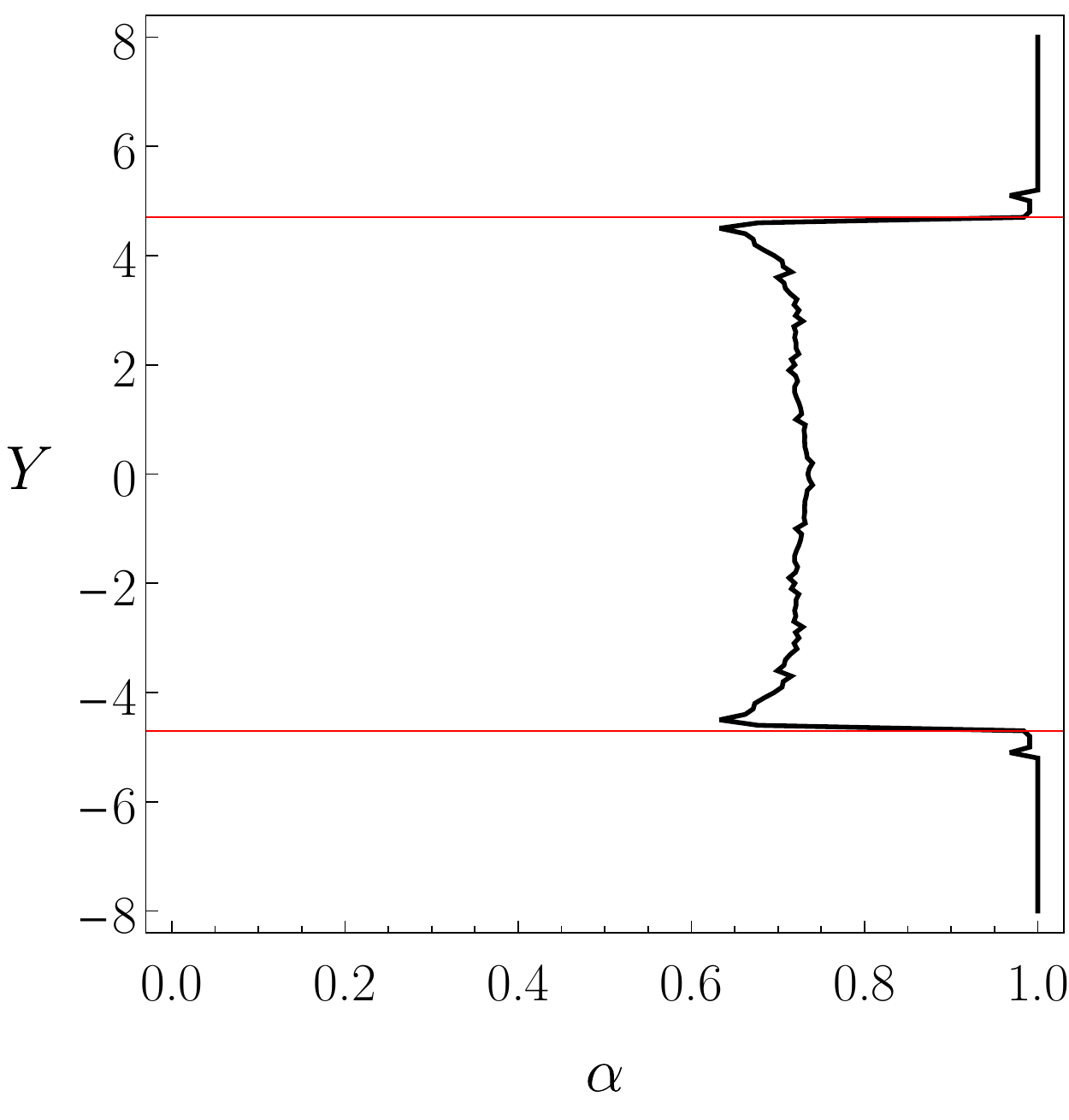} \label{fig:mp_uncertainty_exponent_d2} \hspace{1em}}
\subfigure[$d = 2$]{
\includegraphics[height=0.45\textwidth]{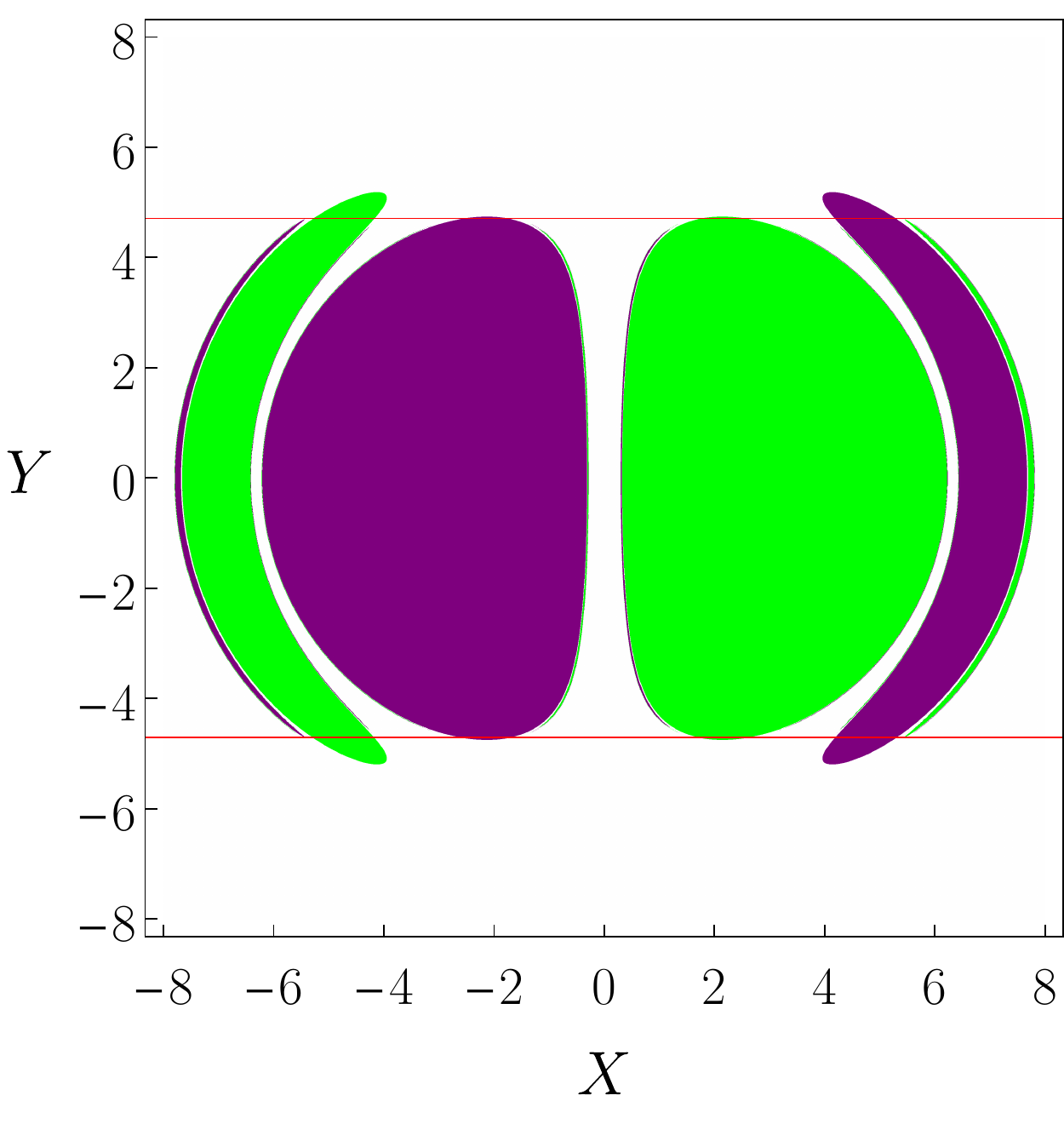}\label{fig:mp_shadow_uncertainty_exponent_d2}}
\subfigure[$d = 1$]{
\includegraphics[height=0.45\textwidth]{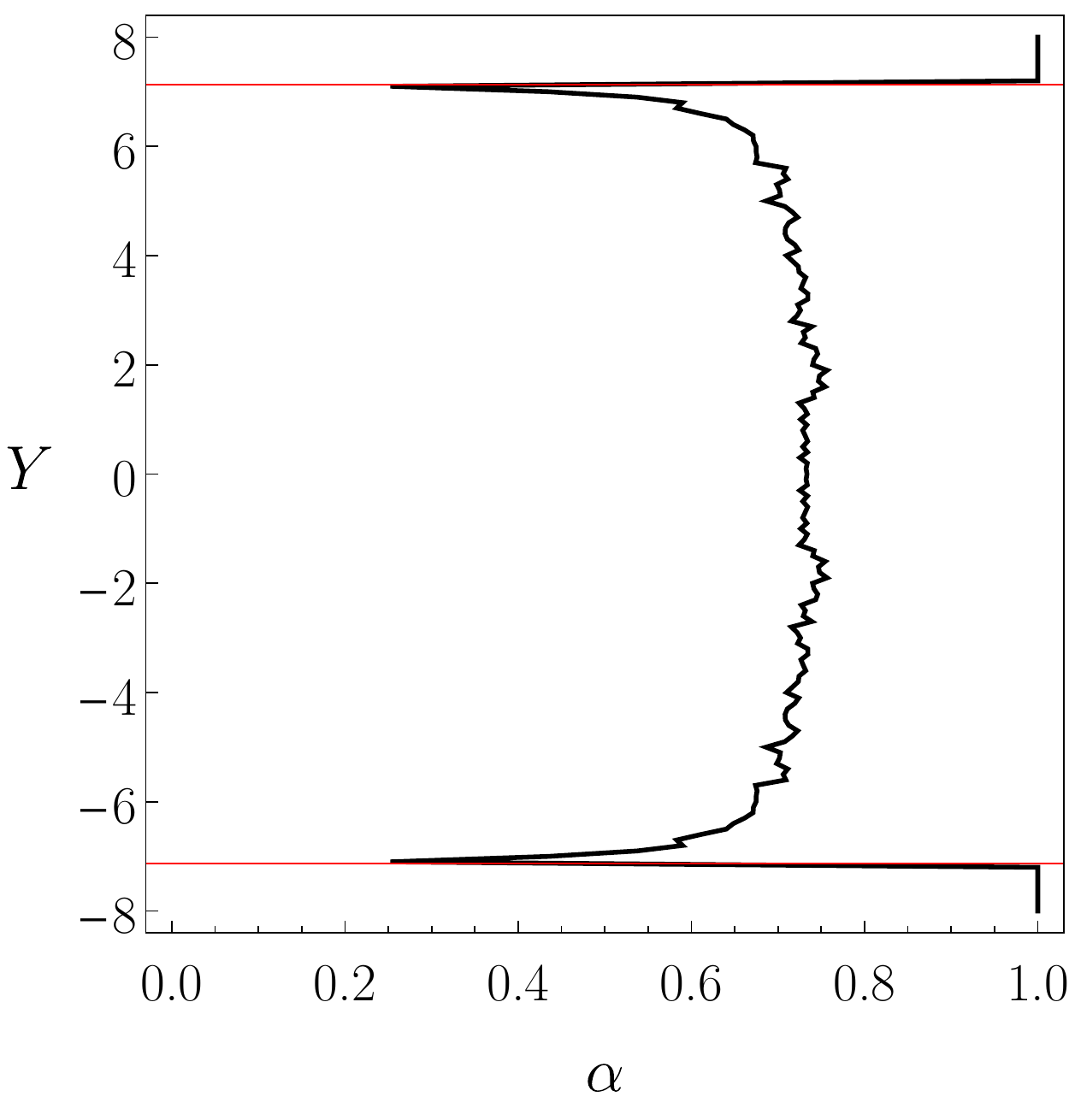} \label{fig:mp_uncertainty_exponent_d1} \hspace{1em}}
\subfigure[$d = 1$]{
\includegraphics[height=0.45\textwidth]{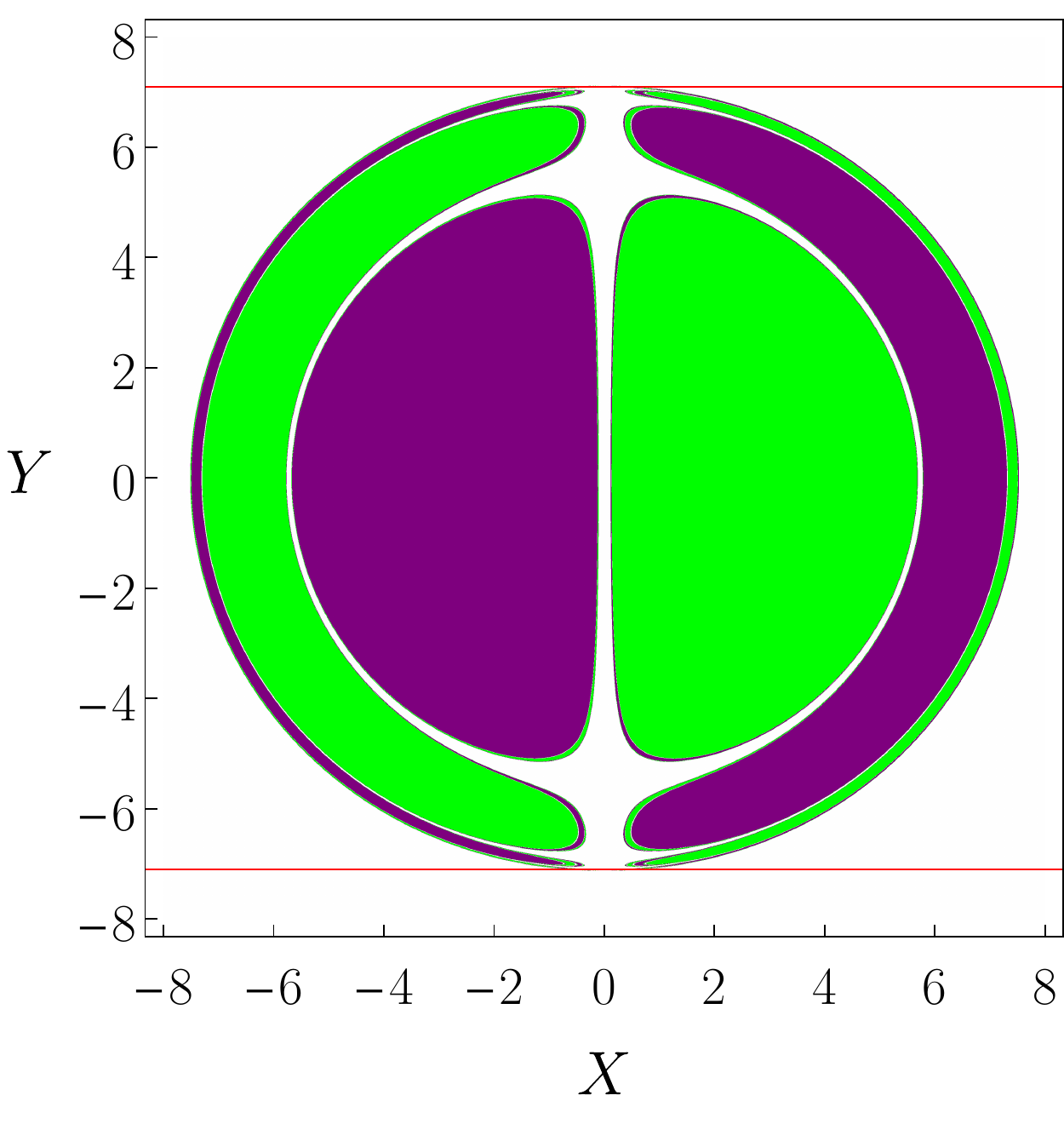}\label{fig:mp_shadow_uncertainty_exponent_d1}}
\caption{Uncertainty exponent $\alpha(Y)$ for one-dimensional slices ($Y = \text{constant}$) of Majumdar--Papapetrou di-hole shadows. (a)--(b) For $d = 2$, the uncertainty exponent is able to distinguish between the regular and fractal (Cantor-like) parts of the shadow. The critical value of $Y$, at which the type II/III fundamental orbits coincide, is shown as a red horizontal line. (c)--(d) In the case $d = 1$, the uncertainty exponent confirms that the entire shadow boundary is fractal. The end of the shadow, corresponding to $p\ind{_{\phi}} = {p\ind{_{\phi}}}^{\ast}$, occurs at $Y = 7.13$. This is shown as a red horizontal line. We have transposed the graphs of $\alpha$ in (a) and (c) to aid comparison with the shadow images in (b) and (d).}
\label{fig:mp_uncertainty_exponent}
\end{center}
\end{figure}

Let us first consider the case $d = 2$, shown in Figures \ref{fig:mp_uncertainty_exponent_d2} and \ref{fig:mp_shadow_uncertainty_exponent_d2}. As outlined in Section \ref{sec:binary_black_hole_shadows}, we observe Cantor-like and regular boundaries in the one-dimensional cross sections of this shadow; see Figure \ref{fig:mp_shadow_decomposition}. We anticipate that the transition from Cantor-like to regular will occur when the outer (type II/III) fundamental orbits coincide. This happens when $p\ind{_{\phi}} = \hat{p}\ind{_{\phi}} \approx 5.09$, which corresponds to $Y \approx 4.71$. Figure \ref{fig:mp_uncertainty_exponent_d2} shows the uncertainty exponent for this case. This figure is transposed, so that $Y$ runs along the vertical axis, to aid comparison with the shadow presented in Figure \ref{fig:mp_shadow_uncertainty_exponent_d2}. The transition from regular to Cantor-like is indicated using a red horizontal line in Figures \ref{fig:mp_uncertainty_exponent_d2} and \ref{fig:mp_shadow_uncertainty_exponent_d2}. We find that $\alpha < 1$ for $Y < 4.71$, so this part of the shadow is fractal; whereas $\alpha \approx 1$ for $Y > 4.71$, which indicates that this part of the shadow is not fractal. This agrees with our understanding of the shadow structure and the fundamental orbits, outlined in Chapter \ref{chap:binary_black_hole_shadows} and Section \ref{sec:mp_black_hole_shadows_wada}. Crucially, the uncertainty exponent is able to distinguish between the fractal and non-fractal parts of the shadow.

We now consider the uncertainty exponent for a di-hole with $d = 1$. Figure \ref{fig:mp_uncertainty_exponent_d1} shows that, in this case, $\alpha < 1$ across the whole shadow, which indicates that the entire shadow boundary is fractal. We observe that the uncertainty exponent attains its minimum value ($\alpha = 0.252$) at $Y = 7.1$; this is the most uncertain (or ``most fractal'') part of the shadow. Beyond this value, there is a discontinuity in the uncertainty exponent; we see that $\alpha = 1$ for $Y > 7.1$, which is the end of the black hole shadow. The end of the shadow corresponds to $p\ind{_{\phi}} = {p\ind{_{\phi}}}^{\ast}$: a ray incident from $r \rightarrow \infty$ with $p\ind{_{\phi}} \geq {p\ind{_{\phi}}}^{\ast}$ is not permitted to fall into the black holes by the critical contour $h(\rho, z) = p\ind{_{\phi}}$. The value of $Y$ corresponding to ${p\ind{_{\phi}}}^{\ast}$ is found to be $Y = 7.13$. This agrees well with the results of Figures \ref{fig:mp_uncertainty_exponent_d1} and \ref{fig:mp_shadow_uncertainty_exponent_d1}.

In the case $d = 1$, the one-dimensional shadows become more uncertain as one increases the value of $Y$ (for $Y < 7.13$). This may be due to the existence of a ``pocket'' with three escapes, leading to the black holes and to spatial infinity; see the discussion in Section \ref{sec:non_planar_rays}. The escapes become narrower as one increases $p\ind{_{\phi}}$ towards ${p\ind{_{\phi}}}^{\ast}$. This results in greater final-state unpredictability, because the ``pocket'' acts as a randomising region.

For the highly symmetric case $d = 1$, as $p\ind{_{\phi}} \rightarrow {p\ind{_{\phi}}}^{\ast}$, the width of the three escapes in phase space tends to zero. In this limit, we also anticipate that $\alpha \rightarrow 0$. In principle, this could be the case; however, the computation of $\alpha$ is difficult in this regime, due to the fact that the volume of the shadow tends to zero. This is similar to the limit $\gamma \rightarrow 0$ for the $\gamma$-Cantor basins, described in Section \ref{sec:uncertainty_definitions_numerical}.
%

\section{The Wada property}
\label{sec:wada_property}

\subsection{Wada basins}
\label{sec:wada_basins}

A collection of three or more open sets $S = \left\{ B_{i} \, | \, i \in \left\{1, 2, \ldots, N \right\}, N \geq 3 \right\}$ are said to satisfy the \emph{Wada property} if each point $p$ on the boundary of any of the $B_{i} \in S$ is on the boundary of \emph{all} $B_{i} \in S$. Such sets are difficult to visualise, and at first seem impossible to realise; however, the first explicit example of sets which possess this property -- the \emph{lakes of Wada} -- was given by Yoneyama in 1917 \cite{Yoneyama1917}.

The lakes of Wada are constructed using the following simple geometrical procedure \cite{Yoneyama1917, HockingYoung1988}. We start with a double annulus; this is taken to be an island in a red ocean with two lakes, one of which is filled with green water, the other with blue. At time $t = 0$, we dig a canal from the red ocean, such that each point on the land is within a distance of $1$ unit of red ocean water. At time $t = \frac{1}{2}$, we dig a canal from the green lake, which brings green water to within a distance $\frac{1}{2}$ from each point on the land. At time $t = \frac{3}{4}$, we dig a canal from the blue lake so that each point on the land is within a distance of $\frac{1}{3}$ from blue water. At time $\frac{7}{8}$, a canal is dug from the first canal so that each point on land is within a distance $\frac{1}{4}$ of every point on land. We continue in this fashion until time $t = 1$, when the remaining land is a continuum in the plane of zero measure which bounds \emph{three} open sets, i.e., the red ocean, the green lake and the blue lake. The boundary of the lakes of Wada is an example of an \emph{indecomposable continuum}, a concept first described by Brouwer \cite{Brouwer1910}.

The Wada property began as a topological curiosity; however, Kennedy and Yorke \cite{KennedyYorke1991} demonstrated that is quite common in dynamics, arising generically in the basins of systems with three or more attractors or escapes. Recall from Section \ref{sec:chaotic_dynamical_systems} that a point $p$ is a \emph{boundary point} of the basin $B$ if every open neighbourhood of $p$ contains at least one point in $B$ and one point not in $B$. If a point $p$ is a boundary point of $N \geq 3$ basins $B_{i}$, $i \in \left\{1, 2, \ldots, N \right\}$, then $p$ is said to be a \emph{Wada point}. If all the boundary points of a basin $B$ are Wada points, then $B$ is called a \emph{Wada basin}. The Wada property may be rephrased as follows: a point $p$ on the basin boundary is a Wada point if and only if every open neighbourhood of $p$ has a non-empty intersection with at least three different basins (open sets).

In their seminal paper, Kennedy and Yorke \cite{KennedyYorke1991} argue that the Poincar\'{e} return map of the forced damped pendulum with four attractors exhibits the Wada property. The Wada property has since been revealed, using a range of techniques (see Section \ref{sec:test_wada_basins}), in a variety of dynamical systems, such as the forced damped pendulum, the three-disc model, the H\'{e}non--Heiles Hamiltonian system, and others \cite{AguirreVianaSanjuan2009}.

Here, we are primarily concerned with open Hamiltonian systems, so the subsequent discussion will refer only to exit basins; however, this could be generalised to include basins of attraction for dissipative systems. When multiple escapes are present in phase space, we typically wish to determine to which basin a point belongs, i.e., through which escape the particle will leave the system when integrated forwards in time. If the basin boundary is sufficiently complicated, a small uncertainty in fixing the initial conditions may result in a large uncertainty in the final state of the system. In situations where the basin boundary is a Wada boundary, one encounters a high level of indeterminacy and an extreme sensitive dependence on initial conditions, despite the underlying system being fully deterministic. The presence of Wada basins in phase space is therefore a hallmark of deterministic chaos in non-linear dynamical systems.

\subsection{The Wada property in Cantor basins}
\label{sec:wada_cantor_symbolic_dynamics}

It is clear from Yoneyama's geometrical construction of the lakes of Wada, reviewed in Section \ref{sec:wada_property}, that the Wada property is intimately related to iteration. Furthermore, we have seen that a Cantor set can be generated using a similar iterative procedure to that which is used to construct the lakes of Wada. In this section, we demonstrate how the Wada property can arise generically in basins with a Cantor-like boundary.

Recall the Cantor basins presented in Section \ref{sec:construction_cantor_like_set}. One can construct a set of $N \geq 3$ basins $\left\{ B_{i} \right\}_{i = 1}^{N}$ from the complement of the middle-thirds Cantor set $\mathcal{C}$ in $I = \left[ 0, 1 \right]$ as follows. Denote by $J_{k}$ the union of the $2^{k - 1}$ open intervals which are \emph{removed} from $I$ at the $k$th iteration in the construction of $\mathcal{C}$. Now, add $J_{k}$ to the basin $B_{i}$, where $k = i \bmod{N}$. The $N$ basins can then be expressed as an infinite union of disjoint open sets:
\begin{equation}
B_{i} = \bigcup_{k = i \bmod{N}} J_{k}, \qquad i \in \left\{1, 2, \ldots, N \right\}.
\end{equation}
Let us colour the open intervals which are removed at each step using $N$ colours, i.e., we assign the intervals a label from the set $\left\{ c_{1}, c_{2}, \ldots, c_{N} \right\}$. The label is chosen in accordance with the iteration: all intervals which belong to the $i$th basin are labelled by the colour $c_{i}$. Clearly, $\partial B_{1} = \partial B_{2} = \ldots = \partial B_{N} = \mathcal{C}$; i.e., all of the basins $B_{i}$ share a common boundary, namely the Cantor set.  Our aim is to demonstrate that the basins $\left\{ B_{i} \right\}_{i = 1}^{N}$ possess the Wada property: every open neighbourhood of a point which belongs to the basin boundary has a non-empty intersection with all $N$ different basins.

As described in Section \ref{sec:construction_cantor_like_set}, the construction of the Cantor basins can be understood in terms of symbolic dynamics for the Cantor set, where each interval is encoded using a string of symbols from a symbolic alphabet $\mathcal{A} = \left\{ 0, 1, 2 \right\}$. At each iteration, the left-hand (closed) interval is assigned the digit $0$; the middle (open) interval is assigned the digit $1$; and the right-hand (closed) interval is assigned the digit $2$. The intervals which are assigned a colour (i.e., which belong to an exit basin) are all those whose symbolic representation ends in the digit $1$. We iterate on the remaining intervals (corresponding to the digits $0$ and $2$) \emph{ad infinitum}.

The boundary points are represented in symbolic dynamics by sequences of infinite length which do not contain the symbol $1$. We note that sequences which terminate in the digit $1$ correspond to \emph{open} intervals, i.e., they do not contain the endpoints. Let $S_{(k)}$ be a sequence of length $k$ which does not contain the symbol $1$. The endpoints of the interval with representation $S_{(k)} 1$ have symbolic representation $S_{(k)} 1 \overline{0}$ and $S_{(k)} 1 \overline{2}$. However, these points also admit alternative representations in symbolic dynamics, namely $S_{k} 0 \overline{2}$ and $S_{k} 2 \overline{0}$, respectively. Clearly, the latter representation shows that these points are boundary points, and should not be included in the interval $S_{(k)} 1$. Thus $S_{(k)} 1$ is open.

The open intervals in the basin $B_{i}$ (which are labelled using the colour $c_{i}$) are represented in symbolic dynamics by sequences of length $l$, with $l = i \bmod{N}$, which terminate in the symbol $1$, and whose first $l - 1$ digits do not contain the symbol $1$. For example, the open interval labelled $020021$ (which can be read as left-right-left-left-right-middle) belongs to the basin $B_{i}$, where $6 = i \mod{N}$.

It is possible to use the symbolic dynamics for the Cantor set to demonstrate that the $N$-colour Cantor basins possess the Wada property. Given any boundary point $p \in \mathcal{C}$, whose symbolic representation is of the form $p_{1} p_{2} p_{3} \ldots$, where each $p_{i} \in \left\{ 0, 2 \right\}$, one can always find a point in \emph{any} of the $N$ basins which is arbitrarily close to $p$. For if we are given a point $q \in B_{i}$, with symbolic representation $p_{1} p_{2} \ldots p_{l} 1$, it is always possible to find a point $q^{\prime} \in B_{i}$ with symbolic representation $p_{1} p_{2} \ldots p_{l^{\prime}} 1$, where $l^{\prime} = l \bmod{N}$ and $l^{\prime} > l$. The point $q^{\prime}$ is closer to $p$ than $q$ is, but is still in the basin $B_{i}$. Since $p$ and $B_{i}$ are arbitrary, this demonstrates that every open neighbourhood of each boundary point $p$ has non-empty intersection with \emph{all} of the $N$ basins $\left\{ B_{1}, B_{2}, \ldots, B_{N} \right\}$. Thus, the $N$-colour Cantor basins (with $N \geq 3$) possess the Wada property. This simple argument illustrates how the Wada property may arise generically in fractal sets with Cantor-like structure, which are typically constructed using iterative processes.

\section{Numerical test for Wada basins}
\label{sec:test_wada_basins}

\subsection{The merging method}
\label{sec:wada_merging_method}

As described in Section \ref{sec:wada_property}, the Wada property arises naturally in non-linear dynamical systems. A number of methods have been proposed to test this striking property. We present a brief overview of these methods here, before reviewing the \emph{merging method} -- a recently developed algorithm which tests whether an exit basin diagram exhibits the Wada property.

In 1996, Nusse and Yorke \cite{NusseYorke1996} established that an unstable manifold which crosses three (or more) basins in phase space could be used to prove the existence of Wada boundaries. The main goal of the work of Nusse and Yorke was to develop criteria for Wada basins which could be verified numerically; Nusse and Yorke applied their method to verify the Wada property in the basins of the H\'{e}non map and the forced damped pendulum \cite{NusseYorke1996}. The so-called \emph{Nusse--Yorke method} has since been used to demonstrate that the basins of Gaspard--Rice three-disc model possess the Wada property \cite{PoonCamposOttEtAl1996}. The Nusse--Yorke method involves the computation of the unstable manifold of a saddle point which lies on the basin boundary. The algorithm therefore requires a detailed knowledge of the dynamical system under consideration, as well as the computation of unstable manifolds; this is often technically and computationally demanding. Indeed, many studies \cite{PoonCamposOttEtAl1996, ToroczkaiKarolyiPentekEtAl1997, AguirreVallejoSanjuan2001, AguirreSanjuan2002} have been devoted to applying the Nusse--Yorke method to a given dynamical system for a particular set of parameters.

In 2015, Daza \emph{et al.} \cite{DazaWagemakersSanjuanEtAl2015} proposed a new computational method to verify the existence of Wada basins in dynamical systems. This approach, referred to as the \emph{grid method}, is based on successive refinements of a grid in phase space. At each stage of the algorithm, further trajectories are computed at higher and higher resolution. This method can be automated in a simple fashion, and does not require a detailed knowledge of the underlying dynamical system. The grid method is therefore easily applied to any dynamical system, and can test for the presence of Wada basins up to finite resolution. One drawback of this approach is that it requires the computation of new trajectories of the system at high resolution. It is therefore computationally expensive.

More recently, with the principal aim of overcoming the drawbacks of the Nusse--Yorke method and the grid method, Daza \emph{et al.} \cite{DazaWagemakersSanjuan2018} developed a novel method to test for Wada basins up to finite resolution. The method, referred to as the \emph{merging method}, does not require a detailed knowledge of the underlying dynamical system, and is computationally inexpensive. Of the three methods outlined here, the merging method is the fastest, and the only one able to provide a reliable test of the existence of Wada basins through a straightforward examination of the basins (at finite resolution), without the need to compute new trajectories of the system or its invariant manifolds.

\begin{figure}[t]
\begin{center}
\subfigure[]{\includegraphics[width=0.22\textwidth]{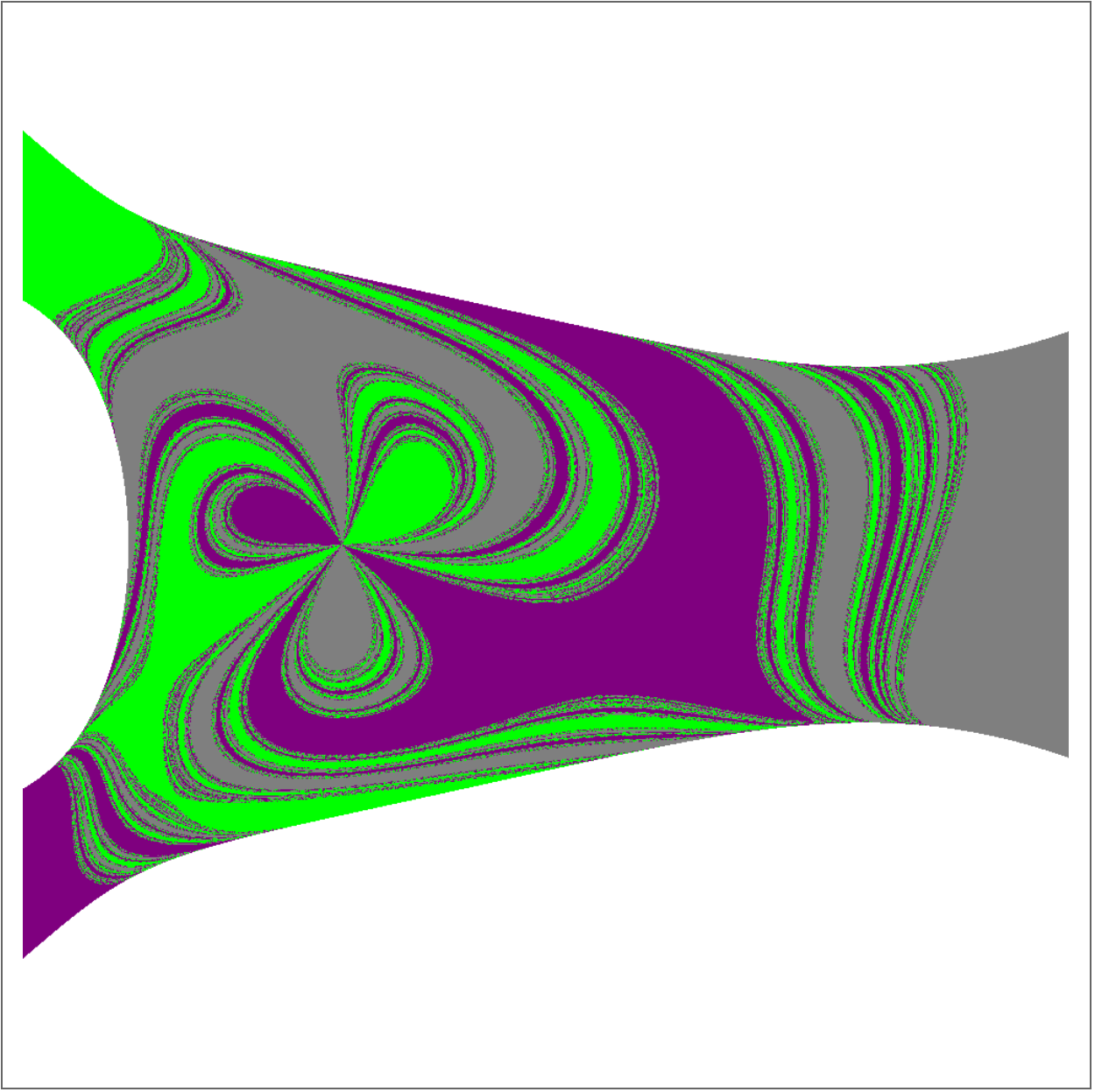} \label{fig:mp_merging_basins_1}}
\subfigure[]{\includegraphics[width=0.22\textwidth]{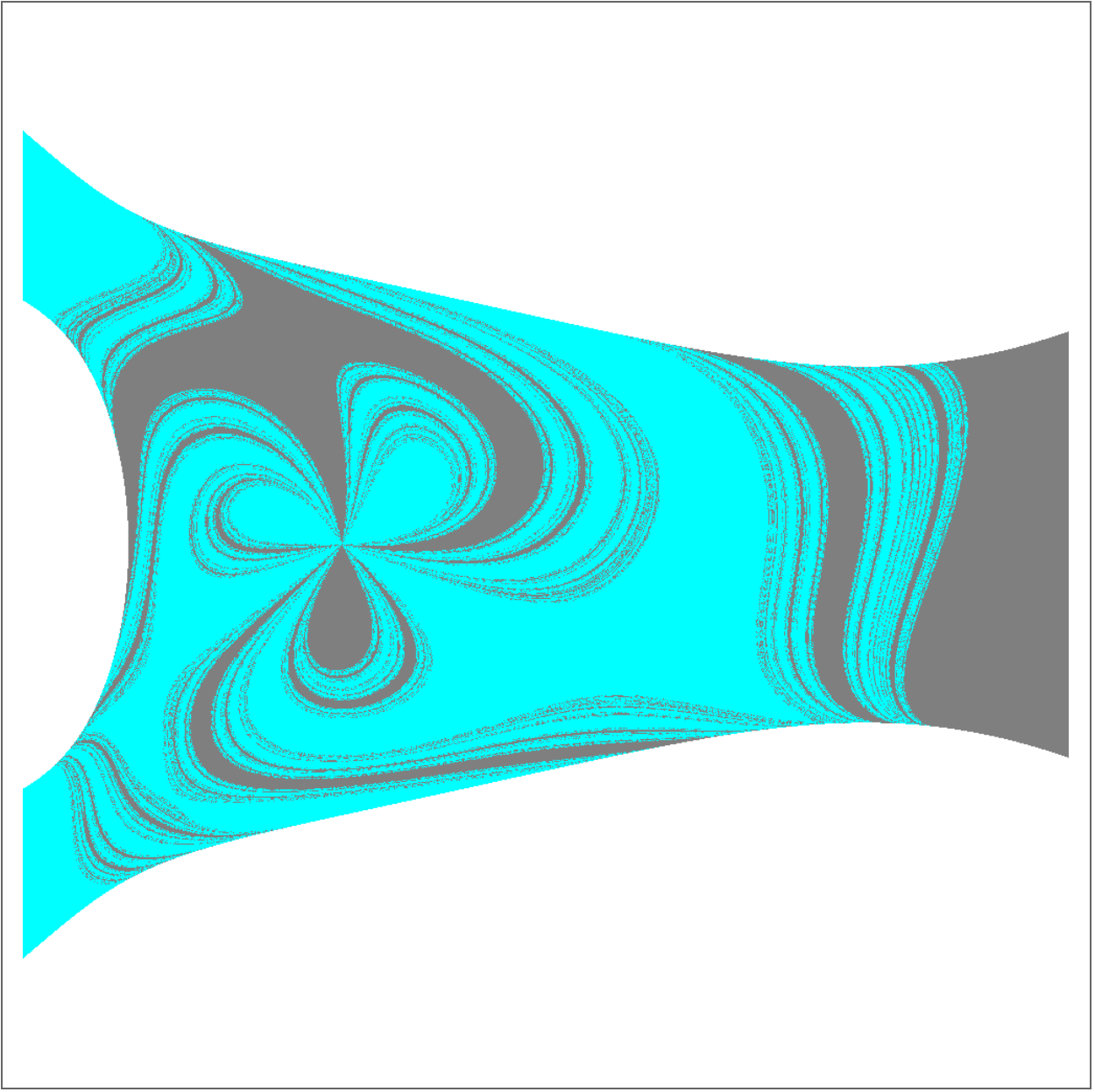} \label{fig:mp_merging_basins_2}}
\subfigure[]{\includegraphics[width=0.22\textwidth]{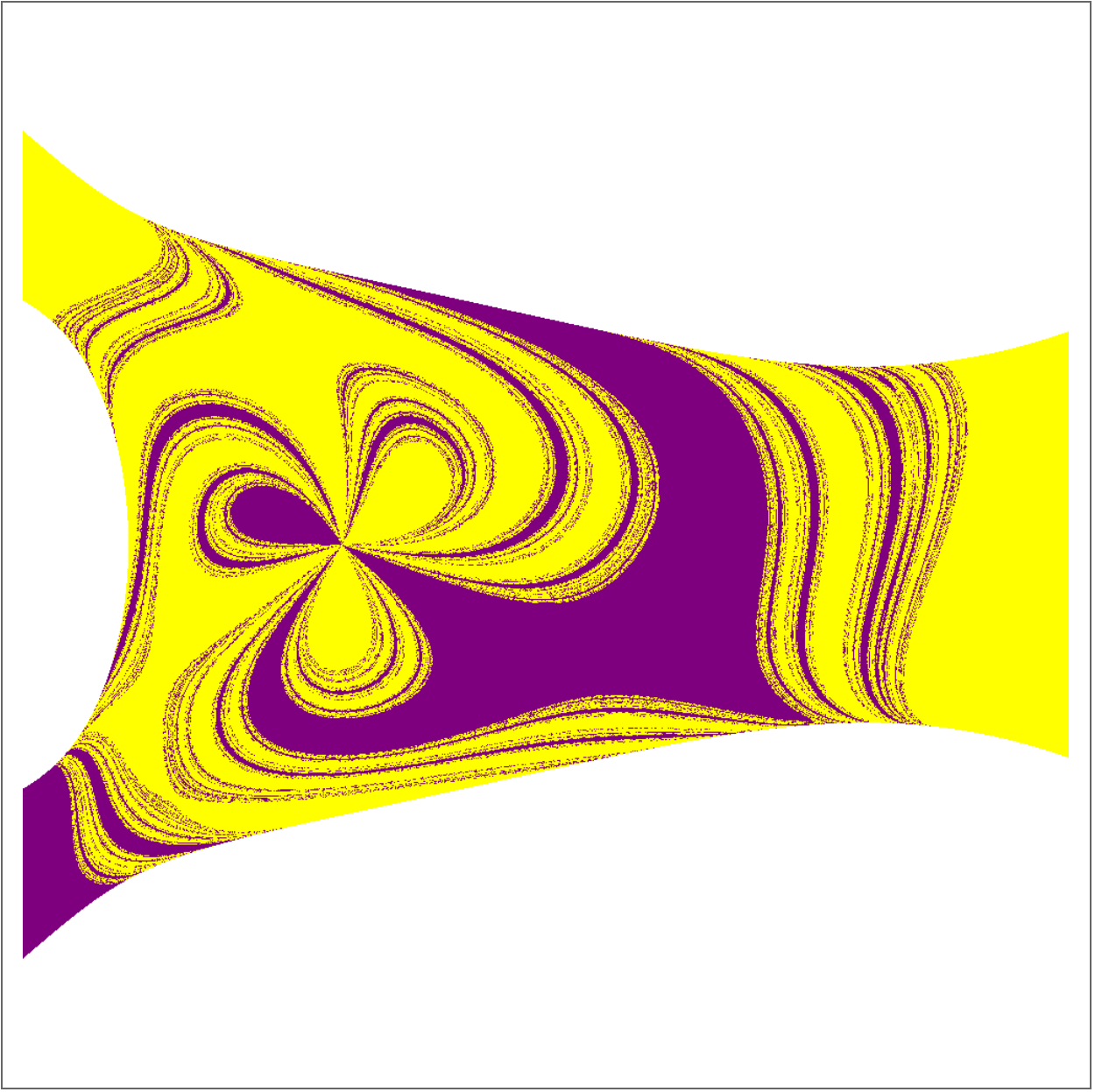} \label{fig:mp_merging_basins_3}}
\subfigure[]{\includegraphics[width=0.22\textwidth]{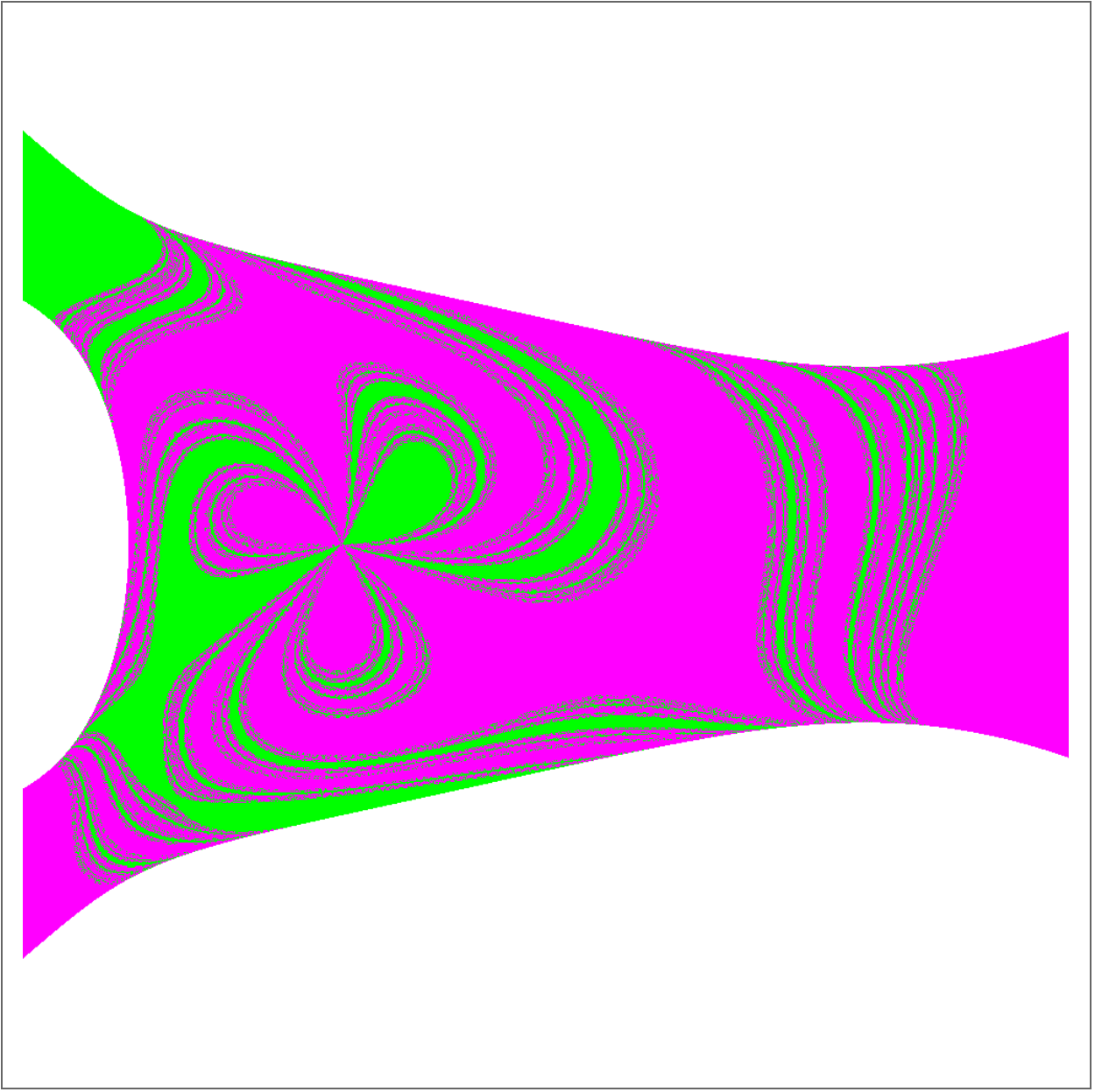} \label{fig:mp_merging_basins_4}}
\subfigure[]{\includegraphics[width=0.22\textwidth]{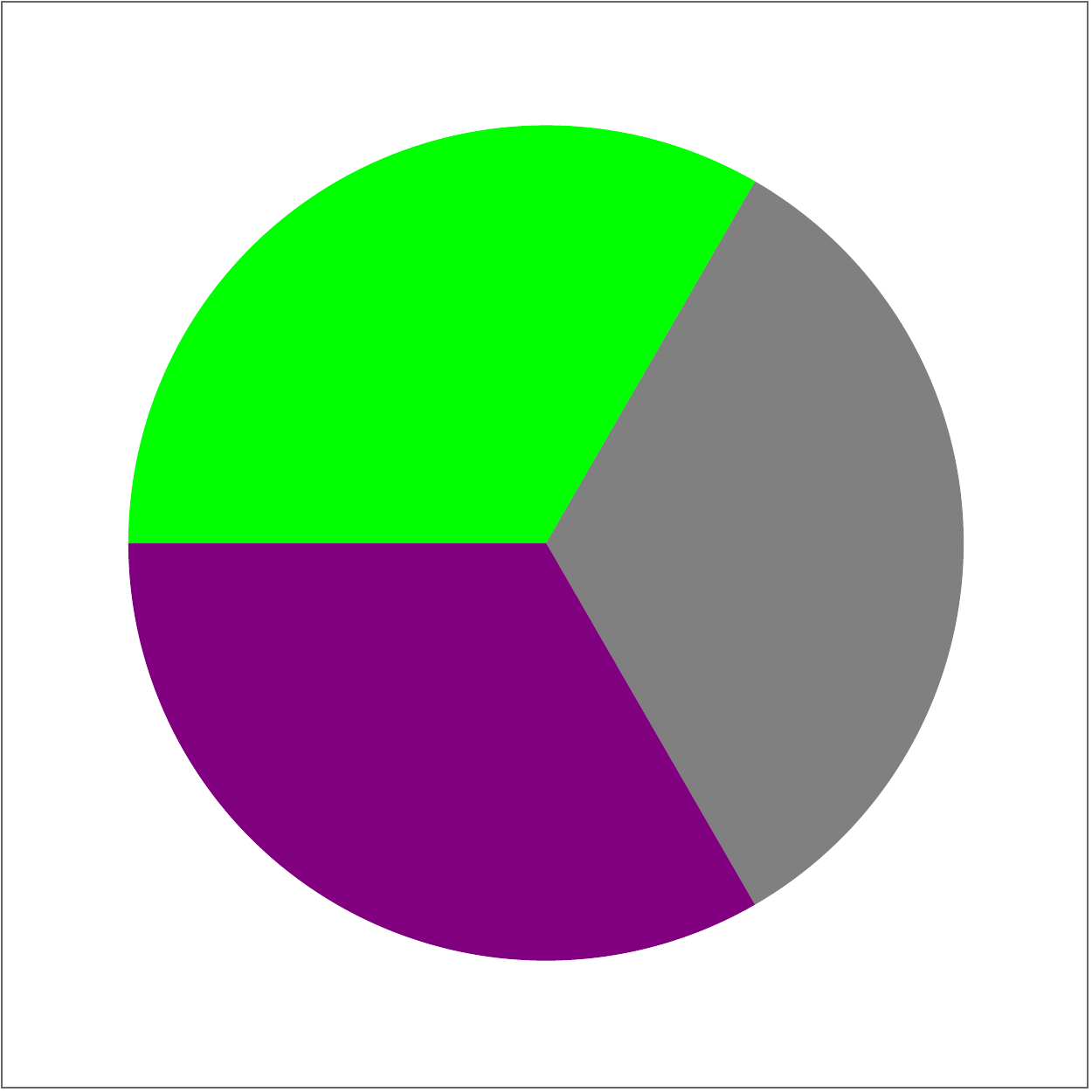} \label{fig:mp_merging_basins_regular_1}}
\subfigure[]{\includegraphics[width=0.22\textwidth]{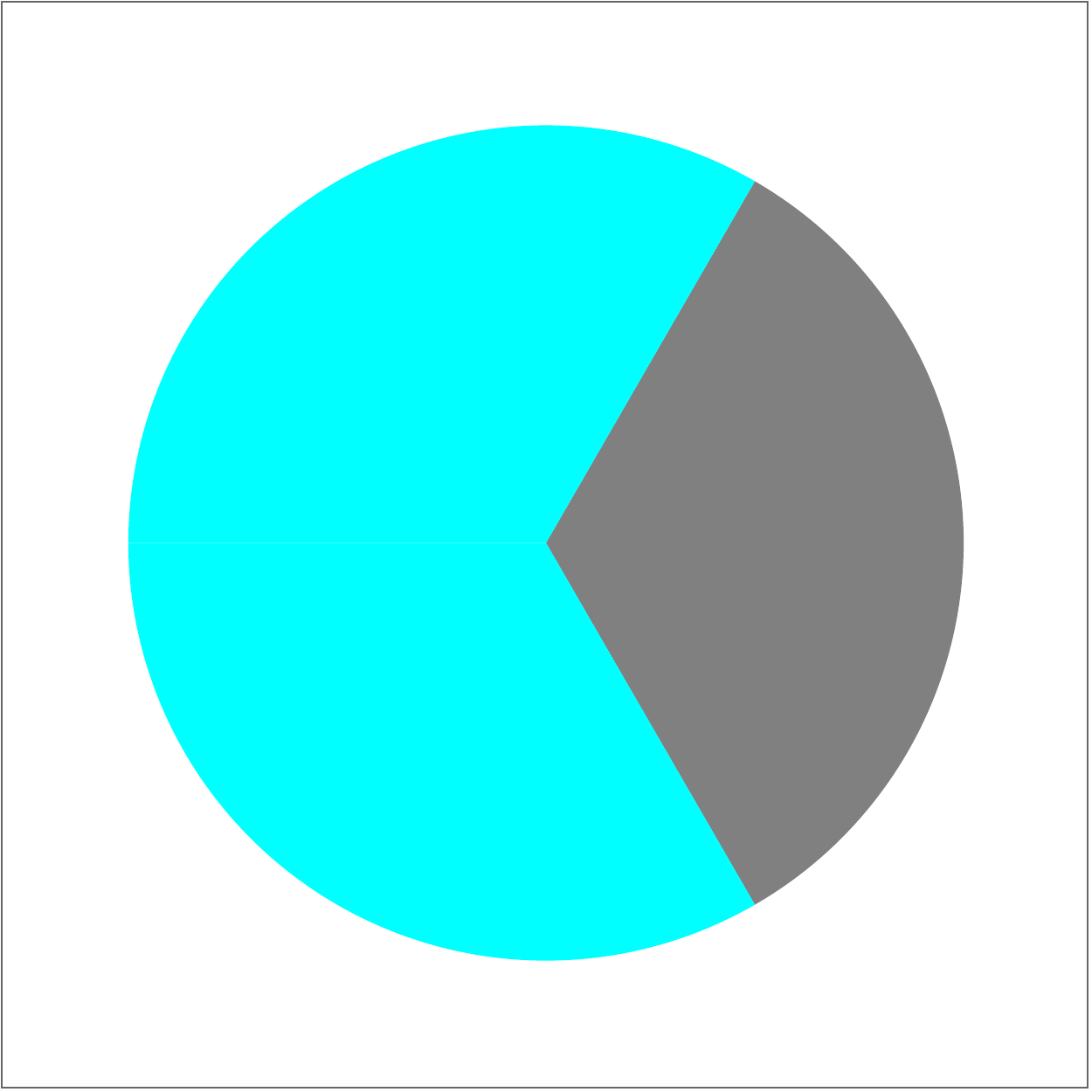} \label{fig:mp_merging_basins_regular_2}}
\subfigure[]{\includegraphics[width=0.22\textwidth]{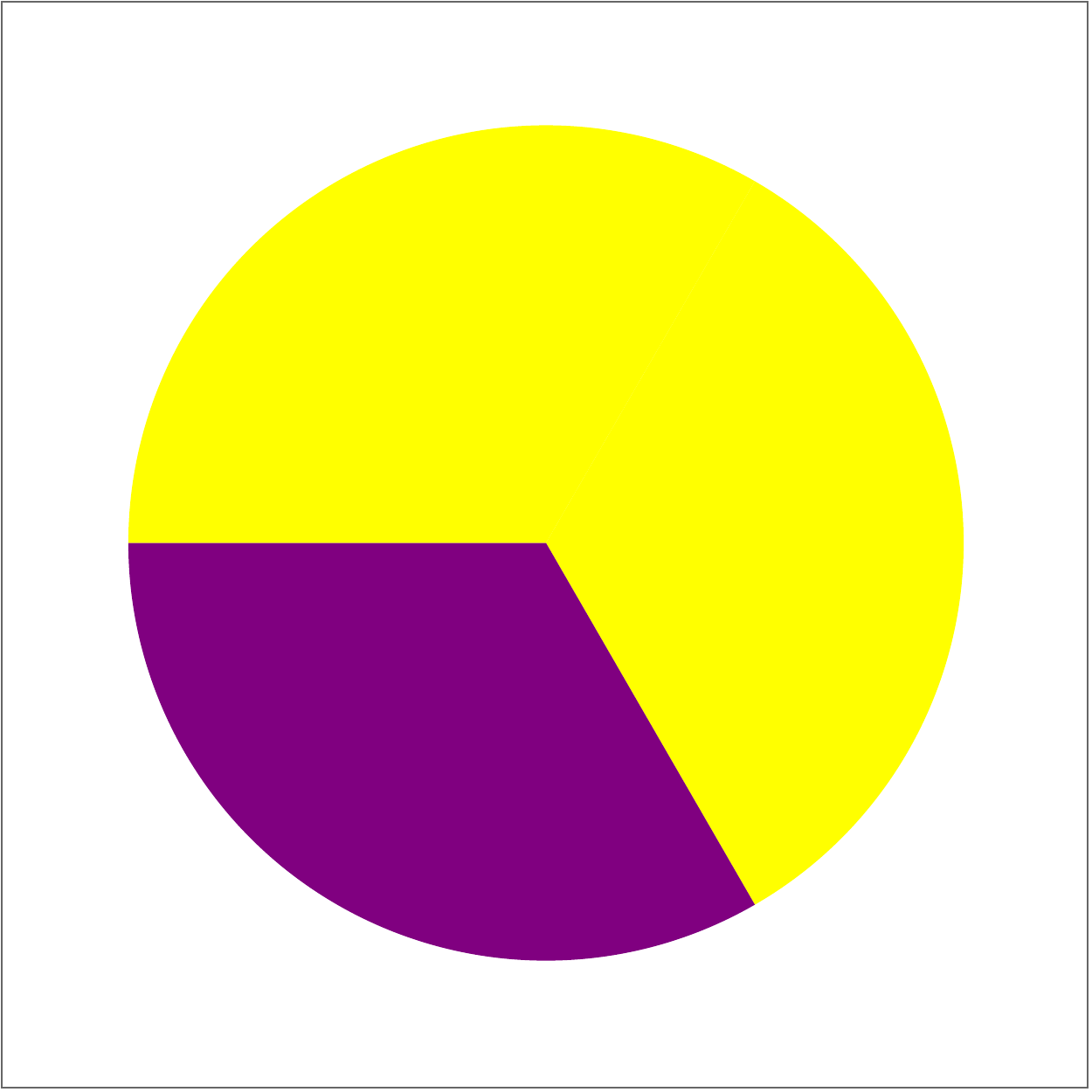} \label{fig:mp_merging_basins_regular_3}}
\subfigure[]{\includegraphics[width=0.22\textwidth]{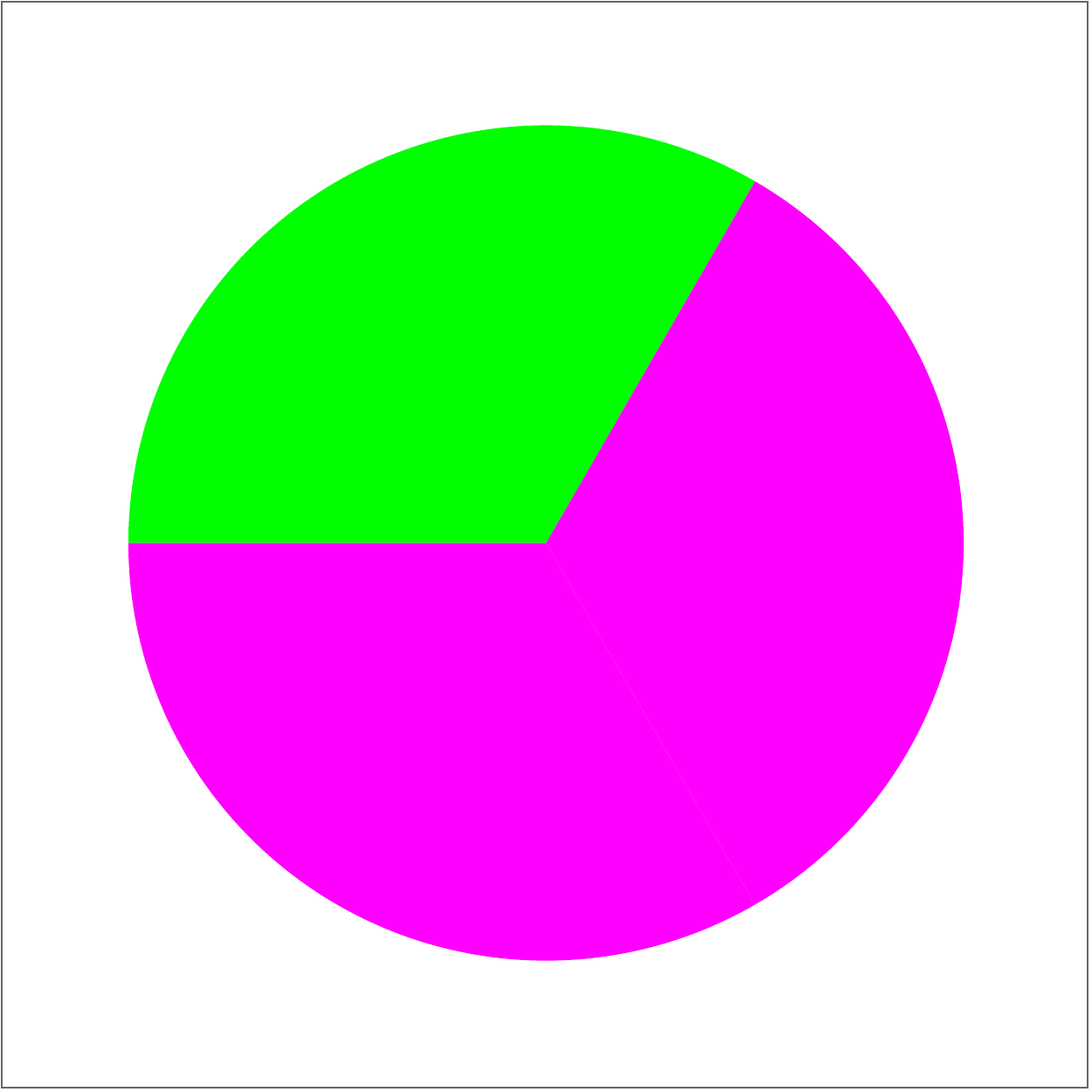} \label{fig:mp_merging_basins_regular_4}}
\caption{Illustration of the merging method for Wada basins and regular basins. (a)--(d) Wada basins for the Majumdar--Papapetrou di-hole system for initial conditions in the $(\rho, z)$-plane with $\Delta p\ind{_{\phi}} = 0.05$. (a) The original three-colour basins: rays which plunge into the upper (lower) black hole are coloured green (purple); rays which escape to infinity are coloured grey. (b) The basins corresponding to the black holes are merged [cyan]. (c) The basins corresponding to the upper black hole and spatial infinity are merged [yellow]. (d) The basins corresponding to the lower black hole and spatial infinity are merged [magenta]. (e)--(h) A simple example of regular basins and their merged versions. In this case, only the centre of the disc, where the three basins meet, is invariant under the pairwise merging of the basins; this is a Wada point. The remaining boundary points are not Wada points.}
\label{fig:mp_merging_basins}
\end{center}
\end{figure}

The method is based on a simple but counterintuitive observation: Wada basin boundaries are invariant under the action of merging any two of the basins together. To see this, consider $N \geq 3$ disjoint open sets (i.e., basins) $\left\{ B_{1}, B_{2}, \ldots, B_{N} \right\}$. Suppose $p \in \partial B_{i}$ for all $i \in \left\{ 1, 2, \ldots, N \right\}$; in other words, $p$ is a Wada point. Define a new basin $B_{1,2} = B_{1} \cup B_{2}$ (that is, \emph{merge} the basins $B_{1}$ and $B_{2}$ together). Clearly $p \in \partial B_{1,2}$ and $p \in \partial B_{j}$ for all $j \in \left\{ 3, \ldots, N \right\}$, so $p$ is still a boundary point of the merged basins. Since the choice of basins $B_{1}$ and $B_{2}$ was arbitrary, the action of merging any two of the basins together leaves the Wada boundary points invariant. In fact, one could continue merging the basins together in this fashion, until one is left with a pair of basins $B_{i}$ and $\bigcup_{j \neq i} B_{j}$ for some $i \in \left\{ 1, 2, \ldots, N \right\}$. Clearly $p \in \partial B_{i}$ and $p \in \partial \left( \bigcup_{j \neq i} B_{j} \right)$. Figure \ref{fig:mp_merging_basins} illustrates this striking property. In Figures \ref{fig:mp_merging_basins_1}--\ref{fig:mp_merging_basins_4}, we show the exit basins in phase space and their merged versions for the Majumdar--Papapetrou di-hole with initial conditions taken in the $(\rho, z)$-plane and an angular momentum parameter $\Delta p\ind{_{\phi}} = {p\ind{_{\phi}}}^{\ast} - p\ind{_{\phi}} = 0.05$ (see Section \ref{sec:exit_basins_phase_space}). On first inspection, it appears that we have simply changed the colours used to depict the basins. However, closer examination reveals that the boundaries in each figure are the same, but that two of the three basins have been merged to form a new basin in each case. It is straightforward to see that non-Wada boundaries do not exhibit this invariance under the pairwise merging of the basins. Figures \ref{fig:mp_merging_basins_regular_1}--\ref{fig:mp_merging_basins_regular_4} depict a simple example of regular (i.e., non-fractal) basins -- constructed by partitioning the unit disc into three equal sectors -- and their merged counterparts. For this example, only one of the boundary points (the centre of the disc) is a Wada point. The non-Wada boundaries are destroyed by the pairwise merging of the basins.

Mathematically, to check that the basins of a system satisfy the Wada property, it suffices to check that the boundary remains unaltered under the merging of the basins. However, this procedure would require either (i) a closed-form expression for the basins and their boundaries; or (ii) an image of the basins up to arbitrarily high resolution.

Typically, the exit basins of a dynamical system are computed using a finite grid of initial conditions in the $\left(u, v\right)$-plane, where $u$ and $v$ are the coordinates of a two-dimensional surface of section in phase space. The basins can, of course, be realised in higher dimensions; however, we restrict our attention to the two-dimensional case for simplicity. Here, a ``pixel'' is taken to be a square of side length $L$ centred on the initial condition $(u_{0}, v_{0})$. The exit basins are realised by colouring the pixels according to the final state of the trajectory with initial condition given by the point $(u_{0}, v_{0})$, i.e., the centre of the pixel.

Due to the issues caused by the finite resolution of the grid used to realise the basins, we \emph{fatten} the basin boundaries by a factor of the \emph{fattening parameter} $r$; that is, we replace each pixel belonging to the (original) \emph{slim boundaries} by itself plus its $r$ nearest neighbours. For a square grid of initial data, this is typically achieved using the Chebyshev metric; however, other grid layouts and metrics can be employed without drastically affecting the method.

Having computed the boundaries and applied the fattening procedure for some fattening parameter $r$, we are left with the set of slim boundaries $\left\{ \partial B_{i} \right\}_{i = 1}^{N}$ and their fattened versions $\left\{ \partial_{r} B_{i} \right\}_{i = 1}^{N}$. The condition of the method states that if all of the (original) slim boundaries are contained in all of the fat boundaries, i.e., $\partial B_{i} \subset \partial_{r} B_{j}$ for all $i, j \in \left\{1, 2, \ldots, N \right\}$, then the basin boundary is a Wada boundary up to the resolution of the fattening parameter $r$. The method is reliant on the choice of parameter $r$; we begin with $r = 1$ and increase its value until either all of the basin boundary points are classified as Wada points, or a halting condition $r > r_{\text{stop}}$ is met. Of course, the merging method only ascertains whether or not a basin is Wada up to a resolution determined by the internal parameters of the method: the grid size $L$, and the fattening parameter $r$.

Let us summarise the steps of the merging method for clarity.
\begin{enumerate}
\item Begin with a finite resolution image of $N$ exit basins $\left\{ B_{i} \right\}_{i = 1}^{N}$. Recall that the method does not require any prior knowledge of the dynamical system; however, this will be necessary to compute the basins in the first place.
\item Choose one basin $B_{i}$ for some $i \in \left\{ 1, 2, \ldots, N \right\}$, and merge all of the other basins $\bigcup_{j \neq i} B_{j}$. This yields a two-colour exit basin diagram. Repeat this procedure for each basin $B_{i}$ in turn. This results in a set of $N$ two-colour basin diagrams (see Figure \ref{fig:mp_merging_basins}).
\item Determine the boundary of each two-colour basin diagram. Practically, this is achieved by identifying pixels which have at least one neighbour of the opposite colour. This yields a set of $N$ \emph{slim boundaries} $\left\{ \partial B_{i} \right\}_{i = 1}^{N}$.
\item Starting with $r = 1$, fatten each slim boundary by a factor of $r$ (i.e., take each boundary point and all other points within a distance $r$) to obtain $N$ \emph{fat boundaries} $\left\{ \partial_{r} B_{i} \right\}_{i = 1}^{N}$.
\item Consider a fat boundary $\partial_{r} B_{j}$ for some $j \in \left\{ 1, 2, \ldots, N \right\}$. Check whether the union of slim boundaries is contained inside the fat boundary. Repeat for each of the $N$ fat boundaries $\partial_{r} B_{j}$ in turn. If the union of the slim boundaries is contained inside \emph{every} fat boundary, then the basins are said to posses the Wada property, up to the resolution determined by the fattening parameter $r$. If this is not the case, increase the value of $r$ ($r \mapsto r + 1$) and return to step 4 of the algorithm. Repeat this process until the halting condition $r > r_{\text{stop}}$ is satisfied. If this condition is met and the union of slim boundaries is not contained in each fat boundary, then the merging method classifies the basins as non-Wada. In the case of partially Wada boundaries (i.e., some boundary points are Wada points, but others are not), the merging method provides a list of all of the boundary points in the original image and their classification as Wada or non-Wada.
\end{enumerate}

Daza \emph{et al.} \cite{DazaWagemakersSanjuan2018} present an overview of the merging method, and provide a detailed analysis of the method. Moreover, the method is used to test for the Wada property in the basins of two well-known systems: the H\'{e}non--Heiles Hamiltonian; and the Newton method to find the $n$th roots of unity in the complex plane. Numerical investigation of these systems demonstrates that the merging method correctly classifies basins which are known to possess the Wada property; and that the classification using a fattening parameter $r$ is independent of (i) the fractal dimension of the boundary, and (ii) the number of basins.

\subsection{The Wada property in Majumdar--Papapetrou exit basins}
\label{sec:results_wada_exit_basins}

As described in Sections \ref{sec:exit_basins_phase_space} and \ref{sec:wada_merging_method}, the exit basins in the phase space of the H\'{e}non--Heiles Hamiltonian system \cite{HenonHeiles1964} are known to possess the Wada property, as demonstrated by Aguirre \emph{et al.} \cite{AguirreVallejoSanjuan2001}, who employed the Nusse--Yorke method \cite{NusseYorke1996}. This has since been confirmed using the grid method \cite{DazaWagemakersSanjuanEtAl2015} and the merging method \cite{DazaShipleyDolanEtAl2018}. Given the striking similarities between the H\'{e}non--Heiles system and the Majumdar--Papapetrou di-hole system (see Section \ref{sec:exit_basins_phase_space} and Appendix \ref{chap:appendix_b}), one would naturally expect the basins of the Majumdar--Papapetrou di-hole -- shown in Figure \ref{fig:mp_exit_basins} -- to exhibit the Wada property. Here, we test this claim using the merging method.

We applied the merging algorithm described in Section \ref{sec:wada_merging_method} to the exit basins of the Majumdar--Papapetrou system with initial data taken in both the $(\rho, z)$-plane and the $(\rho, p\ind{_{\rho}})$-plane (see Figures \ref{fig:mp_rho_z_basin_005}--\ref{fig:mp_rho_z_basin_001} and Figures \ref{fig:mp_rho_p_rho_basin_005}--\ref{fig:mp_rho_p_rho_basin_001}, respectively), where the basins were realised using a grid of $10^{3} \times 10^{3}$ equally spaced initial conditions. We tested all boundary points for the Wada property for a selection of exit basin diagrams with $\Delta p\ind{_{\phi}} \in \left[ 0.02, 0.15 \right]$, where $\Delta p\ind{_{\phi}} = {p\ind{_{\phi}}}^{\ast} - p\ind{_{\phi}}$. For this choice of parameters, the merging algorithm classified \emph{all} boundary points as Wada points for a fattening parameter of $r = 3$.

We remark that, as $\Delta p\ind{_{\phi}} \rightarrow 0$ (i.e., as $p\ind{_{\phi}} \rightarrow {p\ind{_{\phi}}}^{\ast}$), the width of the three escapes in phase space tends to zero. Moreover, KAM islands become dominant -- as shown in Figures \ref{fig:mp_rho_z_basin_001} and \ref{fig:mp_rho_p_rho_basin_001} -- and the escape time for photons which start inside the scattering region blows up \cite{AguirreSanjuan2003}. It is therefore computationally expensive to verify the Wada property for small values of the parameter $\Delta p\ind{_{\phi}}$. Although we have not directly verified the Wada property for Majumdar--Papapetrou basins with $\Delta p\ind{_{\phi}} < 0.02$, we expect all boundary points to retain the Wada property in the limit of small escapes ($\Delta p\ind{_{\phi}} \rightarrow 0$). A detailed discussion of the limit of small escapes in open Hamiltonian systems is given by Aguirre and Sanju\'{a}n \cite{AguirreSanjuan2003}.
%

\subsection{The Wada property in Majumdar--Papapetrou shadows}
\label{sec:results_wada_shadows}

We now turn our attention to the shadows of the Majumdar--Papapetrou di-hole, described in detail in Section \ref{sec:binary_black_hole_shadows}, and reviewed in this chapter in Section \ref{sec:mp_black_hole_shadows_wada}; see Figure \ref{fig:mp_shadow_decomposition} for examples of shadows of equal-mass ($M_{\pm} = M = 1$) black holes separated by distances $d = 1$ and $d = 2$.

We applied the merging method (Section \ref{sec:wada_merging_method}) to test for the Wada property in Majumdar--Papapetrou di-hole shadows for a selection of coordinate separations in the range $d \in \left[ 0, 3 \right]$. In each case, we employed a ray-tracing method (Section \ref{sec:binary_black_hole_shadows}) to generate black hole shadow images for an observer situated at a radial distance $r_{0} = 50 M$ with a viewing angle of $\theta = \frac{\pi}{2}$. Practically, this involved numerically integrating Hamilton's equations for the Hamiltonian \eqref{eqn:mp_hamiltonian_canonical_recall} for a $10^{3} \times 10^{3}$ grid of evenly spaced initial conditions in the $(X, Y)$-plane (the observer's image plane).

\begin{figure}[t]
\begin{center}
\includegraphics[width=0.6\textwidth]{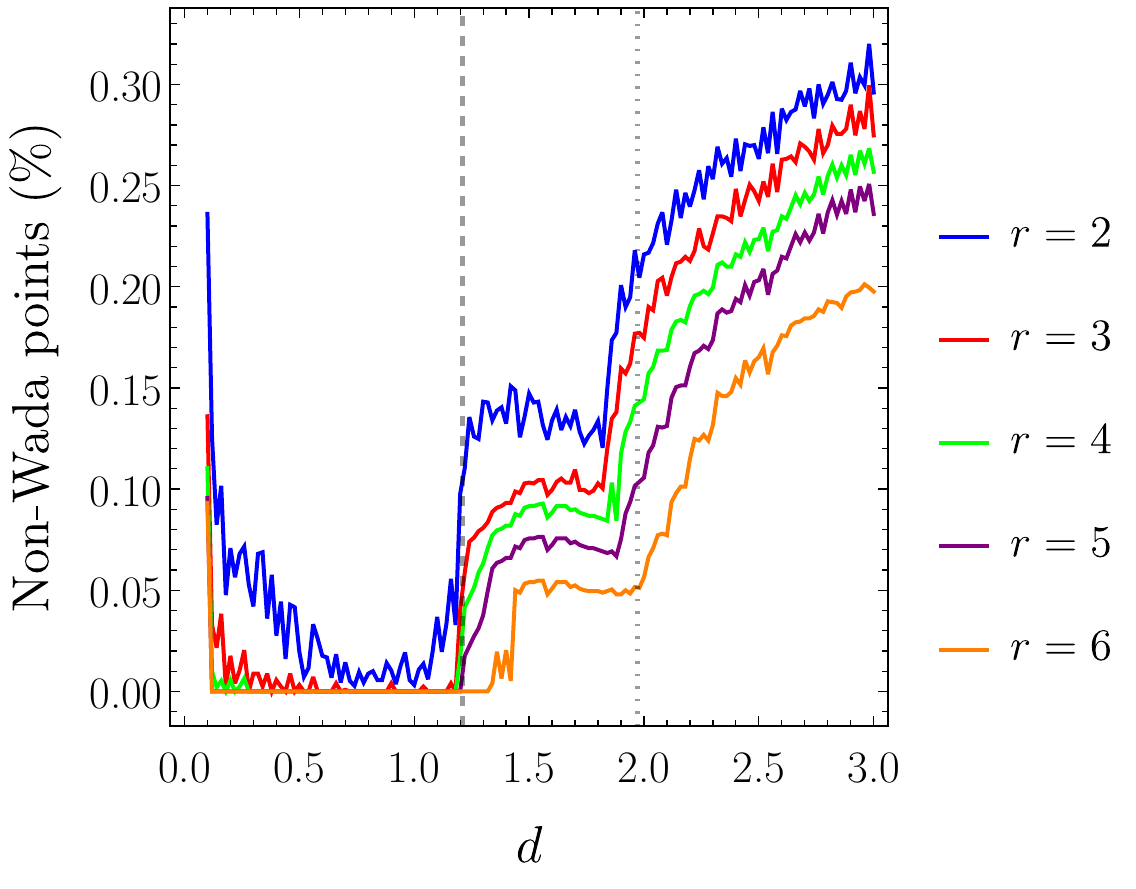}
\caption{Results of the merging method for the equal-mass Majumdar--Papapetrou di-hole with coordinate separation $d$. The numerical algorithm was performed for a range of values of the fattening parameter $r$. The dashed vertical line indicates the critical separation $d = \hat{d} = 1.2085$, below which the shadows exhibit the full Wada property. The dotted vertical line at $d \approx 1.97$ marks a second transition in the nature of the shadow, where there is an increase in the number of non-Wada points.}
\label{fig:mp_wada_test_results}
\end{center}
\end{figure}

The results of the merging algorithm are shown in Figure \ref{fig:mp_wada_test_results}, which depicts the percentage of boundary points which are \emph{not} classified as Wada points by the algorithm, as the coordinate separation $d$ is varied. The results of the algorithm provide evidence that the shadows are fully Wada (i.e., all boundary points are Wada points) for separations in the range $0.1 \lesssim d \lesssim 1.2$. Moreover, the results of Figure \ref{fig:mp_wada_test_results} suggest that there is a qualitative transition in the nature of the shadows at $d \approx 1.2$, after which the shadow is only partially Wada.

Such a qualitative change was anticipated in our study of the fundamental orbits in Sections \ref{sec:mp_wada_photon_orbits} and \ref{sec:critical_separation}. We argued that, for a fixed value of $p\ind{_{\phi}}$, the existence of three types of photon orbits, as shown in Figure \ref{fig:mp_fundamental_orbits_p_phi}, gives rise to Cantor-like structure in the corresponding one-dimensional shadows (see Chapter \ref{chap:binary_black_hole_shadows}). In Section \ref{sec:wada_cantor_symbolic_dynamics}, we presented a simple argument based on symbolic dynamics which illustrates that a set of $N \geq 3$ exit basins which have a Cantor-like boundary will possess the Wada property. Hence, if all three types of fundamental orbits exist for $0 \leq p\ind{_{\phi}} < {p\ind{_{\phi}}}^{\ast}$, then we anticipate that the two-dimensional shadow will be totally Wada. If, on the other hand, there exists some value $\hat{p}\ind{_{\phi}} < {p\ind{_{\phi}}}^{\ast}$ for which the type II and III fundamental orbits cease to exist -- as shown in Figure \ref{fig:mp_fundamental_orbits_p_phi_590} -- then the one-dimensional shadows with $\hat{p}\ind{_{\phi}} < p\ind{_{\phi}} < {p\ind{_{\phi}}}^{\ast}$ will \emph{not} be fractal. As a result, the two-dimensional shadow will only be partially Wada: the boundary points corresponding to $0 \leq p\ind{_{\phi}} \leq \hat{p}\ind{_{\phi}}$ will be Wada points, whereas those corresponding to $\hat{p}\ind{_{\phi}} < p\ind{_{\phi}} < {p\ind{_{\phi}}}^{\ast}$ will not. In Section \ref{sec:critical_separation}, we demonstrated that the latter scenario occurs for sufficiently separated black holes with $d > \hat{d} \approx 1.2085$. This prediction matches the observed transition in the results of the merging algorithm, shown in Figure \ref{fig:mp_wada_test_results}.

Figure \ref{fig:mp_wada_test_results} also indicates that there is a second qualitative change in the nature of the shadow structure at $d \approx 1.9$. It is likely that this transition occurs when the top of the main globular shadow components impinge upon the regular region of the shadow. For $d \gtrsim \hat{d}$, only the tips of the primary eyebrow-like components are regular. As one increases $d$, the regular region of the shadow (i.e., the domain in $Y$ for which all one-dimensional shadows are non-fractal) begins to incorporate the top of the globular features (main shadows). Numerical investigations of the shadow structure and the location of the regular region suggest that this occurs at $d \approx 1.97$, which agrees well with the observed transition in Figure \ref{fig:mp_wada_test_results} (shown as a dotted vertical line).

\begin{figure}[t]
\begin{center}
\subfigure[]{
\includegraphics[height=0.45\textwidth]{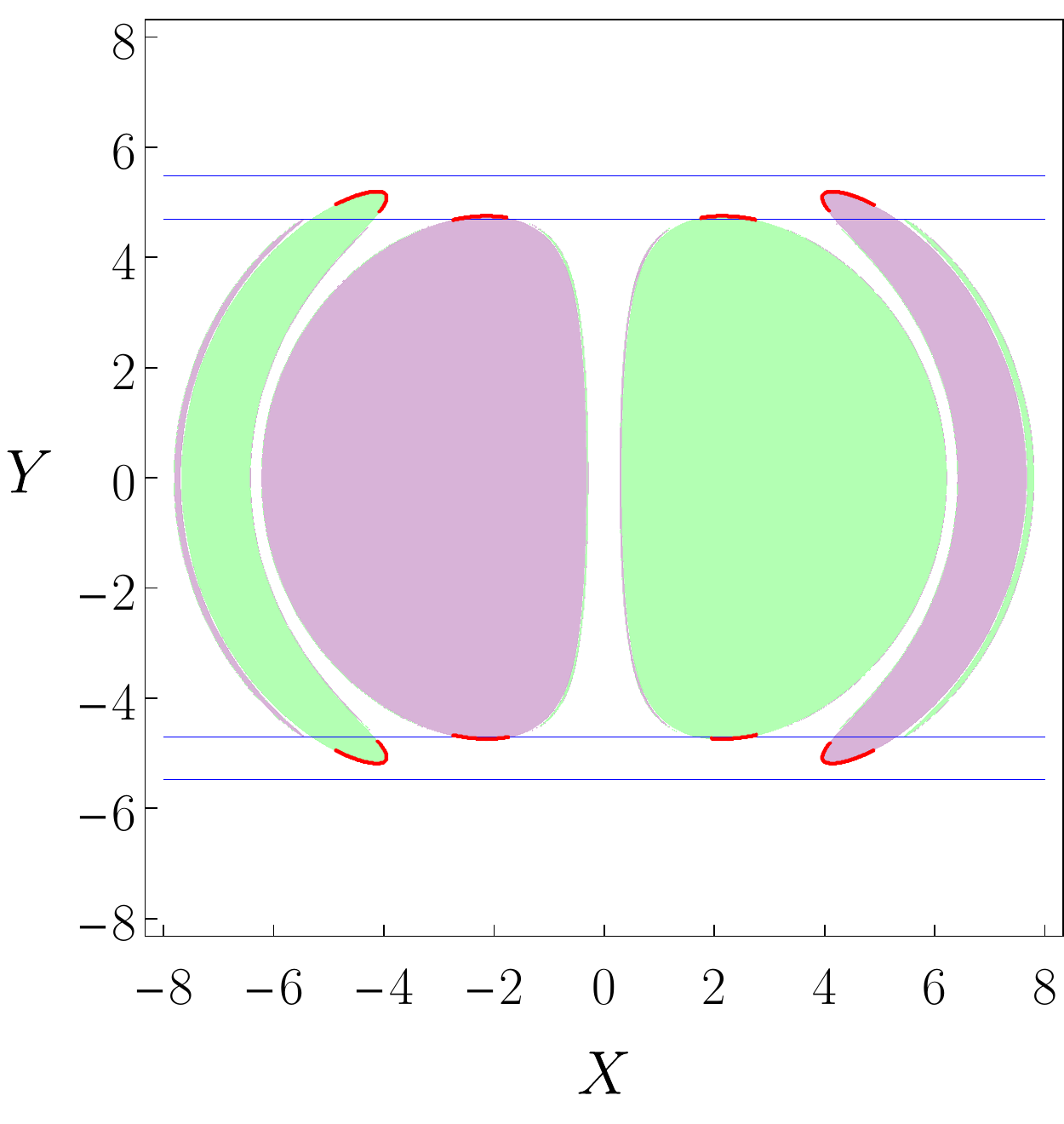} \label{fig:mp_shadow_wada_test} \hspace{1em}}
\subfigure[]{
\includegraphics[height=0.45\textwidth]{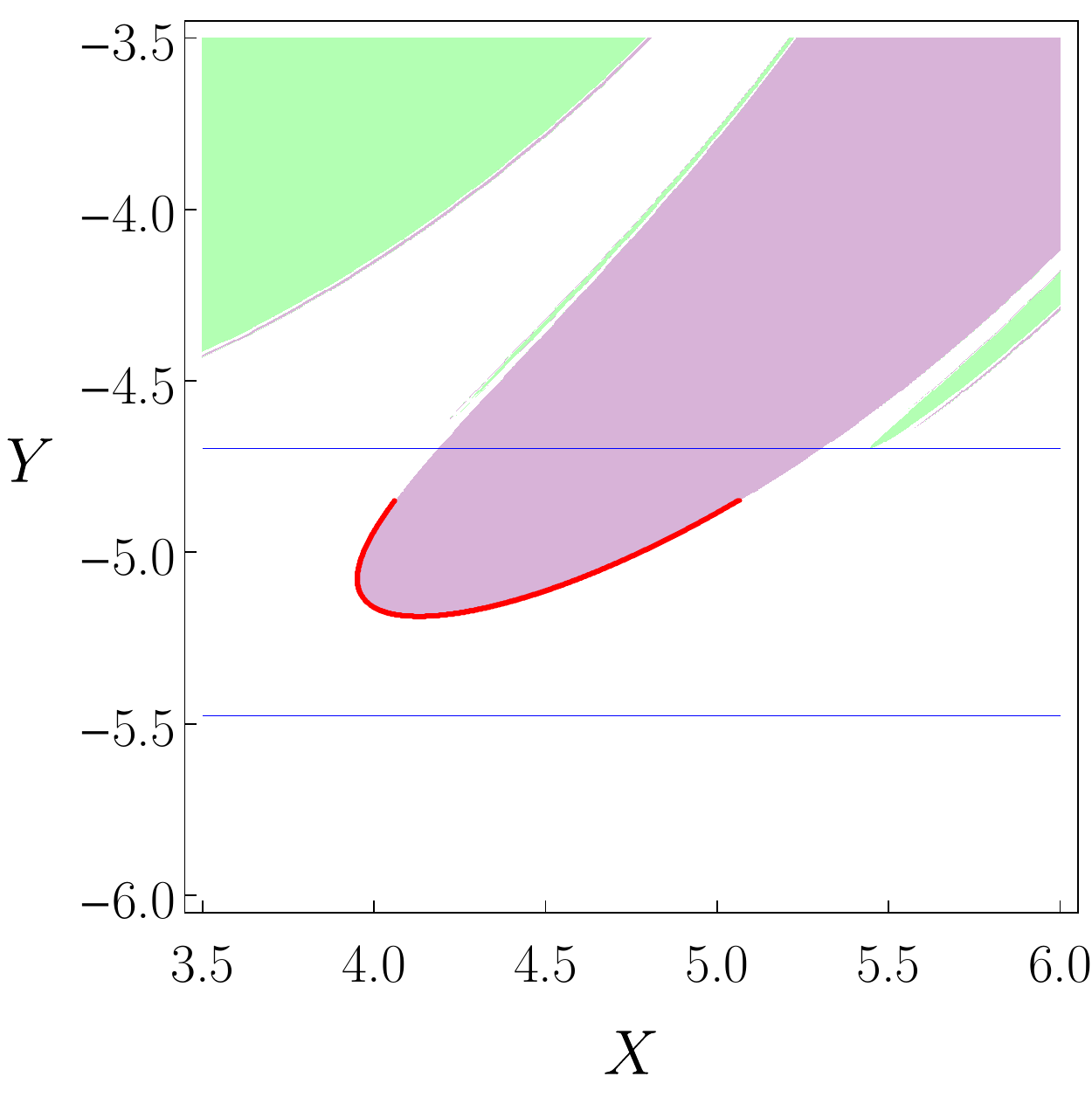} \label{fig:mp_shadow_zoom_wada_test} \hspace{1em}}
\caption{Results of the merging method to test for Wada boundary points in the equal-mass Majumdar--Papapetrou di-hole shadow with coordinate separation $d = 2 M$. (a) Shadow of the Majumdar--Papapetrou di-hole in the image plane with coordinates $(X, Y)$. The shadow of the upper (lower) black hole is shown in green (purple). The non-Wada boundary points found by the merging method (with $r = 5$) are shown in red; critical values of $Y$, which delimit the non-fractal regions, are shown in blue. The latter are obtained by looking at critical values of $p\ind{_{\phi}}$. (b) Magnified region of the shadow (a) which shows good agreement between the non-Wada points detected by the merging method and the non-fractal regions of the shadow.}
\label{fig:mp_shadow_wada_test_full}
\end{center}
\end{figure}

To confirm our interpretation of the Wada property in Majumdar--Papapetrou di-hole shadows, we illustrate the results of the merging method in Figure \ref{fig:mp_shadow_wada_test_full}. We consider a Majumdar--Papapetrou di-hole shadows seen by an observer with a viewing angle of $\theta = \frac{\pi}{2}$, for black holes of mass $M$ separated by coordinate distance $d = 2 M$. The exit basin corresponding to the upper (lower) black hole are plotted in green (purple); the exit basin corresponding to spatial infinity is plotted in white. The non-Wada points detected by the merging method (for a fattening parameter $r = 5$) are shown in red. The regular (non-fractal) region of the shadow is delimited by the blue horizontal lines; this region is determined by considering the values of $Y$ corresponding to $p\ind{_{\phi}} \in ( \hat{p}\ind{_{\phi}}, {p\ind{_{\phi}}}^{\ast} )$. The results of Figure \ref{fig:mp_shadow_wada_test_full} confirm that all of the non-Wada points lie within the regular region. Furthermore, one can see from Figure \ref{fig:mp_shadow_wada_test} that the regular zone has begun to impinge on the primary shadows for $d = 2 > 1.97$.

The agreement between the boundaries of the regular region [blue curves] and the non-Wada points [red] detected by the merging method could be improved by (i) increasing the resolution of the exit basin diagram; and (ii) taking $r_{\textrm{max}} \rightarrow \infty$, where $r_{\textrm{max}}$ is the location of the observer. We caution that implementing these proposals would make the merging method more computationally expensive.

\section{Discussion}
\label{sec:discussion_fractal_structures}

\subsection{Extensions}

Do the shadows cast by real binary black holes, such as the progenitors of gravitational waves detected by LIGO--Virgo, exhibit the Wada property? This remains an open question. Despite the fact that singleton black hole systems have been successfully imaged \cite{EHTC2019a}, there appears to be little prospect of imaging a binary black hole system in the near future. However, realistic simulations of gravitational lensing by dynamical binary black holes have been conducted by Bohn \emph{et al.} \cite{BohnThroweHebertEtAl2015}. Classifying these high-resolution shadow images as fully Wada, partially Wada or otherwise by applying the merging method would certainly be of theoretical interest. One could also apply the algorithm to shadows in other binary systems, such as the imitative models explored in the recent work of Cunha \emph{et al.} \cite{CunhaHerdeiroRodriguez2018}.

Aguirre and Sanju\'{a}n \cite{AguirreSanjuan2003} explore the exit basins of open Hamiltonian systems in the limit that the escape width tends to zero, using the Gaspard--Rice three-disc model and the H\'{e}non--Heiles Hamiltonian system as illustrative cases. They find that, as the width of the escapes decreases, the basins become \emph{totally uncertain} (i.e., the uncertainty exponent approaches zero with the width of the escapes). These totally uncertain basins resemble the case of \emph{riddled basins} in dissipative systems. A basin is said to be \emph{riddled} if all of its points are arbitrarily close to points from another basin \cite{AlexanderYorkeYouEtAl1992}. Riddled basins coincide with their boundary.

In the case of the Majumdar--Papapetrou Hamiltonian system with $d = M$ (highly symmetric case), we have three exits in phase space whose widths approach zero in the limit $\Delta p\ind{_{\phi}} \rightarrow 0$. We anticipate that the exit basins in phase space and the one-dimensional shadows would become totally uncertain in the limit $\Delta p\ind{_{\phi}} \rightarrow 0$.  The basins of Figure \ref{fig:mp_exit_basins}, the black hole shadows of Figure \ref{fig:mp_uncertainty_exponent_d1} and the uncertainty exponent in Figure \ref{fig:mp_shadow_uncertainty_exponent_d1} all support this claim; however, an accurate calculation of the uncertainty exponent in the limit $\Delta p\ind{_{\phi}} \rightarrow 0$ would be computationally expensive.

\subsection{Conclusions}

In this chapter, we have applied techniques from the field of non-linear dynamics to study fractal structures which arise in the shadows of a static binary black hole model -- the Majumdar--Papapetrou di-hole solution -- which is an exact solution to the Einstein--Maxwell equations of gravity and electromagnetism. The solution can be regarded as a surrogate for a more ``realistic'' dynamical binary black hole system, such the progenitors of the gravitational waves detected by the LIGO--Virgo collaboration.

The black hole shadow is intimately related to the scattering of photons, which follow null geodesics on the curved spacetime geometry. The null geodesics of the Majumdar--Papapetrou geometry are governed by a two-dimensional Hamiltonian dynamical system with a conserved parameter for rays (the photon's azimuthal angular momentum). Intriguingly, this Hamiltonian system shares many qualitative features with the well-known H\'{e}non--Heiles system \cite{HenonHeiles1964}, a paradigmatic two-dimensional time-independent model for chaotic scattering in Hamiltonian dynamics.

In Chapter \ref{chap:binary_black_hole_shadows}, we saw that the Majumdar--Papapetrou di-hole system is an example of chaotic scattering: the existence of three dynamically connected fundamental photon orbits gives rise to a fractal set of scattering singularities of measure zero. The chaotic scattering of photons by the two black holes is responsible for Cantor-like fractal structure in the binary black hole shadows. Our perspective is that black hole shadows can be viewed as exit basins defined with respect to the image plane of a distant observer. Black hole shadows are therefore amenable to techniques from non-linear dynamics which have been developed to characterise and quantify fractal structures.

In Section \ref{sec:mp_geometry_and_rays_recap}, we reviewed the key features of the Majumdar--Papapetrou di-hole geometry and the Hamiltonian formalism for null geodesics (first covered in Chapter \ref{chap:binary_black_hole_shadows}). Taking inspiration from the study of the H\'{e}non--Heiles system in \cite{AguirreVallejoSanjuan2001}, we analysed the geodesic dynamics of the Majumdar--Papapetrou di-hole by considering exit basins in phase space in Section \ref{sec:exit_basins_phase_space}. Then, in Section \ref{sec:mp_black_hole_shadows_wada}, we reviewed the black hole shadows of the equal-mass Majumdar--Papapetrou di-hole system for different values of the coordinate separation $d$ between the black holes. The role of fundamental photon orbits on the variation in shadow structure was discussed (Section \ref{sec:mp_wada_photon_orbits}). We outlined a numerical method which could be used to calculate the critical value of the separation parameter $\hat{d}$ (Section \ref{sec:critical_separation}); we conjectured that Majumdar--Papapetrou di-hole shadows with $d < \hat{d}$ will be fully Wada, whereas shadows with $d > \hat{d}$ will only be partially Wada (i.e., they will exhibit both Wada and non-Wada regions). In Section \ref{sec:uncertainty_definitions_numerical}, we reviewed the notion of Hausdorff dimension for fractal sets and the uncertainty exponent, first introduced by Grebogi \emph{et al.} \cite{GrebogiMcDonaldOttEtAl1983} as a quantitative measure for the indeterminacy in final-state prediction when fractal boundaries are present in phase space. We presented a numerical method which can be employed to compute the uncertainty exponent for exit basins. To test and calibrate the numerical method, we calculated the uncertainty exponent for the Cantor basins (first introduced in Section \ref{sec:construction_cantor_like_set}) and compared the numerical data to known analytical results (Section \ref{sec:uncertaint_exponent_cantor_basins}). We then applied the numerical method to one-dimensional Majumdar--Papapetrou di-hole shadows (Section \ref{sec:uncertainty_exponent_shadows}); the results support our understanding of the shadow structure and fundamental photon orbits, outlined in Chapter \ref{chap:binary_black_hole_shadows} and Section \ref{sec:mp_black_hole_shadows_wada}. In Section \ref{sec:wada_basins}, we reviewed the Wada property, with an emphasis on the exit basins of open Hamiltonian systems. Using a simple argument based on ``symbolic dynamics'' for Cantor basins, we demonstrate that the Wada property arises generically in systems with three or more exit basins who share a common Cantor-like fractal basin boundary (Section \ref{sec:wada_cantor_symbolic_dynamics}). We applied the merging method (reviewed in Section \ref{sec:wada_merging_method}) to test for the Wada property in (i) the exit basins in phase space of the Majumdar--Papapetrou di-hole (Section \ref{sec:results_wada_exit_basins}); and (ii) the shadows cast by the black holes, i.e., exit basins on the image plane of an observer (Section \ref{sec:results_wada_shadows}).

Using the merging method, we have shown that the Majumdar--Papapetrou di-hole shadows can exhibit either the partial Wada property or the full Wada property, depending on the separation of centres $d$. In the exit basins of open Hamiltonian systems, the Wada property is associated with high levels of indeterminacy and unpredictability, despite the underlying dynamics being deterministic. In the case of a pair of black holes, the existence of Wada basins implies that a photon whose initial conditions are chosen close to a Wada boundary can end up in one of three final states: it can plunge into either of the black holes of escape to spatial infinity.

An important property of the merging method of \cite{DazaWagemakersSanjuan2018} is that it is agnostic to the underlying dynamical system: no knowledge of the system or its invariant sets is required; all that is necessary is an exit basin diagram at finite resolution. A key result of this work is that the algorithm successfully detected a ``phase transition'' in the black hole shadow, as it changed from fully Wada to partially Wada at a certain value of the separation parameter $d$. The numerical results of the merging method agree well with our theoretical predictions based on fundamental photon orbits (Sections \ref{sec:mp_wada_photon_orbits} and \ref{sec:critical_separation}). In the exit basins of dynamical systems where the underlying dynamics is either unknown or too complicated to study analytically, the merging method can provide novel physical insight.

As described in Section \ref{sec:wada_merging_method}, a number of computational methods have been proposed to test for Wada basins, including the Nusse--Yorke method \cite{NusseYorke1996}. The merging method, employed in this work, offers several advantages over the Nusse--Yorke method. In the latter approach, to verify the Wada property, one must compute an unstable manifold and verify that it crosses all of the exit basins in phase space. In the case of black hole shadows for the Majumdar--Papapetrou di-hole, the image plane mixes phase space and parameter space: the image plane coordinates are dependent on the independent phase space coordinates and the conserved parameter $p\ind{_{\phi}}$. It is not clear how one would construct an unstable manifold on the image plane; as a result, it does not appear possible that the Nusse--Yorke method could be used to test for the Wada property in Majumdar--Papapetrou di-hole shadows.

The work presented in this chapter (based on that of \cite{DazaShipleyDolanEtAl2018}) is, to our knowledge, the first demonstration of the Wada property in the exit basins and/or the black hole shadows of a general relativistic system. Our work demonstrates that tools from the field of non-linear dynamics (e.g.~the uncertainty exponent and the merging method to test for Wada basis) can be exploited to deepen our understanding of scattering processes in general relativity. Furthermore, this investigation of the Majumdar--Papapetrou di-hole system shows that there exist novel dynamical systems in general relativity (and in the wider field of gravitational physics) which are amenable to techniques from non-linear dynamics and chaos theory. 

\chapter{Stable photon orbits in stationary axisymmetric spacetimes} \label{chap:stable_photon_orbits}

\section{Introduction}

The epoch-making gravitational-wave detections by the LIGO--Virgo collaboration provide compelling evidence for the existence of stellar-mass black holes in nature \cite{Abbottothers2018}. Moreover, the characteristic ``chirp'' profiles of these events are consistent with the inspiral, merger and ringdown phases of a binary black hole coalescence, which settles down to an equilibrium state described by Kerr's solution of general relativity \cite{Kerr1963}.

The ringdown phase of a binary black hole coalescence provides strong evidence that the end-product of the merger possesses a \emph{light-ring} \cite{CardosoFranzinPani2016}, i.e., a family of \emph{unstable} photon orbits \cite{Teo2003, Hod2013}. The parameters of the light-ring -- in particular, its orbital frequencies and Lyapunov exponents -- are related to the frequency and decay rate of the ringdown phase of the gravitational-wave signal \cite{Goebel1972, Mashhoon1985, CardosoMirandaBertiEtAl2008, Dolan2010, YangNicholsZhangEtAl2012}. If the merged body is endowed with a light-ring, this will be associated with a range of observational phenomena, including multiple lensing images of distant sources \cite{Perlick2004}; diffraction effects such as glories and orbiting \cite{CrispinoDolanOliveira2009}; and a characteristic spectrum of quasinormal modes \cite{KonoplyaZhidenko2011}.

Intriguingly, the correspondence between the object's complex quasinormal mode oscillations and the existence of an unstable photon orbit suggests that any object with a light-ring will initially vibrate like a black hole \cite{CardosoFranzinPani2016}. As a result, the LIGO--Virgo gravitational-wave detections do not rule out the possibility that the merged body possesses a light-ring but \emph{not} an event horizon. Objects with a light-ring are dubbed \emph{ultra-compact} in the literature \cite{CardosoCrispinoMacedoEtAl2014, CunhaBertiHerdeiro2017}.

Horizonless ultracompact objects -- such as boson stars \cite{LieblingPalenzuela2017}, Proca stars \cite{BritoCardosoHerdeiroEtAl2016}, gravastars \cite{MazurMottola2004, ChirentiRezzolla2016}, and wormholes \cite{DamourSolodukhin2007} -- possess unstable light-rings which are generically accompanied by inner \emph{stable} photon orbits \cite{CardosoCrispinoMacedoEtAl2014}; the latter are associated with a range of phenomenological features. First, the existence of stable photon orbits allows for the trapping and storage of energy in bounded regions; this allows instabilities (e.g.~fragmentation and collapse \cite{CardosoCrispinoMacedoEtAl2014} or ergoregion instabilities \cite{Friedman1978, PaniBarausseBertiEtAl2010}) to flourish. Second, Keir \cite{Keir2016} found that linear waves -- which serve as a model for non-linear perturbations -- decay (no faster than) logarithmically with time, which dominates power-law Price decay \cite{CasalsOttewill2015}; this slow decay suggests that ultra-compact objects are unstable at the non-linear level. Finally, ultra-compact objects with a stable photon orbit exhibit a modified late-time gravitational-wave ringdown which depends on the internal structure of the compact object, and differs from that of a black hole \cite{CardosoFranzinPani2016}. As we enter a new era of precision gravitational-wave experiments, this modified ringdown will be tested and constrained.

Various strands of work suggest that stable photon orbits are relevant not only in horizonless geometries, but also play a key role in black hole spacetimes. In four dimensions, stable photon orbits have been shown to exist within the inner horizons of Kerr--Newman black holes and around naked singularities \cite{Liang1974, CalvaniFeliceNobili1980, Stuchlik1981, BalekBicakStuchlik1989, PuglieseQuevedoRuffini2013, Dokuchaev2011, UlbrichtMeinel2015, KhooOng2016}; and around black holes (or solitons) with a non-zero cosmological constant \cite{StuchlikHledik2002, GrenzebachPerlickLaemmerzahl2014}. In higher-dimensional contexts, stable photon orbits have been revealed in the exterior regions (i.e., outside the event horizon) of five-dimensional black rings \cite{IgataIshiharaTakamori2013} and six-dimensional Myers--Perry black holes \cite{Igata2015}. More recently, stable photon orbits have been revealed around Majumdar--Papapetrou di-holes \cite{WuenschMuellerWeiskopfEtAl2013, ShipleyDolan2016}, which serve as static axisymmetric toy models for dynamical binary black holes; and around hairy black hole models, including Kerr black holes with scalar hair \cite{CunhaGroverHerdeiroEtAl2016} and Proca hair \cite{CunhaHerdeiroRadu2017}. In these scenarios, stable photon orbits have been revealed to play a key role in gravitational lensing; in particular, they are associated with distinctive chaotic features in black hole shadows. Non-planar ``fundamental photon orbits'' (generalisations of light-rings) in generic stationary, axisymmetric spacetimes have been classified -- and their stability explored -- by Cunha \emph{et al.} \cite{CunhaHerdeiroRadu2017}. A review of the relationship between fundamental photon orbits and strong-field lensing effects by ultra-compact objects can be found in \cite{CunhaHerdeiro2018}.

In this chapter, we investigate the existence of stable photon orbits in four-dimensional stationary axisymmetric electrovacuum spacetimes; the majority of the work presented in this chapter is based on \cite{DolanShipley2016}. In Section \ref{sec:spo_photon_orbits}, we review the existence and classification of equatorial circular photon orbits in the charge--spin parameter space for the Kerr--Newman family of spacetimes. In Section \ref{sec:spo_sa_spacetimes}, we employ a Hamiltonian formalism for null geodesics to demonstrate that, for a general stationary axisymmetric (Weyl) spacetime (reviewed in Section \ref{sec:spo_hamiltonian_formalism}), the Einstein--Maxwell equations permit the existence of stable photon orbits in \emph{electrovacuum}, but not in pure vacuum (Section \ref{sec:spo_electrovacuum_case}). In Section \ref{sec:spo_diholes_existence}, we explore the existence of stable photon orbits in the Reissner--Nordstr\"{o}m di-hole class of spacetimes -- a subfamily of the general Bret\'{o}n--Manko--Aguilar two-centre solution \cite{BretonMankoAguilarSanchez1998}, which includes the uncharged Weyl--Bach di-hole and extremal Majumdar--Papapetrou di-hole as special cases. Using Poincar\'{e} sections, we explore the structure of stable photon orbits around the Majumdar--Papapetrou di-hole in Section \ref{sec:spo_diholes_geodesic_structure}. Finally, we conclude with a discussion of this work and possible extensions in Section \ref{sec:spo_discussion}.

\section{Photon orbits in Kerr--Newman spacetime} \label{sec:spo_photon_orbits}

\subsection{Spacetime geometry}

Uniqueness and no-hair theorems \cite{Robinson1975, Mazur1982} support the conjecture that the end-product of gravitational collapse in asymptotically flat electrovacuum is a Kerr--Newman black hole \cite{NewmanJanis1965, NewmanCouchChinnaparedEtAl1965}. The most astrophysically relevant case is that of an uncharged Kerr black hole \cite{Kerr1963}, which has become a cornerstone of Einstein's theory of general relativity. Here, we outline the key features of the Kerr--Newman geometry, paying special attention to its (null) geodesics; for further details, see Sections \ref{sec:black_holes} and \ref{sec:integrability_and_chaos_in_gr}, and references therein.

The Kerr--Newman geometry is characterised by three externally observable parameters: mass $M$, charge $Q$, and spin $a = \frac{J}{M}$ (where $J$ is the angular momentum). In Boyer--Lindquist coordinates $\{t, r, \theta, \phi \}$, the line element takes the form
\begin{align}
\ed s^{2} &= g\ind{_{a b}} \ed x^{a} \ed x^{b} \\
\begin{split}
& = - \left( 1 - \frac{2 M r}{\Sigma} \right) \ed t^{2} + \frac{\Sigma}{\Delta} \, \ed r^{2} + \Sigma \, \ed \theta^{2} + \left(r^{2} + a^{2} + \frac{2 M a^{2} r}{\Sigma} \sin^{2} \theta\right) \sin^{2} \theta \, \ed \phi^{2} \\ & \qquad - \frac{4 M a r \sin^{2} \theta}{\Sigma} \, \ed t \, \ed \phi,
\end{split}
\end{align}
where $\Sigma(r, \theta) = r^{2} + a^{2} \cos^{2}{\theta}$ and $\Delta(r) = r^{2} - 2 M r + a^{2} + Q^{2}$. The event horizons are located at Boyer--Lindquist radii $r = r_{\pm} = M \pm \sqrt{M^{2} - (a^{2} + Q^{2})}$. Introducing the charge ratio $q = \frac{Q}{M}$ and spin ratio $\tilde{a} = \frac{a}{M} = \frac{J}{M^{2}}$, this becomes $r_{\pm} = M \left(1 \pm \sqrt{1 - (\tilde{a}^{2} + q^{2})} \right)$. The black hole subfamily is given by $\tilde{a}^{2} + q^{2} \leq 1$, with equality in the extremal case where there is a single event horizon at $r_{\pm} = M$. The Kerr--Newman family may be extended beyond the black hole case to include naked singularities by considering the region of charge--spin parameter space given by the inequality $\tilde{a}^{2} + q^{2} > 1$.

On Kerr--Newman spacetime, there exist two Killing vector fields, satisfying Killing's equation $\xi\ind{_{(a;b)}} = 0$, which are associated with time translations (${\xi_{(t)}}\ind{^{a}}$), and rotation about the symmetry axis (${\xi_{(\phi)}}\ind{^{a}}$). These isometries are in one-to-one correspondence with constants of motion along the tangent vector $u\ind{^{a}}$ to a geodesic: the energy $E = - {\xi_{(t)}}\ind{^{a}} u\ind{_{a}}$, and azimuthal angular momentum $L\ind{_{z}} = {\xi_{(\phi)}}\ind{^{a}} u\ind{_{a}}$. The existence of a rank-two Killing tensor, which satisfies $K\ind{_{(a b ; c)}} = 0$, is responsible for a further ``hidden'' constant of motion -- the Carter constant, defined as $K = K\ind{_{a b}} u\ind{^{a}} u\ind{^{b}}$. The geodesic equations are separable (and hence Liouville integrable) thanks to the existence of this additional conserved quantity; see Section \ref{sec:integrability_and_chaos_in_gr} for a review.

The radial and latitudinal motion of a massless particle with energy $E$ and angular momentum $L\ind{_{z}}$ is described by the equations \cite{Carter1968a, Chandrasekhar1983}
\begin{align}
\Sigma^{2} \dot{r}^{2} &= \left[ E (r^{2} + a^{2}) - a L\ind{_{z}} \right]^{2} - K \Delta, \label{eqn:kerr_ecpo_radial_geodesic} \\
\Sigma^{2} \dot{\theta}^{2} &= K - \left(a E \sin{\theta} - \frac{L\ind{_{z}}}{\sin{\theta}} \right)^{2}, \label{eqn:kerr_ecpo_latitudinal_geodesic}
\end{align}
where an overdot denotes differentiation with respect to an affine parameter along the ray, say $\lambda$. Rearrangement of \eqref{eqn:kerr_ecpo_latitudinal_geodesic} permits us to write Carter's constant in the form
\begin{equation}
\label{eqn:cartern_constant_kerr_newman_theta_dot}
K = K\ind{_{a b}} u\ind{^{a}} u\ind{^{b}} = \Sigma^{2} \dot{\theta}^{2} + \left(a E \sin{\theta} - \frac{L\ind{_{z}}}{\sin{\theta}} \right)^{2} .
\end{equation}

\subsection{Classification of equatorial circular photon orbits}

We are interested in the problem of classifying equatorial circular photon orbits. Setting $\theta = \frac{\pi}{2}$, we see that \eqref{eqn:cartern_constant_kerr_newman_theta_dot} reduces to $K = (a E - L\ind{_{z}})^{2}$ and \eqref{eqn:kerr_ecpo_radial_geodesic} gives
\begin{equation}
\label{eqn:kerr_ecpo_rdot_r}
\dot{r}^{2} = \frac{E^{2}}{r^{4}} \left[ \left(r^{2} + a^{2} - a b \right)^{2} - (a - b)^{2} \left( r^{2} - 2 M r + a^{2} + Q^{2} \right) \right],
\end{equation}
where $b = \frac{L\ind{_{z}}}{E}$ is the impact parameter. The change of variables $r \mapsto u = \frac{M}{r}$ allows us to express \eqref{eqn:kerr_ecpo_rdot_r} as
\begin{align}
\label{eqn:kerr_ecpo_rdot_u}
\dot{r}^{2} &= E^{2} \mathcal{R}(u), & \mathcal{R}(u) &= 1 - \left( \tilde{b}^{2} - \tilde{a}^{2} \right) u^{2} + \left( \tilde{a} - \tilde{b} \right)^{2} \left( 2 u^{3} - q^{2} u^{4} \right),
\end{align}
where $\tilde{a}$ and $q$ are the spin and charge ratios, respectively, and $\tilde{b} = \frac{b}{M}$. A circular orbit satisfies $\dot{r} = 0 = \ddot{r}$. If $E \neq 0$, the conditions for circular photon orbits can be expressed as $\mathcal{R}(u) = 0 = \mathcal{R}^{\prime}(u)$, where a prime denotes differentiation with respect to $u$. Furthermore, an orbit is stable if $\mathcal{R}^{\prime\prime}(u) < 0$. The problem of classifying equatorial circular photon orbits of non-zero energy therefore reduces to classifying the roots of the quartic $\mathcal{R}(u)$.

We seek the values of the impact parameter $b$ for which $\mathcal{R}$ has repeated roots. Such values may be found by solving $\Delta_{u} (\mathcal{R}) = 0$, where $\Delta_{u}$ denotes the discriminant with respect to $u$. The expression $\Delta_{u} (\mathcal{R})$ is a polynomial in $b$ of degree $10$. Phase boundaries in the $(q^{2}, \tilde{a}^{2})$-plane -- which correspond to repeated roots of the polynomial $\Delta_{u} (\mathcal{R})$ -- are obtained by setting ``the discriminant of the discriminant'' to zero, which factorises as
\begin{equation}
\Delta_{b} \left( \frac{\Delta_{u}(\mathcal{R})}{ ( \tilde{b} - \tilde{a} )^{6}} \right) = 2^{32} \left( 1 - \tilde{a}^{2} - q^{2} \right) \left[ 27 \tilde{a}^{2} - q^{2} \left(9 - 8 q^{2} \right)^{2} \right]^{3}.
\end{equation}
Hence, the phase boundaries in the charge--spin plane are
\begin{align}
\tilde{a}^{2} &= 1 - q^{2}, \label{eqn:spo_kerr_phase_boundary_1} \\
\tilde{a}^{2} &= \frac{1}{27} q^{2} \left(9 - 8 q^{2} \right)^{2}. \label{eqn:spo_kerr_phase_boundary_2}
\end{align}
The first of these is the extremality condition for the Kerr--Newman family: solutions with $\tilde{a}^{2} + q^{2} \leq 1$ are black holes; whereas the converse holds for naked singularities. The phase boundaries intersect at the point $(q, \tilde{a}) = \left(\frac{\sqrt{3}}{2}, \frac{1}{2} \right)$.

\begin{figure}[h]
\begin{center}
\includegraphics[width=0.8\textwidth]{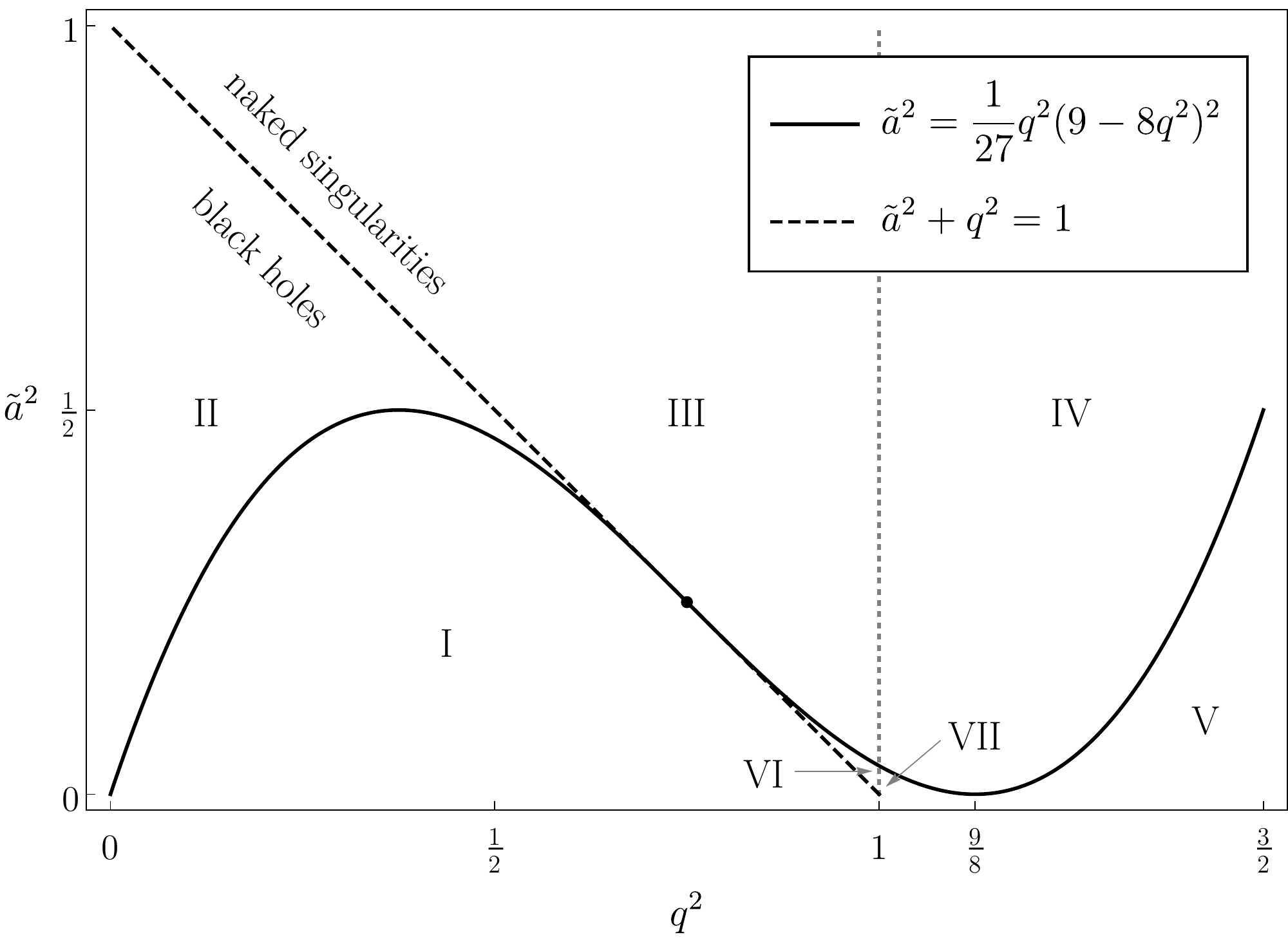}
\end{center}
\caption{Phase diagram for equatorial circular photon orbits of positive Boyer--Lindquist radius ($r > 0$) on Kerr--Newman spacetime in the charge--spin plane $(q^{2}, \tilde{a}^{2})$, where $q = \frac{Q}{M}$ and $\tilde{a} = \frac{a}{M} = \frac{J}{M^{2}}$. The black solid line shows the phase boundary $\tilde{a}^{2} = \frac{1}{27} q^{2} \left(9 - 8 q^{2} \right)^{2}$; the black dashed line is the phase boundary $\tilde{a}^{2} + q^{2} = 1$, which separates black holes (Regions I and II) from naked singularities (Regions III--VII). The phase boundaries meet at the point $(q, \tilde{a}) = \left( \frac{\sqrt{3}}{2}, \frac{1}{2} \right)$. Stable equatorial circular photon orbits exist within the inner horizon for black holes in Region II; on (inside) extremal horizons $r = M$ for $a < \frac{1}{2}$ ($a > \frac{1}{2}$); and around naked singularities in Regions III, IV and VI. (Equivalent phase diagrams are presented in Figure 6 of \cite{Stuchlik1981}, Figure 2.1 of \cite{BalekBicakStuchlik1989}, and Figure 4 of \cite{PuglieseQuevedoRuffini2013}.)}
\label{fig:kerr_ecpos}
\end{figure}

In Figure \ref{fig:kerr_ecpos}, we present the Balek--Bi\v{c}\'{a}k--Stuchl\'{i}k \cite{BalekBicakStuchlik1989} phase diagram in the charge--spin plane for equatorial circular photon orbits of positive Boyer--Lindquist radius ($r > 0$) on Kerr--Newman spacetimes. Here we summarise the classification of equatorial circular photon orbits in Regions I--VII of the phase diagram. Regions I and II correspond to the black hole regime with distinct horizons at $r = r_{\pm}$ ($\tilde{a}^{2} + q^{2} < 1$); Regions III--VIII correspond to naked singularities with no event horizon ($\tilde{a}^{2} + q^{2} > 1$). All orbits are \emph{unstable}, unless stated otherwise. In Region I there are two exterior orbits ($r > r_{+}$). In Region II there are two interior orbits ($r < r_{-}$) and two exterior orbits ($r > r_{+}$); the innermost orbit is stable. Regions III and IV admit two orbits; the innermost is stable. In Region V there are no equatorial circular photon orbits. Finally, Regions VI and VII admit four orbits; the inner pair is stable. We remark that Regions III and VI (naked singularities with $0 < q^{2} < 1$) admit a counter-rotating orbit; whereas Regions IV and VII do not.

We now consider the critical cases, which demarcate the seven (open) regions of the $(q^{2}, \tilde{a}^{2})$-plane considered above. Firstly, for $\tilde{a}^{2} = 0$ and $0 \leq q^{2} < 1$ (Schwarzschild and Reissner--Nordstr\"{o}m black holes), there exists one exterior orbit. For $\tilde{a}^{2} = 0$ and $1 \leq q^{2} < \frac{9}{8}$ (Reissner--Nordstr\"{o}m naked singularities), there are two orbits; the innermost being stable. In the case $q^{2} = 0$, $0 < \tilde{a}^{2} < 1$ (Kerr black holes), there are two exterior orbits and one interior orbit. (Recall from Section \ref{sec:photon_orbits_review} that the Kerr black hole solution admits a pair of unstable light-rings at $r = \hat{r}_{\pm}$; one of which is co-rotating with the spin of the black hole; the other is counter-rotating. In the Schwarzschild limit $a \rightarrow 0$, these light-rings coincide at $r = 3 M$.) In the case of extremal Kerr--Newman black holes, with $\tilde{a}^{2} + q^{2} = 1$, there are three equatorial circular photon orbits, whose stability depends on the value of $\tilde{a}$: for $0 < a < \frac{1}{2}$, there is one stable orbit on the horizon ($r = M$) and two exterior orbits; for $\frac{1}{2} < a < 1$, there is one stable interior orbit ($r < M$), one horizon orbit ($r = M$) and one exterior orbit ($r > M$) \cite{UlbrichtMeinel2015,KhooOng2016}. At the point $(q, \tilde{a}) = \left( \frac{\sqrt{3}}{2}, \frac{1}{2} \right)$, where the phase boundaries \eqref{eqn:spo_kerr_phase_boundary_1} and \eqref{eqn:spo_kerr_phase_boundary_2} intersect, there is a marginally stable orbit ($\mathcal{R}^{\prime\prime}(u) = 0$) on the extremal horizon, and one exterior orbit at $r = 3 M$.

\section{Photon orbits in stationary axisymmetric spacetimes}
\label{sec:spo_sa_spacetimes}

\subsection{Hamiltonian formalism for null geodesics}
\label{sec:spo_hamiltonian_formalism}

Stationary axisymmetric spacetime in electrovacuum can be described in Weyl--Lewis--Papapetrou coordinates $\{ t, \rho, z, \phi \}$ by the line element \cite{StephaniKramerMacCallumEtAl2003,GriffithsPodolsky2009}
\begin{equation}
\label{eqn:wlp_metric}
\ed s^{2} = g\ind{_{a b}} \ed x\ind{^{a}} \ed x\ind{^{b}} = - f \left(\ed t + w \, \ed \phi \right)^{2} + \frac{1}{f} \left[ e^{2 \gamma} \left( \ed \rho^{2} + \ed z^{2} \right) + \rho^{2} \, \ed \phi^{2} \right],
\end{equation}
and electromagnetic one-form potential
\begin{equation}
\label{eqn:wlp_em_four_potential}
A\ind{_{a}} \ed x\ind{^{a}} = A\ind{_{t}} \, \ed t + A\ind{_{\phi}} \, \ed \phi,
\end{equation}
where $f$, $\gamma$, $w$, $A\ind{_{t}}$ and $A\ind{_{\phi}}$ are functions of $\rho$ and $z$ only; see Section \ref{sec:stationary_axisymmetric_solutions} and e.g.~\cite{StephaniKramerMacCallumEtAl2003, GriffithsPodolsky2009} for further details. The geodesics $q\ind{^{a}}(\lambda)$ are the integral curves of Hamilton's equations, with Hamiltonian function $H(q, p) = \frac{1}{2} g\ind{^{a b}} p\ind{_{a}} p\ind{_{b}}$ and canonical momenta $p\ind{_{a}} = g\ind{_{a b}} \dot{q}\ind{^{a}}$. Here, an overdot denotes differentiation with respect to the affine parameter $\lambda$. The metric components are independent of $t$ and $\phi$, so $p\ind{_{t}}$ and $p\ind{_{\phi}}$ are constants of motion; one may set $p\ind{_{t}} = -1$ without loss of generality, by rescaling the affine parameter. Further, $H$ itself is a constant of motion along geodesics, with $H = 0$ in the null case.

As shown in Section \ref{sec:geodesics}, null geodesics are invariant under conformal transformations of the form $g\ind{_{a b}} \mapsto \Omega^{2}(q) g\ind{_{a b}}$, where $\Omega > 0$ is a function of the spacetime coordinates. Employing a conformal transformation with $\Omega^{2} = e^{-2 \gamma} f$, we are able to recast the Hamiltonian associated with \eqref{eqn:wlp_metric} in two-dimensional ``canonical form'' as
\begin{align}
H(\rho, z, p\ind{_{\rho}}, p\ind{_{z}}; p\ind{_{\phi}}) &= \frac{1}{2} \left({p\ind{_{\rho}}}^{2} + {p\ind{_{z}}}^{2} \right) + V(\rho, z; p\ind{_{\phi}}), \label{eqn:wlp_hamiltonian} \\
V(\rho, z; p\ind{_{\phi}}) &= - \frac{e^{2 \gamma}}{2 \rho^{2}} \left[ \frac{\rho^{2}}{f^{2}} - \left( w + p\ind{_{\phi}} \right)^{2} \right], \label{eqn:wlp_potential}
\end{align}
where $p\ind{_{\phi}} = \text{constant}$ is a free parameter. The positivity of the kinetic term in \eqref{eqn:wlp_hamiltonian} and the Hamiltonian constraint $H = 0$ imply that $V \leq 0$. For a photon with fixed azimuthal angular momentum $p\ind{_{\phi}}$, this inequality demarcates the \emph{allowed regions} of configuration space.

Null geodesics of constant $\rho$ and $z$ (stationary points) exist where $V = 0$ and $\bnab V = \mathbf{0}$, where the $\bnab = \left( \partial\ind{_{\rho}}, \partial\ind{_{z}} \right)$ is the standard two-dimensional gradient operator. The stability of such orbits can be determined by considering the extrema of the potential $V$: the orbit is stable if the stationary point is a local minimum of $V$, i.e., if $\det{\mathcal{H}(V)} > 0$ and $\operatorname{tr}{\mathcal{H}(V)} > 0$, where
\begin{equation}
\mathcal{H}(V) =
\begin{bmatrix}
  V\ind{_{, \rho \rho}} & V\ind{_{, \rho z}} \\
  V\ind{_{, \rho z}} & V\ind{_{, z z}} \\
\end{bmatrix}
\end{equation}
is the Hessian matrix for the potential $V$. In general, null geodesics are kinematically bounded if there exists a contour $V(\rho, z) = 0$ which is closed in the $(\rho, z)$-plane that bounds an open region in which the potential is negative.

A drawback of the potential \eqref{eqn:wlp_potential} is that it depends on the orbital parameter $p\ind{_{\phi}}$. One may factorise the potential as
\begin{equation}
\label{eqn:wlp_facotrised_potential}
V = - \frac{e^{2 \gamma}}{2 \rho^{2}} ( h^{+} + p\ind{_{\phi}} ) ( h^{-} - p\ind{_{\phi}} ),
\end{equation}
where we introduce the \emph{height functions} (or \emph{effective potentials})
\begin{equation}
\label{eqn:spo_wlp_effective_potentials}
h^{\pm} (\rho, z) = \frac{\rho}{f} \pm w,
\end{equation}
which are independent of the azimuthal angular momentum $p\ind{_{\phi}}$. (In the case of \emph{static} axisymmetric spacetimes, described by \eqref{eqn:wlp_metric} with $w = 0$, we write $h^{\pm} = h$.) In order to determine the closed contours $V = 0$ (which are necessary for kinematically bounded photon orbits), it is sufficient to seek closed contours $h^{\pm} = \mp p\ind{_{\phi}}$. Since $p\ind{_{\phi}} \in \mathbb{R}$, closed contours $V = 0$ -- and thus stable photon orbits -- exist in the neighbourhood of any local \emph{maximum} of the effective potentials \eqref{eqn:spo_wlp_effective_potentials}.\footnote{Note that local maxima of $h^{\pm}$ correspond to local minima of $V$, due to the minus sign in \eqref{eqn:wlp_facotrised_potential}.} The problem of classifying stable photon orbits in stationary axisymmetric spacetimes reduces to classifying the fixed points of the effective potentials \eqref{eqn:spo_wlp_effective_potentials}, which is achieved by considering $\det{\mathcal{H}(h^{\pm})}$ and $\operatorname{tr}{\mathcal{H}(h^{\pm})}$, where $\mathcal{H}(h^{\pm})$ is the Hessian matrix for $h^{\pm}$. In particular, a fixed point of $h^{\pm}$ is a local extremum (saddle point) if $\det{\mathcal{H}(h^{\pm})} > 0$ ($\det{\mathcal{H}(h^{\pm})} < 0$). In the case of local extrema, we have a maximum (minimum) if $\operatorname{tr}{\mathcal{H}(h^{\pm})} < 0$ ($\operatorname{tr}{\mathcal{H}(h^{\pm})} > 0$).

\subsection{Classification of photon orbits in electrovacuum}
\label{sec:spo_electrovacuum_case}

We now classify the stationary points of the height function \eqref{eqn:spo_wlp_effective_potentials} for a general stationary axisymmetric geometry in electrovacuum, described by the line element \eqref{eqn:wlp_metric} and electromagnetic four-potential \eqref{eqn:wlp_em_four_potential}. In \cite{Ernst1968}, Ernst presents a formulation of the Einstein--Maxwell field equations for the electrovacuum Weyl solution; the equations relevant for our purposes are
\begin{align}
\left( \bnab f \right)^{2} - \rho^{-2} f^{4} \left( \bnab w \right)^{2} + 2 f \left( \bnab A\ind{_{t}} \right)^{2} + 2 \rho^{-2} f^{3} \left( \bnab A\ind{_{\phi}} - w \bnab A\ind{_{t}} \right)^{2} &= f \nabla^{2} f , \label{eqn:ernst_1} \\
\bnab \cdot \left[ \rho^{-2} f^{2} \bnab w - 4 \rho^{-2} f A\ind{_{t}} \left( \bnab A\ind{_{\phi}} - w \bnab A\ind{_{t}} \right) \right] &= 0 , \label{eqn:ernst_2} \\
\bnab \cdot \left[ \rho^{-2} f \left( \bnab A\ind{_{\phi}} - w \bnab A\ind{_{t}} \right) \right] &= 0 . \label{eqn:ernst_3}
\end{align}
These equations may be obtained by considering the trace-reversed Einstein field equations $R\ind{_{a b}} = 8 \pi \overline{T}\ind{_{a b}}$, where $\overline{T}\ind{_{a b}} = T\ind{_{a b}} - \frac{1}{2} T g\ind{_{a b}}$, with $T = 0$ for the electromagnetic field (see Section \ref{sec:einstein_maxwell_field_equations}). In \eqref{eqn:ernst_1}--\eqref{eqn:ernst_1}, the two-gradient operator acting on an arbitrary scalar field $\varphi(\rho, z)$ is $\bnab \varphi = \left(\varphi\ind{_{,\rho}}, \varphi\ind{_{,z}} \right)$; the divergence operator acting on an arbitrary vector field $\mathbf{F}(\rho, z) = \left(F\ind{_{1}}(\rho, z), F\ind{_{2}}(\rho, z) \right)$ is $\bnab \cdot \mathbf{F} = \frac{1}{\rho} {\left( \rho F\ind{_{1}} \right)}\ind{_{,\rho}} + {F\ind{_{2}}}\ind{_{, z}}$; and the Laplacian of a scalar field $\varphi(\rho, z)$ is given by $\nabla^{2} \varphi = \bnab \cdot (\bnab \varphi) = \varphi\ind{_{, \rho \rho}} + \frac{1}{\rho} \varphi\ind{_{,\rho}} + \varphi\ind{_{, z z}}$.

To classify the stationary points of $h^{\pm}$, we seek an expression for $\operatorname{tr}{\mathcal{H}(h^{\pm})} = {h^{\pm}}\ind{_{, \rho \rho}} + {h^{\pm}}\ind{_{, z z}}$. Taking second derivatives of \eqref{eqn:spo_wlp_effective_potentials}, we find
\begin{equation}
\label{eqn:tr_hessian_hpm_1}
\operatorname{tr}{\mathcal{H}(h^{\pm})} = - \rho f^{-3} \left( f \nabla^{2} f \right) - f^{-2} f\ind{_{,\rho}} + 2 \rho f^{-3} \left( \bnab f \right)^{2} \pm \left( w\ind{_{, \rho \rho}} + w\ind{_{, z z}} \right).
\end{equation}
The field equation \eqref{eqn:ernst_2} rearranges to
\begin{equation}
\label{eqn:ernst_2_rearranged}
w\ind{_{, \rho \rho}} + w\ind{_{, z z}} = \rho^{-1} f^{2} w\ind{_{, \rho}} - 2 f^{-1} \bnab f \cdot \bnab w + 4 \rho^{2} f^{-2} \bnab \cdot \left[ \rho^{-2} f A\ind{_{t}} \left( \bnab A\ind{_{\phi}} - w \bnab A\ind{_{t}} \right) \right].
\end{equation}
Inserting \eqref{eqn:ernst_1} and \eqref{eqn:ernst_2_rearranged} into the right-hand side of \eqref{eqn:tr_hessian_hpm_1} yields
\begin{equation}
\label{eqn:tr_hessian_hpm_2}
\begin{split}
\operatorname{tr}{\mathcal{H}(h^{\pm})} &= \rho f^{-3} \left( \bnab f \right)^{2} \mp 2 f^{-1} \bnab f \cdot \bnab w + \rho^{-1} f \left( \bnab w \right)^{2} - f^{-2} f\ind{_{, \rho}} \pm \rho^{-1} w\ind{_{,\rho}} \\
& \qquad - 2 \rho f^{-2} \left( \bnab A\ind{_{t}} \right)^{2} - 2 \rho^{-1} \left( \bnab A\ind{_{\phi}} - w \bnab A\ind{_{t}} \right)^{2} \\
&\qquad \qquad \pm 4 \rho^{2} f^{-2} \bnab \cdot \left[ \rho^{-2} f A\ind{_{t}} \left( \bnab A\ind{_{\phi}} - w \bnab A\ind{_{t}} \right) \right].
\end{split}
\end{equation}

Suppose we have found a stationary point of either $h^{+}$ or $h^{-}$, that is, a point $(\rho, z)$ which satisfies the stationary point conditions ${h^{\pm}}\ind{_{, \rho}} = 0 = {h^{\pm}}\ind{_{, z}}$. Taking partial derivatives of \eqref{eqn:spo_wlp_effective_potentials} and employing the stationary point conditions, we see that the equations
\begin{align}
w\ind{_{, \rho}} &= \pm \left( \rho f^{-2} f\ind{_{, \rho}} - f^{-1} \right), \label{eqn:w_derivative_rho} \\
w\ind{_{, z}} &= \pm \rho f^{-2} f\ind{_{, z}} \label{eqn:w_derivative_z}
\end{align}
must hold at fixed points of $h^{\pm}$. One can use \eqref{eqn:w_derivative_rho} and \eqref{eqn:w_derivative_z} to eliminate terms involving partial derivatives of $w$ from the right-hand side of \eqref{eqn:tr_hessian_hpm_2}; at stationary points of $h^{\pm}$, the relevant terms can be written
\begin{align}
2 f^{-1} \bnab f \cdot \bnab w &= \pm \left[ 2 \rho f^{-3} \left( \bnab f \right)^{2} - 2 f^{-2} f\ind{_{, \rho}} \right], \label{eqn:w_derivative_replace_1} \\
\rho^{-1} f \left( \bnab w \right)^{2} &= \rho f^{-3} \left( \bnab f \right)^{2} - 2 f^{-2} f\ind{_{, \rho}} + \rho^{-1} f^{-1}. \label{eqn:w_derivative_replace_2}
\end{align}

Now, substituting \eqref{eqn:w_derivative_rho}, \eqref{eqn:w_derivative_replace_1} and \eqref{eqn:w_derivative_replace_2} into the right-hand side of \eqref{eqn:tr_hessian_hpm_2}, we see that the first five terms vanish; this leaves us with
\begin{equation}
\label{eqn:tr_hessian_hpm_3}
\begin{split}
\operatorname{tr}{\mathcal{H}(h^{\pm})} &= - 2 \rho f^{-2} \left( \bnab A\ind{_{t}} \right)^{2} - 2 \rho^{-1} \left( \bnab A\ind{_{\phi}} - w \bnab A\ind{_{t}} \right)^{2} \\
& \qquad \pm 4 \rho^{2} f^{-2} \bnab \cdot \left[ \rho^{-2} f A\ind{_{t}} \left( \bnab A\ind{_{\phi}} - w \bnab A\ind{_{t}} \right) \right].
\end{split}
\end{equation}
Consider the final term on the right-hand side of \eqref{eqn:tr_hessian_hpm_3}. Expanding this term using the product rule for the divergence operator, we find
\begin{equation}
\label{eqn:divergence_term_expanded}
\begin{split}
4 \rho^{2} f^{-2} \bnab \cdot \left[ \rho^{-2} f A\ind{_{t}} \left( \bnab A\ind{_{\phi}} - w \bnab A\ind{_{t}} \right) \right] &= 4 f^{-1} \bnab A\ind{_{t}} \cdot \left( \bnab A\ind{_{\phi}} - w \bnab A\ind{_{t}} \right) \\
& \quad + 4 \rho^{2} f^{-2} A\ind{_{t}} \bnab \cdot \left[ \rho^{-2} f \left( \bnab A\ind{_{\phi}} - w \bnab A\ind{_{t}} \right) \right].
\end{split}
\end{equation}
Availing the remaining field equation \eqref{eqn:ernst_3}, we see that the final term on the right-hand side of \eqref{eqn:divergence_term_expanded} vanishes, such that
\begin{equation}
\label{eqn:divergence_term_simplified}
4 \rho^{2} f^{-2} \bnab \cdot \left[ \rho^{-2} f A\ind{_{t}} \left( \bnab A\ind{_{\phi}} - w \bnab A\ind{_{t}} \right) \right] = 4 f^{-1} \bnab A\ind{_{t}} \cdot \left( \bnab A\ind{_{\phi}} - w \bnab A\ind{_{t}} \right).
\end{equation}

Finally, inserting \eqref{eqn:divergence_term_simplified} into the right-hand side of \eqref{eqn:tr_hessian_hpm_3} yields
\begin{equation}
\label{eqn:tr_hessian_hpm_4}
\begin{split}
\operatorname{tr}{\mathcal{H}(h^{\pm})} &= - 2 \rho f^{-2} \left( \bnab A\ind{_{t}} \right)^{2} - 2 \rho^{-1} \left( \bnab A\ind{_{\phi}} - w \bnab A\ind{_{t}} \right)^{2} \\
& \qquad \pm 4 f^{-1} \bnab A\ind{_{t}} \cdot \left( \bnab A\ind{_{\phi}} - w \bnab A\ind{_{t}} \right),
\end{split}
\end{equation}
which factorises to give
\begin{equation}
\label{eqn:tr_hessian_hpm_5}
\operatorname{tr}{\mathcal{H}(h^{\pm})} = - \frac{2}{\rho} \left| \mathbf{W}_{\pm} \right|^{2}, \qquad \mathbf{W}_{\pm} = h^{\pm} \bnab A\ind{_{t}} \mp \bnab A\ind{_{\phi}},
\end{equation}
where $\left| \mathbf{W}_{\pm} \right|^{2} = \mathbf{W}_{\pm} \cdot \mathbf{W}_{\pm}$. Using \eqref{eqn:tr_hessian_hpm_5}, we can replace the terms involving ${h^{\pm}}\ind{_{, z z}}$ in the expression for $\det{\mathcal{H}(h^{\pm})}$; this gives
\begin{equation}
\label{eqn:det_hessian_hpm}
\det{\mathcal{H}(h^{\pm})} = - \left[ {({h^{\pm}}\ind{_{, \rho \rho}})}^{2} + \frac{2}{\rho} \left| \mathbf{W}_{\pm} \right|^{2} {h^{\pm}}\ind{_{, \rho \rho}} + ({h^{\pm}}\ind{_{, \rho z}})^{2} \right].
\end{equation}

In the pure vacuum case, the electromagnetic four-potential \eqref{eqn:wlp_em_four_potential} is zero, so $\mathbf{W}_{\pm} = 0$ and the right-hand side of \eqref{eqn:tr_hessian_hpm_5} vanishes, i.e., $\operatorname{tr}{\mathcal{H}(h^{\pm})} = 0$. Moreover, $\det{\mathcal{H}(h^{\pm})} = - \left[ {({h^{\pm}}\ind{_{, \rho \rho}})}^{2} + ({h^{\pm}}\ind{_{, \rho z}})^{2} \right] \leq 0$. Hence, $h^{\pm}$ cannot possess a first-order local maximum, so generic stable photon orbits are ruled out in pure vacuum.

For $\bnab A\ind{_{a}} \neq \mathbf{0}$, we have $\mathbf{W}_{\pm} \neq \mathbf{0}$, so $\operatorname{tr}{\mathcal{H}(h^{\pm})} < 0$ by \eqref{eqn:tr_hessian_hpm_5}. Therefore local minima of $h^{\pm}$ are forbidden; however, local maxima cannot be ruled out. Local maxima of $h^{\pm}$, and hence stable photon orbits, are possible if the right-hand side of \eqref{eqn:det_hessian_hpm} is positive; this can only be true if ${h^{\pm}}\ind{_{, \rho \rho}} < 0$ and $({h^{\pm}}\ind{_{, \rho z}})^{2} < - {h^{\pm}}\ind{_{, \rho \rho}} \left( {h^{\pm}}\ind{_{, \rho \rho}} + \frac{2}{\rho} \left| \mathbf{W}_{\pm} \right|^{2} \right)$. In the special case of height functions with an equatorial symmetry, the derivative ${h^{\pm}}\ind{_{, \rho z}}$ vanishes, and the stability conditions reduce to the inequality $0 < - {h^{\pm}}\ind{_{, \rho \rho}} < \frac{2}{\rho} \left| \mathbf{W}_{\pm} \right|^{2}$.

The absence of stable photon orbits in static axisymmetric \emph{vacuum} spacetimes has been pointed out by Liang \cite{Liang1974}, who remarked that ``stability within the plane implies instability off the plane''. Equation \eqref{eqn:tr_hessian_hpm_5} demonstrates how Liang's result breaks down in \emph{electrovacuum}, when an electromagnetic field is introduced. Generalisations of the results presented in this section -- to include other fields, matter sources, or a non-zero cosmological constant -- may be considered, and would provide a more complete description of the geodesic dynamics on stationary axisymmetric spacetimes in Einstein's general theory of relativity. This is discussed further in Section \ref{sec:spo_extensions}.
%

\section{Existence of stable photon orbits around di-holes} \label{sec:spo_diholes_existence}

\subsection{Di-hole formalism: the Bret\'{o}n--Manko--Aguilar solution} \label{sec:dihole_formalism_bma}

We consider the $N = 2$ Bret\'{o}n--Manko--Aguilar family of electrostatic solutions to the Einstein--Maxwell equations \cite{PerryCooperstock1997, BretonMankoAguilarSanchez1998, VarzuginChistyakov2002, AlekseevBelinski2007}. In Appendix \ref{chap:appendix_e}, we briefly review the so-called ``physical parametrisation'' of the Bret\'{o}n--Manko--Aguilar solution, first presented by Manko \cite{Manko2007}. In this parametrisation, the Bret\'{o}n--Manko--Aguilar family of solutions is described by five parameters: the black hole masses $M_{\pm}$; the black hole charges $Q_{\pm}$; and the separation of centres $d$.

Here, we restrict our attention to the Reissner--Nordstr\"{o}m di-hole subfamily, in which each of the sources is a black hole (i.e., $Q_{\pm} \leq M_{\pm}$, with equality in the case of extremal black holes); we do not consider naked singularities ($Q_{\pm} > M_{\pm}$). The Reissner--Nordstr\"{o}m di-hole family contains as special cases (i) the Majumdar--Papapetrou di-hole family \cite{Majumdar1947, Papapetrou1947}, in which both black holes are extremal ($Q_{\pm} = M_{\pm}$); (ii) the Weyl--Bach di-hole family \cite{BachWeyl2012}, in which both black holes are uncharged ($Q_{\pm} = 0$). With the exception of the Majumdar--Papapetrou subfamily, the black holes which comprise the Reissner--Nordstr\"{o}m di-hole are held in equilibrium by a ``Weyl strut'' \cite{BachWeyl2012}, and their (outer) event horizons appear as ``rods'' on the symmetry axis of coordinate length $2 \sqrt{M_{\pm}^{2} - Q_{\pm}^{2}}$ in Weyl--Lewis--Papapetrou coordinates. In the Majumdar--Papapetrou case ($Q_{\pm} = M_{\pm}$), the black hole horizons appear as points (of finite proper area) on the $z$-axis.

In this section we consider two special cases of the Reissner--Nordstr\"{o}m di-hole family: (i) the Majumdar--Papapetrou di-hole family, characterised by the mass ratio parameter $\eta = \frac{M_{+} - M_{-}}{M_{+} + M_{-}}$ (Section \ref{sec:spo_mp_dihole_existence}); (ii) the equal-mass, equal-charge RN di-hole, characterised by the charge ratio parameter $q = \frac{Q_{\pm}}{M_{\pm}}$ (Section \ref{sec:spo_rn_dihole}). We analyse the existence of stable photon orbits in both cases; this involves employing a numerical root finder to classify the fixed points of the static height function $h$, given by \eqref{eqn:spo_wlp_effective_potentials} with $w = 0$.

\subsection{Majumdar--Papapetrou di-hole} \label{sec:spo_mp_dihole_existence}

The Majumdar--Papapetrou di-hole comprises a pair of extremal Reissner--Nordstr\"{o}m black holes in static equilibrium \cite{Majumdar1947, Papapetrou1947, Chandrasekhar1989}. The geometry is described by the Weyl line element \eqref{eqn:wlp_metric} with $\gamma = 0 = w$, and $f = \frac{1}{U^{2}}$, where
\begin{equation}
U(\rho, z) = 1 + \frac{M_{+}}{\sqrt{\rho^{2} + (z - z_{+})^{2}}} + \frac{M_{-}}{\sqrt{\rho^{2} + (z - z_{-})^{2}}}.
\end{equation}
The only non-zero component of the electromagnetic four-potential \eqref{eqn:wlp_em_four_potential} is $A\ind{_{t}} = \frac{1}{U}$. We consider the general case, in which the black hole masses are given by $M_{\pm} = \frac{1}{2} (1 \pm \eta) M_{0}$, where $\eta \in (-1, 1)$ parametrises the mass ratio, and $M_{0} = M_{+} + M_{-}$ is the total mass of the system. The black holes are located along the $z$-axis ($\rho = 0$) at $z_{\pm} = \pm \frac{d M_{\mp}}{M_{0}}$, where $d$ denotes the separation between the sources (see Section \ref{sec:mp_dihole_review}). The solution is a special case of the Bret\'{o}n--Manko--Aguilar family (see Section \ref{sec:dihole_formalism_bma}), where each black hole is extremal ($Q_{\pm} = M_{\pm}$). The Majumdar--Papapetrou di-hole was used as a surrogate model to investigate the qualitative features of binary black hole shadows in Chapters \ref{chap:binary_black_hole_shadows} and \ref{chap:fractal_structures}.

For the Majumdar--Papapetrou di-hole, the height function takes the form $h = \rho \, U^{2}$. As described in Section \ref{sec:spo_sa_spacetimes}, the stationary points of $h$ correspond to orbits of constant $\rho$ and $z$; such orbits are stable if the stationary point is a local maximum of $h$. Moreover, the contours of $h$ around a local maximum will be closed curves, permitting kinematically bounded orbits. We now highlight the existence of stable photon orbits around the Majumdar--Papapetrou di-hole with mass parameter $\eta$.

\begin{figure}
\begin{center}
\subfigure[$\eta = 0$, $d = 0.95 M$]{
\includegraphics[height=0.45\textwidth]{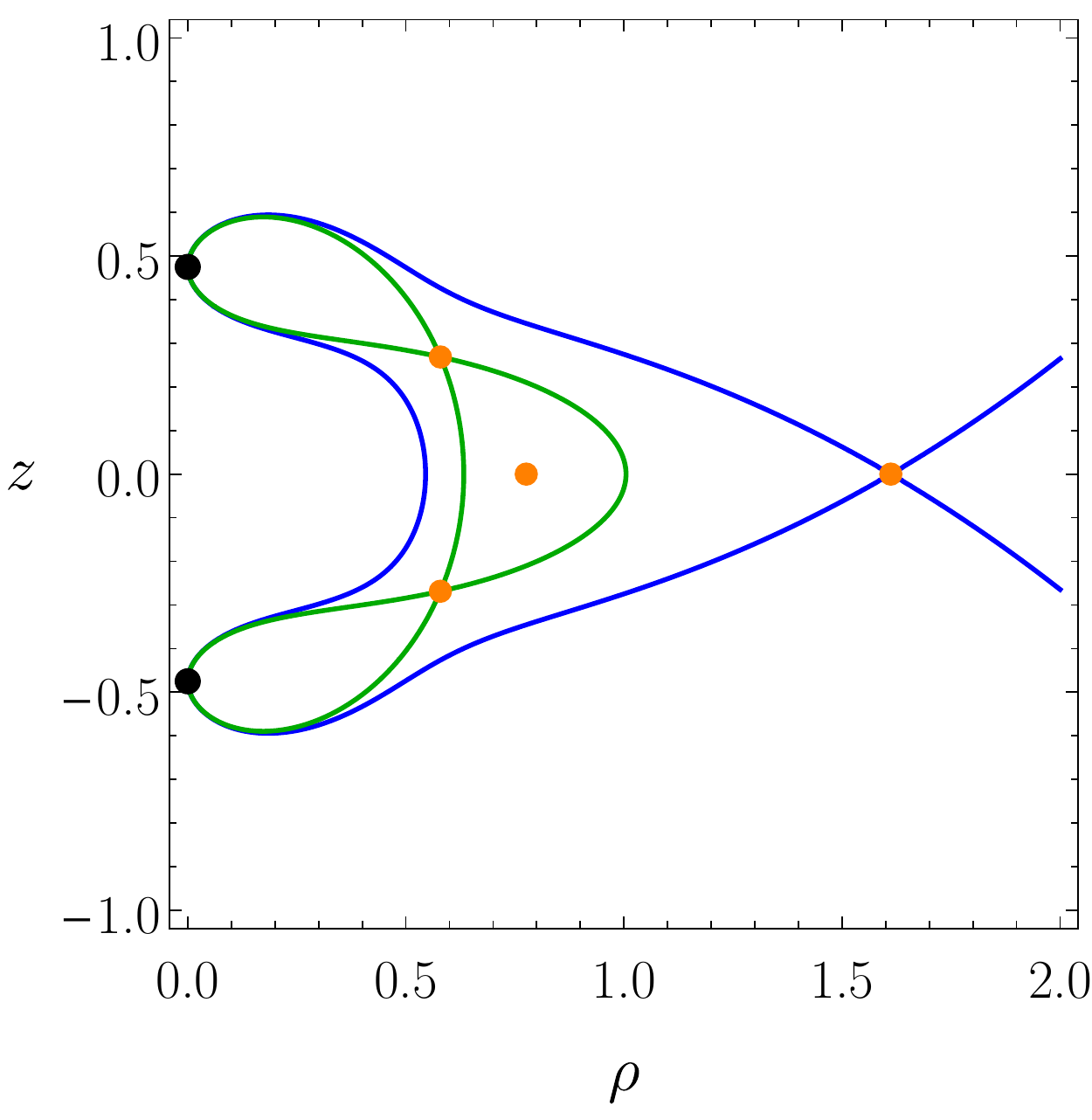} \label{fig:mp_height_function_contours_1} \hspace{1em}}
\hspace{0.5em}
\subfigure[$\eta = 0$, $d = M$]{
\includegraphics[height=0.45\textwidth]{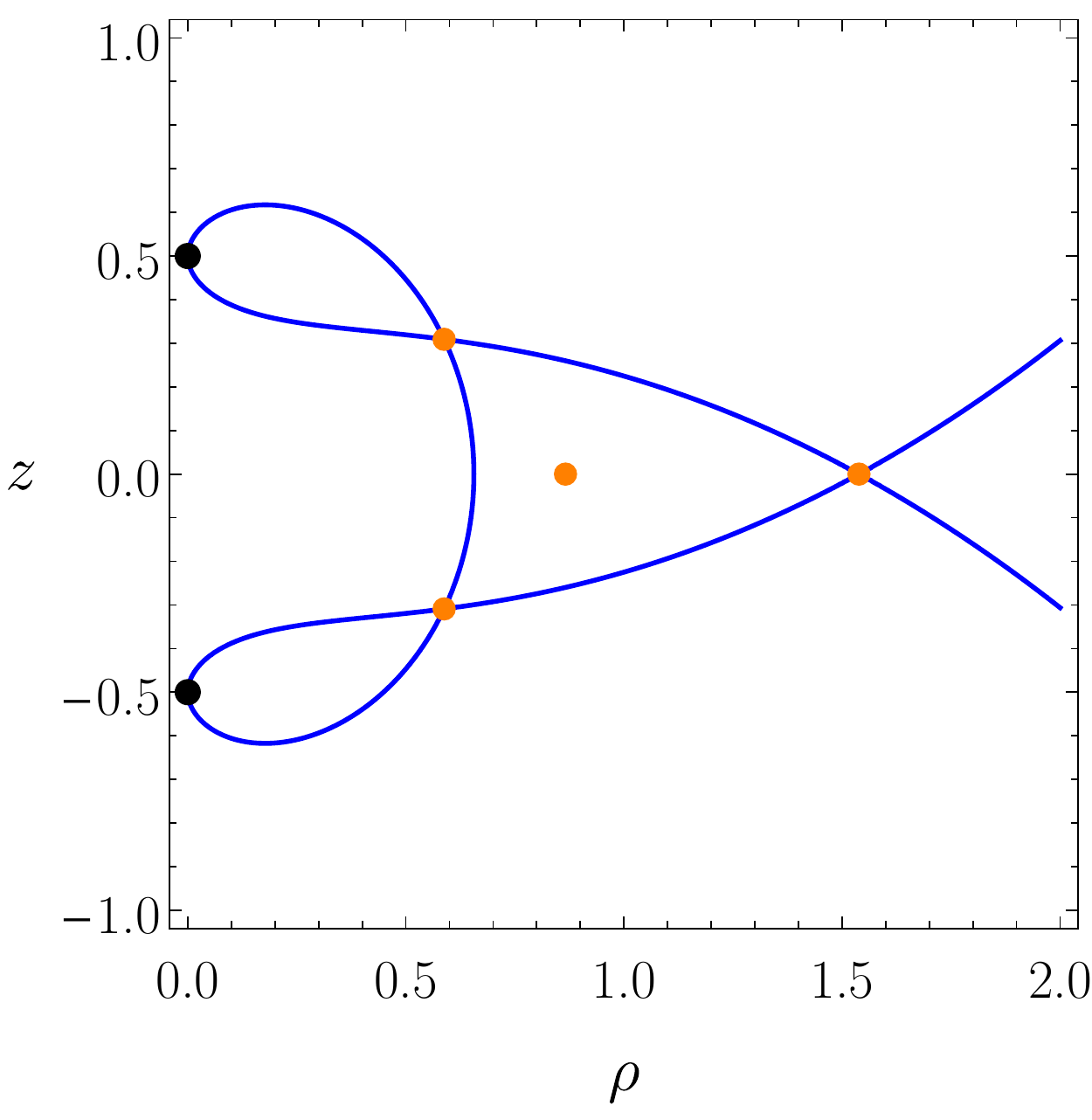} \label{fig:mp_height_function_contours_2} \hspace{1em}}
\subfigure[$\eta = 0$, $d = 1.05 M$]{
\includegraphics[height=0.45\textwidth]{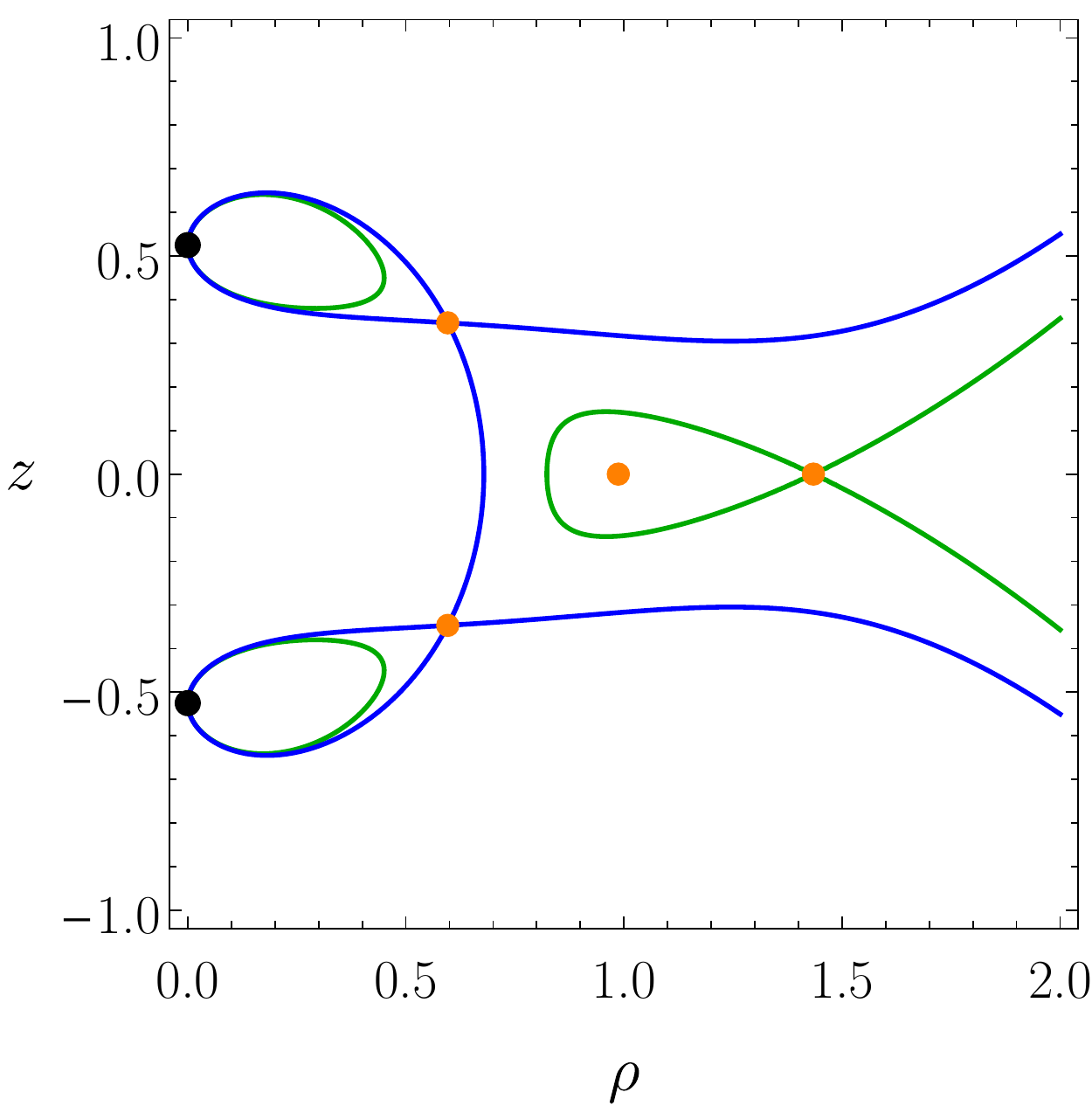} \label{fig:mp_height_function_contours_3} \hspace{1em}}
\hspace{0.5em}
\subfigure[$\eta = 0.05$, $d = \frac{1}{2} M_{0}$]{
\includegraphics[height=0.45\textwidth]{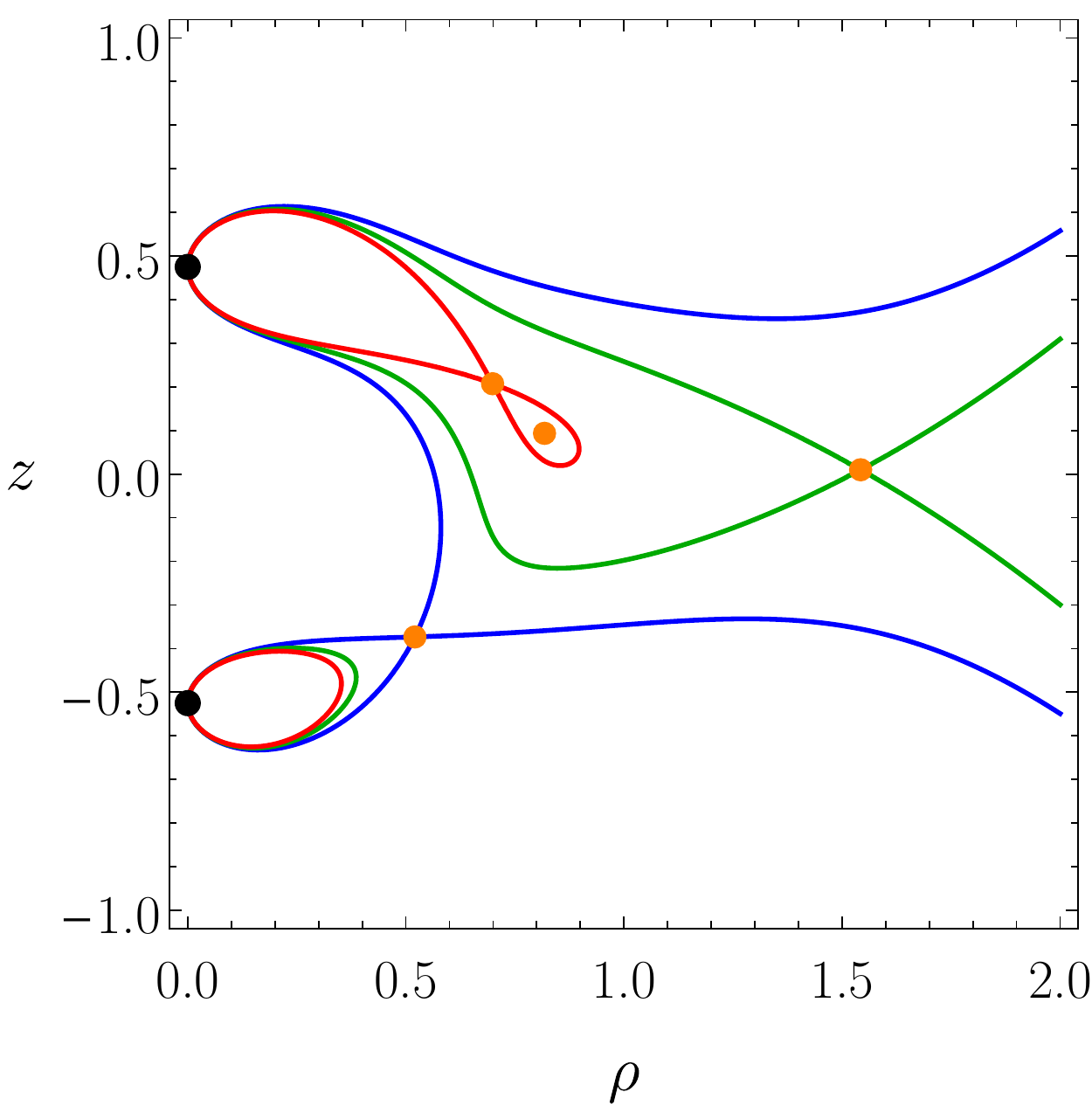} \label{fig:mp_height_function_contours_4} \hspace{1em}}
\end{center}
\caption{Critical contours of the effective potential $h$ for the Majumdar--Papapetrou di-hole, characterised by mass ratio $\eta$ and separation parameter $d$. Filled orange circles indicate stationary points of $h$: three saddle points and one local maximum. Filled black circles on the $z$-axis represent the black hole horizons, which are located at $z = z_{\pm}$. (a)--(c) Equal-mass di-holes with varying separation $d$. (d) An unequal-mass di-hole with mass parameter $\eta = 0.05$, and separation of centres $d = \frac{1}{2} M_{0}$.}
\label{fig:mp_height_function_contours}
\end{figure}

First, consider the equal-mass case, in which $\eta = 0$. We write $M = M_{\pm}$ for the mass of the black holes. The system admits stable photon orbits for separations in the range $\sqrt{\frac{16}{27}} < \frac{d}{M} < \sqrt{\frac{32}{27}}$; this existence region can be determined using the result on stable photon orbits derived in Section \ref{sec:spo_electrovacuum_case}, or an alternative approach; both methods are presented in Appendix \ref{chap:appendix_a}. In this parameter regime, $h$ admits four stationary points: a local maximum in the equatorial plane at $(\rho, z) = (\rho_{(1)}, 0)$, with $h = {p\ind{_{\phi}}}^{(1)}$; a saddle point in the equatorial plane at $(\rho, z) = (\rho_{(2)}, 0)$, with $\rho_{(2)} > \rho_{(1)}$ and $h = {p\ind{_{\phi}}}^{(2)} < {p\ind{_{\phi}}}^{(1)}$; and two saddle points out of the plane at $(\rho, z) = (\rho_{(3)}, \pm z_{(3)})$, with $\rho_{(3)} < \rho_{(1)}$ and $h = {p\ind{_{\phi}}}^{(3)} < {p\ind{_{\phi}}}^{(1)}$. Figures \ref{fig:mp_height_function_contours_1}--\ref{fig:mp_height_function_contours_3} show the stationary points and critical contours of the height function $h$ in the $(\rho, z)$-plane, for a selection of values of $d$ (close to the highly symmetric case $d = M$).

For separations in the range $\sqrt{\frac{16}{27}} M < d < M$, we have ${p\ind{_{\phi}}}^{(2)} < {p\ind{_{\phi}}}^{(3)}$. Null geodesics with ${p\ind{_{\phi}}}^{(2)} \leq p\ind{_{\phi}} < {p\ind{_{\phi}}}^{(3)}$ are permitted to fall into the black holes, but are not able to escape to infinity. Moreover, rays with ${p\ind{_{\phi}}}^{(3)} \leq p\ind{_{\phi}} < {p\ind{_{\phi}}}^{(1)}$ which start in the vicinity of the local maximum are kinematically bounded. This is illustrated Figure \ref{fig:mp_height_function_contours_1}, where we set $d = 0.95 M$.

On the other hand, for $M < d < \sqrt{\frac{32}{27}} M$, we have ${p\ind{_{\phi}}}^{(3)} < {p\ind{_{\phi}}}^{(2)}$, so rays with ${p\ind{_{\phi}}}^{(3)} \leq p\ind{_{\phi}} < {p\ind{_{\phi}}}^{(2)}$ are connected to infinity but not to the black holes. Null rays which start close to the local maximum of $h$ with ${p\ind{_{\phi}}}^{(2)} \leq p\ind{_{\phi}} < {p\ind{_{\phi}}}^{(1)}$ are kinematically bounded. An example of this case is shown in Figure \ref{fig:mp_height_function_contours_3}, in which $d = 1.05 M$.

In the critical case $d = M$, we have ${p\ind{_{\phi}}}^{(2)} = {p\ind{_{\phi}}}^{(3)}$, and we are able to find closed-form expressions for the fixed points of $h$ and the critical contours which pass through them (see Appendix \ref{chap:appendix_a}). There exists a local maximum at $\rho_{(1)} = \frac{\sqrt{3}}{2} M$, with ${p\ind{_{\phi}}}^{(1)} = \frac{9 \sqrt{3}}{2} M$. The maximum is enclosed by a single closed critical contour $h = {p\ind{_{\phi}}}^{(2)} = {p\ind{_{\phi}}}^{(3)} = \frac{1}{2} 5^{5/4} \varphi^{3/2} M$, which connects three saddles at $\rho_{(2)} = \frac{1}{2} 5^{1/4} \varphi^{3/2} M$, $z = 0$ and $\rho_{(3)} = \frac{1}{2} 5^{1/4} \varphi^{- 1/2} M$, $\pm z_{(3)} = \pm \frac{M}{2 \varphi}$, where $\varphi = \frac{1}{2} \left( 1 + \sqrt{5} \right)$ denotes the golden ratio. This special case is illustrated in Figure \ref{fig:mp_height_function_contours_2}.

Figures \ref{fig:mp_critical_contours_d_2} and \ref{fig:mp_critical_contours_d_4} of Section \ref{sec:non_planar_rays_two_dim_shadows} show the critical contours of the effective potential $h = \rho \, U^{2}$ in the extreme cases $d = \sqrt{\frac{16}{27}} M$ and $d = \sqrt{\frac{32}{27}} M$, respectively. When $d = \sqrt{\frac{16}{27}} M$, $h$ admits two saddle points in the equatorial plane. For $d = \sqrt{\frac{32}{27}} M$, there are two non-equatorial saddle points and there is a ``cusp'' in the plane; decreasing $d$ slightly from this marginal value, the cusp becomes a saddle point in the equatorial plane, as shown in Figure \ref{fig:mp_critical_contours_d_3}. For separations in the range $\sqrt{\frac{16}{27}} M < d < \sqrt{\frac{32}{27}} M$, the effective potential admits one equatorial saddle point, two non-equatorial saddles, and a local maximum in the equatorial plane. The latter corresponds to a stable circular photon orbit.

In the more general case of unequal-mass black holes ($\eta \neq 0$), the $\mathbb{Z}_{2}$ symmetry in the equatorial plane is broken: the non-planar saddle points are no longer connected by the same critical contour, and the kinematically bounded ``stable photon orbit region'' is pulled towards the more massive black hole. This is illustrated in Figure \ref{fig:mp_height_function_contours_4}, where the mass ratio is taken to be $\eta = \frac{M_{+} - M_{-}}{M_{+} + M_{-}} = 0.05$ ($M_{+} > M_{-}$), and the separation between the centres is $d = \frac{1}{2} M_{0} = \frac{1}{2} \left( M_{+} + M_{-} \right)$.

We now consider the effect of increasing the mass ratio on the existence of stable photon orbits around Majumdar--Papapetrou di-holes. In general, we are not able to find closed-form expressions for the stable photon orbit existence regions. For a fixed value of $\eta$, we employ a numerical root finder to solve the stationary point conditions $h\ind{_{, \rho}} = 0 = h\ind{_{, z}}$, and the stability condition $\det{\mathcal{H}(h)} = h\ind{_{, \rho \rho}} \, h\ind{_{, z z}} - (h\ind{_{, \rho z}})^{2} = 0$, in order to find the critical values of the dimensionless separation parameter $\tilde{d} = \frac{2 d}{M_{0}}$ for which stable photon orbits exist. In Figure \ref{fig:mp_dihole_spo_region}, we present a plot of the $(\eta, \tilde{d})$ parameter space which highlights the existence region for stable photon orbits. We observe that increasing the mass ratio has the effect of diminishing the stable photon orbit existence region; for $\eta \gtrsim 0.13$ stable photon orbits do not exist. When $\eta = 0$ (equal-mass case), we see that the existence region is given by $\sqrt{\frac{16}{27}} M < d < \sqrt{\frac{32}{27}} M$, which agrees with the analytical results described above.

%

\subsection{Equal-mass, equal-charge Reissner--Nordstr\"{o}m di-hole} \label{sec:spo_rn_dihole}

We now turn our attention to the equal-mass, equal-charge Reissner--Nordstr\"{o}m di-hole. This is a two-parameter subfamily of the five-parameter Bret\'{o}n--Manko--Aguilar solution, presented in Section \ref{sec:dihole_formalism_bma}. This solution comprises a pair of equal-mass black holes ($M_{\pm} = M$), with charges $Q_{\pm} = q M$, where $q \in [0, 1]$ is the charge ratio. The case $q = 0$ corresponds to the uncharged Weyl--Bach di-hole, whilst the extremal $q = 1$ case corresponds to the equal-mass Majumdar--Papapetrou di-hole (described in Section \ref{sec:spo_mp_dihole_existence}). The centres are held in static equilibrium by a Weyl strut, which imparts a force \cite{Manko2007}
\begin{equation}
\mathcal{F} = \frac{M \sigma^{2}}{d^{2} - 4 M^{2} \sigma^{2}},
\end{equation}
where $\sigma = \sqrt{1- q^{2}}$ is a charge parameter. In Weyl--Lewis--Papapetrou coordinates, the black hole event horizons appear as ``rods'' of length $2 \sqrt{M_{\pm}^{2} - Q_{\pm}^{2}} = 2 \sigma$ on the $z$-axis; the coordinate distance between the horizons is given by $d - 2 M \sigma$.

\begin{figure}
\begin{center}
\subfigure[Majumdar--Papapetrou]{
\includegraphics[height=0.45\textwidth]{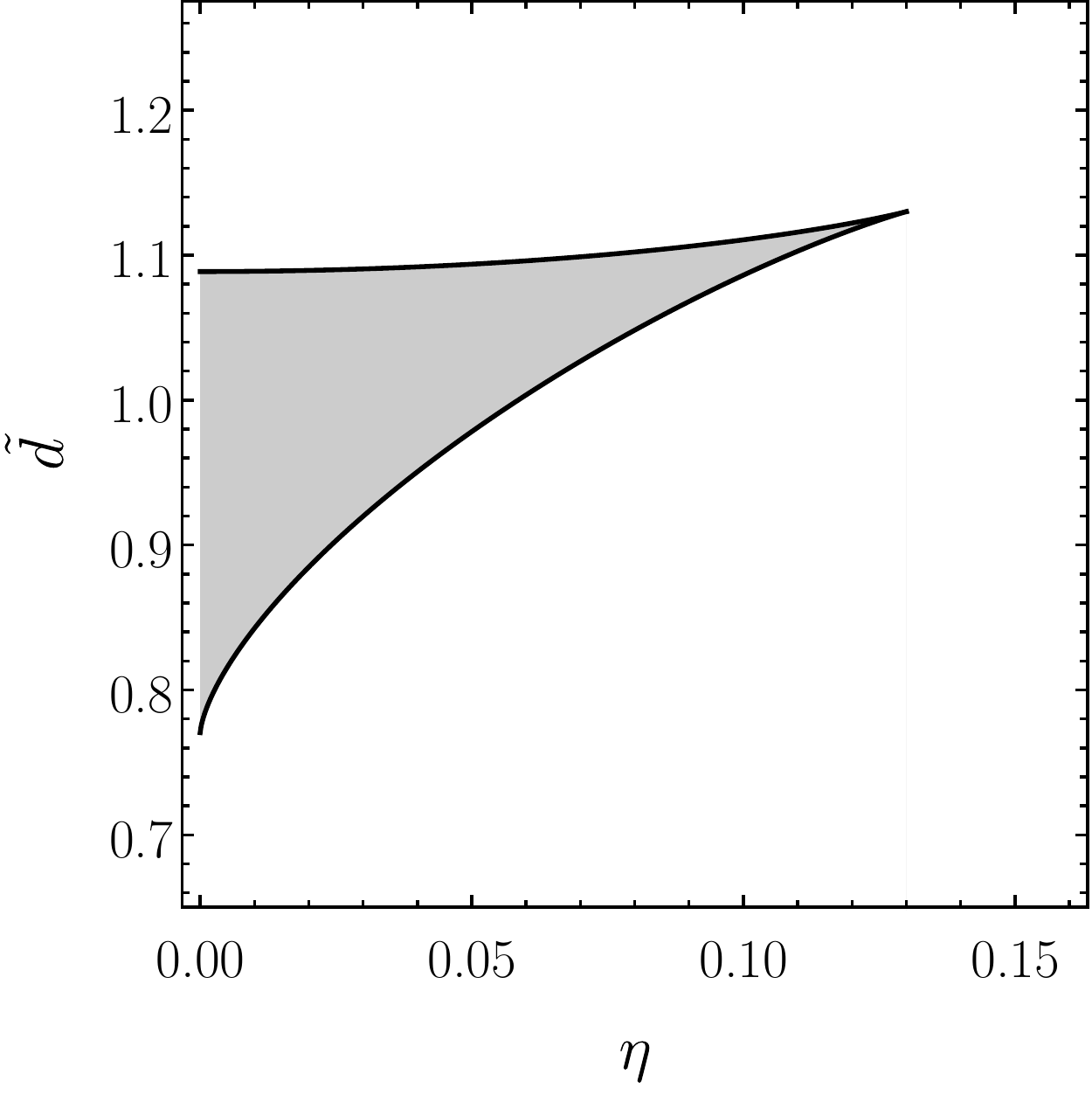} \label{fig:mp_dihole_spo_region} \hspace{1em}}
\subfigure[Reissner--Nordstr\"{o}m]{
\includegraphics[height=0.45\textwidth]{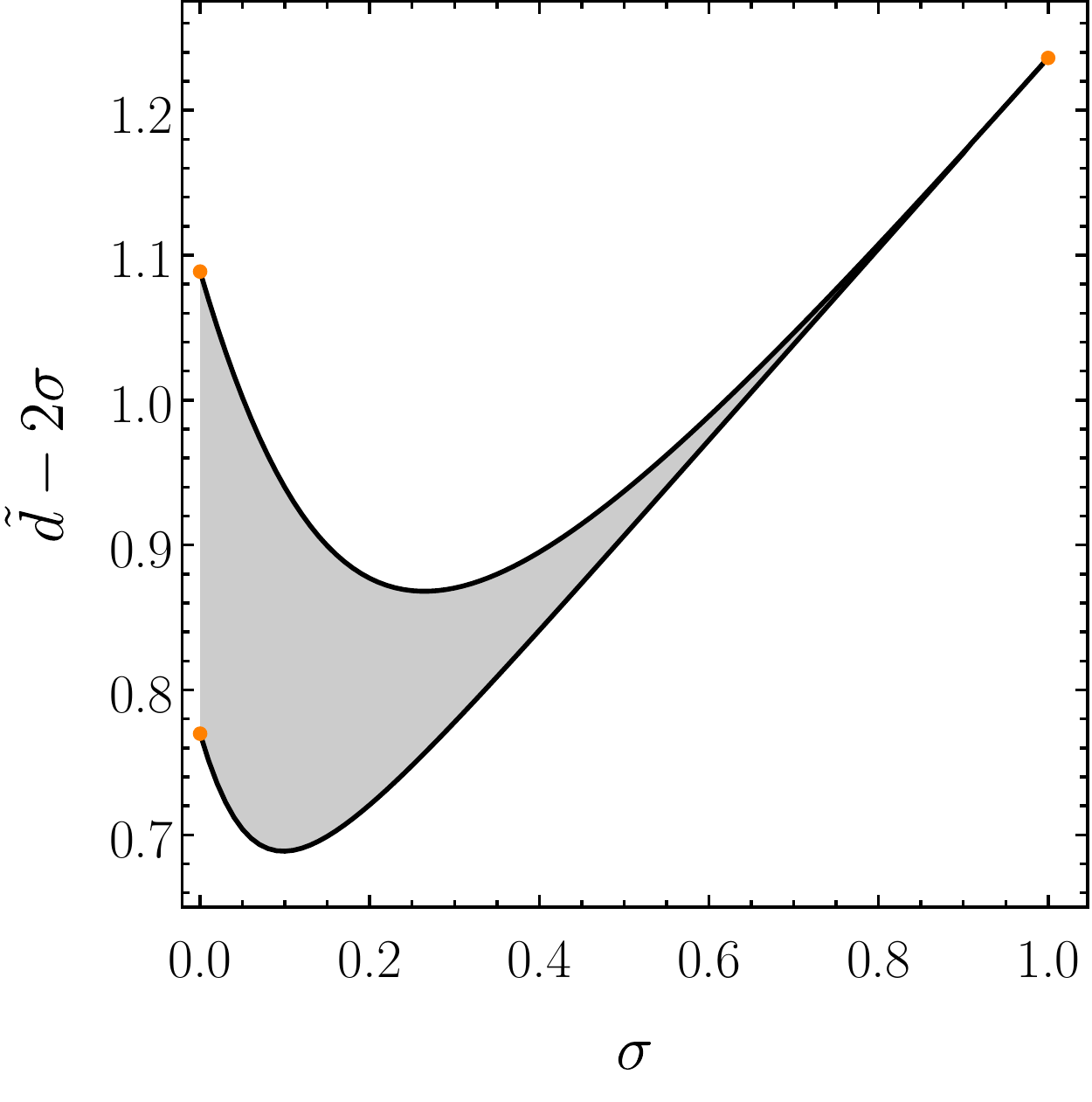} \label{fig:rn_dihole_spo_region} \hspace{1em}}
\end{center}
\caption{Existence region for stable photon orbits around Reissner--Nordstr\"{o}m di-holes, a subfamily of the $N = 2$ Bret\'{o}n--Manko--Aguilar class of electrostatic solutions. (a) Majumdar--Papapetrou di-holes with mass ratio $\eta$ and separation parameter $\tilde{d} = \frac{2 d}{M_{0}}$. (b) Equal-mass Reissner--Nordstr\"{o}m di-holes with charge parameter $\sigma = \sqrt{1 - q^{2}}$; the separation between the event horizons is $d - 2 M \sigma$.}
\label{fig:mp_dihole_parameter_space}
\end{figure}

Employing a similar method to that described in Section \ref{sec:spo_mp_dihole_existence}, we consider the effect of varying the charge parameter $\sigma$ on the existence region for stable photon orbits in the equal-mass, equal-charge Reissner--Nordstr\"{o}m di-hole spacetime. The results of the numerical algorithm are presented in Figure \ref{fig:rn_dihole_spo_region}. We observe that stable photon orbits exist for $0 \leq \sigma < 1$, i.e., up to but not including the uncharged Weyl--Bach limit ($\sigma = 1$) \cite{BachWeyl2012}. (Recall from our key result of Section \ref{sec:spo_electrovacuum_case} that stable photon orbits are permitted in electrovacuum, but not in pure vacuum.) Increasing the charge parameter $\sigma$ has the effect of shrinking the existence region for stable photon orbits. In Figure \ref{fig:rn_dihole_spo_region}, the critical values of the separation parameter for the Majumdar--Papapetrou and Weyl--Bach cases are shown as orange points: the analytical values are $d = \sqrt{\frac{16}{27}} M$ and $d = \sqrt{\frac{32}{27}} M$ (Majumdar--Papapetrou); and $d = 2 \left( \frac{1}{\varphi} + 1 \right) M$ (Weyl--Bach), where $\varphi$ denotes the golden ratio. (The Weyl--Bach case was studied by Coelho and Herdeiro in \cite{CoelhoHerdeiro2009}.)
%

\section{Geodesic structure of stable photon orbits} \label{sec:spo_diholes_geodesic_structure}

%
%

In general, the two-dimensional Hamiltonian system given by the Hamiltonian function \eqref{eqn:wlp_hamiltonian} with potential \eqref{eqn:wlp_potential} is non-integrable. (The stationary axisymmetric geometry \eqref{eqn:wlp_metric} will not admit higher-order Killing tensor, except in special cases.) Thus, stable photon orbits which exist close to local maxima of the effective potential(s) \eqref{eqn:spo_wlp_effective_potentials} will exhibit rich -- possibly chaotic -- behaviour. In this section, we analyse the geodesic dynamics of stable photon orbits for the Majumdar--Papapetrou di-hole, described in Section \ref{sec:spo_mp_dihole_existence}.

Introducing a dimensionless parameter $\mu$, we express the azimuthal angular momentum in the form
\begin{equation}
\label{eqn:mp_p_phi_mu}
p\ind{_{\phi}} = \left( 1 - \mu \right) {p\ind{_{\phi}}}^{\ast} + \mu \, {p\ind{_{\phi}}}^{(1)}, \qquad {p\ind{_{\phi}}}^{\ast} = \max{\left( {p\ind{_{\phi}}}^{(2)}, {p\ind{_{\phi}}}^{(3)} \right)}.
\end{equation}
The contours of $h$ form closed curves on a subset of the $(\rho, z)$-plane for $0 \leq \mu < 1$. Orbits which start in this region are confined to a toroidal region in the three-dimensional space spanned by the coordinates $\{ \rho, z, \phi \}$; such orbits are forbidden from escaping by angular momentum. Thus, kinematically bounded stable photon orbits exist in the regime $0 \leq \mu < 1$.

Close to the local maximum of $h$ (i.e., in the limit $\mu \rightarrow 1$ from below), the bounded region forms a small ellipse in the $(\rho, z)$-plane. At its local minimum $(\rho, z) = (\rho_{0}, z_{0})$, the potential \eqref{eqn:wlp_potential} satisfies $V(\rho_{0}, z_{0}) = 0$, and $V\ind{_{, \rho}}(\rho_{0}, z_{0}) = 0 = V\ind{_{, z}}(\rho_{0}, z_{0})$. Performing a second-order Taylor expansion about this minimum, the potential may be expressed as
\begin{equation}
\label{eqn:anisotropic_simple_harmonic_oscillator}
V \sim \frac{1}{2} \left[ V\ind{_{, \rho \rho}} \left( \rho - \rho_{0} \right)^{2} + 2 V\ind{_{, \rho z}} \left( \rho - \rho_{0} \right) \left( z - z_{0} \right) + V\ind{_{, z z}} \left( z - z_{0} \right)^{2} \right].
\end{equation}
One can then perform a rotation in the $(\rho, z)$-plane to remove the mixed quadratic term in the potential \eqref{eqn:anisotropic_simple_harmonic_oscillator}. The resulting potential resembles that of an anisotropic harmonic oscillator $V = \frac{1}{2} \left[ \omega_{\rho}^{2} \left(\rho - \rho_{0}\right)^{2} + \omega_{z}^{2} \left(z - z_{0}\right)^{2} \right]$, with characteristic frequencies given by $\omega_{\rho} = \sqrt{V\ind{_{, \rho \rho}}}$ and $\omega_{z} = \sqrt{V\ind{_{, z z}}}$. For $d < M$ ($d > M$), we have $\omega_{\rho} > \omega_{z}$ ($\omega_{\rho} < \omega_{z}$). The case $d = M$ is \emph{isotropic}, in the sense that $\omega_{\rho} = \omega_{z}$.

From numerical investigations, we see that generic kinematically bounded photon orbits for $\mu \rightarrow 0$ are precessing ellipses, confined to a small elliptical region of the $(\rho, z)$-plane in the neighbourhood of the local maximum of the effective potential. Even in this regime, higher-order corrections to the potential $V$ cannot be neglected if the characteristic frequencies of the orbit are commensurate, i.e., $\frac{\omega_{\rho}}{\omega_{z}} \approx \frac{n_{1}}{n_{2}}$, where $n_{1}$ and $n_{2}$ are small positive integers. This is because perturbation methods generally break down due to secular terms which appear in the case of low-order resonances \cite{Contopoulos2002}. Interestingly, the isotropic case admits a $1 \colon 1$ resonance, and other low-order resonances.
%

\subsection{Bounded orbits and Poincar\'{e} sections}

The dynamics of kinematically bounded photon orbits can be analysed using Poincar\'{e} sections. For a review of this method, see Section \ref{sec:chaotic_dynamical_systems} and \cite{Berry1978}, for example. In Appendix \ref{chap:appendix_b}, we present a brief, illustrative review of Poincar\'{e} sections for the well-known H\'{e}non--Heiles Hamiltonian system \cite{HenonHeiles1964}. A Poincar\'{e} map allows us to build up a picture of the global dynamics, which is governed by a Hamiltonian flow on a three-dimensional energy hypersurface (here $H = 0$, with $p\ind{_{\phi}} = \text{constant}$), by constructing a discrete dynamical system on a two-dimensional subspace -- the Poincar\'{e} section.

\begin{figure}
\begin{center}
\subfigure[Poincar\'{e} section for $\mu = 0.75$]{
\includegraphics[height=0.45\textwidth]{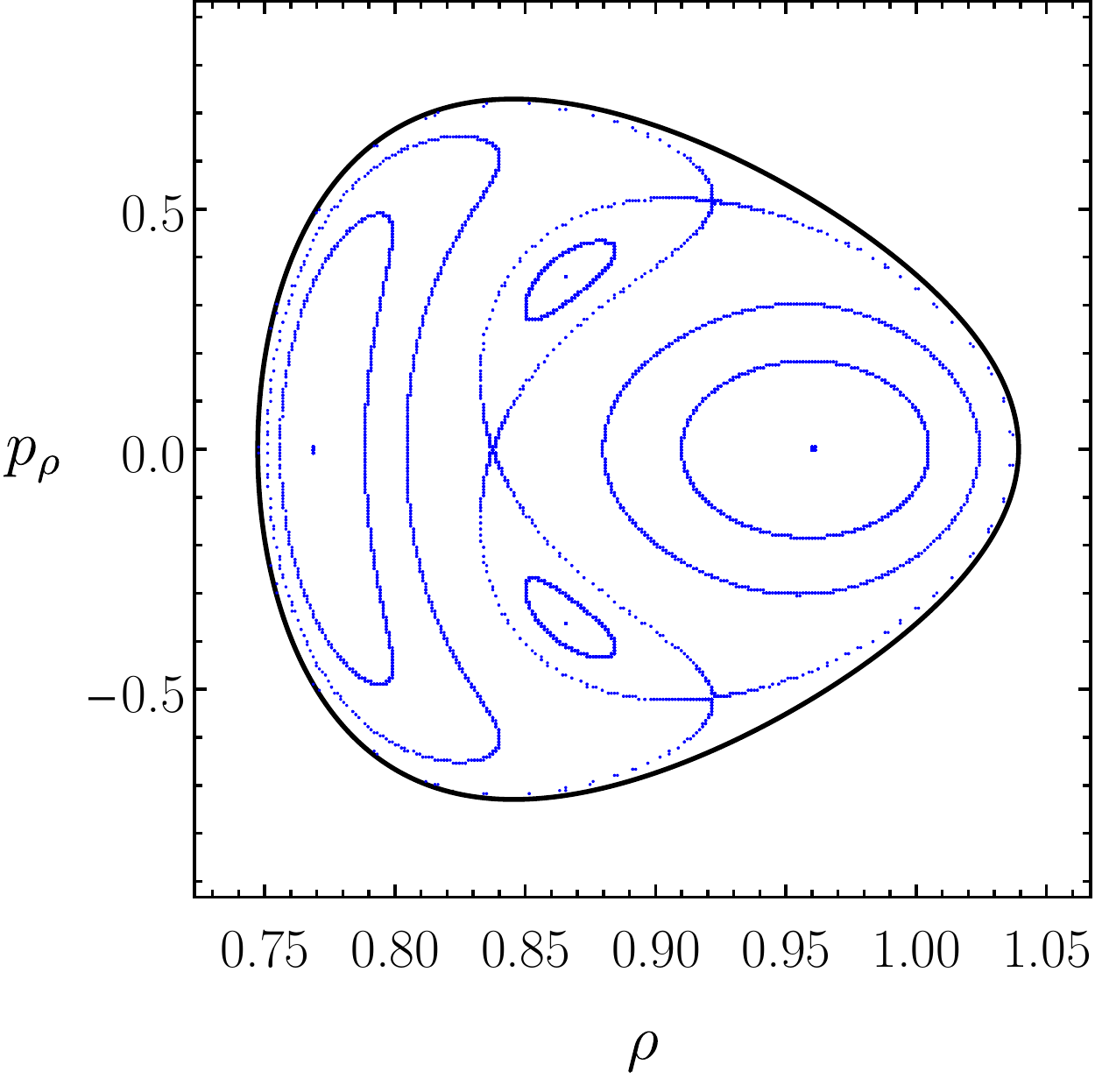} \label{fig:mp_poincare_section_mu_075} \hspace{1.1em}}
\subfigure[Periodic orbits for $\mu = 0.75$]{
\includegraphics[height=0.45\textwidth]{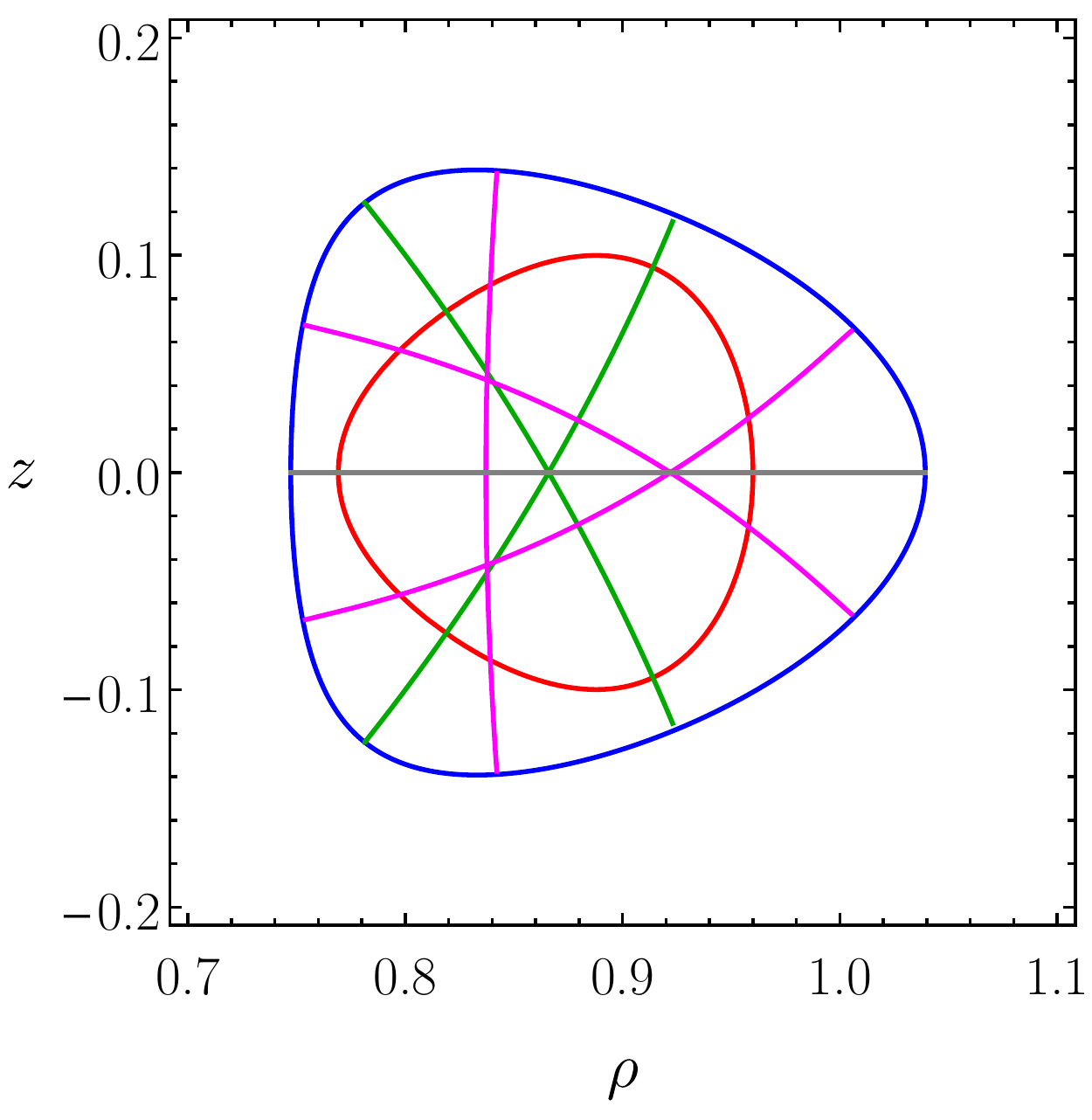} \label{fig:mp_orbits_mu_075} \hspace{1.5em}}
\subfigure[Orbit for $\mu = 0.75$]{
\includegraphics[height=0.45\textwidth]{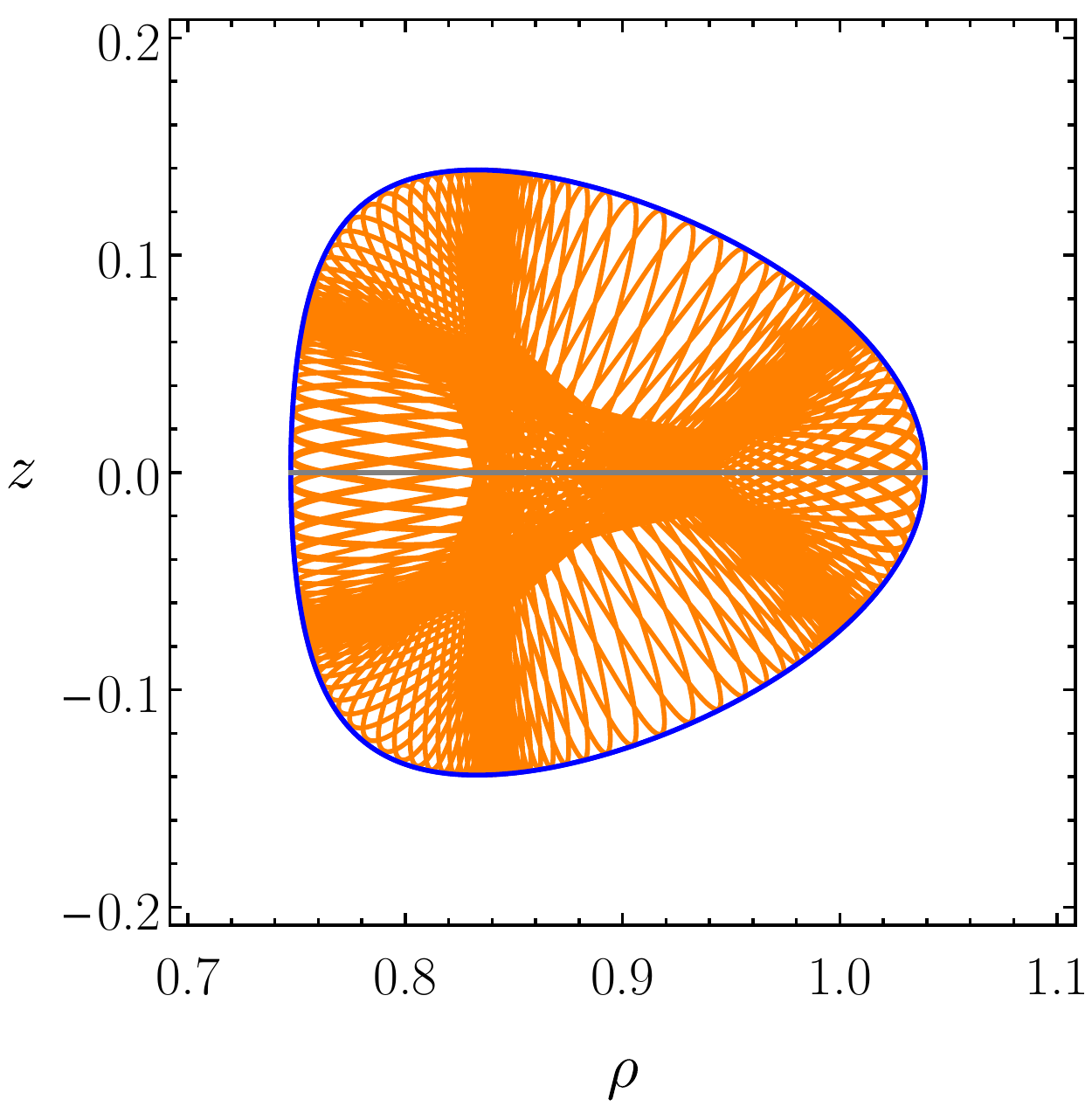} \label{fig:mp_orbits_2_mu_075} \hspace{1em}}
\subfigure[Orbit for $\mu = 0.75$]{
\includegraphics[height=0.45\textwidth]{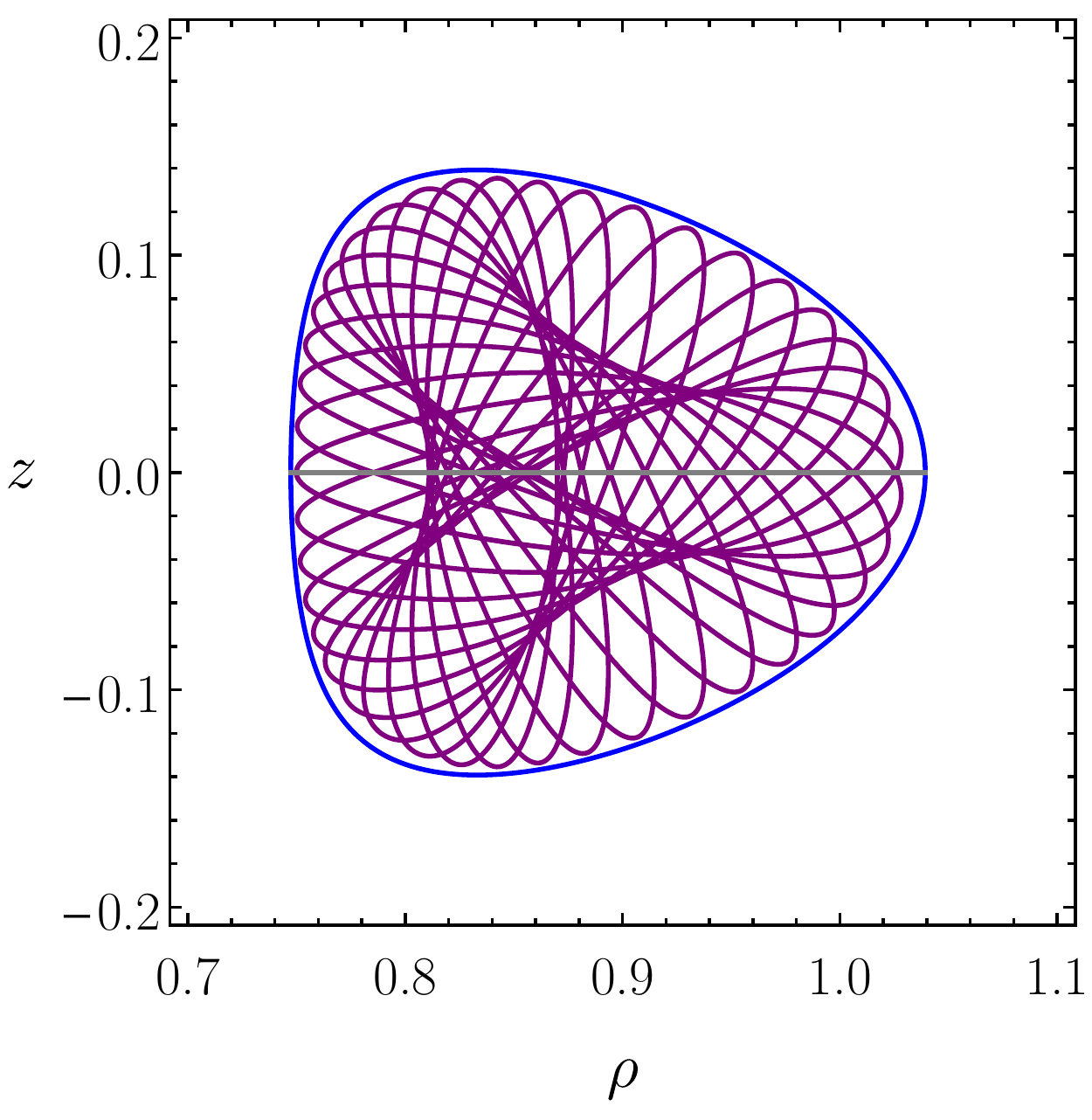} \label{fig:mp_orbits_3_mu_075} \hspace{1em}}
\caption{(a) Poincar\'{e} section in the $(\rho, p\ind{_{\rho}})$-plane for a selection of kinematically bounded photon orbits with dimensionless angular momentum parameter $\mu = 0.75$. (b) Periodic orbits projected onto the $(\rho, z)$-plane: a rotational $1 : 1$ resonant orbit (``loop orbit'') [red], and a pair of librational $1 : 1$ resonant orbits (``linear orbits'') [green], corresponding to elliptic fixed points of the Poincar\'{e} map; three unstable linear orbits [magenta], corresponding to hyperbolic fixed points. (c) An orbit whose initial conditions are perturbed slightly from those of the planar hyperbolic fixed point. (d) A high-order resonance (``box orbit'').}
\label{fig:mp_mu_075}
\end{center}
\end{figure}

\begin{figure}
\begin{center}
\subfigure[Poincar\'{e} section for $\mu = 0.4$]{
\includegraphics[height=0.45\textwidth]{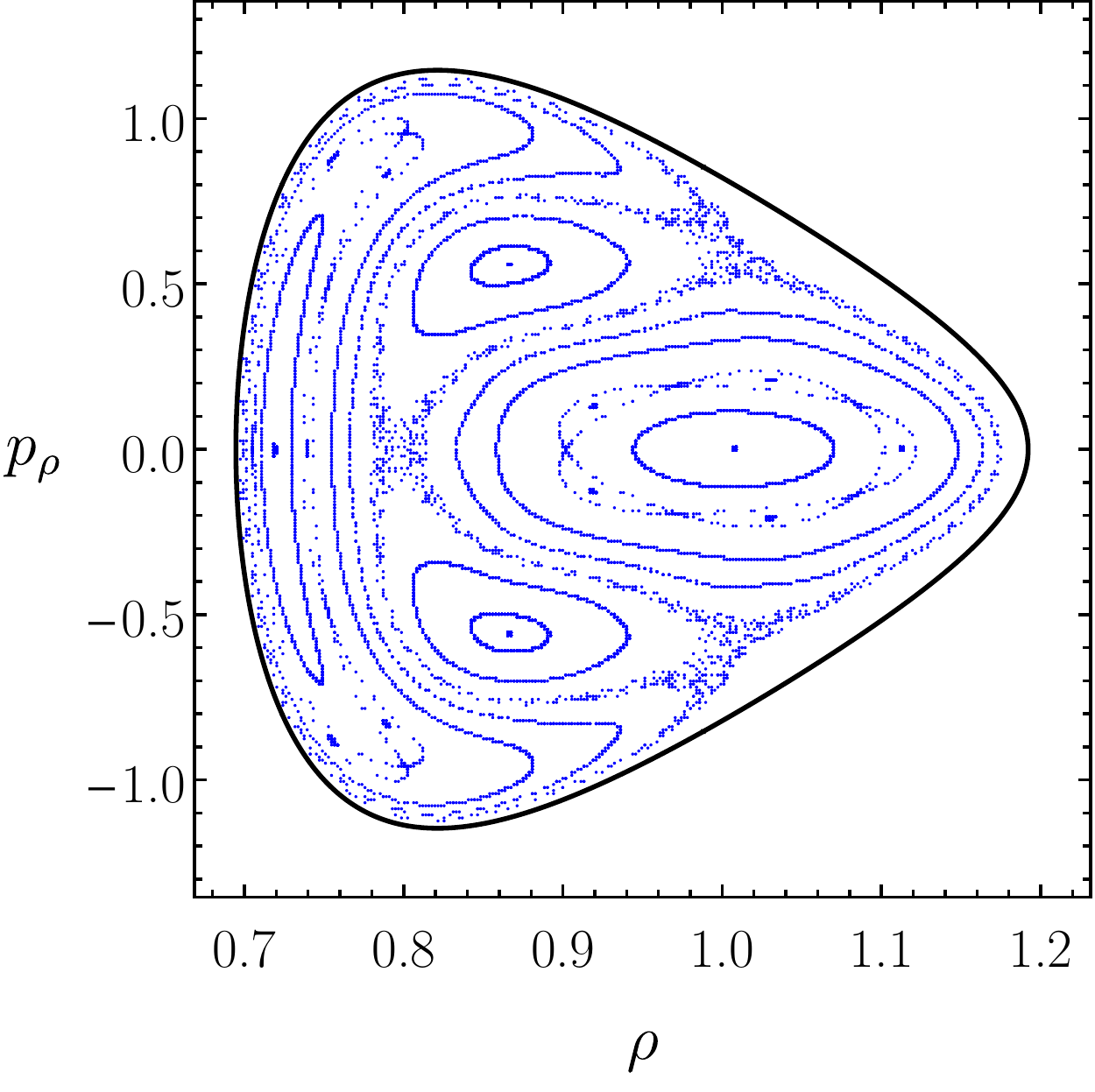} \label{fig:mp_poincare_section_mu_040} \hspace{1.2em}}
\subfigure[Orbits for $\mu = 0.4$]{
\includegraphics[height=0.45\textwidth]{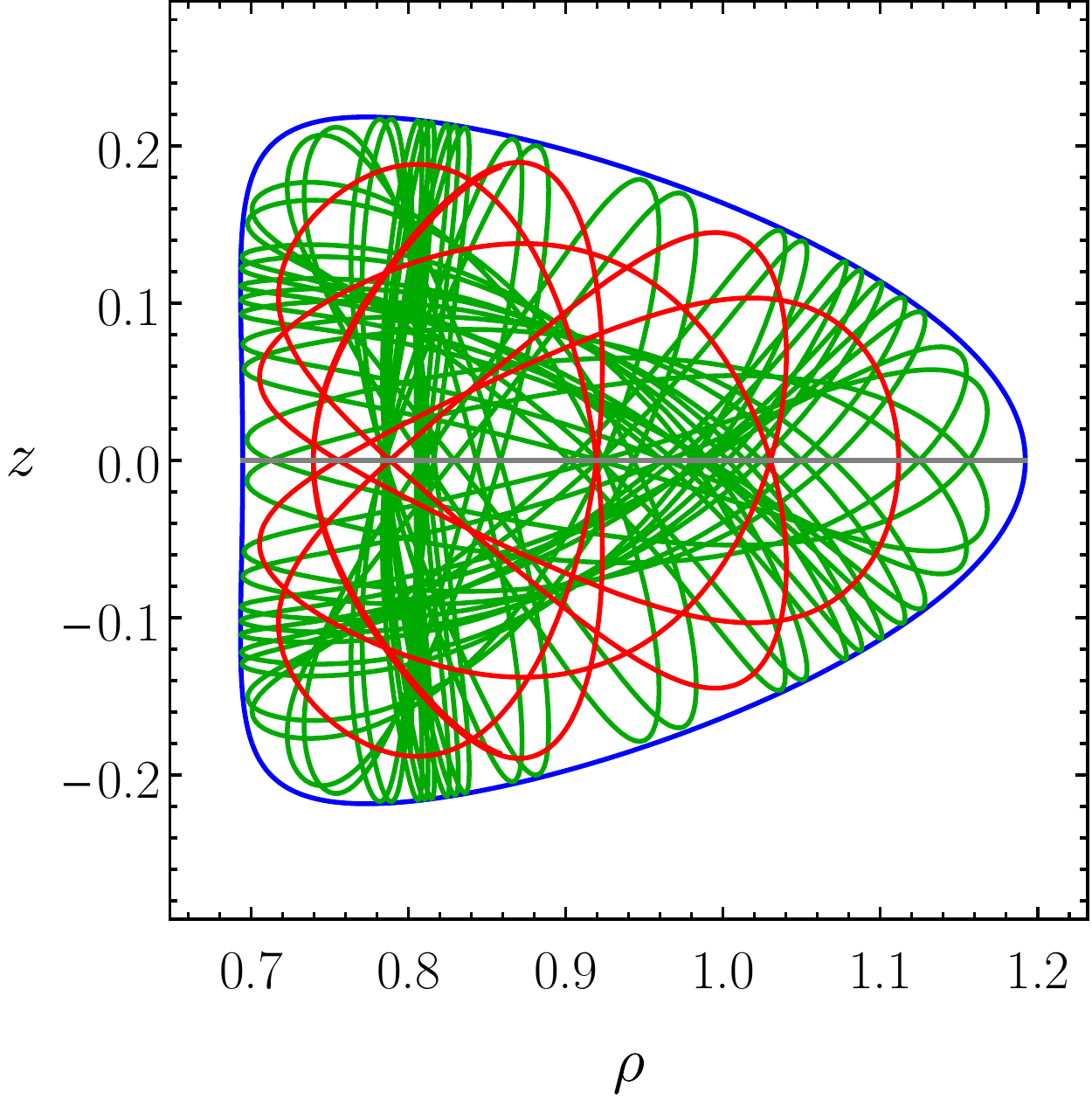} \label{fig:mp_orbits_mu_040} \hspace{1.3em}}
\subfigure[Poincar\'{e} section for $\mu = 0$]{
\includegraphics[height=0.45\textwidth]{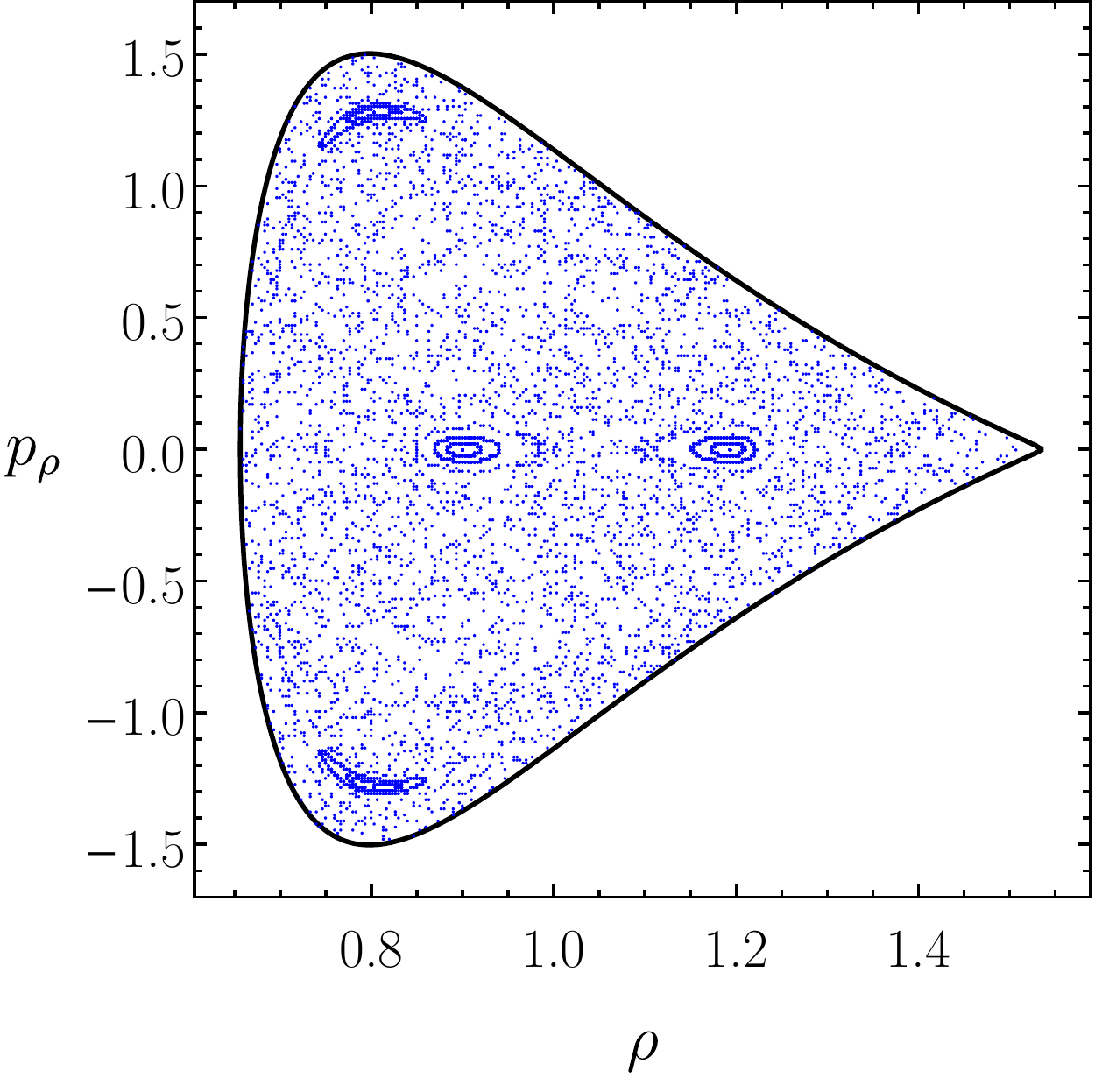} \label{fig:mp_poincare_section_mu_000} \hspace{1em}}
\subfigure[Orbits for $\mu = 0$]{
\includegraphics[height=0.45\textwidth]{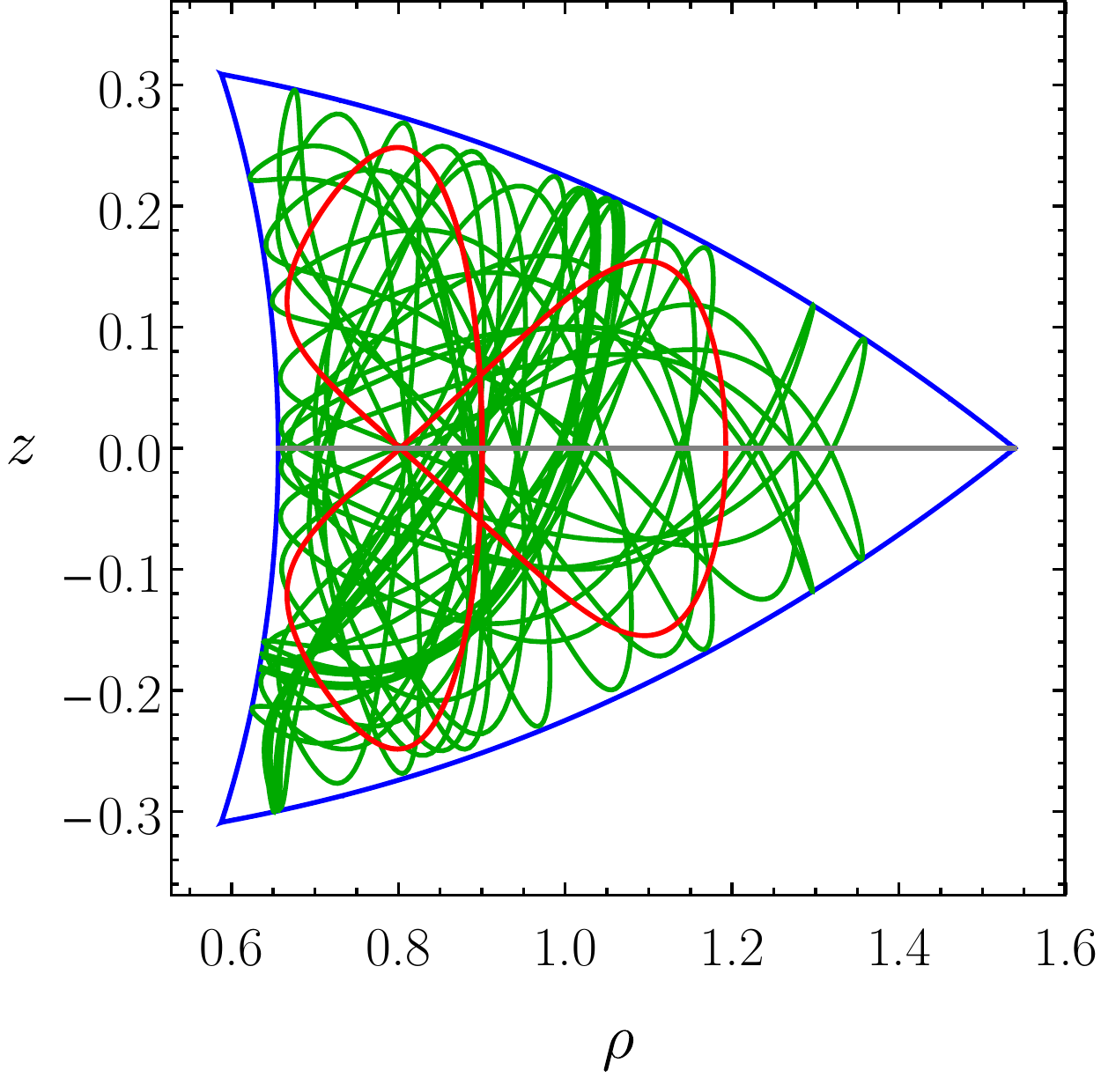} \label{fig:mp_orbits_mu_000} \hspace{1em}}
\caption{(a) Poincar\'{e} section in the $(\rho, p\ind{_{\rho}})$-plane for $\mu = 0.4$. (b) A pair of bounded orbits for $\mu = 0.4$: a low-order resonant orbit [red]; a chaotic orbit [green]. (c) Poincar\'{e} section in the $(\rho, p\ind{_{\rho}})$-plane for $\mu = 0$, with chaos dominant. There exist small ``islands of stability'' in the ``chaotic sea''. (d) A pair of bounded orbits for $\mu = 0$: a $2:2$ resonant orbit [red]; a chaotic orbit [green].}
\label{fig:mp_mu_040_000}
\end{center}
\end{figure}

We let $\mathcal{E} = \left\{ (\rho, z, p\ind{_{\rho}}, p\ind{_{z}})  \,| \, H = 0  \right\}$ be the three-dimensional energy hypersurface, defined by the Hamiltonian constraint. We take the two-dimensional Poincar\'{e} section (or surface of section) to be the surface $\mathcal{S} = \left\{ (\rho, p\ind{_{\rho}})  \,| \, z = 0  \right\}$, which intersects $\mathcal{E}$. The Poincar\'{e} map $P  \colon \mathcal{S} \rightarrow \mathcal{S}$ then maps a trajectory on $\mathcal{E}$ that begins on $\mathcal{S}$ to its next point of intersection with $\mathcal{S}$.
%
%
%

Here, we construct Poincar\'{e} sections for the equal-mass Majumdar--Papapetrou di-hole using a method analogous to that of the H\'{e}non--Heiles system presented in Appendix \ref{chap:appendix_b}. We consider a selection of orbits which start within the closed contour $h = p\ind{_{\phi}}$, where ${p\ind{_{\phi}}}^{\ast} \leq p\ind{_{\phi}} < {p\ind{_{\phi}}}^{(1)}$. The trajectories are numerically integrated using Hamilton's equations. Each time the orbit intersects the surface $z = 0$ (with $p\ind{_{z}} > 0$), the values of $\rho$ and $p\ind{_{\rho}}$ are recorded.

Figures \ref{fig:mp_mu_075} and \ref{fig:mp_mu_040_000} show Poincar\'{e} sections for the isotropic case $d = M$, where we consider the effect of varying the dimensionless angular momentum parameter $\mu$, defined in \eqref{eqn:mp_p_phi_mu}. We also present selected examples of kinematically bounded orbits projected onto the $(\rho, z)$-plane. The orbits fall into three categories: (i) low-order $n_{1} : n_{2}$ resonant orbits; (ii) high-order resonances (or box orbits); and (iii) chaotic orbits. (See \cite{Zotos2015} for a classification of these types of orbits for the H\'{e}non--Heiles system.) The $n_{1} : n_{2}$ families of resonant orbits (where $n_{1}$ and $n_{2}$ are both small positive integers) exhibit an oscillatory pattern: the orbit completes $n_{1}$ oscillations in the $\rho$-direction in the time it takes to complete $n_{2}$ oscillations in the $z$-direction. The ratio $n_{1} : n_{2}$ corresponds to the ratio of the characteristic frequencies of the orbit, which are defined with respect to the frequency of the largest amplitude in each direction; see \cite{Zotos2015} and references therein.

For $\mu = 0.75$, the Poincar\'{e} section, shown in Figure \ref{fig:mp_poincare_section_mu_075}, exhibits four regular regions, delineated by a ``separatrix''. Each of these four regions contains closed, smooth, oval-shaped curves, which surround stable elliptic fixed points of the Poincar\'{e} map. The pair of fixed points in the plane $p\ind{_{\rho}} = 0$ are ``rotational'' orbits which form closed curves in the $(\rho, z)$-plane; whereas the fixed points out of the plane are ``librational'' orbits, which correspond to open curves in the $(\rho, z)$-plane that bounce between two sides of the closed contour $h = p\ind{_{\phi}}$. The separatrix intersects itself three times; the points of intersection are unstable hyperbolic fixed points of the Poincar\'{e} map. The orbits corresponding to the elliptic and hyperbolic fixed points are shown in Figure \ref{fig:mp_orbits_mu_075}: the stable rotational orbit is shown in red; the stable librational orbits are shown in green; and the unstable orbits are shown in magenta. In Figure \ref{fig:mp_orbits_2_mu_075}, we show an orbit whose initial conditions are perturbed slightly from those of the planar unstable hyperbolic fixed point. Figure \ref{fig:mp_orbits_3_mu_075} shows an example of a box orbit (i.e., a higher-order resonance).

\begin{figure}
\begin{center}
\subfigure[Poincar\'{e} section for $d = 0.95 M$, $\mu = 0.2$]{
\includegraphics[height=0.45\textwidth]{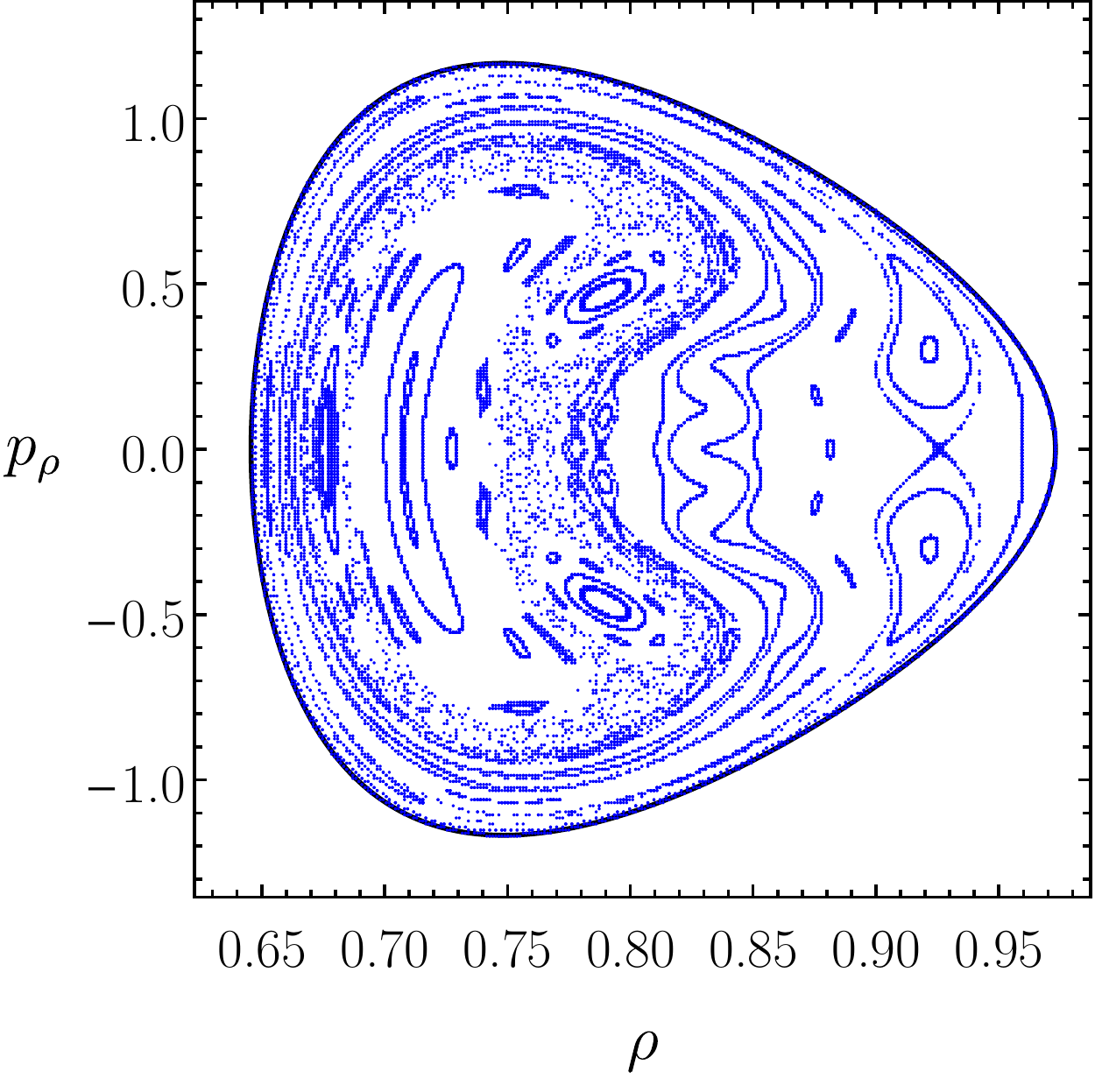} \label{fig:mp_poincare_section_d_095} \hspace{1em}}
\subfigure[Poincar\'{e} section for $d = 1.05 M$, $\mu = 0.05$]{
\includegraphics[height=0.45\textwidth]{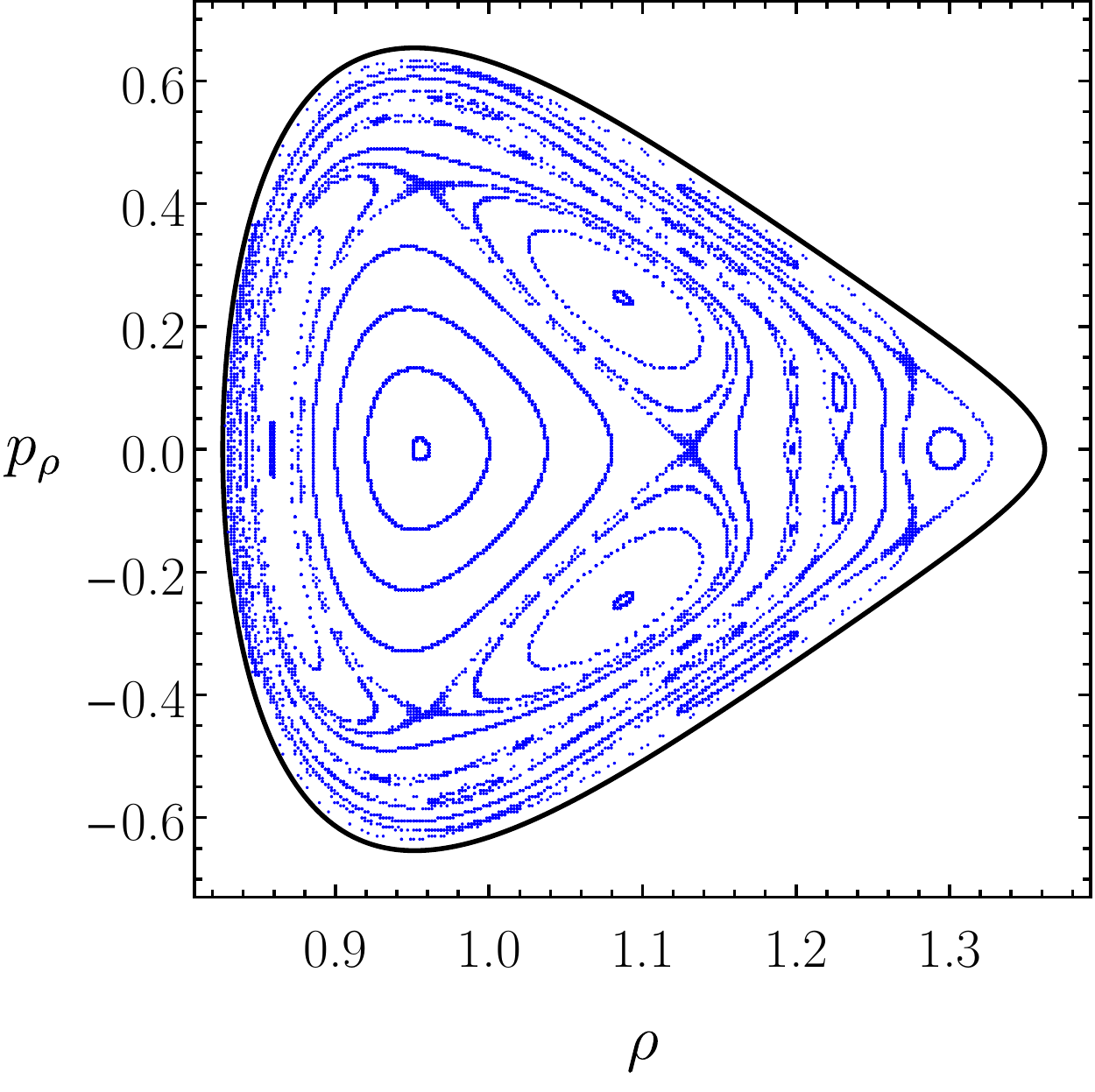} \label{fig:mp_poincare_section_d_105} \hspace{1em}}
\caption{Poincar\'{e} sections for the ``anisotropic'' ($d \neq M$) equal-mass Majumdar--Papapetrou di-hole. (a) An example with $d < M$. The Poincar\'{e} section exhibits rich structure: elliptic and hyperbolic fixed points; low-order resonances; KAM islands; chaotic bands; and novel features, including ``crenulations''. (b) An example with $d > M$ which exhibits regular behaviour.}
\label{fig:mp_poincare_separation}
\end{center}
\end{figure}
\begin{figure}
\begin{center}
\subfigure[$d = 0.95 M$, $\mu = 0.20$]{
\includegraphics[height=0.31\textwidth]{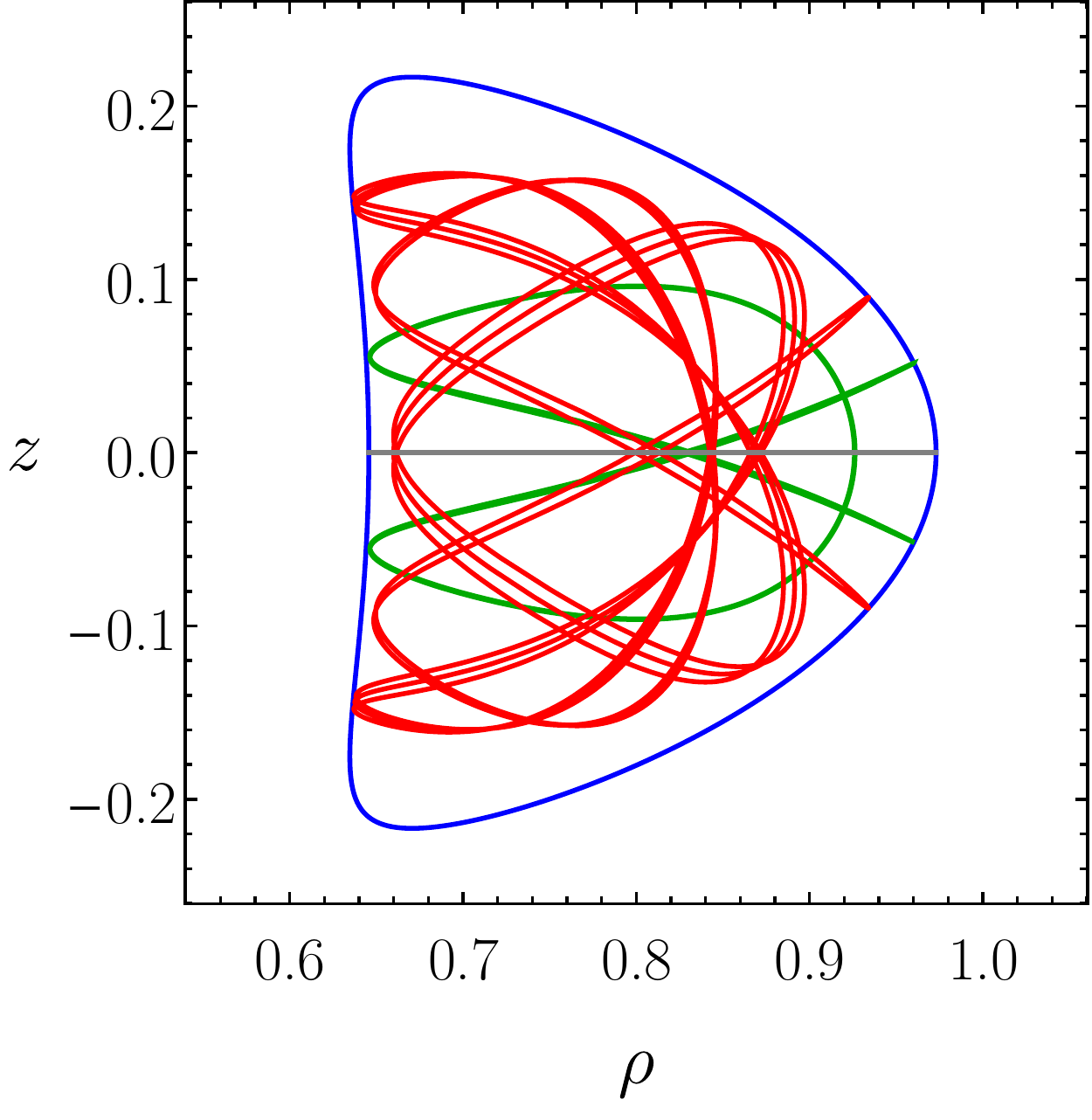} \label{fig:mp_orbits_d_095_mu_020}} \hfill
\subfigure[$d = 1.05 M$, $\mu = 0.05$]{
\includegraphics[height=0.31\textwidth]{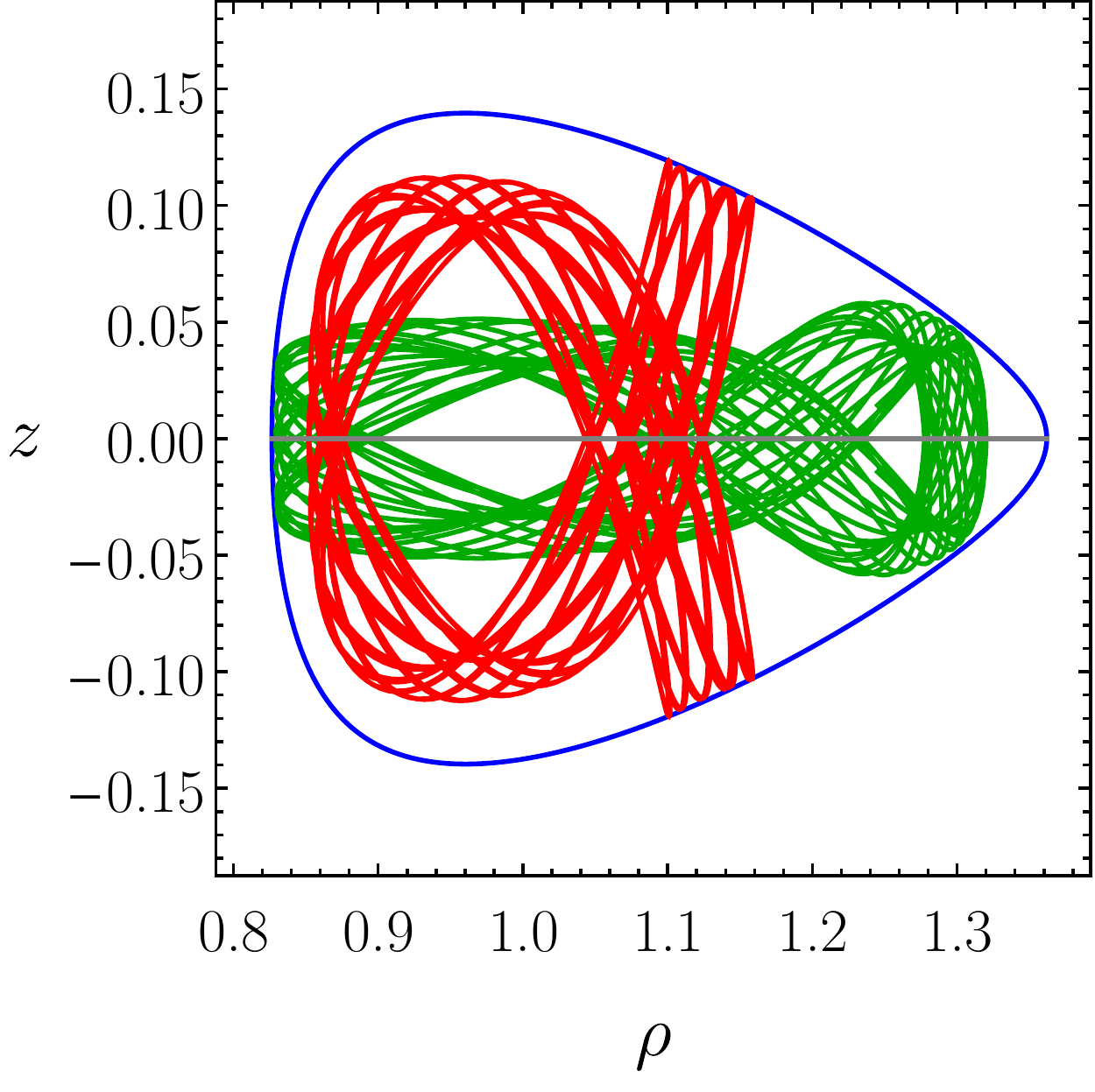} \label{fig:mp_orbits_d_105_mu_005}} \hfill
\subfigure[$d = M$, $\mu = -0.10$]{
\includegraphics[height=0.31\textwidth]{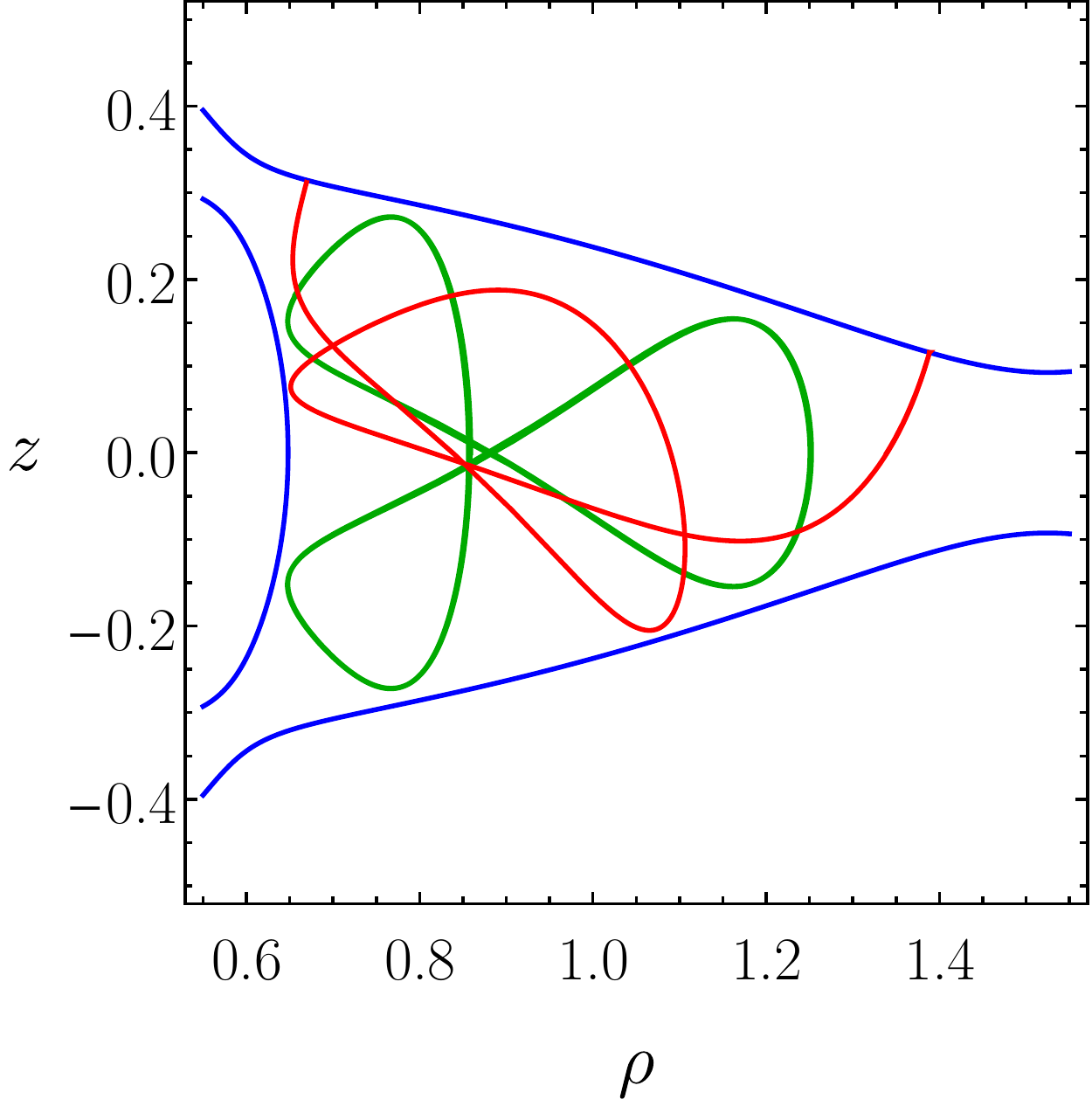} \label{fig:mp_orbits_mu_m010}}
\caption{(a) Regular bounded null orbits for the $d < M$ case, associated with the ``crenulations'' in the Poincar\'{e} section of Figure \ref{fig:mp_poincare_section_d_095}. (b) Regular bounded null orbits for the $d > M$ case; see Figure \ref{fig:mp_poincare_section_d_105} for the corresponding Poincar\'{e} section. (c) Non-escaping orbits in the unbounded regime with $\mu = - 0.1$: an open libration-type orbit which touches the contour $h = p\ind{_{\phi}}$ [red]; a closed rotation-type orbit [green]. These orbits persist, despite the existence of three narrow escapes which connect the scattering region to the black holes and to infinity (cf.~Chapter \ref{chap:fractal_structures}).}
\label{fig:mp_spos_sep_unbounded}
\end{center}
\end{figure}

We now turn our attention to Figure \ref{fig:mp_mu_040_000}, in which we examine the effect of decreasing the parameter $\mu$ (i.e., approaching the threshold value $p\ind{_{\phi}} = {p\ind{_{\phi}}}^{\ast}$). In Figure \ref{fig:mp_poincare_section_mu_040}, we show the Poincar\'{e} section for $\mu = 0.4$. The four regions around the elliptic fixed points containing regular orbits persist; however, the separatrix has degenerated into a chaotic band. Moreover, stable and unstable Kolmogorov--Arnold--Moser (KAM) islands are manifest around resonances \cite{Berry1978}. Figure \ref{fig:mp_orbits_mu_040} shows a $5 : 5$ resonant orbit [red]; and a chaotic orbit [green]. As $\mu$ is decreased, further lower-order resonances bifurcate from the $1 : 1$ resonances exhibited in Figure \ref{fig:mp_orbits_mu_075}; and we observe the transition from order to chaos. At the threshold $\mu = 0$, chaos is dominant -- as shown by the Poincar\'{e} section in Figure \ref{fig:mp_poincare_section_mu_000}. We see that almost all of the Poincar\'{e} section is filled by chaotic orbits, with small ``islands of stability'' manifest. Figure \ref{fig:mp_orbits_mu_000} shows a pair of orbits bounded by the contour $h = {p\ind{_{\phi}}}^{\ast}$. In particular, we show a $2 : 2$ resonant orbit [red], and an example of a chaotic orbit [green].

Intriguingly, the qualitative features of kinematically bounded photon orbits around the isotropic ($d = M$) equal-mass Majumdar--Papapetrou di-hole are shared by the well-known H\'{e}non--Heiles Hamiltonian system \cite{HenonHeiles1964, Berry1978, Zotos2015}, a review of which is given in Appendix \ref{chap:appendix_b}. In particular, compare the Poincar\'{e} sections and bounded orbits in Figures \ref{fig:mp_mu_075} and \ref{fig:mp_mu_040_000} (Majumdar--Papapetrou) with those of Figures \ref{fig:hh_e_1_12} and \ref{fig:hh_e_1_8_and_1_6} (H\'{e}non--Heiles).

Figure \ref{fig:mp_poincare_separation} shows Poincar\'{e} sections for the ``anisotropic'' ($d \neq M$) equal-mass Majumdar--Papapetrou di-hole. We observe a rich variety of features in the $d < M$ case, shown in Figure \ref{fig:mp_poincare_section_d_095}. In addition to elliptic and hyperbolic fixed points (periodic orbits), there exist chaotic domains in which KAM islands of stability are embedded. Moreover, we observe novel features in this Poincar\'{e} section, including the ``crenulations''. The Poincar\'{e} section shown in Figure \ref{fig:mp_poincare_section_d_105} ($d > M$) exhibits regular behaviour, despite the fact that $\mu$ is small (i.e., we are far from the maximum of $h$).

Intriguingly, non-escaping photon orbits survive even in the kinematically unbounded regime where $\mu < 0$ (i.e., $p\ind{_{\phi}} < {p\ind{_{\phi}}}^{(1)}$). Figure \ref{fig:mp_orbits_mu_m010} shows examples of closed (rotation-type) and open (libration-type) periodic orbits in for $\mu = - 0.1$. Recall from Section \ref{sec:exit_basins_phase_space} that, for $\mu \lesssim 0$ there exist KAM tori of quasi-periodic orbits that do not escape the scattering region, despite the system being \emph{open} (i.e., kinematically unbounded); see the exit basins of Figures \ref{fig:mp_rho_z_basin_001} and \ref{fig:mp_rho_p_rho_basin_001}, for example.
%
%

\section{Discussion}
\label{sec:spo_discussion}

\subsection{Extensions}
\label{sec:spo_extensions}
%

\subsubsection{Stable photon orbits in other spacetimes}

Cunha \emph{et al.} \cite{CunhaGroverHerdeiroEtAl2016} explore null geodesic motion around boson stars and Kerr black holes with scalar hair. Two-dimensional effective potentials -- similar to those discussed in Section \ref{sec:spo_hamiltonian_formalism} -- are employed, since the motion is not known (nor expected) to be separable. The authors find that the spacetimes admit stable light-rings, which permit the formation of ``pocket'' features in the two-dimensional configuration space spanned by the Boyer--Lindquist coordinates $\{ r, \theta \}$; these pocket features are delimited by the contours of the effective potential. As in the Majumdar--Papapetrou di-hole case, such features lead to an effective trapping of some null rays. Furthermore, the authors observe that the existence of multiple light-rings (including a stable one) around a horizon is a central ingredient for the existence of multiple disconnected shadows of a single hairy black hole.

Using techniques from the theory of dynamical systems, Grover and Wittig \cite{GroverWittig2017} show how light-rings -- fixed points of the null geodesic potential -- give rise to families of periodic orbits and invariant manifolds in phase space. These invariant structures are shown to define the shape of the black hole shadow as well as a number of key signatures of gravitational lensing. As a case study, the authors analyse chaotic lensing by a Herdeiro--Radu hairy black hole (i.e., a Kerr black hole with scalar hair), similar to those studied by Cunha \emph{et al.} \cite{CunhaHerdeiroRaduEtAl2015, CunhaHerdeiroRaduEtAl2016}.

In a recent study, Jia \emph{et al.} \cite{JiaPangYang2018} explore the existence and stability of null (and timelike) equatorial circular orbits in \emph{static} axisymmetric spacetimes. Using a fixed point approach similar to that considered here, the authors obtain a necessary and sufficient conditions for the (non-)existence of timelike circular orbits; they prove that (stable) timelike circular orbits will always exist at large $\rho$ in an asymptotically flat static axisymmetric spacetime with a positive Arnowitt--Deser--Misner mass. The authors then derive the necessary and sufficient condition for the existence of null circular orbits. They show that the existence of timelike circular orbits in static axisymmetric spacetimes does not, in general, imply the existence of null circular orbits, and vice versa.

To generalise the results for stationary axisymmetric electrovacuum spacetimes (Section \ref{sec:spo_electrovacuum_case}), one could consider (i) including a non-zero cosmological constant; (ii) other fields and matter sources; or (iii) spacetimes with less symmetry. Let us comment on each of these potential generalisations here.

A natural starting point for case (i) would be to consider a cosmological version of the Weyl--Lewis--Papapetrou solution: the spacetime geometry and an Ernst-type formulation of the field equations are presented in vacuum by Charmousis \emph{et al.} \cite{CharmousisLangloisSteerEtAl2007}, and in electrovacuum by Astorino \cite{Astorino2012}. In both cases, the inclusion of a non-zero cosmological constant modifies the metric and field equations in a non-trivial way; a careful generalisation of the results of Section \ref{sec:spo_electrovacuum_case} would therefore be necessary in this case. As a first attempt at case (ii), one may wish to analyse the existence and stability of photon orbits in stationary axisymmetric (electro)vacuum spacetimes which are coupled to a scalar field. Solutions in the case of conformally (and minimally) coupled scalar fields in vacuum are considered by Astorino \cite{Astorino2015}. We caution that, even in the ``simple'' case of massless minimally or conformally couple scalar fields, the metric and field equations are much less tractable than in the case of stationary axisymmetric electrovacuum with no additional matter fields (considered in this work). Finally, we remark that, whilst case (iii) is of interest for dynamical binary black holes (which are expected to be neither stationary nor axisymmetric), the lack of symmetry would mean that studying the existence and stability of photon orbits using two-dimensional effective potentials would not be possible. A special treatment would be required to study the existence and stability of circular photon orbits in dynamical spacetimes. To gain insight into the nature of photon orbits on the spacetimes of dynamical binary black holes, a systematic study of the simulations generated by Bohn \emph{et al.} \cite{BohnThroweHebertEtAl2015}, for example, could repay further investigation.
%

\subsubsection{Light-ring stability for horizonless ultra-compact objects}

In Section \ref{sec:spo_electrovacuum_case}, we considered the stability of photon orbits for (electro)vacuum solutions to the Einstein(--Maxwell) field equations, motivated by the study of black holes. In recent work, Cunha \emph{et al.} \cite{CunhaBertiHerdeiro2017} consider stationary axisymmetric solutions of Einstein's field equations that are formed from classical gravitational collapse of matter which obeys the null energy condition, which are ultra-compact (i.e., they possess a light-ring), but which do not possess an event horizon.

Using a Hamiltonian formalism for null geodesics similar to that of Sections \ref{sec:spo_hamiltonian_formalism} and \ref{sec:spo_electrovacuum_case}, the authors derive a pair of effective potentials which they use to analyse the existence and stability of circular photon orbits. The authors prove that, for ultra-compact objects described by a stationary axisymmetric geometry which is a solution to \emph{any} metric theory of gravity in which photons propagate along null geodesics, light-rings always come in pairs, one of which is a saddle point and the other is a local extremum of an effective potential. The proof depends on a topological argument based on the Brouwer degree of a continuous map. Then, assuming Einstein's equations of general relativity, the authors prove that the local extremum of the effective potential is a \emph{minimum} (i.e., a stable light-ring) if the stress--energy tensor satisfies the null energy condition.

In a follow-up paper, Hod \cite{Hod2018} proves that, whilst the key theorem of Cunha \emph{et al.} \cite{CunhaBertiHerdeiro2017} is generally true, there is an exception in the case of \emph{degenerate light-rings}, which are characterised by the simple relation $8 \pi \left( \varrho + p_{\textrm{T}} \right) r_{\gamma}^{2} = 1$, where $r = r_{\gamma}$ is the location of the light-ring in spherical coordinates, $\varrho = - T\ind{^{t}_{t}}$ is the energy density, and $p_{\textrm{T}} = T\ind{^{\theta}_{\theta}} = T\ind{^{\phi}_{\phi}}$ is the tangential pressure. This is illustrated explicitly by considering the case of a spherically symmetric constant-density stars with dimensionless compactness $\frac{M}{R} = \frac{1}{3}$, where $M$ is the object's mass and $R$ is its surface radius; Hod finds that this system admits a unique \emph{unstable} light-ring.

\subsection{Conclusions}
\label{sec:spo_conclusions}

In this chapter, we have discussed the existence, stability and phenomenology of circular photon orbits (or light-rings) in the context of four-dimensional stationary axisymmetric solutions to the Einstein--Maxwell equations of gravity and electromagnetism.

Geodesic motion on Kerr(--Newman) spacetime, which is perhaps the most astrophysically relevant stationary axisymmetric (electro)vacuum solution, is separable (and hence Liouville integrable) in Boyer--Lindquist coordinates, thanks to the existence of four integrals of motion in involution on phase space. The fourth conserved quantity is the Carter constant, which arises from the ``hidden'' symmetry of Kerr(--Newman) spacetime which is encoded in the existence of a rank-two Killing tensor field (see Section \ref{sec:killing_objects_symmetries_integrability}). The separability of the radial and latitudinal motion in Boyer--Lindquist coordinates means that the problem of classifying the equatorial circular photon orbits of Kerr--Newman spacetime reduces to classifying the roots of the quartic \eqref{eqn:kerr_ecpo_rdot_u} in $u = \frac{M}{r}$, where $r$ is the radial coordinate and $M$ is the mass. We reviewed the Balek--Bi\v{c}\'{a}k--Stuchl\'{i}k phase diagram \cite{BalekBicakStuchlik1989} in the charge--spin plane for equatorial circular photon orbits of positive Boyer--Lindquist radius $r$ (Figure \ref{fig:kerr_ecpos}).

We focussed on stationary axisymmetric spacetimes in four dimensions, which are described in Weyl--Lewis--Papapetrou coordinates $\{ t, \rho, z, \phi \}$ by the line element \eqref{eqn:wlp_metric}. In general, the geodesic motion on such geometries is not expected to be separable (except in special cases, such as the Kerr--Newman spacetime); however, the stationarity and axisymmetry mean that $t$ and $\phi$ are ignorable coordinates, and the motion of null rays is governed by a two-dimensional time-independent Hamiltonian system with conserved parameter $p\ind{_{\phi}}$. We introduced the effective potentials $h^{\pm}(\rho, z)$, given by \eqref{eqn:spo_wlp_effective_potentials}, and demonstrated that the problem of classifying circular photon orbits reduces to that of classifying the stationary points of these effective potentials.

Restricting our attention to the electrovacuum case, we classified the stationary points of the effective potentials by availing a subset of the Einstein--Maxwell field equations. Our key result is that, in the context of four-dimensional stationary axisymmetric spacetimes in general relativity, generic stable photon orbits are forbidden in pure vacuum; however, stable photon orbits may arise in \emph{electrovacuum}, thanks to \eqref{eqn:tr_hessian_hpm_5}.

As a case study, we explored the existence of stable photon orbits around static di-holes in four dimensions. We reviewed the general Bret\'{o}n--Manko--Aguilar family of solutions (Section \ref{sec:dihole_formalism_bma}), focussing on the Reissner--Nordstr\"{o}m di-hole subfamily, which contains the Weyl--Bach and Majumdar--Papapetrou di-holes as special cases. We found that, in the case of the highly symmetric equal-mass Majumdar--Papapetrou di-hole, stable photon orbits exist for separations in the range $\sqrt{\frac{16}{27}} < \frac{d}{M} < \sqrt{\frac{32}{27}}$. In cases with less symmetry (e.g.~the unequal-mass Majumdar--Papapetrou di-hole and the equal-mass, equal-charge Reissner--Nordstr\"{o}m di-hole), we used a numerical method to determine the existence region for stable photon orbits in parameter space (Figure \ref{fig:mp_dihole_parameter_space}). The results of Figure \ref{fig:rn_dihole_spo_region} indicate that stable photon orbits can exist up to (but not including) the vacuum Weyl--Bach case, provided that other parameters are appropriately fine-tuned.

Finally, we explored the rich structure of stable photon orbits for the Majumdar--Papapetrou di-hole, which is non-integrable (Section \ref{sec:spo_diholes_geodesic_structure}). In the isotropic case with separation $d = M$, we found that the system can either be closed or open with three narrow escapes, depending on the value of the conserved azimuthal angular momentum. Close to the minimum of the geodesic potential, the motion is bounded to a small elliptical region. Using Poincar\'{e} sections, we analysed the dynamics of bounded null geodesics, presented some examples of resonant orbits, and commented on the transition from order to chaos as the azimuthal angular momentum parameter is varied. Curiously, in both the bounded regime (this chapter) and the unbounded regime (Chapter \ref{chap:fractal_structures}), the isotropic equal-mass Majumdar--Papapetrou di-hole shares many qualitative features with the H\'{e}non--Heiles Hamiltonian system (Appendix \ref{chap:appendix_b}). In the anisotropic case ($d \neq M$), we observe a variety of rich structure in the Poincar\'{e} sections (e.g.~the ``crenulations'' of Figure \ref{fig:mp_poincare_separation}) and the bounded null geodesics (Figure \ref{fig:mp_spos_sep_unbounded}). 

\chapter{Higher-order geometric optics on Kerr spacetime} \label{chap:geometric_optics_kerr}

\section{Introduction}

The Event Horizon Telescope's observations of the black hole M87$^{\ast}$ \cite{EHTC2019a} and the detections of gravitational waves by LIGO--Virgo \cite{Abbottothers2018} have opened up an exciting new era in gravitational astronomy (see Chapter \ref{chap:introduction}). These events are part of a long line of observations which involve measurements of (typically electromagnetic) waves that have travelled to us over cosmological distances through a dynamical curved spacetime.

Formally, electromagnetic fields are described by wavelike solutions to Maxwell's equations; however, these equations are not typically analysed directly in practice. To describe the effects of gravitational lensing, it is usually sufficient to adopt a leading-order geometric optics approximation \cite{MisnerThorneWheeler1973, SchneiderEhlersFalco1992, Perlick2004}. In this approach, the gradient of the eikonal phase is tangent to a light ray, and light travels along null geodesics in a four-dimensional spacetime, which is typically taken to be a solution to Einstein's field equations of general relativity; variations in the intensity of the radiation are determined by the change in cross-sectional area of an infinitesimal bundle of rays in the geometric optics wavefront; and the polarisation of the wave is parallel-transported along the ray -- the gravitational Faraday effect \cite{IshiharaTakahashiTomimatsu1988}.

Geometric optics does not only apply to electromagnetic waves and the description of gravitational lensing phenomena; in fact, the formalism is widely used in many areas of physics \cite{ThorneBlandford2017}. Geometric optics is an accurate description of wave propagation when the wavelengths and inverse frequency scales are short when compared to other characteristic length and time scales, including the spacetime curvature scale set by the Riemann tensor \cite{MisnerThorneWheeler1973}. In the context of general relativity, the leading-order geometrical optics formalism has been employed to provide a theoretical description of scalar, electromagnetic and gravitational fields on curved spacetime \cite{Dolan2017}.

For observations of gravitational lensing effects, the assumptions on which the geometric optics approximation relies are generally sound. For example, the Event Horizon Telescope employs millimetre and sub-millimetre very-long-baseline interferometry to generate images of supermassive black holes with very large diameters ($\sim 10^{13} \textrm{~m}$ for M87$^{\ast}$). Despite the fact that gravitational waves have much longer wavelengths ($\sim 10^{7} \textrm{~m}$ for GW150914), the geometric optics approximation is still typically safe.

However, there are a number of situations where the geometric optics approximation breaks down. For example, it is well-known that -- even in standard optics -- the geometric optics approximation is not valid at caustics (i.e., transverse self-intersections of the wavefront) \cite{Perlick2004}. There are other scenarios, such as the scattering of waves by Kerr black holes when the wavelength is comparable with the horizon radius \cite{Dolan2008, LeiteDolanCrispino2017}, where the leading-order geometric optics formalism is no longer an accurate description of the wave-optical phenomena. In the latter scenario, one would expect that higher-order corrections would provide a more accurate description, and perhaps provide insight into wavelike effects which are not offered by the leading-order (ray-optics) approximation.

Higher-order extensions to the geometric optics formalism have been studied by Ehlers \cite{Ehlers1967} and Anile \cite{Anile1976}. More recently, modifications to the geometric optics formalism have been given by Dolan \cite{Dolan2018} and Harte \cite{Harte2018, Harte2019}. The work of Dolan \cite{Dolan2018} is motivated by interest in spin--helicity effects, i.e., a coupling between the helicity of a circularly polarised electromagnetic wave (of finite wavelength) and the frame-dragging of spacetime outside a rotating source, such as a Kerr black hole \cite{Mashhoon1974}. It has been demonstrated, for example, that Kerr black holes are able to distinguish and separate waves of opposite helicity \cite{Dolan2008, FrolovShoom2011, FrolovShoom2012}. Moreover, recent analyses of the absorption of planar electromagnetic waves incident on a Kerr black hole along the axis of symmetry demonstrated that waves with a circular polarisation that is rotating in the opposite sense to the black hole are absorbed to a greater degree than co-rotating waves \cite{LeiteDolanCrispino2017}. Dolan \cite{Dolan2018} argued that, in principle, such effects can be captured by an extension to geometric optics at sub-leading order. However, this approach has not yet been employed in any practical calculations of lensing effects by black holes.

In this chapter, we apply the higher-order geometric optics formalism to electromagnetic waves on Kerr spacetime. In Section \ref{sec:go_recap_and_ho_formalism} we review the geometric optics formalism for electromagnetic waves on a four-dimensional curved spacetime. The leading-order formalism was presented in detail in Section \ref{sec:geometric_optics_review}, but is reviewed in Section \ref{sec:leading_order_go_recap} of the present chapter for convenience. In Section \ref{sec:higher_order_go}, we summarise the higher-order geometric optics formalism of Dolan \cite{Dolan2018}. In this work, we are interested in using the formalism to calculate the sub-leading-order corrections to the electromagnetic stress--energy tensor for circularly polarised waves propagating on Kerr spacetime. We review the important properties of Kerr spacetime in Section \ref{sec:go_kerr_spacetime_recap}, including the spacetime geometry, the existence of Killing objects and symmetries, Carter's symmetric tetrad, and the role of spacetime symmetries in the separability of geodesic motion. As described in Section \ref{sec:geometric_optics_review}, (parallel-transported) complex null tetrads play an important role in the analysis of gravitational lensing phenomena through the geometric optics formalism. In Section \ref{sec:null_tetrad_construction}, we use the principal tensor and its Hodge dual to construct a complex null tetrad which is parallel-propagated along the null geodesics of Kerr spacetime; we demonstrate that this tetrad is equivalent to a complex null tetrad which can be built from the legs of a real parallel-propagated tetrad found by Marck \cite{Marck1983b}. Section \ref{sec:np_formalism_and_transport_equations} concerns the Newman--Penrose formalism for our parallel-propagated complex null tetrad: we present the transport equations required to compute the higher-order geometric optics corrections; and we calculate quantities such as the Weyl curvature scalars and their directional derivatives along the legs of the complex null tetrad. In Section \ref{sec:far_field_asymptotics}, we analyse the far-field behaviour of the Weyl scalars, Newman--Penrose scalars, and geometric optics quantities. In Section \ref{sec:conjugate_points}, we review wavefronts and caustics in geometric optics, and analyse the behaviour of (divergent) Newman--Penrose quantities in the neighbourhood of caustic points, where neighbouring rays cross. Finally, in Section \ref{sec:transport_equations_cautics}, we present a practical method for evolving transport equations through caustic points: we introduce a regularisation method in Section \ref{sec:regularisation_transport_equations}; and we present some preliminary numerical results for Kerr spacetime in Section \ref{sec:regularisation_numerical_method_and_results}. We discuss our results and avenues for future work in Section \ref{sec:discussion_go_kerr}.
%

\section{Higher-order geometric optics formalism}
\label{sec:go_recap_and_ho_formalism}

\subsection{Leading-order geometric optics}
\label{sec:leading_order_go_recap}

In this section, we briefly review aspects of the leading-order geometric optics formalism. For further details, see Section \ref{sec:geometric_optics_review} and references therein. The Faraday tensor $F\ind{_{a b}}$ is governed by the Maxwell equations in curved spacetime. One may introduce a complexified Faraday tensor $\mathcal{F}\ind{_{a b}} = F\ind{_{a b}} + i \vp{F}^{\star} F\ind{_{a b}}$, which is a self-dual bivector. Here, $\vp{F}^{\star} F\ind{_{a b}}$ is the Hodge dual of the Faraday tensor, defined in \eqref{eqn:hodge_dual_faraday}. In the absence of sources, the complexified Faraday tensor satisfies the wave equation
\begin{equation}
\label{eqn:complex_faraday_wave_equation}
\Box \mathcal{F}\ind{_{a b}} + 2 R\ind{_{a c b d}} \mathcal{F}\ind{^{c d}} + R\ind{_{a}^{c}} \mathcal{F}\ind{_{b c}} - R\ind{_{b}^{c}} \mathcal{F}\ind{_{a c}} = 0 ,
\end{equation}
which is obtained by taking further covariant derivatives of the source-free Maxwell's equations \eqref{eqn:maxwells_equations_complex}.

The leading-order geometric optics solution to \eqref{eqn:complex_faraday_wave_equation} for a circularly polarised electromagnetic wave is given by
\begin{equation}
\mathcal{F}\ind{_{a b}} = \mathcal{A} \mathfrak{f}\ind{_{a b}} \exp{(i \omega \Phi)} , \qquad \mathfrak{f}\ind{_{a b}} = 2 k\ind{_{ [ a}} m\ind{_{ b ] }} ,
\end{equation}
Here, $k\ind{^{a}}$ is a real null ($k\ind{^{a}} k\ind{_{a}} = 0$) vector field which is defined as the gradient of the eikonal phase ($k\ind{_{a}} = \nabla\ind{_{a}} \Phi$). We denote the directional derivative along $k\ind{^{a}}$ by $D = k\ind{^{a}} \nabla\ind{_{a}}$. From the gradient property, it follows that $k\ind{_{[a ; b]}} = 0$. The fact that $k\ind{^{a}}$ is null and a gradient means $k\ind{^{a}}$ is tangent to a geodesic ($D k\ind{^{a}} = 0$); $k\ind{^{a}}$ is a null generator of constant-phase hypersurfaces ($\Phi = \text{constant}$). The vector $m\ind{^{a}}$ and its complex conjugate $\ol{m}\ind{^{a}}$ are complex null ($m\ind{^{a}} m\ind{_{a}} = 0$) vector fields which are unit $m\ind{^{a}} \ol{m}\ind{_{a}} = 1$; parallel-propagated ($D m\ind{^{a}} = 0 $); transverse to $k\ind{^{a}}$ ($k\ind{^{a}} m\ind{_{a}} = 0$); and hence tangent to constant-phase hypersurfaces ($m\ind{^{a}} \nabla\ind{_{a}} \Phi = 0$). One may introduce an auxiliary real null vector $n\ind{^{a}}$ which is future-pointing and satisfies $k\ind{^{a}} n\ind{_{a}} = - 1$ and $m\ind{^{a}} n\ind{_{a}} = 0$. It follows from these scalar products that, if $k\ind{^{a}}$ and $m\ind{^{a}}$ are parallel-transported, then so too is $n\ind{^{a}}$ ($D n\ind{^{a}} = 0 $). Hence, each leg of the complex null tetrad $\{ k\ind{^{a}}, n\ind{^{a}}, m\ind{^{a}}, \ol{m}\ind{^{a}} \}$ is parallel-propagated along the null generators of the constant-phase hypersurfaces.

We recall from Section \ref{sec:geometric_optics_review} that, at $O(\omega)$, the ambiguity in the definitions of the amplitude $\mathcal{A}$ and the polarisation bivector $\mathfrak{f}\ind{_{a b}}$ can be used to split the transport equations such that
\begin{equation}
D \mathcal{A}^{2} = 2 \rho \mathcal{A}^{2} , \qquad
D \mathfrak{f}\ind{_{a b}} = 0 , \label{eqn:transport_equation_square_amplitude_recap}
\end{equation}
where $\rho = - \nps{m}{k}{\ol{m}}$ is a Newman--Penrose scalar (see Section \ref{sec:newman_penrose_spin_coefficients}). Hereafter, we will use the transport equation for the square-amplitude $\mathcal{A}^{2}$ rather than $\mathcal{A}$ for reasons which shall be discussed in Section \ref{sec:conjugate_points}.

At leading order in $\omega$, the electromagnetic stress--energy tensor takes the form of a null fluid, i.e.,
\begin{equation}
T\ind{_{a b}} = \frac{1}{8 \pi} \mathcal{A}^{2} k\ind{_{a}} k\ind{_{b}} .
\end{equation}

We recall that, for a twist-free null congruence ($k\ind{_{ [ a ; b ] }} = 0$), the cross-sectional area $A$ of an infinitesimal bundle of rays in the geometric-optics wavefront satisfies the transport equation
\begin{equation}
\label{eqn:transport_equation_area_twist_free}
D A = - 2 \rho A ,
\end{equation}
where the Newman--Penrose scalar (or optical scalar) $\rho$, which is responsible for the expansion of the infinitesimal bundle, is real.

\subsection{Higher-order geometric optics}
\label{sec:higher_order_go}

Dolan \cite{Dolan2018} presents a method to extend the geometrical approximation to higher orders in the frequency $\omega$. This is achieved by retaining the standard ansatz \eqref{eqn:complex_faraday_go_ansatz}, and expanding the polarisation bivector as a power series:
\begin{equation}
\label{eqn:polarisation_bovector_expansion}
\mathfrak{f}\ind{_{a b}} = \sum_{j = 0}^{\infty} \omega^{-j} {\mathfrak{f}^{(j)}}\ind{_{a b}}.
\end{equation}
The self-dual bivectors ${\mathfrak{f}^{(j)}}\ind{_{a b}}$ are then expanded in the bivector basis $U\ind{_{a b}} = 2 k\ind{_{[a}} m\ind{_{b]}}$, $V\ind{_{a b}} = 2 \ol{m}\ind{_{[a}} n\ind{_{b]}}$, $W\ind{_{a b}} = 2 ( m\ind{_{[a}} \ol{m}\ind{_{b]}} - k\ind{_{[a}} n\ind{_{b]}} )$, which is constructed from a twist-free parallel-propagated complex null tetrad $\{ k\ind{^{a}}, n\ind{^{a}}, m\ind{^{a}}, \ol{m}\ind{^{a}} \}$ \cite{StephaniKramerMacCallumEtAl2003}, viz.
\begin{equation}
{\mathfrak{f}^{(j)}}\ind{_{a b}} = \mathfrak{u}_{j} U\ind{_{a b}} + \mathfrak{v}_{j} V\ind{_{a b}} + \mathfrak{w}_{j} W\ind{_{a b}},
\end{equation}
where $\mathfrak{u}_{j}$, $\mathfrak{v}_{j}$ and $\mathfrak{w}_{j}$ are complex scalar fields.

Inserting the ansatz \eqref{eqn:polarisation_bovector_expansion} into the first of Maxwell's equations \eqref{eqn:maxwells_equations_complex} and expanding order-by-order in $\omega$ yields a system of equations for the ${\mathfrak{f}^{(j)}}\ind{_{a b}}$. At sub-leading order, the relevant equations are
\begin{align}
D \mathfrak{u}_{1} &= i \left( \frac{1}{\mathcal{A}} \ol{\delta} \delta \mathcal{A} + \frac{\chi}{\mathcal{A}} \ol{\delta} \mathcal{A} + \ol{\delta} \chi - \sigma \lambda \right), \label{eqn:transport_u1} \\
\mathfrak{v}_{1} &= i \sigma, \label{eqn:algebraic_v1} \\
\mathfrak{w}_{1} &= i \left( \frac{1}{\mathcal{A}} \delta \mathcal{A} + \chi \right), \label{eqn:algebraic_w1}
\end{align}
where $\sigma = -\nps{m}{k}{m}$ and $\chi = \nps{\ol{m}}{m}{m}$ are Newman--Penrose scalars (see Sections \ref{sec:newman_penrose} and \ref{sec:np_formalism_and_transport_equations}). The transport equation \eqref{eqn:transport_u1} features second derivatives of the amplitude $\mathcal{A}$ across the wavefront (i.e., in the $m$- and $\ol{m}$-directions).

Using the expression for the electromagnetic stress--energy tensor \eqref{eqn:stress_energy_complexified_faraday}, and identities for the bivector basis $\{ U\ind{_{a b}}, V\ind{_{a b}}, W\ind{_{a b}} \}$, the sub-leading-order correction to the stress--energy tensor is
\begin{equation}
\label{eqn:subleading_stess_energy}
4 \pi T\ind{_{a b}} = \frac{1}{2} \mathcal{A}^{2} k\ind{_{a}} k\ind{_{b}} + \omega^{-1} \mathcal{A}^{2} \on{Re}{ \left( \mathfrak{u}_{1} k\ind{_{a}} k\ind{_{b}} + \mathfrak{v}_{1} \ol{m}\ind{_{a}} \ol{m}\ind{_{b}} - 2 \mathfrak{w}_{1} k\ind{_{(a}} \ol{m}\ind{_{b)}} \right) } + O(\omega^{-2}).
\end{equation}
Since this only features the real part of $\mathfrak{u}_{1}$, one may use \eqref{eqn:transport_u1} and its complex conjugate to obtain a transport equation for $\on{Re}{(\mathfrak{u}_{1})}$. One may then apply the commutator identity $\ol{\delta} \delta - \delta \ol{\delta} = (\ol{\mu} - \mu) D - \ol{\chi} \delta + \chi \ol{\delta}$ to eliminate second derivatives of $\mathcal{A}$ in favour of first derivatives. The relevant transport equation is then
\begin{equation}
\label{eqn:transport_real_u1}
D ( \on{Re}{(\mathfrak{u}_{1})} ) = \frac{i}{2} \left[ \ol{\delta} \chi - \delta \ol{\chi} + \chi \frac{\ol{\delta} \mathcal{A}^{2}}{\mathcal{A}^{2}} - \ol{\chi} \frac{\delta \mathcal{A}^{2}}{\mathcal{A}^{2}} + \rho \left( \ol{\mu} - \mu \right) + \ol{\sigma} \ol{\lambda} - \sigma \lambda \right].
\end{equation}

Our principal aim here is to calculate the sub-dominant correction to the stress--energy tensor \eqref{eqn:subleading_stess_energy}. In principle, the Newman--Penrose scalars which appear in \eqref{eqn:algebraic_v1}, \eqref{eqn:algebraic_w1} and \eqref{eqn:transport_real_u1} may be found along null geodesics by evolving a system of standard transport equations \cite{StephaniKramerMacCallumEtAl2003, Dolan2018} (see Sections \ref{sec:np_field_equations} and \ref{sec:transport_equations_for_np_quantities}). We observe that \eqref{eqn:algebraic_v1}, \eqref{eqn:algebraic_w1} and \eqref{eqn:transport_real_u1} also feature ``higher-order'' Newman--Penrose quantities and geometric optics quantities, e.g.~$\ol{\delta} \chi$ and $\delta \mathcal{A}^{2}$, for which transport equations may also be obtained (see Section \ref{sec:transport_equations_for_np_quantities} and \cite{Dolan2018}).

By considering the asymptotic behaviour of the Newman--Penrose quantities in a flat region of spacetime, Dolan \cite{Dolan2018} finds that, as $s \rightarrow \infty$, $\mathfrak{u}_{j} = O(1)$, $\mathfrak{v}_{j} = O(s^{-2})$ and $\mathfrak{w}_{j} = O(s^{-1})$, where $s$ denotes the affine parameter along a null geodesic. To find the $O(\omega^{-1})$ correction to $T\ind{_{a b}}$ in \eqref{eqn:subleading_stess_energy} far from the black hole, we therefore only need to calculate the quantity $\on{Re}{(\mathfrak{u}_{1})}$ by means of \eqref{eqn:transport_real_u1}. We note that the sub-leading-order correction is proportional to the leading-order piece (i.e., $k\ind{_{a}} k\ind{_{b}}$).
%
%
%
%

\section{Kerr spacetime}
\label{sec:go_kerr_spacetime_recap}

\subsection{Spacetime geometry}
\label{sec:go_kerr_spacetime_geometry}

The Kerr solution to Einstein's field equations describes the spacetime of a rotating black hole \cite{Kerr1963}. The region outside the event horizon may be described in Boyer--Lindquist coordinates $\{ t, r, \theta, \phi \}$ \cite{BoyerLindquist1967}, in which the line element $\ed \bar{s}^{2} = g\ind{_{a b}} \ed x^{a} \ed x^{b}$ takes the form\footnote{Here, we denote the spacetime interval by $\ed \bar{s}$, rather than $\ed s$, to avoid a clash with the notation used for the affine parameter throughout the rest of this chapter.}
\begin{align}
\begin{split}
\ed \bar{s}^{2} &= - \left( 1 - \frac{2 M r}{\Sigma} \right) \ed t^{2} + \frac{\Sigma}{\Delta} \, \ed r^{2} + \Sigma \, \ed \theta^{2} + \left(r^{2} + a^{2} + \frac{2 M a^{2} r}{\Sigma} \sin^{2} \theta\right) \sin^{2} \theta \, \ed \phi^{2} \\ & \qquad - \frac{4 M a r \sin^{2} \theta}{\Sigma} \, \ed t \, \ed \phi,
\label{eqn:kerr_line_element_bl}
\end{split}
\end{align}
where $\Sigma(r, \theta) = r^{2} + a^{2} \cos^{2}{\theta}$ and $\Delta(r) = r^{2} - 2 M r + a^{2}$. Here, $M$ is the mass of the black hole and $a = \frac{J}{M}$ is the spin, where $J$ is the angular momentum. The Kerr spacetime has an outer ($+$) and inner ($-$) horizon at $r_{\pm} = M \pm \sqrt{M^{2} - a^{2} \cos^{2}{\theta}}$; we only consider the exterior region $r > r_{+}$. The non-zero components of the inverse metric $g\ind{^{a b}}$ are
\begin{equation}
g\ind{^{t t}} = -\frac{1}{\Delta} \left( r^{2} + a^{2} + \frac{2 M r a^{2}}{\Sigma} \sin^{2}{\theta} \right), \qquad
g\ind{^{t \phi}} = g\ind{^{\phi t}} = - \frac{2 M a r}{\Delta \Sigma}, \label{eqn:kerr_inverse_metric_1}
\end{equation}
\begin{equation}
g\ind{^{r r}} = \frac{\Delta}{\Sigma}, \qquad
g\ind{^{\theta \theta}} = \frac{1}{\Sigma}, \qquad
g\ind{^{\phi \phi}} = \frac{\Delta - a^{2} \sin^{2}{\theta}}{\Delta \Sigma \sin^{2}{\theta}}. \label{eqn:kerr_inverse_metric_2}
\end{equation}
It is useful to note the identity $\Delta - a^{2} \sin^{2}{\theta} = \Sigma - 2 M r$.

\subsection{Killing objects and symmetries}
\label{sec:symmetries_of_kerr}

In Boyer--Lindquist coordinates, the Kerr solution admits a pair of Killing vectors ${\xi_{(t)}} = \partial\ind{_{t}}$ and ${\xi_{(\phi)}} = \partial\ind{_{\phi}}$, which satisfy Killing's equation,
\begin{equation}
\mathcal{L}_{\xi} g\ind{_{a b}} = \xi\ind{_{a ; b}} + \xi\ind{_{b ; a}} = 2 \xi\ind{_{(a ; b)}} = 0.
\end{equation}
It is well known that the Killing vectors ${\xi_{(t)}}$ and ${\xi_{(\phi)}}$ encode continuous symmetries (isometries) of the metric, corresponding to time-translations and axial rotations, respectively.

In fact, the Kerr metric admits a special geometric object -- the \emph{principal tensor} -- which is the generator of a complete set of explicit and hidden (or implicit) symmetries of the spacetime \cite{FrolovKrtousKubiznak2017}. As described in Section \ref{sec:killing_objects_symmetries_integrability}, the principal tensor $h\ind{_{a b}}$ is a closed ($\ed h = 0$) conformal Killing--Yano two-form, which satisfies the defining equation
\begin{equation}
\label{eqn:ccky_tensor}
h\ind{_{a b ; c}} = g\ind{_{a c}} \xi\ind{_{b}} - g\ind{_{c b}} \xi\ind{_{a}},
\end{equation}
where $\xi\ind{^{a}}$ is an associated vector field. (The terms ``principal tensor'' and ``closed conformal Killing--Yano tensor'' will be used interchangeably.) Contracting \eqref{eqn:ccky_tensor} with $g\ind{^{c b}}$ and using the antisymmetry of $h\ind{_{ab}}$, one may show that
\begin{equation}
\label{eqn:kv_cckyt_relationship}
\xi\ind{_{a}} = \frac{1}{3} h\ind{_{b a}^{; b}}.
\end{equation}
Taking the covariant derivative of \eqref{eqn:kv_cckyt_relationship}, symmetrising, and imposing the Einstein field equations in four dimensions, it is possible to show that \cite{Kashiwada1968, Tachibana1969, KrtousFrolovKubiznak2008}
\begin{equation}
\label{eqn:kv_einstein_spacetime}
\xi\ind{_{(a ; b)}} = \frac{1}{2} R\ind{_{c ( b}} h\ind{_{a )}^{c}}.
\end{equation}
For a spacetime which satisfies Einstein's field equations in vacuum with a cosmological constant, the Ricci tensor is proportional to the metric tensor. Hence, due to the antisymmetry of $h\ind{_{a b}}$, the right-hand side of \eqref{eqn:kv_einstein_spacetime} vanishes, and we see that $\xi\ind{_{(a ; b)}}$ i.e., $\mathcal{L}_{\xi} g\ind{_{a b}} = 0$. In other words, $\xi\ind{^{a}}$ defined through \eqref{eqn:kv_cckyt_relationship} is a Killing vector. In fact, any solution to Einstein's equations which admits a principal tensor $h\ind{_{a b}}$ automatically possesses a Killing vector. This is referred to as the \emph{principal Killing vector}. For Kerr, the principal Killing vector is $\xi\ind{_{(t)}} = \partial\ind{_{t}}$ in Boyer--Lindquist coordinates.

The Hodge dual of the closed conformal Killing--Yano tensor is the \emph{Killing--Yano tensor} $f\ind{_{a b}} = \vp{h}^{\star} h\ind{_{ab}} = \frac{1}{2} \epsilon\ind{_{a b c d}} h\ind{^{c d}}$, which satisfies $f\ind{_{a b ; c}} = f\ind{_{[a b ; c]}}$. The \emph{Killing tensor} $K\ind{_{a b}} = f\ind{_{a}^{c}} f\ind{_{b c}}$ is the ``square'' of the Killing--Yano tensor, and satisfies $K\ind{_{(a b ; c)}} = 0$. The \emph{conformal Killing tensor} $Q\ind{_{a b}} = h\ind{_{a}^{c}} h\ind{_{b c}}$ is the ``square'' of the closed conformal Killing--Yano tensor. This tensor satisfies $Q\ind{_{(a b ; c)}} = g\ind{_{(a b}} q\ind{_{c)}}$, where $q\ind{_{a}} = h\ind{_{a b}}{\xi_{(t)}}\ind{^{b}}$. 

In this chapter, we focus on the null geodesics of Kerr spacetime. Consider an affinely parametrised null geodesic on Kerr spacetime with tangent vector field $k\ind{^{a}}(s) = \dot{x}\ind{^{a}}(s)$, where an overdot denotes differentiation with respect to the affine parameter $s$. In Section \ref{sec:killing_objects_symmetries_integrability}, we saw that each Killing vector $\xi\ind{^{a}}$ yields a conserved quantity along a geodesic, $I = g\ind{_{a b}} \xi\ind{^{a}} k\ind{^{b}}$. On Kerr spacetime, there exists a pair of commuting Killing vectors, $\xi_{(t)}$ and $\xi_{(\phi)}$, which encode the stationarity and axisymmetry of the metric. These Killing vectors yield a pair of conserved quantities, $E = - g\ind{_{a b}} {\xi_{(t)}}\ind{^{a}} k\ind{^{b}}$ and $L\ind{_{z}} = g\ind{_{a b}} {\xi_{(\phi)}}\ind{^{a}} k\ind{^{b}}$, such that $\dot{E} = 0 = \dot{L}\ind{_{z}}$ along rays.

Besides the conserved quantities which arise from Killing vectors, there also exist conserved quantities which are related to higher-rank Killing tensors, which imply the existence of less evident symmetries \cite{FrolovKrtousKubiznak2017}. For Kerr spacetime, define the scalar quantities
\begin{equation}
\label{eqn:kt_conserved_quantities}
K = K\ind{_{a b}} k\ind{^{a}} k\ind{^{b}}, \qquad Q = Q\ind{_{a b}} k\ind{^{a}} k\ind{^{b}},
\end{equation}
along a null geodesic with tangent vector field $k\ind{^{a}}$. Then
\begin{equation}
\dot{K} = k\ind{^{c}} K\ind{_{; c}} = K\ind{_{a b ; c}} k\ind{^{a}} k\ind{^{b}} k\ind{^{c}} = K\ind{_{( a b ; c )}} k\ind{^{a}} k\ind{^{b}} k\ind{^{c}} = 0,
\end{equation}
where the right-hand side vanishes due to the fact that $K\ind{_{( a b ; c )}} = 0$, i.e., $K\ind{_{a b}}$ is a Killing tensor. Similarly,
\begin{equation}
\dot{Q} = k\ind{^{c}} Q\ind{_{; c}} = Q\ind{_{a b ; c}} k\ind{^{a}} k\ind{^{b}} k\ind{^{c}} = Q\ind{_{( a b ; c )}} k\ind{^{a}} k\ind{^{b}} k\ind{^{c}} = g\ind{_{( a b}} q\ind{_{c )}} k\ind{^{a}} k\ind{^{b}} k\ind{^{c}} = k\ind{^{a}} k\ind{_{a}} k\ind{^{c}} q\ind{_{c}} = 0,
\end{equation}
by virtue of the fact that $k\ind{^{a}}$ is null. Hence, the Killing tensor and conformal Killing tensor yield conserved quantities \eqref{eqn:kt_conserved_quantities} along null geodesics. These will be discussed in greater detail in Section \ref{sec:kerr_geodesics}.

\subsection{Carter's canonical tetrad}
\label{sec:carter_tetrad}

Here, we introduce Carter's ``canonical'' (symmetric) tetrad \cite{Carter1968b}, an orthonormal tetrad (see Section \ref{sec:newman_penrose_tetrads}) which is useful when analysing the separability properties of Kerr spacetime \cite{Carter1968a, CarterMcLenaghan1982}. The one-form components of the tetrad $\omega\ind{^{(\alpha)}_{a}}$ are given by
\begin{align}
\omega\ind{^{(0)}_{a}} \, \ed x^{a} &= \sqrt{\frac{\Delta}{\Sigma}} \left( \ed t - a \sin^{2} \theta \, \ed \phi \right), \label{eqn:carter_tetrad_of_0} \\
\omega\ind{^{(1)}_{a}} \, \ed x^{a} &= \sqrt{\frac{\Sigma}{\Delta}} \, \ed r, \label{eqn:carter_tetrad_of_1} \\
\omega\ind{^{(2)}_{a}} \, \ed x^{a} &= \sqrt{\Sigma} \, \ed \theta, \label{eqn:carter_tetrad_of_2} \\
\omega\ind{^{(3)}_{a}} \, \ed x^{a} &= \frac{\sin \theta}{\sqrt{\Sigma}} \left[ a \, \ed t - (r^{2} + a^{2}) \, \ed \phi \right], \label{eqn:carter_tetrad_of_3}
\end{align}
and the vector components of the tetrad $\omega\ind{_{(\alpha)}^{a}}$ are
\begin{align}
\omega\ind{_{(0)}^{a}} \, \partial\ind{_{a}} &= \frac{1}{\sqrt{\Sigma \Delta}} \left[ (r^{2} + a^{2}) \, \partial\ind{_{t}} + a \, \partial\ind{_{\phi}} \right], \label{eqn:carter_tetrad_vec_0} \\
\omega\ind{_{(1)}^{a}} \, \partial\ind{_{a}} &= \sqrt{\frac{\Delta}{\Sigma}} \, \partial\ind{_{r}}, \label{eqn:carter_tetrad_vec_1} \\
\omega\ind{_{(2)}^{a}} \, \partial\ind{_{a}} &= \frac{1}{\sqrt{\Sigma}} \, \partial\ind{_{\theta}}, \label{eqn:carter_tetrad_vec_2} \\
\omega\ind{_{(3)}^{a}} \, \partial\ind{_{a}} &= - \frac{1}{\sqrt{\Sigma} \sin{\theta}} \left( a \sin^{2}{\theta} \, \partial\ind{_{t}} + \partial\ind{_{\phi}} \right). \label{eqn:carter_tetrad_vec_3}
\end{align}
In terms of Carter's orthonormal tetrad, the covariant and contravariant components of the metric tensor are, respectively, given by
\begin{align}
g\ind{_{a b}} &= - \omega\ind{^{(0)}_{a}} \omega\ind{^{(0)}_{b}} + \omega\ind{^{(1)}_{a}} \omega\ind{^{(1)}_{b}} + \omega\ind{^{(2)}_{a}} \omega\ind{^{(2)}_{b}} + \omega\ind{^{(3)}_{a}} \omega\ind{^{(3)}_{b}}, \label{eqn:tetrad_metric} \\
g\ind{^{a b}} &= - \omega\ind{_{(0)}^{a}} \omega\ind{_{(0)}^{b}} + \omega\ind{_{(1)}^{a}} \omega\ind{_{(1)}^{b}} + \omega\ind{_{(2)}^{a}} \omega\ind{_{(2)}^{b}} + \omega\ind{_{(3)}^{a}} \omega\ind{_{(3)}^{b}}. \label{eqn:tetrad_inverse_metric}
\end{align}

Carter's tetrad is closely related to the (normalised) \emph{Darboux basis} $\left\{ \nu, \hat{\nu}, \epsilon, \hat{\epsilon} \right\}$. This is presented by Frolov \emph{et al.} \cite{FrolovKrtousKubiznak2017} in \emph{canonical coordinates} $\{ \tau = t - a \phi, r, y = a \cos{\theta}, \psi = \frac{\phi}{a} \}$, in which the hidden symmetry of Kerr spacetime is more evident. By direct comparison of the Darboux basis in \cite{FrolovKrtousKubiznak2017} with the basis one-forms of Carter's symmetric tetrad \eqref{eqn:carter_tetrad_of_0}--\eqref{eqn:carter_tetrad_of_3}, we see that
\begin{equation}
\nu = \omega^{(1)}, \qquad \hat{\nu} = \omega^{(0)}, \qquad \epsilon = - \omega^{(2)}, \qquad \hat{\epsilon} = \omega^{(3)}.
\end{equation}
Hence, Carter's orthonormal tetrad is equivalent to the Darboux frame (up to a change of sign). One advantage of the Darboux basis (i.e., Carter's tetrad) is that the closed conformal Killing--Yano tensor and related Killing objects take a remarkably simple form \cite{FrolovKrtousKubiznak2017}:
\begin{align}
f\ind{_{a b}} &= - 2 a \cos{\theta} \, \omega\ind{^{(0)}_{{[a}}} \omega\ind{^{(1)}_{{b]}}} - 2 r \, \omega\ind{^{(2)}_{{[a}}} \omega\ind{^{(3)}_{{b]}}}, \label{eqn:cckyt_carter} \\
h\ind{_{a b}} &= 2 r \, \omega\ind{^{(0)}_{{[a}}} \omega\ind{^{(1)}_{{b]}}} - 2 a \cos{\theta} \, \omega\ind{^{(2)}_{{[a}}} \omega\ind{^{(3)}_{{b]}}}, \label{eqn:kyt_carter} \\
K\ind{_{a b}} &= - a^{2} \cos^{2}{\theta} \left( - \omega\ind{^{(0)}_{a}} \omega\ind{^{(0)}_{b}} + \omega\ind{^{(1)}_{a}} \omega\ind{^{(1)}_{b}} \right) + r^{2} \left( \omega\ind{^{(2)}_{a}} \omega\ind{^{(2)}_{b}} + \omega\ind{^{(3)}_{a}} \omega\ind{^{(3)}_{b}} \right), \label{eqn:kt_carter} \\
Q\ind{_{a b}} &= - r^{2} \left( - \omega\ind{^{(0)}_{a}} \omega\ind{^{(0)}_{b}} + \omega\ind{^{(1)}_{a}} \omega\ind{^{(1)}_{b}} \right) + a^{2} \cos^{2}{\theta} \left( \omega\ind{^{(2)}_{a}} \omega\ind{^{(2)}_{b}} + \omega\ind{^{(3)}_{a}} \omega\ind{^{(3)}_{b}} \right). \label{eqn:ckt_carter}
\end{align}

We may use the expressions \eqref{eqn:cckyt_carter}--\eqref{eqn:ckt_carter} to derive some useful identities for the Killing objects on Kerr spacetime. Notice from the form of \eqref{eqn:kt_carter} and \eqref{eqn:ckt_carter} that
\begin{equation}
\label{eqn:kt_minus_ckt}
K\ind{_{a b}} - Q\ind{_{a b}} = \left( r^{2} - a^{2} \cos^{2}{\theta} \right) g\ind{_{a b}},
\end{equation}
where we have used \eqref{eqn:tetrad_metric}. Now, contraction of \eqref{eqn:kt_minus_ckt} with $k\ind{^{a}} k\ind{^{b}}$ gives $K\ind{_{a b}} k\ind{^{a}} k\ind{^{b}} - Q\ind{_{a b}} k\ind{^{a}} k\ind{^{b}} = \left( r^{2} - a^{2} \cos^{2}{\theta} \right) g\ind{_{a b}} k\ind{^{a}} k\ind{^{b}} = 0$, which implies that
\begin{equation}
K = Q.
\end{equation}
Using \eqref{eqn:tetrad_metric} and \eqref{eqn:kt_carter}, we obtain two useful expressions for the Killing tensor:
\begin{align}
K\ind{_{a b}} &= \Sigma \left( \omega\ind{^{(0)}_{a}} \omega\ind{^{(0)}_{b}} - \omega\ind{^{(1)}_{a}} \omega\ind{^{(1)}_{b}} \right) + r^{2} g\ind{_{a b}} \label{eqn:kt_carter_r} \\
&= \Sigma \left( \omega\ind{^{(2)}_{a}} \omega\ind{^{(2)}_{b}} + \omega\ind{^{(3)}_{a}} \omega\ind{^{(3)}_{b}} \right) - a^{2} \cos^{2}{\theta} g\ind{_{a b}}. \label{eqn:kt_carter_theta}
\end{align}

Using \eqref{eqn:cckyt_carter} and \eqref{eqn:kyt_carter} to read off the components of $f\ind{_{(\alpha) (\gamma)}}$ and $h\ind{_{(\alpha) (\gamma)}}$ (i.e., the projection of the conformal Killing--Yano tensor and the Killing--Yano tensor onto Carter's tetrad), one may check that
\begin{equation}
f\ind{_{(\alpha) (\gamma)}} h\ind{_{(\beta)}^{(\gamma)}} = f\ind{_{(\alpha) (\gamma)}} \eta\ind{^{(\gamma) (\delta)}} h\ind{_{(\delta) (\beta)}} = a r \cos{\theta} \, \eta\ind{_{(\alpha) (\gamma)}} ,
\end{equation}
where $\eta\ind{_{(\alpha) (\beta)}}$ are the components of the Minkowski metric. Hence, the ``product'' of $f\ind{_{a b}}$ and $h\ind{_{a b}}$ is
\begin{equation}
\label{eqn:product_cckyt_kyt}
f\ind{_{a c}} h\ind{_{b}^{c}} = h\ind{_{a c}} f\ind{_{b}^{c}} = a r \cos{\theta}\, g\ind{_{a b}}.
\end{equation}


\subsection{Null geodesics: separability and conserved quantities}
\label{sec:kerr_geodesics}

Let us consider an affinely parametrised null geodesic on Kerr spacetime with tangent vector field $k\ind{^{a}}(s) = \dot{x}\ind{^{a}}(s)$ in Boyer--Lindquist coordinates. The geodesic equation may be separated with the aid of the explicit and hidden symmetries of Kerr spacetime. The separation of geodesic motion was first demonstrated by Carter \cite{Carter1968b}.

To demonstrate this separability, let us first consider the explicit symmetries which arise from the existence of the Killing vectors $\xi_{(t)}$ and $\xi_{(\phi)}$. In Boyer--Lindquist coordinates, we have $E = g\ind{_{a b}} {\xi_{(t)}}\ind{^{a}} k\ind{^{b}} = - g\ind{_{t t}} \dot{t} - g\ind{_{t \phi}} \dot{\phi}$ and $L\ind{_{z}} = g\ind{_{a b}} {\xi_{(\phi)}}\ind{^{a}} k\ind{^{b}} = g\ind{_{t \phi}} \dot{t} + g\ind{_{\phi \phi}} \dot{\phi}$. These relationships can be inverted to give
\begin{equation}
\dot{t} = - g\ind{^{t t}} E + g\ind{^{t \phi}} L\ind{_{z}}, \qquad
\dot{\phi} = - g\ind{^{t \phi}} E + g\ind{^{\phi \phi}} L\ind{_{z}}. \label{eqn:t_dot_phi_dot}
\end{equation}

Next, we may use \eqref{eqn:kt_carter_r} and \eqref{eqn:kt_carter_theta} to separate out the $r$- and $\theta$-motion. Contracting these identities with $k\ind{^{a}} k\ind{^{b}}$, we see that the terms involving $g\ind{_{a b}}$ will vanish due to the fact that $k\ind{^{a}}$ is null. We are therefore left with
\begin{align}
K &= \Sigma \left( \omega\ind{^{(0)}_{a}} \omega\ind{^{(0)}_{b}} - \omega\ind{^{(1)}_{a}} \omega\ind{^{(1)}_{b}} \right) k\ind{^{a}} k\ind{^{b}} \label{eqn:carter_const_omegas_0_1} \\
&= \Sigma \left( \omega\ind{^{(2)}_{a}} \omega\ind{^{(2)}_{b}} + \omega\ind{^{(3)}_{a}} \omega\ind{^{(3)}_{b}} \right) k\ind{^{a}} k\ind{^{b}}. \label{eqn:carter_const_omegas_2_3}
\end{align}
Inserting the one-form components of Carter's orthonormal tetrad \eqref{eqn:carter_tetrad_of_0}--\eqref{eqn:carter_tetrad_of_3} into \eqref{eqn:carter_const_omegas_0_1} and \eqref{eqn:carter_const_omegas_2_3}, and inserting the components of the inverse metric \eqref{eqn:kerr_inverse_metric_1} and \eqref{eqn:kerr_inverse_metric_2} into \eqref{eqn:t_dot_phi_dot}, we arrive at the geodesic equations
\begin{align}
\dot{t} &= \frac{1}{\Delta \Sigma} \left\{ E \left[ \left(r^{2} + a^{2}\right)^{2} - \Delta a^{2} \sin^{2} \theta \right] - 2 M a r L_{z} \right\}, \label{eqn:kerr_t_dot} \\
\Sigma^{2} \dot{r}^{2} &= \left[ E \left(r^{2} + a^{2}\right) - a L_{z} \right]^{2} - \Delta K, \label{eqn:kerr_r_dot} \\
\Sigma^{2} \dot{\theta}^{2} &= K - \left( a E \sin \theta - \frac{L_{z}}{\sin \theta} \right)^{2}, \label{eqn:kerr_theta_dot} \\
\dot{\phi} &= \frac{1}{\Delta \Sigma} \left[ 2 M a r E + \frac{(\Sigma - 2 M r) L_{z}}{\sin^{2} \theta}  \right] . \label{eqn:kerr_phi_dot}
\end{align}
The constant $K$, which permits the separation of the radial and latitudinal motion, is the so-called \emph{Carter constant}. This can be written using  \eqref{eqn:kerr_r_dot} or \eqref{eqn:kerr_theta_dot} as
\begin{equation}
K = \frac{1}{\Delta} \left\{ \left[ E \left(r^{2} + a^{2}\right)^{2} - a L_{z} \right]^{2} - \Sigma^{2} \dot{r}^{2} \right\}
= \Sigma^{2} \dot{\theta}^{2} + \left( a E \sin \theta - \frac{L_{z}}{\sin \theta} \right)^{2}.
\end{equation}

Rather than work with the equations \eqref{eqn:kerr_t_dot}--\eqref{eqn:kerr_phi_dot}, we choose to employ the Hamiltonian formalism for null geodesics; see Section \ref{sec:hamiltonian_formulation_geodesics}. The Hamiltonian takes the form $H = \frac{1}{2} g\ind{^{a b}} p\ind{_{a}} p\ind{_{b}}$, where $p\ind{_{a}} = g\ind{_{a b}} \dot{x}\ind{^{b}} = g\ind{_{a b}} k\ind{^{b}}$ are the conjugate momenta, and $g\ind{^{a b}}$ are the contravariant components of the metric tensor given by \eqref{eqn:kerr_inverse_metric_1} and \eqref{eqn:kerr_inverse_metric_2}. The null geodesics are then the integral curves of Hamilton's equations,
\begin{equation}
\dot{x}\ind{^{a}} = \frac{\partial H}{\partial p\ind{_{a}}}, \qquad \dot{p}\ind{_{a}} = - \frac{\partial H}{\partial x\ind{^{a}}}.
\end{equation}

Clearly, the metric \eqref{eqn:kerr_line_element_bl} is independent of the coordinates $t$ and $\phi$, so the momenta $p\ind{_{t}}$ and $p\ind{_{\phi}}$ are constants of the motion. In fact, we have $E = - g\ind{_{a b}} {\xi_{(t)}}\ind{^{a}} k\ind{^{b}} = - {\xi_{(t)}}\ind{^{a}} p\ind{_{a}} = - p\ind{_{t}}$, and $L\ind{_{z}} = g\ind{_{a b}} {\xi_{(\phi)}}\ind{^{a}} k\ind{^{b}} = {\xi_{(\phi)}}\ind{^{a}} p\ind{_{a}} = p\ind{_{\phi}}$. The conserved quantities $E$ and $L\ind{_{z}}$ correspond to the energy and azimuthal angular momentum of the photon, respectively. Since these conserved quantities are linear in the momenta, the symmetries generated by $\xi_{(t)}$ and $\xi_{(\phi)}$ are referred to as \emph{explicit} symmetries \cite{FrolovKrtousKubiznak2017}.

As described in Section \ref{sec:symmetries_of_kerr}, there exist higher-rank Killing tensors on Kerr spacetime which give rise to further conserved quantities along rays. Here, we express these in terms of the Boyer--Lindquist coordinates and their conjugate momenta.

Firstly, we note that the metric itself is a (trivial) rank-two Killing tensor. The quantity $g\ind{^{a b}} p\ind{_{a}} p\ind{_{b}}$ is therefore conserved along geodesics. This means that the Hamiltonian $H = \frac{1}{2} g\ind{^{a b}} p\ind{_{a}} p\ind{_{b}}$ is a constant of the motion, with $H = 0$ in the case of null geodesics.

Furthermore, the Carter constant, which can be written in the form $K = K\ind{_{a b}} k\ind{^{a}} k\ind{^{b}}$ is constant along geodesics. In terms of the Boyer--Lindquist coordinates and their conjugate momenta, the Carter constant can be expressed in the form
\begin{equation}
\label{eqn:carter_constant}
K = K\ind{^{a b}} p\ind{_{a}} p\ind{_{b}} = {p\ind{_{\theta}}}^{2} + \left( a E \sin{\theta} - \frac{L\ind{_{z}}}{\sin{\theta}} \right)^{2}.
\end{equation}
Clearly, the Carter constant is quadratic in the momenta (by virtue of the fact that $K\ind{_{a b}}$ is a rank-two Killing tensor). The Carter constant therefore corresponds to a \emph{hidden} symmetry of Kerr spacetime \cite{FrolovKrtousKubiznak2017}. We also remark here that $K$ is clearly non-negative. (One may arrive at an alternative expression for $K$ by rearranging \eqref{eqn:kerr_r_dot} and noting that $p\ind{_{r}} = \frac{\Sigma}{\Delta} \dot{r}$.)

\section{Construction of complex null tetrads}
\label{sec:null_tetrad_construction}

\subsection{Complex null tetrads from symmetries of Kerr spacetime}
\label{sec:null_tetrads_from_symmetries}

We now use the Killing objects defined in Section \ref{sec:kerr_geodesics} to construct a complex null tetrad $\{ k\ind{^{a}}, n\ind{^{a}}, m\ind{^{a}}, \ol{m}\ind{^{a}} \}$, which is parallel-transported along a null geodesic with tangent vector field $k\ind{^{a}}$. This is achieved by first constructing a non-parallel-transported tetrad $\{ \wt{k}\ind{^{a}}, \wt{n}\ind{^{a}}, \wt{m}\ind{^{a}}, \ol{\wt{m}}\vp{m}\ind{^{a}} \}$, by projecting the tangent vector $\wt{k}$, then performing a Lorentz transformation which keeps $\wt{k}\ind{^{a}}$ fixed; see Section \ref{sec:np_lorentz_transformations}.

Define the complex bivector
\begin{equation}
\label{eqn:complex_bivector}
\mathscr{F}\ind{_{a b}} = f\ind{_{a b}} - i h\ind{_{a b}},
\end{equation}
where $h\ind{_{a b}}$ and $f\ind{_{a b}}$ are the closed conformal Killing--Yano tensor and the Killing--Yano tensor, respectively. Note that $\vp{\mathscr{F}}^{\star} \mathscr{F}\ind{_{a b}} = - i \mathscr{F}\ind{_{a b}}$, so $\mathscr{F}\ind{_{a b}}$ is self-dual. Using the properties of $f\ind{_{a b}}$ and $h\ind{_{a b}}$, one may verify some useful identities for the complex bivector \eqref{eqn:complex_bivector}; see Appendix \ref{chap:appendix_d}. In particular, we have
\begin{align}
\mathscr{F}\ind{_{a c}} \mathscr{F}\ind{_{b}^{c}} &= \left(r - i a \cos{\theta} \right)^{2} g\ind{_{a b}}, \label{eqn:bivector_identity_1} \\
\mathscr{F}\ind{_{a c}} \ol{\mathscr{F}}\vp{\mathscr{F}}\ind{_{b}^{c}} &= K\ind{_{a b}} + Q\ind{_{a b}}. \label{eqn:bivector_identity_2}
\end{align}
We may now use $\mathscr{F}\ind{_{a b}}$ and its geometrical properties to construct the tetrad legs $\{ \wt{n}\ind{^{a}}, \wt{m}\ind{^{a}}, \ol{\wt{m}}\vp{m}\ind{^{a}} \}$, by projecting the tangent vector $\wt{k}\ind{^{a}}$.

First, let
\begin{equation}
\label{eqn:leg_m_sdb}
\wt{m}\ind{^{a}} = \frac{1}{\sqrt{2 K}} \mathscr{F}\ind{^{a}_{b}} \wt{k}\ind{^{b}}.
\end{equation}
The leg $\ol{\wt{m}}\vp{m}\ind{^{a}}$ is obtained by complex conjugation of \eqref{eqn:leg_m_sdb}. Clearly $\wt{m}\ind{^{a}}$ is orthogonal to $\wt{k}\ind{_{a}}$:
\begin{equation}
\wt{k}\ind{^{a}} \wt{m}\ind{_{a}} = \frac{1}{\sqrt{2 K}} \wt{k}\ind{^{a}} \mathscr{F}\ind{_{a b}} \wt{k}\ind{^{b}} = 0,
\end{equation}
by virtue of the fact that $\mathscr{F}\ind{_{a b}}$ is antisymmetric ($\mathscr{F}\ind{_{a b}} = - \mathscr{F}\ind{_{b a}}$). Moreover, using the identity \eqref{eqn:bivector_identity_1}, we find
\begin{equation}
\wt{m}\ind{^{a}} \wt{m}\ind{_{a}} = \frac{1}{2 K} \mathscr{F}\ind{^{a}_{b}} \mathscr{F}\ind{_{a c}} \wt{k}\ind{^{b}} \wt{k}\ind{^{c}} = \frac{1}{2 K} \left(r - i a \cos{\theta} \right)^{2} g\ind{_{b c}} \wt{k}\ind{^{b}} \wt{k}\ind{^{c}} = 0,
\end{equation}
i.e., $\wt{m}\ind{^{a}}$ is null. Finally, using the identity \eqref{eqn:bivector_identity_2}, we see that
\begin{equation}
\wt{m}\ind{^{a}} \ol{\wt{m}}\ind{_{a}} = \frac{1}{2 K} \mathscr{F}\ind{^{a}_{b}} \ol{\mathscr{F}}\ind{_{a c}} \wt{k}\ind{^{b}} \wt{k}\ind{^{c}} = \frac{1}{2 K} \left(K\ind{_{b c}} + Q\ind{_{b c}} \right) \wt{k}\ind{^{b}} \wt{k}\ind{^{c}} = 1,
\end{equation}
where we recall that $K\ind{_{a b}} \wt{k}\ind{^{a}} \wt{k}\ind{^{b}} = K = Q\ind{_{a b}} \wt{k}\ind{^{a}} \wt{k}\ind{^{b}}$.

Applying the derivative operator $\wt{D} = \wt{k}\ind{^{a}} \nabla\ind{_{a}}$ to \eqref{eqn:leg_m_sdb} gives (see Appendix \ref{chap:appendix_d})
\begin{align}
\label{eqn:transport_equation_m_tilde}
\wt{D} \wt{m}\ind{_{a}} = \frac{i E}{\sqrt{2 K}} \wt{k}\ind{_{a}}.
\end{align}
Clearly $\wt{m}\ind{^{a}}$ is \emph{not} parallel-propagated along a null geodesic with tangent vector $\wt{k}\ind{^{a}}$; however, one can construct a parallel-transported vector $m\ind{^{a}}$ by performing a null rotation, as we shall demonstrate.

We define the final leg of our complex null tetrad to be
\begin{equation}
\label{eqn:leg_n_sdb}
\wt{n}\ind{^{a}} = \frac{1}{2 K} \mathscr{F}\ind{^{a b}} \ol{\mathscr{F}}\ind{_{b c}} \wt{k}\ind{^{c}}.
\end{equation}
First, note that
\begin{equation}
\wt{k}\ind{^{a}} \wt{n}\ind{_{a}} = \frac{1}{2 K} \mathscr{F}\ind{_{a}^{b}} \ol{\mathscr{F}}\ind{_{b c}} \wt{k}\ind{^{a}} \wt{k}\ind{^{c}} = - \frac{1}{2 K} \left( K\ind{_{a c}} + Q\ind{_{a c}} \right) \wt{k}\ind{^{a}} \wt{k}\ind{^{c}} = -1,
\end{equation}
where we have used \eqref{eqn:bivector_identity_2} and the fact that $K\ind{_{a b}} \wt{k}\ind{^{a}} \wt{k}\ind{^{b}} = K = Q\ind{_{a b}} \wt{k}\ind{^{a}} \wt{k}\ind{^{b}}$. Next, taking the inner product with $\wt{m}\ind{^{a}}$, we see that
\begin{equation}
\wt{n}\ind{^{a}} \wt{m}\ind{_{a}} = \frac{1}{(2 K)^{3/2}} \mathscr{F}\ind{^{a b}} \ol{\mathscr{F}}\ind{_{b c}} \mathscr{F}\ind{_{a d}} \wt{k}\ind{^{c}} \wt{k}\ind{^{d}} =  \frac{1}{(2 K)^{3/2}} \left(r - i a \cos{\theta} \right)^{2} \ol{\mathscr{F}}\ind{_{c d}} \wt{k}\ind{^{c}} \wt{k}\ind{^{d}} = 0,
\end{equation}
where we have used \eqref{eqn:bivector_identity_1} and the fact that $\ol{\mathscr{F}}\ind{_{a b}}$ is antisymmetric. (It follows from complex conjugation that $\wt{n}\ind{^{a}} \ol{\wt{m}}\ind{_{a}} = 0$, since $\wt{n}\ind{^{a}}$ is real.) Finally,
\begin{equation}
\wt{n}\ind{^{a}} \wt{n}\ind{_{a}} = \frac{1}{(2 K)^{2}} \mathscr{F}\ind{^{a b}} \ol{\mathscr{F}}\ind{_{b c}} \wt{k}\ind{^{c}} \mathscr{F}\ind{_{a}^{d}} \ol{\mathscr{F}}\ind{_{d e}} \wt{k}\ind{^{e}} = \left( \frac{\Sigma}{2 K} \right)^{2} g\ind{_{c e}} \wt{k}\ind{^{c}} \wt{k}\ind{^{e}} = 0,
\end{equation}
where we have used the identity \eqref{eqn:bivector_identity_1} and its complex conjugate, as well as the fact that $\wt{k}\ind{^{a}}$ is null. Hence, $\wt{n}\ind{^{a}}$ is null. We note that $\wt{n}\ind{^{a}}$ is \emph{not} parallel-propagated along null geodesics.
%

We have used the complex self-dual bivector \eqref{eqn:complex_bivector} to construct a complex null tetrad $\{ \wt{k}\ind{^{a}}, \wt{n}\ind{^{a}}, \wt{m}\ind{^{a}}, \ol{\wt{m}}\vp{m}\ind{^{a}} \}$, with all inner products zero except $\wt{k}\ind{^{a}} \wt{n}\ind{_{a}} = -1$ and $\wt{m}\ind{^{a}} \ol{\wt{m}}\ind{_{a}} = 1$. Only the tangent vector $\wt{k}\ind{^{a}}$ is parallel-transported. However, we may construct a parallel-transported tetrad by performing a Lorentz transformation of the form
\begin{equation}
\label{eqn:parallel_transport_lorentz_transformation}
k\ind{^{a}} = \wt{k}\ind{^{a}}, \qquad
m\ind{^{a}} = \wt{m}\ind{^{a}} + B \wt{k}\ind{^{a}}, \qquad
n\ind{^{a}} = \wt{n}\ind{^{a}} + B \ol{\wt{m}}\vp{m}\ind{^{a}} + \ol{B} \wt{m}\ind{^{a}} + B \ol{B} \wt{k}\ind{^{a}}.
\end{equation}
We note that the transformation \eqref{eqn:parallel_transport_lorentz_transformation} preserves the inner product relationships between the tetrad legs, so that $k\ind{^{a}} n\ind{_{a}} = - 1$ and $m\ind{^{a}} \ol{m}\ind{_{a}} = 1$, with all others zero.

Clearly $k\ind{^{a}}$ is parallel-transported. The requirement that $m\ind{^{a}}$ be parallel-transported fixes the null rotation parameter $B$ in \eqref{eqn:parallel_transport_lorentz_transformation} as
\begin{equation}
\label{eqn:null_rotation_parameter}
B = - \frac{i E s}{\sqrt{2 K}} + B_{0},
\end{equation}
where $B_{0} \in \mathbb{C}$ is an arbitrary integration constant. The transport equation \eqref{eqn:transport_equation_m_tilde} then reads
\begin{equation}
\label{eqn:transport_equation_m_tilde_bdot}
D \wt{m}\ind{_{a}} = - \dot{B} k\ind{_{a}}.
\end{equation}
It follows from the fact that each of the legs $\left\{ k\ind{^{a}}, m\ind{^{a}}, \ol{m}\ind{^{a}} \right\}$ are parallel-transported that $n\ind{^{a}}$ is also parallel-propagated along null geodesics.

In summary, taking the complex null tetrad $\{ \wt{k}\ind{^{a}}, \wt{n}\ind{^{a}}, \wt{m}\ind{^{a}}, \ol{\wt{m}}\vp{m}\ind{^{a}} \}$, and performing the Lorentz transformation \eqref{eqn:parallel_transport_lorentz_transformation} with null rotation parameter \eqref{eqn:null_rotation_parameter} yields a complex null tetrad $\{ k\ind{^{a}}, n\ind{^{a}}, m\ind{^{a}}, \ol{m}\ind{^{a}} \}$, which is parallel-transported along null geodesics.

Intriguingly, we can absorb the Lorentz transformation \eqref{eqn:parallel_transport_lorentz_transformation} into the self-dual bivector \eqref{eqn:complex_bivector} by defining a new ``projection operator''
\begin{equation}
\label{eqn:projection_operator_parallel_transport}
\mathscr{G}\ind{_{a b}} = \mathscr{F}\ind{_{a b}} + \sqrt{2 K} B \, g\ind{_{a b}},
\end{equation}
where $B$ is defined in \eqref{eqn:null_rotation_parameter}. (Note that $\mathscr{G}\ind{_{a b}}$ is not a bivector.) Then the definitions \eqref{eqn:leg_m_sdb} and \eqref{eqn:leg_n_sdb} with $\mathscr{F}\ind{_{a b}}$ replaced by $\mathscr{G}\ind{_{a b}}$ (and ``tilded'' quantities replaced by ``untilded'' quantities) yield the parallel-transported legs $m\ind{^{a}}$ and $n\ind{^{a}}$. The procedure outlined above applies to all geometries which admit a principal tensor, i.e., the (off-shell) Kerr--NUT--anti-de Sitter family of spacetimes \cite{FrolovKrtousKubiznak2017}.

We caution here that the construction of these tetrads is valid only for rays with $K > 0$. Constructing a parallel-propagated complex null tetrad along rays with $K = 0$ requires a special treatment.

On Kerr spacetime, the ``tilded'' complex null tetrad derived here can be related to the well-known Kinnersley tetrad by application of a sequence of Lorentz transformations. This is done explicitly in Appendix \ref{chap:appendix_d}.
%

\subsection{Marck's tetrads}
\label{sec:marck_tetrads}

\subsubsection{Quasi-orthonormal tetrad}

Marck \cite{Marck1983b} employs Carter's canonical tetrad (Section \ref{sec:carter_tetrad}) and the symmetries of Kerr spacetime to construct a tetrad which is parallel-propagated along null geodesics in the Kerr(--Newman) spacetime.

This is achieved by first constructing a quasi-orthonormal tetrad $\{ \mqolegup{0}{a}, \mqolegup{1}{a}, \mqolegup{2}{a}, \mqolegup{3}{a} \}$, where $\mqolegup{0}{a}$ and $\mqolegup{3}{a}$ are null, and $\mqolegup{1}{a}$ and $\mqolegup{2}{a}$ are unit spacelike. Marck identifies $\mqolegup{0}{a}$ with the tangent vector $k\ind{^{a}}$, which is clearly parallel-transported. The components of this tetrad leg with respect to Carter's symmetric tetrad $\mqolegup{0}{(\alpha)} = \mqolegup{0}{a} \omega\ind{^{(\alpha)}_{a}}$ are given by
\begin{align}
\mqolegup{0}{(0)} &= \frac{1}{\sqrt{\Delta \Sigma}} \left[ E \left(r^{2} + a^{2}\right) - a L_{z}  \right], \\
\mqolegup{0}{(1)} &= \sqrt{\frac{\Sigma}{\Delta}} \, \dot{r}, \\
\mqolegup{0}{(2)} &= \sqrt{\Sigma} \, \dot{\theta}, \\
\mqolegup{0}{(3)} &= \frac{1}{\sqrt{\Sigma}} \left( a E \sin \theta - \frac{L_{z}}{\sin \theta} \right).
\end{align}

A second parallel-transported vector which is orthogonal to $\mqolegup{0}{(\alpha)}$ arises naturally due to the existence of the non-trivial Killing--Yano tensor $f\ind{_{(\alpha) (\beta)}}$. The properties of this tensor (discussed in Section \ref{sec:symmetries_of_kerr}) ensure that the vector with components $v\ind{^{(\alpha)}} = f\ind{^{(\alpha)}_{(\beta)}} \mqolegup{0}{(\beta)}$ is (i) parallel-propagated along the geodesic with tangent vector $\mqolegup{0}{(\alpha)}$, and (ii) orthogonal to $\mqolegup{0}{(\alpha)}$. The leg $\mqolegup{2}{(\alpha)}$ is chosen to be the unit vector parallel to $v\ind{^{(\alpha)}}$; its components with respect to the symmetric tetrad are
\begin{align}
\mqolegup{2}{(0)} &= \sqrt{\frac{\Sigma}{K \Delta}} a \cos \theta \, \dot{r} = \frac{a \cos{\theta}}{\sqrt{K}} \mqolegup{0}{(1)}, \\
\mqolegup{2}{(1)} &= \frac{a \cos{\theta}}{\sqrt{K \Delta \Sigma}} \left[ E \left(r^{2} + a^{2}\right) - a L_{z}  \right] = \frac{a \cos{\theta}}{\sqrt{K}} \mqolegup{0}{(0)}, \\
\mqolegup{2}{(2)} &= -\frac{r}{\sqrt{K \Sigma}} \left( a E \sin \theta - \frac{L_{z}}{\sin \theta} \right) = - \frac{r}{\sqrt{K}} \mqolegup{0}{(3)}, \\
\mqolegup{2}{(3)} &= \sqrt{\frac{\Sigma}{K}} \, r \, \dot{\theta} = \frac{r}{\sqrt{K}} \mqolegup{0}{(2)}.
\end{align}

There is now a straightforward choice for the vectors $\mqolegup{1}{a}$ and $\mqolegup{3}{a}$, such that the tetrad legs satisfy
\begin{equation}
\mqolegup{0}{(\alpha)} \mqolegdn{3}{(\alpha)} = \mqolegup{1}{(\alpha)} \mqolegdn{1}{(\alpha)} = \mqolegup{2}{(\alpha)} \mqolegdn{2}{(\alpha)} = 1,
\end{equation}
with all other inner products zero. This choice is given by
\begin{align}
\mqolegup{1}{(0)} &= \sqrt{\frac{\Sigma}{K \Delta}} r \, \dot{r} = \frac{r}{\sqrt{K}} \mqolegup{0}{(1)}, \\
\mqolegup{1}{(1)} &= \frac{r}{\sqrt{K \Delta \Sigma}} \left[ E \left(r^{2} + a^{2}\right) - a L_{z}  \right] = \frac{r}{\sqrt{K}} \mqolegup{0}{(0)}, \\
\mqolegup{1}{(2)} &= \frac{a \cos{\theta}}{\sqrt{K \Sigma}} \left( a E \sin \theta - \frac{L_{z}}{\sin \theta} \right) = \frac{a \cos{\theta}}{\sqrt{K}} \mqolegup{0}{(3)}, \\
\mqolegup{1}{(3)} &= \sqrt{\frac{\Sigma}{K}} \, a \cos \theta \, \dot{\theta} = - \frac{a \cos{\theta}}{\sqrt{K}} \mqolegup{0}{(2)},
\end{align}
and
\begin{align}
\mqolegup{3}{(0)} &= -\frac{1}{2K} \sqrt{\frac{\Sigma}{\Delta}} \left[ E \left(r^{2} + a^{2}\right) - a L_{z}  \right] = - \frac{\Sigma}{2 K} \mqolegup{0}{(0)}, \\
\mqolegup{3}{(1)} &= -\frac{1}{2K} \sqrt{\frac{\Sigma^{3}}{\Delta}} \, \dot{r} = - \frac{\Sigma}{2 K} \mqolegup{0}{(1)}, \\
\mqolegup{3}{(2)} &= \frac{1}{2K} \sqrt{\Sigma^{3}} \, \dot{\theta} = \frac{\Sigma}{2 K} \mqolegup{0}{(2)}, \\
\mqolegup{3}{(3)} &= \frac{1}{2K} \sqrt{\Sigma} \left( a E \sin \theta - \frac{L_{z}}{\sin \theta} \right) = \frac{\Sigma}{2 K} \mqolegup{0}{(3)}.
\end{align}
Unlike $\mqolegup{0}{a}$ and $\mqolegup{2}{a}$, the legs $\mqolegup{1}{a}$ and $\mqolegup{3}{a}$ are \emph{not} parallel-propagated along null geodesics with tangent vector $\mqolegup{0}{a}$.

\subsubsection{Parallel-transported tetrad}

Marck \cite{Marck1983b} demonstrates that one may use the tetrad $\{ \mqolegup{0}{a}, \mqolegup{1}{a}, \mqolegup{2}{a}, \mqolegup{3}{a} \}$ to construct a parallel-transported tetrad $\{ \mlegup{0}{a}, \mlegup{1}{a}, \mlegup{2}{a}, \mlegup{3}{a} \}$. First, one may set $\mlegup{0}{a} = \mqolegup{0}{a}$ and $\mlegup{2}{a} = \mqolegup{2}{a}$, as these legs are already parallel-propagated. The remaining legs, $\mlegup{1}{a}$ and $\mlegup{3}{a}$, may be constructed by means of a one-parameter Lorentz transformation of the form
\begin{align}
\mlegup{1}{a} &= \mqolegup{1}{a} - \Psi \mqolegup{0}{a} , &
\mlegup{3}{a} &= \mqolegup{3}{a} + \Psi \mqolegup{1}{a} - \frac{1}{2} \Psi^{2} \mqolegup{0}{a} ,
\end{align}
where $\Psi$ is a real-valued function. The requirement that the legs be parallel-transported yields a first-order ordinary differential equation for $\Psi$ which reads $\dot{\Psi} = \left( k\ind{^{b}} \nabla\ind{_{b}} \mqolegup{1}{a} \right) \mqolegdn{3}{a}$. Integration of this equation permits us to express the parameter $\Psi$ as \cite{Marck1983b}
\begin{equation}
\label{eqn:marck_tetrad_parameter_integral}
\Psi = \frac{1}{\sqrt{K}} \left( \int \frac{E (r^{2} + a^{2})- a L_{z}}{[R(r)]^{1/2}} \ed r - \int a \sin \theta \frac{a E \sin \theta - L_{z} \operatorname{cosec}{\theta}}{[\Theta(\theta)]^{1/2}} \ed \theta \right).
\end{equation}
Here, $R(r) = [ E (r^{2} + a^{2})- a L_{z} ]^{2} - K \Delta$ and $\Theta(\theta) = K - (a E \sin \theta - L_{z} \operatorname{cosec}{\theta})^{2}$, where the signs of $R^{1/2} = \pm \sqrt{R}$ and $\Theta^{1/2} = \pm \sqrt{\Theta}$ are chosen independently and in accordance with the turning points in the $r$- and $\theta$-motion.

Inserting the equations of motion \eqref{eqn:kerr_r_dot} and \eqref{eqn:kerr_theta_dot} into \eqref{eqn:marck_tetrad_parameter_integral}, one can simplify the expression for the tetrad parameter $\Psi$:
\begin{align}
\Psi &= \frac{1}{\sqrt{K}} \int \left( \frac{E (r^{2} + a^{2})- a L_{z}}{[R(r)]^{1/2}} \dot{r} -  a \sin \theta \frac{a E \sin \theta - L_{z} \operatorname{cosec}{\theta}}{[\Theta(\theta)]^{1/2}} \dot{\theta} \right) \ed s \\
&= \frac{1}{\sqrt{K}} \int \frac{E (r^{2} + a^{2} \cos^{2} \theta)}{\Sigma} \, \ed s \\
&= \frac{E}{\sqrt{K}} \int \ed s.
\end{align}
Hence, the Lorentz transformation parameter $\Psi$ takes the form
\begin{equation}
\label{eqn:marck_tetrad_parameter}
\Psi = \frac{E s}{\sqrt{K}} + \psi,
\end{equation}
where $s$ denotes the affine parameter and $\psi$ is an arbitrary constant of integration. One can see from \eqref{eqn:carter_constant} that the Carter constant is non-negative. This construction of the parallel-propagated tetrad is therefore valid for all null rays, apart from those which satisfy $K = 0$. In this degenerate case, the vector $v\ind{^{(\alpha)}} = f\ind{^{(\alpha)}_{(\beta)}} \mqolegup{0}{(\beta)}$ is parallel to $\mlegup{0}{(\alpha)}$; a special treatment would therefore be required to construct a parallel-propagated tetrad along null rays with vanishing Carter constant.

\subsubsection{Complex null tetrads from Marck's tetrads}

One may construct a complex null tetrad from Marck's quasi-orthonormal tetrad via
\begin{align}
\wt{k}\ind{^{a}} &= \mqolegup{0}{a}, &
\wt{n}\ind{^{a}} &= - \mqolegup{3}{a}, &
\wt{m}\ind{^{a}} &= \frac{1}{\sqrt{2}} \left( \mqolegup{2}{a} + i \mqolegup{1}{a} \right), &
\ol{\wt{m}}\vp{m}\ind{^{a}} &= \frac{1}{\sqrt{2}} \left( \mqolegup{2}{a} - i \mqolegup{1}{a} \right). \label{eqn:marck_null_legs}
\end{align}
This tetrad is a Newman--Penrose tetrad: the legs satisfy $\wt{k}\ind{^{a}} \wt{n}\ind{_{a}} = -1$ and $\wt{m}\ind{^{a}} \ol{\wt{m}}\ind{_{a}} = 1$, with all other inner products zero. We note that only $\wt{k}\ind{^{a}}$ is parallel-transported.

One may construct a parallel-transported complex null tetrad by replacing the ``tilded'' legs with parallel-transported legs ($\mqolegup{i}{a} \mapsto {\lambda_{i}}\ind{^{a}}$) in the definitions \eqref{eqn:marck_null_legs}. Alternatively, one may perform a null rotation of the tetrad \eqref{eqn:marck_null_legs} of the form \eqref{eqn:parallel_transport_lorentz_transformation}.
The requirement that the tetrad $\{ k\ind{^{a}}, n\ind{^{a}}, m\ind{^{a}}, \ol{m}\ind{^{a}} \}$ be parallel-transported yields $\dot{B} = - \wt{n}\ind{^{a}} D \wt{m}\ind{_{a}}$. Integrating this ordinary differential equation fixes the null rotation parameter:
\begin{equation}
\label{eqn:null_rotation_parameter_marck}
B = - \frac{i \Psi}{\sqrt{2}},
\end{equation}
where $\Psi$ is Marck's tetrad parameter, given in \eqref{eqn:marck_tetrad_parameter}.

It should be noted that the complex null tetrads built from Marck's tetrads are equivalent to those which were constructed using the complex self-dual bivector $\mathscr{F}\ind{_{a b}}$ in Section \ref{sec:null_tetrads_from_symmetries}. This is best seen by projecting onto Carter's symmetric tetrad, as shown in Appendix \ref{chap:appendix_d}. Moreover, direct substitution of Marck's tetrad parameter \eqref{eqn:marck_tetrad_parameter} into \eqref{eqn:null_rotation_parameter_marck} shows that the null rotation parameters \eqref{eqn:null_rotation_parameter} and \eqref{eqn:null_rotation_parameter_marck} are equivalent.
%

\section{Newman--Penrose formalism and transport equations}
\label{sec:np_formalism_and_transport_equations}

\subsection{Transport equations for Newman--Penrose quantities}
\label{sec:transport_equations_for_np_quantities}

The Newman--Penrose scalars are defined in terms of projections of first derivatives of the complex null tetrad legs \cite{NewmanPenrose1962}; see Section \ref{sec:newman_penrose} for a review. We define directional derivatives along the tetrad legs to be
\begin{equation}
D = k\ind{^{a}} \nabla\ind{_{a}}, \qquad \mathit{\Delta} = n\ind{^{a}} \nabla\ind{_{a}}, \qquad \delta = m\ind{^{a}} \nabla\ind{_{a}}, \qquad \ol{\delta} = \ol{m}\ind{^{a}} \nabla\ind{_{a}}.
\end{equation}
For our parallel-propagated complex null tetrad (Section \ref{sec:null_tetrad_construction}), three of the Newman--Penrose scalars are trivially zero: $\kappa = \pi = \epsilon = 0$. The eight non-zero Newman--Penrose scalars which we use are
\begin{align}
\rho &= - \nps{m}{k}{\ol{m}}, & \sigma &= - \nps{m}{k}{m}, \\
\chi &= - \nps{\ol{m}}{m}{m}, & \tau &= - \nps{m}{k}{n}, \\
\mu &= \nps{\ol{m}}{n}{m} & \lambda &= \nps{\ol{m}}{n}{\ol{m}}, \\
\nu &= \nps{\ol{m}}{n}{n}, & \gamma &= \frac{1}{2} \left( \nps{\ol{m}}{m}{n} - \nps{n}{k}{n} \right).
\end{align}
One can obtain identities for the Newman--Penrose scalars by using the fact that $g\ind{^{a b}} = - k\ind{^{a}} n\ind{^{b}} - n\ind{^{a}} k\ind{^{b}} + m\ind{^{a}} \ol{m}\ind{^{b}} + \ol{m}\ind{^{a}} m\ind{^{b}}$, and using the fact that $k\ind{_{a}} = \Phi\ind{_{; a}}$ is a gradient. For example, it is straightforward to see that the twist-free property of the complex null tetrad (i.e., $k\ind{_{[ a; b]}} = 0$) implies that $\rho$ is a real quantity. In fact, $\rho = - \frac{1}{2} \vartheta$, where $\vartheta = k\ind{^{a}_{;a}}$ is the expansion scalar \cite{Poisson2004}. For our tetrad, $\tau = \beta + \ol{\alpha}$ and $\chi = \beta - \ol{\alpha}$, where $\alpha = \frac{1}{2} ( \nps{k}{n}{\ol{m}} - \nps{m}{\ol{m}}{\ol{m}} )$ and $\beta = \frac{1}{2} ( \nps{\ol{m}}{m}{m} - \nps{n}{k}{m} )$. It is therefore possible to eliminate both $\alpha$ and $\beta$ with the introduction of $\chi$ \cite{Dolan2018}.

Along null geodesics, the Newman--Penrose scalars satisfy a system of transport equations \cite{StephaniKramerMacCallumEtAl2003}. In a Ricci-flat spacetime, the transport equations we require are
\begin{align}
D \rho &= \rho^{2} + \sigma \ol{\sigma}, &
D \sigma &= 2 \rho \sigma + \Psi_{0}, \label{eqn:transport_equation_rho_sigma} \\
D \chi &= \rho \chi - \sigma \ol{\chi} + \Psi_{1}, &
D \tau &= \rho \tau + \sigma \ol{\tau} + \Psi_{1}, \label{eqn:transport_equation_chi_tau} \\
D \mu &= \rho \mu + \sigma \lambda + \Psi_{2}, &
D \lambda &= \rho \lambda + \ol{\sigma} \mu, \label{eqn:transport_equation_mu_lambda} \\
D \nu &= \ol{\tau} \mu + \tau \lambda + \Psi_{3}, &
D \gamma &= \tau \ol{\tau} + \frac{1}{2} \left( \ol{\tau} \chi - \tau \ol{\chi} \right) + \Psi_{2}. \label{eqn:transport_equation_nu_gamma}
\end{align}
Here, $\Psi_{i}$ are the complex Weyl scalars (see Section \ref{sec:newman_penrose_weyl_scalars}). We calculate these for the ``tilded'' (non-parallel-transported) tetrad and the parallel-transported tetrad in Section \ref{sec:go_kerr_weyl}. In this work, we are interested the closed subsystem of equations \eqref{eqn:transport_equation_rho_sigma}--\eqref{eqn:transport_equation_mu_lambda}, satisfied by the Newman--Penrose scalars $\{ \rho, \sigma, \chi, \tau, \mu, \lambda \}$.

In addition to the Newman--Penrose scalars, we wish to calculate higher-order quantities, such as $\delta \mathcal{A}^{2}$ and $\delta \ol{\chi}$, which feature in the transport equations for sub-leading-order geometrical optics quantities (e.g.~$\mathfrak{u}_{1}$); see Section \ref{sec:higher_order_go} and \cite{Dolan2018}. In general, one may obtain transport equations by using the commutator identity $D \delta = \delta D - \tau D + \rho \delta + \sigma \ol{\delta}$ and its complex conjugate \cite{StephaniKramerMacCallumEtAl2003}. The system of transport equations we require is
\begin{align}
D(\delta \mathcal{A}^{2}) &= 3 \rho \, \delta \mathcal{A}^{2} + \sigma \, \ol{\delta} \mathcal{A}^{2} + 2 \mathcal{A}^{2} \left( \delta \rho - \rho \tau \right), \label{eqn:ho_transport_1} \\
D(\ol{\delta} \mathcal{A}^{2}) &= 3 \rho \, \ol{\delta} \mathcal{A}^{2} + \sigma \, \delta \mathcal{A}^{2} + 2 \mathcal{A}^{2} \left( \ol{\delta} \rho - \rho \ol{\tau} \right), \label{eqn:ho_transport_2} \\
D(\delta \rho) &= 3 \rho \, \delta \rho + \ol{\sigma} \, \delta \sigma + \sigma \, \left( \ol{\delta} \sigma \right)^{\ast} - \left( \rho^{2} + \sigma \ol{\sigma} \right) \tau + \sigma \, \ol{\delta} \rho, \label{eqn:ho_transport_3} \\
D(\ol{\delta} \rho) &= 3 \rho \, \ol{\delta} \rho + \ol{\sigma} \, \ol{\delta} \sigma + \sigma \, \left( \delta \sigma \right)^{\ast} - \left( \rho^{2} + \sigma \ol{\sigma} \right) \ol{\tau} + \ol{\sigma} \, \delta \rho, \label{eqn:ho_transport_4} \\
D(\delta \sigma) &= 3 \rho \, \delta \sigma + 2 \sigma \, \delta \rho - 2 \rho \sigma \tau + \sigma \, \ol{\delta} \sigma - \tau \Psi_{0} + \delta \Psi_{0}, \label{eqn:ho_transport_5} \\
D(\ol{\delta} \sigma) &= 3 \rho \, \ol{\delta} \sigma + 2 \sigma \, \ol{\delta} \rho - 2 \rho \sigma \ol{\tau} + \ol{\sigma} \, \delta \sigma - \ol{\tau} \Psi_{0} + \ol{\delta} \Psi_{0}, \label{eqn:ho_transport_6} \\
D(\delta \chi) &= 2 \rho \, \delta \chi + \chi \, \delta \rho + \sigma \ol{\delta} \chi - \sigma \left( \ol{\delta} \chi \right)^{\ast} - \ol{\chi} \, \delta \sigma - \tau \left( \rho \chi - \sigma \ol{\chi} + \Psi_{1} \right) + \delta \Psi_{1}, \label{eqn:ho_transport_7} \\
D(\ol{\delta} \chi) &= 2 \rho \, \ol{\delta} \chi + \chi \, \ol{\delta} \rho + \ol{\sigma} \, \delta \chi - \sigma \, \left( \delta \chi \right)^{\ast} - \ol{\chi} \, \ol{\delta} \sigma - \ol{\tau} \left( \rho \chi - \sigma \ol{\chi} + \Psi_{1} \right) + \ol{\delta} \Psi_{1} , \qquad \label{eqn:ho_transport_8}
\end{align}
where for extra clarity we use an asterisk (as well as an overbar) to denote complex conjugation. This system of transport equations features directional derivatives of the first two Weyl scalars along the legs $m$ and $\overline{m}$: $\{ \delta \Psi_{0}, \ol{\delta} \Psi_{0}, \delta \Psi_{1}, \ol{\delta} \Psi_{1} \}$. These quantities are calculated in Section \ref{sec:go_kerr_d_d_weyl}.


\subsection{Weyl scalars}
\label{sec:go_kerr_weyl}

The Weyl tensor $C\ind{_{a b c d}}$, which in this case is identical to the Riemann tensor due to the fact that Kerr spacetime is Ricci-flat, can be expressed in terms of Carter's symmetric tetrad as \cite{Carter1973, Marck1983a}
\begin{align}
\Omega\ind{^{(0)}_{(1)}} &= 2 I_{1} \, \omega\ind{^{(0)}} \wedge \omega\ind{^{(1)}} + 2 I_{2} \, \omega\ind{^{(2)}} \wedge \omega\ind{^{(3)}}, \\
\Omega\ind{^{(0)}_{(2)}} &= - I_{1} \, \omega\ind{^{(0)}} \wedge \omega\ind{^{(2)}} + I_{2} \, \omega\ind{^{(1)}} \wedge \omega\ind{^{(3)}}, \\
\Omega\ind{^{(0)}_{(3)}} &= - I_{1} \, \omega\ind{^{(0)}} \wedge \omega\ind{^{(3)}} - I_{2} \, \omega\ind{^{(1)}} \wedge \omega\ind{^{(2)}}, \\
\Omega\ind{^{(1)}_{(2)}} &= - I_{1} \, \omega\ind{^{(1)}} \wedge \omega\ind{^{(2)}} + I_{2} \, \omega\ind{^{(0)}} \wedge \omega\ind{^{(3)}}, \\
\Omega\ind{^{(3)}_{(1)}} &= I_{1} \, \omega\ind{^{(1)}} \wedge \omega\ind{^{(3)}} + I_{2} \, \omega\ind{^{(0)}} \wedge \omega\ind{^{(2)}}, \\
\Omega\ind{^{(3)}_{(2)}} &= - 2 I_{1} \, \omega\ind{^{(2)}} \wedge \omega\ind{^{(3)}} + 2 I_{2} \, \omega\ind{^{(0)}} \wedge \omega\ind{^{(1)}},
\end{align}
where $\Omega\ind{^{(\alpha)}_{(\beta)}} = \frac{1}{2} C\ind{^{(\alpha)}_{(\beta) (\gamma) (\delta)}} \omega\ind{^{(\gamma)}} \wedge \omega\ind{^{(\delta)}}$ is the \emph{curvature two-form}, and the functions $I_{1}$ and $I_{2}$ are defined to be
\begin{equation}
I_{1} = \frac{M r}{\Sigma^{3}} \left( r^{2} - 3 a^{2} \cos^{2} \theta \right), \qquad
I_{2} = \frac{M a \cos \theta}{\Sigma^{3}} \left( 3 r^{2} - a^{2} \cos^{2} \theta \right).
\end{equation}
These quantities satisfy the identity $I_{1} + i I_{2} = \frac{M}{ \left( r - i a \cos{\theta} \right)^{3} }$.

Using the components of the Weyl tensor in terms of Carter's symmetric tetrad, we calculate the complex Weyl scalars in the ``tilded'' complex null tetrad. We find
\begin{alignat}{2}
\wt{\Psi}_{0} &= C\ind{_{\wt{k} \wt{m} \wt{k} \wt{m}}} &&= \frac{3 K M}{(r + i a \cos \theta)^{5}}, \label{eqn:psi_0_tilde} \\
\wt{\Psi}_{1} &= C\ind{_{\wt{k} \wt{n} \wt{k} \wt{m}}} &&= 0, \label{eqn:psi_1_tilde} \\
\wt{\Psi}_{2} &= C\ind{_{\wt{k} \wt{m} \ol{\wt{m}} \wt{n}}} &&= \frac{M}{2(r + i a \cos \theta)^{3}}, \label{eqn:psi_2_tilde} \\
\wt{\Psi}_{3} &= C\ind{_{\wt{k} \wt{n} \ol{\wt{m}} \wt{n}}} &&= 0, \label{eqn:psi_3_tilde} \\
\wt{\Psi}_{4} &= C\ind{_{\ol{\wt{m}} \wt{n} \ol{\wt{m}} \wt{n}}} &&= \frac{3 M}{4 K (r + i a \cos \theta)}, \label{eqn:psi_4_tilde}
\end{alignat}
where we employ the notation $C\ind{_{\wt{k} \wt{m} \wt{k} \wt{m}}} = C\ind{_{a b c d}} \wt{k}\ind{^{a}} \wt{m}\ind{^{b}} \wt{k}\ind{^{c}} \wt{m}\ind{^{d}}$ for brevity.

Recall that the ``tilded'' complex null tetrad $\{ \wt{k}, \wt{n}, \wt{m}, \ol{\wt{m}} \}$ and the parallel-transported complex null tetrad $\{ k, n, m, \ol{m} \}$ are related by the Lorentz transformation \eqref{eqn:parallel_transport_lorentz_transformation} with \eqref{eqn:null_rotation_parameter}. The Weyl scalars in the parallel-propagated frame can be expressed in terms of \eqref{eqn:psi_0_tilde}--\eqref{eqn:psi_4_tilde} as \cite{StephaniKramerMacCallumEtAl2003}
\begin{alignat}{2}
\Psi_{0} &= C\ind{_{k m k m}} &&= \wt{\Psi}_{0}, \label{eqn:psi_0} \\
\Psi_{1} &= C\ind{_{k n k m}} &&= \wt{\Psi}_{1} + \ol{B} \wt{\Psi}_{0}, \label{eqn:psi_1} \\
\Psi_{2} &= C\ind{_{k m \ol{m} n}} &&= \wt{\Psi}_{2} + 2 \ol{B} \wt{\Psi}_{1} + \ol{B}^{2} \wt{\Psi}_{0}, \label{eqn:psi_2} \\
\Psi_{3} &= C\ind{_{k n \ol{m} n}} &&= \wt{\Psi}_{3} + 3 \ol{B} \wt{\Psi}_{2} + 3 \ol{B}^{2} \wt{\Psi}_{1} + \ol{B}^{3} \wt{\Psi}_{0}, \label{eqn:psi_3} \\
\Psi_{4} &= C\ind{_{\ol{m} n \ol{m} n}} &&= \wt{\Psi}_{4} + 4 \ol{B} \wt{\Psi}_{3} + 6 \ol{B}^{2} \wt{\Psi}_{2} + 4 \ol{B}^{3} \wt{\Psi}_{1} + \ol{B}^{4} \wt{\Psi}_{0} , \label{eqn:psi_4}
\end{alignat}
with $B$ given by \eqref{eqn:null_rotation_parameter}.

\subsection{Directional derivatives of Weyl scalars}
\label{sec:go_kerr_d_d_weyl}

The transport equations in the higher-order geometric optics formalism feature the quantities $\left\{ \delta \Psi_{0}, \ol{\delta} \Psi_{0}, \delta \Psi_{1}, \ol{\delta} \Psi_{1} \right\}$, i.e., directional derivatives along $m$ and $\ol{m}$ of the Weyl scalars $\Psi_{0}$ and $\Psi_{1}$. The calculation of these quantities is more involved than the calculation of the Weyl scalars.

First, note that the Weyl scalars of interest are $\Psi_{0} = C\ind{_{k m k m}} = C\ind{_{a b c d}} k\ind{^{a}} m\ind{^{b}} k\ind{^{c}} m\ind{^{d}}$ and $\Psi_{1} = C\ind{_{k m \ol{m} m}} = C\ind{_{a b c d}} k\ind{^{a}} m\ind{^{b}} \ol{m}\ind{^{c}} m\ind{^{d}}$. Applying the differential operators $\delta$ and $\ol{\delta}$ to these quantities and using the Leibniz rule, we see that we will require directional derivatives (along $m$ and $\ol{m}$) of the tetrad legs $k$, $m$ and $\ol{m}$. These are given by
\begin{align}
\delta k\ind{^{a}} &= \tau k\ind{^{a}} - \rho m\ind{^{a}} - \sigma \ol{m}\ind{^{a}}, &
\ol{\delta} k\ind{^{a}} &= \ol{\tau} k\ind{^{a}} - \ol{\sigma} m\ind{^{a}} - \rho \ol{m}\ind{^{a}}, \label{eqn:delta_deltabar_k} \\
\delta m\ind{^{a}} &= \ol{\lambda} k\ind{^{a}} - \sigma n\ind{^{a}} + \chi m\ind{^{a}}, &
\ol{\delta} m\ind{^{a}} &= \lambda k\ind{^{a}} - \ol{\sigma} n\ind{^{a}} + \ol{\chi} \ol{m}\ind{^{a}}. \label{eqn:delta_deltabar_m}
\end{align}

The directional derivatives of the Weyl scalars can therefore be obtained by applying the derivative operators $\delta$ and $\ol{\delta}$ to the expressions $\Psi_{0} = C\ind{_{a b c d}} k\ind{^{a}} m\ind{^{b}} k\ind{^{c}} m\ind{^{d}}$ and $\Psi_{1} = C\ind{_{a b c d}} k\ind{^{a}} m\ind{^{b}} \ol{m}\ind{^{c}} m\ind{^{d}}$, and using the identities \eqref{eqn:delta_deltabar_k}--\eqref{eqn:delta_deltabar_m}. This yields
\begin{align}
\delta \Psi_{0} &= 2 \left( \tau + \chi \right) \Psi_{0} - 4 \sigma \Psi_{1} + \left( \delta C \right)\ind{_{k m k m}}, \label{eqn:d_psi_0} \\
\ol{\delta} \Psi_{0} &= 2 \left( \ol{\tau} - \ol{\chi} \right) \Psi_{0} - 4 \sigma \Psi_{1} + \left( \ol{\delta} C \right)\ind{_{k m k m}}, \label{eqn:dbar_psi_0} \\
\delta \Psi_{1} &= \mu \Psi_{0} + \left( \tau + \chi \right) \Psi_{1} - 3 \sigma \Psi_{2} + \left( \delta C \right)\ind{_{k m \ol{m} m}}, \label{eqn:d_psi_1} \\
\ol{\delta} \Psi_{1} &=  \lambda \Psi_{0} + \left( \ol{\tau} - \ol{\chi} \right) \Psi_{1} - 3 \rho \Psi_{2} + \left( \ol{\delta} C \right)\ind{_{k m \ol{m} m}}, \label{eqn:dbar_psi_1}
\end{align}
where $\left( \delta C \right)\ind{_{k m k m}} = \left( m\ind{^{e}} \nabla\ind{_{e}} C\ind{_{a b c d}} \right) k\ind{^{a}} m\ind{^{b}} k\ind{^{c}} m\ind{^{d}}$, for example. The quantities $\left( \delta C \right)\ind{_{k m k m}}$, $\left( \ol{\delta} C \right)\ind{_{k m k m}}$, $\left( \delta C \right)\ind{_{k m \ol{m} m}}$ and $\left( \ol{\delta} C \right)\ind{_{k m \ol{m} m}}$ can be calculated explicitly by transforming between the Kinnersley tetrad and the parallel-transported complex null tetrad; this calculation is presented in Appendix \ref{chap:appendix_d}.

\subsection{Constraints from Newman--Penrose field equations}
\label{sec:constraints_np_field_equations}

When projected onto a complex null tetrad, the Ricci identities give rise to a system of Newman--Penrose field equations, which relate the Newman--Penrose scalars, the Weyl scalars and the Ricci scalars; the full system of equations can be found in (e.g.) \cite{StephaniKramerMacCallumEtAl2003}. The Newman--Penrose field equations relevant for our purposes are
\begin{align}
Z = \delta \rho - \ol{\delta} \sigma - \rho \tau + \sigma \left( \ol{\tau} - 2 \ol{\chi} \right) + \Psi_{1} &= 0, \label{eqn:ricci_constraint_1} \\
W = \ol{\delta} \chi + \left( \ol{\delta} \chi \right)^{\ast} + \rho \left( \mu + \ol{\mu} \right) - \sigma \lambda - \ol{\sigma} \ol{\lambda} + 2 \chi \ol{\chi} - \Psi_{2} - \ol{\Psi}_{2} &= 0. \label{eqn:ricci_constraint_2}
\end{align}
We note that $W$ is real. The identities \eqref{eqn:ricci_constraint_1} and \eqref{eqn:ricci_constraint_2} can be viewed as algebraic constraints which must be satisfied along null rays by the Newman--Penrose scalars, higher-order Newman--Penrose quantities, and Weyl curvature scalars.

With some work, it is possible to use the transport equations of Section \ref{sec:transport_equations_for_np_quantities} to demonstrate that the quantities $Z$ and $W$ satisfy a coupled system of transport equations along rays, namely
\begin{equation}
D Z = 3 \rho Z - \ol{\sigma} Z, \qquad D W = 2 \rho W + \ol{\chi} Z + \chi \ol{Z}.
\end{equation}
These equations indicate that, if $Z \neq 0$ or $W \neq 0$ initially (perhaps due to numerical error), then the quantities $Z$ and $W$ will grow, which could cause propagation of errors when numerically integrating transport equations along rays. It is therefore important to keep track of the constraints (i.e., Ricci identities) along rays.

\section{Far-field asymptotics}
\label{sec:far_field_asymptotics}

We are interested in the special case of an initially parallel null geodesic congruence which starts ar $r = \infty$. In practice -- i.e., when numerically integrating our transport equations -- we must start at some finite radius $r = r_{0}$. We therefore wish to determine the far-field behaviour of the Newman--Penrose scalars and higher-order quantities as power series in $r^{-1}$.

We first note that, for $r \rightarrow \infty$, the geodesic equation \eqref{eqn:kerr_r_dot} is $\dot{r}^{2} \sim E^{2}$. This has solution $r \sim \pm E s + r_{0}$, where $r_{0} = r(0)$. Here, the upper ($+$) sign corresponds to \emph{outgoing} rays (with $\dot{r} > 0$), and the lower ($-$) sign corresponds to \emph{ingoing} rays (with $\dot{r} < 0$). We consider the latter case, as we are interested in the far-field behaviour of the system \emph{before} scattering has occurred.

Recall that the Weyl scalars in the parallel-transported basis are related to those of the ``tilded'' tetrad by a parabolic Lorentz transformation, which features the complex parameter \eqref{eqn:null_rotation_parameter}, which can be rewritten in the form
\begin{equation}
\label{eqn:tetrad_parameter_far_field_1}
B = - \frac{i (E s + b_{0})}{\sqrt{2 K}},
\end{equation}
where $b_{0} \in \mathbb{C}$ is an arbitrary integration constant. For rays which are ingoing from $r = r_{0} \gg 1$, we have $E s \sim r_{0} - r$. Making the replacement $E s \mapsto r - r_{0}$, and choosing $b_{0} = - r_{0}$ to eliminate any dependence on $r_{0}$ in \eqref{eqn:tetrad_parameter_far_field_1} yields
\begin{equation}
\label{eqn:tetrad_parameter_far_field_2}
B = \frac{i r}{\sqrt{2 K}}.
\end{equation}
%

\subsection{Far-field behaviour of the Weyl scalars}

Recall the Weyl scalars in the parallel-propagated frame \eqref{eqn:psi_0}--\eqref{eqn:psi_4}. Inserting the far-field null rotation parameter \eqref{eqn:tetrad_parameter_far_field_2} into these expressions, and expanding to leading order in $r^{-1}$, we find
\begin{align}
\Psi_{0} &= 3 M K r^{-5} + O(r^{-6}) , \label{eqn:weyl_scalars_far_0} \\
\Psi_{1} &= - 3 i M \sqrt{K} r^{-4} + O(r^{-5}) , \label{eqn:weyl_scalars_far_1} \\
\Psi_{2} &= -M r^{-3} + O(r^{-4}) , \label{eqn:weyl_scalars_far_2} \\
\Psi_{3} &= \frac{3 M a \cos{\theta}}{\sqrt{2 K}} r^{-3} + O(r^{-4}) , \label{eqn:weyl_scalars_far_3} \\
\Psi_{4} &= -\frac{3 M a^{2} \cos^{2}{\theta}}{K} r^{-3} + O(r^{-4}) . \label{eqn:weyl_scalars_far_4}
\end{align}

We see that the first three Weyl scalars \eqref{eqn:weyl_scalars_far_0}--\eqref{eqn:weyl_scalars_far_2} exhibit a peeling behaviour at infinity: $\Psi_{i} = O( r^{i - 5})$ for $i \in \left\{ 0, 1, 2 \right\}$. Moreover, the leading-order terms for these quantities are independent of the black hole spin $a$. The leading-order solutions for the remaining Weyl scalars \eqref{eqn:weyl_scalars_far_3} and \eqref{eqn:weyl_scalars_far_4} depend on the spin of the black hole $a$, and the polar angle $\theta$. The Weyl scalars $\Psi_{3}$ and $\Psi_{4}$ therefore exhibit a first-order post-Newtonian effect.
%

\subsection{Far-field behaviour of the Newman--Penrose scalars}
\label{sec:far_field_np_scalars}

The far-field behaviour of the Newman--Penrose scalars may be determined using the transport equations \eqref{eqn:transport_equation_rho_sigma}--\eqref{eqn:transport_equation_mu_lambda}, and the far-field expressions for the Weyl scalars \eqref{eqn:weyl_scalars_far_0}--\eqref{eqn:weyl_scalars_far_4}. We note that, in the far-field limit, the derivative operator which appears on the left-hand side of the transport equations may be recast in terms of the radial coordinate using the chain rule: $D = \frac{\ed}{\ed s} = \dot{r} \frac{\ed}{\ed r} \sim - E \frac{\ed}{\ed r}$, for ingoing rays.

The Newman--Penrose transport equations are hierarchical, so we consider them in order. We first focus on the smallest closed subset, which are the Sachs equations \eqref{eqn:transport_equation_rho_sigma}. We make a leading-order ansatz of the form $\rho = \rho_{0} r^{n_{\rho}}$, $\sigma = \sigma_{0} r^{n_{\sigma}}$. Substitution of this ansatz and the leading-order expression for $\Psi_{0}$, given in \eqref{eqn:weyl_scalars_far_0}, into the Sachs equations yields
\begin{align}
- n_{\rho} E \rho_{0} r^{n_{\rho} - 1} &= \rho_{0}^{2} r^{2 n_{\rho}} + \sigma_{0} \ol{\sigma}_{0} r^{2 n_{\sigma}}, \label{eqn:far_field_expansion_rho} \\
- n_{\sigma} E \sigma_{0} r^{n_{\sigma} - 1} &= 2 \rho_{0} \sigma_{0} r^{n_{\rho} + n_{\sigma}} + 3 M K r^{-5}. \label{eqn:far_field_expansion_sigma}
\end{align}
We assume that the term on the right-hand side of \eqref{eqn:far_field_expansion_sigma} at $O(r^{n_{\rho} + n_{\sigma}})$ is sub-dominant, and can be neglected at leading order. The left-hand side of \eqref{eqn:far_field_expansion_sigma} (i.e., the change in $\sigma$) is therefore sourced by the curvature term $\Psi_{0} = O(r^{-5})$. Balancing these terms forces $n_{\sigma} = - 4$. Rearrangement then fixes the leading-order coefficient $\sigma_{0} = \frac{3 K M}{4 E}$. The shear generated by spacetime curvature through \eqref{eqn:far_field_expansion_sigma}, in turn, generates expansion $\rho$ through \eqref{eqn:far_field_expansion_rho}. Assuming that the term at $O(r^{2 n_{\rho}})$ on the right-hand side of \eqref{eqn:far_field_expansion_rho} is sub-dominant, we find $n_{\rho} = - 7$. Moreover, the coefficient $\rho_{0}$ is fixed in terms of $\sigma_{0}$: $\rho_{0} = \frac{1}{7 E} \sigma_{0} \ol{\sigma}_{0}$.

Going through the full system of transport equations in this way, we find that the far-field leading-order solutions for the six relevant Newman--Penrose scalars are
\begin{align}
\rho &= \frac{1}{7 E} \left( \frac{3 K M}{4 E} \right)^{2} r^{-7} + O(r^{-8}) , &
\sigma &= \frac{3 K M}{4 E} r^{-4} + O(r^{-5}) , \label{eqn:far_field_solutions_rho_sigma} \\
\chi &= - \frac{i M \sqrt{K}}{\sqrt{2} E} r^{-3} + O(r^{-4}) , &
\tau &= - \frac{i M \sqrt{K}}{\sqrt{2} E} r^{-3} + O(r^{-4}) , \label{eqn:far_field_solutions_chi_tau} \\
\mu &= - \frac{M}{2 E} r^{-2} + O(r^{-3}) , &
\lambda &= - \frac{3 K M^{2}}{40 E^{3}} r^{-5} + O(r^{-6}) . \label{eqn:far_field_solutions_mu_lambda}
\end{align}

\subsection{Far-field behaviour of higher-order quantities}

We now consider the far-field behaviour of the directional derivatives of Weyl scalars \eqref{eqn:d_psi_0}--\eqref{eqn:dbar_psi_1}, which appear on the right-hand side of the transport equations \eqref{eqn:ho_transport_1}--\eqref{eqn:ho_transport_8}. Inserting the leading-order solutions for the Newman--Penrose scalars, the Weyl scalars and the null rotation parameter, the leading-order solutions for the directional derivatives of the first two Weyl scalars are
\begin{align}
\delta \Psi_{0} &= - \frac{15 E M \sqrt{K} a \cos{\theta}}{\sqrt{2}} r^{-6} + O(r^{-7}) , &
\overline{\delta} \Psi_{0} &= - 6 i \sqrt{2 K} E M r^{-5} + O(r^{-6}) , \\
\delta \Psi_{1} &= 6 i E M a \cos{\theta} r^{-5} + O(r^{-6}) , &
\overline{\delta} \Psi_{1} &= - 3 E M r^{-4} + O(r^{-5}) .
\end{align}

Now, one may combine the results for the Newman--Penrose scalars, the Weyl scalars and their directional derivatives to calculate leading-order solutions for the higher-order quantities $\left\{ \delta \mathcal{A}^{2}, \ol{\delta} \mathcal{A}^{2}, \delta \rho, \ol{\delta} \rho, \delta \sigma, \ol{\delta} \sigma, \delta \chi, \ol{\delta} \chi \right\}$. Employing a method similar to that of Section \ref{sec:far_field_np_scalars} (i.e., substituting a leading-order ansatz into the transport equations and equating powers of $r^{-1}$), we find
\begin{align}
\delta \rho &= \frac{9 M^{2} \sqrt{2 K^{3}} i}{56 E^{2}} r^{-7} + O(r^{-8}) , &
\ol{\delta} \rho &= - \frac{9 M^{2} \sqrt{2 K^{3}} i}{56 E^{2}} r^{-7} + O(r^{-8}) , \label{eqn:far_field_delta_quantities_1}
\\
\delta \sigma &= - \frac{3 M \sqrt{2 K} a \cos{\theta}}{2} r^{-5} + O(r^{-6}) , &
\ol{\delta} \sigma &= - \frac{3 M \sqrt{2 K} i}{2} r^{-4} + O(r^{-5}) ,
\label{eqn:far_field_delta_quantities_2}
\\
\delta \chi &= \frac{3 M a \cos{\theta}}{2} r^{-4} + O(r^{-5}) , &
\ol{\delta} \chi &= - M r^{-3} + O(r^{-4}) ,
\label{eqn:far_field_delta_quantities_3}
\\
\delta \mathcal{A}^{2} &= \frac{3 M^{2} \sqrt{2 K^{3}} i}{56 E^{3}} r^{-6} + O(r^{-7}) . &
\ol{\delta} \mathcal{A}^{2} &= \frac{3 M^{2} \sqrt{2 K^{3}} i}{56 E^{3}} r^{-6} + O(r^{-7}) .
\label{eqn:far_field_delta_quantities_4}
\end{align}
We note that $\ol{\delta} \rho = \left( \delta \rho \right)^{\ast}$ and $\ol{\delta} \mathcal{A}^{2} = \left( \delta \mathcal{A}^{2} \right)^{\ast}$, since both $\rho$ and $\mathcal{A}$ are real for a twist-free tetrad.

\subsection{Far-field behaviour of geometric optics quantities}

The geometric optics quantity of interest is $\on{Re}{(\mathfrak{u}_{1})}$ which satisfies the transport equation \eqref{eqn:transport_real_u1}. The leading-order terms on the right-hand side of this transport equation come from $\ol{\delta} \chi$ and its complex conjugate, which are of order $r^{-3}$, as given in \eqref{eqn:far_field_delta_quantities_3}. The coefficient of the leading-order term in the far-field expansion for $\ol{\delta} \chi$ is $- M$, which is real. The leading-order terms in the transport equation \eqref{eqn:transport_real_u1} will therefore cancel. In order to determine the leading-order behaviour of $\on{Re}{(\mathfrak{u}_{1})}$, it is necessary to consider the sub-leading-order behaviour of the terms on the right-hand side of \eqref{eqn:transport_real_u1}.

To find the sub-leading-order part of $\ol{\delta} \chi$ from its transport equation \eqref{eqn:ho_transport_8}, we require the sub-leading-order pieces of the terms on the right-hand side. The highest-order terms are the $O(r^{-4})$ and $O(r^{-5})$ pieces of $\ol{\delta} \Psi_{1}$. Using \eqref{eqn:dbar_psi_1}, we see that these are, respectively, the $O(r^{-4})$ and $O(r^{-5})$ parts of $\left( \ol{\delta} C \right)\ind{_{k m \ol{m} m}}$; an expression for the latter quantity is given in \eqref{eqn:delta_derivative_weyl_4} of Appendix \ref{chap:appendix_d}. Performing a series expansion of this quantity in powers of $r^{-1}$, we find
\begin{equation}
\left( \ol{\delta} C \right)\ind{_{k m \ol{m} m}} = - 3 E M r^{-4} + 24 i E M a \cos{\theta} r^{-5} + O(r^{-6}) .
\end{equation}
Using this expression, we find that
\begin{equation}
\label{eqn:far_field_delta_bar_chi_slo}
\ol{\delta} \chi = - M r^{-3} + 6 i M a \cos{\theta} r^{-4} + O(r^{-5}) .
\end{equation}
Crucially, the $O(r^{-4})$ coefficient in \eqref{eqn:far_field_delta_bar_chi_slo} has a non-zero imaginary part. The term $\ol{\delta} \chi - \delta \ol{\chi}$ will therefore contribute to the right-hand side of the transport equation \eqref{eqn:transport_real_u1} at $O(r^{-4})$. Using this transport equation, we see that the leading-order far-field series solution for $\on{Re}{(\mathfrak{u}_{1})}$ is
\begin{equation}
\on{Re}{(\mathfrak{u}_{1})} = - \frac{2 M a \cos{\theta}}{E} r^{-3} + O(r^{-2}) .
\end{equation}
We remark that the leading-order piece depends on the spin of the black hole $a$, and vanishes in the case of a Schwarzschild black hole ($a = 0$).
%

\section{Wavefronts, caustic points and transport equations}
\label{sec:conjugate_points}

\subsection{Wavefronts and caustics in geometric optics}

We are interested in the propagation of electromagnetic waves on Kerr spacetime using the (higher-order) geometric optics approximation. Here, we review the important concept of a wavefront in general relativity for a general spacetime $(\mathcal{M}, g\ind{_{a b}})$; we refer the reader to \cite{Perlick2004} and references therein for a more detailed overview.

A \emph{wavefront} is a subset of spacetime that can be constructed as follows \cite{Perlick2004}. First, choose an orientable spacelike two-surface $\mathcal{S}$. Next, at each point $p \in \mathcal{S}$, choose a null direction orthogonal to $\mathcal{S}$ that depends smoothly on $p$. Finally, consider all rays which are tangent to the chosen null directions. These null geodesics are the \emph{generators} of the wavefront. The wavefront $\mathcal{W}$ is the union of all generators. If $\mathcal{S}$ is a two-sphere of radius $\varepsilon$ centred on $p \in \mathcal{M}$, then in the limit $\varepsilon \rightarrow 0$, the wavefront is the light cone with vertex $p$.

One may describe a wavefront using a coordinate system which is well-adapted to the behaviour of the generators \cite{Poisson2004}. We allow the affine parameter $s$ to be one of the coordinates, and we introduce two additional (local) coordinates $\left\{ \theta\ind{^{1}}, \theta\ind{^{2}} \right\}$ to label the null generators; the latter are constant along each generator, and span the two-surface $\mathcal{S}$ (which is transverse to the generators). Each null generator $\gamma(s)$ is chosen to be affinely parametrised such that $\gamma(s_{0}) \in \mathcal{S}$ and the tangent vector field $k\ind{^{a}}$ depends smoothly on the point $\gamma(s_{0})$. The wavefront $\mathcal{W}$ is then the image of a map
\begin{equation}
\label{eqn:wavefront_coordinate_representation}
(s, \theta\ind{^{1}}, \theta\ind{^{2}}) \mapsto W\ind{^{a}} (s, \theta\ind{^{1}}, \theta\ind{^{2}}), \qquad a \in \left\{ 0, 1, 2, 3 \right\} .
\end{equation}
In the neighbourhood of the two-surface $\mathcal{S}$ (i.e., close to $s = s_{0}$), the map \eqref{eqn:wavefront_coordinate_representation} is an embedding, so $\mathcal{W}$ is a three-dimensional (null) submanifold of $\mathcal{M}$.\footnote{A submanifold is said to be \emph{null} (or \emph{lightlike}) if its induced metric is degenerate. Since $k\ind{^{a}}$ is orthogonal to $\mathcal{S}$, it follows that the induced metric on the wavefront close to $\mathcal{S}$ is degenerate \cite{Poisson2004}. It is therefore a null submanifold.} However, away from $\mathcal{S}$, neighbouring null generators may intersect each other and the wavefront may fail to be a submanifold.

The \emph{caustic} of the wavefront is defined as the set of all points where \eqref{eqn:wavefront_coordinate_representation} fails to be an immersion; in other words, the rank of the differential of the map \eqref{eqn:wavefront_coordinate_representation} is not maximal (i.e., it is strictly less than $3$) \cite{Perlick2004}. Heuristically, a caustic is a transverse self-intersection of the wavefront. We note that the derivative of the map \eqref{eqn:wavefront_coordinate_representation} with respect to the affine parameter $s$ is always non-zero, so the rank of the map cannot be zero. If the rank is $3 - 1 = 2$, then the caustic point is said to have \emph{multiplicity one}. If, on the other hand, the rank is $3 - 2 = 1$, the caustic point is said to be of \emph{multiplicity two}. In the former case, the cross-section of an infinitesimally thin bundle of null rays (generators) in the instantaneous wavefront collapses down to a line at a caustic point. In the latter case, the bundle of generators collapses down to a single point.

If the wavefront is a light cone with vertex $p$, caustic points are said to be \emph{conjugate} to the point $p$ along a generator (i.e., a null geodesic) \cite{Perlick2004}. For an arbitrary wavefront, a caustic point is said to be \emph{conjugate} to a spacelike two-surface in the initial wavefront $\mathcal{S}$. The terms \emph{conjugate point} and \emph{caustic point} are therefore used interchangeably.

Away from caustic points, a wavefront is a hypersurface (i.e., a three-dimensional submanifold), which can be expressed as a function of the spacetime coordinates in the form $S(x\ind{^{a}}) = \textrm{constant}$, where $S$ satisfies the \emph{Hamilton--Jacobi equation} (or \emph{eikonal equation}) $g\ind{^{a b}} S\ind{_{, a}} S\ind{_{,b}} = 0$. However, at caustic points, a wavefront will typically exhibit complicated structure, with ``cuspidal edges'', or ``vertices''. Caustics can be classified locally using the singularity theory of Arnold \cite{Arnold1989, EhlersNewman2000}.

Recall from Section \ref{sec:geometric_optics_review} that the lensing dynamics of an infinitesimal bundle of rays in a wavefront $\mathcal{S}$ can be understood by analysing the geodesic deviation equation (or Jacobi equation), given in \eqref{eqn:null_geodesic_deviation_equation}. The cross-section of a bundle of rays is described by the shape parameters $d_{\pm}$. For an infinitesimal twist-free bundle of generators, a point where only one of the two shape parameters vanishes is a caustic point of multiplicity one; a point where both of the shape parameters vanish is a caustic point of multiplicity two \cite{Perlick2004}. It is clear from this definition that, at a caustic point, the cross-sectional area of the bundle $A = \pi d_{+} d_{-}$ vanishes.

The behaviour of the optical scalars in the neighbourhood of caustic points can be determined by considering the transport equations \eqref{eqn:transport_equation_rho_sigma}, i.e., the Sachs equations. Suppose there is a caustic point at $s = s_{0}$. For a caustic point of multiplicity one, the optical scalars are \cite{Perlick2004, SeitzSchneiderEhlers1994}
\begin{align}
\label{eqn:optical_scalars_caustic_point_multiplicity_one}
\rho(s) &= - \frac{1}{2} (s - s_{0})^{-1} + O\left( s - s_{0} \right), &
\left| \sigma(s) \right| &= \frac{1}{2} (s - s_{0})^{-1} + O\left( s - s_{0} \right) .
\end{align}
For a caustic point of multiplicity two, the optical scalars are given by \cite{Perlick2004, SeitzSchneiderEhlers1994}
\begin{align}
\rho(s) &= (s - s_{0})^{-1} + O\left( s - s_{0} \right), &
\sigma(s) &= \frac{1}{3} \Psi_{0}(s_{0}) (s - s_{0}) + O\left( (s - s_{0})^{2} \right) .
\end{align}
Caustic points of multiplicity one are somewhat more generic than caustic points of multiplicity two. Hereafter, when we use the term caustic point, we mean a caustic point of multiplicity one, unless stated otherwise.
%

\subsection{Behaviour of Newman--Penrose scalars at caustic points}
\label{sec:np_scalars_caustic_point}

It is clear from \eqref{eqn:optical_scalars_caustic_point_multiplicity_one} that the optical scalars diverge like $(s-s_{0})^{-1}$ at a caustic point. In the vicinity of a caustic point, the behaviour of the full set of complex Newman--Penrose scalars can be deduced from the system of transport equations \eqref{eqn:transport_equation_rho_sigma}--\eqref{eqn:transport_equation_mu_lambda} as follows.

In particular, let us assume that we have a caustic point of multiplicity one at $s = 0$, where $s$ is the affine parameter. (The affine parameter can always be redefined such that the caustic point occurs at $s = 0$.) We suppose that each Newman--Penrose scalar $z \in \{ \rho, \sigma, \chi, \tau, \mu, \lambda \}$ can be expanded as a power series of the form
\begin{equation}
\label{eqn:newman_penrose_scalar_power_series}
z(s) = s^{n_{z}} \sum_{j = 0}^{\infty} z_{j} s^{j},
\end{equation}
where $n_{z}$ is some exponent to be determined for each Newman--Penrose scalar $z$, and the coefficients $z_{j}$ are complex. Moreover, we expand the Weyl scalars, which appear on the right-hand side of the Newman--Penrose transport equations \eqref{eqn:transport_equation_rho_sigma}--\eqref{eqn:transport_equation_mu_lambda}, as a power series of the form
\begin{equation}
\label{eqn:weyl_scalar_power_series}
\Psi_{i}(s) = \sum_{j = 0}^{\infty} \psi_{i}^{(j)} s^{j},
\end{equation}
where $\psi_{i}^{(j)}$ are complex coefficients. (We note that the Weyl scalars are regular at caustic points.)

We insert the ansatz \eqref{eqn:newman_penrose_scalar_power_series} and \eqref{eqn:weyl_scalar_power_series} into the hierarchical system of transport equations \eqref{eqn:transport_equation_rho_sigma}--\eqref{eqn:transport_equation_mu_lambda} and consider each closed subsystem in turn (see the method of Section \ref{sec:far_field_np_scalars}). Balancing the leading order terms on both sides of the transport equations fixes the exponents $n_{z} = -1$ for all $z \in \{ \rho, \sigma, \chi, \tau, \lambda, \mu \}$. We find that, close to a caustic point of multiplicity one at $s = 0$, the behaviour of the Newman--Penrose scalars through sub-leading order is given by
\begin{align}
\rho(s) &= -\frac{1}{2} s^{-1} + \rho_{1} + O(s) , &
\sigma(s) &= - \frac{1}{2} e^{2 i \varphi} s^{-1} - e^{2 i \varphi} \rho_{1} + O(s) , \label{eqn:power_series_solutions_rho_sigma} \\
\chi(s) &= \left| \chi_{0} \right| i e^{i \varphi} s^{-1} + \left| \chi_{1} \right| e^{i \varphi} + O(s) , &
\tau(s) &= \left| \tau_{0} \right| e^{i \varphi} s^{-1} + \left| \tau_{1} \right| i e^{i \varphi} + O(s) , \label{eqn:power_series_solutions_chi_tau} \\
\mu(s) &= \mu_{0} s^{-1} + \mu_{1} + O(s) , &
\lambda(s) &= \mu_{0} e^{- 2 i \varphi} s^{-1} - \mu_{1} e^{- 2 i \varphi} + O(s) , \label{eqn:power_series_solutions_lambda_mu}
\end{align}
where $\varphi \in [0, 2 \pi)$, $\rho_{1}, \left| \chi_{0} \right|, \left| \chi_{1} \right|, \left| \tau_{0} \right|, \left| \tau_{1} \right| \in \mathbb{R}$ and $\mu_{0}, \mu_{1} \in \mathbb{C}$ are free parameters determined by the ray along which the transport equations \eqref{eqn:transport_equation_rho_sigma}--\eqref{eqn:transport_equation_mu_lambda} are evolved. The Weyl scalars \eqref{eqn:weyl_scalar_power_series}, which are $O(1)$ quantities, do not contribute to the series solutions \eqref{eqn:power_series_solutions_rho_sigma}--\eqref{eqn:power_series_solutions_lambda_mu} until $O(s)$, i.e., sub-sub-leading order. Hence, close to a caustic point of multiplicity one, spacetime curvature does not play a significant role in the behaviour of the Newman--Penrose scalars. The leading-order results for the optical scalars, given in \eqref{eqn:power_series_solutions_rho_sigma}, are equivalent to those presented by Perlick \cite{Perlick2004} and Seitz \emph{et al.} \cite{SeitzSchneiderEhlers1994}, cf.~\eqref{eqn:optical_scalars_caustic_point_multiplicity_one}.

The cross-sectional area $A$ and geometric optics square-amplitude $\mathcal{A}^{2}$ satisfy the transport equations $D A = - 2 \rho A$ and $D \mathcal{A}^{2} = 2 \rho \mathcal{A}^{2}$. Close to the caustic point, we expand these quantities as generalised power series of the form
\begin{equation}
\label{eqn:ansatz_area_amplitude}
A(s) = \sum_{j = 0}^{\infty} A_{j} s^{j}, \qquad \mathcal{A}^{2}(s) = s^{n} \sum_{j = 0}^{\infty} a_{j} s^{j},
\end{equation}
where $n$ is some index to be determined. Inserting \eqref{eqn:ansatz_area_amplitude} into the above transport equations, and noting that the cross-sectional area vanishes at the caustic point (i.e., $A(0) = 0$), we find the sub-leading-order solutions
\begin{align}
A(s) &= A_{1} s - 2 \rho_{1} A_{1} s^{2} + O(s^{3}) , \label{eqn:area_slo} \\
\mathcal{A}^{2}(s) &= a_{0} s^{-1} + 2 \rho_{1} a_{0} + O(s) , \label{eqn:amp_squared_slo}
\end{align}
where $A_{1}, a_{0} \in \mathbb{R}$ are free parameters. The square-amplitude $\mathcal{A}^{2}$ therefore diverges like $O(s^{-1})$ close to a caustic point of multiplicity one. We use the square of the amplitude (rather than the amplitude $\mathcal{A}$) to avoid fractional powers in the generalised power series expansions close to the caustic point.
%

\subsection{Behaviour of higher-order quantities at caustic points}
\label{sec:caustic_higher_order_np_quantities}

We now turn our attention to the quantities $\{ \delta \rho, \ol{\delta} \rho, \delta \sigma, \ol{\delta} \sigma, \delta \chi, \ol{\delta} \chi \}$, which are governed by the transport equations \eqref{eqn:ho_transport_3}--\eqref{eqn:ho_transport_8}. At first glance, this system of equations appears to be non-linear. However, we note that, if the Newman--Penrose scalars are known, then the transport equations \eqref{eqn:ho_transport_3}--\eqref{eqn:ho_transport_8} may be coupled to their complex conjugates to yield a linear system of first-order ordinary differential equations for the unknown quantities and their complex conjugates. This system may be expressed as a linear inhomogeneous system of ordinary differential equations of the form
\begin{equation}
\label{eqn:linear_higher_order_transport}
D \mathbf{x} = M \mathbf{x} + \mathbf{S} + \mathbf{W},
\end{equation}
where
\begin{equation}
\label{eqn:ho_np_dependent_variables}
\mathbf{x} = \left[
\delta \rho ,
(\delta \rho)^{\ast} ,
\delta \sigma ,
(\delta \sigma)^{\ast} ,
\overline{\delta} \sigma ,
(\overline{\delta} \sigma)^{\ast} ,
\delta \chi ,
(\delta \chi)^{\ast} ,
\overline{\delta} \chi ,
(\overline{\delta} \chi)^{\ast}
\right],
\end{equation}
is a vector which contains all of the dependent variables and their complex conjugates. We recall that $(\delta \rho)^{\ast} = \ol{\delta} \rho$, since $\rho$ is real. In \eqref{eqn:linear_higher_order_transport}, $M$ is a coefficient matrix of Newman--Penrose scalars, which is given by
\begin{equation}
\label{eqn:ho_np_matrix}
M =
\left[
\begin{array}{c c c c c c c c c c}
3 \rho & \sigma & \overline{\sigma} & 0 & 0 & \sigma & 0 & 0 & 0 & 0 \\
\overline{\sigma} & 3 \rho & 0 & \sigma & \overline{\sigma} & 0 & 0 & 0 & 0 & 0 \\
2 \sigma & 0 & 3 \rho & 0 & \sigma & 0 & 0 & 0 & 0 & 0 \\
0 & 2 \overline{\sigma} & 0 & 3 \rho & 0 &  \overline{\sigma} & 0 & 0 & 0 & 0 \\
0 & 2 \sigma & \overline{\sigma} & 0 & 3 \rho & 0 & 0 & 0 & 0 & 0 \\
2 \overline{\sigma} & 0 & 0 & \sigma & 0 & 3 \rho & 0 & 0 & 0 & 0 \\
\chi & 0 & - \overline{\chi} & 0 & 0 & 0 & 2 \rho & 0 & \sigma & - \sigma \\
0 & \overline{\chi} & 0 & - \chi & 0 & 0 & 0 & 2 \rho & - \overline{\sigma} & \overline{\sigma} \\
0 & \chi & 0 & 0 & - \overline{\chi} & 0 & \overline{\sigma} & - \sigma & 2 \rho & 0 \\
\overline{\chi} & 0 & 0 & 0 & 0 & - \chi & - \overline{\sigma} & \sigma & 0 & 2 \rho
\end{array}
\right].
\end{equation}
The components of the inhomogeneous terms which appear on the right-hand side of \eqref{eqn:linear_higher_order_transport} are given by
\begin{align}
\mathbf{S} &=
\left[
(\rho^{2} + \sigma \overline{\sigma}) \tau ,
(\ast),
2 \rho \sigma \tau ,
(\ast) ,
2 \rho \sigma \overline{\tau} ,
(\ast) ,
\tau (\rho \chi - \sigma \overline{\chi}) ,
(\ast) ,
\overline{\tau} (\rho \chi - \sigma \overline{\chi}) ,
(\ast)
\right],
\label{eqn:inhomogeneous_terms_np}
\\
\mathbf{W} &=
\left[
0 ,
(\ast) ,
\delta \Psi_{0} - \tau \Psi_{0} ,
(\ast) ,
\overline{\delta} \Psi_{0} - \overline{\tau} \Psi_{0} ,
(\ast) ,
\delta \Psi_{1} - \tau \Psi_{1} ,
(\ast) ,
\overline{\delta} \Psi_{1} - \overline{\tau} \Psi_{1} ,
(\ast)
\right],
\label{eqn:inhomogeneous_terms_weyl}
\end{align}
where an asterisk in parentheses $(\ast)$ is used for brevity to denote the complex conjugate of the preceding component of the vector. For example, the second component of $\mathbf{S}$ is $\left( (\rho^{2} + \sigma \overline{\sigma}) \tau \right)^{\ast} = (\rho^{2} + \sigma \overline{\sigma}) \ol{\tau}$. The vector $\mathbf{S}$ contains all of the Newman--Penrose scalars, and $\mathbf{W}$ contains all of the curvature terms, i.e., the Weyl curvature scalars and their directional derivatives.

Our aim is to find a generalised power series solution for the higher-order Newman--Penrose quantities $\{ \delta \rho, \ol{\delta} \rho, \delta \sigma, \ol{\delta} \sigma, \delta \chi, \ol{\delta} \chi \}$ through sub-leading order, in the vicinity of a caustic point of multiplicity one. To do this, we seek a generalised power series solution to the linear inhomogeneous system \eqref{eqn:linear_higher_order_transport}, with the additional constraint that the solution must satisfy the complex conjugacy property $x_{2k - 1} = \ol{x}_{2 k}$ for all $k \in \{ 1, 2, 3, 4, 5\}$, where $x_{k}$ are the coefficients of the solution $\mathbf{x}$. We expand $\mathbf{x}$ as a generalised power series of the form
\begin{equation}
\mathbf{x}(s) = s^{n} \sum_{j = 0}^{\infty} \mathbf{x}_{j} s^{j},
\end{equation}
where $\mathbf{x}_{j}$ are vectors of complex coefficients and the index $n$ is to be determined from analysis of the system \eqref{eqn:linear_higher_order_transport}. We note here that since the solution $\mathbf{x}$ must satisfy the complex conjugacy property (by construction), each of the coefficients $\mathbf{x}_{j}$ at $O(s^{j+n})$ must also satisfy the same property. Upon substitution of \eqref{eqn:power_series_solutions_rho_sigma}--\eqref{eqn:power_series_solutions_lambda_mu} into \eqref{eqn:ho_np_matrix}--\eqref{eqn:inhomogeneous_terms_weyl}, we see that $M = O(s^{-1})$, $\mathbf{S} = O(s^{-3})$ and $\mathbf{W} = O(s^{-1})$. We therefore write
\begin{align}
M(s) &= s^{-1} \sum_{j = 0}^{\infty} M_{j} s^{j}, &
\mathbf{S}(s) &= s^{-3} \sum_{j = 0}^{\infty} \mathbf{S}_{j} s^{j}, &
\mathbf{W}(s) &= s^{-3} \sum_{j = 0}^{\infty} \mathbf{W}_{j} s^{j}.
\end{align}
Here, $M_{j}$ are matrices given by \eqref{eqn:ho_np_matrix} with the Newman--Penrose scalars replaced by their $O(s^{j})$ coefficients from the series expansions \eqref{eqn:newman_penrose_scalar_power_series}. Similarly, $\mathbf{S}_{j}$ and $\mathbf{W}_{j}$ are vectors given by \eqref{eqn:inhomogeneous_terms_np} and \eqref{eqn:inhomogeneous_terms_weyl}, respectively, with the Newman--Penrose scalars and curvature terms replaced by their $O(s^{j})$ coefficients.

At leading order, the system of equations \eqref{eqn:linear_higher_order_transport} is
\begin{equation}
\label{eqn:eigenvalue_equation_leading_order}
M_{0} \mathbf{x}_{0} = n \mathbf{x}_{0},
\end{equation}
where we have assumed that the $O(s^{-3})$ terms are sub-dominant. This is an eigenvalue problem for the matrix $M_{0}$. The characteristic polynomial for the matrix $M_{0}$ is
\begin{equation}
\label{eqn:characteristic_polynomial}
n^{2} \left( n + 1 \right)^{4} \left( n + 2 \right)^{3} \left( n + 3 \right) = 0 ,
\end{equation}
from which the eigenvalues and their multiplicities can be read off straightforwardly. The smallest eigenvalue of $M_{0}$ is $n = -3$, so $\mathbf{x} = O(s^{-3})$. The eigenvector of the matrix $M_{0}$ corresponding to the eigenvalue $n = -3$ is
\begin{equation}
\mathbf{v}^{(-3)} =
\left[
e^{i \varphi} ,
(\ast) ,
e^{3 i \varphi} ,
(\ast) ,
e^{i \varphi}  ,
(\ast) ,
- 2 \left| \chi_{0} \right| i e^{2 i \varphi} ,
(\ast) ,
- 2 i \left| \chi_{0} \right| ,
(\ast)
\right] ,
\label{eqn:eigenvector_nm3}
\end{equation}
where we have again used the notation $(\ast)$ to denote the complex conjugate of the previous term. The eigenvector \eqref{eqn:eigenvector_nm3} clearly satisfies the complex conjugacy property. The leading-order solution is therefore any (complex) scalar multiple of this eigenvector: $\mathbf{x}_{0} = \alpha_{0} \mathbf{v}^{(-3)}$. Demanding that the leading-order solution $\mathbf{x}_{0}$ satisfies the complex conjugacy property means that $\alpha_{0}$ must be a real quantity. To leading order, the solutions to the transport equations \eqref{eqn:ho_transport_3}--\eqref{eqn:ho_transport_8} close to a caustic point of multiplicity one are therefore given by
\begin{align}
\delta \rho(s) &= \alpha_{0} e^{i \varphi} s^{-3} + O(s^{-2}), &
\ol{\delta} \rho(s) &= \alpha_{0} e^{i \varphi} s^{-3} + O(s^{-2}), \label{eqn:np_lo_delta_rho} \\
\delta \sigma(s) &= \alpha_{0} e^{3 i \varphi} s^{-3} + O(s^{-2}), &
\ol{\delta} \sigma(s) &= \alpha_{0} e^{i \varphi} s^{-3} + O(s^{-2}), \label{eqn:np_lo_delta_sigma} \\
\delta \chi(s) &= - 2 \alpha_{0} \left| \chi_{0} \right| i e^{2 i \varphi} s^{-3} + O(s^{-2}), &
\ol{\delta} \chi(s) &= - 2 \alpha_{0} \left| \chi_{0} \right| i s^{-3} + O(s^{-2}). \label{eqn:np_lo_delta_chi}
\end{align}

The sub-leading-order terms in \eqref{eqn:linear_higher_order_transport} yield an inhomogeneous system,
\begin{equation}
\label{eqn:eigenvalue_equation_subleading_order}
\left( M_{0} + 2 I \right) \mathbf{x}_{1} = - \left( \mathbf{S_{0}} + M_{1} \mathbf{x}_{0} \right),
\end{equation}
where $I$ denotes the identity matrix. Interestingly, we find that $\mathbf{x}_{0} \in \on{Null}{(M_{1})}$, where $\on{Null}{(M_{1})}$ denotes the null space of the matrix $M_{1}$. Hence, the term $M_{1} \mathbf{x}_{0}$ on the right-hand side of \eqref{eqn:eigenvalue_equation_subleading_order} vanishes. The matrix $M_{0} + 2 I$ is singular since $n = -2$ is an eigenvalue of $M_{0}$. One can see from \eqref{eqn:characteristic_polynomial} that this eigenvalue has multiplicity three. Solving the system \eqref{eqn:eigenvalue_equation_subleading_order} using Gaussian elimination, we arrive at the sub-leading-order solution
\begin{equation}
\label{eqn:inhomogeneous_solution_subleading_order}
\mathbf{x}_{1} = \sum_{k = 1}^{3} w_{k} \mathbf{v}_{k}^{(- 2)} + \mathbf{y}_{1},
\end{equation}
where $w_{k} = \xi_{k} + i \eta_{k}$ are arbitrary complex coefficients. The three eigenvectors of $M_{0}$ corresponding to the eigenvalue $n = -2$ are
\begin{align}
\mathbf{v}_{1}^{(-2)} &=
\left[
i e^{i \varphi} ,
- i  e^{- i \varphi} ,
3 i e^{3 i \varphi} ,
- 3 i e^{- 3 i \varphi} ,
i e^{i \varphi}  ,
- i e^{- i \varphi}  ,
4 \left| \chi_{0} \right| e^{2 i \varphi}  ,
4 \left| \chi_{0} \right| e^{- 2 i \varphi} ,
0 ,
0
\right] ,
\label{eqn:eigenvector_v_nm2_1}
\\
\mathbf{v}_{2}^{(-2)} &=
\left[
0 ,
0 ,
i e^{3 i \varphi} ,
- i e^{- 3 i \varphi} ,
i e^{i \varphi}  ,
- i e^{- i \varphi}  ,
2 \left| \chi_{0} \right| e^{2 i \varphi}  ,
0 ,
2 \left| \chi_{0} \right| ,
0
\right] ,
\label{eqn:eigenvector_v_nm2_2}
\\
\mathbf{v}_{3}^{(-2)} &=
\left[
i e^{i \varphi} ,
- i  e^{- i \varphi} ,
i e^{3 i \varphi} ,
- i e^{- 3 i \varphi} ,
- i e^{i \varphi}  ,
i e^{- i \varphi}  ,
4 \left| \chi_{0} \right| e^{2 i \varphi}  ,
0 ,
0 ,
- 4 \left| \chi_{0} \right|
\right] ,
\label{eqn:eigenvector_v_nm2_3}
\end{align}
which are clearly linearly independent. The particular solution to the inhomogeneous system \eqref{eqn:eigenvalue_equation_subleading_order} is given by
\begin{equation}
\mathbf{y}_{1} =
- \frac{1}{2} \left| \tau_{0} \right|
\left[
e^{i \varphi} ,
e^{- i \varphi} ,
e^{3 i \varphi} ,
e^{-3 i \varphi} ,
e^{i \varphi} ,
e^{-i \varphi} ,
0 ,
0 ,
0 ,
0
\right].
\label{eqn:particular_solution_nm2}
\end{equation}
We note here that the particular solution \eqref{eqn:particular_solution_nm2} satisfies the complex conjugacy property; however, the eigenvectors \eqref{eqn:eigenvector_v_nm2_2} and \eqref{eqn:eigenvector_v_nm2_3} do not. Demanding that the components of the sub-leading-order solution \eqref{eqn:inhomogeneous_solution_subleading_order} satisfy the complex conjugacy property gives a relationship between the complex coefficients $w_{k}$:
\begin{align}
w_{1} &= \xi_{1} + i \eta_{1}, & w_{2} &= \xi_{2} - i \eta_{1}, & w_{3} &= - 2 \xi_{2} - 2 i \eta_{1},
\end{align}
where $\xi_{1}, \xi_{2}, \eta_{1} \in \mathbb{R}$ are free parameters.

The generalised power series solutions for the higher-order Newman--Penrose quantities through sub-leading order are given by
\begin{align}
\delta \rho(s) &= \alpha_{0} e^{i \varphi} s^{-3} + \left( \xi_{1} + \xi_{2} + \frac{i}{2} \left| \tau_{0} \right| \right) i e^{i \varphi} s^{-2} + O(s^{-1}) , \label{eqn:ho_np_slo_1} \\
\ol{\delta} \rho(s) &= \alpha_{0} e^{i \varphi} s^{-3} - \left( \xi_{1} + \xi_{2} - \frac{i}{2} \left| \tau_{0} \right| \right) i e^{- i \varphi} s^{-2} + O(s^{-1}) , \\
\delta \sigma(s) &= \alpha_{0} e^{3 i \varphi} s^{-3} + \left( 3 \xi_{1} - \xi_{2} - \frac{i}{2} \left| \tau_{0} \right| \right) i e^{3 i \varphi} s^{-2} + O(s^{-1}) , \\
\ol{\delta} \sigma(s) &= \alpha_{0} e^{i \varphi} s^{-3} + \left( \xi_{1} - 3 \xi_{2} - \frac{i}{2} \left| \tau_{0} \right| \right) i e^{i \varphi} s^{-2} + O(s^{-1}) , \\
\delta \chi(s) &= - 2 \alpha_{0} \left| \chi_{0} \right| i e^{2 i \varphi} s^{-3} + 4 \left| \chi_{0} \right| \left( \xi_{1} - i \eta_{1} \right) e^{2 i \varphi} s^{-2} + O(s^{-1}) , \\
\ol{\delta} \chi(s) &= - 2 \alpha_{0} \left| \chi_{0} \right| i s^{-3} - 4 \left| \chi_{0} \right| \left( \xi_{2} + i \eta_{1} \right) s^{-2} + O(s^{-1}) . \label{eqn:ho_np_slo_6}
\end{align}
These solutions feature three free parameters ($\varphi, \left| \chi_{0} \right|, \left| \tau_{0} \right| \in \mathbb{R}$) which appear in the leading-order solutions for the Newman--Penrose scalars; one free parameter ($\alpha_{0} \in \mathbb{R}$) which comes from the leading-order part; and three free parameters ($\xi_{1}, \xi_{2}, \eta_{1} \in \mathbb{R}$) which arise from the sub-leading-order part.

As described in Section \ref{sec:constraints_np_field_equations}, we have the pair of algebraic constraints $Z = 0$ and $W = 0$, given in \eqref{eqn:ricci_constraint_1} and \eqref{eqn:ricci_constraint_1}, which must be satisfied by the Newman--Penrose scalars, the higher-order Newman--Penrose quantities, and the Weyl scalars. We now impose these constraints on the solutions \eqref{eqn:ho_np_slo_1}--\eqref{eqn:ho_np_slo_6} through $O(s^{-2})$. The dominant terms in \eqref{eqn:ricci_constraint_1} are the $O(s^{-3})$ pieces of $\delta \rho$ and $\ol{\delta} \sigma$. Remarkably, these are equal, so the constraint $Z = 0$ is satisfied at leading order. Similarly, the dominant terms in \eqref{eqn:ricci_constraint_2} are the $O(s^{-3})$ pieces of $\ol{\delta} \chi$ and its complex conjugate. Since these quantities are purely imaginary, their sum vanishes, and the constraint $W = 0$ is satisfied at leading order. In order for the constraints \eqref{eqn:ricci_constraint_1} and \eqref{eqn:ricci_constraint_2} to be satisfied at $O(s^{-2})$, we require
\begin{equation}
\label{eqn:ricci_constraints_slo}
\xi_{2} = \frac{1}{4} \left| \chi_{0} \right|.
\end{equation}
Both $Z = 0$ and $W = 0$ yield the same condition. Replacing the parameter $\xi_{2}$ in \eqref{eqn:ho_np_slo_1}--\eqref{eqn:ho_np_slo_6} using \eqref{eqn:ricci_constraints_slo} yields the generalised power series solutions through sub-leading order for the quantities $\{ \delta \rho, \ol{\delta} \rho, \delta \sigma, \ol{\delta} \sigma, \delta \chi, \ol{\delta} \chi \}$.

Using the sub-leading-order solutions for the Newman--Penrose scalars and the higher-order Newman--Penrose quantities, we may also look for a generalised series solution to the transport equations \eqref{eqn:ho_transport_1} and \eqref{eqn:ho_transport_2} close to a caustic point of multiplicity one. Noting that $\ol{\delta} \mathcal{A}^{2} = (\delta \mathcal{A}^{2})^{\ast}$, we seek a solution of the form
\begin{equation}
\label{eqn:power_series_delta_amp_2}
\delta \mathcal{A}^{2}(s) = s^{n} \sum_{j = 0}^{\infty} \mathfrak{a}_{j} s^{j} ,
\end{equation}
where $\mathfrak{a}_{j}$ are complex coefficients, and the index $n$ is to be determined by balancing terms in \eqref{eqn:ho_transport_1} and \eqref{eqn:ho_transport_2}. Inserting \eqref{eqn:power_series_delta_amp_2} and its complex conjugate into the transport equations \eqref{eqn:ho_transport_1} and \eqref{eqn:ho_transport_2}, we find $n = -3$; hence, $\delta \mathcal{A}^{2}$ diverges like $s^{-3}$ at a caustic point of multiplicity one.

At $O(s^{-3})$, the equations \eqref{eqn:ho_transport_1} and \eqref{eqn:ho_transport_2} lead to
\begin{equation}
\left[
\begin{array}{c c}
3 \rho_{0} + 3 & \sigma_{0} \\
\ol{\sigma}_{0} & 3 \rho_{0} + 3
\end{array}
\right]
\left[
\begin{array}{c}
\mathfrak{a}_{0} \\
\ol{\mathfrak{a}}_{0}
\end{array}
\right]
=
- 2 a_{0} \alpha_{0}
\left[
\begin{array}{c}
e^{i \varphi} \\
e^{- i \varphi}
\end{array}
\right] .
\end{equation}
The matrix on the left-hand side of this system is invertible, and it follows that $\mathfrak{a}_{0} = - 2 a_{0} \alpha_{0} e^{i \varphi}$. Hence $\delta \mathcal{A}^{2} = - 2 a_{0} \alpha_{0} e^{i \varphi} s^{-3} + O(s^{-2})$. At $O(s^{-2})$, the equations \eqref{eqn:ho_transport_1} and \eqref{eqn:ho_transport_2} yield
\begin{equation}
\left[
\begin{array}{c c}
3 \rho_{0} + 2 & \sigma_{0} \\
\ol{\sigma}_{0} & 3 \rho_{0} + 2
\end{array}
\right]
\left[
\begin{array}{c}
\mathfrak{a}_{1} \\
\ol{\mathfrak{a}}_{1}
\end{array}
\right]
=
\frac{1}{2} a_{0} \left( 4 \xi_{1} + \left| \chi_{0} \right| \right)
\left[
\begin{array}{c}
- i e^{i \varphi} \\
i e^{- i \varphi}
\end{array}
\right] .
\end{equation}
This system is non-invertible; however, it may be solved using Gaussian elimination, for example. The solution is $\mathfrak{a}_{1} = \left[ \zeta_{1} + \frac{i a_{0}}{2} \left( 4 \xi_{1} + \left| \chi_{0} \right| \right) \right] e^{i \varphi}$, where $\zeta_{1} \in \mathbb{R}$ is a free parameter. The general solution to \eqref{eqn:ho_transport_1} and \eqref{eqn:ho_transport_2} through sub-leading order is therefore
\begin{equation}
\label{eqn:delta_amp_squared_slo}
\delta \mathcal{A}^{2}(s) = - 2 a_{0} \alpha_{0} e^{i \varphi} s^{-3} + \left[ \zeta_{1} + \frac{i a_{0}}{2} \left( 4 \xi_{1} + \left| \chi_{0} \right| \right) \right] e^{i \varphi} s^{-2} + O(s^{-1}) ,
\end{equation}
with $\ol{\delta} \mathcal{A}^{2}$ given by the complex conjugate of \eqref{eqn:delta_amp_squared_slo}.

Inserting all of the relevant sub-leading-order series solutions into the right-hand side of the transport equation \eqref{eqn:transport_real_u1}, we find
\begin{equation}
D (\on{Re}{(\mathfrak{u}_{1})} ) = 4 \alpha_{0} \left| \chi_{0} \right| s^{-3} - \left[ 4 \left|\chi_{0} \right| \left(\alpha_{0} \rho_{1} - \eta_{1} \right) + \on{Im}{(\mu_{0})} + \frac{\zeta_{1} \left|\chi_{0} \right|}{a_{0}} \right] s^{-2} + O(s^{-1}) .
\end{equation}
Integration of this transport equation with respect to $s$ gives
\begin{equation}
\label{eqn:power_series_re_u1}
\on{Re}{(\mathfrak{u}_{1}(s))} = - 2 \alpha_{0} \left| \chi_{0} \right| s^{-2} + \left[ 4 \left|\chi_{0} \right| \left(\alpha_{0} \rho_{1} - \eta_{1} \right) + \on{Im}{(\mu_{0})} + \frac{\zeta_{1} \left|\chi_{0} \right|}{a_{0}} \right] s^{-1}  + O(1) .
\end{equation}
Hence, the quantity $\on{Re}{(\mathfrak{u}_{1})}$ diverges like $s^{-2}$ at a caustic point of multiplicity one. The $O(s^{-2})$ part of the divergence is controlled by the parameters $\left| \chi_{0} \right|$ and $\alpha_{0}$; the former arises in the leading-order solution for the Newman--Penrose scalars \eqref{eqn:power_series_solutions_rho_sigma}--\eqref{eqn:power_series_solutions_lambda_mu}, whilst the latter comes from the leading-order series solution for the higher-order Newman--Penrose quantities \eqref{eqn:np_lo_delta_rho}--\eqref{eqn:np_lo_delta_chi}. The $O(s^{-1})$ piece features the parameters $\rho_{1}$, $\left| \chi_{0} \right|$ and $\mu_{0}$, which come from the leading-order solutions \eqref{eqn:power_series_solutions_rho_sigma}--\eqref{eqn:power_series_solutions_lambda_mu}; $a_{0}$, which comes from the leading-order part of \eqref{eqn:amp_squared_slo}; $\alpha_{0}$ and $\eta_{1}$, which arise from the leading- and sub-leading-order solutions \eqref{eqn:ho_np_slo_1}--\eqref{eqn:ho_np_slo_6}, respectively; and $\zeta_{1}$, which appears in the sub-leading-order piece of \eqref{eqn:delta_amp_squared_slo}.

We remark here that a quantity with spin weight $\mathfrak{s}$ admits a power series expansion close to the caustic point whose coefficients are proportional to the unit complex number $e^{i \mathfrak{s} \pi}$. (We refer the reader to Section \ref{sec:ghp_formalism} for a brief review of the Geroch--Held--Penrose calculus and the notion of spin weight.)

The transport equations for which we have obtained series solutions come from the higher-order geometric optics formalism of Dolan \cite{Dolan2018}, which assumes only that the spacetime is Ricci-flat. The results obtained in Section \ref{sec:np_scalars_caustic_point} and the present section are general, in that they apply to any Ricci-flat manifold. The curvature terms (i.e., Weyl scalars and their derivatives) do not play a role until sub-sub-leading order.
%

\section{Evolving transport equations through caustic points}
\label{sec:transport_equations_cautics}

Our primary aim in this chapter is to calculate higher-order geometrical optics corrections to the stress--energy tensor (see Section \ref{sec:higher_order_go}). This involves calculating the quantity $\on{Re}{(\mathfrak{u}_{1})}$, for which we have a transport equation involving Newman--Penrose scalars and higher-order Newman--Penrose quantities (e.g.~$\ol{\delta} \chi$). When evolving the relevant system of transport equations for an infinitesimal bundle of rays, we will generically encounter caustic points, where neighbouring rays cross. As demonstrated in Section \ref{sec:conjugate_points}, the Newman--Penrose scalars, higher-order Newman--Penrose quantities and geometric optics quantities diverge at caustic points; we therefore require a method which allows us to evolve the transport equations through caustic points, so that the quantities of interest, such as $\on{Re}{(\mathfrak{u}_{1})}$, can be calculated after scattering has occurred (at $r \rightarrow \infty$).

As reviewed in Section \ref{sec:geometric_optics_review}, Dolan \cite{Dolan2017, Dolan2018} demonstrated that certain Newman--Penrose scalars -- namely $\left\{ \rho, \sigma, \chi, \tau, \mu, \lambda \right\}$ -- can be calculated by decomposing the geodesic deviation equation \eqref{eqn:null_geodesic_deviation_equation} and the differential precession equation \eqref{eqn:differential_precession_equation} in a parallel-transported null tetrad $\{ k, n, m, \ol{m} \}$. This approach yields a system of first- and second-order ordinary differential equations which \emph{can} be evolved through caustic points. The Newman--Penrose scalars can then be found by inverting a system of simultaneous equations for a pair of linearly independent solutions to the aforementioned differential equations. This inversion breaks down at caustic points; however, it is well-defined everywhere else along the central ray of the bundle. The Newman--Penrose scalars may therefore be calculated \emph{after} the first caustic point using this method.

Whilst it is possible that a similar formalism could be used to find \emph{all} of the Newman--Penrose and geometrical optics quantities along an entire ray (i.e., past the first caustic point), it is not immediately obvious which equations (cf.~geodesic deviation and differential precession) should be considered for this purpose. We therefore proceed by proposing a more pragmatic approach to evolving the full system of transport equations through caustic points, which relies on ``regularising'' the divergent quantities. In Section \ref{sec:regularisation_transport_equations}, we introduce the regularisation method, which could be applied to any parallel-propagated twist-free complex null tetrad on a Ricci-flat spacetime. In Section \ref{sec:regularisation_numerical_method_and_results}, we apply the numerical method to Kerr spacetime, and present some preliminary results. We comment on these results in Section \ref{sec:discussion_go_kerr}.


\subsection{Regularisation of transport equations}
\label{sec:regularisation_transport_equations}

In this section, we assume that there is a caustic point (of multiplicity one) at $s = s_{0}$, where $s$ is the affine parameter along the central ray. We note first that, close to a caustic point, the cross-sectional area scales like $(s - s_{0})$; see \eqref{eqn:area_slo}.

Consider the Newman--Penrose scalars $\left\{ \rho, \sigma, \chi, \tau, \mu, \lambda \right\}$, which obey the transport equations \eqref{eqn:transport_equation_rho_sigma}--\eqref{eqn:transport_equation_mu_lambda}, and which admit generalised power series solutions in the neighbourhood of the caustic point given by \eqref{eqn:power_series_solutions_rho_sigma}--\eqref{eqn:power_series_solutions_lambda_mu} through sub-leading order. Each Newman--Penrose scalar diverges like $(s - s_{0})^{-1}$ as we approach the caustic point. We choose to regularise the Newman--Penrose scalars by multiplying each one by the cross-sectional area, e.g.~$\wh{\rho} = A \rho$, to cancel out the $O\left( (s - s_{0})^{-1} \right)$ divergence. The regularised Newman--Penrose scalars, which do not diverge at caustic points, satisfy the system of transport equations
\begin{align}
D \wh{\rho} &= \frac{1}{A} \left( \wh{\sigma} \ol{\wh{\sigma}} - \wh{\rho}^{2} \right), &
D \wh{\sigma} &= A \Psi_{0}, \label{eqn:regularised_transport_rho_sigma} \\
D \wh{\chi} &= - \frac{1}{A} \left( \wh{\rho} \wh{\chi} + \wh{\sigma} \ol{\wh{\chi}} \right) + A \Psi_{1}, &
D \wh{\tau} &= \frac{1}{A} \left( \wh{\sigma} \ol{\wh{\tau}} - \wh{\rho} \wh{\tau} \right) + A \Psi_{1}, \\
D \wh{\lambda} &= \frac{1}{A} \left( \ol{\wh{\sigma}} \wh{\mu} - \wh{\rho} \wh{\lambda} \right), &
D \wh{\mu} &= \frac{1}{A} \left( \wh{\sigma} \wh{\lambda} - \wh{\rho} \wh{\mu} \right) + A \Psi_{2}. \label{eqn:regularised_transport_lambda_mu}
\end{align}
We note that the cross-sectional area now satisfies the transport equation
\begin{equation}
\label{eqn:regularised_transport_equation_area}
D A = - 2 \wh{\rho}.
\end{equation}
As in the case of the Newman--Penrose scalars, the square-amplitude $\mathcal{A}^{2}$ diverges like $(s - s_{0})^{-1}$, so we regularise via $\wh{\mathcal{A}}^{2} = A \mathcal{A}^{2}$. The regularised square-amplitude then satisfies
\begin{equation}
D \wh{\mathcal{A}}^{2} = 0,
\end{equation}
i.e., it is constant. We remark that the notation used here is somewhat misleading: $\mathcal{A}^{2}$ is not necessarily a non-negative quantity. One can see from the series solution \eqref{eqn:area_slo} that $A$ will change sign at a caustic point; since $\mathcal{A}^{2}$ is proportional to $A^{-1}$, this quantity will also change sign at a caustic point.

Now consider the higher-order quantities $\{ \delta \rho, \ol{\delta} \rho, \delta \sigma, \ol{\delta} \sigma, \delta \chi, \ol{\delta} \chi, \delta \mathcal{A}^{2}, \ol{\delta} \mathcal{A}^{2} \}$. The generalised series solutions for these quantities are given in \eqref{eqn:ho_np_slo_1}--\eqref{eqn:ho_np_slo_6} and \eqref{eqn:delta_amp_squared_slo}. Each of these quantities diverges like $(s - s_{0})^{-3}$ at a caustic point. We may therefore regularise these quantities by multiplying by $A^{3}$, e.g.~$\wh{\delta \rho} = A^{3} \delta \rho$. The rescaled quantities satisfy the system of equations
\begin{align}
D(\wh{\delta \rho}) &= \frac{1}{A} \left( - 3 \, \wh{\rho} \, \wh{\delta \rho} + \wh{\sigma} \, \wh{\ol{\delta} \rho} + \ol{\wh{\sigma}} \, \wh{\delta \sigma} + \wh{\sigma} \, ( \wh{ \ol{\delta} \sigma } )^{\ast} \right) - \left( \wh{\rho}^{2} + \wh{\sigma} \, \ol{\wh{\sigma}} \right) \wh{\tau} ,  \label{eqn:regularised_transport_delta_rho} \\
D(\wh{\ol{\delta} \rho}) &= \frac{1}{A} \left( -3 \wh{\rho} \, \wh{\ol{\delta} \rho} + \ol{\wh{\sigma}} \, \wh{\delta \rho} + \ol{\wh{\sigma}} \, \wh{\ol{\delta} \sigma} + \wh{\sigma} \, ( \wh{\delta \sigma})^{\ast} \right) - \left( \wh{\rho}^{2} + \wh{\sigma} \ol{\wh{\sigma}} \right) \ol{\wh{\tau}} , \\
D(\wh{\delta \sigma}) &= \frac{1}{A} \left( -3 \wh{\rho} \, \wh{\delta \sigma} + 2 \wh{\sigma} \, \wh{\delta \rho} + \wh{\sigma} \, \wh{\ol{\delta} \sigma} \right) - 2 \wh{\rho} \, \wh{\sigma} \wh{\tau} - A^{2} \wh{\tau} \Psi_{0} + A^{3} \delta \Psi_{0} , \\
D(\wh{\ol{\delta} \sigma}) &= \frac{1}{A} \left( -3 \wh{\rho} \, \wh{\ol{\delta} \sigma} + 2 \wh{\sigma} \, \wh{\ol{\delta} \rho} + \ol{\wh{\sigma}} \, \wh{\delta \sigma} \right) - 2 \wh{\rho} \, \wh{\sigma} \ol{\wh{\tau}} - A^{2} \ol{\wh{\tau}} \Psi_{0} + A^{3} \ol{\delta} \Psi_{0},
\\
\begin{split}
D(\wh{\delta \chi}) &=
\frac{1}{A} \left( - 4 \wh{\rho} \, \wh{\delta \chi} + \wh{\chi} \, \wh{\delta \rho} + \wh{\sigma} \, \wh{\ol{\delta} \chi} - \wh{\sigma} (\wh{\ol{\delta} \chi})^{\ast} - \ol{\wh{\chi}} \, \wh{\delta \sigma} \right) - \left( \wh{\rho} \wh{\chi} - \wh{\sigma} \ol{\wh{\chi}} \right) \wh{\tau} \\
& \qquad - A^{2} \wh{\tau} \Psi_{1} + A^{3} \delta \Psi_{1},
\end{split}
\\
\begin{split}
D(\wh{\ol{\delta} \chi}) &=
\frac{1}{A} \left( - 4 \wh{\rho} \, \wh{\ol{\delta} \chi} + \wh{\chi} \, \wh{\ol{\delta} \rho} + \ol{\wh{\sigma}} \, \wh{\delta \chi} - \wh{\sigma} (\wh{\delta \chi})^{\ast} - \ol{\wh{\chi}} \, \wh{\ol{\delta} \sigma} \right) - \left( \wh{\rho} \wh{\chi} - \wh{\sigma} \ol{\wh{\chi}} \right) \ol{\wh{\tau}} \\
& \qquad - A^{2} \ol{\wh{\tau}} \Psi_{1} + A^{3} \ol{\delta} \Psi_{1} ,
\end{split}
\label{eqn:regularised_transport_deltabar_chi}
\\
D ( \wh{\delta \mathcal{A}^{2}} ) &= \frac{1}{A} \left( 2 \, \wh{\mathcal{A}^{2}} \, \wh{\delta \rho} - 3 \, \wh{\rho} \, \wh{\delta \mathcal{A}^{2}} + \wh{\sigma} \, \wh{ \ol{\delta} \mathcal{A}^{2} } \right) - 2 \, \wh{ \mathcal{A}^{2} } \, \wh{\rho} \, \wh{\tau} , \label{eqn:regularised_transport_delta_amp_sq} \\
D ( \wh{\ol{\delta} \mathcal{A}^{2}} ) &= \frac{1}{A} \left( 2 \, \wh{\mathcal{A}^{2}} \, \wh{\ol{\delta} \rho} - 3 \, \wh{\rho} \, \wh{\ol{\delta} \mathcal{A}^{2}} + \ol{\wh{\sigma}} \, \wh{ \delta \mathcal{A}^{2} } \right) - 2 \, \wh{ \mathcal{A}^{2} } \, \wh{\rho} \, \ol{\wh{\tau}} , \label{eqn:regularised_transport_deltabar_amp_sq}
\end{align}
where the divergent quantities are to be replaced by their regularised counterparts in $\delta \Psi_{0}$, $\ol{\delta} \Psi_{0}$, $\delta \Psi_{1}$ and $\ol{\delta} \Psi_{1}$, which are defined in \eqref{eqn:d_psi_0}--\eqref{eqn:dbar_psi_1}.

Finally, we see from \eqref{eqn:power_series_re_u1} that the quantity $\on{Re}{(\mathfrak{u}_{1})}$ scales like $(s - s_{0})^{-2}$. We therefore introduced the regularised quantity $\on{Re}{(\wh{\mathfrak{u}}_{1})} = A^{2} \on{Re}{(\mathfrak{u}_{1})}$, which is not divergent at the caustic point. This quantity satisfies the transport equation
\begin{equation}
\label{eqn:regularised_transport_re_u1}
\begin{split}
D\left( \operatorname{Re}(\wh{\mathfrak{u}}_{1}) \right) =& \frac{1}{A} \left[ - 4 \wh{\rho} \operatorname{Re}(\wh{\mathfrak{u}}_{1}) + \frac{i}{2} \left( \wh{\ol{\delta} \chi} - (\wh{\ol{\delta} \chi} )^{\ast} + \wh{\chi} \frac{\wh{\ol{\delta} \mathcal{A}^{2}}}{\wh{\mathcal{A}^{2}}} - \ol{\wh{\chi}} \frac{\wh{\delta \mathcal{A}^{2}}}{\wh{\mathcal{A}^{2}}} \right) \right] \\
& \qquad + \frac{i}{2} \left[ \wh{\rho} \left( \ol{\wh{\mu}} - \wh{\mu} \right) + \ol{\wh{\sigma}} \ol{\wh{\lambda}} - \wh{\sigma} \wh{\lambda} \right] .
\end{split}
\end{equation}

We note that the regularisation operation (denoted by a caret) commutes with complex conjugation (denoted by an overline or an asterisk) because the cross-sectional area $A$ is real. (Recall that a divergent quantity $z$ which scales like $(s - s_{0})^{n}$ at the caustic point is regularised by taking $\wh{z} = A^{-n} z$. Then $\ol{\wh{z}} = A^{-n} \ol{z} = \wh{\ol{z}}$.)

\subsection{Numerical method and results}
\label{sec:regularisation_numerical_method_and_results}

Our principal aim is to consider spin--helicity effects for circularly polarised electromagnetic waves on Kerr spacetime. In particular, we shall consider a circularly polarised electromagnetic wave which begins far from the system, interacts with the black hole, and is then scattered to infinity. We employ the higher-order geometric optics formalism to consider the lensing effects of an initially circular infinitesimal bundle of rays in the electromagnetic wavefront. We shall evolve the transport equations of Section \ref{sec:regularisation_transport_equations} along null geodesics of Kerr spacetime, in order to calculate $\on{Re}{(\mathfrak{u}_{1})}$ along a ray. (We consider the case of a rapidly rotating Kerr black hole of mass $M = 1$ with spin parameter $a = 0.9$.)

\subsubsection{Null geodesic equations and initial conditions}

We use the Hamiltonian formalism for null geodesics in Boyer--Lindquist coordinates. In particular, we evolve the full set of Hamilton's equations, which is a system of eight first-order ordinary differential equations for the spacetime coordinates and their conjugate momenta. We recall that there are four conserved quantities along null rays in the Kerr spacetime: the energy $E = - p_{t}$, the azimuthal angular momentum $L\ind{_{z}} = p\ind{_{\phi}}$; the Hamiltonian $H = 0$; and the Carter constant $K = {p\ind{_{\theta}}}^{2} + \left( a E \sin{\theta} - L\ind{_{z}} \on{cosec}{\theta} \right)^{2}$.

Following Johannsen and Psaltis \cite{JohannsenPsaltis2010}, we consider a ray which begins on an ``image plane'' whose centre is a distance $D_{0}$ away from the origin of the coordinate system centred on the black hole, at an inclination angle $\iota$, with its centre in the $(x, z)$-plane; see Appendix of \cite{JohannsenPsaltis2010}. Since the ray begins far from the black hole, the initial data can be expressed using standard Euclidean geometry. The image plane coordinates $(X, Y)$ are related to the Cartesian coordinates $(x, y, z)$ centred on the black hole via
\begin{align}
x &= - Y \cos{\iota} + D_{0} \sin{\iota}, &
y &= X, &
z &= Y \sin{\iota} + D_{0} \cos{\iota} .
\end{align}
A photon which begins on the image plane at $(X, Y)$ with initial three-momentum orthogonal to the image plane has initial conditions
\begin{align}
r &= \sqrt{X^{2} + Y^{2} + D_{0}^{2}} , &
p\ind{_{r}} &= - \frac{D_{0}}{r} , \label{eqn:coordinates_image_plane_r} \\
\theta &= \arccos{ \left( \frac{Y \sin{\iota} + D_{0} \cos{\iota} }{r} \right) } , &
p\ind{_{\theta}} &= \frac{\cos{\iota} - D_{0} \left( Y \sin{\iota} + D_{0} \cos{\iota} \right)}{r^{2} \sqrt{X^{2} + \left( D_{0} \sin{\iota} - Y \cos{\iota} \right)^{2}} } , \label{eqn:coordinates_image_plane_theta} \\
\phi &= \arctan{ \left( \frac{X}{D_{0} \sin{\iota} - Y \cos{\iota}} \right) } , &
p\ind{_{\phi}} &= \frac{X \sin{\iota}}{X^{2} + \left( D\sin{\iota} - Y\cos{\iota} \right)^{2}} . \label{eqn:coordinates_image_plane_phi}
\end{align}
The $t$-component of the four-momentum is calculated from the three-momentum so that the norm $g\ind{^{a b}} p\ind{_{a}} p\ind{_{b}}$ vanishes. In practice, we use the fact that $g\ind{_{a b}} \rightarrow \eta\ind{_{a b}}$ as $r \rightarrow \infty$ to calculate $p\ind{_{t}}$. In this limit, the choice of normalisation of the three-momentum means that we have $p\ind{_{t}} = -1$.
%

\subsubsection{Transport equations and far-field initial data}

In order to calculate $\on{Re}{(\mathfrak{u}_{1})}$, one must evolve the system of transport equations \eqref{eqn:transport_equation_area_twist_free} for the cross-sectional area; \eqref{eqn:transport_equation_square_amplitude_recap} for the square-amplitude; \eqref{eqn:transport_equation_rho_sigma}--\eqref{eqn:transport_equation_mu_lambda} for the Newman--Penrose scalars; \eqref{eqn:ho_transport_1}--\eqref{eqn:ho_transport_8} for the higher-order Newman--Penrose and geometrical optics quantities; and \eqref{eqn:transport_real_u1} for $\on{Re}{(\mathfrak{u}_{1})}$ itself. These must be evolved along a central null geodesic, whose initial conditions are determined using the set-up presented in the previous section.

We wish to consider an (initially circular) infinitesimal bundle of rays which starts at infinity; however, in practice, the numerical integration of the transport equations must begin at some finite (but large) value of $r$. One may use the far-field solutions to the transport equations given in Section \ref{sec:far_field_asymptotics} to determine the leading-order-in-$\frac{1}{r}$ corrections to the initial conditions which arise from the fact that $r(0) = r_{0} < \infty$.

For example, at infinity, one would choose the expansion $\rho$ and shear $\sigma$ to be zero initially. One can see from the Sachs equations \eqref{eqn:transport_equation_rho_sigma} that, even in regions of spacetime where the curvature is small, shear will be generated through the transport equation for $\sigma$, which will generate expansion through the transport equation for $\rho$. The leading-order correction to the initial conditions for $\rho$ and $\sigma$ are determined by \eqref{eqn:far_field_solutions_rho_sigma}, i.e.,
\begin{equation}
\rho(0) = \frac{1}{7 E} \left( \frac{3 K M}{4 E} \right)^{2} \frac{1}{r_{0}^{7}} , \qquad
\sigma(0) = \frac{3 K M}{4 E} \frac{1}{r_{0}^{4}} ,
\end{equation}
where $r_{0} \gg 1$. The initial conditions for the other Newman--Penrose quantities can be determined using the leading-order far-field power series solutions given in \eqref{eqn:far_field_solutions_rho_sigma}--\eqref{eqn:far_field_solutions_mu_lambda}.

We test the numerical method by integrating the transport equations along a ray with initial conditions
\begin{align}
\label{eqn:photon_initial_conditions_bundle}
X(0) &= 5, & Y(0) &= 10, & D_{0}(0) &= 10^{5}, & \iota(0) = \frac{17 \pi}{18} .
\end{align}
The initial values of the spacetime coordinates and their conjugate momenta can be determined through \eqref{eqn:coordinates_image_plane_r} and \eqref{eqn:coordinates_image_plane_theta}. In particular, $r(0) = \sqrt{X(0)^{2} + Y(0)^{2} + D_{0}(0)^{2}}$. This value of $r(0)$ is then used to determine the initial data for the remaining quantities, using the far-field solutions of Section \ref{sec:far_field_asymptotics}. We also choose $A(0) = 1$.

The transport equations can be evolved along the ray until we reach (the neighbourhood of) the first caustic point at $s = s_{0}$, where the Newman--Penrose quantities diverge; see Section \ref{sec:conjugate_points}. In practice, we can never reach the caustic point at $s = s_{0}$, so we must impose a halting condition: the numerical integration is stopped at $s = s_{- \varepsilon} = s_{0} - \varepsilon$ (where $\varepsilon > 0$) when
\begin{equation}
\label{eqn:caustic_point_halting_condition}
A(s_{- \varepsilon}) = A_{- \varepsilon}, \qquad 0 < A_{- \varepsilon} \ll 1;
\end{equation}
typically, we take $A_{- \varepsilon} = 10^{-5}$. (Recall that $A(s_{0}) = 0$ at the caustic point, where neighbouring rays cross, and one of the two shape parameters $d_{\pm}$ vanishes.)

In order to calibrate the numerical method and check the solutions to the transport equations, we fit a generalised power series solution to each of the variables in the vicinity of the caustic point, and compare with the results of Section \ref{sec:conjugate_points}. Here, we briefly outline the steps involved.

The output of the numerical integration is an interpolating function for each of the (complex) scalar quantities described above (i.e., the Newman--Penrose scalars, higher-order Newman--Penrose quantities, and geometric optics quantities). We take as numerical data the interpolating function evaluated at a selection of $N$ equally spaced values of $s \in \left[ s_{- \varepsilon} - s_{\ast}, s_{-\varepsilon}  \right]$ within a distance $s_{\ast}$ of the stopping value $s_{-\varepsilon}$. Here, $s_{\ast}$ is taken to be some sufficiently small value, so that the generalised power series solution through sub-leading order is a good approximation for the true solution. (Typically, we take $s_{\ast} = 0.1$ and $N = 10^{3}$.) For a quantity which diverges like $\left( s - s_{0} \right)^{n}$ at leading order, we find a least-squares fit to the numerical data as a linear combination of the leading- and sub-leading-order functions, $\left( s - s_{0} \right)^{n}$ and $\left( s - s_{0} \right)^{n + 1}$. (In practice, this is achieved using Mathematica's \texttt{Fit} function.)

Upon inspection of plots in the neighbourhood of the caustic point, we see good agreement between the solutions to the transport equations and the least-squares fit through sub-leading order for all of the quantities under consideration.

Recall that, there are a number of free parameters in the sub-leading-order generalised power series solutions obtained in Section \ref{sec:conjugate_points}. The free parameters which arise from the leading- and sub-leading-order solutions to the transport equations for the Newman--Penrose scalars $\{ \rho, \sigma, \chi, \tau, \mu, \lambda \}$ are $\varphi \in [0, 2 \pi)$, $\rho_{1}, \left| \chi_{0} \right|, \left| \chi_{1} \right|, \left| \tau_{0} \right|, \left| \tau_{1} \right| \in \mathbb{R}$ and $\mu_{0}, \mu_{1} \in \mathbb{C}$. The cross-sectional area $A$ features one free parameter $A_{1} \in \mathbb{R}$ which appears at leading order. The square-amplitude $\mathcal{A}^{2}$ also features one free parameter $a_{0} \in \mathbb{R}$ which arises from the leading-order piece. The free parameters which appear in the higher-order Newman--Penrose quantities $\{ \delta \rho, \ol{\delta} \rho, \delta \sigma, \ol{\delta} \sigma, \delta \chi, \ol{\delta} \chi \}$ are $\alpha_{0} \in \mathbb{R}$ which comes from the leading-order part, and $\xi_{1}, \eta_{1} \in \mathbb{R}$ which arise from the sub-leading-order part. Finally, in the higher-order geometric optics quantity $\delta \mathcal{A}^{2}$, there is one free parameter $\zeta_{1} \in \mathbb{R}$ which arises at sub-leading order. For each of the quantities described above, we may extract the free parameters which appear in the coefficients of the generalised power series solutions using the least-squares fits through sub-leading order. We see consistency between the free parameters which are extracted from the numerical solutions.
%

\subsubsection{Evolving regularised transport equations through caustic points}

In order to calculate quantities along a ray beyond the first caustic point, we use the regularised transport equations of Section \ref{sec:regularisation_transport_equations}. Here, we briefly outline the numerical method, and illustrate this by evolving the regularised transport equations along a reference trajectory.

The full system of equations we evolve is given by the geodesic equations (i.e., Hamilton's equations) for the four spacetime coordinates and their conjugate momenta, as well as the set of regularised transport equations \eqref{eqn:regularised_transport_rho_sigma}--\eqref{eqn:regularised_transport_re_u1}. The right-hand sides of the regularised equations feature denominators which are singular at the caustic point. It is therefore necessary to integrate just short of the caustic point, and take a ``step'' over the caustic point, before resuming the numerical integration on the other side.

The regularised quantities are all related to the original quantities by a factor of $A$ to some power. Since we choose to set $A(0) = 1$, the initial data for the regularised quantities is the same as that for the original (divergent) quantities described above. We integrate the full system of ordinary differential equations along a central ray with initial data given by \eqref{eqn:photon_initial_conditions_bundle} until we reach the neighbourhood of the first caustic point. In practice, we integrate from $s = 0$ up to $s = s_{- \varepsilon} = s_{0} - \varepsilon$, where $A(s_{- \varepsilon}) = A_{- \varepsilon}$ for some predetermined tolerance parameter $0 < A_{-\varepsilon} \ll 1$; see the halting condition \eqref{eqn:caustic_point_halting_condition}. Performing a Taylor expansion of $A(s)$ in the neighbourhood of the stopping point $s = s_{- \varepsilon}$, we find
\begin{equation}
\label{eqn:area_taylor_series_stopping_point}
A(s) \sim A_{- \varepsilon} - 2 \wh{\rho}_{- \varepsilon} (s - s_{\varepsilon}),
\end{equation}
where $\wh{\rho}_{- \varepsilon} = \wh{\rho}(s_{-\varepsilon})$, and we have used the regularised transport equation \eqref{eqn:regularised_transport_equation_area} to replace the coefficient at $O(s - s_{\varepsilon})$. Evaluating \eqref{eqn:area_taylor_series_stopping_point} at the caustic point $s = s_{0}$, noting that $A(s_{0}) = 0$ by definition, and rearranging, we find
\begin{equation}
\label{eqn:epsilon_caustic_point_distance}
\varepsilon = \frac{A_{- \varepsilon}}{\wh{\rho}_{- \varepsilon}} .
\end{equation}
Hence, given the location of the stopping point $s_{- \varepsilon}$, one can approximate the location of the caustic point $s_{0} = s_{\varepsilon} + \varepsilon$ using information about $A(s)$ and its first derivative -- i.e., $\wh{\rho}(s)$ -- at the stopping point.

To evolve the transport equations ``through'' the caustic point, we take a step of size $2 \varepsilon$ from the stopping point to $s_{\varepsilon} = s_{0} + \varepsilon$ and restart the numerical integration of the regularised transport equations and Hamilton's equations. Beyond the caustic point, at $s = s_{\varepsilon}$, the new initial data are determined using information about the first derivative of each quantity at the stopping point. As an illustrative example, consider the evolution of $\wh{\rho}(s)$, which satisfies the transport equation $D \wh{\rho} =  A^{-1} ( \wh{\sigma} \ol{\wh{\sigma}} - \wh{\rho}^{2} )$. The value of $\wh{\rho}$ at the stopping point is $\wh{\rho}_{- \varepsilon} = \wh{\rho}(s_{- \varepsilon})$. Using the Taylor series expansion of $\wh{\rho}(s)$ about the stopping point, one may write
\begin{equation}
\label{eqn:new_id_rho_hat_1}
\wh{\rho}(s) \sim \wh{\rho}_{- \varepsilon} + \left. {D \wh{\rho}} \, \right|_{s = s_{- \varepsilon}} (s - s_{- \varepsilon}) ,
\end{equation}
where it is understood that the term $D \wh{\rho}$ is to be replaced by the right-hand side of the transport equation for $\wh{\rho}$. Evaluating \eqref{eqn:new_id_rho_hat_1} at $s = s_{\varepsilon}$, the (first-order approximation to the) new initial condition after the caustic point is
\begin{equation}
\label{eqn:new_id_rho_hat_2}
\wh{\rho}(s_{\varepsilon}) = \wh{\rho}_{- \varepsilon} + \frac{2 \varepsilon}{A_{- \varepsilon}} \left( \wh{\sigma}_{-\varepsilon} \ol{\wh{\sigma}}_{-\varepsilon} - \wh{\rho}_{-\varepsilon}^{2} \right) ,
\end{equation}
where $\wh{\sigma}_{- \varepsilon} = \wh{\sigma}(s_{- \varepsilon})$ and $\varepsilon$ is determined by \eqref{eqn:epsilon_caustic_point_distance}. The initial data for all other quantities of interest may be determined in a similar fashion. Once the new initial data are known, the ordinary differential equations may be evolved for $s \geq s_{\varepsilon}$. In practice, we carry out the numerical integration until either
$\Delta(r) = r^{2} - 2 M r + a^{2} \leq \varepsilon_{\textrm{H}} \ll 1$, or $r \geq R \gg 1$. The former case corresponds to a ray which approaches the outer event horizon; the latter case corresponds to a ray which reaches infinity. (The particular ray we consider here satisfies the second condition.)

Once the regularised quantities have been obtained by numerically solving the transport equations \eqref{eqn:regularised_transport_rho_sigma}--\eqref{eqn:regularised_transport_re_u1}, the divergent quantities can be obtained straightforwardly, e.g.~$\rho = A^{-1} \wh{\rho}$. In order to check the validity of the numerical solutions to the transport equations, one may compare them to the generalised power series solutions through sub-leading order. The latter are valid on \emph{both} sides of the caustic point (for $\left| s - s_{0} \right|$ sufficiently small), so a comparison between the two solutions permits us to check whether the evolution through the caustic point has been successful.

\begin{figure}[t]
\begin{center}
\subfigure[$\rho = A^{-1} \wh{\rho}$]{
\includegraphics[height=0.31\textwidth]{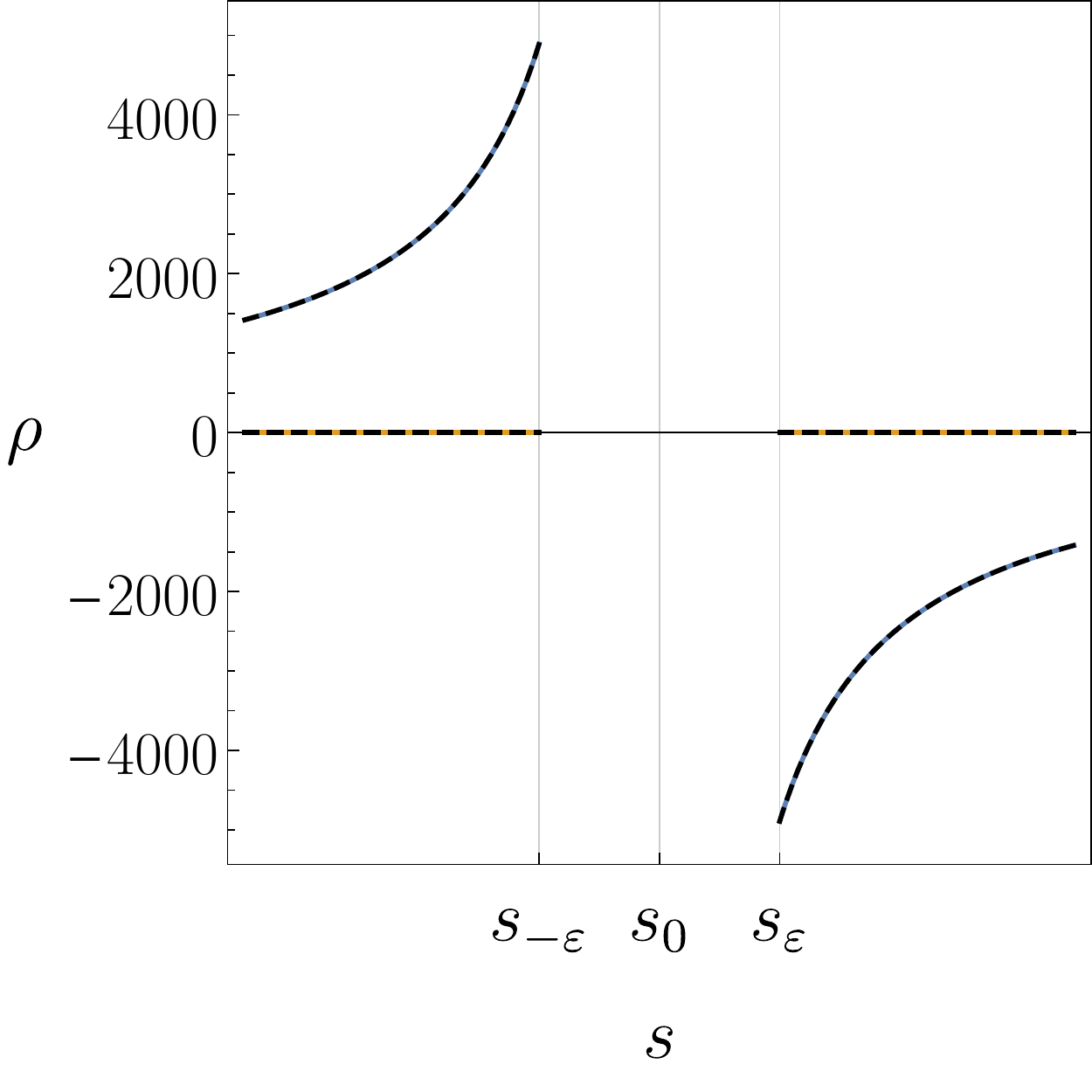} \label{fig:np_rho_caustic_fit_figure}} \hfill
\subfigure[$\sigma = A^{-1} \wh{\sigma}$]{
\includegraphics[height=0.31\textwidth]{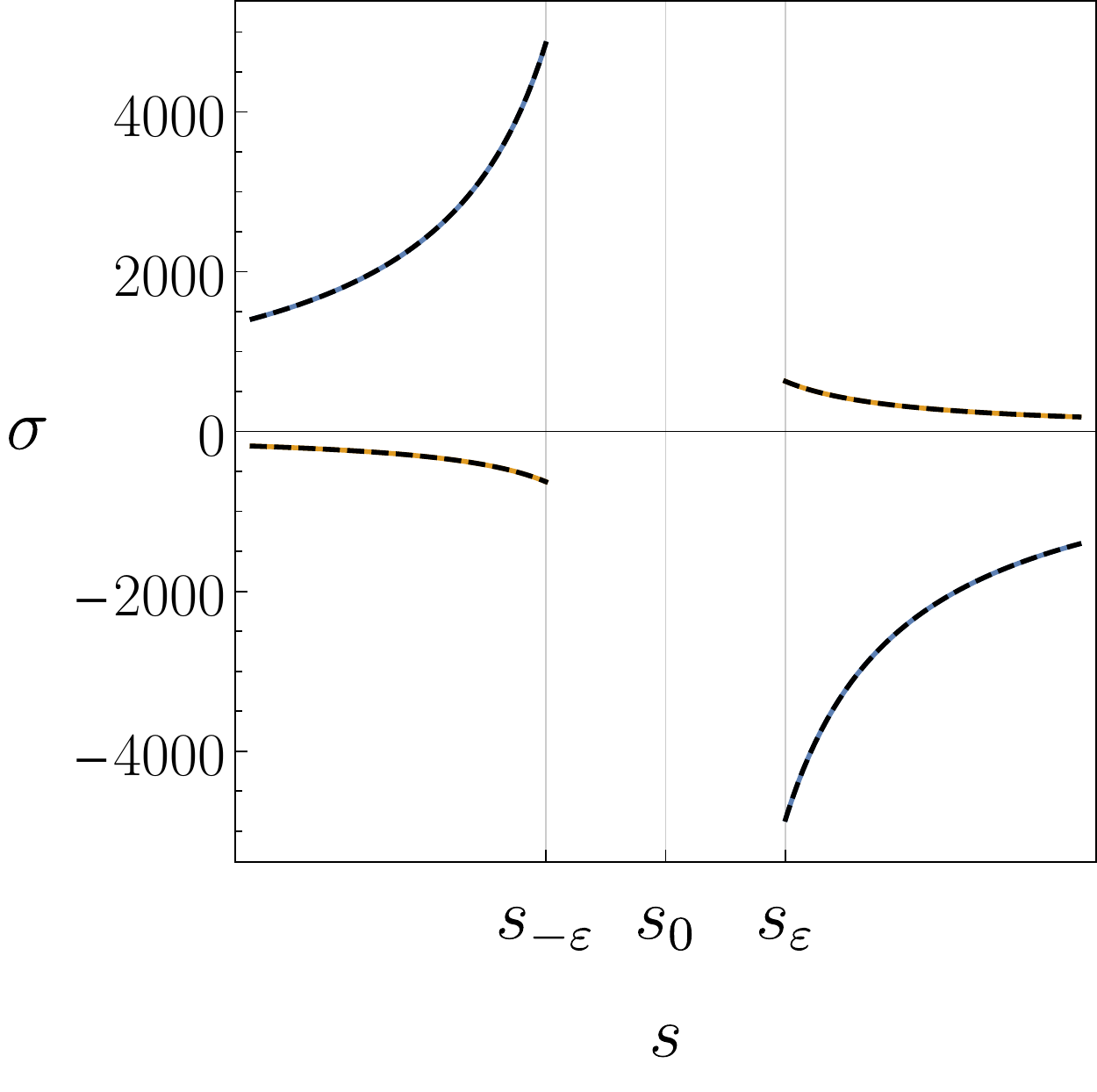} \label{fig:np_sigma_caustic_fit_figure}} \hfill
\subfigure[$\chi = A^{-1} \wh{\chi}$]{
\includegraphics[height=0.31\textwidth]{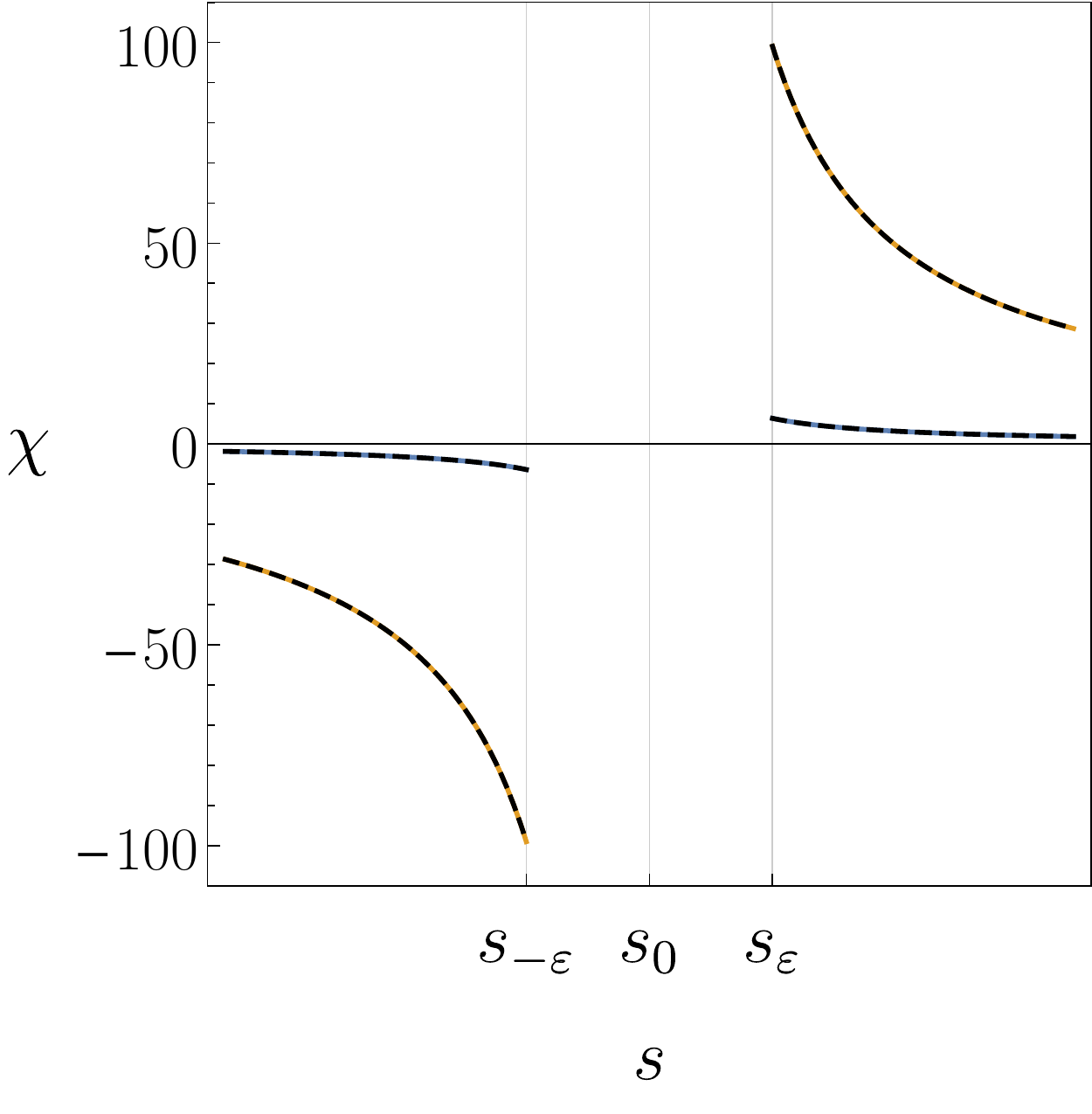} \label{fig:np_chi_caustic_fit_figure}}
\subfigure[$\tau = A^{-1} \wh{\tau}$]{
\includegraphics[height=0.31\textwidth]{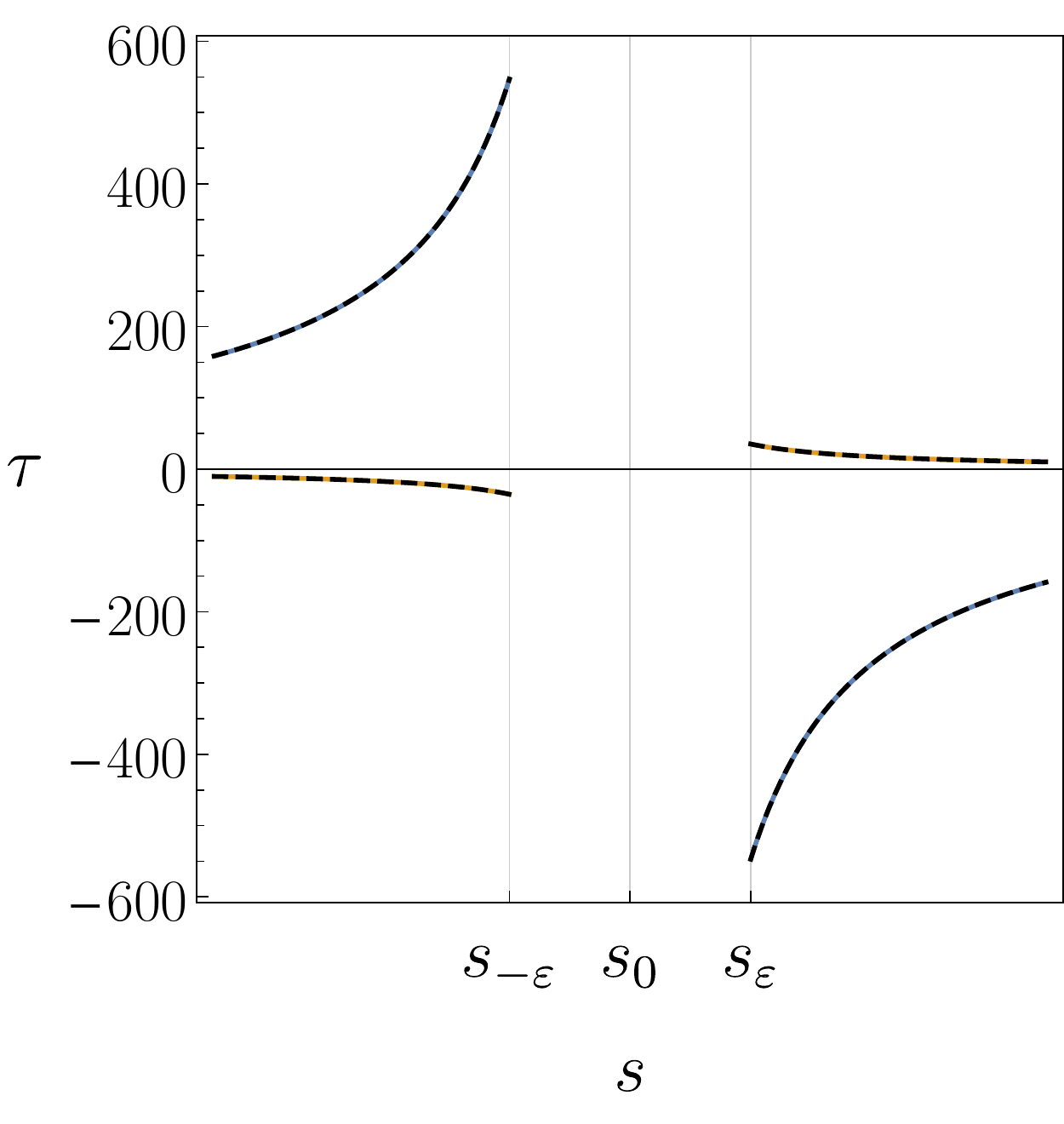} \label{fig:np_tau_caustic_fit_figure}} \hfill
\subfigure[$\mu = A^{-1} \wh{\mu}$]{
\includegraphics[height=0.31\textwidth]{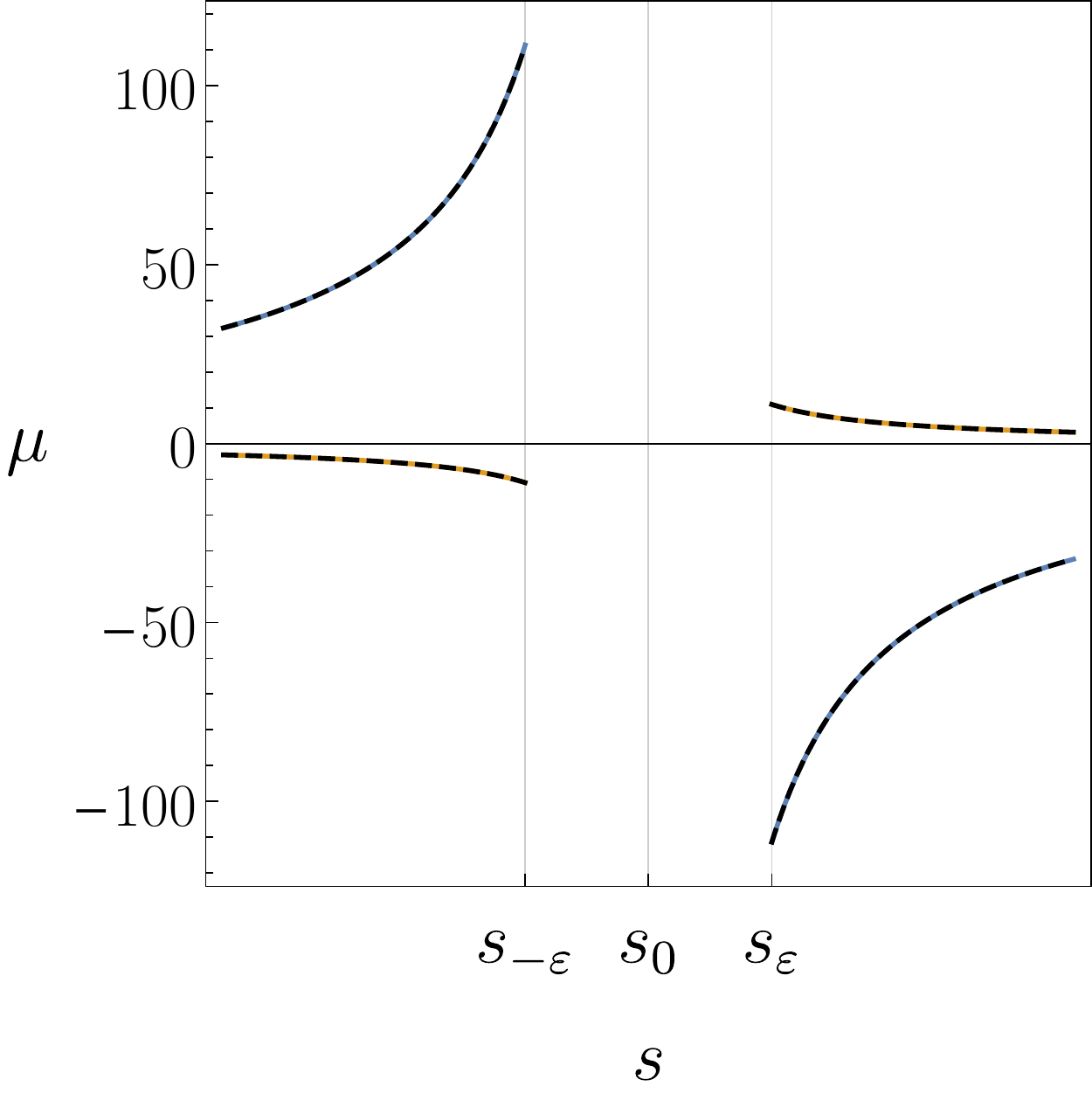} \label{fig:np_mu_caustic_fit_figure}} \hfill
\subfigure[$\lambda = A^{-1} \wh{\lambda}$]{
\includegraphics[height=0.31\textwidth]{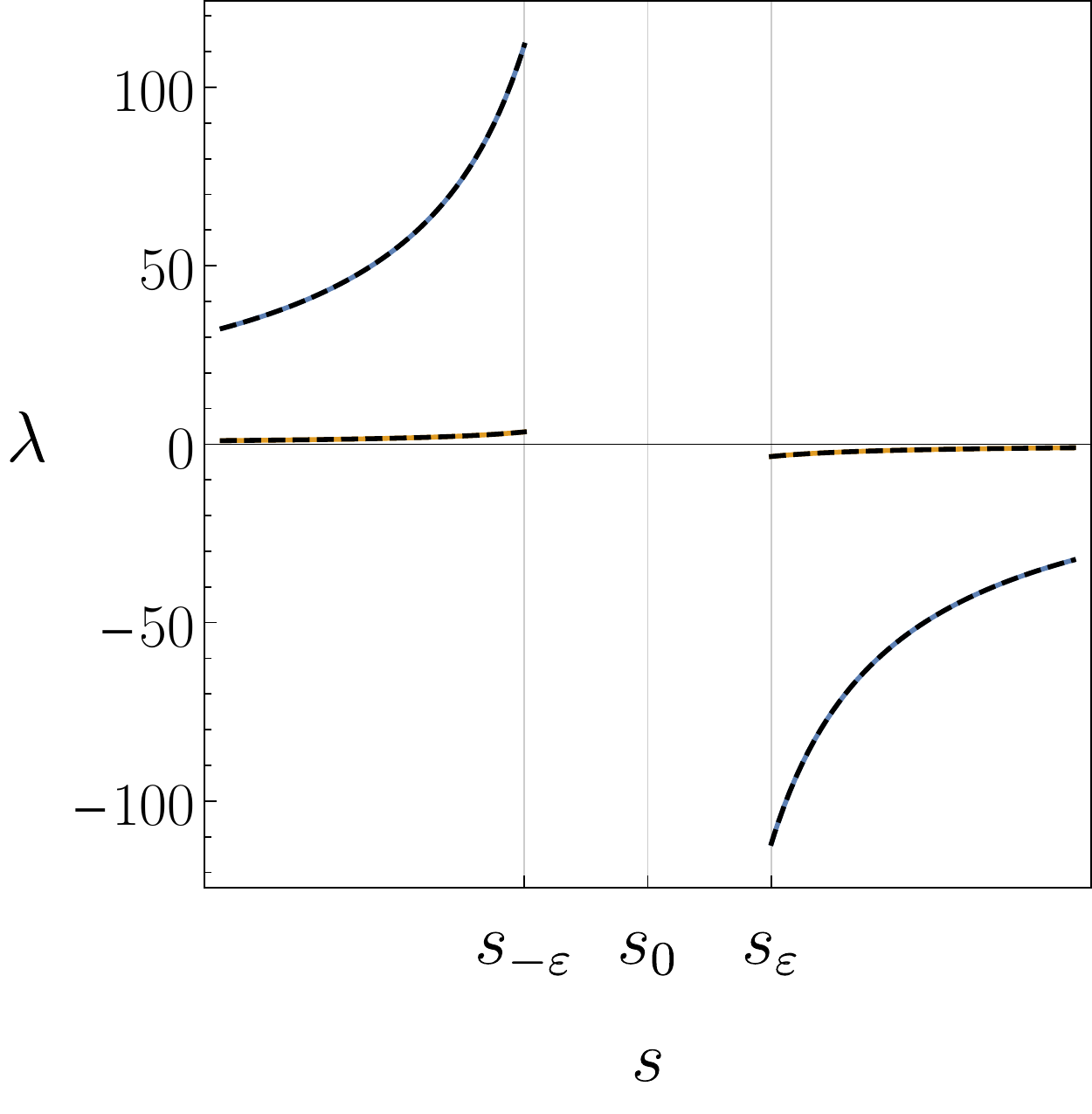} \label{fig:np_lambda_caustic_fit_figure}}
\caption{Newman--Penrose scalars in the neighbourhood of the caustic point of multiplicity one at $s = s_{0}$, where $s$ denotes the affine parameter. The real (imaginary) part of each Newman--Penrose scalar is shown in blue (orange). The generalised power series solutions through sub-leading order are overlaid as black dashed curves. We see good agreement between the interpolating functions and the generalised power series expansions through sub-leading order on both sides of the caustic point. (The absolute relative error is always less than $10^{-4}$.)}
\label{fig:np_scalars_caustic_point}
\end{center}
\end{figure}

\begin{figure}
\begin{center}
\subfigure[$\delta \rho = A^{-3} \wh{\delta \rho}$]{
\includegraphics[height=0.275\textwidth]{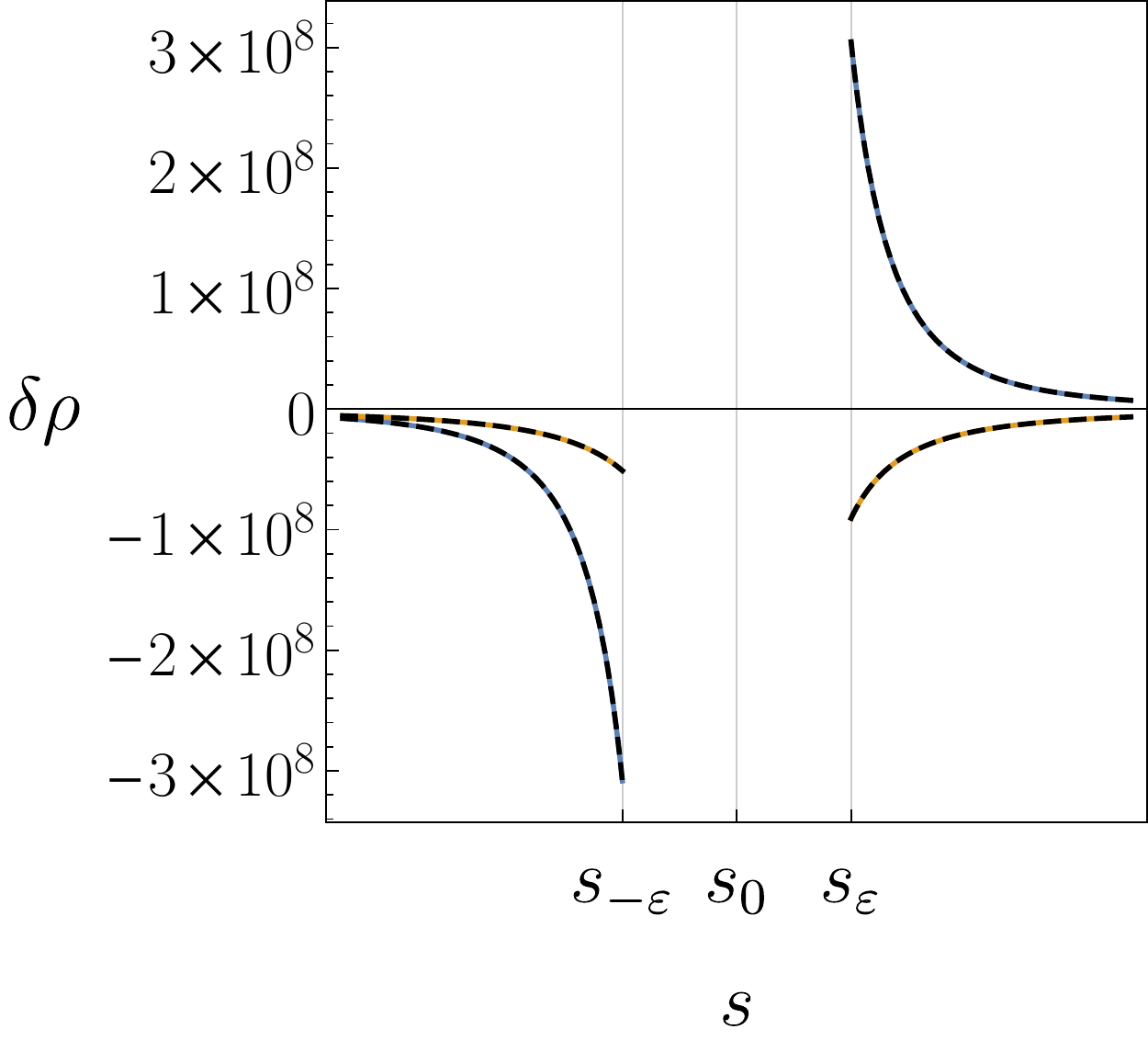} \label{fig:np_delta_rho_caustic_fit_figure}} \hfill
\subfigure[$\ol{\delta} \rho = A^{-3} \wh{\ol{\delta} \rho}$]{
\includegraphics[height=0.275\textwidth]{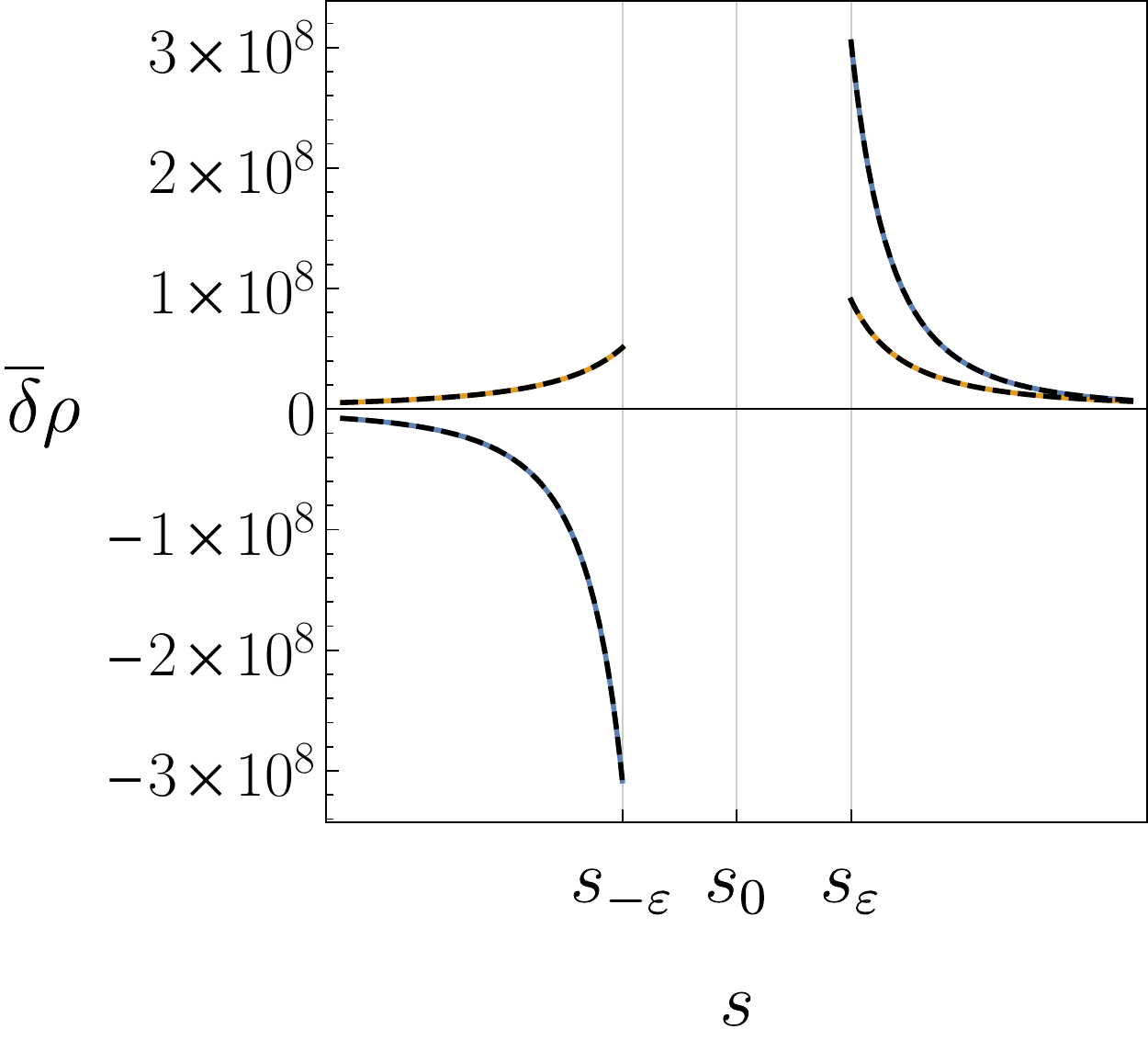} \label{fig:np_deltabar_rho_caustic_fit_figure}} \hfill
\subfigure[$\delta \sigma = A^{-3} \wh{\delta \sigma}$]{
\includegraphics[height=0.275\textwidth]{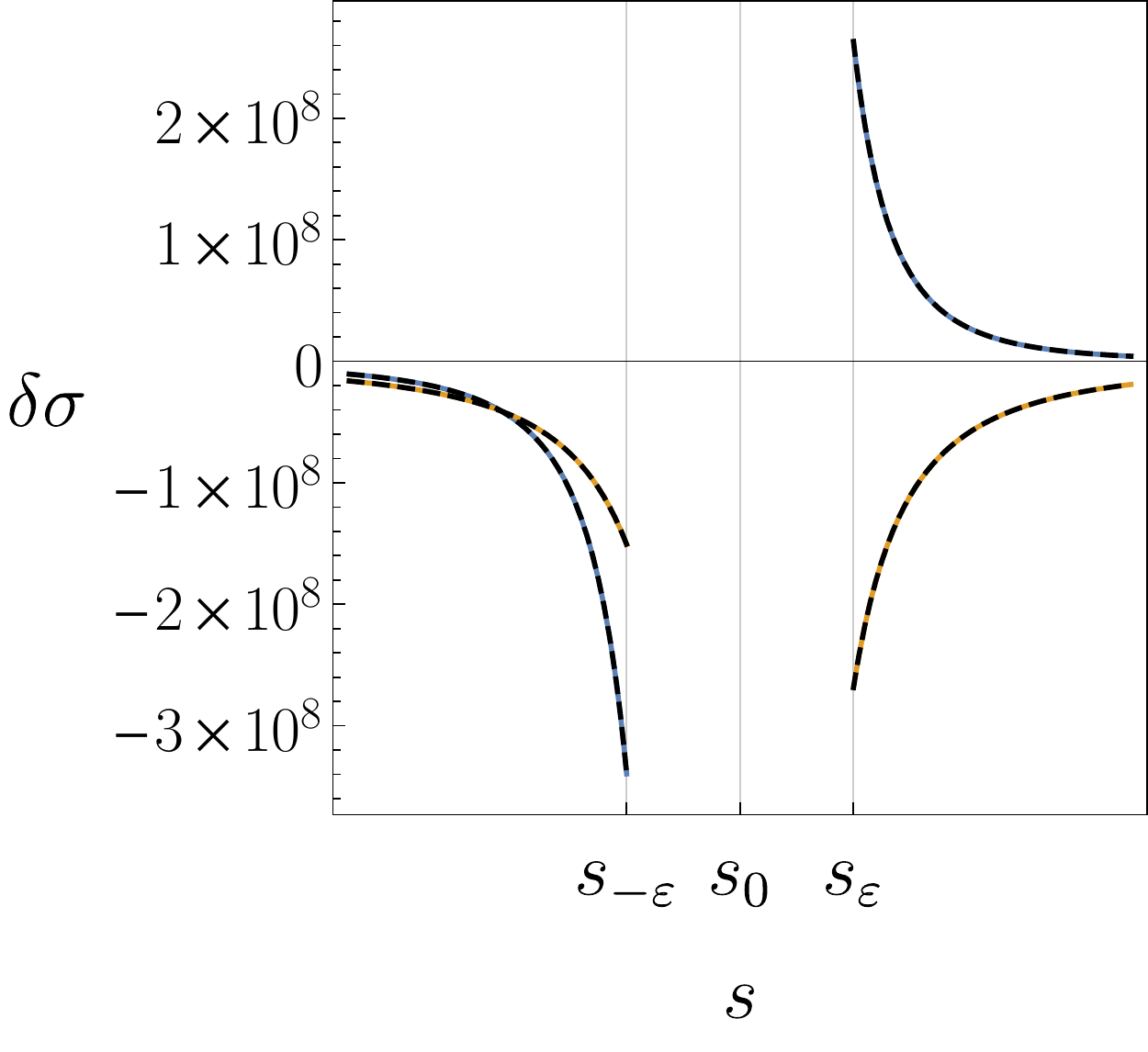} \label{fig:np_delta_sigma_caustic_fit_figure}}
\subfigure[$\ol{\delta} \sigma = A^{-3} \ol{\delta} \sigma$]{
\includegraphics[height=0.275\textwidth]{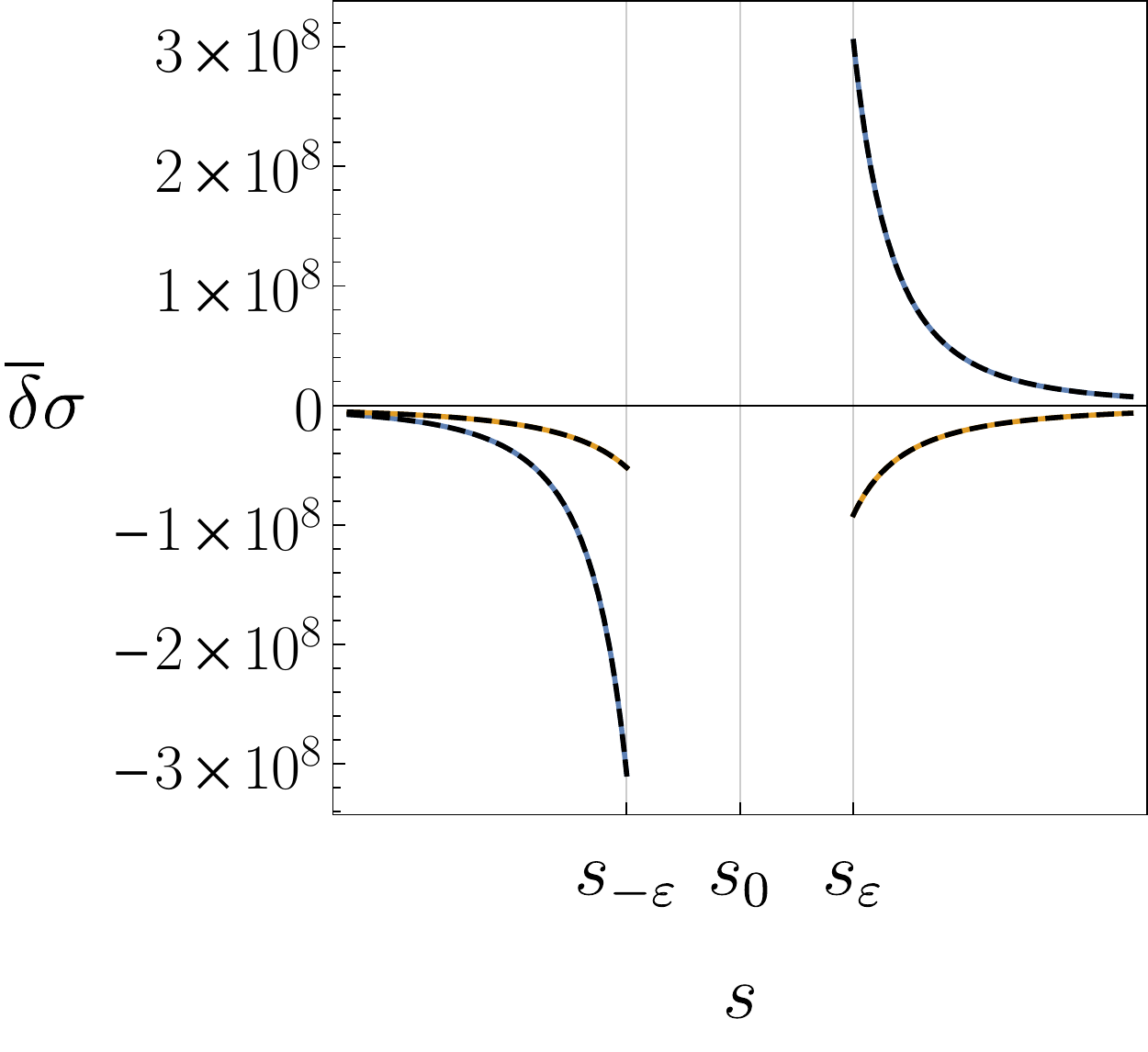} \label{fig:np_deltabar_sigma_caustic_fit_figure}} \hfill
\subfigure[$\delta \chi = A^{-3} \wh{\delta \chi}$]{
\includegraphics[height=0.275\textwidth]{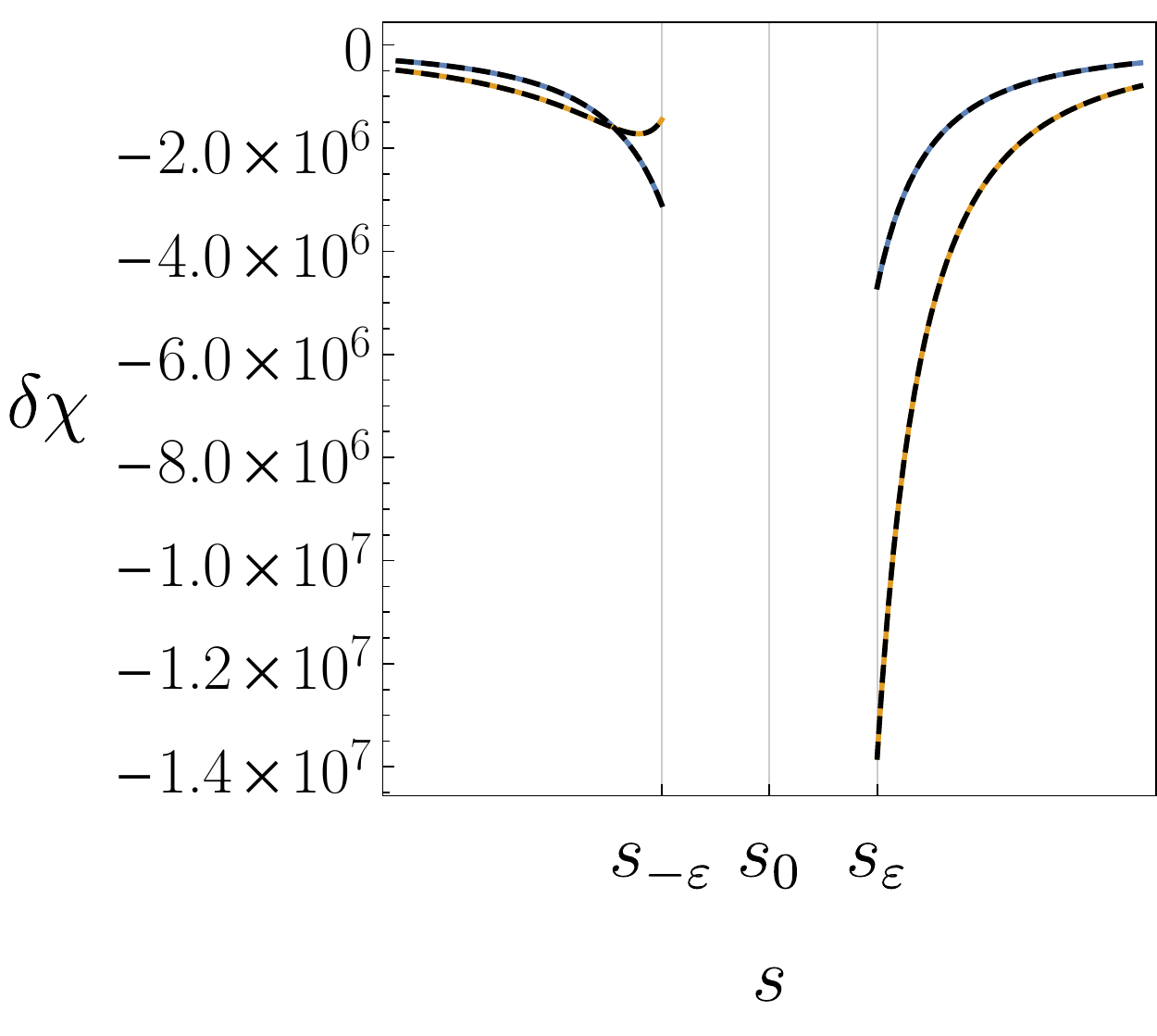} \label{fig:np_delta_chi_caustic_fit_figure}} \hfill
\subfigure[$\ol{\delta} \chi = A^{-3} \wh{\ol{\delta} \chi}$]{
\includegraphics[height=0.275\textwidth]{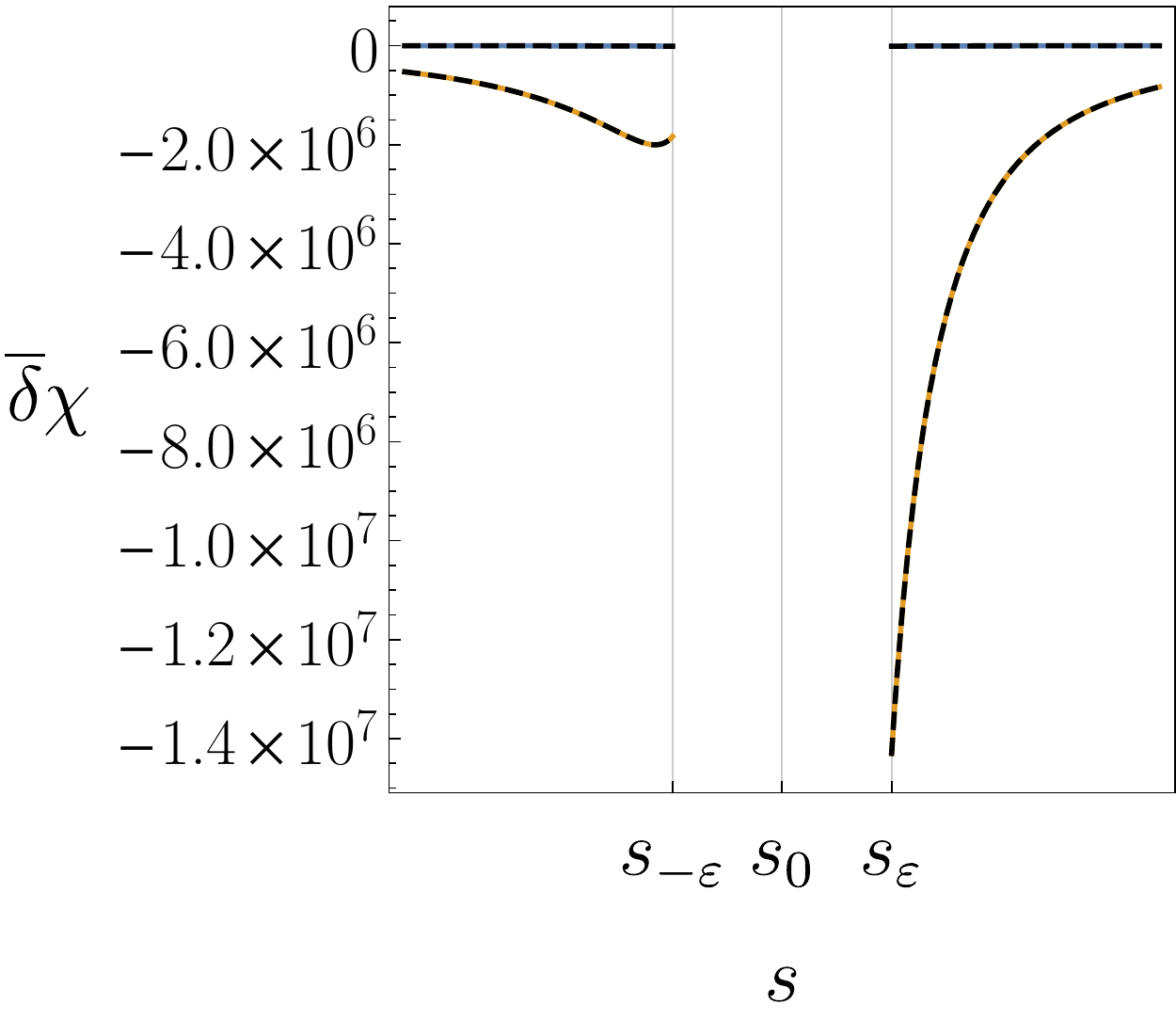} \label{fig:np_deltabar_chi_caustic_fit_figure}}
\caption{Higher-order Newman--Penrose quantities in the neighbourhood of the caustic point of multiplicity one at $s = s_{0}$. The real (imaginary) part of each quantity is shown in blue (orange). The generalised power series solutions through sub-leading order are overlaid as black dashed curves. There is good agreement between the interpolating functions and the generalised power series expansions through sub-leading order on both sides of the caustic point; the absolute relative error is always less than $10^{-4}$.}
\label{fig:np_higher_order_caustic_point}
\end{center}
\end{figure}

\begin{figure}
\begin{center}
\subfigure[$\mathcal{A}^{2} = A^{-1} \wh{\mathcal{A}}^{2}$]{
\includegraphics[height=0.275\textwidth]{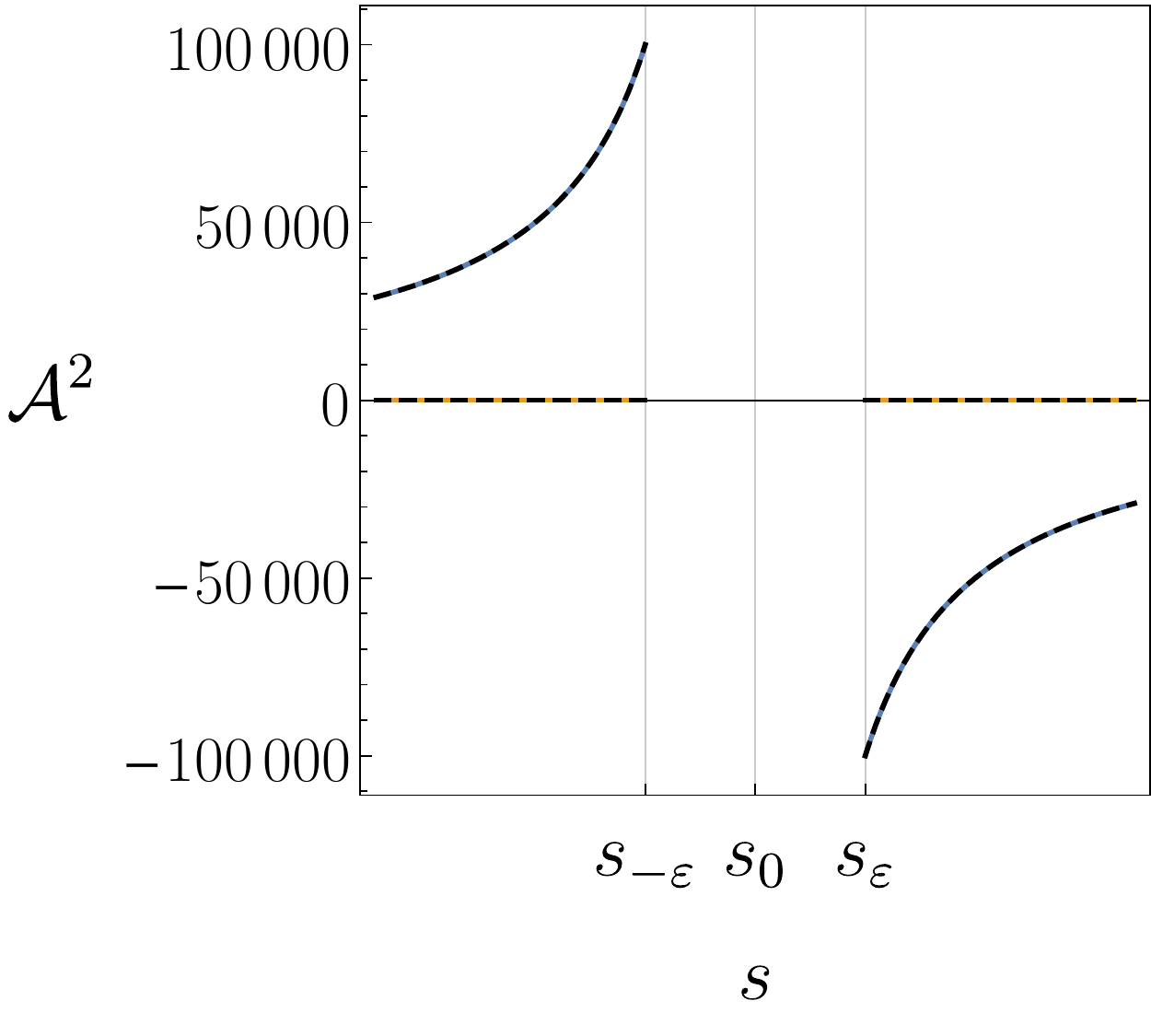} \label{fig:np_amp_sq_caustic_fit_figure}} \hfill
\subfigure[$\delta \mathcal{A}^{2} = A^{-3} \wh{ \delta \mathcal{A}^{2}}$]{
\includegraphics[height=0.275\textwidth]{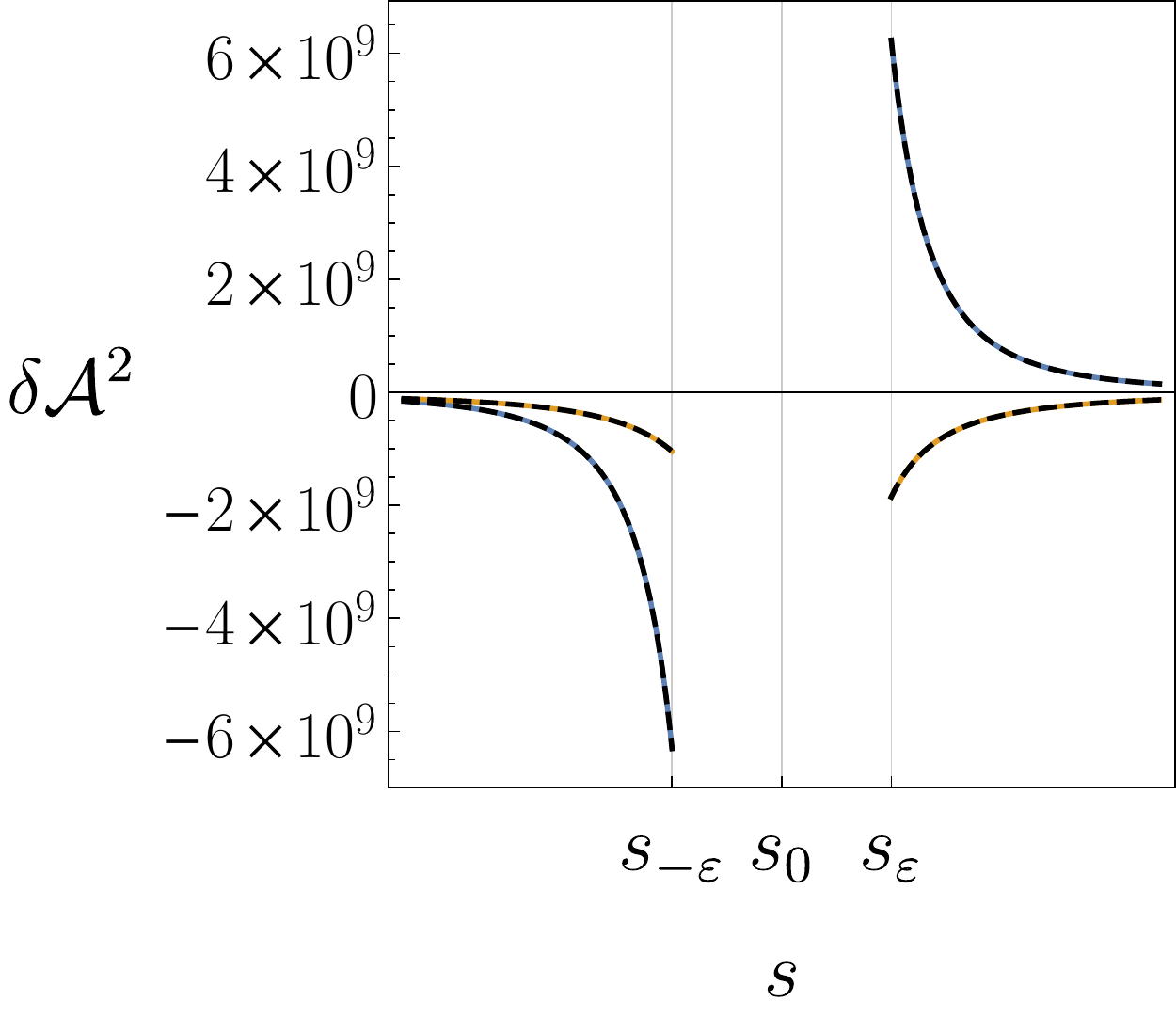} \label{fig:np_delta_amp_sq_caustic_fit_figure}} \hfill
\subfigure[$\on{Re}{(\mathfrak{u}_{1})} = A^{-2} \on{Re}{(\wh{\mathfrak{u}}_{1})}$]{
\includegraphics[height=0.275\textwidth]{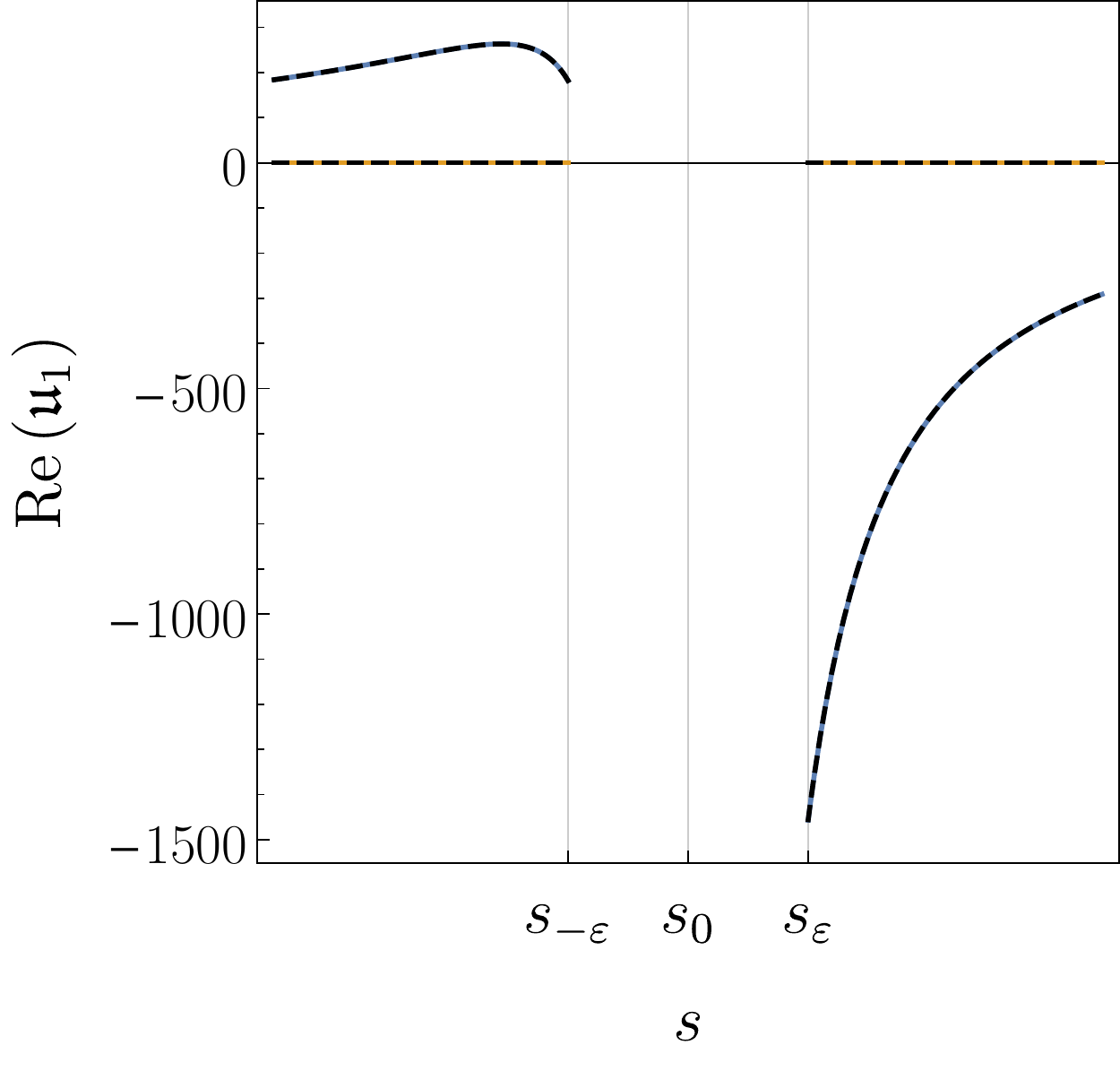} \label{fig:np_re_u1_caustic_fit_figure}}
\caption{Geometric optics quantities in the neighbourhood of the caustic point of multiplicity one at $s = s_{0}$. The real (imaginary) part of each quantity is shown in blue (orange). The generalised power series solutions through sub-leading order are overlaid as black dashed curves. We observe good agreement between the interpolating functions and the generalised power series expansions through sub-leading order on both sides of the caustic point.}
\label{fig:go_higher_order_caustic_point}
\end{center}
\end{figure}

In Figures \ref{fig:np_scalars_caustic_point}--\ref{fig:go_higher_order_caustic_point}, we present the numerical results obtained by evolving the regularised transport equations alongside the sub-leading-order generalised power series solutions in the vicinity of the caustic point. For the particular ray we consider, we set $A_{- \varepsilon} = 10^{-5}$. The value of the affine parameter when the halting condition is imposed is $s_{- \varepsilon} = 100023.582612$, and the (estimated) value of the affine parameter at the caustic is $s_{0} = 100023.582715$. Let us now describe the results of Figures \ref{fig:np_scalars_caustic_point}--\ref{fig:go_higher_order_caustic_point} in more detail.

Figure \ref{fig:np_scalars_caustic_point} shows the Newman--Penrose scalars $\{ \rho = A^{-1} \wh{\rho}, \sigma = A^{-1} \wh{\sigma}, \chi = A^{-1} \wh{\chi}, \tau = A^{-1} \wh{\chi}, \mu = A^{-1} \wh{\mu}, \lambda = A^{-1} \wh{\lambda} \}$, which are built from the regularised quantities, in the neighbourhood of the caustic point at $s = s_{0}$. The real (imaginary) part of the numerical solution to the transport equations is plotted in blue (orange). As described above, the quantities are evolved from $s = 0$ until $s = s_{-\varepsilon}$, where $A(s) = A_{- \varepsilon} \ll 1$. The location of the caustic point, $s = s_{0}$, is estimated by extrapolation of $A(s)$. The numerical integration is then restarted at $s = s_{\varepsilon} = s_{- \varepsilon} + 2 \varepsilon$ using the initial conditions described above. The generalised power series solutions \eqref{eqn:power_series_solutions_rho_sigma}--\eqref{eqn:power_series_solutions_lambda_mu} are shown as black dashed curves. We see good agreement between the numerical data and the least-squares fits of the generalised power series through sub-leading order on both sides of the caustic point: the absolute value of the relative error between the numerical solution and the least-squares fit is always less than $10^{-4}$. We note that the solutions are odd about $s = s_{0}$ because the leading-order terms at $O \left( (s - s_{0})^{-1} \right)$ are equal and opposite on either side of $s = s_{0}$.

In Figure \ref{fig:np_higher_order_caustic_point} we present numerical results for the higher-order Newman--Penrose quantities $\{ \delta \rho = A^{-3} \wh{\delta \rho}, \ol{\delta} \rho = A^{-3} \wh{\ol{\delta} \rho}, \delta \sigma = A^{-3} \wh{\delta \sigma}, \ol{\delta} \sigma = A^{-3} \ol{\delta} \sigma, \delta \chi = A^{-3} \wh{\delta \chi}, \ol{\delta} \chi = A^{-3} \wh{\ol{\delta} \chi} \}$ in the neighbourhood of the caustic point at $s = s_{0}$. Again, the real (imaginary) parts of the solutions to the regularised transport equations are shown in blue (orange). The least-squares fits of the sub-leading-order generalised power series solutions \eqref{eqn:ho_np_slo_1}--\eqref{eqn:ho_np_slo_6} are shown as black dashed curves. We again see good agreement with the numerical solutions to the transport equations; the absolute value of the relative error is always less than $10^{-4}$. We clearly see that there is an asymmetry in the solutions about $s = s_{0}$, due to the presence of competing terms at leading and sub-leading order, i.e., $O \left( (s - s_{0})^{-3} \right)$ and $O \left( (s - s_{0})^{-2} \right)$. Intriguingly, we find that, for certain quantities, the sub-leading-order terms dominate until very close to the caustic point, due to the fact that the leading-order coefficients are small in magnitude.

Figure \ref{fig:go_higher_order_caustic_point} shows the geometric optics quantities $\mathcal{A}^{2} = A^{-1} \wh{\mathcal{A}}^{2}$, $\delta \mathcal{A}^{2} = A^{-3} \wh{ \delta \mathcal{A}^{2}}$ and $\on{Re}{(\mathfrak{u}_{1})} = A^{-2} \on{Re}{(\wh{\mathfrak{u}}_{1})}$. We again see good agreement between the interpolating functions built from numerical solutions to the regularised transport equations and the sub-leading order generalised power series solutions \eqref{eqn:amp_squared_slo} and \eqref{eqn:delta_amp_squared_slo}. For each of these quantities, the absolute value of the relative error is less than $10^{-4}$ in the neighbourhood of the caustic point.

Recall that the regularisation method involves halting the numerical integrator just short of the caustic point at $s = s_{-\varepsilon}$, where $A(s_{-\varepsilon}) = A_{- \varepsilon}$. We estimate the value of the affine parameter at the caustic point ($s = s_{0}$) via extrapolation; see \eqref{eqn:area_taylor_series_stopping_point} and \eqref{eqn:epsilon_caustic_point_distance}. Dolan \cite{Dolan2017, Dolan2018} demonstrated that certain Newman--Penrose quantities (e.g.~$\rho$ and $\sigma$) may be calculated beyond caustic points using a second-order formalism; see Section \ref{sec:geometric_optics_review} for a review. The complex scalar quantity $c$, which corresponds to a point in the instantaneous geometric optics wavefront, satisfies the second-order equation \eqref{eqn:c_double_dot} which is well-behaved at caustic points. We recall that the shape parameters of the bundle satisfy $d_{+} d_{-} = 2 \left| \on{Im}{\left( c_{1} \ol{c}_{2} \right)} \right|$, where $c_{1}$ and $c_{2}$ are any pair of linearly independent solutions to \eqref{eqn:c_double_dot}. At a caustic point, one of the shape parameters $d_{\pm}$ vanishes. One may therefore calculate the value of the affine parameter at the caustic point independently by solving the equation $\on{Im}{\left( c_{1} \ol{c}_{2} \right)} = 0$ for $s$. Doing this for our reference ray, we find that the absolute value of the relative error between the value estimated via extrapolation of the regularised transport equations and the value obtained using the second-order equation \eqref{eqn:c_double_dot} is approximately $10^{-11}$.

\subsubsection{Sensitivity to internal parameters}

Let us now quantify the sensitivity of the numerical results to changes in the numerical method's internal parameters, such as the caustic-handling parameter $A_{- \varepsilon}$. To do this, we twice solve the full system of transport equations subject to the initial data described above, comparing the results obtained for $A_{- \varepsilon}^{(1)} = 10^{-5}$ and $A_{- \varepsilon}^{(2)} = 2 \times 10^{-5}$.

We find that the Newman--Penrose scalars are insensitive to changes in the caustic-handling parameter $A_{- \varepsilon}$ beyond the caustic point. The absolute value of the relative error between the two runs is less than approximately $10^{-5}$ for all regularised Newman--Penrose scalars (e.g.~$\wh{\rho}$ and $\wh{\sigma}$) beyond the caustic point. We can therefore be confident that the regularisation method and its numerical implementation have allowed us to calculate the Newman--Penrose scalars beyond the caustic point, far from the black hole.

In addition, we observe good agreement for the higher-order quantities, such as $\wh{\delta \rho}$, $\wh{\delta \chi}$ and $\on{Re}{(\wh{\mathfrak{u}}_{1})}$, in the neighbourhood of the caustic point: within a distance $s^{\prime} = s - s_{0} \sim 0.005$ of the caustic point, the absolute value of the relative error between the two runs is $O(10^{-4})$ or smaller. However, we find that the higher-order quantities become highly sensitive to changes in the caustic-handling parameter downstream from the caustic point (i.e., as $s^{\prime}$ is increased). A consequence is that we cannot be confident in the numerical results for the higher-order quantities far from the black hole. For example, when $A_{- \varepsilon} = A_{- \varepsilon}^{(1)}$, we find $\lim_{s \rightarrow \infty} \on{Re}{(\mathfrak{u}_{1})} \sim 200$; however, when $A_{- \varepsilon} = A_{- \varepsilon}^{(2)}$, we obtain $\lim_{s \rightarrow \infty} \on{Re}{(\mathfrak{u}_{1})} \sim -30$.

\begin{figure}
\begin{center}
\subfigure[Neighbourhood of the caustic point]{
\includegraphics[height=0.38\textwidth]{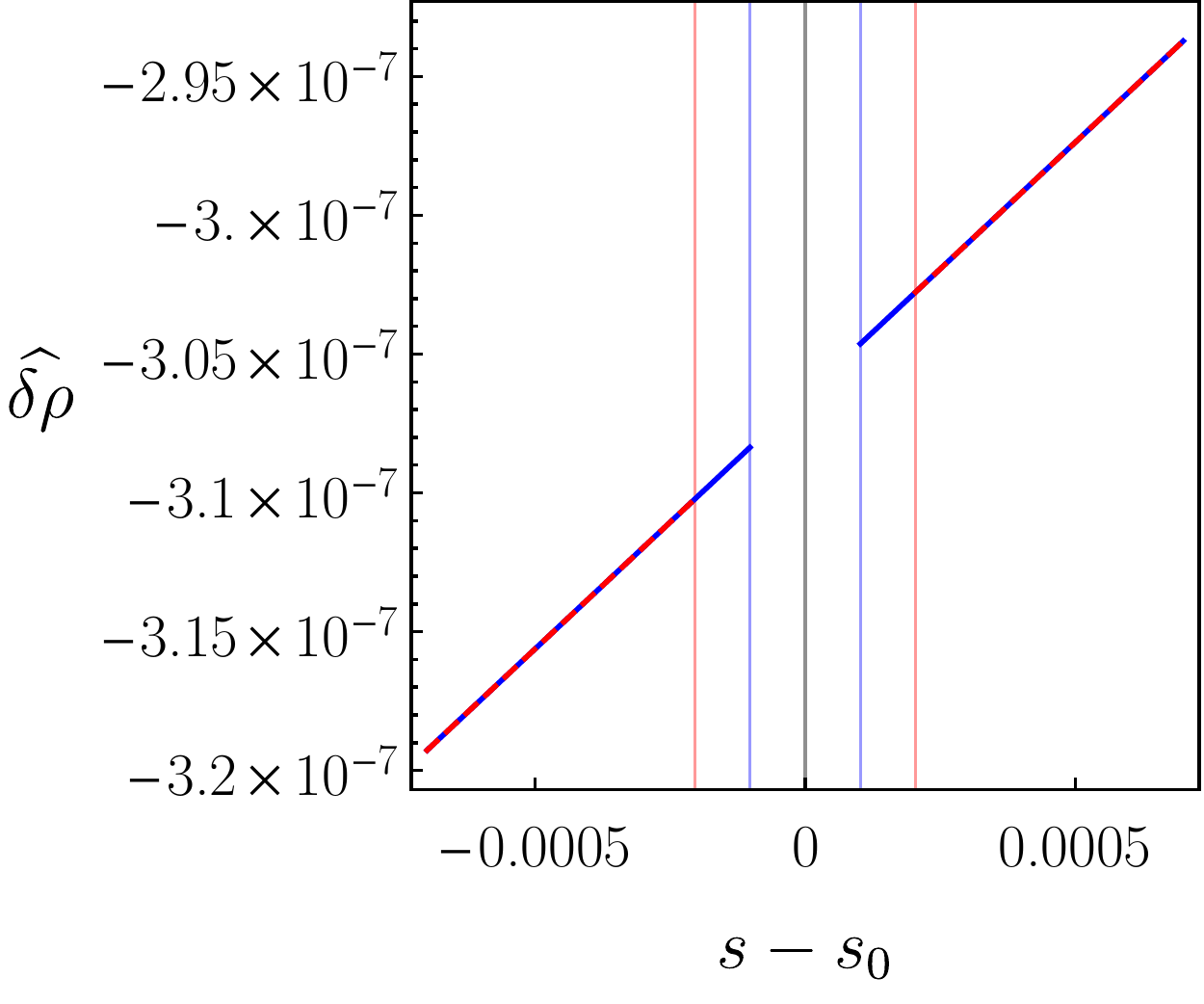} \label{fig:delta_rho_sensitivity}}
\subfigure[Beyond the caustic point]{
\includegraphics[height=0.38\textwidth]{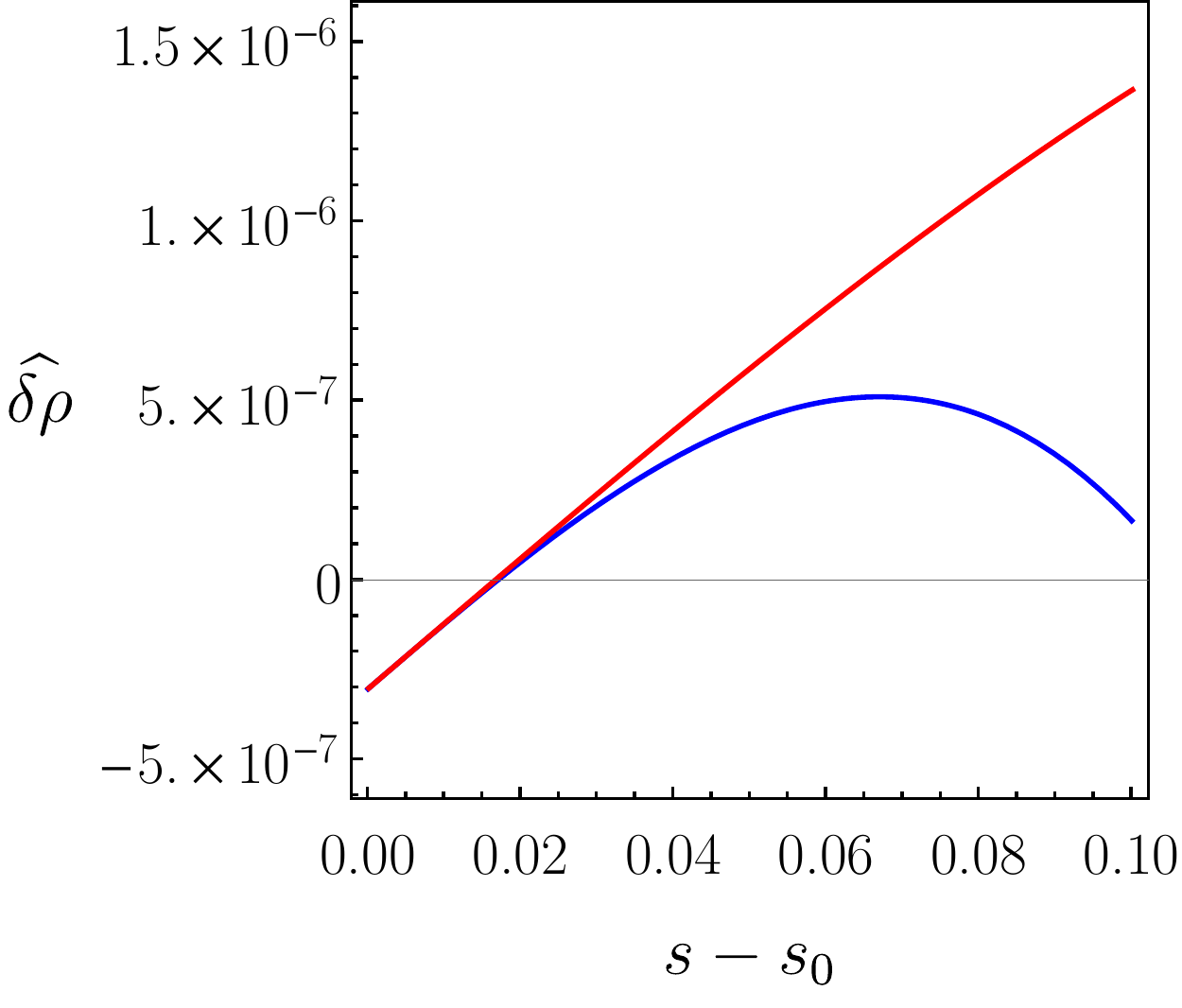} \label{fig:delta_rho_disagreement}}
\caption{Numerical results for $\on{Re}{ ( \wh{\delta \rho} ) }$ for two runs using $A_{- \varepsilon}^{(1)} = 10^{-5}$ [blue] and $A_{- \varepsilon}^{(2)} = 2 \times 10^{-5}$ [red]. The horizontal axis is rescaled, so that the caustic point occurs at $s^{\prime} = s - s_{0} = 0$. (a) Numerical solutions for the two runs in the vicinity of the caustic point. We observe good agreement between the curves obtained from the first [blue] and second [red, dashed] runs. (b) Disagreement between the two solutions beyond the caustic point. The curves begin to diverge for $s^{\prime} = s - s_{0} \sim 0.02$.}
\label{fig:sensitivity_figure}
\end{center}
\end{figure}

Figure \ref{fig:sensitivity_figure} shows the numerical results for $\on{Re}{ ( \wh{\delta \rho} ) }$ -- chosen as an example of a typical (regularised) higher-order Newman--Penrose quantity -- with $A_{- \varepsilon} = A_{- \varepsilon}^{(1)}$ [blue] and $A_{- \varepsilon} = A_{- \varepsilon}^{(2)}$ [red]. In both figures, the horizontal axis has been rescaled ($s \mapsto s^{\prime} = s - s_{0}$) so that the caustic point occurs at $s^{\prime} = 0$. In Figure \ref{fig:delta_rho_sensitivity}, we compare the results in the vicinity of the caustic point. For the first run, the numerical integration is halted when $A = A_{- \varepsilon}^{(1)}$; the corresponding value of $s_{-\varepsilon}^{(1)}$ is shown as a blue vertical line. The value $s_{0}^{(1)} = s_{-\varepsilon}^{(1)} + \varepsilon^{(1)}$, which is estimated via extrapolation using \eqref{eqn:epsilon_caustic_point_distance}, is shown as a vertical grey line. The numerical integration is resumed at $s_{\varepsilon}^{(1)} = s_{0}^{(1)} + \varepsilon$, shown using a second vertical blue line. The second run is halted when $A = A_{- \varepsilon}^{(2)}$. The corresponding values of $s_{\pm \varepsilon}^{(2)}$ are shown as red vertical lines. The estimated value of the affine parameter at the caustic point, denoted $s_{0}^{(2)}$, is \emph{not} identical to the value obtained from the first run; however, the absolute difference between these two quantities is $O(10^{-10})$. The numerical results for $\on{Re}{ ( \wh{\delta \rho} ) }$ are in good agreement on both sides of the caustic point: on the domain shown in Figure \ref{fig:delta_rho_sensitivity}, the absolute value of the relative error between the numerical data obtained from the two runs is less than $O(10^{-5})$. Figure \ref{fig:delta_rho_disagreement} shows the numerical results for $\on{Re}{ ( \wh{\delta \rho} ) }$ beyond the caustic point (which is again located at $s^{\prime} = 0$). One can clearly see that for $s^{\prime} \sim 0.02$, the curves begin to diverge from one another. We observe similar behaviour for all other higher-order quantities. This demonstrates that the numerical solutions to the transport equations beyond the caustic point are highly sensitive to small changes in the caustic-handling parameter $A_{- \varepsilon}$. In Section \ref{sec:extensions_go_kerr}, we comment on potential ways that the issues described here could be alleviated.

\section{Discussion}
\label{sec:discussion_go_kerr}

\subsection{Extensions}
\label{sec:extensions_go_kerr}

\subsubsection{Construction of parallel-transported complex null tetrads}

In Section \ref{sec:null_tetrads_from_symmetries}, we constructed a complex null tetrad by availing the properties of the principal tensor $h\ind{_{a b}}$ and its Hodge dual $f\ind{_{a b}}$. The construction of the parallel-transported complex null tetrad should therefore apply to any geometry which admits a principal tensor, i.e., the (off-shell) Kerr--NUT--(anti-)de Sitter spacetime. The parallel-transported complex null tetrad constructed here can be of practical use in other scenarios beyond gravitational lensing by Kerr black holes.

We recall that the construction of the complex null tetrads of Section \ref{sec:null_tetrads_from_symmetries} breaks down for rays with $K = 0$. This is due to the fact that the tangent vector $k\ind{^{a}}$ and the vector $f\ind{^{a}_{b}} k\ind{^{b}}$ are collinear in this degenerate case. One could extend the work of Section \ref{sec:null_tetrads_from_symmetries} by paying special attention to the construction of a parallel-propagated complex null tetrad along null geodesics with vanishing Carter constant.

\subsubsection{Modified geometric optics, transport equations and caustic points}

The results of Figure \ref{fig:sensitivity_figure} demonstrate that the numerical solutions to the transport equations of modified geometric optics are sensitive to small changes in the caustic-handling parameter $A_{- \varepsilon}$ beyond the caustic point. This hints at a potential flaw in either (i) the modified geometric optics formalism (Section \ref{sec:go_recap_and_ho_formalism}); (ii) its implementation in terms of first-order transport equations (Section \ref{sec:np_formalism_and_transport_equations}); or (iii) the numerical method used to evolve transport equations through caustic points (Section \ref{sec:transport_equations_cautics}).

Let us first consider point (i). The geometric optics approximation relies on the assumption that the wavelength is small in comparison to other physically relevant length scales (e.g.~spacetime curvature scales). This assumption breaks down at caustic points. The geometric optics approximation becomes valid again after passing through a caustic point, when the rays in the bundle are sufficiently separated. However, the passage through the caustic point gives rise to a change in sign of the Jacobian of the lensing map \cite{Roemer2005, ZenginogluGalley2012}. The (complex) geometric optics amplitude will therefore change phase by $\pm \frac{\pi}{2}$ through a caustic, becoming purely imaginary ($\mathcal{A} \mapsto \pm i \mathcal{A}$). Zengino\u{g}lu and Galley \cite{ZenginogluGalley2012} remark that the sign of the phase shift cannot be computed within the geometric optics approximation; instead, this must be determined by matching to the full solution from numerical simulations, or by some alternative approach. It is clear that a better understanding of caustic points and the passage through them is required in order to overcome the issues with the modified geometric optics formalism.

Now consider point (ii) from the above list. The leading-order geometric optics formalism (Section \ref{sec:leading_order_go_recap}) and its sub-leading-order modification (Section \ref{sec:higher_order_go}) are formulated in terms of first-order transport equations for quantities which diverge at caustic points. This issue could perhaps be alleviated by adopting an alternative approach. Recall that a subset of the Newman--Penrose scalars can be obtained beyond caustic points by solving a system of first- and second-order ordinary differential equations along rays; see Section \ref{sec:geometric_optics_review} and \cite{Dolan2017, Dolan2018}. The quantities which satisfy these differential equations do not diverge, and can be evolved through caustic points without issue. It is not clear whether there exists a similar system of equations satisfied by well-behaved quantities from which higher-order Newman--Penrose quantities (e.g.~$\delta \rho$) and geometric optics quantities (e.g.~$\mathfrak{u}_{1}$) may be obtained.

Finally, let us consider point (iii) from the above list. There are a number of ways that the accuracy of the numerical method presented in Section \ref{sec:regularisation_numerical_method_and_results} could be improved. First, recall that we use extrapolation to step over the caustic point, based on a first-order Taylor expansion of the dependent variables; see \eqref{eqn:area_taylor_series_stopping_point} and \eqref{eqn:new_id_rho_hat_1}, for example. For improved accuracy, one could extend this Taylor expansion to higher orders; this would involve taking further derivatives of the dependent variables in order to work out the new initial data after the caustic point. Alternatively, one could use a ``midpoint method'' to calculate the initial data at $s = s_{\varepsilon}$, taking an average of the values obtained by performing (first-order) Taylor series expansions about $s = s_{- \varepsilon}$ and $s = s_{\varepsilon}$. Second, recall that we obtained generalised power series solutions through sub-leading order for all of the Newman--Penrose scalars and higher-order Newman--Penrose quantities in the vicinity of a caustic point of multiplicity one (Section \ref{sec:conjugate_points}). These generalised power series solutions are valid on both sides of the caustic point; we therefore used them as a consistency check of the numerical solution obtained by regularisation beyond the caustic point; see Figures \ref{fig:np_scalars_caustic_point}--\ref{fig:go_higher_order_caustic_point}. In addition to employing these solutions as a consistency check, one could use them to perform a more sophisticated evolution of the divergent quantities through the caustic point, by performing ``matched asymptotic expansions'' in the neighbourhood of the caustic point.

In a recent paper, Dolan \cite{Dolan2019} presented a method to calculate the vector potential for electromagnetic waves on Kerr spacetime in Lorenz gauge. To further test the modified geometric optics formalism, one could perform a direct comparison between the results obtained from solving wave equations in Lorenz gauge, and those obtained by evolving systems of transport equations (considered here).

\subsection{Conclusions}

In this chapter, we have employed the higher-order geometric optics formalism of Dolan \cite{Dolan2018} (see Section \ref{sec:go_recap_and_ho_formalism}) to understand aspects of gravitational lensing by Kerr black holes in general relativity. Our principal aim was to calculate the $O(\omega^{-1})$ correction to the electromagnetic stress--energy tensor \eqref{eqn:subleading_stess_energy}, which involves the quantities $\mathfrak{u}_{1}$, $\mathfrak{v}_{1}$ and $\mathfrak{w}_{1}$. In Section IV of \cite{Dolan2018}, Dolan enumerates several practical hurdles which one faces when applying the higher-order geometric optics formalism, including: (1) finding a suitable parallel-transported complex null tetrad; (2) calculating key quantities, such as the Weyl curvature scalars and their derivatives; (3) solving transport equations numerically or otherwise; and (4) handling ray-crossings and caustic points.

For Kerr spacetime (reviewed in Section \ref{sec:go_kerr_spacetime_recap}), we used the complex bivector \eqref{eqn:complex_bivector} to construct a complex null tetrad $\{ k\ind{^{a}}, \wt{n}\ind{^{a}}, \wt{m}\ind{^{a}}, \ol{\wt{m}}\vp{m}\ind{^{a}} \}$ along null geodesics by projecting along the tangent vector $k\ind{^{a}}$. Performing a Lorentz transformation \eqref{eqn:parallel_transport_lorentz_transformation} with complex rotation parameter $B$ given by \eqref{eqn:null_rotation_parameter}, we mapped the ``tilded'' tetrad onto a complex null tetrad $\{ k\ind{^{a}}, n\ind{^{a}}, m\ind{^{a}}, \ol{m}\vp{m}\ind{^{a}} \}$ which is parallel-transported along the null geodesics of Kerr spacetime. Thus we have overcome hurdle (1) from the above list. Intriguingly, we were able to arrive at this tetrad by defining a new operator \eqref{eqn:projection_operator_parallel_transport} and projecting along $k\ind{^{a}}$, bypassing the ``tilded'' tetrad altogether. Moreover, we demonstrated that the parallel-propagated tetrad constructed here is equivalent to a complex null tetrad constructed from the legs of Marck's (real) parallel-transported tetrad \cite{Marck1983b}.

In order to calculate the quantities $\mathfrak{u}_{1}$, $\mathfrak{v}_{1}$ and $\mathfrak{w}_{1}$ in the sub-leading-order correction to the electromagnetic stress--energy tensor, it is necessary to evolve a system of transport equations for Newman--Penrose quantities along rays (see Section \ref{sec:transport_equations_for_np_quantities}). These transport equations feature the complex Weyl scalars $\Psi_{i}$ and their derivatives along the wavefront (e.g.~$\delta \Psi_{0}$). The Weyl scalars take a simple form in the ``tilded'' tetrad, and can be transformed to the parallel-transported tetrad straightforwardly by means of the Lorentz transformation \eqref{eqn:parallel_transport_lorentz_transformation}; see Section \ref{sec:go_kerr_weyl}. The calculation of the quantities $\{ \delta \Psi_{0}, \ol{\delta} \Psi_{0}, \delta \Psi_{1}, \ol{\delta} \Psi_{1} \}$ is more involved; however, progress can be made by transforming to the symmetrised Kinnersley tetrad. Explicit expressions for these quantities are given for the parallel-propagated tetrad in Appendix \ref{chap:appendix_d}. Hence, we have successfully calculated closed-form expressions for the key quantities described in point (2) of the above list. In Section \ref{sec:far_field_asymptotics}, we analysed the far-field behaviour of these quantities, and used the results to understand the asymptotic behaviour of the Newman--Penrose scalars for $r \rightarrow \infty$, which is useful in gravitational lensing calculations.

Points (3) and (4) from the list pose more of an issue. In principle, the relevant transport equations (Section \ref{sec:transport_equations_for_np_quantities}) can be calculated numerically along a ray. However, on Kerr spacetime (and indeed in other situations), caustic points arise generically. The geometric optics formalism breaks down at caustic points, and quantities like $\rho$ (expansion) and $\sigma$ (shear) diverge there. In Sections \ref{sec:np_scalars_caustic_point} and \ref{sec:caustic_higher_order_np_quantities}, we used the full system of Newman--Penrose transport equations (Section \ref{sec:transport_equations_for_np_quantities}) to deduce the behaviour of Newman--Penrose scalars and higher-order geometric optics quantities in the neighbourhood of a caustic point of multiplicity one. In particular, we obtained these quantities as generalised power series in $(s - s_{0})$ through sub-leading order, where $s = s_{0}$ is the value of the affine parameter at the caustic point, as measured along the central ray of the bundle. The results of Section \ref{sec:np_scalars_caustic_point} are general, in that they are valid for any Ricci-flat manifold.

In Section \ref{sec:transport_equations_cautics}, we presented a practical method to evolve the full system of Newman--Penrose transport equations through caustic points. The method involves regularising divergent quantities at caustic points: a complex scalar quantity $z$ which scales like $(s - s_{0})^{n}$ at a caustic point is regularised by taking $\wh{z} = A^{-n} z$. The regularised variables are finite at caustic points. However, the transport equations feature denominators which are singular at the caustic point. We therefore integrate the transport equations just short of the caustic point, then use a first-order Taylor expansion of the regularised quantities to ``step over'' the caustic point. The integration is then resumed on the other side of the caustic point. We presented some preliminary numerical results for a reference trajectory on Kerr spacetime (with $M = 1$, $a = 0.9$). In Figures \ref{fig:np_scalars_caustic_point}--\ref{fig:go_higher_order_caustic_point}, we observe good agreement between the numerical solution to the transport equations and a least-squares fit of the sub-leading-order generalised power series solutions in the neighbourhood of the caustic point. Moreover, we find that the regularised Newman--Penrose scalars (e.g.~$\wh{\rho}$ and $\wh{\sigma}$) are insensitive to changes in caustic-handling parameter $A_{- \varepsilon}$ downstream of the caustic point. This suggests that the results obtained for the Newman--Penrose scalars are valid far from the black hole, after scattering has occurred. However, we find that the higher-order quantities (e.g.~$\wh{\delta \rho}$ and $\on{Re}{ ( \wh{ \mathfrak{u} }_{1} ) }$) are sensitive to small changes in $A_{- \varepsilon}$ beyond the caustic point, as shown in Figure \ref{fig:sensitivity_figure}. We comment on this issue in Section \ref{sec:extensions_go_kerr}.

%


\begin{appendices}

\addtocontents{toc}{\protect\setcounter{tocdepth}{1}}
\makeatletter
\addtocontents{toc}{%
	\begingroup
	\let\protect\l@chapter\protect\l@section
	\let\protect\l@section\protect\l@subsection
}
\makeatother

\chapter{Translation between symbolic codes} \label{chap:appendix_c}

Recall from Chapter \ref{chap:binary_black_hole_shadows} that the symbolic dynamics of Cornish and Gibbons \cite{CornishGibbons1997}, which we refer to as \emph{collision dynamics}, may be used to describe the perpetual orbits of the Majumdar--Papapetrou di-hole in the meridian plane. This is achieved using the decision dynamics alphabet $\mathcal{A}_{\text{CG}} = \left\{ +1, 0, -1 \right\}$. In Section \ref{sec:symbolic_dynamics_mp}, we present an alternative symbolic code -- \emph{decision dynamics} -- to describe trajectories in the Majumdar--Papapetrou di-hole geometry. This is achieved using the symbolic alphabet $\mathcal{A} = \left\{ 0, 1, 2, 3, 4 \right\}$. The perpetual orbits can be described using infinitely long sequences from the reduced alphabet $\mathcal{A}_{\textrm{R}} = \left\{ 0, 2, 4 \right\}$.

Consider a perpetual orbit described by the sequence $a = a_{1} a_{2} a_{3} \cdots$, with $a_{i} \in \mathcal{A}_{\text{CG}}$, $\forall i \in \mathbb{N}$, as described by Cornish and Gibbons's collision dynamics. Given the sequence $a$ in collision dynamics, we would like to be able to translate to a sequence $\mathcal{X} = \mathcal{X}_{1} \mathcal{X}_{2} \mathcal{X}_{3} \cdots$, with $\mathcal{X}_{i} \in \mathcal{A}_{\textrm{R}}$, $\forall i \in \mathbb{N}$, as described by decision dynamics. This can be achieved using the following algorithm.

We note first that, in decision dynamics, a ``decision point'' is preceded by the symbol $\pm 1$ in collision dynamics. Such points are referred to as \emph{pivots}, and are central to the translation between the two symbolic codes. We assume that $a_{1} = \pm 1$ is our first pivot. Consider the first decision point, which follows the pivot $a_{1} = \pm 1$. If  $\left| a_{1} - a_{2} \right| = 2$ (i.e., $a_{1}$ and $a_{2}$ have opposite signs), then replace $a_{1}$ with $\mathcal{X}_{1} = 0 \in \mathcal{A}_{\textrm{R}}$ in decision dynamics. We then use $a_{2} = \mp 1$ as the new pivot. On the other hand, if $\left| a_{1} - a_{2} \right| = 1$ (i.e., $a_{2} = 0$), we must consider the third element $a_{3}$ of the collision dynamics sequence. If $\left| a_{1} - a_{3} \right| = 2$ (i.e., $a_{1}$ and $a_{3}$ have opposite signs), then we replace $a_{1} a_{2}$ with $\mathcal{X}_{1} = 2 \in \mathcal{A}_{\textrm{R}}$ in decision dynamics. Alternatively, if $\left| a_{1} - a_{3} \right| = 0$ (i.e., $a_{1}$ and $a_{3}$ have the same sign), then we replace $a_{1} a_{2}$ with $\mathcal{X}_{1} = 4 \in \mathcal{A}_{\textrm{R}}$ in decision dynamics. In both of these cases, we use $a_{3} = \pm 1$ as the new pivot. The whole sequence can be translated from collision dynamics to decision dynamics by continuing in this fashion, all the way along the sequence from left to right.

\begin{table}
\begin{center}
\begin{tabular}{c c c c}
\hline \hline
$a_{n}$ & $a_{n+1}$ & $a_{n + 2}$ & $\mathcal{X}_{n}$ \\ \hline
$\pm 1$ & {} & {} & {} \\
{} & $\mp 1$ & {} & 0 \\
{} & $0$ & $\mp 1$ & 2 \\
{} & $0$ & $\pm 1$ & 4 \\ \hline \hline
\end{tabular}
\caption{Summary of the translation between collision dynamics and decision dynamics. Beginning with the pivot $a_{n} = \pm 1$, we replace a finite substring in collision dynamics  -- either $a_{n}$ or $a_{n} a_{n + 1}$ -- with a symbol $\mathcal{X}_{n}$ from decision dynamics. If $a_{n}$ and $a_{n + 1}$ have opposite signs, then we make the replacement $a_{n} \mapsto \mathcal{X}_{n} = 0$, and we use $a_{n + 1} = \mp 1$ as the new pivot. If $a_{n + 1} = 0$, then we consider $a_{n + 2}$. In this case, if $a_{n}$  and $a_{n + 2}$ have the opposite (same) sign, then we make the replacement $a_{n} a_{n + 1} \mapsto \mathcal{X}_{n} = 2$ ($a_{n} a_{n + 1} \mapsto \mathcal{X}_{n} = 4$), and we use $a_{n + 2} = \pm 1$ as the new pivot. \label{table:translation_symbolic_dynamics}}
\end{center}
\end{table}

The algorithm is summarised in Table \ref{table:translation_symbolic_dynamics}. The leftmost column, denoted $a_{n}$, represents the pivot under consideration; this is always $a_{n} = \pm 1$ in collision dynamics. The second and third columns correspond to the subsequent elements, denoted by $a_{n + 1}$ and $a_{n + 2}$, of the finite substring we wish to translate. The final column, labelled $\mathcal{X}_{n}$, gives the corresponding symbol from the reduced decision dynamics alphabet $\mathcal{A}_{\textrm{R}}$. In each case, the final element of the finite substring under consideration -- be that $a_{n + 1}$ or $a_{n + 2}$ -- becomes the new pivot; this always takes the value $\pm 1$.

\begin{figure}[h]
\begin{center}
\includegraphics[width=0.5\textwidth]{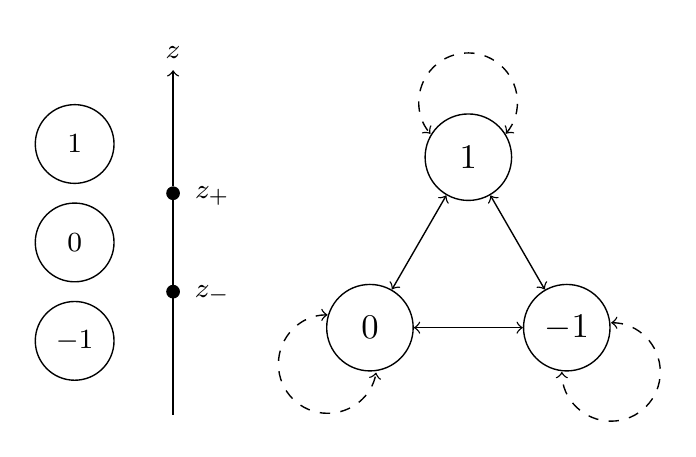}
\caption{Collision dynamics for the Majumdar--Papapetrou di-hole. The left-hand figure shows a schematic diagram of the symbolic coding. The right-hand figure shows the transition diagram for this symbolic dynamics: allowed transitions are shown using a solid line; forbidden transitions are shown using a dashed line. (Adapted from Figure 5 of \cite{CornishGibbons1997}.)}
\label{fig:cornish_gibbons_symbolic_dynamics}
\end{center}
\end{figure}

Figure \ref{fig:cornish_gibbons_symbolic_dynamics} shows a schematic diagram of the collision dynamics symbolic coding of Cornish and Gibbons \cite{CornishGibbons1997}. The two black holes are located along the $z$-axis at $z = z_{\pm}$. Each time a null geodesic passes through $z \in (z_{+}, \infty)$ we assign the digit $1$; when the ray passes through $z \in (z_{-}, z_{+})$ we assign the symbol $0$; and when the ray passes through $z \in (- \infty, z_{-})$ we assign the digit $-1$. The transition diagram on the right-hand side of Figure \ref{fig:cornish_gibbons_symbolic_dynamics} encodes the fact that no digit can follow itself. This highlights explicitly the correspondence with the Gaspard--Rice three disc model (see Section \ref{sec:symbolic_dynamics_gaspard_rice}).
%

\chapter{Circular photon orbits of the Majumdar--Papapetrou di-hole} \label{chap:appendix_a}

In this section, we derive some of the key results of Chapter \ref{chap:binary_black_hole_shadows} and Chapter \ref{chap:stable_photon_orbits} on the existence of stable photon orbits around the equal-mass Majumdar--Papapetrou di-hole. These results were originally presented in \cite{ShipleyDolan2016}.

In this analysis, we use units in which the individual black hole masses are set to unity, i.e., $M_{\pm} = M = 1$. This is equivalent to a rescaling of the coordinates $q\ind{^{a}} \mapsto \frac{1}{M} q\ind{^{a}}$, momenta $p\ind{_{a}} \mapsto \frac{1}{M} p\ind{_{a}}$, and the separation $d \mapsto \frac{d}{M}$.

\section{Equatorial circular photon orbits}

As described in Sections \ref{sec:mp_dihole_hamiltonian_formalism} and \ref{sec:spo_hamiltonian_formalism}, null geodesic motion for the equal-mass Majumdar--Papapetrou di-hole is governed by the two-dimensional effective potential
\begin{equation}
h(\rho, z) = \rho \, U^{2}, \qquad U(\rho, z) = 1 + \frac{1}{\sqrt{\rho^{2} + \left( z - \frac{d}{2} \right)^{2}}} + \frac{1}{\sqrt{\rho^{2} + \left( z + \frac{d}{2} \right)^{2}}}.
\end{equation}
The equal-mass di-hole is symmetric about the equatorial ($z = 0$) plane. Motion confined to this surface is governed by a one-dimensional effective potential $\hat{h}(\rho) = h(\rho, 0) = \rho \, \hat{U}^{2}$, where $\hat{U}(\rho) = 1 + \frac{2}{R}$, where we have introduced a new radial coordinate $R(\rho) = \sqrt{\rho^{2} + \frac{d^{2}}{4}}$. (Note that $R > \frac{d^{2}}{4} > 0$ and $\hat{U} > 0$.) Circular null geodesics exist where $\hat{h}\ind{_{,\rho}} = 0$. Differentiation of the effective potential gives
\begin{equation}
\label{eqn:mp_spo_existence_h_rho}
\hat{h}\ind{_{,\rho}} = \frac{\hat{U}}{R^{3}} P(R), \qquad P(R) = R^{3} - 2 R^{2} + d^{2}.
\end{equation}
The positivity of $\hat{U}$ and $R$ mean that the task of locating circular photon orbits reduces to finding the roots of the cubic $P(R) = R^{3} - 2 R^{2} + d^{2}$: circular orbits exist where $P(R)$ has roots. The discriminant of this polynomial is
\begin{equation}
\Delta_{R} (P) = - d^{2} \left( 27 d^{2} - 32 \right).
\end{equation}
For $d > d_{2} = \sqrt{\frac{32}{27}}$, $\Delta_{R} (P) < 0$, so $P(R)$ has one real root. It is straightforward to show that, in this regime, there are no roots with $R > 0$, and hence no equatorial circular photon orbits. If $d = d_{2}$, $\Delta_{R} (P) = 0$, and there is a repeated root with $R > 0$. In this case there is one circular photon orbit. For $0 < d < d_{2}$, we have $\Delta_{R} (P) > 0$, so $P(R)$ has three distinct real roots, two of which have $R > 0$. Thus, there are two circular photon orbits in the equatorial plane.

The stability of these orbits can be determined by considering $\hat{h}\ind{_{, \rho \rho}}$: orbits are stable (unstable) under radial perturbations if $\hat{h}\ind{_{, \rho \rho}} < 0$ ($\hat{h}\ind{_{, \rho \rho}} > 0$). In the case where $d < d_{2}$, the inner (outer) orbit is stable (unstable) under radial perturbations. In the critical case $d = d_{2}$, the unique circular photon orbit is marginally stable ($\hat{h}\ind{_{, \rho \rho}} = 0$).

One may also consider the stability of equatorial circular photon orbits under perturbations in the $z$-direction. The equatorial symmetry of the equal-mass Majumdar--Papapetrou di-hole implies that $U\ind{_{, z}}(\rho, 0) = 0$, thus $h\ind{_{, z}}(\rho, 0) = 0$. Taking a second derivative, we find
\begin{equation}
\label{eqn:mp_height_function_zz_derivative}
h\ind{_{, z z}} (\rho, 0) = 2 \rho \, \hat{U} \, U\ind{_{, z z}}(\rho, 0), \qquad U\ind{_{, z z}}(\rho, 0) = \frac{3 d^{2} - 4 R^{2}}{2 R^{5}}.
\end{equation}
One can deduce the stability of equatorial circular photon orbits under perturbations in the $z$-direction by considering the sign of the quadratic $3 d^{2} - 4 R^{2}$. Using the fact that stable orbits satisfy $P(R) = 0$, we deduce that $d^{2} \left( d - \sqrt{\frac{16}{27}} \right) > 0$. The inner orbit is therefore stable under perturbations out of the plane if $d > d_{1} = \sqrt{\frac{16}{27}}$ (and it exists if $d < d_{2}$).

Summarising these results, we have demonstrated that stable equatorial photon orbits exist for equal-mass Majumdar--Papapetrou di-holes with coordinate separations in the range $d_{1} < d < d_{2}$. We remark that the upper bound of this result ($d < d_{2}$) is consistent with Coelho and Herdeiro's results \cite{CoelhoHerdeiro2009}; and with those of W\"{u}nsch \emph{et al.} \cite{WuenschMuellerWeiskopfEtAl2013}.

We now focus on the special case $d = 1$. In this case, the polynomial $P(R)$ can be factorised as $P(R) = \left( R - 1 \right) \left( R^{2} - R - 1 \right)$. The (positive) roots are therefore $R = 1$ ($\rho = \frac{\sqrt{3}}{2}$) and $R = \varphi$ ($\rho = \sqrt{\varphi^{2} - \frac{1}{4}} = \frac{1}{2} 5^{1/4} \varphi^{3/2}$), where $\varphi = \frac{1}{2} \left( 1 + \sqrt{5} \right)$ is the golden ratio. Under radial perturbations, the inner (outer) orbit at $R = 1$ ($R = \varphi$) is stable (unstable).

In Section \ref{sec:spo_electrovacuum_case}, we derive a condition for the existence of stable photon orbits in stationary axisymmetric electrovacuum spacetimes. We may employ this as a consistency check for the results derived above in the case of an equal-mass Majumdar--Papapetrou di-hole. A key result of Section \ref{sec:spo_electrovacuum_case} is that \eqref{eqn:tr_hessian_hpm_5} must be satisfied at stationary points, i.e., stationary points of $h$ must satisfy
\begin{equation}
\label{eqn:mp_spo_existence_condition}
\operatorname{tr}{\mathcal{H}(h)} = - \frac{2}{\rho} \left| \mathbf{W} \right|^{2}, \qquad \mathbf{W} = h \bnab A\ind{_{t}} - \bnab A\ind{_{\phi}},
\end{equation}
where $A\ind{_{t}} = \frac{1}{U}$ and $A\ind{_{\phi}} = 0$ for the Majumdar--Papapetrou solution. Introducing the radial coordinate $R = \sqrt{\rho^{2} + \frac{d^{2}}{4}}$, we find that the existence condition \eqref{eqn:mp_spo_existence_condition} is satisfied if $P(R) = R^{3} - 2 R^{2} + d^{2} = 0$, which is equivalent to condition \eqref{eqn:mp_spo_existence_h_rho}. At a stationary point, the black hole separation can therefore be expressed in terms of the radial coordinate $R$ as $d^{2} = 2 R^{2} - R^{3}$.

In the case of an equal-mass di-hole, there is an equatorial symmetry. The stability condition for equatorial circular orbits derived in Section \ref{sec:spo_electrovacuum_case} therefore reduces to $0 < - h\ind{_{, \rho \rho}} < \frac{2}{\rho} \left| \mathbf{W} \right|^{2}$. Rewriting this condition in terms of the radial coordinate $R$, and making the replacement $d^{2} = 2 R^{2} - R^{3}$, we arrive at the inequality $\frac{2}{3} < R < \frac{4}{3}$. Noting that $d = \sqrt{2 R^{2} - R^{3}}$ is monotonically increasing on the domain $R \in \left[\frac{2}{3}, \frac{4}{3}\right]$, the inequality on $R$ may be recast in terms of the separation as $\sqrt{\frac{16}{27}} < d < \sqrt{\frac{32}{27}}$. Hence, we have recovered the existence condition for stable equatorial circular null geodesics in the equal-mass Majumdar--Papapetrou di-hole spacetime.


\section{Non-planar circular photon orbits}

Locating the stationary points of $h$ which do not lie in the equatorial plane poses more of a challenge; however, it is possible to make progress through the introduction of elliptic coordinates $(\xi, \eta)$, which are related to $(\rho, z)$ via
\begin{equation}
\label{eqn:cylindrical_to_elliptical_coordinates}
\rho = \frac{d}{2} \sinh{\xi} \sin{\eta}, \qquad z = \frac{d}{2} \cosh{\xi} \cos{\eta}.
\end{equation}
Making the further substitutions $X = \cosh{\xi}$ and $Y = \cos{\eta}$, \eqref{eqn:cylindrical_to_elliptical_coordinates} may be recast as
\begin{equation}
\label{eqn:cylindrical_to_elliptical_coordinates_2}
\rho = \frac{d}{2} \sqrt{X^{2} - 1} \sqrt{1 - Y^{2}}, \qquad z = \frac{d}{2} X Y.
\end{equation}
To ensure that $\rho > 0$, we must have $X \in \left(1, \infty\right)$ and $Y \in \left(-1, 1\right)$. In these coordinates, we have
\begin{equation}
U = 1 + \frac{4 X}{d \left( X^{2} - Y^{2} \right)},
\end{equation}
and the effective potential is
\begin{equation}
h = \frac{d}{2} \sqrt{X^{2} - 1} \sqrt{1 - Y^{2}} \left(  1 + \frac{4 X}{d \left( X^{2} - Y^{2} \right)} \right)^{2},
\end{equation}

Stationary points satisfy the pair of equations $h\ind{_{, X}} = 0 = h\ind{_{, Y}}$. It is quick to establish that these conditions yield the system of equations
\begin{align}
d X^{5} - 2 d X^{3} Y^{2} + d X Y^{4} + 4 X^{4} + 12 X^{2} Y^{2} - 16 X^{2} &= 0, \label{eqn:elliptic_stationary_point_condition_1} \\
d X^{5} - 2 d X^{3} Y^{2} + d X Y^{4} - 4 X^{4} - 12 X^{2} Y^{2} + 8 X^{2} + 8 Y^{2} &= 0. \label{eqn:elliptic_stationary_point_condition_2}
\end{align}
Subtracting \eqref{eqn:elliptic_stationary_point_condition_2} from \eqref{eqn:elliptic_stationary_point_condition_1} and simplifying yields
\begin{equation}
X^{4} + 3 X^{2} Y^{2} - 3 X^{2} - Y^{2} = 0.
\end{equation}
This relationship allows us to express $Y^{2} = \cos^{2}{\eta}$ as a function of $X = \cosh{\xi}$:
\begin{equation}
\label{eqn:elliptic_stationary_point_relationship}
Y^{2} = \frac{X^{2} \left( X^{2} - 3 \right)}{1 - 3 X^{2}}.
\end{equation}
Remarkably, this relationship is independent of the value of the black hole separation $d$. Now, addition of \eqref{eqn:elliptic_stationary_point_condition_2} and \eqref{eqn:elliptic_stationary_point_condition_1} results in
\begin{equation}
\label{eqn:sum_elliptic_stationary_point_conditions}
d X^{5} - 2 d X^{3} Y^{2} + d X Y^{4} - 4 X^{2} + 4 Y^{2} = 0.
\end{equation}
Eliminating $Y$ from \eqref{eqn:sum_elliptic_stationary_point_conditions} using \eqref{eqn:elliptic_stationary_point_relationship} leads to a quintic in the $X$-coordinate:
\begin{equation}
\label{eqn:elliptic_stationary_point_quintic}
d X^{5} - d X^{3} - 3 X^{2} + 1 = 0.
\end{equation}

For general values of the separation $d$, one is not able to factorise the left-hand side of \eqref{eqn:elliptic_stationary_point_quintic}, or to find closed-form expressions for its roots in terms of $d$. However, when $d = 1$, the quintic factorises as $X^{5} - X^{3} - 3 X^{2} + 1 = (X^{3} + X^{2} + X - 1)(X^{2} - X - 1)$. The cubic factor has no roots with $X > 1$ (i.e., $\rho > 0$). The quadratic factor has one such root at $X = \varphi$. Using the relationship \eqref{eqn:elliptic_stationary_point_relationship}, $X = \varphi$ implies that $Y = \pm \frac{1}{\varphi^{2}}$. In the case $d = 1$, the stationary point conditions are satisfied by $\cosh \xi = \varphi$, $\cos \eta = \pm \frac{1}{\varphi^{2}}$, i.e., $(\rho, z) = (\frac{1}{2} 5^{1/4} \varphi^{-1/2}, \pm \frac{1}{2 \varphi})$. Inserting these values into the effective potential $h = \rho \, U^{2}$, it is quick to verify that these two non-planar saddle points lie on the same contour as the equatorial saddle point, given by $h = p\ind{_{\phi}} = \frac{1}{2} 5^{5/4} \varphi^{3/2}$.

The linearity of \eqref{eqn:elliptic_stationary_point_quintic} in $d$ allows us to express the separation in terms of the $X$-coordinate as
\begin{equation}
\label{eqn:relationship_separation_elliptic}
d = \frac{3 X^{2} - 1}{X^{3} \left( X^{2} - 1 \right)}.
\end{equation}
Therefore, it is possible to substitute a closed-form value of $X$ into the right-hand side of \eqref{eqn:relationship_separation_elliptic} to find the corresponding value of $d$. Recalling that we require $0 < Y^{2} < 1$, the relation \eqref{eqn:elliptic_stationary_point_relationship} may then be used as a consistency check. For example, $X = \sqrt{2}$ is a solution to \eqref{eqn:elliptic_stationary_point_quintic} if $d = \frac{5 \sqrt{2}}{4}$. Using \eqref{eqn:elliptic_stationary_point_relationship}, it is quick to check that $Y^{2} = \frac{2}{5}$, as required.

Consider \eqref{eqn:relationship_separation_elliptic} for a general value of $d$. The conditions $X > 1$ and $0 < Y^{2} < 1$ together imply that we require $X^{2} < 3$ for a solution. The critical value $X = \sqrt{3}$ corresponds to a separation of $d = d_{1}$. For $d < d_{1}$, there are no circular photon orbits out of the equatorial plane.

Finally, let us consider the limit $X \rightarrow 1$ (i.e., $\rho \rightarrow 0$). Performing a first order perturbative expansion with $X \sim 1 + \varepsilon$, where $\varepsilon \ll 1$, we have $X^{2} \sim 1 + 2 \varepsilon$. Then, through \eqref{eqn:elliptic_stationary_point_relationship},
\begin{equation}
Y^{2} = \frac{(1 + 2 \varepsilon) (2 - 2 \varepsilon)}{2 + 6 \varepsilon} \sim 1 - 2 \varepsilon.
\end{equation}
Hence, $Y^{2} \sim X^{-2}$ as $X \rightarrow 1$. For widely separated black holes with $d \gg 1$, $X \sim 1 + \frac{1}{d}$ and $Y \sim 1 - \frac{1}{d}$. In this case, the equal-mass Majumdar--Papapetrou di-hole will resemble a pair of isolated extremal Reissner--Nordstr\"{o}m black holes, with unstable circular photon orbits around each of the sources at $\rho \sim 1$ and $z \sim \pm \frac{d}{2}$. (For a single Reissner--Nordstr\"{o}m black hole of mass $M = 1$, there is an unstable equatorial orbit at $\rho = 1$.) 

\chapter{H\'{e}non--Heiles Hamiltonian system} \label{chap:appendix_b}

In the early 1960s, H\'{e}non and Heiles \cite{HenonHeiles1964} investigated the non-linear motion of stars about the galactic centre. The main purpose of this work was to obtain a third integral of motion -- in addition to the energy and angular momentum -- in galactic dynamics. To do this, H\'{e}non and Heiles consider a simple two-dimensional Hamiltonian, described in phase space coordinates $\{x, y, p\ind{_{x}}, p\ind{_{y}} \}$ by
\begin{align}
H &= \frac{1}{2} \left( {p\ind{_{x}}}^{2} + {p\ind{_{y}}}^{2} \right) + V(x, y) , \label{eqn:hh_hamiltonian} \\
V &= \frac{1}{2} k \left( x^{2} + y^{2} \right) + \lambda \left( x^{2} y - \frac{1}{3} y^{3} \right), \label{eqn:hh_potential}
\end{align}
where $k, \lambda \in \mathbb{R}$ are parameters. If $\lambda$ is small, the potential resembles that of a two-dimensional isotropic harmonic oscillator, with two cubic perturbation terms.

The system exhibits a threefold rotational symmetry in the $(x, y)$-plane. This is seen most easily by transforming to planar polar coordinates $x = r \cos{\theta}$, $y = r \sin{\theta}$, in which the potential reads $V = \frac{1}{2} k r^{2} + \frac{1}{3} \lambda r^{3} \sin{3 \theta}$. The invariance of the potential under transformations of the form $\theta \mapsto \theta + \frac{2 n\pi}{3}$ ($n \in \mathbb{Z}$) is now manifest.

For simplicity, we consider the case $k = 1 = \lambda$. The Hamiltonian $H$ is a constant of the motion, say $E$, which corresponds to the total energy of the system. The equations of motion, i.e., Hamilton's equations, read
\begin{align}
\dot{x} &= p\ind{_{x}}, & \dot{y} &= p\ind{_{y}},
& \dot{p\ind{_{x}}} &= - x - 2 x y, & \dot{p\ind{_{y}}} &= - y - x^{2} + y^{2}, \label{eqn:hh_hamiltons_equations}
\end{align}
where an overdot denotes differentiation with respect to some time parameter, say $t$. The coupled non-linear system \eqref{eqn:hh_hamiltons_equations} is not Liouville integrable, and has no closed-form solution. One should therefore anticipate rich dynamics.

The full phase space of the H\'{e}non--Heiles system is described by two pairs of conjugate variables $\{x, y, p\ind{_{x}}, p\ind{_{y}}\}$, and an energy constraint $H = E$. For a fixed value of the energy $E$, we must have
\begin{align}
\frac{1}{2} \left( {p\ind{_{x}}}^{2} + {p\ind{_{y}}}^{2} \right) &\leq E, \label{eqn:hh_kinetic_inequality} \\
V(x, y) &\leq E. \label{eqn:hh_potential_inequality}
\end{align}
The inequality \eqref{eqn:hh_kinetic_inequality} indicates that the kinetic energy of the particle is bounded above; the inequality \eqref{eqn:hh_potential_inequality} demarcates the allowed regions of configuration space which are accessible by the particle.

Orbits of constant $x$ and $y$ satisfy $\bnab V = \left( V\ind{_{, x}}, V\ind{_{, y}} \right) = \left(0, 0 \right)$, with $V = E$; in other words, such orbits are stationary points of the potential. Locating these stationary points is tantamount to solving the system \eqref{eqn:hh_hamiltons_equations} for $x$ and $y$, with $p\ind{_{x}} = 0 = p\ind{_{y}}$. A stationary point is stable if it is a local minimum of the potential $V$, i.e., if $\det{\mathcal{H}(V)} > 0$ and $\operatorname{tr}{\mathcal{H}(V)} > 0$, where $\mathcal{H}(V)$ denotes the Hessian matrix of the potential $V$. The H\'{e}non--Heiles potential admits a local minimum $(x, y) = (0, 0)$, where $V = 0$. Moreover, there are three saddle points which lie on the vertices of an equilateral triangle with coordinates $(0, 1)$ and $\left( \pm \frac{\sqrt{3}}{2}, - \frac{1}{2} \right)$. The three saddles are connected by a single contour $V = E^{\ast} = \frac{1}{6}$ which encloses the local minimum.

\begin{figure}[t]
\begin{center}
\includegraphics[height=0.45\textwidth]{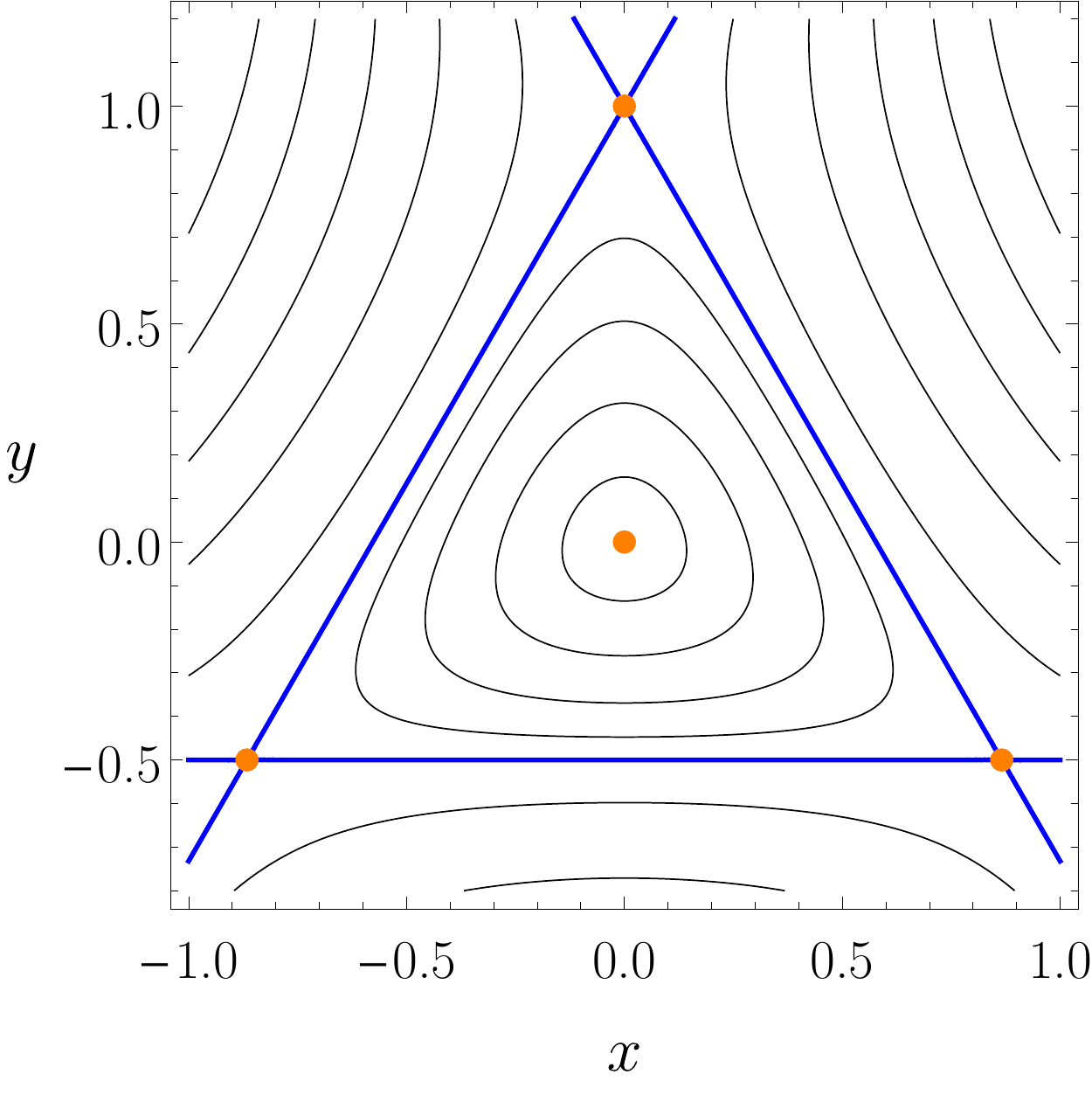}
\caption{Equipotential curves of the two-dimensional H\'{e}non--Heiles potential for a selection of values of the conserved energy $E$. The stationary points of $V$ (three saddle points and a local minimum) are represented by orange points. The local minimum is at the origin $(0, 0)$, and has $V = 0$. The three saddles are located at $(0, 1)$ and $\left( \pm \frac{\sqrt{3}}{2}, - \frac{1}{2} \right)$, and are connected by the critical contour $V = E^{\ast}$. For energies with $0 < E \leq E^{\ast}$ the contours form closed curves; whereas for $E > E^{\ast}$ the contours are open, and a particle can access spatial infinity through one of three escapes.}
\label{fig:hh_contours}
\end{center}
\end{figure}

In Figure \ref{fig:hh_contours}, we  show the equipotential curves for the H\'{e}non--Heiles potential \eqref{eqn:hh_potential} (i.e., curves of constant $V$) in the $(x, y)$-plane. For $E > E^{\ast}$ ($E \leq E^{\ast}$), the H\'{e}non--Heiles equipotential lines are open (closed). The former case is related to the investigation of Wada structures in binary black hole shadows, presented in Chapter \ref{chap:fractal_structures}. The latter case is related to the study of bounded stable photon orbits around Majumdar--Papapetrou di-holes, discussed in Chapter \ref{chap:stable_photon_orbits}. We discuss each case separately below.
%

\section{Open system}

For $E > E^{\ast}$, the H\'{e}non--Heiles system is open: orbits are permitted to escape to spatial infinity. For energies close to the threshold $E \gtrsim E^{\ast}$, there are three narrow channels which connect the scattering region to spatial infinity. In this regime, the H\'{e}non--Heiles system is an example of an open Hamiltonian system with three escapes \cite{AguirreVianaSanjuan2009}.

In an open Hamiltonian system with multiple escapes, an exit basin is defined as the set of initial conditions which lead to a certain exit in phase space (see Section \ref{sec:chaotic_dynamical_systems}). As described above, the H\'{e}non--Heiles system is a two-dimensional time-independent Hamiltonian system: the phase space is described by two pairs of conjugate variables $\{x, y, p\ind{_{x}}, p\ind{_{y}}\}$, and an energy constraint $H = E$. One must fix three initial conditions to define a trajectory: the ``physical'' phase space is therefore three-dimensional. In order to visualise the exit basins of such a Hamiltonian system, one can employ a surface of section (or Poincar\'{e} section) to reduce the dimensionality of the phase space by one; see Section \ref{sec:chaotic_dynamical_systems} for a review. In \cite{AguirreVallejoSanjuan2001}, the authors employ two choices of initial conditions which define a two-dimensional surface of section; these initial conditions are chosen so that the exit basins contain a \emph{Lyapunov orbit} -- an unstable periodic orbit which exists for all energies above the threshold value, such that any trajectory which crosses the Lyapunov orbit with its velocity vector pointing outwards is forced to escape without returning to the scattering region.

\begin{figure}
\begin{center}
\subfigure[$(x, y)$-space]{
\includegraphics[width=0.451\textwidth]{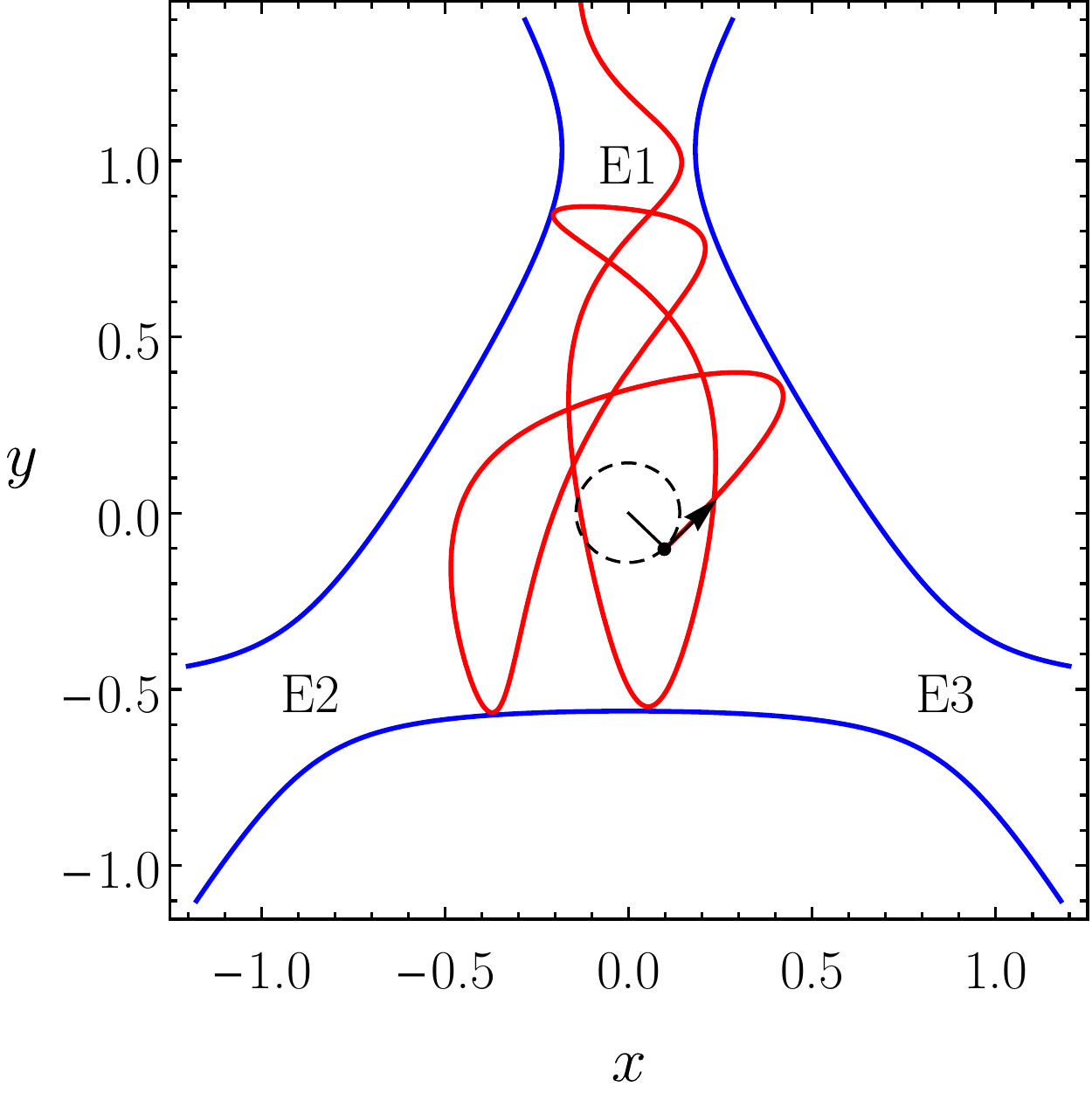} \label{fig:hh_x_y_basin_set_up}} \hspace{0.5em}
\subfigure[$(y, p\ind{_{y}})$-space]{
\includegraphics[width=0.45\textwidth]{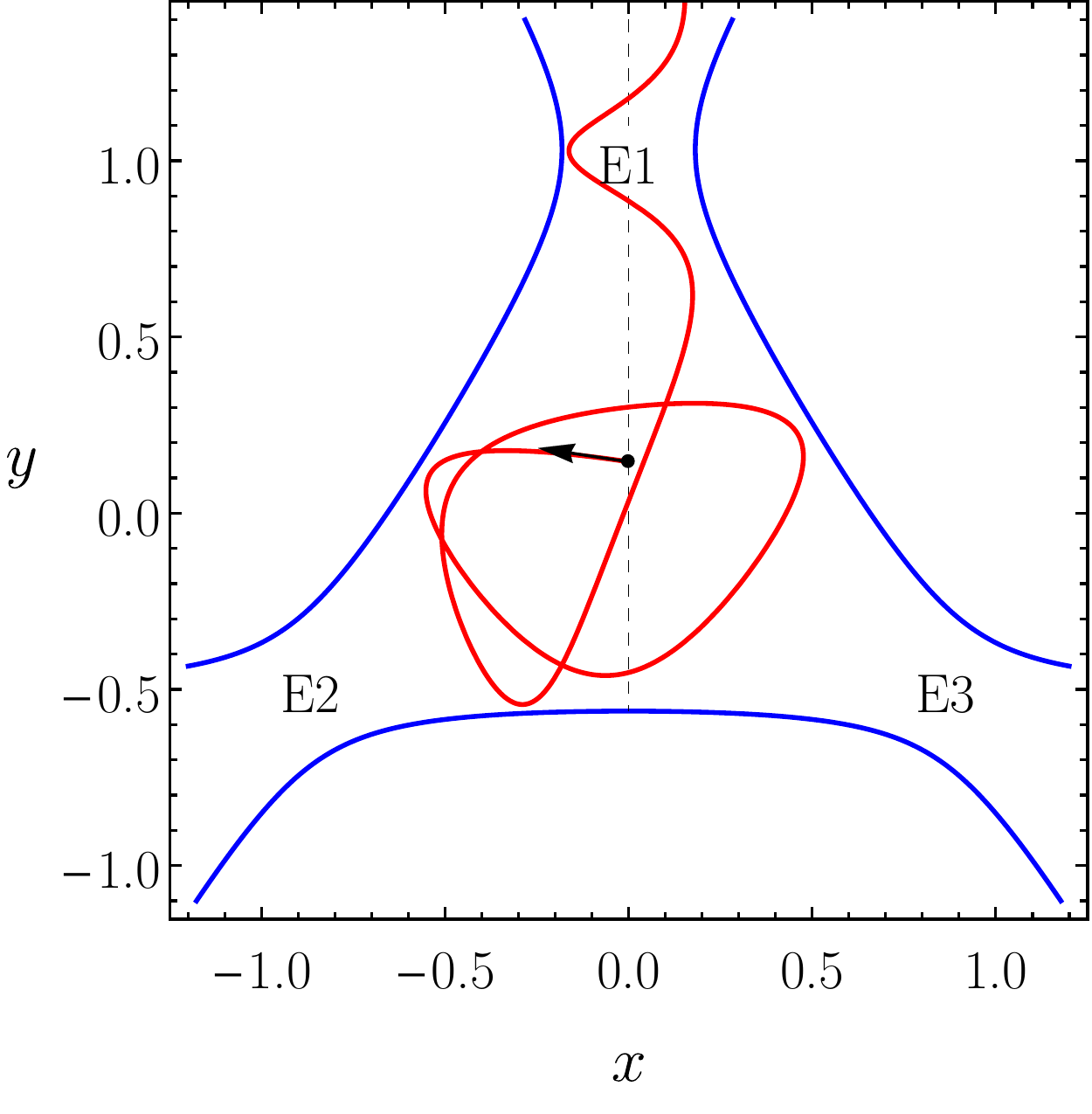} \label{fig:hh_y_py_basin_set_up}}
\caption{Initial conditions for two-dimensional exit basins in the H\'{e}non--Heiles open Hamiltonian system in (a) the $(x, y)$-plane; and (b) the $(y, p\ind{_{y}})$-plane. In both cases, the contour $V = E^{\ast} + \Delta E$ is shown in blue, and a trajectory which escapes the scattering region through Exit 1 is shown in red. \label{fig:hh_exit_basins_set_up}}
\end{center}
\end{figure}

The first choice of initial conditions is defined as follows. For each pair of coordinates $(x, y)$ which satisfy the inequality \eqref{eqn:hh_potential_inequality} (i.e., which lie in the allowed region of configuration space), we choose the initial momentum $(p\ind{_{x}}, p\ind{_{y}})$ to be tangent (in the anti-clockwise sense) to the circle centred on $(0, 0)$ which passes through the initial condition $(x, y)$. In other words, the two-dimensional surface of section is defined by the constraint $x p\ind{_{x}} + y p\ind{_{y}} = 0$, with $x p\ind{_{y}} - y p\ind{_{x}} > 0$. (See Figure \ref{fig:hh_x_y_basin_set_up} for the set-up of these initial conditions.)

For our second choice of initial conditions, we fix $x = 0$ and vary the initial values of $y$ and $p\ind{_{y}}$; the initial value of the momentum $p\ind{_{x}}$ is determined by rearranging the Hamiltonian constraint $H = E$. (The set-up of these initial conditions is shown in Figure \ref{fig:hh_y_py_basin_set_up}.)

\begin{figure}[t]
\begin{center}
\subfigure[$\Delta E = 0.05$]{
\includegraphics[width=0.31\textwidth]{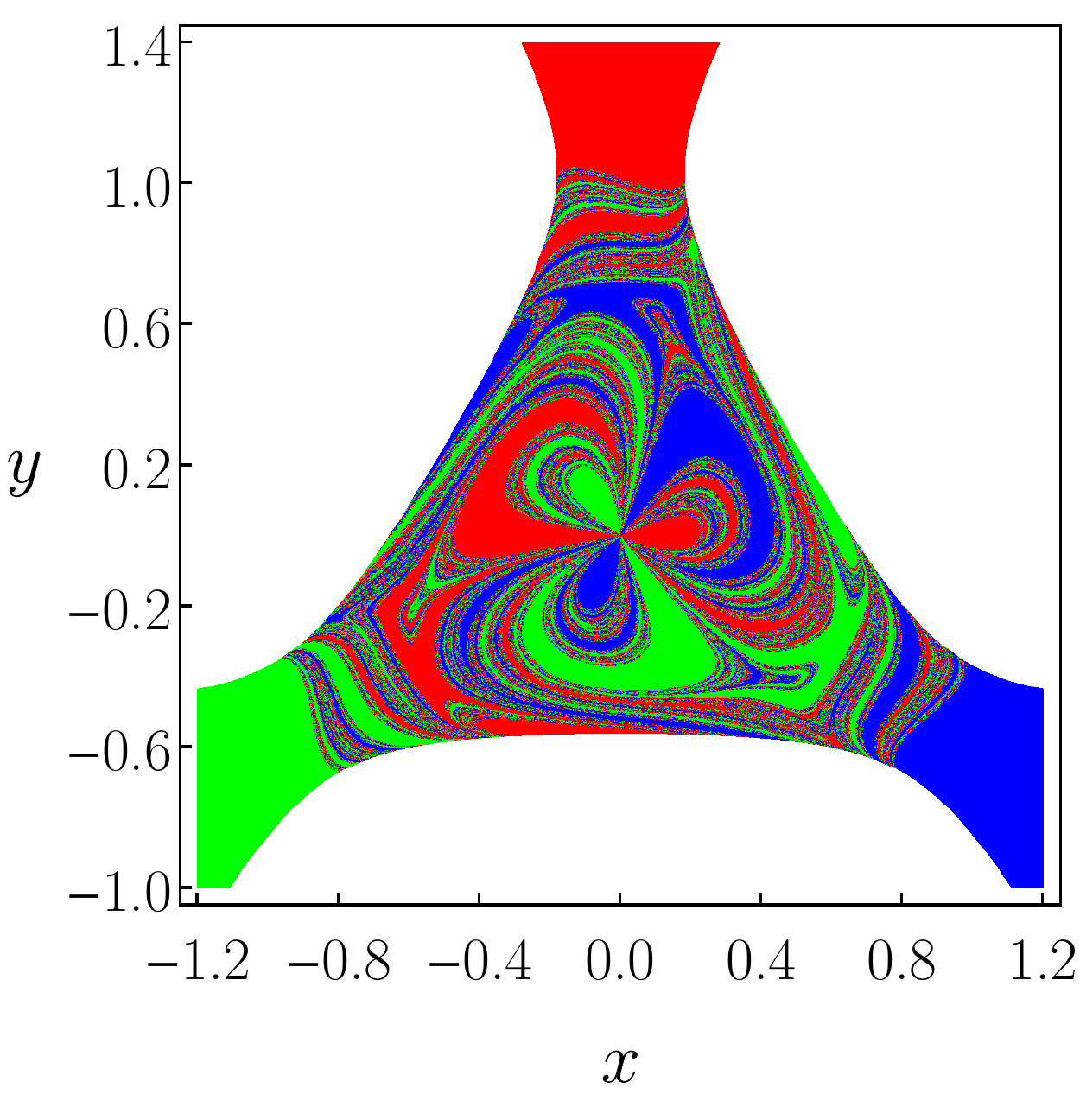} \label{fig:hh_x_y_basins_005}} \hfill
\subfigure[$\Delta E = 0.03$]{
\includegraphics[width=0.31\textwidth]{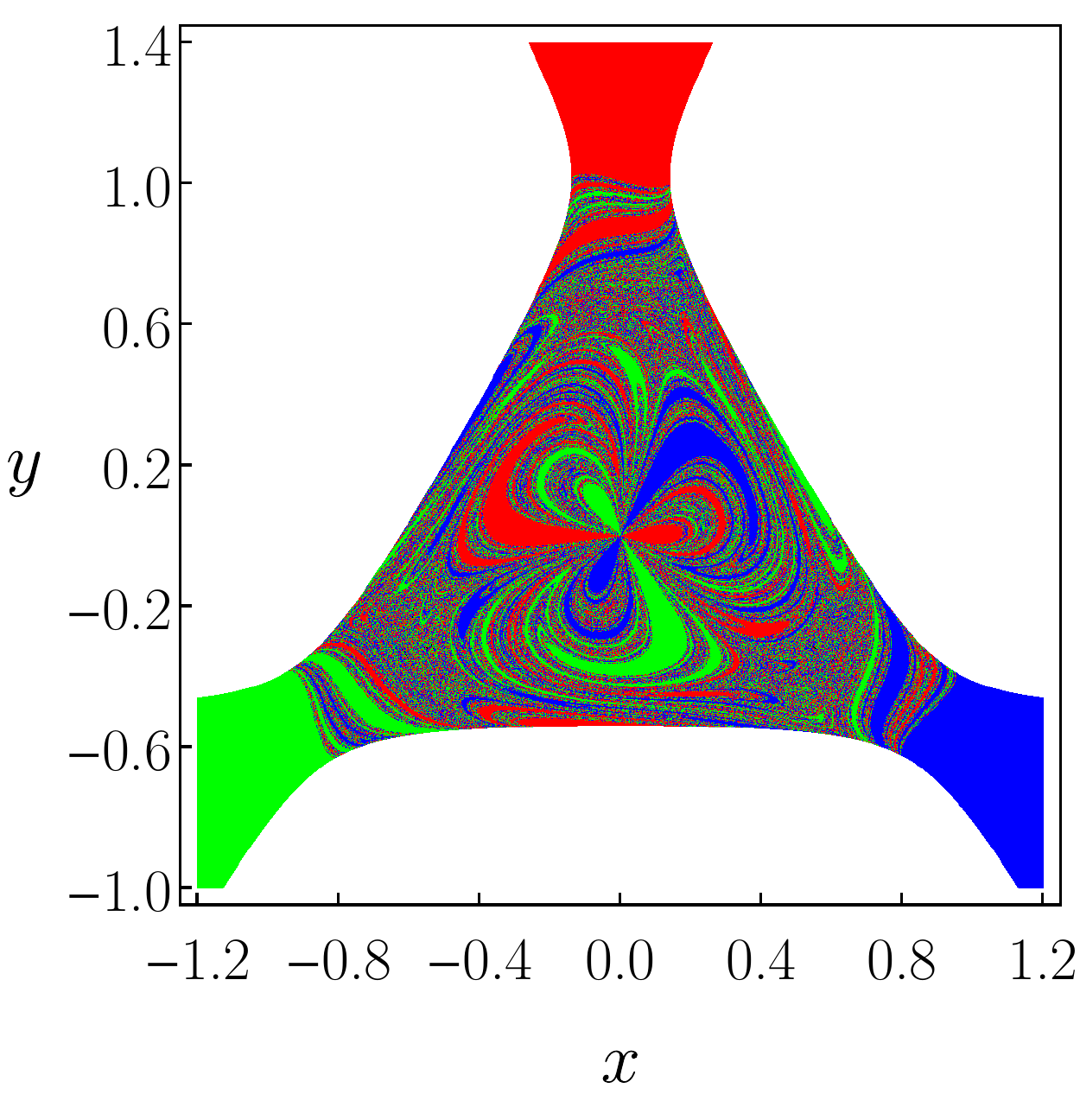} \label{fig:hh_x_y_basins_003}} \hfill
\subfigure[$\Delta E = 0.01$]{
\includegraphics[width=0.31\textwidth]{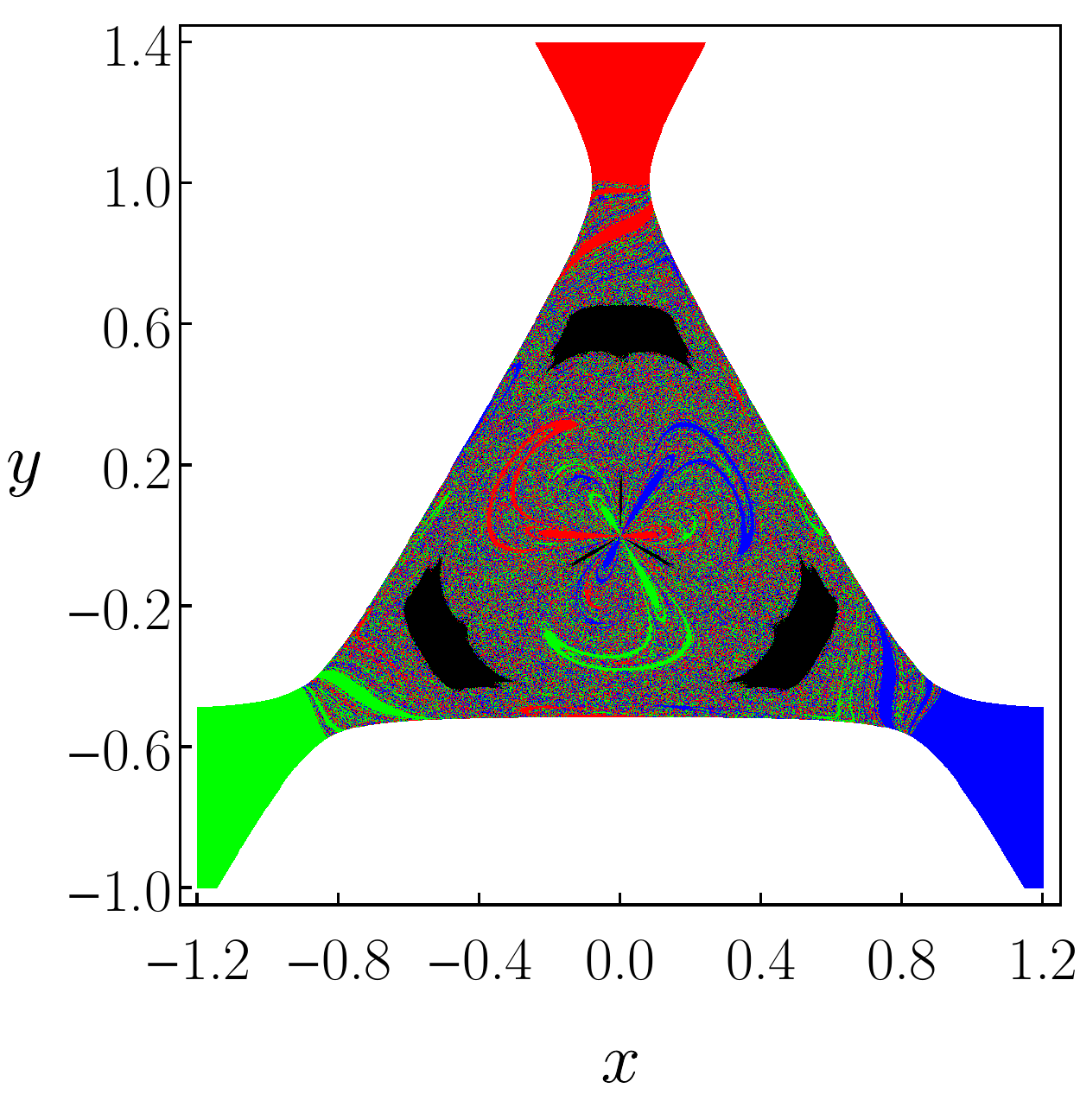} \label{fig:hh_x_y_basins_001}}
\subfigure[$\Delta E = 0.05$]{
\includegraphics[width=0.31\textwidth]{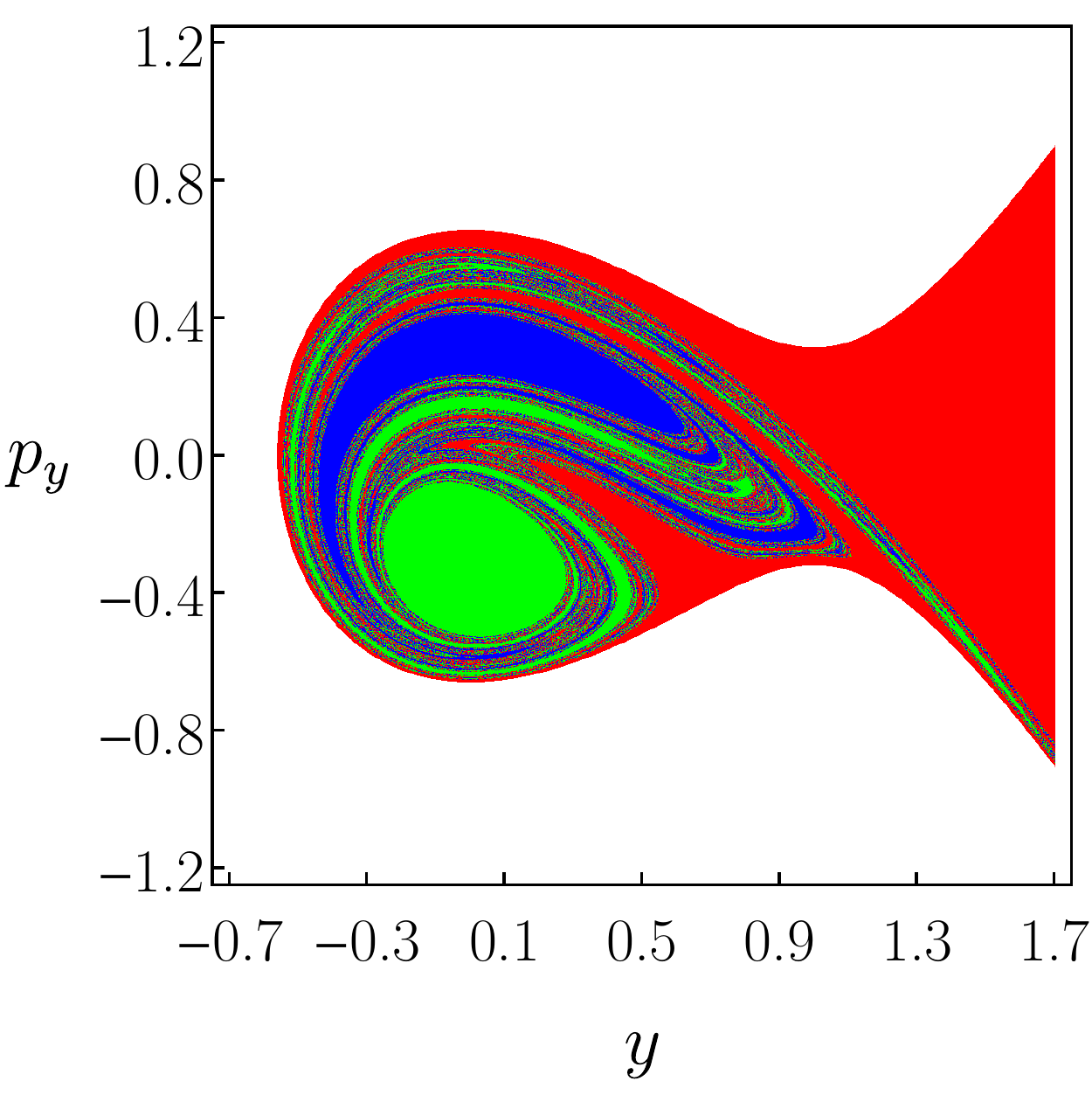} \label{fig:hh_y_py_basins_005}} \hfill
\subfigure[$\Delta E = 0.03$]{
\includegraphics[width=0.31\textwidth]{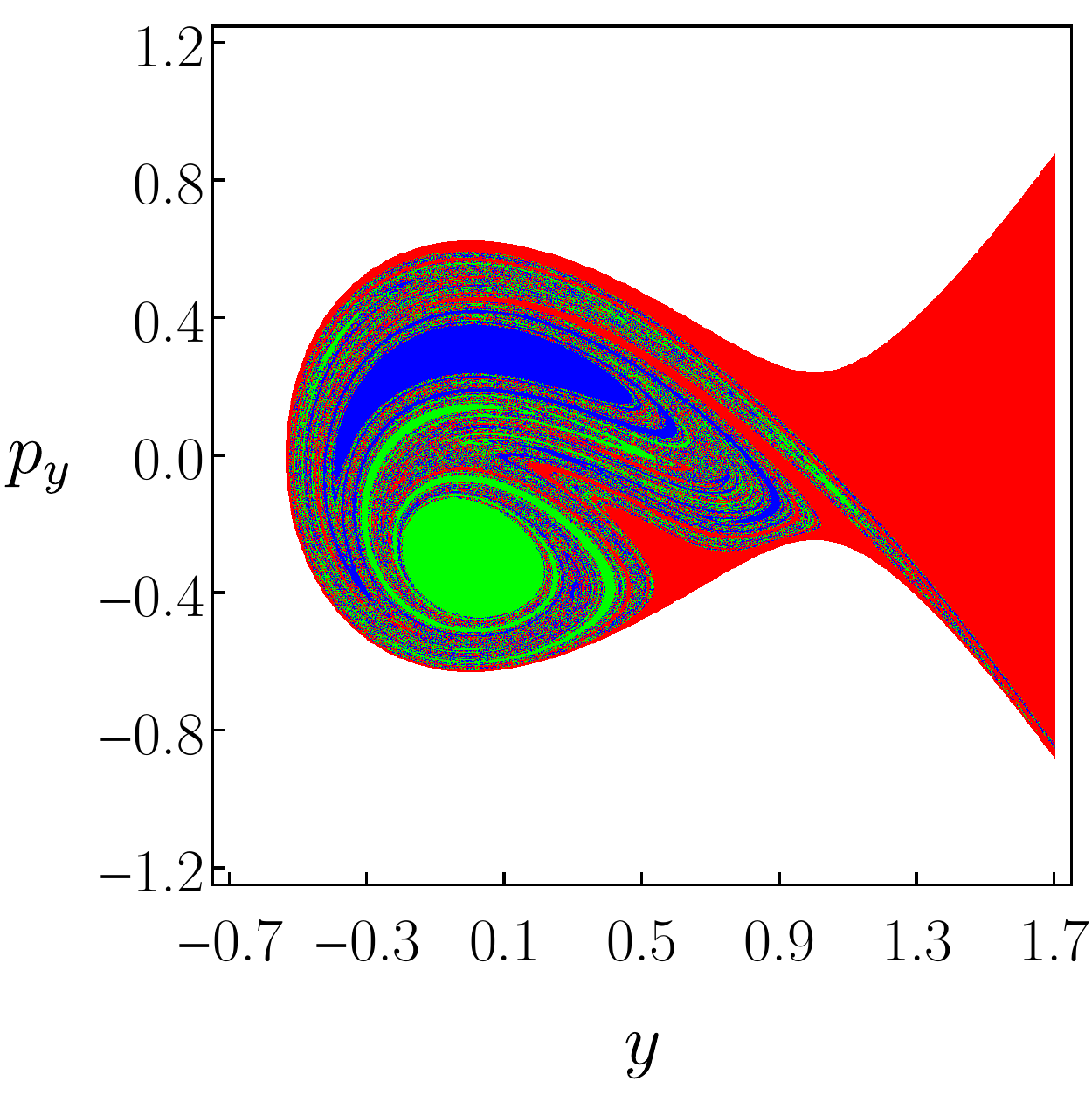} \label{fig:hh_y_py_basins_003}} \hfill
\subfigure[$\Delta E = 0.01$]{
\includegraphics[width=0.31\textwidth]{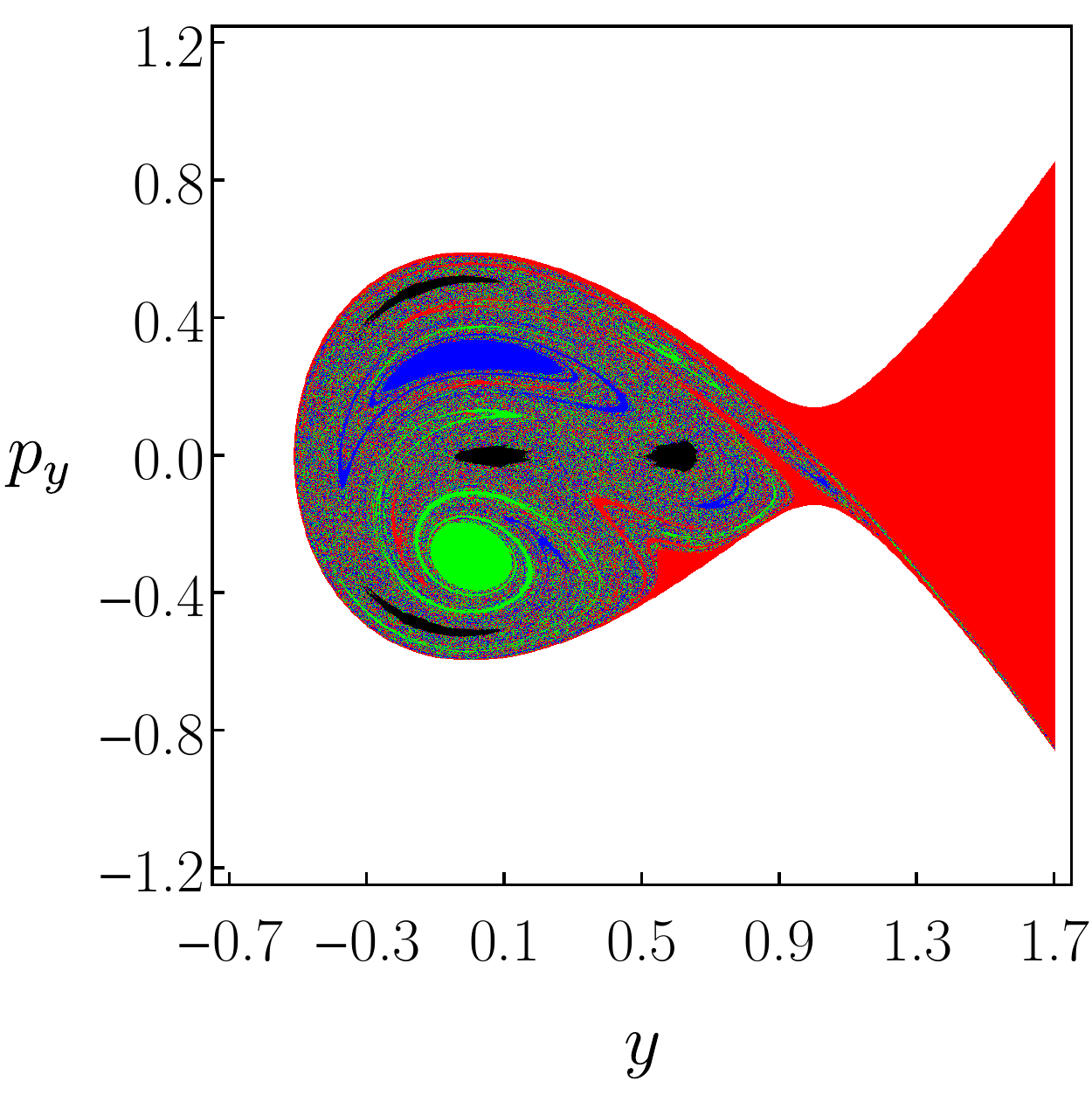} \label{fig:hh_y_py_basins_001}}
\caption{Exit basins in the H\'{e}non--Heiles open Hamiltonian system for (a)--(c) initial conditions in the $(x, y)$-plane; and (d)--(f) initial conditions in $(y, p\ind{_{y}})$-plane. The basins corresponding to Exits 1, 2, and 3 are plotted in red, green, and blue, respectively. \label{fig:hh_exit_basins}}
\end{center}
\end{figure}

In Figure \ref{fig:hh_exit_basins}, we show the exit basins for the H\'{e}non--Heiles system for a selection of values of the energy parameter $\Delta E = E - E^{\ast}$, using the two types of initial conditions described above. In practice, the exit basins are realised by numerically integrating Hamilton's equations for a fine grid of initial conditions in either the $(x, y)$-plane or the $(y, p\ind{_{y}})$-plane. The initial conditions are coloured red, green or blue according to whether they leave the scattering region through Exit 1, Exit 2 or Exit 3, respectively.

Figure \ref{fig:hh_exit_basins} shows the effect of varying the energy $\Delta E$ on the structure of the basins. As $\Delta E \rightarrow 0$ (i.e., as $E \rightarrow E^{\ast} = \frac{1}{6}$ from above), the basins become more ``fractalised'': the three (disconnected) basins are intertwined in a complicated fashion, and their boundary is highly intricate. Moreover, Kolmogorov--Arnold--Moser islands of stability (depicted in black) remain as ``deterministic islands'' in a ``random sea''. The exit basins of the H\'{e}non--Heiles system have been explore in detail by Aguirre \emph{et al.} \cite{AguirreVallejoSanjuan2001}, and in the limit of small escapes by Aguirre and Sanju\'{a}n \cite{AguirreSanjuan2003}.

The exit basins of the H\'{e}non--Heiles system possess the Wada property: any point on the boundary of one basin is on the boundary of the other two basins (see Section \ref{sec:wada_property}). This was verified numerically using the Nusse--Yorke method by Aguirre \emph{et al.} in 2001 \cite{AguirreVallejoSanjuan2001}. Since then, a range of numerical algorithms have been proposed to test for the Wada property in the exit basins of open Hamiltonian systems, using the H\'{e}non--Heiles system as a standard example; these include the grid method \cite{DazaWagemakersSanjuanEtAl2015}, the merging method \cite{DazaWagemakersSanjuan2018}, and the saddle--straddle method \cite{WagemakersDazaSanjuan2019}. In practice, the existence of Wada basins in phase space mean that the system exhibits extreme sensitive dependence on initial conditions: a small uncertainty in fixing the initial conditions in the neighbourhood of a Wada boundary results in high levels of indeterminism in the final state of the system, despite the underlying dynamics being fully deterministic. This is discussed in more depth throughout Chapter \ref{chap:fractal_structures}.

Additional studies of the fractal structures which arise in the exit basins of the H\'{e}non--Heiles system have been conducted by Barrio \emph{et al.} \cite{BarrioBlesaSerrano2008} and Zotos \cite{Zotos2017}.
%

\section{Closed system}

For conserved energies in the range $0 < E \leq \frac{1}{6}$, the H\'{e}non--Heiles system is closed: orbits are kinematically bounded by the energy, and the motion is restricted to an open subset of the $(x, y)$-plane. The inequality \eqref{eqn:hh_kinetic_inequality} indicates that the kinetic energy of the particle is bounded above; the inequality \eqref{eqn:hh_potential_inequality} demarcates the allowed regions of configuration space which are accessible by the particle. Any trajectory which starts inside a closed contour $V = E$ will remain so for all $t > 0$.

In order to visualise the complex nature of trajectories in phase space, one may use a Poincar\'{e} section (see Section \ref{sec:chaotic_dynamical_systems}). A rearrangement of the Hamiltonian constraint $H = E$ allows us to express one of the phase space coordinates, say $p\ind{_{x}}$, in terms of the other three. The dimensionality of the ``physical'' phase space is therefore reduced by one, and the motion is confined to a three-dimensional energy hypersurface. Since the motion is kinematically bounded, the orbits will repeatedly intersect any embedded two-dimensional surface, e.g.~the $(y, p\ind{_{y}})$-plane ($x = 0$). If there exists an additional conserved quantity involving $y$ and $p\ind{_{y}}$, then it could be used to express $p\ind{_{y}} = p\ind{_{y}}(y)$; the points of intersection between the trajectory and the surface $x = 0$ will therefore lie on a smooth curve.

To illustrate the utility of the Poincar\'{e} section, let us consider the dynamics of bounded trajectories the H\'{e}non--Heiles system for a selection of energies $0 < E \leq \frac{1}{6}$, where the surface of section is taken to be $x = 0$, i.e., the $(y, p\ind{_{y}})$-plane (see Section \ref{sec:chaotic_dynamical_systems}). The Hamiltonian constraint $H = E$, coupled with the fact that ${p\ind{_{x}}}^{2} \geq 0$ imply that motion is restricted to a subset of the $(y, p\ind{_{y}})$-plane given by
\begin{equation}
\label{eqn:hh_poincare_section_boundary}
\frac{1}{2} {p\ind{_{y}}}^{2} + \frac{1}{2} y^{2} - \frac{1}{3} y^{3} \leq E.
\end{equation}
The extreme values of $p\ind{_{y}}$ occur where $y = 0$, i.e., $p\ind{_{y}} = \pm \sqrt{2 E}$. Similarly, the extreme values of $y$ occur where $p\ind{_{y}} = 0$, i.e., at the two real roots of the cubic $Q(y) = \frac{1}{3} y^{3} - \frac{1}{2} y^{2} + E$ which satisfy the inequality \eqref{eqn:hh_potential_inequality} with $x = 0$. The cubic $Q(y)$ has discriminant $\Delta_{y}(Q) = \frac{1}{2} E \left( 1 - 6 E \right)$; two such roots are therefore guaranteed to exist for energies in the range $0 < E \leq \frac{1}{6}$.

On the surface $x = 0$, we can express $p\ind{_{x}}$ as a function of $y$ and $p\ind{_{y}}$ as
\begin{equation}
\label{eqn:hh_px_expression}
p\ind{_{x}} = \pm \sqrt{2 E - {p\ind{_{y}}}^{2} - y^{2} + \frac{2}{3} y^{3} }.
\end{equation}
In practice, the construction of a Poincar\'{e} section involves numerically integrating trajectories in the three-dimensional phase space for a selection of initial conditions, recording the values of $y$ and $p\ind{_{y}}$ each time the trajectory intersects $x = 0$ with $p\ind{_{x}} > 0$, i.e., taking the upper ($+$) sign in \eqref{eqn:hh_px_expression}.

\begin{figure}
\begin{center}
\subfigure[Poincar\'{e} section for $E = \frac{1}{12}$]{
\includegraphics[height=0.45\textwidth]{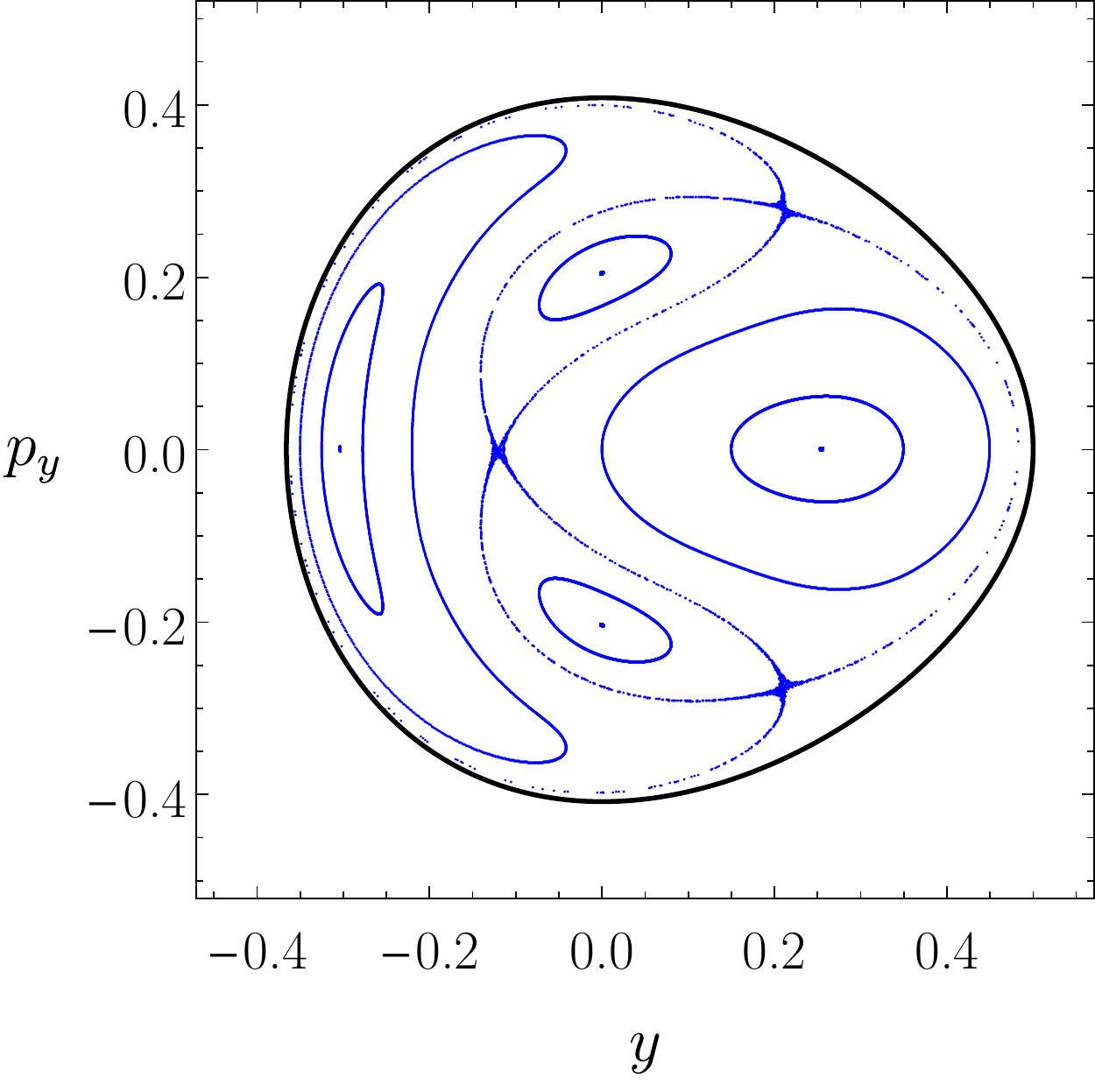} \label{fig:hh_poincare_section_e_1_12} \hspace{1em}}
\subfigure[Orbits for $E = \frac{1}{12}$]{
\includegraphics[height=0.45\textwidth]{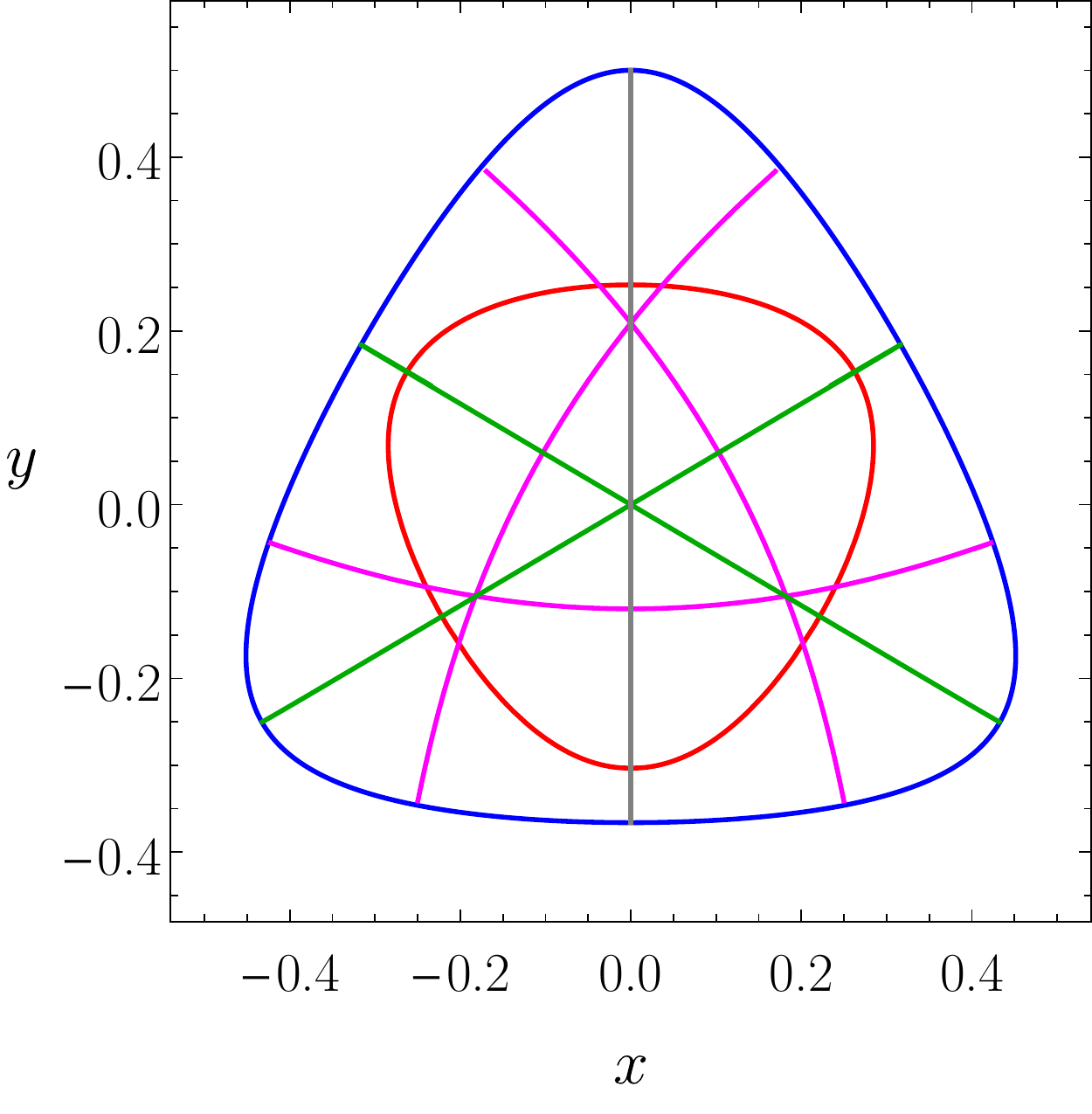} \label{fig:hh_orbits_e_1_12} \hspace{1em}}
\caption{(a) Poincar\'{e} section in the $(y, p\ind{_{y}})$-plane for the H\'{e}non--Heiles Hamiltonian system with conserved total energy $E = \frac{1}{12}$. (b) A selection of (stable and unstable) periodic orbits which lie within the closed contour $V = E = \frac{1}{12}$ [blue]: a rotational $1 : 1$ resonant orbit (or ``loop orbit'') [red]; a pair of librational $1 : 1$ resonant orbits (or ``linear orbits'') [green]; and three unstable linear orbits [magenta].}
\label{fig:hh_e_1_12}
\end{center}
\end{figure}

\begin{figure}
\begin{center}
\subfigure[Poincar\'{e} section for $E = \frac{1}{8}$]{
\includegraphics[height=0.45\textwidth]{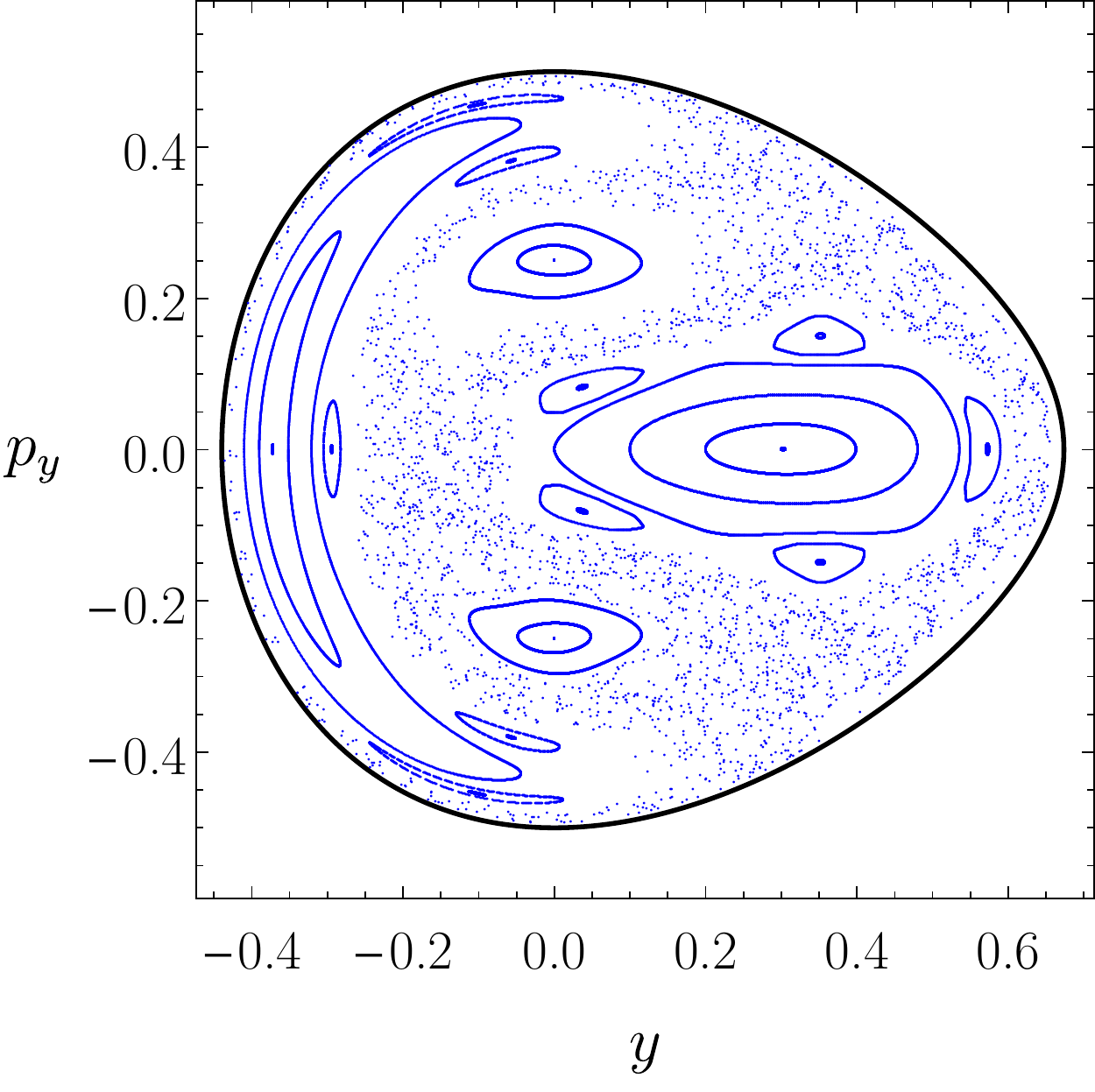} \label{fig:hh_poincare_section_e_1_8} \hspace{1em}}
\subfigure[Orbits for $E = \frac{1}{8}$]{
\includegraphics[height=0.45\textwidth]{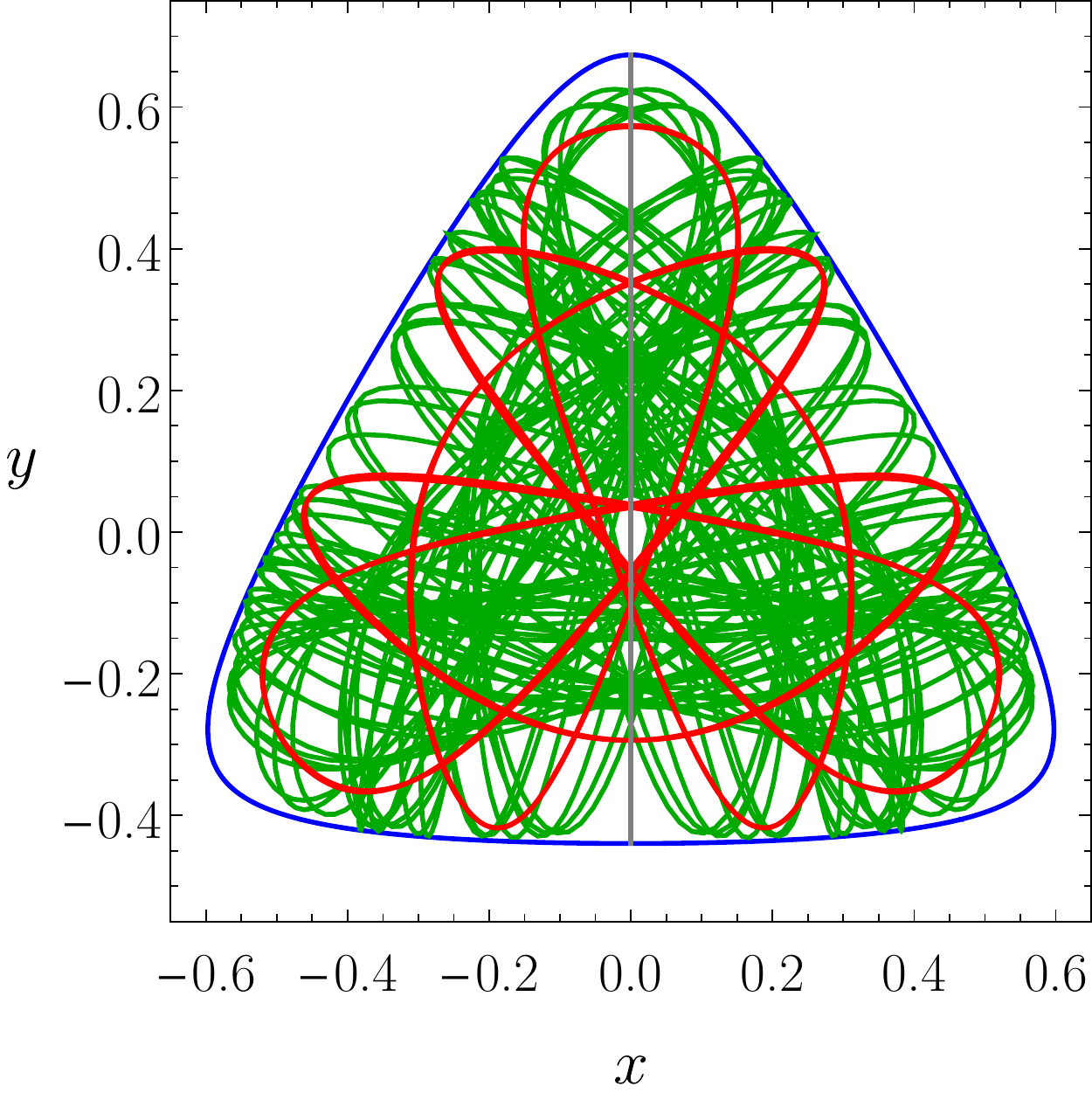} \label{fig:hh_orbits_e_1_8} \hspace{1em}}
\subfigure[Poincar\'{e} section for $E = \frac{1}{6}$]{
\includegraphics[height=0.45\textwidth]{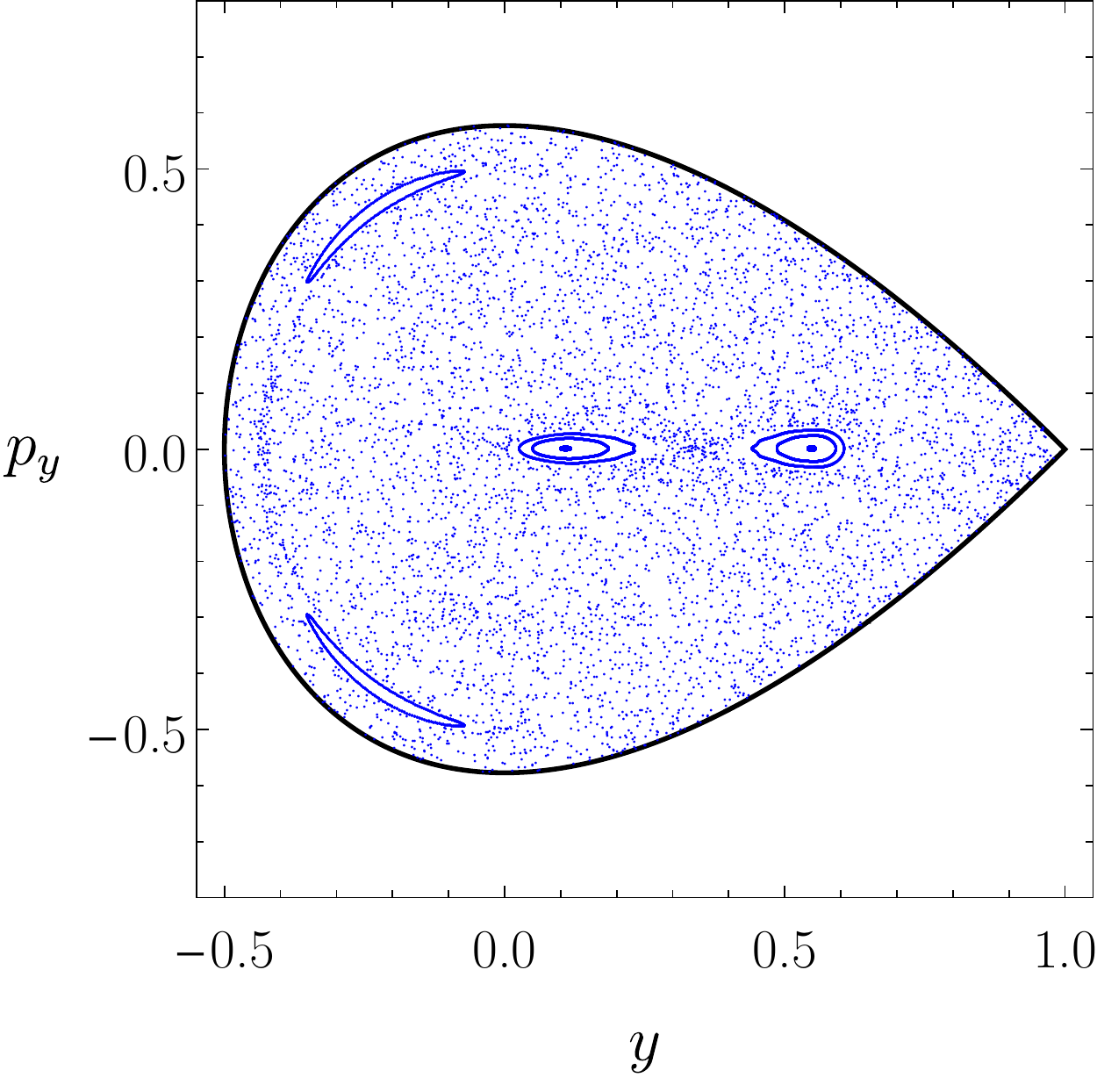} \label{fig:hh_poincare_section_e_1_6} \hspace{1em}}
\subfigure[Orbits for $E = \frac{1}{6}$]{
\includegraphics[height=0.45\textwidth]{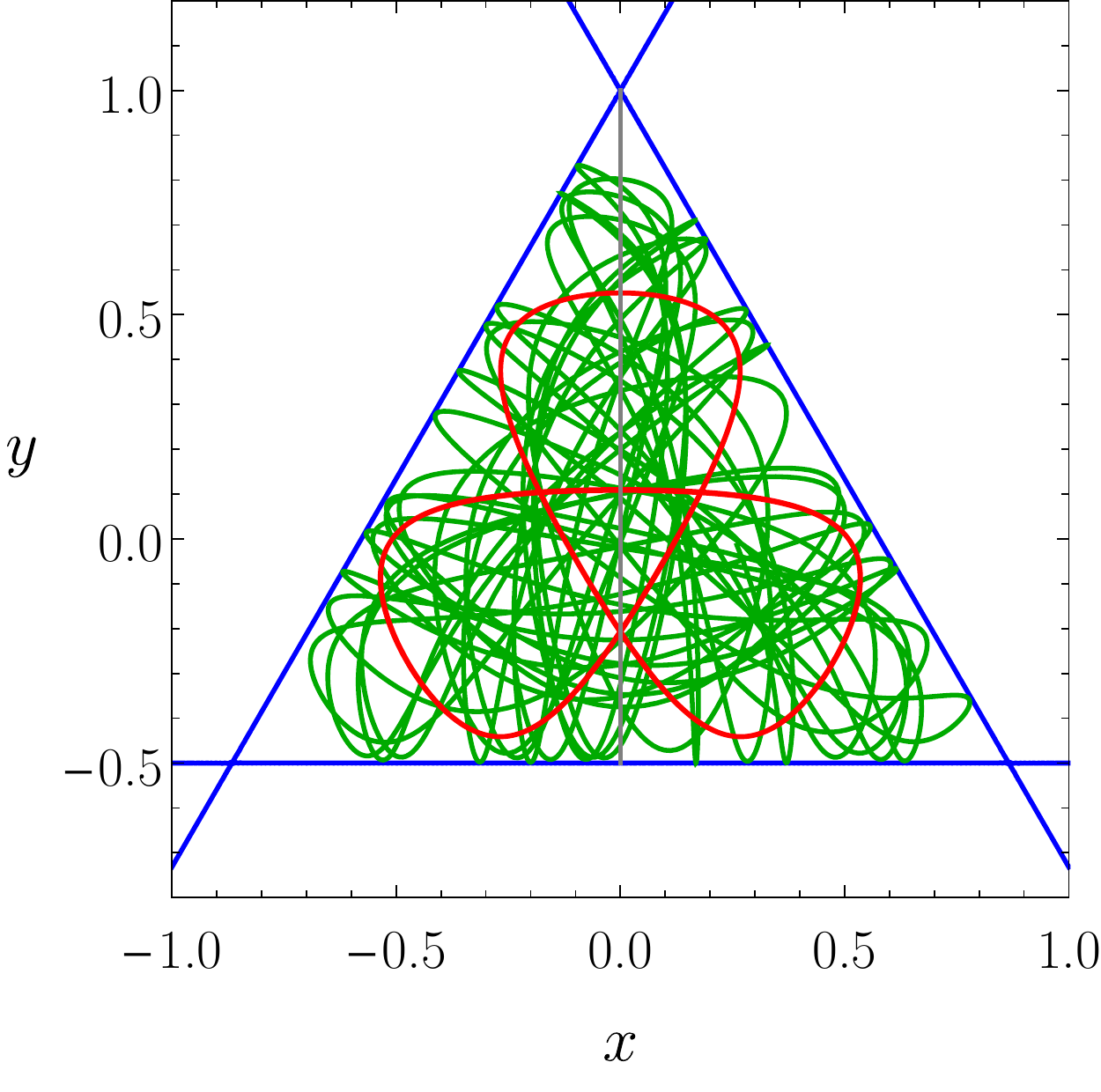} \label{fig:hh_orbits_e_1_6} \hspace{1em}}
\caption{(a) Poincar\'{e} section in the $(y, p\ind{_{y}})$-plane for $E = \frac{1}{8}$. We observe regions of regular motion around elliptic fixed points, and chaotic motion. (b) Orbits bounded by the closed contour $V = E = \frac{1}{8}$ [blue]: a $5 : 5$ resonant orbit [red]; and a chaotic orbit [green]. (c) Poincar\'{e} section for $E = \frac{1}{6}$ with chaos dominant. (d) Orbits which lie within the closed contour $V = E = \frac{1}{6}$ [blue]: a $2 : 2$ resonant orbit [red]; and a chaotic orbit [green], which wanders through the strange attractor region of phase space.}
\label{fig:hh_e_1_8_and_1_6}
\end{center}
\end{figure}

Figure \ref{fig:hh_poincare_section_e_1_12} shows an example of a Poincar\'{e} section for the H\'{e}non--Heiles system with $E = \frac{1}{12}$. There are four regions which contain closed oval-shaped curves, surrounding four (stable) elliptic fixed points. These four regions are separated by a continuous curve, which has self-intersections at three (unstable) hyperbolic fixed points. (The Poincar\'{e} section is clearly symmetric about the $y$-axis, due to the fact that the Hamiltonian \eqref{eqn:hh_hamiltonian} is invariant under the transformation $p\ind{_{y}} \mapsto - p\ind{_{y}}$.) In Figure \ref{fig:hh_orbits_e_1_12}, we present a selection of periodic orbits, bounded by the closed contour $V = E = \frac{1}{12}$, shown in blue. The red curve is a ``rotational'' orbit, corresponding to the pair of elliptic fixed points in the plane $p\ind{_{y}} = 0$. The green curves are ``librational'' orbits, corresponding to the non-planar elliptic fixed points. The rotational and librational orbits are partitioned by a ``separatrix'', which intersects itself at the three elliptic fixed points; the corresponding unstable periodic orbits are shown as magenta curves.

We now consider the effect of increasing the energy $E$, towards the threshold $E = \frac{1}{6}$. In Figure \ref{fig:hh_e_1_8_and_1_6}, we present a selection of Poincar\'{e} sections and orbits for $E \in \left\{ \frac{1}{8}, \frac{1}{6} \right\}$.

Figure \ref{fig:hh_poincare_section_e_1_8} shows that, when $E = \frac{1}{8}$, there exist regions containing closed oval-shaped curves which surround elliptic fixed points. However, these regular ``islands'' are no longer demarcated by a smooth separatrix: the intermediate regions are filled with irregular orbits. In fact, all of the points in the ``speckled'' region of Figure \ref{fig:hh_poincare_section_e_1_8} are generated by the repeated intersection of a single orbit with the surface of section. In Figure \ref{fig:hh_orbits_e_1_8}, we present a selection of orbits for $E = \frac{1}{8}$. A resonant periodic orbit is shown in red; and an irregular (chaotic) orbit is shown in green.

Increasing the energy to its maximum value $E = \frac{1}{6}$, we see that chaos becomes dominant. Figure \ref{fig:hh_poincare_section_e_1_6} shows the Poincar\'{e} section for $E = \frac{1}{6}$. We see that, in this case, chaotic orbits fill almost all of the available phase space; however, there exist very small regular islands, which surround elliptic fixed points. We show a selection of orbits for $E = \frac{1}{6}$ in Figure \ref{fig:hh_orbits_e_1_6}. The stable periodic orbit -- corresponding to the elliptic fixed point of the Poincar\'{e} map -- is shown in red. We also show an example of a chaotic orbit in green.

More comprehensive discussions of bounded motion in the H\'{e}non--Heiles Hamiltonian system as well as a detailed description of the Poincar\'{e} sections can be found in \cite{Berry1978, GoldsteinPooleSafko2002, Zotos2015}, for example.
%

\chapter{Bret\'{o}n--Manko--Aguilar di-hole solution} \label{chap:appendix_e}

The $N = 2$ Bret\'{o}n--Manko--Aguilar \cite{BretonMankoAguilarSanchez1998} solution describes a pair of aligned Reissner--Nordstr\"{o}m particles in static equilibrium. The geometry is a solution to the Einstein--Maxwell equations, and is described by the static Weyl line element
\begin{equation}
\ed s^{2} = - f \ed t^{2} + \frac{1}{f} \left[ e^{2 \gamma} \left( \ed \rho^{2} + \ed z^{2} \right) + \rho^{2} \ed \phi^{2} \right],
\end{equation}
where $f = f(\rho, z)$, $\gamma = \gamma(\rho, z)$, and the only non-zero component of the electromagnetic gauge potential is $A\ind{_{t}}(\rho, z)$.

Manko \cite{Manko2007} presents the general five-parameter Bret\'{o}n--Manko--Aguilar solution in terms of the ``physical parameters'': the black hole masses $M_{\pm}$; the black hole charges $Q_{\pm}$; and the separation between the centres $d$. In the so-called physical parametrisation, the metric functions and electrostatic potential take the form
\begin{equation}
f = \frac{A^{2} - B^{2} + C^{2}}{\left( A + B \right)^{2}}, \qquad
e^{2 \gamma} = \frac{A^{2} - B^{2} + C^{2}}{16 \sigma_{+}^{2} \sigma_{-}^{2} (\nu + 2 \kappa)^{2} r_{1} r_{2} r_{3} r_{4}}, \qquad
A\ind{_{t}} = \frac{C}{A + B},
\end{equation}
where the functions $A$, $B$ and $C$ are given by
\begin{align}
\begin{split}
A &= \sigma_{+} \sigma_{-} \left[ \nu \left(r_{1} + r_{2}\right) \left(r_{3} + r_{4}\right) + 4 \kappa \left(r_{1} r_{2} + r_{3} r_{4}\right) \right] \\
& \qquad - \left( \mu^{2} \nu - 2 \kappa^{2} \right) \left(r_{1} - r_{2}\right) \left(r_{3} - r_{4}\right),
\end{split}
\\
\begin{split}
B &= 2 \sigma_{+} \sigma_{-} \left[ \left( \nu M_{+} + 2 \kappa M_{-} \right) \left(r_{1} + r_{2}\right) + \left( \nu M_{-} + 2 \kappa M_{+} \right) \left(r_{3} + r_{4}\right) \right] \\
& \qquad - 2 \sigma_{+} \left[ \nu \mu \left(Q_{-} + \mu\right) + 2 \kappa \left( d M_{-} + \mu Q_{+} - \mu^2 \right) \right] \left(r_{1} - r_{2}\right) \\
& \qquad \qquad - 2 \sigma_{-} \left[ \nu \mu \left(Q_{+} - \mu\right) - 2 \kappa \left( d M_{+} - \mu Q_{-} - \mu^2 \right) \right] \left(r_{3} - r_{4}\right),
\end{split}
\\
\begin{split}
C &=
2 \sigma_{+} \sigma_{-} \left\{ \left[ \nu (Q_{+} - \mu) +
2 \kappa (Q_{-} + \mu) \right] (r_{1} + r_{2}) \right. \\
& \qquad \left. + \left[ \nu (Q_{-} + \mu) +
2 \kappa (Q_{+} - \mu) \right] (r_{3} + r_{4}) \right\} \\
& \qquad \qquad - 2 \sigma_{+} (\mu \nu M_{-} +
2 \kappa (\mu M_{+} + d Q_{-} + \mu d)) (r_{1} - r_{2}) \\
& \qquad \qquad \qquad - 2 \sigma_{-} \left[ \mu \nu M_{+} +
2 \kappa (\mu M_{-} - d Q_{+} + \mu d) \right] (r_{3} - r_{4}) ,
\end{split}
\end{align}
with
\begin{align}
\nu &= d^{2} - \sigma_{+}^{2} - \sigma_{-}^{2} + 2 \mu^{2}, &
\kappa &= M_{+} M_{-} - (Q_{+} - \mu) (Q_{-} + \mu), \\
\mu &= \frac{M_{-} Q_{+} - M_{+} Q_{-}}{M_{+} + M_{-} + d}, &
\sigma_{\pm} &= \sqrt{M_{\pm}^2 - Q_{\pm}^2 \pm 2\mu Q_{\pm}}, \\
r_{1} &= \sqrt{\rho^{2} + \left(z - \frac{d}{2} - \sigma_{-} \right)^{2}}, &
r_{2} &= \sqrt{\rho^{2} + \left(z - \frac{d}{2} + \sigma_{-} \right)^{2}}, \\
r_{3} &= \sqrt{\rho^{2} + \left(z + \frac{d}{2} - \sigma_{+} \right)^{2}}, &
r_{4} &= \sqrt{\rho^{2} + \left(z + \frac{d}{2} + \sigma_{+} \right)^{2}} .
\end{align}

In \cite{Manko2007}, Manko derives a concise formula for the force imparted by the Weyl strut which holds the sources in static equilibrium:
\begin{equation}
\mathcal{F} = \frac{\kappa}{\nu - 2 \kappa} = \frac{M_{+} M_{-} - (Q_{+} - \mu) (Q_{-} + \mu)}{d^{2} - \left( M_{+} + M_{-} \right)^{2} + \left( Q_{+} + Q_{-} \right)^{2} }.
\end{equation} 

\chapter{Tetrad calculations} \label{chap:appendix_d}

\section{Identities for Killing objects and tetrads}

Using the properties of the Killing objects from Chapter \ref{chap:geometric_optics_kerr}, we prove the identities  \eqref{eqn:bivector_identity_1} and \eqref{eqn:bivector_identity_2} for the self-dual bivector $\mathscr{F}\ind{_{a b}} = f\ind{_{a b}} - i h\ind{_{a b}}$, where $f\ind{_{a b}}$ is the Killing--Yano tensor and $h\ind{_{a b}}$ is the closed conformal Killing--Yano tensor.

Firstly, recall that $f\ind{_{a c}} h\ind{_{b}^{c}} = h\ind{_{a c}} f\ind{_{b}^{c}} = a r \cos{\theta} g\ind{_{a b}}$, and $K\ind{_{a b}} - Q\ind{_{a b}} = \left( r^{2} - a^{2} \cos^{2}{\theta} \right) g\ind{_{a b}}$. The ``square'' of $\mathscr{F}\ind{_{a b}}$ is then
\begin{align}
\mathscr{F}\ind{_{a c}} \mathscr{F}\ind{_{b}^{c}}
&= \left( f\ind{_{a c}} - i h\ind{_{a c}} \right) \left( f\ind{_{b}^{c}} - i h\ind{_{b}^{c}} \right) \\
&= f\ind{_{a c}} f\ind{_{b}^{c}} - h\ind{_{a c}} h\ind{_{b}^{c}} - i \left( f\ind{_{a c}} h\ind{_{b}^{c}} + h\ind{_{a c}} f\ind{_{b}^{c}} \right) \\
&= K\ind{_{a b}} - Q\ind{_{a b}} - 2 i a r \cos{\theta} g\ind{_{a b}} \\
&= \left( r^{2} - a^{2} \cos^{2}{\theta} - 2 i a r \cos{\theta} \right) g\ind{_{a b}} \\
&= \left( r - i a \cos{\theta} \right)^{2} g\ind{_{a b}}.
\end{align}
This proves \eqref{eqn:bivector_identity_1}. Similarly, the product of $\mathscr{F}\ind{_{a b}}$ with its complex conjugate is
\begin{align}
\mathscr{F}\ind{_{a c}} \overline{\mathscr{F}}\vp{\mathscr{F}}\ind{_{b}^{c}}
&= \left( f\ind{_{a c}} - i h\ind{_{a c}} \right) \left( f\ind{_{b}^{c}} + i h\ind{_{b}^{c}} \right) \\
&= f\ind{_{a c}} f\ind{_{b}^{c}} + h\ind{_{a c}} h\ind{_{b}^{c}} + i \left( f\ind{_{a c}} h\ind{_{b}^{c}} - h\ind{_{a c}} f\ind{_{b}^{c}} \right) \\
&= K\ind{_{a b}} + Q\ind{_{a b}}.
\end{align}
This proves result \eqref{eqn:bivector_identity_2}.

We now derive the transport equation \eqref{eqn:transport_equation_m_tilde}  by applying the derivative operator $\wt{D} = \wt{k}\ind{^{a}} \nabla\ind{_{a}}$ to the tetrad legs $\wt{m}\ind{^{a}}$ and $\wt{n}\ind{^{a}}$, which are defined in terms of the self-dual bivector $\mathscr{F}\ind{_{a b}}$ as
\begin{equation}
\wt{m}\ind{^{a}} = \frac{1}{\sqrt{2 K}} \mathscr{F}\ind{^{a}_{b}} \wt{k}\ind{^{b}}, \qquad
\wt{n}\ind{^{a}} = \frac{1}{2 K} \mathscr{F}\ind{^{a b}} \ol{\mathscr{F}}\ind{_{b c}} \wt{k}\ind{^{c}}.
\end{equation}
First, recall that $f\ind{_{a b ; c}} = f\ind{_{[ a b ; c ]}}$, and $h\ind{_{a b ; c}} = g\ind{_{a c}} \xi\ind{_{b}} - g\ind{_{c b}} \xi\ind{_{a}}$, where $\xi\ind{^{a}} = {\xi_{(t)}}\ind{^{a}}$ is the primary Killing vector field. We also note that $\wt{k}\ind{^{a}} \xi\ind{_{a}} = - E$, where $E$ is the photon's conserved energy. Using these results, the transport equation for $\wt{m}\ind{^{a}}$ is then
\begin{align}
\wt{D} \wt{m}\ind{_{a}}
&= \frac{1}{\sqrt{2 K}} \wt{k}\ind{^{c}} \left( \mathscr{F}\ind{_{a b}} \wt{k}\ind{^{b}} \right)\ind{_{; c}} \\
&= \frac{1}{\sqrt{2 K}} \mathscr{F}\ind{_{a b ; c}} \wt{k}\ind{^{b}} \wt{k}\ind{^{c}} \\
&= \frac{1}{\sqrt{2 K}} \left( f\ind{_{a b ; c}} - i h\ind{_{a b ; c}} \right) \wt{k}\ind{^{b}} \wt{k}\ind{^{c}} \\
&= \frac{1}{\sqrt{2 K}} \left[ f\ind{_{[ a b ; c ]}} - i \left( g\ind{_{a c}} \xi\ind{_{b}} - g\ind{_{c b}} \xi\ind{_{a}} \right) \right] \wt{k}\ind{^{b}} \wt{k}\ind{^{c}} \\
&= - \frac{i}{\sqrt{2 K}} \left( \wt{k}\ind{^{b}} \xi\ind{_{b}} \right) \wt{k}\ind{_{a}} \\
&= \frac{i E}{\sqrt{2 K}} \wt{k}\ind{_{a}}.
\end{align}
This is precisely \eqref{eqn:transport_equation_m_tilde}.

\section{Relationship with Marck's tetrads}

In Section \ref{sec:null_tetrad_construction} we claim that the complex null tetrad $\{ \wt{k}\ind{^{a}}, \wt{n}\ind{^{a}}, \wt{m}\ind{^{a}}, \ol{\wt{m}}\vp{m}\ind{^{a}} \}$ constructed from the symmetries of Kerr spacetime is equivalent to that constructed using the legs of Marck's quasi-orthonormal tetrad. In particular,
\begin{equation}
\label{eqn:equivalent_marck_tetrad}
\wt{n}\ind{^{a}} = \frac{1}{2 K} \mathscr{F}\ind{^{a b}} \ol{\mathscr{F}}\ind{_{b c}} \wt{k}\ind{^{c}} = - \mqolegup{3}{a}, \qquad \wt{m}\ind{^{a}} = \frac{1}{\sqrt{2 K}} \mathscr{F}\ind{^{a}_{b}} \wt{k}\ind{^{b}} = \frac{1}{\sqrt{2}} \left( \mqolegup{2}{a} + i \mqolegup{1}{a} \right).
\end{equation}
This is best demonstrated by projecting onto Carter's tetrad as follows. First, we note that $\wt{k}\ind{^{a}} = \mqolegup{0}{a}$ by definition. For notational simplicity, we introduce the coordinate $y = a \cos{\theta}$.\footnote{In fact, $y = a \cos{\theta}$ is one of the so-called \emph{canonical coordinates} \cite{FrolovKrtousKubiznak2017}.} When projected onto Carter's symmetric tetrad, the Killing objects take the form \cite{FrolovKrtousKubiznak2017}
\begin{align}
f\ind{_{(\alpha) (\beta)}} &=
\left[
\begin{array}{*4{C{1.1em}}}
0 & - y & 0 & 0 \\
y & 0 & 0 & 0 \\
0 & 0 & 0 & - r \\
0 & 0 & r & 0
\end{array}
\right]
,
&
h\ind{_{(\alpha) (\beta)}} &=
\left[
\begin{array}{*4{C{1.1em}}}
0 & r & 0 & 0 \\
- r & 0 & 0 & 0 \\
0 & 0 & 0 & - y \\
0 & 0 & y & 0
\end{array}
\right]
,
\\
K\ind{_{(\alpha) (\beta)}} &=
\left[
\begin{array}{*4{C{1.1em}}}
y^{2} & 0 & 0 & 0 \\
0 & - y^{2} & 0 & 0 \\
0 & 0 & r^{2} & 0 \\
0 & 0 & 0 & r^{2}
\end{array}
\right]
,
&
Q\ind{_{(\alpha) (\beta)}} &=
\left[
\begin{array}{*4{C{1.1em}}}
r^{2} & 0 & 0 & 0 \\
0 & - r^{2} & 0 & 0 \\
0 & 0 & y^{2} & 0 \\
0 & 0 & 0 & y^{2}
\end{array}
\right]
.
\end{align}

Recall that $\Sigma = r^{2} + a^{2} \cos^{2}{\theta} = r^{2} + y^{2}$. Using the definition \eqref{eqn:equivalent_marck_tetrad} and the identity $\mathscr{F}\ind{_{a c}} \ol{\mathscr{F}}\vp{\mathscr{F}}\ind{_{b}^{c}} = K\ind{_{a b}} + Q\ind{_{a b}}$, we find that the leg $n\ind{^{a}}$, when projected onto Carter's symmetric tetrad, takes the form
\begin{align}
\wt{n}\ind{^{(\alpha)}}
&= - \frac{1}{2 K} \left( K\ind{^{(\alpha)}_{(\beta)}} + Q\ind{^{(\alpha)}_{(\beta)}} \right) \mqolegup{0}{(\beta)} \\
&=
- \frac{\Sigma}{2 K}
\left[
\begin{array}{*4{C{1.1em}}}
-1 & 0 & 0 & 0 \\
0 & -1 & 0 & 0 \\
0 & 0 & 1 & 0 \\
0 & 0 & 0 & 1
\end{array}
\right]
\left[
\begin{array}{c}
\mqolegup{0}{(0)} \\
\mqolegup{0}{(1)} \\
\mqolegup{0}{(2)} \\
\mqolegup{0}{(3)}
\end{array}
\right]
\\
&=
- \frac{\Sigma}{2 K}
\left[
\begin{array}{c}
- \mqolegup{0}{(0)} \\
- \mqolegup{0}{(1)} \\
\mqolegup{0}{(2)} \\
\mqolegup{0}{(3)}
\end{array}
\right].
\label{eqn:n_equals_lambda_3_last}
\end{align}
By comparing \eqref{eqn:n_equals_lambda_3_last} with the components of $\mqolegup{3}{(\alpha)}$ given in Section \ref{sec:marck_tetrads}, we see that $n\ind{^{a}} = - \mqolegup{3}{a}$, as claimed.

Similarly, when projected onto Carter's tetrad, we find that the leg $\wt{m}\ind{^{a}}$ may be expressed in the form
\begin{align}
\wt{m}\ind{^{(\alpha)}}
&= \frac{1}{\sqrt{2 K}} \left( f\ind{^{(\alpha)}_{(\beta)}} - i h\ind{^{(\alpha)}_{(\beta)}} \right) \mqolegup{0}{(\beta)} \\
&=
\frac{1}{\sqrt{2 K}}
\left(
\left[
\begin{array}{*4{C{1.1em}}}
0 & y & 0 & 0 \\
- y & 0 & 0 & 0 \\
0 & 0 & 0 & - r \\
0 & 0 &  r & 0
\end{array}
\right]
+ i
\left[
\begin{array}{*4{C{1.1em}}}
0 & r & 0 & 0 \\
r & 0 & 0 & 0 \\
0 & 0 & 0 & y \\
0 & 0 &  -y & 0
\end{array}
\right]
\right)
\left[
\begin{array}{c}
\mqolegup{0}{(0)} \\
\mqolegup{0}{(1)} \\
\mqolegup{0}{(2)} \\
\mqolegup{0}{(3)}
\end{array}
\right]
\\
&=
\frac{1}{\sqrt{2}}
\left(
\frac{1}{\sqrt{K}}
\left[
\begin{array}{c}
y \mqolegup{0}{(1)} \\
y \mqolegup{0}{(0)} \\
- r \mqolegup{0}{(3)} \\
r \mqolegup{0}{(2)}
\end{array}
\right]
+
\frac{i}{\sqrt{K}}
\left[
\begin{array}{c}
r \mqolegup{0}{(1)} \\
r \mqolegup{0}{(0)} \\
y \mqolegup{0}{(3)} \\
- y \mqolegup{0}{(2)}
\end{array}
\right]
\right).
\label{eqn:m_equals_lambda_1_2_last}
\end{align}
Comparison of \eqref{eqn:m_equals_lambda_1_2_last} with the legs $\mqolegup{1}{(\alpha)}$ and $\mqolegup{1}{(\alpha)}$ of Marck's quasi-orthonormal tetrad (Section \ref{sec:marck_tetrads}) shows that $\wt{m}\ind{^{a}} = \frac{1}{\sqrt{2}} \left( \mqolegup{2}{a} + i \mqolegup{1}{a} \right)$. Considering the real and imaginary parts of $\wt{m}\ind{^{a}}$ separately, we see that
\begin{equation}
\mqolegup{2}{a} = \frac{1}{\sqrt{K}} f\ind{^{a}_{b}} k\ind{^{b}}, \qquad \mqolegup{1}{a} = - \frac{1}{\sqrt{K}} h\ind{^{a}_{b}} k\ind{^{b}}.
\end{equation}
The first of these identities is unsurprising: Marck \cite{Marck1983b} defines the leg $\mqolegup{2}{a}$ to be the unit spacelike vector which is parallel to the vector $v\ind{^{a}} = f\ind{^{a}_{b}} k\ind{^{a}}$.

\section{Relationship with Kinnersley's tetrad}
\label{sec:kinnersley_marck}

We begin with the well-known \emph{Kinnersley tetrad} $\{ \mathsfit{k}\ind{^{a}}, \mathsfit{n}\ind{^{a}}, \mathsfit{m}\ind{^{a}}, \overline{\mathsfit{m}}\ind{^{a}} \}$ \cite{Kinnersley1969}, which in Boyer--Lindquist coordinates has components
\begin{align}
{\mathsfit{k}}\ind{^{a}} \partial\ind{_{a}} &= \frac{1}{\Delta} \left[ \left(r^2 + a^2 \right) \partial\ind{_{t}} + \Delta \, \partial\ind{_{r}} + a \, \partial\ind{_{\phi}} \right] , \\
{\mathsfit{n}}\ind{^{a}} \partial\ind{_{a}} &= \frac{1}{2 \Sigma} \left[ \left(r^2 + a^2 \right) \partial\ind{_{t}} - \Delta \, \partial\ind{_{r}} + a \, \partial\ind{_{\phi}} \right] , \\
{\mathsfit{m}}\ind{^{a}} \partial\ind{_{a}} &= \frac{1}{\sqrt{2} \left( r + i a \cos{\theta} \right)} \left( i a \sin{\theta} \, \partial\ind{_{t}} + \partial\ind{_{\theta}} + \frac{i}{\sin{\theta}} \partial\ind{_{\phi}} \right), \\
\ol{{\mathsfit{m}}}\ind{^{a}} \partial\ind{_{a}} &= \frac{1}{\sqrt{2} \left( r - i a \cos{\theta} \right)} \left( - i a \sin{\theta} \, \partial\ind{_{t}} + \partial\ind{_{\theta}} - \frac{i}{\sin{\theta}} \partial\ind{_{\phi}} \right),
\end{align}
where $\Sigma(r, \theta) = r^{2} + a^{2} \cos^{2}{\theta}$ and $\Delta(r) = r^{2} - 2 M r + a^{2}$. All inner products of the Kinnersley tetrad are zero, except $\mathsfit{k}\ind{^{a}} \mathsfit{n}\ind{_{a}} = -1$ and $\mathsfit{m}\ind{^{a}} \overline{\mathsfit{m}}\ind{_{a}} = 1$, i.e., it is a complex null tetrad.

We define the \emph{symmetrised Kinnersley tetrad} as
\begin{equation}
\label{eqn:kinnersley_symmetrised}
\hat{k}\ind{^{a}} = \sqrt{\frac{\Delta}{2 \Sigma}} \mathsfit{k}\ind{^{a}}, \qquad \hat{n}\ind{^{a}} = \sqrt{\frac{2 \Sigma}{\Delta}} \mathsfit{n}\ind{^{a}}, \qquad \hat{m}\ind{^{a}} = \mathsfit{m}\ind{^{a}}, \qquad
\overline{\hat{m}}\vp{m}\ind{^{a}} = \overline{\mathsfit{m}}\ind{^{a}}.
\end{equation}
This is a Lorentz boost in the $(\mathsfit{k}, \mathsfit{n})$-plane; see the transformation \eqref{eqn:lt_k_n} in Chapter \ref{chap:general_relativity}. Clearly, the symmetrised null tetrad satisfies the same inner product relationships as the standard Kinnersley tetrad.

One may express the ``tilded'' complex null tetrad $\{ k\ind{^{a}}, \widetilde{n}\ind{^{a}}, \widetilde{m}\ind{^{a}}, \overline{\widetilde{m}}\vp{m}\ind{^{a}} \}$ (see Chapter \ref{chap:geometric_optics_kerr}) in terms of the symmetrised Kinnersley tetrad as
\begin{align}
k\ind{^{a}} &= A \left( e^{+\gamma} \hat{k}\ind{^{a}} + e^{-\gamma} \hat{n}\ind{^{a}} + e^{- i \beta} \hat{m}\ind{^{a}} + e^{+ i \beta} \overline{\hat{m}}\vp{m}\ind{^{a}} \right), \label{eqn:kin_marck_k} \\
\widetilde{n}\ind{^{a}} &= \frac{1}{4 A} \left( e^{+\gamma} \hat{k}\ind{^{a}} + e^{-\gamma} \hat{n}\ind{^{a}} - e^{- i \beta} \hat{m}\ind{^{a}} - e^{+ i \beta} \overline{\hat{m}}\vp{m}\ind{^{a}} \right), \label{eqn:kin_marck_n} \\
\widetilde{m}\ind{^{a}} &= \frac{i e^{i \alpha}}{2} \left( e^{+\gamma} \hat{k}\ind{^{a}} - e^{-\gamma} \hat{n}\ind{^{a}} + e^{- i \beta} \hat{m}\ind{^{a}} - e^{+ i \beta} \overline{\hat{m}}\vp{m}\ind{^{a}} \right), \label{eqn:kin_marck_m} \\
\ol{\widetilde{m}}\vp{m}\ind{^{a}} &= -\frac{i e^{- i \alpha}}{2} \left( e^{+\gamma} \hat{k}\ind{^{a}} - e^{-\gamma} \hat{n}\ind{^{a}} - e^{- i \beta} \hat{m}\ind{^{a}} + e^{+ i \beta} \overline{\hat{m}}\vp{m}\ind{^{a}} \right), \label{eqn:kin_marck_mbar}
\end{align}
where we have made the definitions
\begin{align}
A &= \sqrt{\frac{K}{2 \Sigma}}, \label{eqn:kinn_capital_a} \\
e^{\pm \gamma} &= \frac{\Sigma}{\sqrt{K \Delta}} \left[ \frac{E \left( r^{2} + a^{2} \right) - a L\ind{_{z}}}{\Sigma} \pm \dot{r} \right], \\
e^{i \beta} &= e^{i \alpha} \sqrt{\frac{Z}{\overline{Z}}}, \label{eqn:kinn_exp_i_beta} \\
e^{i \alpha} &= \sqrt{\frac{r - i a \sin{\theta}}{r + i a \sin{\theta}}} = \frac{r - i a \sin{\theta}}{\sqrt{\Sigma}} = \frac{\sqrt{\Sigma}}{r + i a \sin{\theta}}, \label{eqn:kinn_exp_i_alpha} \\
Z &= \dot{\theta} + \frac{i}{\Sigma} \left( \frac{L\ind{_{z}}}{\sin{\theta}} - a E \sin{\theta} \right),
\end{align}
and $K = K\ind{_{a b}} k\ind{^{a}} k\ind{^{b}}$ is Carter's constant. It is quick to show that $Z \overline{Z} = \frac{K}{\Sigma^{2}}$, so $\frac{Z}{\overline{Z}} = \frac{\Sigma^{2} Z^{2}}{K}$. We can therefore write \eqref{eqn:kinn_exp_i_beta} as
\begin{equation}
e^{i \beta} = \frac{e^{i \alpha}}{\sqrt{K}} \left[ \Sigma \dot{\theta} + i \left( \frac{L\ind{_{z}}}{\sin{\theta}} - a E \sin{\theta} \right) \right].
\end{equation}

The transformation \eqref{eqn:kin_marck_k}--\eqref{eqn:kin_marck_m} can be achieved by performing a series of Lorentz transformations of the symmetrised Kinnersley tetrad $\{ \hat{k}, \hat{n}, \hat{m}, \ol{\hat{m}} \}$. First, perform a boost in the $(\hat{k}, \hat{n})$-plane and a spatial rotation in the $(\hat{m}, \ol{\hat{m}})$-plane, such that $k_{1} = e^{+ \gamma} \hat{k}$, $n_{1} = e^{- \gamma} \hat{n}$ and $m_{1} = e^{+ i \beta} \ol{\hat{m}}$. (Note that we have changed the handedness in the $(\hat{{m}}, \ol{\hat{m}})$-plane.) Second, perform a null rotation which keeps $n_{1}$ fixed of the form $k_{2} = k_{1} + \ol{\mathcal{E}} m_{1} + \mathcal{E} \ol{m}_{1} + \mathcal{E} \ol{\mathcal{E}} n_{1}$, $n_{2} = n_{1}$, $m_{2} = m_{1} + \mathcal{E} n_{1}$, with $\mathcal{E} = 1$. Third, perform a null rotation which keeps $k_{2}$ fixed of the form $k_{3} = k_{2}$, $n_{3} = n_{2} + \ol{\mathcal{B}} m_{1} + \mathcal{B} \ol{m}_{1} + \mathcal{B} \ol{\mathcal{B}} k_{1}$, $m_{3} = m_{2} + \mathcal{B} k_{2}$, with $\mathcal{B} = - \frac{1}{2}$. Finally, perform a boost in the $(k_{3}, n_{3})$-plane and a spatial rotation in the $(m_{3}, \ol{m}_{3})$-plane, such that $\wt{k} = A k_{3}$, $\wt{n} = \frac{1}{A} n_{3}$ and $\wt{m} = - i e^{i \alpha} m_{3}$, where $A$ and $e^{i \alpha}$ are defined in \eqref{eqn:kinn_capital_a} and \eqref{eqn:kinn_exp_i_alpha}, respectively.

The relationships \eqref{eqn:kin_marck_k}--\eqref{eqn:kin_marck_m} permit us to calculate quantities in terms of the symmetrised Kinnersley tetrad $\{ \hat{k}, \hat{n}, \hat{m}, \ol{\hat{m}} \}$, and map them to the ``tilded'' Marck tetrad $\{ k, \widetilde{n}, \widetilde{m}, \ol{\widetilde{m}} \}$. Transforming from the ``tilded'' tetrad to the parallel-transported tetrad using the Lorentz transformation \eqref{eqn:parallel_transport_lorentz_transformation} with null rotation parameter \eqref{eqn:null_rotation_parameter} is then straightforward.

As an example, let us calculate the Weyl scalars in the ``tilded'' frame by transforming from the Kinnersley tetrad. It is well known that Kerr spacetime is Petrov type D, i.e., it has two double principal null directions; these are precisely the legs $\hat{k}$ and $\hat{n}$ of the symmetrised Kinnersley tetrad. This means that, in the symmetrised Kinnersley tetrad, there is only one non-vanishing Weyl scalar:
\begin{equation}
\label{eqn:weyl_kinnersley_psi_2}
\hat{\Psi}_{2} = C\ind{_{a b c d}} \hat{k}\ind{^{a}} \hat{m}\ind{^{b}} \ol{\hat{m}}\vp{m}\ind{^{c}} \hat{n}\ind{^{d}} = -\frac{M}{ \left(r - i a \cos{\theta} \right)^{3} }.
\end{equation}
One may now employ the sequence of Lorentz transformations described above to calculate the Weyl scalars in the ``tilded'' frame. We note that, in changing from the frame $\{ \hat{k}, \hat{n}, \hat{m}, \ol{\hat{m}} \}$ to the frame $\{ k_{1}, n_{1}, m_{1}, \ol{m}_{1} \}$ we interchanged $m_{1}$ and $\ol{m}_{1}$; the only non-vanishing Weyl scalar in the latter frame is therefore $\Psi_{2}^{(1)} = C\ind{_{a b c d}} {k_{1}}\ind{^{a}} {m_{1}}\ind{^{b}} {\ol{m}_{1}}\ind{^{c}} {n_{1}}\ind{^{d}} = C\ind{_{a b c d}} \hat{k}\ind{^{a}} \ol{\hat{m}}\vp{m}\ind{^{b}} \hat{m}\ind{^{c}} \hat{n}\ind{^{d}} = \ol{\hat{\Psi}}_{2}$, which is the complex conjugate of \eqref{eqn:weyl_kinnersley_psi_2}. Applying the sequence of Lorentz transformations, we arrive at the Weyl scalars in the tilded basis:
\begin{alignat}{2}
\wt{\Psi}_{0} &= C\ind{_{a b c d}} \wt{k}\ind{^{a}} \wt{m}\ind{^{b}} \wt{k}\ind{^{c}} \wt{m}\ind{^{d}} &&= - A^{2} e^{2 i \alpha} \left( 6 \mathcal{E}^{2} \right) \ol{\hat{\Psi}}_{2}, \label{eqn:tilde_weyl_kinnersley_transformation_0} \\
\wt{\Psi}_{1} &= C\ind{_{a b c d}} \wt{k}\ind{^{a}} \wt{n}\ind{^{b}} \wt{k}\ind{^{c}} \wt{m}\ind{^{d}} &&= - A i e^{i \alpha} \left( 3 \mathcal{E} + 6 \mathcal{E}^{2} \ol{\mathcal{B}} \right) \ol{\hat{\Psi}}_{2} , \\
\wt{\Psi}_{2} &= C\ind{_{a b c d}} \wt{k}\ind{^{a}} \wt{m}\ind{^{b}} \ol{\wt{m}}\vp{m}\ind{^{c}} \wt{n}\ind{^{d}} &&= \left( 1 + 6 \mathcal{E} \ol{\mathcal{B}} + 6 \mathcal{E}^{2} \ol{\mathcal{B}}\vp{\mathcal{B}}^{2} \right) \ol{\hat{\Psi}}_{2} , \\
\wt{\Psi}_{3} &= C\ind{_{a b c d}} \wt{k}\ind{^{a}} \wt{n}\ind{^{b}} \ol{\wt{m}}\vp{m}\ind{^{c}} \wt{n}\ind{^{d}} &&= \frac{i e^{- i \alpha}}{A} \left( 3 \ol{\mathcal{B}} + 9 \mathcal{E} \ol{\mathcal{B}}\vp{\mathcal{B}}^{2} + 6 \mathcal{E}^{2} \ol{\mathcal{B}}\vp{\mathcal{B}}^{3} \right) \ol{\hat{\Psi}}_{2} , \\
\wt{\Psi}_{4} &= C\ind{_{a b c d}} \ol{\wt{m}}\vp{m}\ind{^{a}} \wt{n}\ind{^{b}} \ol{\wt{m}}\vp{m}\ind{^{c}} \wt{n}\ind{^{d}} &&= - \frac{e^{- 2 i \alpha}}{A^{2}} \left( 6 \ol{\mathcal{B}}\vp{\mathcal{B}}^{2} + 12 \mathcal{E} \ol{\mathcal{B}}\vp{\mathcal{B}}^{3} + 6 \mathcal{E}^{2} \ol{\mathcal{B}}\vp{\mathcal{B}}^{4} \right) \ol{\hat{\Psi}}_{2} , \label{eqn:tilde_weyl_kinnersley_transformation_4}
\end{alignat}
where $\mathcal{E} = 1$ and $\mathcal{B} = - \frac{1}{2}$ are the null rotation parameters defined above. Upon simplification, the expressions agree \eqref{eqn:tilde_weyl_kinnersley_transformation_0}--\eqref{eqn:tilde_weyl_kinnersley_transformation_4} with \eqref{eqn:psi_0_tilde}--\eqref{eqn:psi_4_tilde}, which were computed by contracting the ``tilded'' tetrad legs with the Weyl tensor. Using the expressions for the Weyl scalars in the ``tilded'' frame, one may obtain the Weyl scalars in the parallel-transported basis using by applying the Lorentz transformation outlined in Section \ref{sec:go_kerr_weyl}.

\section{Calculating directional derivatives of Weyl scalars}

In Section \ref{sec:go_kerr_d_d_weyl}, we provide expressions for directional derivatives of the Weyl scalars $\Psi_{0}$ and $\Psi_{1}$ along the legs $m\ind{^{a}}$ and $\ol{m}\ind{^{a}}$ of our parallel-transported null tetrad. These involve directional derivatives of the Weyl tensor projected along tetrad legs, e.g.~\eqref{eqn:d_psi_0} features the quantity $\left( \delta C \right)\ind{_{k m k m}} = \left( m\ind{^{e}} \nabla\ind{_{e}} C\ind{_{a b c d}} \right) k\ind{^{a}} m\ind{^{b}} k\ind{^{c}} m\ind{^{d}}$. In this section, we calculate these quantities explicitly, by first calculating analogous quantities in the Kinnersley tetrad, then transforming to the ``tilded'' complex null tetrad using the transformation \eqref{eqn:kin_marck_k}--\eqref{eqn:kin_marck_m} of Section \ref{sec:kinnersley_marck}.

Let us denote directional derivatives along Kinnersley's symmetrised tetrad \eqref{eqn:kinnersley_symmetrised} by
\begin{equation}
\hat{D} = \hat{k}\ind{^{a}} \nabla\ind{_{a}}, \qquad
\mathit{\hat{\Delta}} = \hat{n}\ind{^{a}} \nabla\ind{_{a}}, \qquad
\hat{\delta} = \hat{m}\ind{^{a}} \nabla\ind{_{a}}, \qquad
\ol{\hat{\delta}} = \ol{\hat{m}}\vp{m}\ind{^{a}} \nabla\ind{_{a}}.
\end{equation}
Using a symbolic algebra package, we find
\begin{align}
( \hat{D} C\ind{_{a b c d}} ) k\ind{^{a}} \wt{m}\ind{^{b}} k\ind{^{c}} \wt{m}\ind{^{d}} &= \frac{6 i K M a \sin{\theta}}{\sqrt{2} \varrho^{7}} e^{- \gamma - i \beta} - \frac{9 K M}{\varrho^{6}} \sqrt{\frac{\Delta}{2 \Sigma}} , \\
( \mathit{\hat{\Delta}} C\ind{_{a b c d}} ) k\ind{^{a}} \wt{m}\ind{^{b}} k\ind{^{c}} \wt{m}\ind{^{d}} &= \frac{6 i K M a \sin{\theta}}{\sqrt{2} \varrho^{6} \ol{\varrho}} e^{+ \gamma + i \beta} + \frac{9 K M}{\varrho^{6}} \sqrt{\frac{\Delta}{2 \Sigma}} , \\
( \hat{\delta} C\ind{_{a b c d}} ) k\ind{^{a}} \wt{m}\ind{^{b}} k\ind{^{c}} \wt{m}\ind{^{d}} &= \frac{9 i K M a \sin{\theta}}{\sqrt{2} \varrho^{7}} - \frac{6 K M e^{+ \gamma + i \beta}}{\varrho^{6}} \sqrt{\frac{\Delta}{2 \Sigma}} , \\
( \ol{\hat{\delta}} C\ind{_{a b c d}} ) k\ind{^{a}} \wt{m}\ind{^{b}} k\ind{^{c}} \wt{m}\ind{^{d}} &= \frac{9 i K M a \sin{\theta}}{\sqrt{2} \varrho^{6} \ol{\varrho}} + \frac{6 K M e^{- \gamma - i \beta}}{\varrho^{6}} \sqrt{\frac{\Delta}{2 \Sigma}} , \\
( \hat{D} C\ind{_{a b c d}} ) k\ind{^{a}} \wt{m}\ind{^{b}} \ol{m}\ind{^{c}} \wt{m}\ind{^{d}} &= - \frac{3 A M a \sin{\theta}}{\sqrt{2} \varrho^{5}} e^{i \alpha - \gamma - i \beta} , \\
( \mathit{\hat{\Delta}} C\ind{_{a b c d}} ) k\ind{^{a}} \wt{m}\ind{^{b}} \ol{m}\ind{^{c}} \wt{m}\ind{^{d}} &= \frac{3 A M a \sin{\theta}}{\sqrt{2} \varrho^{4} \ol{\varrho}} e^{i \alpha + \gamma + i \beta} , \\
( \hat{\delta} C\ind{_{a b c d}} ) k\ind{^{a}} \wt{m}\ind{^{b}} \ol{m}\ind{^{c}} \wt{m}\ind{^{d}} &= \frac{3 i A M}{\varrho^{4}} \sqrt{\frac{\Delta}{2 \Sigma}} e^{i \alpha + \gamma + i \beta} , \\
( \ol{\hat{\delta}} C\ind{_{a b c d}} ) k\ind{^{a}} \wt{m}\ind{^{b}} \ol{m}\ind{^{c}} \wt{m}\ind{^{d}} &= \frac{3 i A M}{\varrho^{4}} \sqrt{\frac{\Delta}{2 \Sigma}} e^{i \alpha - \gamma - i \beta} ,
\end{align}
where we have introduced the quantity $\varrho = r + i a \cos{\theta}$ for notational simplicity. Transforming to the ``tilded'' complex null tetrad via \eqref{eqn:kin_marck_k}--\eqref{eqn:kin_marck_m}, we find
\begin{align}
( D C\ind{_{a b c d}} ) k\ind{^{a}} \wt{m}\ind{^{b}} k\ind{^{c}} \wt{m}\ind{^{d}} &= - \frac{15 K M}{\varrho^{6}} \left( \dot{r} - i \dot{\theta} a \sin{\theta} \right) = D \Psi_{0} , \label{eqn:derivative_weyl_kmkm_1} \\
( \wt{\delta} C\ind{_{a b c d}} ) k\ind{^{a}} \wt{m}\ind{^{b}} k\ind{^{c}} \wt{m}\ind{^{d}} &= \frac{15 i M \sqrt{2 K}}{2 \varrho^{7}} \left[ 2 a \left( L\ind{_{z}} - a E \right) - E \left( r^{2} - a^{2} \cos^{2}{\theta} \right) \right] , \\
( \ol{\wt{\delta}} C\ind{_{a b c d}} ) k\ind{^{a}} \wt{m}\ind{^{b}} k\ind{^{c}} \wt{m}\ind{^{d}} &= \frac{3 i E M \sqrt{2 K}}{2 \varrho^{5}} , \\
( D C\ind{_{a b c d}} ) k\ind{^{a}} \wt{m}\ind{^{b}} \ol{\wt{m}}\ind{^{c}} \wt{m}\ind{^{d}} &= \frac{3 i E M \sqrt{2 K}}{2 \varrho^{5}} , \\
( \wt{\delta} C\ind{_{a b c d}} ) k\ind{^{a}} \wt{m}\ind{^{b}} \ol{\wt{m}}\ind{^{c}} \wt{m}\ind{^{d}} &= - \frac{3 M \ol{\varrho}}{2 \varrho^{5}} \left( \dot{r} + i \dot{\theta} a \sin{\theta} \right) , \\
( \ol{\wt{\delta}} C\ind{_{a b c d}} ) k\ind{^{a}} \wt{m}\ind{^{b}} \ol{\wt{m}}\ind{^{c}} \wt{m}\ind{^{d}} &= - \frac{3 M }{2 \varrho^{4}} \left( \dot{r} - i \dot{\theta} a \sin{\theta} \right) = D \Psi_{2} . \label{eqn:derivative_weyl_kmmbarm_3}
\end{align}

Finally, we may calculate the desired quantities by using a null rotation to transform to the parallel-transported complex null tetrad. Under this transformation, the leg $\wt{m}\ind{^{a}}$ transforms as $m\ind{^{a}} = \wt{m}\ind{^{a}} + B k\ind{^{a}}$, where the null rotation parameter $B$ is defined in \eqref{eqn:null_rotation_parameter}. Using this, we find
\begin{align}
( \delta C )\ind{_{k m k m}} &= ( \wt{\delta} C )\ind{_{k \wt{m} k \wt{m}}} + B ( D C )\ind{_{k \wt{m} k \wt{m}}} , \label{eqn:delta_derivative_weyl_1} \\
( \ol{\delta} C )\ind{_{k m k m}} &= ( \ol{\delta} C )\ind{_{k \wt{m} k \wt{m}}} + B ( D C )\ind{_{k \wt{m} k \wt{m}}} , \label{eqn:delta_derivative_weyl_2} \\
( \delta C )\ind{_{k m \ol{m} m}} &= ( \wt{\delta} C )\ind{_{k \wt{m} \ol{\wt{m}} \wt{m}}} + B ( D C )\ind{_{k \wt{m} \ol{\wt{m}} \wt{m}}} + \ol{B} ( \wt{\delta} C )\ind{_{k \wt{m} k \wt{m}}} + B \ol{B} ( D C )\ind{_{k \wt{m} k \wt{m}}} , \label{eqn:delta_derivative_weyl_3} \\
( \ol{\delta} C )\ind{_{k m \ol{m} m}} &= ( \ol{\wt{\delta}} C )\ind{_{k \wt{m} \ol{\wt{m}} \wt{m}}} + \ol{B} ( D C )\ind{_{k \wt{m} \ol{\wt{m}} \wt{m}}} + \ol{B} ( \ol{\wt{\delta}} C )\ind{_{k \wt{m} k \wt{m}}} + \ol{B}^{2} ( D C )\ind{_{k \wt{m} k \wt{m}}} , \label{eqn:delta_derivative_weyl_4}
\end{align}
where explicit expressions for the quantities on the right-hand side are given in \eqref{eqn:derivative_weyl_kmkm_1}--\eqref{eqn:derivative_weyl_kmmbarm_3}. The directional derivatives \eqref{eqn:delta_derivative_weyl_1}--\eqref{eqn:delta_derivative_weyl_4} can now be inserted into \eqref{eqn:d_psi_0}--\eqref{eqn:dbar_psi_1} to complete the derivation of $\delta \Psi_{0}$, $\ol{\delta} \Psi_{0}$, $\delta \Psi_{1}$ and $\ol{\delta} \Psi_{1}$.

\addtocontents{toc}{\endgroup} 
\end{appendices}


\pagestyle{plain}
\bibliographystyle{ieeetr}
\bibliography{References/thesis_refs}
\addcontentsline{toc}{chapter}{Bibliography}

\end{document}